\documentclass{article}[10pt,letterpaper]
\usepackage{fancyhdr}
\usepackage{supertabular}
\renewcommand{\sectionmark}[1]{\markboth{#1}{}}
\usepackage{color}
\usepackage{amsmath}
\usepackage{pbox}
\usepackage{amssymb}
\usepackage{amsbsy}
\usepackage{scalefnt}
\usepackage{ifpdf}
\usepackage{graphicx}
\usepackage{multicol}
\usepackage[colorlinks=true, linkcolor=blue, linktocpage=true,urlcolor=blue]{hyperref}
\usepackage{longtable}
\usepackage{needspace}
\usepackage{eso-pic}
\usepackage{mathptmx}
\usepackage{multirow}
\usepackage{lscape}
\usepackage{textcomp}
\usepackage{changepage}
\usepackage{xfrac}
\usepackage{txfonts}
\usepackage{pdfcomment}
\usepackage[absolute]{textpos}
\ifpdf
\fi
\makeatletter
\renewcommand{\l@section}{\@dottedtocline{2}{1em}{0em}}
\renewcommand{\l@subsection}{\@dottedtocline{3}{2.5cm}{0em}}
\renewcommand{\hrulefill}{\leavevmode \leaders \hrule \@height 1pt \hfill \kern\z@}
\makeatother
\renewcommand{\underline}[1]{\begin{tabular}{@{\extracolsep{\fill}}c@{\extracolsep{\fill}}}#1\\[-0.2cm]\hrulefill\end{tabular}}
\renewcommand{\footrulewidth}{1pt}
\renewcommand{\headrulewidth}{1pt}
\newcounter{chappage}
\AddToShipoutPicture{\stepcounter{chappage}}
\fancypagestyle{plain}{\lhead{}\rhead{}}
\fancypagestyle{single}{\lhead{\leftmark \ \\}\rhead{\leftmark \ \\}}
\fancypagestyle{bob}{\lhead{\leftmark -\thechappage \ \\}\rhead{\leftmark -\thechappage \ \\}}
\begin{document}
\fontsize{9}{10}\selectfont
\fontdimen2\font=1.3\fontdimen2\font
\thispagestyle{empty}
\setcounter{page}{1}
\begin{center}

\vspace{0.5cm}
{ \huge Energy Levels of \ensuremath{^{\textnormal{18}}}Ne*}\\
\vspace{1.0cm}
{ \normalsize K. Setoodehnia\ensuremath{^{\textnormal{1,2}}}, J. H. Kelley\ensuremath{^{\textnormal{1,3}}} and C. G. Sheu\ensuremath{^{\textnormal{1,2}}}}\\
\vspace{0.2in}
{ \small \it \ensuremath{^{\textnormal{1}}}Triangle Universities Nuclear Laboratory, Duke University,\\
  Durham North Carolina 27708, USA.\\
  \ensuremath{^{\textnormal{2}}}Department of Physics, Duke University, Durham, North\\
  Carolina 27708, USA.\\
  \ensuremath{^{\textnormal{3}}}Department of Physics, North Carolina State University\\
  Raleigh, North Carolina 27607, USA}\\
\vspace{0.2in}
\end{center}

\setlength{\parindent}{-0.5cm}
\addtolength{\leftskip}{2cm}
\addtolength{\rightskip}{2cm}
{\bf Abstract: }
In this document, experimental nuclear structure data are evaluated for \ensuremath{^{\textnormal{18}}}Ne. \ensuremath{^{\textnormal{18}}}Ne was first identified by (\href{https://www.nndc.bnl.gov/nsr/nsrlink.jsp?1954Go17,B}{1954Go17}), see (\href{https://www.nndc.bnl.gov/nsr/nsrlink.jsp?2012Th01,B}{2012Th01}). The details of each reaction and decay experiment populating \ensuremath{^{\textnormal{18}}}Ne levels are compiled and evaluated. The combined results provide a set of adopted values that include level energies, spins and parities, level half-lives, \ensuremath{\gamma}-ray energies, decay types and branching ratios, and other nuclear properties. This work supersedes the earlier work by Ron Tilley (\href{https://www.nndc.bnl.gov/nsr/nsrlink.jsp?1995Ti07,B}{1995Ti07}) published in Nuclear Physics A \textbf{595} (1995) 1. The earlier evaluations were published by Fay Ajzenberg-Selove in (\href{https://www.nndc.bnl.gov/nsr/nsrlink.jsp?1959Aj76,B}{1959Aj76}, \href{https://www.nndc.bnl.gov/nsr/nsrlink.jsp?1972Aj02,B}{1972Aj02}, \href{https://www.nndc.bnl.gov/nsr/nsrlink.jsp?1978Aj03,B}{1978Aj03}, \href{https://www.nndc.bnl.gov/nsr/nsrlink.jsp?1983Aj01,B}{1983Aj01}, and \href{https://www.nndc.bnl.gov/nsr/nsrlink.jsp?1987Aj02,B}{1987Aj02}).\\

{\bf Cutoff Date: }
Literature available up to July 31, 2024 has been considered; the primary bibliographic source, the NSR database (\href{https://www.nndc.bnl.gov/nsr/nsrlink.jsp?2011Pr03,B}{2011Pr03}) available at Brookhaven National Laboratory web page: www.nndc.bnl.gov/nsr/.\\

{\bf General Policies and Organization of Material: }
See the April 2025 issue of the {\it Nuclear Data Sheets} or \\https://www.nndc.bnl.gov/nds/docs/NDSPolicies.pdf. \\

{\bf Acknowledgements: }
The authors expresses her gratitude to personnel at the National Nuclear Data Center (NNDC) at Brookhaven National Laboratory for facilitating this work.\\

\vfill

* This work is supported by the Office of Nuclear Physics, Office of Science, U.S. Department of Energy under contracts: DE-FG02-97ER41042 {\textminus} North Carolina State University and DE-FG02-97ER41033 {\textminus} Duke University\\

\setlength{\parindent}{+0.5cm}
\addtolength{\leftskip}{-2cm}
\addtolength{\rightskip}{-2cm}
\newpage
\pagestyle{plain}
\setlength{\columnseprule}{1pt}
\setlength{\columnsep}{1cm}
\begin{center}
\underline{\normalsize Index for A=18}
\end{center}
\hspace{.3cm}\raggedright\underline{Nuclide}\hspace{1cm}\underline{Data Type\mbox{\hspace{2.3cm}}}\hspace{2cm}\underline{Page}\hspace{1cm}
\raggedright\underline{Nuclide}\hspace{1cm}\underline{Data Type\mbox{\hspace{2.3cm}}}\hspace{2cm}\underline{Page}
\begin{adjustwidth}{}{0.05\textwidth}
\begin{multicols}{2}
\setcounter{tocdepth}{3}
\renewcommand{\contentsname}{\protect\vspace{-0.8cm}}
\tableofcontents
\end{multicols}
\end{adjustwidth}
\clearpage
\thispagestyle{empty}
\mbox{}
\clearpage
\clearpage
\pagestyle{bob}
\begin{center}
\section[\ensuremath{^{18}_{10}}Ne\ensuremath{_{8}^{~}}]{ }
\vspace{-30pt}
\setcounter{chappage}{1}
\subsection[\hspace{-0.2cm}Adopted Levels, Gammas]{ }
\vspace{-20pt}
\vspace{0.3cm}
\hypertarget{NE0}{{\bf \small \underline{Adopted \hyperlink{18NE_LEVEL}{Levels}, \hyperlink{18NE_GAMMA}{Gammas}}}}\\
\vspace{4pt}
\vspace{8pt}
\parbox[b][0.3cm]{17.7cm}{\addtolength{\parindent}{-0.2in}Q(\ensuremath{\beta^-})=$-$19720 {\it 90}; S(n)=19254.2 {\it 5}; S(p)=3923.1 {\it 4}; Q(\ensuremath{\alpha})=$-$5115.1 {\it 4}\hspace{0.2in}\href{https://www.nndc.bnl.gov/nsr/nsrlink.jsp?2021Wa16,B}{2021Wa16}}\\
\parbox[b][0.3cm]{17.7cm}{\addtolength{\parindent}{-0.2in}S\ensuremath{_{\textnormal{2n}}}=34812 keV \textit{20}, S\ensuremath{_{\textnormal{2p}}}=4523.3 keV \textit{4} (\href{https://www.nndc.bnl.gov/nsr/nsrlink.jsp?2021Wa16,B}{2021Wa16}).}\\

\vspace{0.385cm}
\parbox[b][0.3cm]{17.7cm}{\addtolength{\parindent}{-0.2in}\ensuremath{^{\textnormal{18}}}Ne was observed for the first time by Gow and Alvarez in 1954 (\href{https://www.nndc.bnl.gov/nsr/nsrlink.jsp?1954Go17,B}{1954Go17}) using the \ensuremath{^{\textnormal{19}}}F(p,2n)\ensuremath{^{\textnormal{18}}}Ne reaction. This study was}\\
\parbox[b][0.3cm]{17.7cm}{conducted at Berkeley linear accelerator facility (see the associated dataset). The decay curve of \ensuremath{^{\textnormal{18}}}Ne\ensuremath{_{\textnormal{g.s.}}} was measured. The}\\
\parbox[b][0.3cm]{17.7cm}{positron end point energy and the \ensuremath{^{\textnormal{18}}}Ne\ensuremath{_{\textnormal{g.s.}}} half-life were deduced.}\\
\parbox[b][0.3cm]{17.7cm}{\addtolength{\parindent}{-0.2in}\ensuremath{^{\textnormal{18}}}Ne\ensuremath{_{\textnormal{g.s.}}}(J\ensuremath{^{\ensuremath{\pi}}}=0\ensuremath{^{\textnormal{+}}}) \ensuremath{\beta}\ensuremath{^{\textnormal{+}}} decays to: (1) \ensuremath{^{\textnormal{18}}}F\ensuremath{_{\textnormal{g.s.}}}(1\ensuremath{^{\textnormal{+}}}) with Q\ensuremath{_{\textnormal{EC}}}(g.s.)=4444.21 keV \textit{68} (\href{https://www.nndc.bnl.gov/nsr/nsrlink.jsp?2020Ha30,B}{2020Ha30}) and BR=92.25\% \textit{13} (weighted}\\
\parbox[b][0.3cm]{17.7cm}{average of 92.5\% \textit{2} (\href{https://www.nndc.bnl.gov/nsr/nsrlink.jsp?1970As06,B}{1970As06}); 92.11\% \textit{21} (\href{https://www.nndc.bnl.gov/nsr/nsrlink.jsp?1975Ha21,B}{1975Ha21}); and 92.12\% \textit{21} (\href{https://www.nndc.bnl.gov/nsr/nsrlink.jsp?2013Gr03,B}{2013Gr03})). (2) \ensuremath{^{\textnormal{18}}}F*(1041.55 keV, 0\ensuremath{^{\textnormal{+}}}) {\textminus} superallowed}\\
\parbox[b][0.3cm]{17.7cm}{Fermi decay with Q\ensuremath{_{\textnormal{EC}}}=3402.66 keV \textit{69} (\href{https://www.nndc.bnl.gov/nsr/nsrlink.jsp?2020Ha30,B}{2020Ha30}) and BR=7.70\% \textit{21} (weighted average of 9\% \textit{3} (\href{https://www.nndc.bnl.gov/nsr/nsrlink.jsp?1963Fr10,B}{1963Fr10}); and 7.69\% \textit{21}}\\
\parbox[b][0.3cm]{17.7cm}{(\href{https://www.nndc.bnl.gov/nsr/nsrlink.jsp?2020Ha30,B}{2020Ha30})). (3) \ensuremath{^{\textnormal{18}}}F*(1700.81 keV, 1\ensuremath{^{\textnormal{+}}}) with BR=0.1866\% \textit{32} (weighted average of 0.17\% \textit{5} (\href{https://www.nndc.bnl.gov/nsr/nsrlink.jsp?1970As06,B}{1970As06}); 0.23\% \textit{3} (\href{https://www.nndc.bnl.gov/nsr/nsrlink.jsp?1975Ha21,B}{1975Ha21});}\\
\parbox[b][0.3cm]{17.7cm}{0.19\% \textit{4} (\href{https://www.nndc.bnl.gov/nsr/nsrlink.jsp?1981Ad01,B}{1981Ad01} and E. G. Adelberger private communication to Fay Ajzenberg-Selove, see (\href{https://www.nndc.bnl.gov/nsr/nsrlink.jsp?1983Aj01,B}{1983Aj01})); 0.183\% \textit{6}}\\
\parbox[b][0.3cm]{17.7cm}{(\href{https://www.nndc.bnl.gov/nsr/nsrlink.jsp?1982He04,B}{1982He04}); 0.188\% \textit{6} (\href{https://www.nndc.bnl.gov/nsr/nsrlink.jsp?1983Ad03,B}{1983Ad03}); and 0.187\% \textit{5} (\href{https://www.nndc.bnl.gov/nsr/nsrlink.jsp?2013Gr03,B}{2013Gr03})). (4) \ensuremath{^{\textnormal{18}}}F*(1080.54 keV, 0\ensuremath{^{-}}) {\textminus} the first forbidden decay with}\\
\parbox[b][0.3cm]{17.7cm}{BR=0.00219\% \textit{11} (weighted average of 0.00228\% \textit{17} (\href{https://www.nndc.bnl.gov/nsr/nsrlink.jsp?1975Ha21,B}{1975Ha21}); 0.00214\% \textit{26} (\href{https://www.nndc.bnl.gov/nsr/nsrlink.jsp?1982He04,B}{1982He04}); 0.00207\% \textit{28} (\href{https://www.nndc.bnl.gov/nsr/nsrlink.jsp?1983Ad03,B}{1983Ad03}); and}\\
\parbox[b][0.3cm]{17.7cm}{0.00214\% \textit{25} (\href{https://www.nndc.bnl.gov/nsr/nsrlink.jsp?2013Gr03,B}{2013Gr03})). (\href{https://www.nndc.bnl.gov/nsr/nsrlink.jsp?1968Go05,B}{1968Go05}) estimated an upper limit of BR\ensuremath{<}1.5\% for the \ensuremath{^{\textnormal{18}}}Ne(\ensuremath{\beta}\ensuremath{^{\textnormal{+}}})\ensuremath{^{\textnormal{18}}}F*(2100.61 keV, 2\ensuremath{^{-}}) decay but}\\
\parbox[b][0.3cm]{17.7cm}{this branch is not observed and may not exist, see (\href{https://www.nndc.bnl.gov/nsr/nsrlink.jsp?1982He04,B}{1982He04}).}\\
\parbox[b][0.3cm]{17.7cm}{\addtolength{\parindent}{-0.2in}\ensuremath{^{\textnormal{18}}}Ne (T\ensuremath{_{\textnormal{z}}}={\textminus}1) is a superallowed \ensuremath{\beta} emitter. If its \textit{Ft} value can be estimated with sub-1\% precision, the theoretical isospin symmetry}\\
\parbox[b][0.3cm]{17.7cm}{breaking correction can be constrained. The uncertainty in \textit{Ft} for \ensuremath{^{\textnormal{18}}}Ne, evaluated most recently by (\href{https://www.nndc.bnl.gov/nsr/nsrlink.jsp?2020Ha30,B}{2020Ha30}), is dominated by the}\\
\parbox[b][0.3cm]{17.7cm}{2.7\% fractional uncertainty in the branching ratio of the superallowed Fermi decay branch measured by (\href{https://www.nndc.bnl.gov/nsr/nsrlink.jsp?1975Ha21,B}{1975Ha21}). This study}\\
\parbox[b][0.3cm]{17.7cm}{provided the normalization ratio of 7.83\% \textit{21}, with which the absolute branching ratios mentioned earlier are deduced. The}\\
\parbox[b][0.3cm]{17.7cm}{branching ratio of the superallowed Fermi decay branch was remeasured in 2012 as 7.29\% \textit{5} (H. Bouzomita-Zran, \textit{Mesure de}}\\
\parbox[b][0.3cm]{17.7cm}{\textit{pr\'{e}cision de la d\'{e}croissance super-premise de \ensuremath{^{18}}Ne}, Ph.D. Thesis, Universit\'{e} de Caen Normandie, 2015, unpublished). The}\\
\parbox[b][0.3cm]{17.7cm}{uncertainty was reduced to 0.7\% (unpublished). A proposal (Tech. Rep. CERN-INTC-2016-051 INTC-P-481) was submitted in}\\
\parbox[b][0.3cm]{17.7cm}{2016 to ISOLDE and Neutron Time-of-Flight Committee at CERN to improve this measurement by a factor of \ensuremath{\sim}2 in order to}\\
\parbox[b][0.3cm]{17.7cm}{achieve the required \ensuremath{\pm}0.3\% precision. To reduce the contributions of the \ensuremath{^{\textnormal{18}}}Ne and \ensuremath{^{\textnormal{18}}}F ground-state masses to the uncertainty of}\\
\parbox[b][0.3cm]{17.7cm}{the superallowed Q\ensuremath{_{\textnormal{EC}}}, another proposal (Tech. Rep. CERN-INTC-2018-015 INTC-P-545) was submitted in 2018 to measure the}\\
\parbox[b][0.3cm]{17.7cm}{Q\ensuremath{_{\textnormal{EC}}}(\ensuremath{^{\textnormal{18}}}Ne\ensuremath{_{\textnormal{g.s.}}}\ensuremath{\rightarrow}\ensuremath{^{\textnormal{18}}}F\ensuremath{_{\textnormal{g.s.}}}) to a precision of about 20 eV, and thus improving the current uncertainty (\href{https://www.nndc.bnl.gov/nsr/nsrlink.jsp?2021Wa16,B}{2021Wa16}) by a factor 30.}\\
\vspace{0.385cm}
\parbox[b][0.3cm]{17.7cm}{\addtolength{\parindent}{-0.2in}Measurements:}\\
\parbox[b][0.3cm]{17.7cm}{\addtolength{\parindent}{-0.2in}\textit{Mass}: \href{https://www.nndc.bnl.gov/nsr/nsrlink.jsp?1960Aj01,B}{1960Aj01}, \href{https://www.nndc.bnl.gov/nsr/nsrlink.jsp?1961To03,B}{1961To03}, \href{https://www.nndc.bnl.gov/nsr/nsrlink.jsp?1961Du02,B}{1961Du02}, \href{https://www.nndc.bnl.gov/nsr/nsrlink.jsp?1963Fr10,B}{1963Fr10}, \href{https://www.nndc.bnl.gov/nsr/nsrlink.jsp?1994Ma14,B}{1994Ma14}, \href{https://www.nndc.bnl.gov/nsr/nsrlink.jsp?2008Ge07,B}{2008Ge07}, \href{https://www.nndc.bnl.gov/nsr/nsrlink.jsp?2017Se09,B}{2017Se09}. See also \href{https://www.nndc.bnl.gov/nsr/nsrlink.jsp?2004Bl20,B}{2004Bl20}, A. Kellerbauer {et al}., AIP}\\
\parbox[b][0.3cm]{17.7cm}{Conf. Proc. 831 (2006) 49.}\\
\parbox[b][0.3cm]{17.7cm}{\addtolength{\parindent}{-0.2in}\textit{B(E2;0\ensuremath{^{+}_{\textnormal{g.s.}}}\ensuremath{\rightarrow}2\ensuremath{^{\textnormal{+}}_{\textnormal{1}}})}: \href{https://www.nndc.bnl.gov/nsr/nsrlink.jsp?1976Mc02,B}{1976Mc02}, \href{https://www.nndc.bnl.gov/nsr/nsrlink.jsp?1988Se04,B}{1988Se04}, \href{https://www.nndc.bnl.gov/nsr/nsrlink.jsp?2000Ri15,B}{2000Ri15}.}\\
\parbox[b][0.3cm]{17.7cm}{\addtolength{\parindent}{-0.2in}\textit{RMS charge radius}: \href{https://www.nndc.bnl.gov/nsr/nsrlink.jsp?2000Ne11,B}{2000Ne11}, \href{https://www.nndc.bnl.gov/nsr/nsrlink.jsp?2008Ge07,B}{2008Ge07}, \href{https://www.nndc.bnl.gov/nsr/nsrlink.jsp?2011Ma48,B}{2011Ma48}, \href{https://www.nndc.bnl.gov/nsr/nsrlink.jsp?2013An02,B}{2013An02}: the average \textit{rms} charge radius as defined by the evaluation of}\\
\parbox[b][0.3cm]{17.7cm}{(\href{https://www.nndc.bnl.gov/nsr/nsrlink.jsp?2013An02,B}{2013An02}) is R=2.9714 fm \textit{76}.}\\
\parbox[b][0.3cm]{17.7cm}{\addtolength{\parindent}{-0.2in}\textit{Half-life}: \href{https://www.nndc.bnl.gov/nsr/nsrlink.jsp?1954Go17,B}{1954Go17}: 1.6 s \textit{2}; \href{https://www.nndc.bnl.gov/nsr/nsrlink.jsp?1959Du81,B}{1959Du81}: 1.25 s \textit{20} (as cited in \href{https://www.nndc.bnl.gov/nsr/nsrlink.jsp?1961Bu05,B}{1961Bu05}); \href{https://www.nndc.bnl.gov/nsr/nsrlink.jsp?1961Bu05,B}{1961Bu05}: 1.46 s \textit{7}; \href{https://www.nndc.bnl.gov/nsr/nsrlink.jsp?1961Ec02,B}{1961Ec02}: 1.7 s \textit{4}; \href{https://www.nndc.bnl.gov/nsr/nsrlink.jsp?1963Fr10,B}{1963Fr10},}\\
\parbox[b][0.3cm]{17.7cm}{\href{https://www.nndc.bnl.gov/nsr/nsrlink.jsp?1965Fr09,B}{1965Fr09}: 1.47 s \textit{10}; \href{https://www.nndc.bnl.gov/nsr/nsrlink.jsp?1970Al11,B}{1970Al11}: 1.67 s \textit{2}; \href{https://www.nndc.bnl.gov/nsr/nsrlink.jsp?1970As06,B}{1970As06}: 1.69 s \textit{4}; \href{https://www.nndc.bnl.gov/nsr/nsrlink.jsp?1972Ha58,B}{1972Ha58}: 1.655 s \textit{25}; \href{https://www.nndc.bnl.gov/nsr/nsrlink.jsp?1975Al27,B}{1975Al27}: 1.669 s \textit{4}; \href{https://www.nndc.bnl.gov/nsr/nsrlink.jsp?1975Ha21,B}{1975Ha21}: 1.687 s \textit{9};}\\
\parbox[b][0.3cm]{17.7cm}{\href{https://www.nndc.bnl.gov/nsr/nsrlink.jsp?1975Ha45,B}{1975Ha45}: 1.682 s \textit{15} (the average value); \href{https://www.nndc.bnl.gov/nsr/nsrlink.jsp?2007Gr18,B}{2007Gr18}: 1.6656 s \textit{19}; \href{https://www.nndc.bnl.gov/nsr/nsrlink.jsp?2013Gr03,B}{2013Gr03}: 1.6648 s \textit{11}; \href{https://www.nndc.bnl.gov/nsr/nsrlink.jsp?2015La19,B}{2015La19} and A. Laffoley, \textit{High}}\\
\parbox[b][0.3cm]{17.7cm}{\textit{precision half-life measurements for the superallowed Fermi beta emitters oxygen-14 and neon-18}, Ph.D. Thesis, University of}\\
\parbox[b][0.3cm]{17.7cm}{Guelph, (2015): 1.66400 s \textit{+57{\textminus}48}. See also compilations and evaluations: \href{https://www.nndc.bnl.gov/nsr/nsrlink.jsp?1969Ka41,B}{1969Ka41}, \href{https://www.nndc.bnl.gov/nsr/nsrlink.jsp?1970Mc23,B}{1970Mc23}, \href{https://www.nndc.bnl.gov/nsr/nsrlink.jsp?1973To04,B}{1973To04}, \href{https://www.nndc.bnl.gov/nsr/nsrlink.jsp?1975Ra37,B}{1975Ra37}, \href{https://www.nndc.bnl.gov/nsr/nsrlink.jsp?1975Ha45,B}{1975Ha45},}\\
\parbox[b][0.3cm]{17.7cm}{\href{https://www.nndc.bnl.gov/nsr/nsrlink.jsp?2005Ha27,B}{2005Ha27}, \href{https://www.nndc.bnl.gov/nsr/nsrlink.jsp?2005Ha65,B}{2005Ha65}, \href{https://www.nndc.bnl.gov/nsr/nsrlink.jsp?2009Ha12,B}{2009Ha12}, \href{https://www.nndc.bnl.gov/nsr/nsrlink.jsp?2012Ha46,B}{2012Ha46}, \href{https://www.nndc.bnl.gov/nsr/nsrlink.jsp?2015Ha07,B}{2015Ha07}, \href{https://www.nndc.bnl.gov/nsr/nsrlink.jsp?2016Pr01,B}{2016Pr01}, \href{https://www.nndc.bnl.gov/nsr/nsrlink.jsp?2020Ha30,B}{2020Ha30}, \href{https://www.nndc.bnl.gov/nsr/nsrlink.jsp?2023Se01,B}{2023Se01}.}\\
\parbox[b][0.3cm]{17.7cm}{\addtolength{\parindent}{-0.2in}\textit{Superallowed \ensuremath{\beta}\ensuremath{^{+}} transition}: M. Eibach and M. Mougeot, \textit{High-precision measurement of the \ensuremath{^{18}}Ne superallowed \ensuremath{\beta}-decay} \textit{Q-value},}\\
\parbox[b][0.3cm]{17.7cm}{Tech. Rep. CERNINTC-2018-015. INTC-P-545 (CERN, Geneva, 2018).}\\
\parbox[b][0.3cm]{17.7cm}{\addtolength{\parindent}{-0.2in}\textit{\ensuremath{\beta}\ensuremath{^{+}} decay}: \href{https://www.nndc.bnl.gov/nsr/nsrlink.jsp?1954Go17,B}{1954Go17}, \href{https://www.nndc.bnl.gov/nsr/nsrlink.jsp?1960Bu03,B}{1960Bu03}, \href{https://www.nndc.bnl.gov/nsr/nsrlink.jsp?1980LoZY,B}{1980LoZY}, \href{https://www.nndc.bnl.gov/nsr/nsrlink.jsp?1981Ad01,B}{1981Ad01}, \href{https://www.nndc.bnl.gov/nsr/nsrlink.jsp?1981Ba67,B}{1981Ba67}, \href{https://www.nndc.bnl.gov/nsr/nsrlink.jsp?1981BaZJ,B}{1981BaZJ}, \href{https://www.nndc.bnl.gov/nsr/nsrlink.jsp?1981BoZY,B}{1981BoZY}, \href{https://www.nndc.bnl.gov/nsr/nsrlink.jsp?1981VoZY,B}{1981VoZY}, \href{https://www.nndc.bnl.gov/nsr/nsrlink.jsp?1982DaZZ,B}{1982DaZZ}, \href{https://www.nndc.bnl.gov/nsr/nsrlink.jsp?1982He04,B}{1982He04},}\\
\parbox[b][0.3cm]{17.7cm}{\href{https://www.nndc.bnl.gov/nsr/nsrlink.jsp?1982HeZW,B}{1982HeZW}, \href{https://www.nndc.bnl.gov/nsr/nsrlink.jsp?1982VoZY,B}{1982VoZY}, \href{https://www.nndc.bnl.gov/nsr/nsrlink.jsp?1983Ad03,B}{1983Ad03}, \href{https://www.nndc.bnl.gov/nsr/nsrlink.jsp?1988La01,B}{1988La01}, \href{https://www.nndc.bnl.gov/nsr/nsrlink.jsp?1994Re24,B}{1994Re24}, \href{https://www.nndc.bnl.gov/nsr/nsrlink.jsp?1997Eg02,B}{1997Eg02}, \href{https://www.nndc.bnl.gov/nsr/nsrlink.jsp?1998Br23,B}{1998Br23}, \href{https://www.nndc.bnl.gov/nsr/nsrlink.jsp?1999Og03,B}{1999Og03}, \href{https://www.nndc.bnl.gov/nsr/nsrlink.jsp?2001Be67,B}{2001Be67}, \href{https://www.nndc.bnl.gov/nsr/nsrlink.jsp?2002Vo11,B}{2002Vo11}, \href{https://www.nndc.bnl.gov/nsr/nsrlink.jsp?2009BaZT,B}{2009BaZT}.}\\
\vspace{0.385cm}
\parbox[b][0.3cm]{17.7cm}{\addtolength{\parindent}{-0.2in}Theory:}\\
\parbox[b][0.3cm]{17.7cm}{\addtolength{\parindent}{-0.2in}\textit{Mass excess, particle separation and binding energies}: \href{https://www.nndc.bnl.gov/nsr/nsrlink.jsp?1965Ry01,B}{1965Ry01}, \href{https://www.nndc.bnl.gov/nsr/nsrlink.jsp?1977Sh13,B}{1977Sh13}, \href{https://www.nndc.bnl.gov/nsr/nsrlink.jsp?1999Si13,B}{1999Si13}, \href{https://www.nndc.bnl.gov/nsr/nsrlink.jsp?2002Gu10,B}{2002Gu10}, \href{https://www.nndc.bnl.gov/nsr/nsrlink.jsp?2010La06,B}{2010La06}, \href{https://www.nndc.bnl.gov/nsr/nsrlink.jsp?2013Ho01,B}{2013Ho01},}\\
\parbox[b][0.3cm]{17.7cm}{\href{https://www.nndc.bnl.gov/nsr/nsrlink.jsp?2013Xu15,B}{2013Xu15}, \href{https://www.nndc.bnl.gov/nsr/nsrlink.jsp?2015Mo16,B}{2015Mo16}, \href{https://www.nndc.bnl.gov/nsr/nsrlink.jsp?2018Fo04,B}{2018Fo04} (evaluation), \href{https://www.nndc.bnl.gov/nsr/nsrlink.jsp?2018Is03,B}{2018Is03} (evaluation), \href{https://www.nndc.bnl.gov/nsr/nsrlink.jsp?2021Ma33,B}{2021Ma33}, \href{https://www.nndc.bnl.gov/nsr/nsrlink.jsp?2022Zo01,B}{2022Zo01}, \href{https://www.nndc.bnl.gov/nsr/nsrlink.jsp?2023Di08,B}{2023Di08}.}\\
\parbox[b][0.3cm]{17.7cm}{\addtolength{\parindent}{-0.2in}\textit{2p decay}: \href{https://www.nndc.bnl.gov/nsr/nsrlink.jsp?1996Be27,B}{1996Be27}, \href{https://www.nndc.bnl.gov/nsr/nsrlink.jsp?2002Br22,B}{2002Br22}, \href{https://www.nndc.bnl.gov/nsr/nsrlink.jsp?2003Ba99,B}{2003Ba99}, \href{https://www.nndc.bnl.gov/nsr/nsrlink.jsp?2004Ro26,B}{2004Ro26}, \href{https://www.nndc.bnl.gov/nsr/nsrlink.jsp?2005Pf02,B}{2005Pf02}, \href{https://www.nndc.bnl.gov/nsr/nsrlink.jsp?2005Ro23,B}{2005Ro23}, \href{https://www.nndc.bnl.gov/nsr/nsrlink.jsp?2005Ro41,B}{2005Ro41}, \href{https://www.nndc.bnl.gov/nsr/nsrlink.jsp?2005Pf01,B}{2005Pf01} (evaluation), \href{https://www.nndc.bnl.gov/nsr/nsrlink.jsp?2007Be54,B}{2007Be54},}\\
\parbox[b][0.3cm]{17.7cm}{\href{https://www.nndc.bnl.gov/nsr/nsrlink.jsp?2008Ch31,B}{2008Ch31}, \href{https://www.nndc.bnl.gov/nsr/nsrlink.jsp?2008Bl03,B}{2008Bl03}, \href{https://www.nndc.bnl.gov/nsr/nsrlink.jsp?2009Yu04,B}{2009Yu04}.}\\
\parbox[b][0.3cm]{17.7cm}{\addtolength{\parindent}{-0.2in}\textit{Ground state properties}: \href{https://www.nndc.bnl.gov/nsr/nsrlink.jsp?1968Be82,B}{1968Be82}, \href{https://www.nndc.bnl.gov/nsr/nsrlink.jsp?1968Ar02,B}{1968Ar02}, \href{https://www.nndc.bnl.gov/nsr/nsrlink.jsp?1971KhZT,B}{1971KhZT}, \href{https://www.nndc.bnl.gov/nsr/nsrlink.jsp?1977Br03,B}{1977Br03}, \href{https://www.nndc.bnl.gov/nsr/nsrlink.jsp?1979Va07,B}{1979Va07}, \href{https://www.nndc.bnl.gov/nsr/nsrlink.jsp?1980Zh01,B}{1980Zh01}, \href{https://www.nndc.bnl.gov/nsr/nsrlink.jsp?1981OvZZ,B}{1981OvZZ}, \href{https://www.nndc.bnl.gov/nsr/nsrlink.jsp?1982Ov01,B}{1982Ov01}, \href{https://www.nndc.bnl.gov/nsr/nsrlink.jsp?1982Se15,B}{1982Se15},}\\
\parbox[b][0.3cm]{17.7cm}{\href{https://www.nndc.bnl.gov/nsr/nsrlink.jsp?1983AnZQ,B}{1983AnZQ}, \href{https://www.nndc.bnl.gov/nsr/nsrlink.jsp?1984Sa37,B}{1984Sa37}, \href{https://www.nndc.bnl.gov/nsr/nsrlink.jsp?1987Po01,B}{1987Po01}, \href{https://www.nndc.bnl.gov/nsr/nsrlink.jsp?1994Ci02,B}{1994Ci02}, \href{https://www.nndc.bnl.gov/nsr/nsrlink.jsp?1996Go38,B}{1996Go38}, \href{https://www.nndc.bnl.gov/nsr/nsrlink.jsp?1996Gr21,B}{1996Gr21}, \href{https://www.nndc.bnl.gov/nsr/nsrlink.jsp?1996Re03,B}{1996Re03}, \href{https://www.nndc.bnl.gov/nsr/nsrlink.jsp?1997Pa38,B}{1997Pa38}, \href{https://www.nndc.bnl.gov/nsr/nsrlink.jsp?1997Ch09,B}{1997Ch09}, \href{https://www.nndc.bnl.gov/nsr/nsrlink.jsp?1998La02,B}{1998La02}, \href{https://www.nndc.bnl.gov/nsr/nsrlink.jsp?1998Re07,B}{1998Re07},}\\
\parbox[b][0.3cm]{17.7cm}{\href{https://www.nndc.bnl.gov/nsr/nsrlink.jsp?2001Oz04,B}{2001Oz04}, \href{https://www.nndc.bnl.gov/nsr/nsrlink.jsp?2002Fo11,B}{2002Fo11}, \href{https://www.nndc.bnl.gov/nsr/nsrlink.jsp?2002Gu10,B}{2002Gu10}, \href{https://www.nndc.bnl.gov/nsr/nsrlink.jsp?2002Mi14,B}{2002Mi14}, \href{https://www.nndc.bnl.gov/nsr/nsrlink.jsp?2003St22,B}{2003St22}, \href{https://www.nndc.bnl.gov/nsr/nsrlink.jsp?2004Sa58,B}{2004Sa58}, \href{https://www.nndc.bnl.gov/nsr/nsrlink.jsp?2004Ge02,B}{2004Ge02}, \href{https://www.nndc.bnl.gov/nsr/nsrlink.jsp?2004La24,B}{2004La24}, \href{https://www.nndc.bnl.gov/nsr/nsrlink.jsp?2004Sa58,B}{2004Sa58}, \href{https://www.nndc.bnl.gov/nsr/nsrlink.jsp?2005Ch71,B}{2005Ch71}, \href{https://www.nndc.bnl.gov/nsr/nsrlink.jsp?2005Ma98,B}{2005Ma98},}\\
\parbox[b][0.3cm]{17.7cm}{\href{https://www.nndc.bnl.gov/nsr/nsrlink.jsp?2005Sa59,B}{2005Sa59}, \href{https://www.nndc.bnl.gov/nsr/nsrlink.jsp?2005Sa63,B}{2005Sa63}, \href{https://www.nndc.bnl.gov/nsr/nsrlink.jsp?2006Ma17,B}{2006Ma17}, \href{https://www.nndc.bnl.gov/nsr/nsrlink.jsp?2006Sa29,B}{2006Sa29}, \href{https://www.nndc.bnl.gov/nsr/nsrlink.jsp?2008Sc02,B}{2008Sc02}, \href{https://www.nndc.bnl.gov/nsr/nsrlink.jsp?2008Wi11,B}{2008Wi11}, \href{https://www.nndc.bnl.gov/nsr/nsrlink.jsp?2010Zh45,B}{2010Zh45}, \href{https://www.nndc.bnl.gov/nsr/nsrlink.jsp?2011Eb02,B}{2011Eb02}, \href{https://www.nndc.bnl.gov/nsr/nsrlink.jsp?2011Gu03,B}{2011Gu03}, \href{https://www.nndc.bnl.gov/nsr/nsrlink.jsp?2011Ro50,B}{2011Ro50}, \href{https://www.nndc.bnl.gov/nsr/nsrlink.jsp?2011Su14,B}{2011Su14},}\\
\clearpage
\vspace{0.3cm}
{\bf \small \underline{Adopted \hyperlink{18NE_LEVEL}{Levels}, \hyperlink{18NE_GAMMA}{Gammas} (continued)}}\\
\vspace{0.3cm}
\parbox[b][0.3cm]{17.7cm}{\href{https://www.nndc.bnl.gov/nsr/nsrlink.jsp?2012Ma12,B}{2012Ma12}, \href{https://www.nndc.bnl.gov/nsr/nsrlink.jsp?2013FeZW,B}{2013FeZW}, \href{https://www.nndc.bnl.gov/nsr/nsrlink.jsp?2013Xu15,B}{2013Xu15}, \href{https://www.nndc.bnl.gov/nsr/nsrlink.jsp?2014Ch39,B}{2014Ch39}, \href{https://www.nndc.bnl.gov/nsr/nsrlink.jsp?2014Va13,B}{2014Va13}, \href{https://www.nndc.bnl.gov/nsr/nsrlink.jsp?2015Mo16,B}{2015Mo16}, \href{https://www.nndc.bnl.gov/nsr/nsrlink.jsp?2015Si12,B}{2015Si12}, \href{https://www.nndc.bnl.gov/nsr/nsrlink.jsp?2016Wa14,B}{2016Wa14}, \href{https://www.nndc.bnl.gov/nsr/nsrlink.jsp?2016Ja03,B}{2016Ja03}, \href{https://www.nndc.bnl.gov/nsr/nsrlink.jsp?2016St12,B}{2016St12}, \href{https://www.nndc.bnl.gov/nsr/nsrlink.jsp?2017Ah08,B}{2017Ah08},}\\
\parbox[b][0.3cm]{17.7cm}{\href{https://www.nndc.bnl.gov/nsr/nsrlink.jsp?2019Ra09,B}{2019Ra09}, \href{https://www.nndc.bnl.gov/nsr/nsrlink.jsp?2019Sa38,B}{2019Sa38}, \href{https://www.nndc.bnl.gov/nsr/nsrlink.jsp?2019Sa58,B}{2019Sa58}, \href{https://www.nndc.bnl.gov/nsr/nsrlink.jsp?2020An13,B}{2020An13}, \href{https://www.nndc.bnl.gov/nsr/nsrlink.jsp?2020Pa38,B}{2020Pa38}, \href{https://www.nndc.bnl.gov/nsr/nsrlink.jsp?2020No10,B}{2020No10}, \href{https://www.nndc.bnl.gov/nsr/nsrlink.jsp?2021Am03,B}{2021Am03}, \href{https://www.nndc.bnl.gov/nsr/nsrlink.jsp?2022Yu02,B}{2022Yu02}, \href{https://www.nndc.bnl.gov/nsr/nsrlink.jsp?2023Sa22,B}{2023Sa22}.}\\
\parbox[b][0.3cm]{17.7cm}{\addtolength{\parindent}{-0.2in}\textit{Calculated levels}: \href{https://www.nndc.bnl.gov/nsr/nsrlink.jsp?1968Va24,B}{1968Va24}, \href{https://www.nndc.bnl.gov/nsr/nsrlink.jsp?1969Ra28,B}{1969Ra28}, \href{https://www.nndc.bnl.gov/nsr/nsrlink.jsp?1970El08,B}{1970El08}, \href{https://www.nndc.bnl.gov/nsr/nsrlink.jsp?1970El23,B}{1970El23}, \href{https://www.nndc.bnl.gov/nsr/nsrlink.jsp?1972Ka01,B}{1972Ka01}, \href{https://www.nndc.bnl.gov/nsr/nsrlink.jsp?1982Zh01,B}{1982Zh01}, \href{https://www.nndc.bnl.gov/nsr/nsrlink.jsp?1987Ra01,B}{1987Ra01}, \href{https://www.nndc.bnl.gov/nsr/nsrlink.jsp?1987Wi11,B}{1987Wi11}, \href{https://www.nndc.bnl.gov/nsr/nsrlink.jsp?1989SuZT,B}{1989SuZT}, \href{https://www.nndc.bnl.gov/nsr/nsrlink.jsp?1989Fu01,B}{1989Fu01},}\\
\parbox[b][0.3cm]{17.7cm}{\href{https://www.nndc.bnl.gov/nsr/nsrlink.jsp?1990Av06,B}{1990Av06}, \href{https://www.nndc.bnl.gov/nsr/nsrlink.jsp?1998Sh35,B}{1998Sh35}, \href{https://www.nndc.bnl.gov/nsr/nsrlink.jsp?2001Ra27,B}{2001Ra27}, \href{https://www.nndc.bnl.gov/nsr/nsrlink.jsp?2003Fo13,B}{2003Fo13}, \href{https://www.nndc.bnl.gov/nsr/nsrlink.jsp?2003Jh01,B}{2003Jh01}, \href{https://www.nndc.bnl.gov/nsr/nsrlink.jsp?2004Sa58,B}{2004Sa58}, \href{https://www.nndc.bnl.gov/nsr/nsrlink.jsp?2011Fo12,B}{2011Fo12}, \href{https://www.nndc.bnl.gov/nsr/nsrlink.jsp?2011YuZY,B}{2011YuZY}, \href{https://www.nndc.bnl.gov/nsr/nsrlink.jsp?2012GoZU,B}{2012GoZU}, \href{https://www.nndc.bnl.gov/nsr/nsrlink.jsp?2012Su28,B}{2012Su28}, \href{https://www.nndc.bnl.gov/nsr/nsrlink.jsp?2013Ho01,B}{2013Ho01},}\\
\parbox[b][0.3cm]{17.7cm}{\href{https://www.nndc.bnl.gov/nsr/nsrlink.jsp?2013Ja13,B}{2013Ja13}, \href{https://www.nndc.bnl.gov/nsr/nsrlink.jsp?2013Sc14,B}{2013Sc14}, \href{https://www.nndc.bnl.gov/nsr/nsrlink.jsp?2014Ca21,B}{2014Ca21}, \href{https://www.nndc.bnl.gov/nsr/nsrlink.jsp?2014Sa30,B}{2014Sa30}, \href{https://www.nndc.bnl.gov/nsr/nsrlink.jsp?2014Yu02,B}{2014Yu02}, \href{https://www.nndc.bnl.gov/nsr/nsrlink.jsp?2016Wa14,B}{2016Wa14}, \href{https://www.nndc.bnl.gov/nsr/nsrlink.jsp?2019Oi01,B}{2019Oi01}, \href{https://www.nndc.bnl.gov/nsr/nsrlink.jsp?2019Pa38,B}{2019Pa38}, \href{https://www.nndc.bnl.gov/nsr/nsrlink.jsp?2019Wu12,B}{2019Wu12}, \href{https://www.nndc.bnl.gov/nsr/nsrlink.jsp?2023Ya06,B}{2023Ya06}.}\\
\parbox[b][0.3cm]{17.7cm}{\addtolength{\parindent}{-0.2in}\textit{Shell model and ab initio calculations}: \href{https://www.nndc.bnl.gov/nsr/nsrlink.jsp?1965En02,B}{1965En02}, \href{https://www.nndc.bnl.gov/nsr/nsrlink.jsp?1969Be94,B}{1969Be94}, \href{https://www.nndc.bnl.gov/nsr/nsrlink.jsp?1969Zu03,B}{1969Zu03}, \href{https://www.nndc.bnl.gov/nsr/nsrlink.jsp?1970El23,B}{1970El23}, \href{https://www.nndc.bnl.gov/nsr/nsrlink.jsp?1970Ha49,B}{1970Ha49}, T. Engeland \textit{ et al}., Nucl. Phys. A}\\
\parbox[b][0.3cm]{17.7cm}{181 (1972) 368, \href{https://www.nndc.bnl.gov/nsr/nsrlink.jsp?1972Ka01,B}{1972Ka01}, \href{https://www.nndc.bnl.gov/nsr/nsrlink.jsp?1977He18,B}{1977He18}, J. A. Harvey, J. Phys. Soc. Jpn. Suppl. 44 (1978) 127, A. M. Bernstein, V. R. Brown, and}\\
\parbox[b][0.3cm]{17.7cm}{V. A. Madsen, Phys. Rev. Lett. 42 (1979) 425, \href{https://www.nndc.bnl.gov/nsr/nsrlink.jsp?1982Br24,B}{1982Br24}, \href{https://www.nndc.bnl.gov/nsr/nsrlink.jsp?1983Br29,B}{1983Br29}, \href{https://www.nndc.bnl.gov/nsr/nsrlink.jsp?1984Wi17,B}{1984Wi17}, \href{https://www.nndc.bnl.gov/nsr/nsrlink.jsp?1985Br29,B}{1985Br29}, D. J. Rowe, Rept. Prog. Phys. 48}\\
\parbox[b][0.3cm]{17.7cm}{(1985) 1419, \href{https://www.nndc.bnl.gov/nsr/nsrlink.jsp?1986An07,B}{1986An07}, \href{https://www.nndc.bnl.gov/nsr/nsrlink.jsp?1986An10,B}{1986An10}, \href{https://www.nndc.bnl.gov/nsr/nsrlink.jsp?1996Pa28,B}{1996Pa28}, \href{https://www.nndc.bnl.gov/nsr/nsrlink.jsp?1997Bo47,B}{1997Bo47}, \href{https://www.nndc.bnl.gov/nsr/nsrlink.jsp?1997Mi08,B}{1997Mi08}, \href{https://www.nndc.bnl.gov/nsr/nsrlink.jsp?1999Si13,B}{1999Si13}, \href{https://www.nndc.bnl.gov/nsr/nsrlink.jsp?1999Va03,B}{1999Va03}, \href{https://www.nndc.bnl.gov/nsr/nsrlink.jsp?2006Gu07,B}{2006Gu07}, \href{https://www.nndc.bnl.gov/nsr/nsrlink.jsp?2006Or01,B}{2006Or01}, \href{https://www.nndc.bnl.gov/nsr/nsrlink.jsp?2007Ma54,B}{2007Ma54},}\\
\parbox[b][0.3cm]{17.7cm}{\href{https://www.nndc.bnl.gov/nsr/nsrlink.jsp?2007Wa30,B}{2007Wa30}, \href{https://www.nndc.bnl.gov/nsr/nsrlink.jsp?2008Ne13,B}{2008Ne13}, \href{https://www.nndc.bnl.gov/nsr/nsrlink.jsp?2008NeZX,B}{2008NeZX}, \href{https://www.nndc.bnl.gov/nsr/nsrlink.jsp?2012Po15,B}{2012Po15}, \href{https://www.nndc.bnl.gov/nsr/nsrlink.jsp?2013La15,B}{2013La15}, \href{https://www.nndc.bnl.gov/nsr/nsrlink.jsp?2015La10,B}{2015La10}, \href{https://www.nndc.bnl.gov/nsr/nsrlink.jsp?2015Wu07,B}{2015Wu07}, \href{https://www.nndc.bnl.gov/nsr/nsrlink.jsp?2018Ji07,B}{2018Ji07}, \href{https://www.nndc.bnl.gov/nsr/nsrlink.jsp?2019Mi22,B}{2019Mi22}, \href{https://www.nndc.bnl.gov/nsr/nsrlink.jsp?2020Wi12,B}{2020Wi12}, \href{https://www.nndc.bnl.gov/nsr/nsrlink.jsp?2020Ma25,B}{2020Ma25},}\\
\parbox[b][0.3cm]{17.7cm}{\href{https://www.nndc.bnl.gov/nsr/nsrlink.jsp?2021Ka42,B}{2021Ka42}, \href{https://www.nndc.bnl.gov/nsr/nsrlink.jsp?2021Sa49,B}{2021Sa49}, \href{https://www.nndc.bnl.gov/nsr/nsrlink.jsp?2021Ya26,B}{2021Ya26}, \href{https://www.nndc.bnl.gov/nsr/nsrlink.jsp?2022St03,B}{2022St03}, \href{https://www.nndc.bnl.gov/nsr/nsrlink.jsp?2022Zh57,B}{2022Zh57}, \href{https://www.nndc.bnl.gov/nsr/nsrlink.jsp?2023Sa22,B}{2023Sa22}, J. G. Li, K. H. Li, N. Michel, H. H. Li, and W. Zuo,}\\
\parbox[b][0.3cm]{17.7cm}{\textit{Mechanisms of mirror energy difference for states exhibiting Thomas-Ehrman shift: Gamow shell model case studies of \ensuremath{^{18}}Ne/\ensuremath{^{\textnormal{18}}}O}}\\
\parbox[b][0.3cm]{17.7cm}{\textit{and \ensuremath{^{19}}Na/\ensuremath{^{\textnormal{19}}}O}, arXiv:2407.00884v1 [nucl-th] (July-2024) (unpublished).}\\
\parbox[b][0.3cm]{17.7cm}{\addtolength{\parindent}{-0.2in}\textit{Superallowed \ensuremath{\beta} transition}: \href{https://www.nndc.bnl.gov/nsr/nsrlink.jsp?1978Sz03,B}{1978Sz03}, \href{https://www.nndc.bnl.gov/nsr/nsrlink.jsp?2002To19,B}{2002To19}, \href{https://www.nndc.bnl.gov/nsr/nsrlink.jsp?2005Ha15,B}{2005Ha15}, \href{https://www.nndc.bnl.gov/nsr/nsrlink.jsp?2005Ha65,B}{2005Ha65}, \href{https://www.nndc.bnl.gov/nsr/nsrlink.jsp?2006Ha12,B}{2006Ha12}, \href{https://www.nndc.bnl.gov/nsr/nsrlink.jsp?2008To03,B}{2008To03}, \href{https://www.nndc.bnl.gov/nsr/nsrlink.jsp?2012Ha46,B}{2012Ha46}, \href{https://www.nndc.bnl.gov/nsr/nsrlink.jsp?2012Sa50,B}{2012Sa50}, G. F. Grinyer}\\
\parbox[b][0.3cm]{17.7cm}{\textit{et} \textit{al}., Nucl. Instr. Meth. Phys. Res. A 741 (2014) 18 (GEANT4 simulation for \ensuremath{^{\textnormal{18}}}Ne decay), \href{https://www.nndc.bnl.gov/nsr/nsrlink.jsp?2015To02,B}{2015To02}, \href{https://www.nndc.bnl.gov/nsr/nsrlink.jsp?2022Xa01,B}{2022Xa01}.}\\
\parbox[b][0.3cm]{17.7cm}{\addtolength{\parindent}{-0.2in}\textit{\ensuremath{\beta}\ensuremath{^{+}} decay}: \href{https://www.nndc.bnl.gov/nsr/nsrlink.jsp?1976Lo01,B}{1976Lo01}, \href{https://www.nndc.bnl.gov/nsr/nsrlink.jsp?1977Az02,B}{1977Az02}, \href{https://www.nndc.bnl.gov/nsr/nsrlink.jsp?1977To11,B}{1977To11}, \href{https://www.nndc.bnl.gov/nsr/nsrlink.jsp?1981Ha06,B}{1981Ha06}, E. G. Adelberger, Comments Nucl. Part. Phys. 11 (1983) 189, P.}\\
\parbox[b][0.3cm]{17.7cm}{G.Bizzeti, Riv. Nuovo Cim. vol 6, issue 12 (1983) 1-64, \href{https://www.nndc.bnl.gov/nsr/nsrlink.jsp?1984Ha58,B}{1984Ha58}, E. G. Adelberger and W. C. Haxton, Ann. Rev. Nucl. Part.}\\
\parbox[b][0.3cm]{17.7cm}{Sci. 35 (1985) 501, \href{https://www.nndc.bnl.gov/nsr/nsrlink.jsp?1986KiZR,B}{1986KiZR}, I. S. Towner, Ann. Rev. Nucl. Part. Sci. 36 (1986) 115, \href{https://www.nndc.bnl.gov/nsr/nsrlink.jsp?1987Ki03,B}{1987Ki03}, \href{https://www.nndc.bnl.gov/nsr/nsrlink.jsp?1991Ja13,B}{1991Ja13}, \href{https://www.nndc.bnl.gov/nsr/nsrlink.jsp?1992Ba22,B}{1992Ba22},}\\
\parbox[b][0.3cm]{17.7cm}{\href{https://www.nndc.bnl.gov/nsr/nsrlink.jsp?1992He12,B}{1992He12}, \href{https://www.nndc.bnl.gov/nsr/nsrlink.jsp?1996Ka04,B}{1996Ka04}, \href{https://www.nndc.bnl.gov/nsr/nsrlink.jsp?1999Va08,B}{1999Va08}, \href{https://www.nndc.bnl.gov/nsr/nsrlink.jsp?1999Va25,B}{1999Va25}, \href{https://www.nndc.bnl.gov/nsr/nsrlink.jsp?2002Va14,B}{2002Va14}, \href{https://www.nndc.bnl.gov/nsr/nsrlink.jsp?2004Mc05,B}{2004Mc05}, \href{https://www.nndc.bnl.gov/nsr/nsrlink.jsp?2009Ha12,B}{2009Ha12}, \href{https://www.nndc.bnl.gov/nsr/nsrlink.jsp?2009Li22,B}{2009Li22}, \href{https://www.nndc.bnl.gov/nsr/nsrlink.jsp?2010Li52,B}{2010Li52}, \href{https://www.nndc.bnl.gov/nsr/nsrlink.jsp?2015El06,B}{2015El06}, \href{https://www.nndc.bnl.gov/nsr/nsrlink.jsp?2016Sa34,B}{2016Sa34},}\\
\parbox[b][0.3cm]{17.7cm}{\href{https://www.nndc.bnl.gov/nsr/nsrlink.jsp?2020Oh01,B}{2020Oh01}, \href{https://www.nndc.bnl.gov/nsr/nsrlink.jsp?2023Se05,B}{2023Se05}.}\\
\parbox[b][0.3cm]{17.7cm}{\addtolength{\parindent}{-0.2in}\textit{Mirror and analog states}: \href{https://www.nndc.bnl.gov/nsr/nsrlink.jsp?1969Mu09,B}{1969Mu09}, \href{https://www.nndc.bnl.gov/nsr/nsrlink.jsp?1971Bl12,B}{1971Bl12}, \href{https://www.nndc.bnl.gov/nsr/nsrlink.jsp?1987Ki03,B}{1987Ki03}, \href{https://www.nndc.bnl.gov/nsr/nsrlink.jsp?2020Br14,B}{2020Br14}, \href{https://www.nndc.bnl.gov/nsr/nsrlink.jsp?2021Am03,B}{2021Am03}, \href{https://www.nndc.bnl.gov/nsr/nsrlink.jsp?2023Li03,B}{2023Li03}.}\\
\parbox[b][0.3cm]{17.7cm}{\addtolength{\parindent}{-0.2in}\textit{Giant multipole resonances}: \href{https://www.nndc.bnl.gov/nsr/nsrlink.jsp?1994StZY,B}{1994StZY}, \href{https://www.nndc.bnl.gov/nsr/nsrlink.jsp?2012Ma12,B}{2012Ma12}, \href{https://www.nndc.bnl.gov/nsr/nsrlink.jsp?2017Lv02,B}{2017Lv02}.}\\
\parbox[b][0.3cm]{17.7cm}{\addtolength{\parindent}{-0.2in}\textit{Other nuclear structure information}: \href{https://www.nndc.bnl.gov/nsr/nsrlink.jsp?1972En03,B}{1972En03}, \href{https://www.nndc.bnl.gov/nsr/nsrlink.jsp?1972Ra08,B}{1972Ra08}, \href{https://www.nndc.bnl.gov/nsr/nsrlink.jsp?1976Sh04,B}{1976Sh04}, \href{https://www.nndc.bnl.gov/nsr/nsrlink.jsp?1977Sz03,B}{1977Sz03}, \href{https://www.nndc.bnl.gov/nsr/nsrlink.jsp?1979En05,B}{1979En05} (compilation), \href{https://www.nndc.bnl.gov/nsr/nsrlink.jsp?1979Sa31,B}{1979Sa31}, \href{https://www.nndc.bnl.gov/nsr/nsrlink.jsp?1982Ri04,B}{1982Ri04} (E2}\\
\parbox[b][0.3cm]{17.7cm}{strengths), \href{https://www.nndc.bnl.gov/nsr/nsrlink.jsp?1982La26,B}{1982La26} (compilation), \href{https://www.nndc.bnl.gov/nsr/nsrlink.jsp?1985Al21,B}{1985Al21}, \href{https://www.nndc.bnl.gov/nsr/nsrlink.jsp?1985An28,B}{1985An28} (compilation), B. G. Harvey, J. Phys. (Paris) 47 (1986) C4-29, \href{https://www.nndc.bnl.gov/nsr/nsrlink.jsp?1986St15,B}{1986St15}}\\
\parbox[b][0.3cm]{17.7cm}{(transition densities), \href{https://www.nndc.bnl.gov/nsr/nsrlink.jsp?1988Tr02,B}{1988Tr02}, \href{https://www.nndc.bnl.gov/nsr/nsrlink.jsp?1992Ch50,B}{1992Ch50} (astrophysical resonance properties), \href{https://www.nndc.bnl.gov/nsr/nsrlink.jsp?2004Su26,B}{2004Su26} (B(E2) strengths), \href{https://www.nndc.bnl.gov/nsr/nsrlink.jsp?2005Li48,B}{2005Li48}}\\
\parbox[b][0.3cm]{17.7cm}{(Gamow-Teller transitions), \href{https://www.nndc.bnl.gov/nsr/nsrlink.jsp?2005Sa63,B}{2005Sa63} (B(E2) strengths), \href{https://www.nndc.bnl.gov/nsr/nsrlink.jsp?2008He15,B}{2008He15}, \href{https://www.nndc.bnl.gov/nsr/nsrlink.jsp?2008Ti09,B}{2008Ti09} (spectroscopic factors and ANCs), \href{https://www.nndc.bnl.gov/nsr/nsrlink.jsp?2009PeZW,B}{2009PeZW}}\\
\parbox[b][0.3cm]{17.7cm}{(B(E1) strengths), \href{https://www.nndc.bnl.gov/nsr/nsrlink.jsp?2009Yu04,B}{2009Yu04} (Fadeev approach), \href{https://www.nndc.bnl.gov/nsr/nsrlink.jsp?2011Ma18,B}{2011Ma18} (B(E1) distributions), \href{https://www.nndc.bnl.gov/nsr/nsrlink.jsp?2012Ok02,B}{2012Ok02} (ANCs), \href{https://www.nndc.bnl.gov/nsr/nsrlink.jsp?2017Pr04,B}{2017Pr04} (evaluated B(E2)),}\\
\parbox[b][0.3cm]{17.7cm}{\href{https://www.nndc.bnl.gov/nsr/nsrlink.jsp?2018Mi22,B}{2018Mi22} (M1 strengths), \href{https://www.nndc.bnl.gov/nsr/nsrlink.jsp?2019Mu05,B}{2019Mu05} (ANCs), \href{https://www.nndc.bnl.gov/nsr/nsrlink.jsp?2022Oh01,B}{2022Oh01}, \href{https://www.nndc.bnl.gov/nsr/nsrlink.jsp?2022Gu11,B}{2022Gu11}, \href{https://www.nndc.bnl.gov/nsr/nsrlink.jsp?2022Su17,B}{2022Su17}, \href{https://www.nndc.bnl.gov/nsr/nsrlink.jsp?2023Fo05,B}{2023Fo05}, \href{https://www.nndc.bnl.gov/nsr/nsrlink.jsp?2023Xu10,B}{2023Xu10}, \href{https://www.nndc.bnl.gov/nsr/nsrlink.jsp?2024Ga05,B}{2024Ga05} (charge radius).}\\
\vspace{0.385cm}
\parbox[b][0.3cm]{17.7cm}{\addtolength{\parindent}{-0.2in}Other topics:}\\
\parbox[b][0.3cm]{17.7cm}{\addtolength{\parindent}{-0.2in}\textit{Decay of trapped ions}: \href{https://www.nndc.bnl.gov/nsr/nsrlink.jsp?2020Oh01,B}{2020Oh01}.}\\
\parbox[b][0.3cm]{17.7cm}{\addtolength{\parindent}{-0.2in}\textit{Other reactions that populated \ensuremath{^{18}}Ne}:}\\
\parbox[b][0.3cm]{17.7cm}{\addtolength{\parindent}{-0.2in}\ensuremath{^{\textnormal{12}}}C(\ensuremath{^{\textnormal{18}}}O,\ensuremath{^{\textnormal{18}}}Ne)\ensuremath{^{\textnormal{12}}}Be: M. Takaki \textit{et al}., Proc. Conf. Advances in Radioactive Isotope Science (ARIS2014) JPS Conf. Proc. 6 (2015)}\\
\parbox[b][0.3cm]{17.7cm}{020038.}\\
\parbox[b][0.3cm]{17.7cm}{\addtolength{\parindent}{-0.2in}\ensuremath{^{\textnormal{18}}}O(\ensuremath{^{\textnormal{12}}}C,\ensuremath{^{\textnormal{12}}}Be(0\ensuremath{^{\textnormal{+}}_{\textnormal{2}}}))\ensuremath{^{\textnormal{18}}}Ne: \href{https://www.nndc.bnl.gov/nsr/nsrlink.jsp?2015TaZY,B}{2015TaZY}: populated the ground state of \ensuremath{^{\textnormal{18}}}Ne.}\\
\parbox[b][0.3cm]{17.7cm}{\addtolength{\parindent}{-0.2in}\ensuremath{^{\textnormal{40}}}Ca(\ensuremath{^{\textnormal{18}}}O,\ensuremath{^{\textnormal{18}}}Ne*)\ensuremath{^{\textnormal{40}}}Ar: \href{https://www.nndc.bnl.gov/nsr/nsrlink.jsp?2015Ca24,B}{2015Ca24}, \href{https://www.nndc.bnl.gov/nsr/nsrlink.jsp?2016Ca12,B}{2016Ca12}. These studies populated the \ensuremath{^{\textnormal{18}}}Ne*(1887 keV, 2\ensuremath{^{\textnormal{+}}}) state. It was observed as}\\
\parbox[b][0.3cm]{17.7cm}{unresolved and mixed with the \ensuremath{^{\textnormal{40}}}Ar\ensuremath{_{\textnormal{g.s.}}} and \ensuremath{^{\textnormal{40}}}Ar*(1.46 MeV) state.}\\
\parbox[b][0.3cm]{17.7cm}{\addtolength{\parindent}{-0.2in}\ensuremath{^{\textnormal{116}}}Sn(\ensuremath{^{\textnormal{18}}}O,\ensuremath{^{\textnormal{18}}}Ne)\ensuremath{^{\textnormal{116}}}Cd: \href{https://www.nndc.bnl.gov/nsr/nsrlink.jsp?2019CaZW,B}{2019CaZW}.}\\
\vspace{0.385cm}
\parbox[b][0.3cm]{17.7cm}{\addtolength{\parindent}{-0.2in}Previous \ensuremath{^{\textnormal{18}}}Ne evaluations: \href{https://www.nndc.bnl.gov/nsr/nsrlink.jsp?1959Aj76,B}{1959Aj76}, \href{https://www.nndc.bnl.gov/nsr/nsrlink.jsp?1972Aj02,B}{1972Aj02}, \href{https://www.nndc.bnl.gov/nsr/nsrlink.jsp?1978Aj03,B}{1978Aj03}, \href{https://www.nndc.bnl.gov/nsr/nsrlink.jsp?1981AjZY,B}{1981AjZY}, \href{https://www.nndc.bnl.gov/nsr/nsrlink.jsp?1983Aj01,B}{1983Aj01}, \href{https://www.nndc.bnl.gov/nsr/nsrlink.jsp?1987Aj02,B}{1987Aj02}, \href{https://www.nndc.bnl.gov/nsr/nsrlink.jsp?1978Aj03,B}{1978Aj03}, \href{https://www.nndc.bnl.gov/nsr/nsrlink.jsp?1987Co31,B}{1987Co31}, \href{https://www.nndc.bnl.gov/nsr/nsrlink.jsp?1995Ti07,B}{1995Ti07}.}\\
\vspace{0.385cm}
\vspace{12pt}
\hypertarget{18NE_LEVEL}{\underline{$^{18}$Ne Levels}}\\
\vspace{0.34cm}
\parbox[b][0.3cm]{17.7cm}{\addtolength{\parindent}{-0.254cm}When possible, measured resonance energies in the center-of-mass have been used to deduced the corresponding E\ensuremath{_{\textnormal{x}}} using either}\\
\parbox[b][0.3cm]{17.7cm}{S\ensuremath{_{\textnormal{p}}}=3923.1 keV \textit{4}, or S\ensuremath{_{\ensuremath{\alpha}}}=5115.1 keV \textit{4} from (\href{https://www.nndc.bnl.gov/nsr/nsrlink.jsp?2021Wa16,B}{2021Wa16}). The given resonant reaction (A(b,c):res) determines which separation}\\
\parbox[b][0.3cm]{17.7cm}{energy is used. These deduced E\ensuremath{_{\textnormal{x}}} values are included in the recommended weighted average excitation energies given in the \ensuremath{^{\textnormal{18}}}Ne}\\
\parbox[b][0.3cm]{17.7cm}{Adopted Levels.}\\
\vspace{0.34cm}
\begin{longtable}[c]{llllll}
\multicolumn{6}{c}{\underline{Cross Reference (XREF) Flags}}\\
 \\
\hyperlink{AL1}{\texttt{A }}& \ensuremath{^{\textnormal{22}}}Al \ensuremath{\varepsilon}\ensuremath{\alpha} decay:91.1 ms & \hyperlink{NE11}{\texttt{K }}& \ensuremath{^{\textnormal{9}}}Be(\ensuremath{^{\textnormal{20}}}Mg,X) & \hyperlink{NE21}{\texttt{U }}& \ensuremath{^{\textnormal{19}}}F(p,2n)\\
\hyperlink{NA2}{\texttt{B }}& \ensuremath{^{\textnormal{19}}}Na p decay & \hyperlink{NE12}{\texttt{L }}& \ensuremath{^{\textnormal{9}}}Be(\ensuremath{^{\textnormal{34}}}Ar,\ensuremath{^{\textnormal{18}}}Ne\ensuremath{\gamma}) & \hyperlink{NE22}{\texttt{V }}& \ensuremath{^{\textnormal{20}}}Ne(p,t)\\
\hyperlink{NE3}{\texttt{C }}& \ensuremath{^{\textnormal{1}}}H(\ensuremath{^{\textnormal{17}}}F,\ensuremath{\gamma}) & \hyperlink{NE13}{\texttt{M }}& \ensuremath{^{\textnormal{12}}}C(\ensuremath{^{\textnormal{12}}}C,\ensuremath{^{\textnormal{6}}}He) & \hyperlink{NE23}{\texttt{W }}& \ensuremath{^{\textnormal{27}}}Al(\ensuremath{\pi}\ensuremath{^{-}},\ensuremath{^{\textnormal{18}}}Ne\ensuremath{\gamma})\\
\hyperlink{NE4}{\texttt{D }}& \ensuremath{^{\textnormal{1}}}H(\ensuremath{^{\textnormal{17}}}F,p),(\ensuremath{^{\textnormal{17}}}F,2p),(\ensuremath{^{\textnormal{17}}}F,\ensuremath{\alpha}):res & \hyperlink{NE14}{\texttt{N }}& \ensuremath{^{\textnormal{14}}}N(\ensuremath{^{\textnormal{17}}}F,\ensuremath{^{\textnormal{18}}}Ne) & \hyperlink{NE24}{\texttt{X }}& \ensuremath{^{\textnormal{27}}}Al(p,X\ensuremath{\gamma})\\
\hyperlink{NE5}{\texttt{E }}& \ensuremath{^{\textnormal{1}}}H(\ensuremath{^{\textnormal{18}}}Ne,p\ensuremath{'}) & \hyperlink{NE15}{\texttt{O }}& \ensuremath{^{\textnormal{16}}}O(\ensuremath{^{\textnormal{3}}}He,n) & \hyperlink{NE25}{\texttt{Y }}& \ensuremath{^{\textnormal{27}}}Al(\ensuremath{\alpha},\ensuremath{^{\textnormal{18}}}Ne\ensuremath{\gamma}),\ensuremath{^{\textnormal{28}}}Si(\ensuremath{\alpha},\ensuremath{^{\textnormal{18}}}Ne\ensuremath{\gamma})\\
\end{longtable}
\vspace{-0.5cm}
\begin{longtable}{ccccccc@{\extracolsep{\fill}}c}
\multicolumn{2}{c}{E(level)$^{}$}&J$^{\pi}$$^{}$&\multicolumn{2}{c}{T\ensuremath{_{\textnormal{1/2}}} or \ensuremath{\Gamma}$^{}$}&XREF&Comments&\\[-.2cm]
\multicolumn{2}{c}{\hrulefill}&\hrulefill&\multicolumn{2}{c}{\hrulefill}&\hrulefill&\hrulefill&
\endfirsthead
\multicolumn{1}{r@{}}{0}&\multicolumn{1}{@{}l}{}&\multicolumn{1}{l}{0\ensuremath{^{+}}\ensuremath{^{{\hyperlink{NE0LEVEL3}{d}}}}}&\multicolumn{1}{r@{}}{1}&\multicolumn{1}{@{.}l}{66427 s {\it 72}}&\multicolumn{1}{l}{\texttt{\hyperlink{AL1}{A}\hyperlink{NA2}{B}\hyperlink{NE3}{C}\ \hyperlink{NE5}{E}\hyperlink{NE6}{F}\ \ \ \ \hyperlink{NE11}{K}\hyperlink{NE12}{L}\hyperlink{NE13}{M}\hyperlink{NE14}{N}\hyperlink{NE15}{O}\hyperlink{NE16}{P}\ \ \hyperlink{NE19}{S}\hyperlink{NE20}{T}\hyperlink{NE21}{U}\hyperlink{NE22}{V}\hyperlink{NE23}{W}\hyperlink{NE24}{X}\hyperlink{NE25}{Y}\hyperlink{NE26}{Z}} }&\parbox[t][0.3cm]{8.336241cm}{\raggedright XREF: Others: \hyperlink{NE27}{\texttt{AA}}\vspace{0.1cm}}&\\
&&&&&&\parbox[t][0.3cm]{8.336241cm}{\raggedright \%\ensuremath{\varepsilon}+\%\ensuremath{\beta}\ensuremath{^{\textnormal{+}}}=100 (\href{https://www.nndc.bnl.gov/nsr/nsrlink.jsp?1981Ad01,B}{1981Ad01})\vspace{0.1cm}}&\\
&&&&&&\parbox[t][0.3cm]{8.336241cm}{\raggedright XREF: X(?)\vspace{0.1cm}}&\\
&&&&&&\parbox[t][0.3cm]{8.336241cm}{\raggedright T=1 (\href{https://www.nndc.bnl.gov/nsr/nsrlink.jsp?1961Bu05,B}{1961Bu05}, \href{https://www.nndc.bnl.gov/nsr/nsrlink.jsp?1963Fr10,B}{1963Fr10}, \href{https://www.nndc.bnl.gov/nsr/nsrlink.jsp?1974Ne04,B}{1974Ne04}, \href{https://www.nndc.bnl.gov/nsr/nsrlink.jsp?1975Ha21,B}{1975Ha21},\vspace{0.1cm}}&\\
&&&&&&\parbox[t][0.3cm]{8.336241cm}{\raggedright {\ }{\ }{\ }\href{https://www.nndc.bnl.gov/nsr/nsrlink.jsp?1979Se08,B}{1979Se08},\href{https://www.nndc.bnl.gov/nsr/nsrlink.jsp?1981Ne09,B}{1981Ne09}, \href{https://www.nndc.bnl.gov/nsr/nsrlink.jsp?1981Ad01,B}{1981Ad01}).\vspace{0.1cm}}&\\
&&&&&&\parbox[t][0.3cm]{8.336241cm}{\raggedright T\ensuremath{_{\textnormal{z}}}={\textminus}1 (\href{https://www.nndc.bnl.gov/nsr/nsrlink.jsp?1968Gi09,B}{1968Gi09}, \href{https://www.nndc.bnl.gov/nsr/nsrlink.jsp?1975Ha21,B}{1975Ha21}).\vspace{0.1cm}}&\\
\end{longtable}
\begin{textblock}{29}(0,27.3)
Continued on next page (footnotes at end of table)
\end{textblock}
\clearpage
\begin{longtable}{ccccccc@{\extracolsep{\fill}}c}
\\[-.4cm]
\multicolumn{8}{c}{{\bf \small \underline{Adopted \hyperlink{18NE_LEVEL}{Levels}, \hyperlink{18NE_GAMMA}{Gammas} (continued)}}}\\
\multicolumn{8}{c}{~}\\
\multicolumn{8}{c}{\underline{\ensuremath{^{18}}Ne Levels (continued)}}\\
\multicolumn{8}{c}{~}\\
\multicolumn{2}{c}{E(level)$^{}$}&J$^{\pi}$$^{}$&\multicolumn{2}{c}{T\ensuremath{_{\textnormal{1/2}}} or \ensuremath{\Gamma}$^{}$}&XREF&Comments&\\[-.2cm]
\multicolumn{2}{c}{\hrulefill}&\hrulefill&\multicolumn{2}{c}{\hrulefill}&\hrulefill&\hrulefill&
\endhead
&&&&&&\parbox[t][0.3cm]{8.828441cm}{\raggedright {\ }{\ }{\ }it was adopted by those authors from the \ensuremath{^{\textnormal{18}}}Ne evaluation of\vspace{0.1cm}}&\\
&&&&&&\parbox[t][0.3cm]{8.828441cm}{\raggedright {\ }{\ }{\ }(\href{https://www.nndc.bnl.gov/nsr/nsrlink.jsp?1978Aj03,B}{1978Aj03}).\vspace{0.1cm}}&\\
&&&&&&\parbox[t][0.3cm]{8.828441cm}{\raggedright T\ensuremath{_{1/2}}: From \ensuremath{\tau}=0.68 ps \textit{6}, which is the weighted average (with\vspace{0.1cm}}&\\
&&&&&&\parbox[t][0.3cm]{8.828441cm}{\raggedright {\ }{\ }{\ }external errors and assuming statistical uncertainties) of \ensuremath{\tau}=0.49\vspace{0.1cm}}&\\
&&&&&&\parbox[t][0.3cm]{8.828441cm}{\raggedright {\ }{\ }{\ }ps \textit{+17{\textminus}9} (\href{https://www.nndc.bnl.gov/nsr/nsrlink.jsp?1968Gi09,B}{1968Gi09}, \href{https://www.nndc.bnl.gov/nsr/nsrlink.jsp?1969Ro08,B}{1969Ro08}); \ensuremath{\tau}=0.67 ps \textit{6} (\href{https://www.nndc.bnl.gov/nsr/nsrlink.jsp?1976Mc02,B}{1976Mc02});\vspace{0.1cm}}&\\
&&&&&&\parbox[t][0.3cm]{8.828441cm}{\raggedright {\ }{\ }{\ }and \ensuremath{\tau}=0.77 ps \textit{+9{\textminus}7} (\href{https://www.nndc.bnl.gov/nsr/nsrlink.jsp?2003Ri08,B}{2003Ri08}). See also \ensuremath{\tau}=0.63 ps \textit{13} (sys.)\vspace{0.1cm}}&\\
&&&&&&\parbox[t][0.3cm]{8.828441cm}{\raggedright {\ }{\ }{\ }(\href{https://www.nndc.bnl.gov/nsr/nsrlink.jsp?1974Mc17,B}{1974Mc17}), which is excluded (see the \ensuremath{^{\textnormal{16}}}O(\ensuremath{^{\textnormal{3}}}He,n\ensuremath{\gamma}) dataset\vspace{0.1cm}}&\\
&&&&&&\parbox[t][0.3cm]{8.828441cm}{\raggedright {\ }{\ }{\ }for discussion).\vspace{0.1cm}}&\\
\end{longtable}
\begin{textblock}{29}(0,27.3)
Continued on next page (footnotes at end of table)
\end{textblock}
\clearpage
\begin{longtable}{ccccccc@{\extracolsep{\fill}}c}
\\[-.4cm]
\multicolumn{8}{c}{{\bf \small \underline{Adopted \hyperlink{18NE_LEVEL}{Levels}, \hyperlink{18NE_GAMMA}{Gammas} (continued)}}}\\
\multicolumn{8}{c}{~}\\
\multicolumn{8}{c}{\underline{\ensuremath{^{18}}Ne Levels (continued)}}\\
\multicolumn{8}{c}{~}\\
\multicolumn{2}{c}{E(level)$^{}$}&J$^{\pi}$$^{}$&\multicolumn{2}{c}{T\ensuremath{_{\textnormal{1/2}}} or \ensuremath{\Gamma}$^{}$}&XREF&Comments&\\[-.2cm]
\multicolumn{2}{c}{\hrulefill}&\hrulefill&\multicolumn{2}{c}{\hrulefill}&\hrulefill&\hrulefill&
\endhead
&&&&&&\parbox[t][0.3cm]{8.098701cm}{\raggedright {\ }{\ }{\ }(\href{https://www.nndc.bnl.gov/nsr/nsrlink.jsp?1968Gi09,B}{1968Gi09}, \href{https://www.nndc.bnl.gov/nsr/nsrlink.jsp?1969Ro08,B}{1969Ro08}) and may explain the\vspace{0.1cm}}&\\
&&&&&&\parbox[t][0.3cm]{8.098701cm}{\raggedright {\ }{\ }{\ }inconsistencies between those values and the lifetime\vspace{0.1cm}}&\\
&&&&&&\parbox[t][0.3cm]{8.098701cm}{\raggedright {\ }{\ }{\ }deduced by (\href{https://www.nndc.bnl.gov/nsr/nsrlink.jsp?1972Gi01,B}{1972Gi01}).\vspace{0.1cm}}&\\
&&&&&&\parbox[t][0.3cm]{8.098701cm}{\raggedright J\ensuremath{^{\pi}}: From the n\ensuremath{\gamma} angular correlation measurement of\vspace{0.1cm}}&\\
&&&&&&\parbox[t][0.3cm]{8.098701cm}{\raggedright {\ }{\ }{\ }(\href{https://www.nndc.bnl.gov/nsr/nsrlink.jsp?1970Sh04,B}{1970Sh04}: studied the \ensuremath{^{\textnormal{16}}}O(\ensuremath{^{\textnormal{3}}}He,n)\ensuremath{^{\textnormal{18}}}Ne*\ensuremath{\rightarrow}\ensuremath{^{\textnormal{18}}}Ne\ensuremath{_{\textnormal{g.s.}}}+\ensuremath{\gamma}\vspace{0.1cm}}&\\
&&&&&&\parbox[t][0.3cm]{8.098701cm}{\raggedright {\ }{\ }{\ }reaction), which yielded a unique spin-parity assignment\vspace{0.1cm}}&\\
&&&&&&\parbox[t][0.3cm]{8.098701cm}{\raggedright {\ }{\ }{\ }of J\ensuremath{^{\ensuremath{\pi}}}=4\ensuremath{^{\textnormal{+}}}. This J\ensuremath{^{\ensuremath{\pi}}} assignment is supported by various\vspace{0.1cm}}&\\
&&&&&&\parbox[t][0.3cm]{8.098701cm}{\raggedright {\ }{\ }{\ }DWBA analyses and mirror states analyses (see\vspace{0.1cm}}&\\
\end{longtable}
\begin{textblock}{29}(0,27.3)
Continued on next page (footnotes at end of table)
\end{textblock}
\clearpage
\begin{longtable}{ccccccc@{\extracolsep{\fill}}c}
\\[-.4cm]
\multicolumn{8}{c}{{\bf \small \underline{Adopted \hyperlink{18NE_LEVEL}{Levels}, \hyperlink{18NE_GAMMA}{Gammas} (continued)}}}\\
\multicolumn{8}{c}{~}\\
\multicolumn{8}{c}{\underline{\ensuremath{^{18}}Ne Levels (continued)}}\\
\multicolumn{8}{c}{~}\\
\multicolumn{2}{c}{E(level)$^{}$}&J$^{\pi}$$^{}$&\multicolumn{2}{c}{T\ensuremath{_{\textnormal{1/2}}} or \ensuremath{\Gamma}$^{}$}&XREF&Comments&\\[-.2cm]
\multicolumn{2}{c}{\hrulefill}&\hrulefill&\multicolumn{2}{c}{\hrulefill}&\hrulefill&\hrulefill&
\endhead
&&&&&&\parbox[t][0.3cm]{8.770661cm}{\raggedright T\ensuremath{_{1/2}}: From \ensuremath{\tau}=0.063 ps \textit{+30{\textminus}20} (\href{https://www.nndc.bnl.gov/nsr/nsrlink.jsp?1969Ro08,B}{1969Ro08}). See also \ensuremath{\tau}=0.06 ps\vspace{0.1cm}}&\\
&&&&&&\parbox[t][0.3cm]{8.770661cm}{\raggedright {\ }{\ }{\ }\textit{+3{\textminus}2} (\href{https://www.nndc.bnl.gov/nsr/nsrlink.jsp?1968Gi09,B}{1968Gi09}).\vspace{0.1cm}}&\\
&&&&&&\parbox[t][0.3cm]{8.770661cm}{\raggedright \ensuremath{\Gamma}=1.05\ensuremath{\times}10\ensuremath{^{\textnormal{$-$2}}} eV (\href{https://www.nndc.bnl.gov/nsr/nsrlink.jsp?1969Ro08,B}{1969Ro08}: from \ensuremath{\tau}=0.063 ps).\vspace{0.1cm}}&\\
&&&&&&\parbox[t][0.3cm]{8.770661cm}{\raggedright J\ensuremath{^{\pi}}: From the \ensuremath{^{\textnormal{20}}}Ne(p,t) measurement by (\href{https://www.nndc.bnl.gov/nsr/nsrlink.jsp?1981Ne09,B}{1981Ne09}: see Fig.\vspace{0.1cm}}&\\
&&&&&&\parbox[t][0.3cm]{8.770661cm}{\raggedright {\ }{\ }{\ }14), which deduced J\ensuremath{^{\ensuremath{\pi}}}=2\ensuremath{^{\textnormal{+}}} and L=2 through comparison of\vspace{0.1cm}}&\\
&&&&&&\parbox[t][0.3cm]{8.770661cm}{\raggedright {\ }{\ }{\ }the triton angular distribution corresponding to this state with\vspace{0.1cm}}&\\
&&&&&&\parbox[t][0.3cm]{8.770661cm}{\raggedright {\ }{\ }{\ }known L=2 transitions in the similar \ensuremath{^{\textnormal{20}}}Ne(p,t) measurements\vspace{0.1cm}}&\\
&&&&&&\parbox[t][0.3cm]{8.770661cm}{\raggedright {\ }{\ }{\ }(\href{https://www.nndc.bnl.gov/nsr/nsrlink.jsp?1970Le08,B}{1970Le08}, \href{https://www.nndc.bnl.gov/nsr/nsrlink.jsp?1970Fa17,B}{1970Fa17}); mirror levels analysis; two-nucleon\vspace{0.1cm}}&\\
\end{longtable}
\begin{textblock}{29}(0,27.3)
Continued on next page (footnotes at end of table)
\end{textblock}
\clearpage
\begin{longtable}{ccccccc@{\extracolsep{\fill}}c}
\\[-.4cm]
\multicolumn{8}{c}{{\bf \small \underline{Adopted \hyperlink{18NE_LEVEL}{Levels}, \hyperlink{18NE_GAMMA}{Gammas} (continued)}}}\\
\multicolumn{8}{c}{~}\\
\multicolumn{8}{c}{\underline{\ensuremath{^{18}}Ne Levels (continued)}}\\
\multicolumn{8}{c}{~}\\
\multicolumn{2}{c}{E(level)$^{}$}&J$^{\pi}$$^{}$&\multicolumn{2}{c}{T\ensuremath{_{\textnormal{1/2}}} or \ensuremath{\Gamma}$^{}$}&XREF&Comments&\\[-.2cm]
\multicolumn{2}{c}{\hrulefill}&\hrulefill&\multicolumn{2}{c}{\hrulefill}&\hrulefill&\hrulefill&
\endhead
&&&&&&\parbox[t][0.3cm]{7.79482cm}{\raggedright {\ }{\ }{\ }known L=1 transitions in the similar \ensuremath{^{\textnormal{20}}}Ne(p,t)\vspace{0.1cm}}&\\
&&&&&&\parbox[t][0.3cm]{7.79482cm}{\raggedright {\ }{\ }{\ }measurements (\href{https://www.nndc.bnl.gov/nsr/nsrlink.jsp?1970Le08,B}{1970Le08}, \href{https://www.nndc.bnl.gov/nsr/nsrlink.jsp?1970Fa17,B}{1970Fa17}); mirror levels\vspace{0.1cm}}&\\
&&&&&&\parbox[t][0.3cm]{7.79482cm}{\raggedright {\ }{\ }{\ }analysis; two-nucleon Coulomb shift analysis; and\vspace{0.1cm}}&\\
&&&&&&\parbox[t][0.3cm]{7.79482cm}{\raggedright {\ }{\ }{\ }guided by the DWBA analysis of the \ensuremath{^{\textnormal{20}}}Ne(p,t) data of\vspace{0.1cm}}&\\
&&&&&&\parbox[t][0.3cm]{7.79482cm}{\raggedright {\ }{\ }{\ }(\href{https://www.nndc.bnl.gov/nsr/nsrlink.jsp?1970Fa17,B}{1970Fa17}).\vspace{0.1cm}}&\\
&&&&&&\parbox[t][0.3cm]{7.79482cm}{\raggedright J\ensuremath{^{\pi}}: This assignment is supported by various DWBA\vspace{0.1cm}}&\\
&&&&&&\parbox[t][0.3cm]{7.79482cm}{\raggedright {\ }{\ }{\ }analyses, Hauser-Feshbach calculations, isobaric analog\vspace{0.1cm}}&\\
&&&&&&\parbox[t][0.3cm]{7.79482cm}{\raggedright {\ }{\ }{\ }states analysis, and Coulomb shift analysis (see\vspace{0.1cm}}&\\
\end{longtable}
\begin{textblock}{29}(0,27.3)
Continued on next page (footnotes at end of table)
\end{textblock}
\clearpage
\begin{longtable}{ccccccc@{\extracolsep{\fill}}c}
\\[-.4cm]
\multicolumn{8}{c}{{\bf \small \underline{Adopted \hyperlink{18NE_LEVEL}{Levels}, \hyperlink{18NE_GAMMA}{Gammas} (continued)}}}\\
\multicolumn{8}{c}{~}\\
\multicolumn{8}{c}{\underline{\ensuremath{^{18}}Ne Levels (continued)}}\\
\multicolumn{8}{c}{~}\\
\multicolumn{2}{c}{E(level)$^{}$}&J$^{\pi}$$^{}$&\multicolumn{2}{c}{T\ensuremath{_{\textnormal{1/2}}} or \ensuremath{\Gamma}$^{}$}&XREF&Comments&\\[-.2cm]
\multicolumn{2}{c}{\hrulefill}&\hrulefill&\multicolumn{2}{c}{\hrulefill}&\hrulefill&\hrulefill&
\endhead
&&&&&&\parbox[t][0.3cm]{8.779221cm}{\raggedright {\ }{\ }{\ }(\href{https://www.nndc.bnl.gov/nsr/nsrlink.jsp?1981Ne09,B}{1981Ne09}: (p,t), this result supersedes \ensuremath{\Gamma}\ensuremath{\leq}40 keV reported\vspace{0.1cm}}&\\
&&&&&&\parbox[t][0.3cm]{8.779221cm}{\raggedright {\ }{\ }{\ }in an earlier (p,t) measurement by (\href{https://www.nndc.bnl.gov/nsr/nsrlink.jsp?1974Ne04,B}{1974Ne04})).\vspace{0.1cm}}&\\
&&&&&&\parbox[t][0.3cm]{8.779221cm}{\raggedright See also \ensuremath{\Gamma}\ensuremath{_{\textnormal{p}}}=1 keV (\href{https://www.nndc.bnl.gov/nsr/nsrlink.jsp?1991Ga03,B}{1991Ga03}): deduced for the state they\vspace{0.1cm}}&\\
&&&&&&\parbox[t][0.3cm]{8.779221cm}{\raggedright {\ }{\ }{\ }observed at E\ensuremath{_{\textnormal{x}}}=4589 keV. This width was obtained using the\vspace{0.1cm}}&\\
&&&&&&\parbox[t][0.3cm]{8.779221cm}{\raggedright {\ }{\ }{\ }experimentally determined proton spectroscopic factor of the\vspace{0.1cm}}&\\
&&&&&&\parbox[t][0.3cm]{8.779221cm}{\raggedright {\ }{\ }{\ }mirror level in \ensuremath{^{\textnormal{18}}}O (\href{https://www.nndc.bnl.gov/nsr/nsrlink.jsp?1976Li01,B}{1976Li01}) and the single-particle reduced\vspace{0.1cm}}&\\
&&&&&&\parbox[t][0.3cm]{8.779221cm}{\raggedright {\ }{\ }{\ }width determined from an optical model calculation by\vspace{0.1cm}}&\\
&&&&&&\parbox[t][0.3cm]{8.779221cm}{\raggedright {\ }{\ }{\ }(\href{https://www.nndc.bnl.gov/nsr/nsrlink.jsp?1991Ga03,B}{1991Ga03}). \ensuremath{\Gamma}\ensuremath{_{\textnormal{p}}}=1 keV was used by those authors to\vspace{0.1cm}}&\\
\end{longtable}
\begin{textblock}{29}(0,27.3)
Continued on next page (footnotes at end of table)
\end{textblock}
\clearpage
\begin{longtable}{ccccccc@{\extracolsep{\fill}}c}
\\[-.4cm]
\multicolumn{8}{c}{{\bf \small \underline{Adopted \hyperlink{18NE_LEVEL}{Levels}, \hyperlink{18NE_GAMMA}{Gammas} (continued)}}}\\
\multicolumn{8}{c}{~}\\
\multicolumn{8}{c}{\underline{\ensuremath{^{18}}Ne Levels (continued)}}\\
\multicolumn{8}{c}{~}\\
\multicolumn{2}{c}{E(level)$^{}$}&J$^{\pi}$$^{}$&\multicolumn{2}{c}{T\ensuremath{_{\textnormal{1/2}}} or \ensuremath{\Gamma}$^{}$}&XREF&Comments&\\[-.2cm]
\multicolumn{2}{c}{\hrulefill}&\hrulefill&\multicolumn{2}{c}{\hrulefill}&\hrulefill&\hrulefill&
\endhead
&&&&&&\parbox[t][0.3cm]{8.837001cm}{\raggedright {\ }{\ }{\ }\href{https://www.nndc.bnl.gov/nsr/nsrlink.jsp?2011He09,B}{2011He09}) are based on an assumption that the width is\vspace{0.1cm}}&\\
&&&&&&\parbox[t][0.3cm]{8.837001cm}{\raggedright {\ }{\ }{\ }entirely due to L=0 proton emission to the \ensuremath{^{\textnormal{17}}}F\ensuremath{_{\textnormal{g.s.}}}, which is\vspace{0.1cm}}&\\
&&&&&&\parbox[t][0.3cm]{8.837001cm}{\raggedright {\ }{\ }{\ }confirmed by (\href{https://www.nndc.bnl.gov/nsr/nsrlink.jsp?2005De15,B}{2005De15}). See also \ensuremath{\Gamma}\ensuremath{\leq}80 keV (\href{https://www.nndc.bnl.gov/nsr/nsrlink.jsp?1974Ne04,B}{1974Ne04}:\vspace{0.1cm}}&\\
&&&&&&\parbox[t][0.3cm]{8.837001cm}{\raggedright {\ }{\ }{\ }(p,t)); and \ensuremath{\Gamma}\ensuremath{\leq}60 keV (\href{https://www.nndc.bnl.gov/nsr/nsrlink.jsp?1974Ne04,B}{1974Ne04}, \href{https://www.nndc.bnl.gov/nsr/nsrlink.jsp?1981Ne09,B}{1981Ne09}: (\ensuremath{^{\textnormal{3}}}He,n)).\vspace{0.1cm}}&\\
&&&&&&\parbox[t][0.3cm]{8.837001cm}{\raggedright J\ensuremath{^{\pi}}: From R-matrix analysis of (\href{https://www.nndc.bnl.gov/nsr/nsrlink.jsp?2000Ga50,B}{2000Ga50}) and the zero-range\vspace{0.1cm}}&\\
&&&&&&\parbox[t][0.3cm]{8.837001cm}{\raggedright {\ }{\ }{\ }DWBA analysis of (\href{https://www.nndc.bnl.gov/nsr/nsrlink.jsp?1999Pa07,B}{1999Pa07}) using L=2 (\ensuremath{^{\textnormal{20}}}Ne(p,t)\vspace{0.1cm}}&\\
&&&&&&\parbox[t][0.3cm]{8.837001cm}{\raggedright {\ }{\ }{\ }reaction).\vspace{0.1cm}}&\\
&&&&&&\parbox[t][0.3cm]{8.837001cm}{\raggedright J\ensuremath{^{\pi}}: This assignment is supported by the R-matrix analyses of\vspace{0.1cm}}&\\
\end{longtable}
\begin{textblock}{29}(0,27.3)
Continued on next page (footnotes at end of table)
\end{textblock}
\clearpage
\begin{longtable}{ccccccc@{\extracolsep{\fill}}c}
\\[-.4cm]
\multicolumn{8}{c}{{\bf \small \underline{Adopted \hyperlink{18NE_LEVEL}{Levels}, \hyperlink{18NE_GAMMA}{Gammas} (continued)}}}\\
\multicolumn{8}{c}{~}\\
\multicolumn{8}{c}{\underline{\ensuremath{^{18}}Ne Levels (continued)}}\\
\multicolumn{8}{c}{~}\\
\multicolumn{2}{c}{E(level)$^{}$}&J$^{\pi}$$^{}$&\multicolumn{2}{c}{T\ensuremath{_{\textnormal{1/2}}} or \ensuremath{\Gamma}$^{}$}&XREF&Comments&\\[-.2cm]
\multicolumn{2}{c}{\hrulefill}&\hrulefill&\multicolumn{2}{c}{\hrulefill}&\hrulefill&\hrulefill&
\endhead
&&&&&&\parbox[t][0.3cm]{8.438961cm}{\raggedright {\ }{\ }{\ }and the result of (\href{https://www.nndc.bnl.gov/nsr/nsrlink.jsp?1996Ha26,B}{1996Ha26}), where J\ensuremath{^{\ensuremath{\pi}}}=(2\ensuremath{^{-}}) was assigned\vspace{0.1cm}}&\\
&&&&&&\parbox[t][0.3cm]{8.438961cm}{\raggedright {\ }{\ }{\ }from \ensuremath{^{\textnormal{12}}}C(\ensuremath{^{\textnormal{12}}}C,\ensuremath{^{\textnormal{6}}}He) based on the \ensuremath{^{\textnormal{6}}}He angular distribution\vspace{0.1cm}}&\\
&&&&&&\parbox[t][0.3cm]{8.438961cm}{\raggedright {\ }{\ }{\ }which peaked at backward angle. This was taken as a\vspace{0.1cm}}&\\
&&&&&&\parbox[t][0.3cm]{8.438961cm}{\raggedright {\ }{\ }{\ }signature for an un-natural parity assignment by\vspace{0.1cm}}&\\
&&&&&&\parbox[t][0.3cm]{8.438961cm}{\raggedright {\ }{\ }{\ }(\href{https://www.nndc.bnl.gov/nsr/nsrlink.jsp?1996Ha26,B}{1996Ha26}).\vspace{0.1cm}}&\\
&&&&&&\parbox[t][0.3cm]{8.438961cm}{\raggedright J\ensuremath{^{\pi}}: This assignment is supported by the result of (\href{https://www.nndc.bnl.gov/nsr/nsrlink.jsp?1996Ha26,B}{1996Ha26}:\vspace{0.1cm}}&\\
&&&&&&\parbox[t][0.3cm]{8.438961cm}{\raggedright {\ }{\ }{\ }(\ensuremath{^{\textnormal{3}}}He,n)), where a combination of penetrability\vspace{0.1cm}}&\\
&&&&&&\parbox[t][0.3cm]{8.438961cm}{\raggedright {\ }{\ }{\ }considerations, level widths, mirror levels analysis, and\vspace{0.1cm}}&\\
\end{longtable}
\begin{textblock}{29}(0,27.3)
Continued on next page (footnotes at end of table)
\end{textblock}
\clearpage
\begin{longtable}{ccccccc@{\extracolsep{\fill}}c}
\\[-.4cm]
\multicolumn{8}{c}{{\bf \small \underline{Adopted \hyperlink{18NE_LEVEL}{Levels}, \hyperlink{18NE_GAMMA}{Gammas} (continued)}}}\\
\multicolumn{8}{c}{~}\\
\multicolumn{8}{c}{\underline{\ensuremath{^{18}}Ne Levels (continued)}}\\
\multicolumn{8}{c}{~}\\
\multicolumn{2}{c}{E(level)$^{}$}&J$^{\pi}$$^{}$&\multicolumn{2}{c}{T\ensuremath{_{\textnormal{1/2}}} or \ensuremath{\Gamma}$^{}$}&XREF&Comments&\\[-.2cm]
\multicolumn{2}{c}{\hrulefill}&\hrulefill&\multicolumn{2}{c}{\hrulefill}&\hrulefill&\hrulefill&
\endhead
&&&&&&\parbox[t][0.3cm]{8.837001cm}{\raggedright {\ }{\ }{\ }leading to the \ensuremath{^{\textnormal{16}}}O\ensuremath{_{\textnormal{g.s.}}} 31\% \textit{7} of the time (\href{https://www.nndc.bnl.gov/nsr/nsrlink.jsp?2008Ra12,B}{2008Ra12},\vspace{0.1cm}}&\\
&&&&&&\parbox[t][0.3cm]{8.837001cm}{\raggedright {\ }{\ }{\ }\href{https://www.nndc.bnl.gov/nsr/nsrlink.jsp?2010Gi05,B}{2010Gi05}, \href{https://www.nndc.bnl.gov/nsr/nsrlink.jsp?2010Ra14,B}{2010Ra14}); (2) 2p democratic three-body decay\vspace{0.1cm}}&\\
&&&&&&\parbox[t][0.3cm]{8.837001cm}{\raggedright {\ }{\ }{\ }66\% \textit{9} of the time (\href{https://www.nndc.bnl.gov/nsr/nsrlink.jsp?2008Ra12,B}{2008Ra12}, \href{https://www.nndc.bnl.gov/nsr/nsrlink.jsp?2010Gi05,B}{2010Gi05}, \href{https://www.nndc.bnl.gov/nsr/nsrlink.jsp?2010Ra14,B}{2010Ra14}); and (3)\vspace{0.1cm}}&\\
&&&&&&\parbox[t][0.3cm]{8.837001cm}{\raggedright {\ }{\ }{\ }2p virtual sequential decay via the \ensuremath{^{\textnormal{17}}}F*(3.1 MeV, 1/2\ensuremath{^{-}}) state\vspace{0.1cm}}&\\
&&&&&&\parbox[t][0.3cm]{8.837001cm}{\raggedright {\ }{\ }{\ }3\% \textit{2} of the time (\href{https://www.nndc.bnl.gov/nsr/nsrlink.jsp?2008Ra12,B}{2008Ra12}, \href{https://www.nndc.bnl.gov/nsr/nsrlink.jsp?2010Gi05,B}{2010Gi05}, \href{https://www.nndc.bnl.gov/nsr/nsrlink.jsp?2010Ra14,B}{2010Ra14}).\vspace{0.1cm}}&\\
&&&&&&\parbox[t][0.3cm]{8.837001cm}{\raggedright J\ensuremath{^{\pi}}: From the R-matrix analysis of (\href{https://www.nndc.bnl.gov/nsr/nsrlink.jsp?2014Hu16,B}{2014Hu16}: L=1 from the\vspace{0.1cm}}&\\
&&&&&&\parbox[t][0.3cm]{8.837001cm}{\raggedright {\ }{\ }{\ }\ensuremath{^{\textnormal{1}}}H(\ensuremath{^{\textnormal{17}}}F,p) reaction).\vspace{0.1cm}}&\\
&&&&&&\parbox[t][0.3cm]{8.837001cm}{\raggedright J\ensuremath{^{\pi}}: Supported by R-matrix analyses of (\href{https://www.nndc.bnl.gov/nsr/nsrlink.jsp?2001Ga18,B}{2001Ga18}, \href{https://www.nndc.bnl.gov/nsr/nsrlink.jsp?2001Go01,B}{2001Go01},\vspace{0.1cm}}&\\
\end{longtable}
\begin{textblock}{29}(0,27.3)
Continued on next page (footnotes at end of table)
\end{textblock}
\clearpage
\begin{longtable}{ccccccc@{\extracolsep{\fill}}c}
\\[-.4cm]
\multicolumn{8}{c}{{\bf \small \underline{Adopted \hyperlink{18NE_LEVEL}{Levels}, \hyperlink{18NE_GAMMA}{Gammas} (continued)}}}\\
\multicolumn{8}{c}{~}\\
\multicolumn{8}{c}{\underline{\ensuremath{^{18}}Ne Levels (continued)}}\\
\multicolumn{8}{c}{~}\\
\multicolumn{2}{c}{E(level)$^{}$}&J$^{\pi}$$^{}$&\multicolumn{2}{c}{T\ensuremath{_{\textnormal{1/2}}} or \ensuremath{\Gamma}$^{}$}&XREF&Comments&\\[-.2cm]
\multicolumn{2}{c}{\hrulefill}&\hrulefill&\multicolumn{2}{c}{\hrulefill}&\hrulefill&\hrulefill&
\endhead
&&&&&&\parbox[t][0.3cm]{7.7349005cm}{\raggedright {\ }{\ }{\ }(\href{https://www.nndc.bnl.gov/nsr/nsrlink.jsp?2019Ch16,B}{2019Ch16}) reported the doublet as E\ensuremath{_{\textnormal{x}}}\ensuremath{\sim}6.3 MeV with\vspace{0.1cm}}&\\
&&&&&&\parbox[t][0.3cm]{7.7349005cm}{\raggedright {\ }{\ }{\ }J\ensuremath{^{\ensuremath{\pi}}}=(2\ensuremath{^{-}},3\ensuremath{^{-}}).\vspace{0.1cm}}&\\
&&&&&&\parbox[t][0.3cm]{7.7349005cm}{\raggedright J\ensuremath{^{\pi}}: From R-matrix analysis of (\href{https://www.nndc.bnl.gov/nsr/nsrlink.jsp?2014Hu16,B}{2014Hu16}: L=1 from\vspace{0.1cm}}&\\
&&&&&&\parbox[t][0.3cm]{7.7349005cm}{\raggedright {\ }{\ }{\ }the \ensuremath{^{\textnormal{1}}}H(\ensuremath{^{\textnormal{17}}}F,p) reaction) guided by the resonance\vspace{0.1cm}}&\\
&&&&&&\parbox[t][0.3cm]{7.7349005cm}{\raggedright {\ }{\ }{\ }parameters from (\href{https://www.nndc.bnl.gov/nsr/nsrlink.jsp?1996Ha26,B}{1996Ha26}). These parameters were\vspace{0.1cm}}&\\
&&&&&&\parbox[t][0.3cm]{7.7349005cm}{\raggedright {\ }{\ }{\ }deduced by (\href{https://www.nndc.bnl.gov/nsr/nsrlink.jsp?1996Ha26,B}{1996Ha26}) from a combination of\vspace{0.1cm}}&\\
&&&&&&\parbox[t][0.3cm]{7.7349005cm}{\raggedright {\ }{\ }{\ }Coulomb penetrability considerations, level widths,\vspace{0.1cm}}&\\
&&&&&&\parbox[t][0.3cm]{7.7349005cm}{\raggedright {\ }{\ }{\ }mirror levels analysis, and Woods-Saxon calculations\vspace{0.1cm}}&\\
\end{longtable}
\begin{textblock}{29}(0,27.3)
Continued on next page (footnotes at end of table)
\end{textblock}
\clearpage
\begin{longtable}{ccccccc@{\extracolsep{\fill}}c}
\\[-.4cm]
\multicolumn{8}{c}{{\bf \small \underline{Adopted \hyperlink{18NE_LEVEL}{Levels}, \hyperlink{18NE_GAMMA}{Gammas} (continued)}}}\\
\multicolumn{8}{c}{~}\\
\multicolumn{8}{c}{\underline{\ensuremath{^{18}}Ne Levels (continued)}}\\
\multicolumn{8}{c}{~}\\
\multicolumn{2}{c}{E(level)$^{}$}&J$^{\pi}$$^{}$&\multicolumn{2}{c}{T\ensuremath{_{\textnormal{1/2}}} or \ensuremath{\Gamma}$^{}$}&XREF&Comments&\\[-.2cm]
\multicolumn{2}{c}{\hrulefill}&\hrulefill&\multicolumn{2}{c}{\hrulefill}&\hrulefill&\hrulefill&
\endhead
&&&&&&\parbox[t][0.3cm]{8.70555cm}{\raggedright {\ }{\ }{\ }with L=2 and L=3, respectively, from the \ensuremath{^{\textnormal{1}}}H(\ensuremath{^{\textnormal{17}}}F,p)\vspace{0.1cm}}&\\
&&&&&&\parbox[t][0.3cm]{8.70555cm}{\raggedright {\ }{\ }{\ }reaction). However, (\href{https://www.nndc.bnl.gov/nsr/nsrlink.jsp?2014Hu16,B}{2014Hu16}) prefers the J\ensuremath{^{\ensuremath{\pi}}}=0\ensuremath{^{\textnormal{+}}}\vspace{0.1cm}}&\\
&&&&&&\parbox[t][0.3cm]{8.70555cm}{\raggedright {\ }{\ }{\ }assignment guided by the potential evidence of this state\vspace{0.1cm}}&\\
&&&&&&\parbox[t][0.3cm]{8.70555cm}{\raggedright {\ }{\ }{\ }being probably observed by (\href{https://www.nndc.bnl.gov/nsr/nsrlink.jsp?2004No14,B}{2004No14}, \href{https://www.nndc.bnl.gov/nsr/nsrlink.jsp?2004No18,B}{2004No18}) in a\vspace{0.1cm}}&\\
&&&&&&\parbox[t][0.3cm]{8.70555cm}{\raggedright {\ }{\ }{\ }direct \ensuremath{\alpha}(\ensuremath{^{\textnormal{14}}}O,p) reaction measurement, in which case, the\vspace{0.1cm}}&\\
&&&&&&\parbox[t][0.3cm]{8.70555cm}{\raggedright {\ }{\ }{\ }state most likely is of the natural-parity nature.\vspace{0.1cm}}&\\
&&&&&&\parbox[t][0.3cm]{8.70555cm}{\raggedright If this state is a 0\ensuremath{^{\textnormal{+}}} state, it would be the head of the 6p-4h\vspace{0.1cm}}&\\
&&&&&&\parbox[t][0.3cm]{8.70555cm}{\raggedright {\ }{\ }{\ }band (\href{https://www.nndc.bnl.gov/nsr/nsrlink.jsp?2003Fo13,B}{2003Fo13}, \href{https://www.nndc.bnl.gov/nsr/nsrlink.jsp?2011Fo12,B}{2011Fo12}).\vspace{0.1cm}}&\\
\end{longtable}
\begin{textblock}{29}(0,27.3)
Continued on next page (footnotes at end of table)
\end{textblock}
\clearpage
\begin{longtable}{cccccc@{\extracolsep{\fill}}c}
\\[-.4cm]
\multicolumn{7}{c}{{\bf \small \underline{Adopted \hyperlink{18NE_LEVEL}{Levels}, \hyperlink{18NE_GAMMA}{Gammas} (continued)}}}\\
\multicolumn{7}{c}{~}\\
\multicolumn{7}{c}{\underline{\ensuremath{^{18}}Ne Levels (continued)}}\\
\multicolumn{7}{c}{~}\\
\multicolumn{2}{c}{E(level)$^{}$}&\multicolumn{2}{c}{T\ensuremath{_{\textnormal{1/2}}} or \ensuremath{\Gamma}$^{}$}&XREF&Comments&\\[-.2cm]
\multicolumn{2}{c}{\hrulefill}&\multicolumn{2}{c}{\hrulefill}&\hrulefill&\hrulefill&
\endhead
&&&&&\parbox[t][0.3cm]{9.44581cm}{\raggedright {\ }{\ }{\ }of (\href{https://www.nndc.bnl.gov/nsr/nsrlink.jsp?1998Sh35,B}{1998Sh35}) and the Coulomb shift analysis between the mirror\vspace{0.1cm}}&\\
&&&&&\parbox[t][0.3cm]{9.44581cm}{\raggedright {\ }{\ }{\ }levels, which ruled out the J\ensuremath{^{\ensuremath{\pi}}}=1\ensuremath{^{-}} assignment made by (\href{https://www.nndc.bnl.gov/nsr/nsrlink.jsp?1999Ha14,B}{1999Ha14}:\vspace{0.1cm}}&\\
&&&&&\parbox[t][0.3cm]{9.44581cm}{\raggedright {\ }{\ }{\ }for the state observed at E\ensuremath{_{\textnormal{x}}}=7.16 MeV \textit{15}).\vspace{0.1cm}}&\\
&&&&&\parbox[t][0.3cm]{9.44581cm}{\raggedright The sequential 2p-decay via an intermediate \ensuremath{^{\textnormal{17}}}F state is\vspace{0.1cm}}&\\
&&&&&\parbox[t][0.3cm]{9.44581cm}{\raggedright {\ }{\ }{\ }energetically possible for this state (\href{https://www.nndc.bnl.gov/nsr/nsrlink.jsp?2008Ra12,B}{2008Ra12}, \href{https://www.nndc.bnl.gov/nsr/nsrlink.jsp?2010Ra14,B}{2010Ra14}). The\vspace{0.1cm}}&\\
&&&&&\parbox[t][0.3cm]{9.44581cm}{\raggedright {\ }{\ }{\ }decay via \ensuremath{^{\textnormal{16}}}O+2p channel is also observed by (\href{https://www.nndc.bnl.gov/nsr/nsrlink.jsp?2007Ra36,B}{2007Ra36},\vspace{0.1cm}}&\\
&&&&&\parbox[t][0.3cm]{9.44581cm}{\raggedright {\ }{\ }{\ }\href{https://www.nndc.bnl.gov/nsr/nsrlink.jsp?2008SfZZ,B}{2008SfZZ}, \href{https://www.nndc.bnl.gov/nsr/nsrlink.jsp?2008Ra12,B}{2008Ra12}, \href{https://www.nndc.bnl.gov/nsr/nsrlink.jsp?2010Gi05,B}{2010Gi05}, and \href{https://www.nndc.bnl.gov/nsr/nsrlink.jsp?2010Ra14,B}{2010Ra14}). (\href{https://www.nndc.bnl.gov/nsr/nsrlink.jsp?2008Ra12,B}{2008Ra12})\vspace{0.1cm}}&\\
&&&&&\parbox[t][0.3cm]{9.44581cm}{\raggedright {\ }{\ }{\ }mentions that the mode of 2p-decay is either via sequential decay\vspace{0.1cm}}&\\
\end{longtable}
\begin{textblock}{29}(0,27.3)
Continued on next page (footnotes at end of table)
\end{textblock}
\clearpage
\begin{longtable}{ccccccc@{\extracolsep{\fill}}c}
\\[-.4cm]
\multicolumn{8}{c}{{\bf \small \underline{Adopted \hyperlink{18NE_LEVEL}{Levels}, \hyperlink{18NE_GAMMA}{Gammas} (continued)}}}\\
\multicolumn{8}{c}{~}\\
\multicolumn{8}{c}{\underline{\ensuremath{^{18}}Ne Levels (continued)}}\\
\multicolumn{8}{c}{~}\\
\multicolumn{2}{c}{E(level)$^{}$}&J$^{\pi}$$^{}$&\multicolumn{2}{c}{T\ensuremath{_{\textnormal{1/2}}} or \ensuremath{\Gamma}$^{}$}&XREF&Comments&\\[-.2cm]
\multicolumn{2}{c}{\hrulefill}&\hrulefill&\multicolumn{2}{c}{\hrulefill}&\hrulefill&\hrulefill&
\endhead
\multicolumn{1}{r@{}}{7352}&\multicolumn{1}{@{ }l}{{\it 13}}&\multicolumn{1}{l}{(2\ensuremath{^{+}},1\ensuremath{^{-}})}&\multicolumn{1}{r@{}}{70}&\multicolumn{1}{@{ }l}{keV {\it 60}}&\multicolumn{1}{l}{\texttt{\ \ \ \hyperlink{NE4}{D}\ \ \hyperlink{NE7}{G}\ \ \ \ \ \hyperlink{NE13}{M}\ \hyperlink{NE15}{O}\ \ \ \ \ \ \ \ \ \ \ } }&\parbox[t][0.3cm]{7.9724407cm}{\raggedright E(level): Weighted average of 7350 keV \textit{20} (\href{https://www.nndc.bnl.gov/nsr/nsrlink.jsp?1996Ha26,B}{1996Ha26}:\vspace{0.1cm}}&\\
&&&&&&\parbox[t][0.3cm]{7.9724407cm}{\raggedright {\ }{\ }{\ }\ensuremath{^{\textnormal{12}}}C(\ensuremath{^{\textnormal{12}}}C,\ensuremath{^{\textnormal{6}}}He)); 7350 keV \textit{18} (\href{https://www.nndc.bnl.gov/nsr/nsrlink.jsp?1996Ha26,B}{1996Ha26}: (\ensuremath{^{\textnormal{3}}}He,n)); and\vspace{0.1cm}}&\\
&&&&&&\parbox[t][0.3cm]{7.9724407cm}{\raggedright {\ }{\ }{\ }7.40 MeV \textit{6} (\href{https://www.nndc.bnl.gov/nsr/nsrlink.jsp?2002Ha15,B}{2002Ha15}: from \ensuremath{^{\textnormal{1}}}H(\ensuremath{^{\textnormal{17}}}F,p):res E\ensuremath{_{\textnormal{c.m.}}}=3.48\vspace{0.1cm}}&\\
&&&&&&\parbox[t][0.3cm]{7.9724407cm}{\raggedright {\ }{\ }{\ }MeV \textit{6}).\vspace{0.1cm}}&\\
&&&&&&\parbox[t][0.3cm]{7.9724407cm}{\raggedright E(level): Note that (\href{https://www.nndc.bnl.gov/nsr/nsrlink.jsp?1999Ha14,B}{1999Ha14}) reported E\ensuremath{_{\textnormal{x}}}=7.37 MeV \textit{6}\vspace{0.1cm}}&\\
&&&&&&\parbox[t][0.3cm]{7.9724407cm}{\raggedright {\ }{\ }{\ }corresponding to a resonance at E\ensuremath{_{\textnormal{c.m.}}}=3.48 MeV above\vspace{0.1cm}}&\\
&&&&&&\parbox[t][0.3cm]{7.9724407cm}{\raggedright {\ }{\ }{\ }the proton threshold in \ensuremath{^{\textnormal{18}}}Ne. The reported excitation\vspace{0.1cm}}&\\
&&&&&&\parbox[t][0.3cm]{7.9724407cm}{\raggedright {\ }{\ }{\ }energy is erroneous because assuming the use of\vspace{0.1cm}}&\\
\end{longtable}
\begin{textblock}{29}(0,27.3)
Continued on next page (footnotes at end of table)
\end{textblock}
\clearpage
\begin{longtable}{ccccccc@{\extracolsep{\fill}}c}
\\[-.4cm]
\multicolumn{8}{c}{{\bf \small \underline{Adopted \hyperlink{18NE_LEVEL}{Levels}, \hyperlink{18NE_GAMMA}{Gammas} (continued)}}}\\
\multicolumn{8}{c}{~}\\
\multicolumn{8}{c}{\underline{\ensuremath{^{18}}Ne Levels (continued)}}\\
\multicolumn{8}{c}{~}\\
\multicolumn{2}{c}{E(level)$^{}$}&J$^{\pi}$$^{}$&\multicolumn{2}{c}{T\ensuremath{_{\textnormal{1/2}}} or \ensuremath{\Gamma}$^{}$}&XREF&Comments&\\[-.2cm]
\multicolumn{2}{c}{\hrulefill}&\hrulefill&\multicolumn{2}{c}{\hrulefill}&\hrulefill&\hrulefill&
\endhead
&&&&&&\parbox[t][0.3cm]{7.79605cm}{\raggedright {\ }{\ }{\ }featureless in this energy region and suffer from poor\vspace{0.1cm}}&\\
&&&&&&\parbox[t][0.3cm]{7.79605cm}{\raggedright {\ }{\ }{\ }statistics. Therefore, a J\ensuremath{^{\ensuremath{\pi}}}=1\ensuremath{^{-}} assignment may not be\vspace{0.1cm}}&\\
&&&&&&\parbox[t][0.3cm]{7.79605cm}{\raggedright {\ }{\ }{\ }improbable from these data (see Fig. 9). This\vspace{0.1cm}}&\\
&&&&&&\parbox[t][0.3cm]{7.79605cm}{\raggedright {\ }{\ }{\ }assignment is also consistent with what was proposed\vspace{0.1cm}}&\\
&&&&&&\parbox[t][0.3cm]{7.79605cm}{\raggedright {\ }{\ }{\ }by the theoretical work of (\href{https://www.nndc.bnl.gov/nsr/nsrlink.jsp?2000Fo19,B}{2000Fo19}). Note that the\vspace{0.1cm}}&\\
&&&&&&\parbox[t][0.3cm]{7.79605cm}{\raggedright {\ }{\ }{\ }results from (\href{https://www.nndc.bnl.gov/nsr/nsrlink.jsp?2002Ha15,B}{2002Ha15}) are not based on an R-matrix\vspace{0.1cm}}&\\
&&&&&&\parbox[t][0.3cm]{7.79605cm}{\raggedright {\ }{\ }{\ }fit. They were deduced by comparison of the excitation\vspace{0.1cm}}&\\
&&&&&&\parbox[t][0.3cm]{7.79605cm}{\raggedright {\ }{\ }{\ }functions with R-matrix calculations for various spin\vspace{0.1cm}}&\\
\end{longtable}
\begin{textblock}{29}(0,27.3)
Continued on next page (footnotes at end of table)
\end{textblock}
\clearpage
\begin{longtable}{cccccc@{\extracolsep{\fill}}c}
\\[-.4cm]
\multicolumn{7}{c}{{\bf \small \underline{Adopted \hyperlink{18NE_LEVEL}{Levels}, \hyperlink{18NE_GAMMA}{Gammas} (continued)}}}\\
\multicolumn{7}{c}{~}\\
\multicolumn{7}{c}{\underline{\ensuremath{^{18}}Ne Levels (continued)}}\\
\multicolumn{7}{c}{~}\\
\multicolumn{2}{c}{E(level)$^{}$}&\multicolumn{2}{c}{T\ensuremath{_{\textnormal{1/2}}} or \ensuremath{\Gamma}$^{}$}&XREF&Comments&\\[-.2cm]
\multicolumn{2}{c}{\hrulefill}&\multicolumn{2}{c}{\hrulefill}&\hrulefill&\hrulefill&
\endhead
&&&&&\parbox[t][0.3cm]{9.361121cm}{\raggedright E(level): Weighted average (with external errors) of 7915 keV \textit{12}\vspace{0.1cm}}&\\
&&&&&\parbox[t][0.3cm]{9.361121cm}{\raggedright {\ }{\ }{\ }(\href{https://www.nndc.bnl.gov/nsr/nsrlink.jsp?1974Ne04,B}{1974Ne04}, \href{https://www.nndc.bnl.gov/nsr/nsrlink.jsp?1981Ne09,B}{1981Ne09}: (\ensuremath{^{\textnormal{3}}}He,n)); 7903 keV \textit{15} (\href{https://www.nndc.bnl.gov/nsr/nsrlink.jsp?1977Ev01,B}{1977Ev01}); and\vspace{0.1cm}}&\\
&&&&&\parbox[t][0.3cm]{9.361121cm}{\raggedright {\ }{\ }{\ }7940 keV \textit{10} (\href{https://www.nndc.bnl.gov/nsr/nsrlink.jsp?1996Ha26,B}{1996Ha26}: (\ensuremath{^{\textnormal{3}}}He,n)). Evaluator notes that most\vspace{0.1cm}}&\\
&&&&&\parbox[t][0.3cm]{9.361121cm}{\raggedright {\ }{\ }{\ }likely there is a close lying doublet in this region. One of these\vspace{0.1cm}}&\\
&&&&&\parbox[t][0.3cm]{9.361121cm}{\raggedright {\ }{\ }{\ }members seems to be populated via the \ensuremath{^{\textnormal{16}}}O(\ensuremath{^{\textnormal{3}}}He,n) reaction,\vspace{0.1cm}}&\\
&&&&&\parbox[t][0.3cm]{9.361121cm}{\raggedright {\ }{\ }{\ }while the other via \ensuremath{^{\textnormal{20}}}Ne(p,t). (\href{https://www.nndc.bnl.gov/nsr/nsrlink.jsp?1996Ha26,B}{1996Ha26}) populated both states\vspace{0.1cm}}&\\
&&&&&\parbox[t][0.3cm]{9.361121cm}{\raggedright {\ }{\ }{\ }using these reactions. The two states were kept separate in the\vspace{0.1cm}}&\\
&&&&&\parbox[t][0.3cm]{9.361121cm}{\raggedright {\ }{\ }{\ }discussions (see section V.E) but were combined into one state\vspace{0.1cm}}&\\
\end{longtable}
\begin{textblock}{29}(0,27.3)
Continued on next page (footnotes at end of table)
\end{textblock}
\clearpage
\begin{longtable}{ccccccc@{\extracolsep{\fill}}c}
\\[-.4cm]
\multicolumn{8}{c}{{\bf \small \underline{Adopted \hyperlink{18NE_LEVEL}{Levels}, \hyperlink{18NE_GAMMA}{Gammas} (continued)}}}\\
\multicolumn{8}{c}{~}\\
\multicolumn{8}{c}{\underline{\ensuremath{^{18}}Ne Levels (continued)}}\\
\multicolumn{8}{c}{~}\\
\multicolumn{2}{c}{E(level)$^{}$}&J$^{\pi}$$^{}$&\multicolumn{2}{c}{T\ensuremath{_{\textnormal{1/2}}} or \ensuremath{\Gamma}$^{}$}&XREF&Comments&\\[-.2cm]
\multicolumn{2}{c}{\hrulefill}&\hrulefill&\multicolumn{2}{c}{\hrulefill}&\hrulefill&\hrulefill&
\endhead
&&&&&&\parbox[t][0.3cm]{7.6826305cm}{\raggedright {\ }{\ }{\ }(\href{https://www.nndc.bnl.gov/nsr/nsrlink.jsp?1974Ne04,B}{1974Ne04}, \href{https://www.nndc.bnl.gov/nsr/nsrlink.jsp?1981Ne09,B}{1981Ne09}: (p,t)), and 75 keV (\href{https://www.nndc.bnl.gov/nsr/nsrlink.jsp?2022Ba39,B}{2022Ba39}:\vspace{0.1cm}}&\\
&&&&&&\parbox[t][0.3cm]{7.6826305cm}{\raggedright {\ }{\ }{\ }from \ensuremath{\Gamma}\ensuremath{_{\ensuremath{\alpha}}}+\ensuremath{\Gamma}\ensuremath{_{\textnormal{p}}} for J\ensuremath{^{\ensuremath{\pi}}}=2\ensuremath{^{\textnormal{+}}}).\vspace{0.1cm}}&\\
\multicolumn{1}{r@{}}{8098}&\multicolumn{1}{@{ }l}{{\it 9}}&&\multicolumn{1}{r@{}}{$\leq$30}&\multicolumn{1}{@{ }l}{keV}&\multicolumn{1}{l}{\texttt{\ \ \ \ \ \ \hyperlink{NE7}{G}\ \ \ \ \ \hyperlink{NE13}{M}\ \hyperlink{NE15}{O}\ \ \ \ \ \ \ \ \ \ \ } }&\parbox[t][0.3cm]{7.6826305cm}{\raggedright \ensuremath{\Gamma}\ensuremath{\alpha}/\ensuremath{\Gamma}=21 \textit{13} (\href{https://www.nndc.bnl.gov/nsr/nsrlink.jsp?2012Al11,B}{2012Al11})\vspace{0.1cm}}&\\
&&&&&&\parbox[t][0.3cm]{7.6826305cm}{\raggedright \ensuremath{\Gamma}\ensuremath{_{\textnormal{p0}}}/\ensuremath{\Gamma}=65 \textit{12} (\href{https://www.nndc.bnl.gov/nsr/nsrlink.jsp?2012Al11,B}{2012Al11})\vspace{0.1cm}}&\\
&&&&&&\parbox[t][0.3cm]{7.6826305cm}{\raggedright \ensuremath{\Gamma}\ensuremath{_{\textnormal{p1}}}/\ensuremath{\Gamma}=14 \textit{12} (\href{https://www.nndc.bnl.gov/nsr/nsrlink.jsp?2012Al11,B}{2012Al11})\vspace{0.1cm}}&\\
&&&&&&\parbox[t][0.3cm]{7.6826305cm}{\raggedright \ensuremath{\Gamma}\ensuremath{_{\textnormal{p1}}}/\ensuremath{\Gamma}\ensuremath{_{\textnormal{p0}}}=0.22 (\href{https://www.nndc.bnl.gov/nsr/nsrlink.jsp?2012Al11,B}{2012Al11})\vspace{0.1cm}}&\\
&&&&&&\parbox[t][0.3cm]{7.6826305cm}{\raggedright \ensuremath{\Gamma}\ensuremath{\alpha}/\ensuremath{\Gamma}\ensuremath{_{\textnormal{p}}}=0.32 (\href{https://www.nndc.bnl.gov/nsr/nsrlink.jsp?2012Al11,B}{2012Al11})\vspace{0.1cm}}&\\
\end{longtable}
\begin{textblock}{29}(0,27.3)
Continued on next page (footnotes at end of table)
\end{textblock}
\clearpage
\begin{longtable}{ccccccc@{\extracolsep{\fill}}c}
\\[-.4cm]
\multicolumn{8}{c}{{\bf \small \underline{Adopted \hyperlink{18NE_LEVEL}{Levels}, \hyperlink{18NE_GAMMA}{Gammas} (continued)}}}\\
\multicolumn{8}{c}{~}\\
\multicolumn{8}{c}{\underline{\ensuremath{^{18}}Ne Levels (continued)}}\\
\multicolumn{8}{c}{~}\\
\multicolumn{2}{c}{E(level)$^{}$}&J$^{\pi}$$^{}$&\multicolumn{2}{c}{T\ensuremath{_{\textnormal{1/2}}} or \ensuremath{\Gamma}$^{}$}&XREF&Comments&\\[-.2cm]
\multicolumn{2}{c}{\hrulefill}&\hrulefill&\multicolumn{2}{c}{\hrulefill}&\hrulefill&\hrulefill&
\endhead
&&&&&&\parbox[t][0.3cm]{8.254011cm}{\raggedright J\ensuremath{^{\pi}}: From the R-matrix analyses of (1) (\href{https://www.nndc.bnl.gov/nsr/nsrlink.jsp?2008Fu07,B}{2008Fu07}): deduced\vspace{0.1cm}}&\\
&&&&&&\parbox[t][0.3cm]{8.254011cm}{\raggedright {\ }{\ }{\ }J\ensuremath{^{\ensuremath{\pi}}}=(1\ensuremath{^{-}}, 0\ensuremath{^{\textnormal{+}}}) for the 8.7-MeV state by taking into account\vspace{0.1cm}}&\\
&&&&&&\parbox[t][0.3cm]{8.254011cm}{\raggedright {\ }{\ }{\ }the data of (\href{https://www.nndc.bnl.gov/nsr/nsrlink.jsp?2007Bu01,B}{2007Bu01}). (\href{https://www.nndc.bnl.gov/nsr/nsrlink.jsp?2008Fu07,B}{2008Fu07}) preferred the J\ensuremath{^{\ensuremath{\pi}}}=1\ensuremath{^{-}}\vspace{0.1cm}}&\\
&&&&&&\parbox[t][0.3cm]{8.254011cm}{\raggedright {\ }{\ }{\ }assignment. Inclusion of J\ensuremath{^{\ensuremath{\pi}}}=0\ensuremath{^{\textnormal{+}}} could not be excluded. (2)\vspace{0.1cm}}&\\
&&&&&&\parbox[t][0.3cm]{8.254011cm}{\raggedright {\ }{\ }{\ }(\href{https://www.nndc.bnl.gov/nsr/nsrlink.jsp?2010Ha15,B}{2010Ha15}): deduced J\ensuremath{^{\ensuremath{\pi}}}=(1\ensuremath{^{-}}). Fitted the data using\vspace{0.1cm}}&\\
&&&&&&\parbox[t][0.3cm]{8.254011cm}{\raggedright {\ }{\ }{\ }J\ensuremath{^{\ensuremath{\pi}}}=0\ensuremath{^{\textnormal{+}}}, which caused a large energy shift for the strong\vspace{0.1cm}}&\\
&&&&&&\parbox[t][0.3cm]{8.254011cm}{\raggedright {\ }{\ }{\ }resonance populated at 9.2 MeV. Therefore, (\href{https://www.nndc.bnl.gov/nsr/nsrlink.jsp?2010Ha15,B}{2010Ha15})\vspace{0.1cm}}&\\
&&&&&&\parbox[t][0.3cm]{8.254011cm}{\raggedright {\ }{\ }{\ }suggested that the J\ensuremath{^{\ensuremath{\pi}}}=1\ensuremath{^{-}} assignment is more probable and\vspace{0.1cm}}&\\
\end{longtable}
\begin{textblock}{29}(0,27.3)
Continued on next page (footnotes at end of table)
\end{textblock}
\clearpage
\begin{longtable}{ccccccc@{\extracolsep{\fill}}c}
\\[-.4cm]
\multicolumn{8}{c}{{\bf \small \underline{Adopted \hyperlink{18NE_LEVEL}{Levels}, \hyperlink{18NE_GAMMA}{Gammas} (continued)}}}\\
\multicolumn{8}{c}{~}\\
\multicolumn{8}{c}{\underline{\ensuremath{^{18}}Ne Levels (continued)}}\\
\multicolumn{8}{c}{~}\\
\multicolumn{2}{c}{E(level)$^{}$}&J$^{\pi}$$^{}$&\multicolumn{2}{c}{T\ensuremath{_{\textnormal{1/2}}} or \ensuremath{\Gamma}$^{}$}&XREF&Comments&\\[-.2cm]
\multicolumn{2}{c}{\hrulefill}&\hrulefill&\multicolumn{2}{c}{\hrulefill}&\hrulefill&\hrulefill&
\endhead
&&&&&&\parbox[t][0.3cm]{7.5102005cm}{\raggedright {\ }{\ }{\ }strong \ensuremath{\alpha}-cluster structure (\href{https://www.nndc.bnl.gov/nsr/nsrlink.jsp?2019Ch16,B}{2019Ch16}), this level\vspace{0.1cm}}&\\
&&&&&&\parbox[t][0.3cm]{7.5102005cm}{\raggedright {\ }{\ }{\ }presents a strong \ensuremath{\alpha}-cluster structure as reported by\vspace{0.1cm}}&\\
&&&&&&\parbox[t][0.3cm]{7.5102005cm}{\raggedright {\ }{\ }{\ }(\href{https://www.nndc.bnl.gov/nsr/nsrlink.jsp?2022Ba39,B}{2022Ba39}).\vspace{0.1cm}}&\\
\multicolumn{1}{r@{}}{9197}&\multicolumn{1}{@{}l}{\ensuremath{^{{\hyperlink{NE0LEVEL1}{b}}{\hyperlink{NE0LEVEL2}{c}}}} {\it 8}}&\multicolumn{1}{l}{(2\ensuremath{^{+}},3\ensuremath{^{-}})}&\multicolumn{1}{r@{}}{490}&\multicolumn{1}{@{ }l}{keV {\it 21}}&\multicolumn{1}{l}{\texttt{\ \ \ \ \ \ \hyperlink{NE7}{G}\hyperlink{NE8}{H}\ \ \ \ \hyperlink{NE13}{M}\ \ \ \ \ \ \ \ \hyperlink{NE22}{V}\ \ \ \ } }&\parbox[t][0.3cm]{7.5102005cm}{\raggedright XREF: G(9.14E3)M(9.18E3)\vspace{0.1cm}}&\\
&&&&&&\parbox[t][0.3cm]{7.5102005cm}{\raggedright E(level): Weighted average of 9170 keV \textit{30}\vspace{0.1cm}}&\\
&&&&&&\parbox[t][0.3cm]{7.5102005cm}{\raggedright {\ }{\ }{\ }(\href{https://www.nndc.bnl.gov/nsr/nsrlink.jsp?1970Fa17,B}{1970Fa17}); 9215 keV \textit{20} (\href{https://www.nndc.bnl.gov/nsr/nsrlink.jsp?1972Pa02,B}{1972Pa02}); 9198 keV \textit{10}\vspace{0.1cm}}&\\
&&&&&&\parbox[t][0.3cm]{7.5102005cm}{\raggedright {\ }{\ }{\ }(\href{https://www.nndc.bnl.gov/nsr/nsrlink.jsp?1974Ne04,B}{1974Ne04}, \href{https://www.nndc.bnl.gov/nsr/nsrlink.jsp?1981Ne09,B}{1981Ne09}: (p,t)); 9180 keV \textit{20}\vspace{0.1cm}}&\\
&&&&&&\parbox[t][0.3cm]{7.5102005cm}{\raggedright {\ }{\ }{\ }(\href{https://www.nndc.bnl.gov/nsr/nsrlink.jsp?1996Ha26,B}{1996Ha26}: \ensuremath{^{\textnormal{12}}}C(\ensuremath{^{\textnormal{12}}}C,\ensuremath{^{\textnormal{6}}}He)); 9.22 MeV \textit{4}\vspace{0.1cm}}&\\
\end{longtable}
\begin{textblock}{29}(0,27.3)
Continued on next page (footnotes at end of table)
\end{textblock}
\clearpage
\begin{longtable}{ccccccc@{\extracolsep{\fill}}c}
\\[-.4cm]
\multicolumn{8}{c}{{\bf \small \underline{Adopted \hyperlink{18NE_LEVEL}{Levels}, \hyperlink{18NE_GAMMA}{Gammas} (continued)}}}\\
\multicolumn{8}{c}{~}\\
\multicolumn{8}{c}{\underline{\ensuremath{^{18}}Ne Levels (continued)}}\\
\multicolumn{8}{c}{~}\\
\multicolumn{2}{c}{E(level)$^{}$}&J$^{\pi}$$^{}$&\multicolumn{2}{c}{T\ensuremath{_{\textnormal{1/2}}} or \ensuremath{\Gamma}$^{}$}&XREF&Comments&\\[-.2cm]
\multicolumn{2}{c}{\hrulefill}&\hrulefill&\multicolumn{2}{c}{\hrulefill}&\hrulefill&\hrulefill&
\endhead
&&&&&&\parbox[t][0.3cm]{8.21426cm}{\raggedright {\ }{\ }{\ }what was observed by (\href{https://www.nndc.bnl.gov/nsr/nsrlink.jsp?1974Ne04,B}{1974Ne04}, \href{https://www.nndc.bnl.gov/nsr/nsrlink.jsp?1981Ne09,B}{1981Ne09}, \href{https://www.nndc.bnl.gov/nsr/nsrlink.jsp?1970Fa17,B}{1970Fa17},\vspace{0.1cm}}&\\
&&&&&&\parbox[t][0.3cm]{8.21426cm}{\raggedright {\ }{\ }{\ }\href{https://www.nndc.bnl.gov/nsr/nsrlink.jsp?1972Pa02,B}{1972Pa02}, \href{https://www.nndc.bnl.gov/nsr/nsrlink.jsp?1996Ha26,B}{1996Ha26}) measurements.\vspace{0.1cm}}&\\
&&&&&&\parbox[t][0.3cm]{8.21426cm}{\raggedright J\ensuremath{^{\pi}}: From the R-matrix analyses of (\href{https://www.nndc.bnl.gov/nsr/nsrlink.jsp?2008Fu07,B}{2008Fu07}, \href{https://www.nndc.bnl.gov/nsr/nsrlink.jsp?2010Ha15,B}{2010Ha15},\vspace{0.1cm}}&\\
&&&&&&\parbox[t][0.3cm]{8.21426cm}{\raggedright {\ }{\ }{\ }\href{https://www.nndc.bnl.gov/nsr/nsrlink.jsp?2015Ki07,B}{2015Ki07}: J\ensuremath{^{\ensuremath{\pi}}}=3\ensuremath{^{-}}); and (\href{https://www.nndc.bnl.gov/nsr/nsrlink.jsp?2022Ba39,B}{2022Ba39}: J\ensuremath{^{\ensuremath{\pi}}}=2\ensuremath{^{\textnormal{+}}}). The\vspace{0.1cm}}&\\
&&&&&&\parbox[t][0.3cm]{8.21426cm}{\raggedright {\ }{\ }{\ }R-matrix fits of (\href{https://www.nndc.bnl.gov/nsr/nsrlink.jsp?2022Ba39,B}{2022Ba39}) seems to be poorer than\vspace{0.1cm}}&\\
&&&&&&\parbox[t][0.3cm]{8.21426cm}{\raggedright {\ }{\ }{\ }those of (\href{https://www.nndc.bnl.gov/nsr/nsrlink.jsp?2008Fu07,B}{2008Fu07}, \href{https://www.nndc.bnl.gov/nsr/nsrlink.jsp?2010Ha15,B}{2010Ha15}, \href{https://www.nndc.bnl.gov/nsr/nsrlink.jsp?2015Ki07,B}{2015Ki07}). Also note that\vspace{0.1cm}}&\\
&&&&&&\parbox[t][0.3cm]{8.21426cm}{\raggedright {\ }{\ }{\ }in the R-matrix analysis of (\href{https://www.nndc.bnl.gov/nsr/nsrlink.jsp?2015Ki07,B}{2015Ki07}), J\ensuremath{^{\ensuremath{\pi}}}=3\ensuremath{^{-}} was fixed\vspace{0.1cm}}&\\
\end{longtable}
\begin{textblock}{29}(0,27.3)
Continued on next page (footnotes at end of table)
\end{textblock}
\clearpage
\begin{longtable}{ccccccc@{\extracolsep{\fill}}c}
\\[-.4cm]
\multicolumn{8}{c}{{\bf \small \underline{Adopted \hyperlink{18NE_LEVEL}{Levels}, \hyperlink{18NE_GAMMA}{Gammas} (continued)}}}\\
\multicolumn{8}{c}{~}\\
\multicolumn{8}{c}{\underline{\ensuremath{^{18}}Ne Levels (continued)}}\\
\multicolumn{8}{c}{~}\\
\multicolumn{2}{c}{E(level)$^{}$}&J$^{\pi}$$^{}$&\multicolumn{2}{c}{T\ensuremath{_{\textnormal{1/2}}} or \ensuremath{\Gamma}$^{}$}&XREF&Comments&\\[-.2cm]
\multicolumn{2}{c}{\hrulefill}&\hrulefill&\multicolumn{2}{c}{\hrulefill}&\hrulefill&\hrulefill&
\endhead
\multicolumn{1}{r@{}}{10.07\ensuremath{\times10^{3}}}&\multicolumn{1}{@{ }l}{{\it 4}}&\multicolumn{1}{l}{(1\ensuremath{^{-}},2\ensuremath{^{+}})}&\multicolumn{1}{r@{}}{302}&\multicolumn{1}{@{ }l}{keV {\it 32}}&\multicolumn{1}{l}{\texttt{\ \ \ \ \ \ \hyperlink{NE7}{G}\hyperlink{NE8}{H}\ \ \ \ \ \ \ \ \ \ \ \ \ \ \ \ \ \ } }&\parbox[t][0.3cm]{6.98804cm}{\raggedright \ensuremath{\Gamma}\ensuremath{_{\textnormal{p}}}=94 keV \textit{10}\vspace{0.1cm}}&\\
&&&&&&\parbox[t][0.3cm]{6.98804cm}{\raggedright \ensuremath{\Gamma}\ensuremath{\alpha}=207 keV \textit{30}\vspace{0.1cm}}&\\
&&&&&&\parbox[t][0.3cm]{6.98804cm}{\raggedright E(level): Weighted average of 10.07 MeV \textit{4}\vspace{0.1cm}}&\\
&&&&&&\parbox[t][0.3cm]{6.98804cm}{\raggedright {\ }{\ }{\ }(\href{https://www.nndc.bnl.gov/nsr/nsrlink.jsp?2010Ha15,B}{2010Ha15}: from \ensuremath{^{\textnormal{4}}}He(\ensuremath{^{\textnormal{14}}}O,\ensuremath{\alpha}):res E\ensuremath{_{\textnormal{c.m.}}}=4.95\vspace{0.1cm}}&\\
&&&&&&\parbox[t][0.3cm]{6.98804cm}{\raggedright {\ }{\ }{\ }MeV \textit{4}); and 10.07 MeV \textit{8} (\href{https://www.nndc.bnl.gov/nsr/nsrlink.jsp?2015Ki07,B}{2015Ki07}). See also\vspace{0.1cm}}&\\
&&&&&&\parbox[t][0.3cm]{6.98804cm}{\raggedright {\ }{\ }{\ }10.12 MeV (\href{https://www.nndc.bnl.gov/nsr/nsrlink.jsp?2007Fu09,B}{2007Fu09}, \href{https://www.nndc.bnl.gov/nsr/nsrlink.jsp?2008FuZZ,B}{2008FuZZ}); and 10.1\vspace{0.1cm}}&\\
&&&&&&\parbox[t][0.3cm]{6.98804cm}{\raggedright {\ }{\ }{\ }MeV (\href{https://www.nndc.bnl.gov/nsr/nsrlink.jsp?2008Fu07,B}{2008Fu07}: associated with E\ensuremath{_{\textnormal{c.m.}}}=5.0\vspace{0.1cm}}&\\
&&&&&&\parbox[t][0.3cm]{6.98804cm}{\raggedright {\ }{\ }{\ }MeV from \ensuremath{^{\textnormal{4}}}He(\ensuremath{^{\textnormal{14}}}O,\ensuremath{\alpha}):res).\vspace{0.1cm}}&\\
\end{longtable}%
\parbox[b][0.3cm]{17.7cm}{\makebox[1ex]{\ensuremath{^{\hypertarget{NE0LEVEL0}{a}}}} The data for the states located near the low- or high-energy edge of the excitation function measured by (\href{https://www.nndc.bnl.gov/nsr/nsrlink.jsp?2022Ba39,B}{2022Ba39}) are}\\
\parbox[b][0.3cm]{17.7cm}{{\ }{\ }incomplete. Also, due to the high thresholds in some detectors used in the experiment of (\href{https://www.nndc.bnl.gov/nsr/nsrlink.jsp?2022Ba39,B}{2022Ba39}), the angular distributions in}\\
\parbox[b][0.3cm]{17.7cm}{{\ }{\ }the excitation energy range from 7 to 8 MeV are incomplete. Therefore, (\href{https://www.nndc.bnl.gov/nsr/nsrlink.jsp?2022Ba39,B}{2022Ba39}) report these states in parentheses.}\\
\parbox[b][0.3cm]{17.7cm}{\makebox[1ex]{\ensuremath{^{\hypertarget{NE0LEVEL1}{b}}}} This level has a pronounced \ensuremath{\alpha}-cluster structure reported by (\href{https://www.nndc.bnl.gov/nsr/nsrlink.jsp?2022Ba39,B}{2022Ba39}), which is demonstrated with the large (\ensuremath{>}0.1)}\\
\parbox[b][0.3cm]{17.7cm}{{\ }{\ }dimensionless reduced width (\ensuremath{\theta}\ensuremath{_{\ensuremath{\alpha}}^{\textnormal{2}}}).}\\
\parbox[b][0.3cm]{17.7cm}{\makebox[1ex]{\ensuremath{^{\hypertarget{NE0LEVEL2}{c}}}} All the evidence presented here suggests that this state could be a superposition of more than one states.}\\
\parbox[b][0.3cm]{17.7cm}{\makebox[1ex]{\ensuremath{^{\hypertarget{NE0LEVEL3}{d}}}} For a direct, one-step transfer via the (\ensuremath{^{\textnormal{3}}}He,n) or (p,t) reaction, in which compound nuclear processes can be neglected, on a}\\
\parbox[b][0.3cm]{17.7cm}{{\ }{\ }target nucleus whose J\ensuremath{^{\ensuremath{\pi}}} assignment is 0\ensuremath{^{\textnormal{+}}} (such as \ensuremath{^{\textnormal{16}}}O or \ensuremath{^{\textnormal{20}}}Ne), the double stripping selection rules require J=L and \ensuremath{\pi}=({\textminus}1)\ensuremath{^{\textnormal{L}}}.}\\
\parbox[b][0.3cm]{17.7cm}{{\ }{\ }Therefore, the identification of L determines the spin and parity of the final state, particularly when the final state is strongly}\\
\parbox[b][0.3cm]{17.7cm}{{\ }{\ }populated indicating the transfer of a diproton or a dineutron in a relative \textit{s}=0 state.}\\
\vspace{0.5cm}
\clearpage
\vspace{0.3cm}
\vspace*{-0.5cm}
{\bf \small \underline{Adopted \hyperlink{18NE_LEVEL}{Levels}, \hyperlink{18NE_GAMMA}{Gammas} (continued)}}\\
\vspace{0.3cm}
\hypertarget{18NE_GAMMA}{\underline{$\gamma$($^{18}$Ne)}}\\
\begin{longtable}{ccccccccc@{}ccccc@{\extracolsep{\fill}}c}
\multicolumn{2}{c}{E\ensuremath{_{i}}(level)}&J\ensuremath{^{\pi}_{i}}&\multicolumn{2}{c}{E\ensuremath{_{\gamma}}}&\multicolumn{2}{c}{I\ensuremath{_{\gamma}}\ensuremath{^{\hyperlink{NE0GAMMA0}{a}}}}&\multicolumn{2}{c}{E\ensuremath{_{f}}}&J\ensuremath{^{\pi}_{f}}&Mult.&\multicolumn{2}{c}{\ensuremath{\delta}}&Comments&\\[-.2cm]
\multicolumn{2}{c}{\hrulefill}&\hrulefill&\multicolumn{2}{c}{\hrulefill}&\multicolumn{2}{c}{\hrulefill}&\multicolumn{2}{c}{\hrulefill}&\hrulefill&\hrulefill&\multicolumn{2}{c}{\hrulefill}&\hrulefill&
\endfirsthead
\multicolumn{1}{r@{}}{1887}&\multicolumn{1}{@{.}l}{4}&\multicolumn{1}{l}{2\ensuremath{^{+}}}&\multicolumn{1}{r@{}}{1887}&\multicolumn{1}{@{.}l}{3 {\it 2}}&\multicolumn{1}{r@{}}{100}&\multicolumn{1}{@{}l}{}&\multicolumn{1}{r@{}}{0}&\multicolumn{1}{@{}l}{}&\multicolumn{1}{@{}l}{0\ensuremath{^{+}}}&\multicolumn{1}{l}{E2}&&&\parbox[t][0.3cm]{7.998101cm}{\raggedright B(E2)(W.u.)=18.0 \textit{+17{\textminus}14}\vspace{0.1cm}}&\\
&&&&&&&&&&&&&\parbox[t][0.3cm]{7.998101cm}{\raggedright E\ensuremath{_{\gamma}}: From (\href{https://www.nndc.bnl.gov/nsr/nsrlink.jsp?1968Gi09,B}{1968Gi09}, \href{https://www.nndc.bnl.gov/nsr/nsrlink.jsp?1969Ro08,B}{1969Ro08}).\vspace{0.1cm}}&\\
&&&&&&&&&&&&&\parbox[t][0.3cm]{7.998101cm}{\raggedright I\ensuremath{_{\gamma}}: From (\href{https://www.nndc.bnl.gov/nsr/nsrlink.jsp?1968Gi09,B}{1968Gi09}, \href{https://www.nndc.bnl.gov/nsr/nsrlink.jsp?1969Ro08,B}{1969Ro08}, \href{https://www.nndc.bnl.gov/nsr/nsrlink.jsp?1969Ro22,B}{1969Ro22}).\vspace{0.1cm}}&\\
&&&&&&&&&&&&&\parbox[t][0.3cm]{7.998101cm}{\raggedright Mult.: From (\href{https://www.nndc.bnl.gov/nsr/nsrlink.jsp?1974Mc17,B}{1974Mc17}).\vspace{0.1cm}}&\\
&&&&&&&&&&&&&\parbox[t][0.3cm]{7.998101cm}{\raggedright \ensuremath{\delta}=0 (mixing ratio): from (\href{https://www.nndc.bnl.gov/nsr/nsrlink.jsp?1969Ro08,B}{1969Ro08}).\vspace{0.1cm}}&\\
\multicolumn{1}{r@{}}{3376}&\multicolumn{1}{@{.}l}{4}&\multicolumn{1}{l}{4\ensuremath{^{+}}}&\multicolumn{1}{r@{}}{1488}&\multicolumn{1}{@{.}l}{9 {\it 3}}&\multicolumn{1}{r@{}}{100}&\multicolumn{1}{@{}l}{}&\multicolumn{1}{r@{}}{1887}&\multicolumn{1}{@{.}l}{4 }&\multicolumn{1}{@{}l}{2\ensuremath{^{+}}}&\multicolumn{1}{l}{E2}&&&\parbox[t][0.3cm]{7.998101cm}{\raggedright B(E2)(W.u.)=9.1 \textit{+14{\textminus}11}\vspace{0.1cm}}&\\
&&&&&&&&&&&&&\parbox[t][0.3cm]{7.998101cm}{\raggedright E\ensuremath{_{\gamma}}: From (\href{https://www.nndc.bnl.gov/nsr/nsrlink.jsp?1968Gi09,B}{1968Gi09}, \href{https://www.nndc.bnl.gov/nsr/nsrlink.jsp?1969Ro08,B}{1969Ro08}).\vspace{0.1cm}}&\\
&&&&&&&&&&&&&\parbox[t][0.3cm]{7.998101cm}{\raggedright Mult.: From (\href{https://www.nndc.bnl.gov/nsr/nsrlink.jsp?1972Gi01,B}{1972Gi01}). See also E2/M3 (\href{https://www.nndc.bnl.gov/nsr/nsrlink.jsp?1969Ro08,B}{1969Ro08},\vspace{0.1cm}}&\\
&&&&&&&&&&&&&\parbox[t][0.3cm]{7.998101cm}{\raggedright {\ }{\ }{\ }\href{https://www.nndc.bnl.gov/nsr/nsrlink.jsp?1969Ro22,B}{1969Ro22}, \href{https://www.nndc.bnl.gov/nsr/nsrlink.jsp?1970Sh04,B}{1970Sh04}) with a mixing ratio of \ensuremath{\delta}=0.04 \textit{3}\vspace{0.1cm}}&\\
&&&&&&&&&&&&&\parbox[t][0.3cm]{7.998101cm}{\raggedright {\ }{\ }{\ }from the weighted average of +0.06 \textit{7} (\href{https://www.nndc.bnl.gov/nsr/nsrlink.jsp?1969Ro08,B}{1969Ro08}: see\vspace{0.1cm}}&\\
&&&&&&&&&&&&&\parbox[t][0.3cm]{7.998101cm}{\raggedright {\ }{\ }{\ }Table 2 assuming J=4); +0.00 \textit{4} (\href{https://www.nndc.bnl.gov/nsr/nsrlink.jsp?1969Ro22,B}{1969Ro22}: see Fig. 3\vspace{0.1cm}}&\\
&&&&&&&&&&&&&\parbox[t][0.3cm]{7.998101cm}{\raggedright {\ }{\ }{\ }assuming J=4); and +0.12 \textit{7} (\href{https://www.nndc.bnl.gov/nsr/nsrlink.jsp?1970Sh04,B}{1970Sh04}: see Fig. 1).\vspace{0.1cm}}&\\
&&&&&&&&&&&&&\parbox[t][0.3cm]{7.998101cm}{\raggedright {\ }{\ }{\ }Note that mixing ratios deduced by (\href{https://www.nndc.bnl.gov/nsr/nsrlink.jsp?1970Sh04,B}{1970Sh04}) and\vspace{0.1cm}}&\\
&&&&&&&&&&&&&\parbox[t][0.3cm]{7.998101cm}{\raggedright {\ }{\ }{\ }(\href{https://www.nndc.bnl.gov/nsr/nsrlink.jsp?1972Gi01,B}{1972Gi01}) are based on the phase convention of\vspace{0.1cm}}&\\
&&&&&&&&&&&&&\parbox[t][0.3cm]{7.998101cm}{\raggedright {\ }{\ }{\ }(\href{https://www.nndc.bnl.gov/nsr/nsrlink.jsp?1967Ro21,B}{1967Ro21}).\vspace{0.1cm}}&\\
&&&&&&&&&&&&&\parbox[t][0.3cm]{7.998101cm}{\raggedright I\ensuremath{_{\gamma}}: From (\href{https://www.nndc.bnl.gov/nsr/nsrlink.jsp?1968Gi09,B}{1968Gi09}, \href{https://www.nndc.bnl.gov/nsr/nsrlink.jsp?1969Ro08,B}{1969Ro08}, \href{https://www.nndc.bnl.gov/nsr/nsrlink.jsp?1969Ro22,B}{1969Ro22}: see Fig. 3,\vspace{0.1cm}}&\\
&&&&&&&&&&&&&\parbox[t][0.3cm]{7.998101cm}{\raggedright {\ }{\ }{\ }\href{https://www.nndc.bnl.gov/nsr/nsrlink.jsp?1970Sh04,B}{1970Sh04}: see Fig. 1).\vspace{0.1cm}}&\\
&&&&&&&&&&&&&\parbox[t][0.3cm]{7.998101cm}{\raggedright (\href{https://www.nndc.bnl.gov/nsr/nsrlink.jsp?1970Sh04,B}{1970Sh04}) found that at \ensuremath{\theta}\ensuremath{_{\ensuremath{\gamma}\textnormal{,lab}}}=90\ensuremath{^\circ} the ratio of the\vspace{0.1cm}}&\\
&&&&&&&&&&&&&\parbox[t][0.3cm]{7.998101cm}{\raggedright {\ }{\ }{\ }yield of the 1.87-MeV to the 1.49-MeV \ensuremath{\gamma} rays is 1.01\vspace{0.1cm}}&\\
&&&&&&&&&&&&&\parbox[t][0.3cm]{7.998101cm}{\raggedright {\ }{\ }{\ }\textit{4}. This is consistent with J\ensuremath{^{\ensuremath{\pi}}}=4\ensuremath{^{\textnormal{+}}} assignment for the\vspace{0.1cm}}&\\
&&&&&&&&&&&&&\parbox[t][0.3cm]{7.998101cm}{\raggedright {\ }{\ }{\ }3376.4-keV level because both of the \ensuremath{\gamma}-ray transitions\vspace{0.1cm}}&\\
&&&&&&&&&&&&&\parbox[t][0.3cm]{7.998101cm}{\raggedright {\ }{\ }{\ }involved in the 4\ensuremath{^{\textnormal{+}}_{\textnormal{1}}}\ensuremath{\rightarrow}2\ensuremath{^{\textnormal{+}}_{\textnormal{1}}}\ensuremath{\rightarrow}0\ensuremath{^{\textnormal{+}}_{\textnormal{1}}} cascade have the same\vspace{0.1cm}}&\\
&&&&&&&&&&&&&\parbox[t][0.3cm]{7.998101cm}{\raggedright {\ }{\ }{\ }angular correlation with respect to the corresponding\vspace{0.1cm}}&\\
&&&&&&&&&&&&&\parbox[t][0.3cm]{7.998101cm}{\raggedright {\ }{\ }{\ }neutron groups independent of the reaction mechanism.\vspace{0.1cm}}&\\
&&&&&&&&&&&&&\parbox[t][0.3cm]{7.998101cm}{\raggedright For the unobserved decay branch from \ensuremath{^{\textnormal{18}}}Ne*(3376\vspace{0.1cm}}&\\
&&&&&&&&&&&&&\parbox[t][0.3cm]{7.998101cm}{\raggedright {\ }{\ }{\ }keV)\ensuremath{\rightarrow}\ensuremath{^{\textnormal{18}}}Ne\ensuremath{_{\textnormal{g.s.}}}: I\ensuremath{_{\ensuremath{\gamma}}}\ensuremath{<}1\% (\href{https://www.nndc.bnl.gov/nsr/nsrlink.jsp?1969Ro22,B}{1969Ro22}: see Fig. 3); I\ensuremath{_{\ensuremath{\gamma}}}\ensuremath{<}1\%\vspace{0.1cm}}&\\
&&&&&&&&&&&&&\parbox[t][0.3cm]{7.998101cm}{\raggedright {\ }{\ }{\ }(\href{https://www.nndc.bnl.gov/nsr/nsrlink.jsp?1970Sh04,B}{1970Sh04}: see Fig. 1); and I\ensuremath{_{\ensuremath{\gamma}}}\ensuremath{<}4\% (\href{https://www.nndc.bnl.gov/nsr/nsrlink.jsp?1968Gi09,B}{1968Gi09},\vspace{0.1cm}}&\\
&&&&&&&&&&&&&\parbox[t][0.3cm]{7.998101cm}{\raggedright {\ }{\ }{\ }\href{https://www.nndc.bnl.gov/nsr/nsrlink.jsp?1969Ro08,B}{1969Ro08}). Therefore, we adopt I\ensuremath{_{\ensuremath{\gamma}}}\ensuremath{<}1\% for this branch.\vspace{0.1cm}}&\\
\multicolumn{1}{r@{}}{3576}&\multicolumn{1}{@{.}l}{3}&\multicolumn{1}{l}{0\ensuremath{^{+}}}&\multicolumn{1}{r@{}}{1689}&\multicolumn{1}{@{ }l}{{\it 2}}&\multicolumn{1}{r@{}}{100}&\multicolumn{1}{@{}l}{}&\multicolumn{1}{r@{}}{1887}&\multicolumn{1}{@{.}l}{4 }&\multicolumn{1}{@{}l}{2\ensuremath{^{+}}}&\multicolumn{1}{l}{E2}&&&\parbox[t][0.3cm]{7.998101cm}{\raggedright B(E2)(W.u.)=5.3 \textit{+44{\textminus}18}\vspace{0.1cm}}&\\
&&&&&&&&&&&&&\parbox[t][0.3cm]{7.998101cm}{\raggedright E\ensuremath{_{\gamma}}: From (\href{https://www.nndc.bnl.gov/nsr/nsrlink.jsp?1968Gi09,B}{1968Gi09}, \href{https://www.nndc.bnl.gov/nsr/nsrlink.jsp?1969Ro08,B}{1969Ro08}).\vspace{0.1cm}}&\\
&&&&&&&&&&&&&\parbox[t][0.3cm]{7.998101cm}{\raggedright I\ensuremath{_{\gamma}}: From (\href{https://www.nndc.bnl.gov/nsr/nsrlink.jsp?1968Gi09,B}{1968Gi09}, \href{https://www.nndc.bnl.gov/nsr/nsrlink.jsp?1969Ro08,B}{1969Ro08}, \href{https://www.nndc.bnl.gov/nsr/nsrlink.jsp?1969Ro22,B}{1969Ro22}: see Fig. 3).\vspace{0.1cm}}&\\
&&&&&&&&&&&&&\parbox[t][0.3cm]{7.998101cm}{\raggedright Mult.: From (\href{https://www.nndc.bnl.gov/nsr/nsrlink.jsp?1972Gi01,B}{1972Gi01}).\vspace{0.1cm}}&\\
&&&&&&&&&&&&&\parbox[t][0.3cm]{7.998101cm}{\raggedright The \ensuremath{^{\textnormal{18}}}Ne*(3576.3 keV)\ensuremath{\rightarrow}\ensuremath{^{\textnormal{18}}}Ne\ensuremath{_{\textnormal{g.s.}}} decay branch remains\vspace{0.1cm}}&\\
&&&&&&&&&&&&&\parbox[t][0.3cm]{7.998101cm}{\raggedright {\ }{\ }{\ }unobserved. The branching ratio is estimated to be\vspace{0.1cm}}&\\
&&&&&&&&&&&&&\parbox[t][0.3cm]{7.998101cm}{\raggedright {\ }{\ }{\ }I\ensuremath{_{\ensuremath{\gamma}}}\ensuremath{<}5\% (\href{https://www.nndc.bnl.gov/nsr/nsrlink.jsp?1969Ro22,B}{1969Ro22}: see Fig. 3); and I\ensuremath{_{\ensuremath{\gamma}}}\ensuremath{<}17\%\vspace{0.1cm}}&\\
&&&&&&&&&&&&&\parbox[t][0.3cm]{7.998101cm}{\raggedright {\ }{\ }{\ }(\href{https://www.nndc.bnl.gov/nsr/nsrlink.jsp?1968Gi09,B}{1968Gi09}, \href{https://www.nndc.bnl.gov/nsr/nsrlink.jsp?1969Ro08,B}{1969Ro08}). We, therefore, adopt I\ensuremath{_{\ensuremath{\gamma}}}\ensuremath{<}5\% for\vspace{0.1cm}}&\\
&&&&&&&&&&&&&\parbox[t][0.3cm]{7.998101cm}{\raggedright {\ }{\ }{\ }this branch.\vspace{0.1cm}}&\\
\multicolumn{1}{r@{}}{3616}&\multicolumn{1}{@{.}l}{5}&\multicolumn{1}{l}{2\ensuremath{^{+}}}&\multicolumn{1}{r@{}}{1729}&\multicolumn{1}{@{.}l}{2 {\it 5}}&\multicolumn{1}{r@{}}{100}&\multicolumn{1}{@{ }l}{{\it 3}}&\multicolumn{1}{r@{}}{1887}&\multicolumn{1}{@{.}l}{4 }&\multicolumn{1}{@{}l}{2\ensuremath{^{+}}}&\multicolumn{1}{l}{M1+E2}&\multicolumn{1}{r@{}}{0}&\multicolumn{1}{@{.}l}{05 {\it 7}}&\parbox[t][0.3cm]{7.998101cm}{\raggedright B(M1)(W.u.)=0.088 \textit{+45{\textminus}31}; B(E2)(W.u.)\ensuremath{<}5.8\vspace{0.1cm}}&\\
&&&&&&&&&&&&&\parbox[t][0.3cm]{7.998101cm}{\raggedright E\ensuremath{_{\gamma}}: From (\href{https://www.nndc.bnl.gov/nsr/nsrlink.jsp?1968Gi09,B}{1968Gi09}, \href{https://www.nndc.bnl.gov/nsr/nsrlink.jsp?1969Ro08,B}{1969Ro08}).\vspace{0.1cm}}&\\
&&&&&&&&&&&&&\parbox[t][0.3cm]{7.998101cm}{\raggedright I\ensuremath{_{\gamma}}: I\ensuremath{_{\ensuremath{\gamma}}}=90.9\% \textit{27} is the weighted average (with external\vspace{0.1cm}}&\\
&&&&&&&&&&&&&\parbox[t][0.3cm]{7.998101cm}{\raggedright {\ }{\ }{\ }errors) of 93\% \textit{2} (\href{https://www.nndc.bnl.gov/nsr/nsrlink.jsp?1969Ro22,B}{1969Ro22}: see Fig. 3); and 87.5\% \textit{25}\vspace{0.1cm}}&\\
&&&&&&&&&&&&&\parbox[t][0.3cm]{7.998101cm}{\raggedright {\ }{\ }{\ }(\href{https://www.nndc.bnl.gov/nsr/nsrlink.jsp?1972Gi01,B}{1972Gi01}). The resulting weighted average is\vspace{0.1cm}}&\\
&&&&&&&&&&&&&\parbox[t][0.3cm]{7.998101cm}{\raggedright {\ }{\ }{\ }normalized to 100\%.\vspace{0.1cm}}&\\
&&&&&&&&&&&&&\parbox[t][0.3cm]{7.998101cm}{\raggedright Mult.: From (\href{https://www.nndc.bnl.gov/nsr/nsrlink.jsp?1969Ro08,B}{1969Ro08}, \href{https://www.nndc.bnl.gov/nsr/nsrlink.jsp?1970Sh04,B}{1970Sh04}, \href{https://www.nndc.bnl.gov/nsr/nsrlink.jsp?1972Gi01,B}{1972Gi01}).\vspace{0.1cm}}&\\
&&&&&&&&&&&&&\parbox[t][0.3cm]{7.998101cm}{\raggedright {\ }{\ }{\ }(\href{https://www.nndc.bnl.gov/nsr/nsrlink.jsp?1969Ro08,B}{1969Ro08}) calculated the matrix elements for E2 and\vspace{0.1cm}}&\\
&&&&&&&&&&&&&\parbox[t][0.3cm]{7.998101cm}{\raggedright {\ }{\ }{\ }M1 transitions to be \ensuremath{\vert}M\ensuremath{\vert}\ensuremath{^{\textnormal{2}}}\ensuremath{<}190 W.u. and\vspace{0.1cm}}&\\
&&&&&&&&&&&&&\parbox[t][0.3cm]{7.998101cm}{\raggedright {\ }{\ }{\ }0.03\ensuremath{<}\ensuremath{\vert}M\ensuremath{\vert}\ensuremath{^{\textnormal{2}}}\ensuremath{<}0.1, respectively.\vspace{0.1cm}}&\\
&&&&&&&&&&&&&\parbox[t][0.3cm]{7.998101cm}{\raggedright \ensuremath{\delta}: Weighted average (with external errors) of {\textminus}1.1 \textit{+10{\textminus}3}\vspace{0.1cm}}&\\
&&&&&&&&&&&&&\parbox[t][0.3cm]{7.998101cm}{\raggedright {\ }{\ }{\ }(\href{https://www.nndc.bnl.gov/nsr/nsrlink.jsp?1969Ro08,B}{1969Ro08}: see Table 2 assuming for J=2); {\textminus}0.9 \textit{7}\vspace{0.1cm}}&\\
&&&&&&&&&&&&&\parbox[t][0.3cm]{7.998101cm}{\raggedright {\ }{\ }{\ }(\href{https://www.nndc.bnl.gov/nsr/nsrlink.jsp?1969Ro22,B}{1969Ro22}: see Fig. 3); +0.09 \textit{7} (\href{https://www.nndc.bnl.gov/nsr/nsrlink.jsp?1970Sh04,B}{1970Sh04}: see Fig. 1);\vspace{0.1cm}}&\\
&&&&&&&&&&&&&\parbox[t][0.3cm]{7.998101cm}{\raggedright {\ }{\ }{\ }and +0.03 \textit{9} (\href{https://www.nndc.bnl.gov/nsr/nsrlink.jsp?1972Gi01,B}{1972Gi01}). See also \ensuremath{\delta}=+0.06 \textit{6} computed\vspace{0.1cm}}&\\
&&&&&&&&&&&&&\parbox[t][0.3cm]{7.998101cm}{\raggedright {\ }{\ }{\ }by (\href{https://www.nndc.bnl.gov/nsr/nsrlink.jsp?1972Gi01,B}{1972Gi01}) as the weighted average of the above\vspace{0.1cm}}&\\
&&&&&&&&&&&&&\parbox[t][0.3cm]{7.998101cm}{\raggedright {\ }{\ }{\ }mentioned values.\vspace{0.1cm}}&\\
&&&\multicolumn{1}{r@{}}{3614}&\multicolumn{1}{@{ }l}{{\it 3}}&\multicolumn{1}{r@{}}{10}&\multicolumn{1}{@{ }l}{{\it 3}}&\multicolumn{1}{r@{}}{0}&\multicolumn{1}{@{}l}{}&\multicolumn{1}{@{}l}{0\ensuremath{^{+}}}&\multicolumn{1}{l}{E2}&&&\parbox[t][0.3cm]{7.998101cm}{\raggedright B(E2)(W.u.)=0.68 \textit{+38{\textminus}28}\vspace{0.1cm}}&\\
\end{longtable}
\begin{textblock}{29}(0,27.3)
Continued on next page (footnotes at end of table)
\end{textblock}
\clearpage
\begin{longtable}{ccccc@{\extracolsep{\fill}}c}
\\[-.4cm]
\multicolumn{6}{c}{{\bf \small \underline{Adopted \hyperlink{18NE_LEVEL}{Levels}, \hyperlink{18NE_GAMMA}{Gammas} (continued)}}}\\
\multicolumn{6}{c}{~}\\
\multicolumn{6}{c}{\underline{$\gamma$($^{18}$Ne) (continued)}}\\
\multicolumn{6}{c}{~~~}\\
\multicolumn{2}{c}{E\ensuremath{_{i}}(level)}&\multicolumn{2}{c}{E\ensuremath{_{\gamma}}}&Comments&\\[-.2cm]
\multicolumn{2}{c}{\hrulefill}&\multicolumn{2}{c}{\hrulefill}&\hrulefill&
\endhead
&&&&\parbox[t][0.3cm]{15.402cm}{\raggedright E\ensuremath{_{\gamma}}: From (\href{https://www.nndc.bnl.gov/nsr/nsrlink.jsp?1969Ro22,B}{1969Ro22}).\vspace{0.1cm}}&\\
&&&&\parbox[t][0.3cm]{15.402cm}{\raggedright I\ensuremath{_{\gamma}}: I\ensuremath{_{\ensuremath{\gamma}}}=9.2\% \textit{27} is the weighted average (with external errors) of 7\% \textit{2} (\href{https://www.nndc.bnl.gov/nsr/nsrlink.jsp?1969Ro22,B}{1969Ro22}: see Fig. 3) and 12.5\% \textit{25}\vspace{0.1cm}}&\\
&&&&\parbox[t][0.3cm]{15.402cm}{\raggedright {\ }{\ }{\ }(\href{https://www.nndc.bnl.gov/nsr/nsrlink.jsp?1972Gi01,B}{1972Gi01}). See also BR\ensuremath{<}9\% (\href{https://www.nndc.bnl.gov/nsr/nsrlink.jsp?1968Gi09,B}{1968Gi09}, \href{https://www.nndc.bnl.gov/nsr/nsrlink.jsp?1969Ro22,B}{1969Ro22}); BR\ensuremath{<}3\% (\href{https://www.nndc.bnl.gov/nsr/nsrlink.jsp?1970Sh04,B}{1970Sh04}: see Fig. 1). The resulting weighted\vspace{0.1cm}}&\\
&&&&\parbox[t][0.3cm]{15.402cm}{\raggedright {\ }{\ }{\ }average mentioned above is renormalized to deduce I\ensuremath{_{\ensuremath{\gamma}}}=10\% \textit{3} given here.\vspace{0.1cm}}&\\
&&&&\parbox[t][0.3cm]{15.402cm}{\raggedright Mult.: From (\href{https://www.nndc.bnl.gov/nsr/nsrlink.jsp?1972Gi01,B}{1972Gi01}).\vspace{0.1cm}}&\\
\end{longtable}
\parbox[b][0.3cm]{17.7cm}{\makebox[1ex]{\ensuremath{^{\hypertarget{NE0GAMMA0}{a}}}} Since (\href{https://www.nndc.bnl.gov/nsr/nsrlink.jsp?1968Gi09,B}{1968Gi09}, \href{https://www.nndc.bnl.gov/nsr/nsrlink.jsp?1969Ro08,B}{1969Ro08}, \href{https://www.nndc.bnl.gov/nsr/nsrlink.jsp?1969Ro22,B}{1969Ro22}) considered the branching ratio of each observed branch as 100\% and estimated an upper}\\
\parbox[b][0.3cm]{17.7cm}{{\ }{\ }limit for the branching ratio of the unobserved branch for each level above the 1.89 MeV state, the evaluator renormalized the}\\
\parbox[b][0.3cm]{17.7cm}{{\ }{\ }branching ratios from each level so they add to 100\%.}\\
\vspace{0.5cm}
\begin{figure}[h]
\begin{center}
\includegraphics{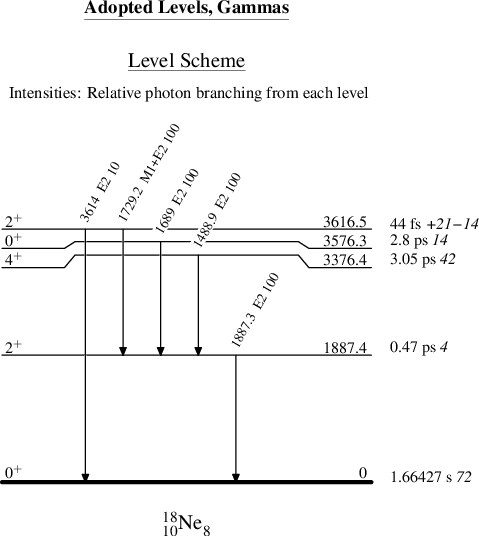}\\
\end{center}
\end{figure}
\clearpage
\subsection[\hspace{-0.2cm}\ensuremath{^{\textnormal{22}}}Al \ensuremath{\varepsilon}\ensuremath{\alpha} decay:91.1 ms]{ }
\vspace{-27pt}
\vspace{0.3cm}
\hypertarget{AL1}{{\bf \small \underline{\ensuremath{^{\textnormal{22}}}Al \ensuremath{\varepsilon}\ensuremath{\alpha} decay:91.1 ms\hspace{0.2in}\href{https://www.nndc.bnl.gov/nsr/nsrlink.jsp?1997Bl03,B}{1997Bl03},\href{https://www.nndc.bnl.gov/nsr/nsrlink.jsp?2021Wu10,B}{2021Wu10}}}}\\
\vspace{4pt}
\vspace{8pt}
\parbox[b][0.3cm]{17.7cm}{\addtolength{\parindent}{-0.2in}Parent: $^{22}$Al: E=0; J$^{\pi}$=(4)\ensuremath{^{+}}; T$_{1/2}$=91.1 ms {\it 5}; Q(\ensuremath{\varepsilon}\ensuremath{\alpha})=10.46\ensuremath{\times10^{3}} {\it syst}; \%\ensuremath{\varepsilon}\ensuremath{\alpha} decay=0.038 {\it 17}

}\\
\parbox[b][0.3cm]{17.7cm}{\addtolength{\parindent}{-0.2in}\ensuremath{^{22}}Al-T$_{1/2}$: In (\href{https://www.nndc.bnl.gov/nsr/nsrlink.jsp?2006Ac04,B}{2006Ac04}), the \ensuremath{^{\textnormal{22}}}Al half-life was measured based on 3 different analyses, two of which are prone to errors}\\
\parbox[b][0.3cm]{17.7cm}{(according to the authors). They recommended T\ensuremath{_{\textnormal{1/2}}}=91.1 ms \textit{5} based on the error-weighted average half-life deduced from fits of}\\
\parbox[b][0.3cm]{17.7cm}{decay time spectra in coincidence with various different proton groups. The recommended value is adopted in the latest ENSDF}\\
\parbox[b][0.3cm]{17.7cm}{evaluation of \ensuremath{^{\textnormal{22}}}Al (\href{https://www.nndc.bnl.gov/nsr/nsrlink.jsp?2015Ba27,B}{2015Ba27}).}\\
\parbox[b][0.3cm]{17.7cm}{\addtolength{\parindent}{-0.2in}\ensuremath{^{22}}Al-J$^{\pi}$: Based on T=2 isospin multiplet (\href{https://www.nndc.bnl.gov/nsr/nsrlink.jsp?1982Ca16,B}{1982Ca16}) and from comparison of the measured \ensuremath{\gamma}-ray intensity, summed Gamow-Teller}\\
\parbox[b][0.3cm]{17.7cm}{strength, and the delayed proton branching ratios accompanied by shell model calculations presented in (\href{https://www.nndc.bnl.gov/nsr/nsrlink.jsp?2006Ac04,B}{2006Ac04}), the evaluator}\\
\parbox[b][0.3cm]{17.7cm}{assigned J=(4), which is consistent with the adopted (by \href{https://www.nndc.bnl.gov/nsr/nsrlink.jsp?2015Ba27,B}{2015Ba27}) J\ensuremath{^{\ensuremath{\pi}}} value in ENSDF and the J\ensuremath{^{\ensuremath{\pi}}} assignments presented in}\\
\parbox[b][0.3cm]{17.7cm}{(\href{https://www.nndc.bnl.gov/nsr/nsrlink.jsp?1983Ca01,B}{1983Ca01}, \href{https://www.nndc.bnl.gov/nsr/nsrlink.jsp?1984Ca29,B}{1984Ca29}, \href{https://www.nndc.bnl.gov/nsr/nsrlink.jsp?1985Ja07,B}{1985Ja07}, and \href{https://www.nndc.bnl.gov/nsr/nsrlink.jsp?2021Wu10,B}{2021Wu10}).}\\
\parbox[b][0.3cm]{17.7cm}{\addtolength{\parindent}{-0.2in}\ensuremath{^{22}}Al-Q(\ensuremath{\varepsilon}\ensuremath{\alpha}): From Q(\ensuremath{\varepsilon})(\ensuremath{^{\textnormal{22}}}Al)=18600 (uncertainty is systematic: \href{https://www.nndc.bnl.gov/nsr/nsrlink.jsp?2021Wa16,B}{2021Wa16}) and S\ensuremath{_{\ensuremath{\alpha}}}(\ensuremath{^{\textnormal{22}}}Mg\ensuremath{_{\textnormal{g.s.}}})=8142.5 \textit{4} (\href{https://www.nndc.bnl.gov/nsr/nsrlink.jsp?2021Wa16,B}{2021Wa16}).}\\
\parbox[b][0.3cm]{17.7cm}{\addtolength{\parindent}{-0.2in}\ensuremath{^{22}}Al-\%\ensuremath{\varepsilon}\ensuremath{\alpha} decay: From Adopted Levels of \ensuremath{^{\textnormal{22}}}Al in (\href{https://www.nndc.bnl.gov/nsr/nsrlink.jsp?2015Ba27,B}{2015Ba27}) and taken from (\href{https://www.nndc.bnl.gov/nsr/nsrlink.jsp?2006Ac04,B}{2006Ac04}). See also \ensuremath{\varepsilon}\ensuremath{\alpha}= 0.31 \% \textit{9} (\href{https://www.nndc.bnl.gov/nsr/nsrlink.jsp?1997Bl03,B}{1997Bl03}),}\\
\parbox[b][0.3cm]{17.7cm}{and \ensuremath{\varepsilon}\ensuremath{\alpha}=0.12 \% \textit{1} (\href{https://www.nndc.bnl.gov/nsr/nsrlink.jsp?2021Wu10,B}{2021Wu10}). The branching ratio deduced in (\href{https://www.nndc.bnl.gov/nsr/nsrlink.jsp?2006Ac04,B}{2006Ac04}) was obtained from the intensity of the 1887-keV \ensuremath{\gamma} ray}\\
\parbox[b][0.3cm]{17.7cm}{(from the \ensuremath{\gamma}-decay of the first excited state of \ensuremath{^{\textnormal{18}}}Ne to \ensuremath{^{\textnormal{18}}}Ne\ensuremath{_{\textnormal{g.s.}}}) observed in coincidence with the \ensuremath{\beta}-delayed \ensuremath{\alpha}-particle from the}\\
\parbox[b][0.3cm]{17.7cm}{decay of \ensuremath{^{\textnormal{22}}}Mg*(IAS) to \ensuremath{^{\textnormal{18}}}Ne*(1887 keV).}\\
\vspace{0.385cm}
\parbox[b][0.3cm]{17.7cm}{\addtolength{\parindent}{-0.2in}\href{https://www.nndc.bnl.gov/nsr/nsrlink.jsp?1997Bl03,B}{1997Bl03}, \href{https://www.nndc.bnl.gov/nsr/nsrlink.jsp?1997Cz02,B}{1997Cz02}: \ensuremath{^{\textnormal{22}}}Al(\ensuremath{\beta}\ensuremath{^{\textnormal{+}}})\ensuremath{^{\textnormal{22}}}Mg*(\ensuremath{\alpha})\ensuremath{^{\textnormal{18}}}Ne* E=74 MeV/nucleon; implanted \ensuremath{^{\textnormal{22}}}Al beam into a Si detector used as an entrance}\\
\parbox[b][0.3cm]{17.7cm}{window (integrated in the cathode) of a Micro-Strip Gas Counter (MSGC). The particle-recoil and particle-particle coincidences}\\
\parbox[b][0.3cm]{17.7cm}{from the \ensuremath{^{\textnormal{22}}}Al decay were measured using these detectors as well as a Si detector downstream the MSGC. After collection of}\\
\parbox[b][0.3cm]{17.7cm}{sufficient activity, the beam was switched off for 100 ms. Measured \ensuremath{\beta}-delayed protons-, two-protons-, and \ensuremath{\alpha}-emission spectra and}\\
\parbox[b][0.3cm]{17.7cm}{E\ensuremath{_{\textnormal{p}}} and E\ensuremath{_{\ensuremath{\alpha}}} for the \ensuremath{\beta}-delayed p, 2p, and \ensuremath{\alpha}-emissions. Deduced the half-life of \ensuremath{^{\textnormal{22}}}Al; discussed the J\ensuremath{^{\ensuremath{\pi}}} assignment of \ensuremath{^{\textnormal{22}}}Al\ensuremath{_{\textnormal{g.s.}}};}\\
\parbox[b][0.3cm]{17.7cm}{deduced \ensuremath{^{\textnormal{18}}}Ne*(1887 keV) level and its associated \ensuremath{\beta}-delayed \ensuremath{\alpha}-branching ratio; and measured the \ensuremath{\beta}-delayed \ensuremath{\alpha}-emission to the}\\
\parbox[b][0.3cm]{17.7cm}{\ensuremath{^{\textnormal{18}}}Ne*(2\ensuremath{^{\textnormal{+}}_{\textnormal{1}}}) state from the isobaric analog state in \ensuremath{^{\textnormal{22}}}Mg. The weaker \ensuremath{\beta}-delayed \ensuremath{\alpha}-decay branches were hindered because of the}\\
\parbox[b][0.3cm]{17.7cm}{stronger transitions from contaminants.}\\
\parbox[b][0.3cm]{17.7cm}{\addtolength{\parindent}{-0.2in}\href{https://www.nndc.bnl.gov/nsr/nsrlink.jsp?2006Ac04,B}{2006Ac04}: \ensuremath{^{\textnormal{22}}}Al(\ensuremath{\beta}\ensuremath{^{\textnormal{+}}})\ensuremath{^{\textnormal{22}}}Mg*(\ensuremath{\alpha})\ensuremath{^{\textnormal{18}}}Ne* E=48 MeV/nucleon; implanted \ensuremath{^{\textnormal{22}}}Al beam into a Si telescope; identified the charged particles}\\
\parbox[b][0.3cm]{17.7cm}{from the \ensuremath{^{\textnormal{22}}}Al decay using a downstream set of three silicon detectors; detected \ensuremath{\gamma}-rays from the decay events by the EXOGAM}\\
\parbox[b][0.3cm]{17.7cm}{HPGe clover detector in close geometry. Measured E\ensuremath{_{\ensuremath{\gamma}}}, I\ensuremath{_{\ensuremath{\gamma}}}, \ensuremath{\beta}, \ensuremath{\beta}-\ensuremath{\gamma} and (\ensuremath{\beta}-delayed particles)-\ensuremath{\gamma} coincidences, branching ratios, and}\\
\parbox[b][0.3cm]{17.7cm}{the \ensuremath{^{\textnormal{22}}}Al half-life in beam-on/beam-off mode (120 ms of implantation, 300 ms beam-off). A GEANT Monte Carlo simulation, the}\\
\parbox[b][0.3cm]{17.7cm}{\ensuremath{^{\textnormal{22}}}Al decay scheme, comparisons with shell model calculations and with previous experimental results are presented. Observed the}\\
\parbox[b][0.3cm]{17.7cm}{\ensuremath{\beta}-2p transition to the \ensuremath{^{\textnormal{20}}}Ne*(2\ensuremath{^{\textnormal{+}}_{\textnormal{1}}}) and the \ensuremath{\beta}-\ensuremath{\alpha} decay to the \ensuremath{^{\textnormal{18}}}Ne*(2\ensuremath{^{\textnormal{+}}_{\textnormal{1}}}) state. To obtain the branching ratios for these transitions,}\\
\parbox[b][0.3cm]{17.7cm}{\ensuremath{\gamma}-ray coincidence events were used. Deduced the mass excess of \ensuremath{^{\textnormal{22}}}Al and discussed the J\ensuremath{^{\ensuremath{\pi}}} assignments of the measured \ensuremath{^{\textnormal{22}}}Mg*}\\
\parbox[b][0.3cm]{17.7cm}{states.}\\
\parbox[b][0.3cm]{17.7cm}{\addtolength{\parindent}{-0.2in}\href{https://www.nndc.bnl.gov/nsr/nsrlink.jsp?2021Wu10,B}{2021Wu10}: \ensuremath{^{\textnormal{22}}}Al(\ensuremath{\beta}\ensuremath{^{\textnormal{+}}})\ensuremath{^{\textnormal{22}}}Mg*(\ensuremath{\alpha})\ensuremath{^{\textnormal{18}}}Ne* E not given; implanted \ensuremath{^{\textnormal{22}}}Al into a position sensitive segmented Si detector array that}\\
\parbox[b][0.3cm]{17.7cm}{composed of 3 DSSDs with different thicknesses. Downstream this array, 3 other silicon detectors identified the \ensuremath{\beta}-rays from the}\\
\parbox[b][0.3cm]{17.7cm}{decay and vetoed the light beam contaminants. Measured the \ensuremath{\beta}-delayed \ensuremath{\gamma} rays by a \ensuremath{\gamma}-ray detector array consisting of 5 HPGe}\\
\parbox[b][0.3cm]{17.7cm}{clover detectors surrounding the position sensitive Si array. Measured the \ensuremath{\beta}-delayed protons, 2p, and \ensuremath{\alpha}- particles, E\ensuremath{_{\ensuremath{\beta}}}, E\ensuremath{_{\ensuremath{\gamma}}}, I\ensuremath{_{\ensuremath{\gamma}}}, \ensuremath{\beta}\ensuremath{\gamma}}\\
\parbox[b][0.3cm]{17.7cm}{and \ensuremath{\alpha}-\ensuremath{\gamma} coincidences, and decay curve of \ensuremath{^{\textnormal{22}}}Al; deduced level energies, J\ensuremath{^{\ensuremath{\pi}}} values, \ensuremath{^{\textnormal{22}}}Al half-life, and branching ratios. The}\\
\parbox[b][0.3cm]{17.7cm}{measured absolute branching ratios are summed to 91.6\% \textit{51}. However, there are 8 unassigned transitions, which contribute to the}\\
\parbox[b][0.3cm]{17.7cm}{missing branching ratios. These 8 transitions are, however, not in coincidence with any known \ensuremath{\gamma} rays in \ensuremath{^{\textnormal{21}}}Na (from \ensuremath{\beta}p decay of}\\
\parbox[b][0.3cm]{17.7cm}{\ensuremath{^{\textnormal{22}}}Al), \ensuremath{^{\textnormal{20}}}Ne (from \ensuremath{\beta}2p decay of \ensuremath{^{\textnormal{22}}}Al), or \ensuremath{^{\textnormal{18}}}Ne (from \ensuremath{\beta}\ensuremath{\alpha} decay of \ensuremath{^{\textnormal{22}}}Al). Moreover, based on energy summation, these}\\
\parbox[b][0.3cm]{17.7cm}{transitions cannot be connecting to the ground state in the above mentioned nuclei. The authors tentatively assigned them to \ensuremath{\beta}p}\\
\parbox[b][0.3cm]{17.7cm}{decay of \ensuremath{^{\textnormal{22}}}Mg* states to \ensuremath{^{\textnormal{21}}}Na\ensuremath{_{\textnormal{g.s.}}}. Comparisons with shell model calculations are presented.}\\
\vspace{0.385cm}
\parbox[b][0.3cm]{17.7cm}{\addtolength{\parindent}{-0.2in}\textit{Previously Known \ensuremath{^{22}}Al Decay Schemes}:}\\
\parbox[b][0.3cm]{17.7cm}{\addtolength{\parindent}{-0.2in}\href{https://www.nndc.bnl.gov/nsr/nsrlink.jsp?1983Ca01,B}{1983Ca01}, \href{https://www.nndc.bnl.gov/nsr/nsrlink.jsp?1983CaZU,B}{1983CaZU}, \href{https://www.nndc.bnl.gov/nsr/nsrlink.jsp?1983CaZT,B}{1983CaZT}, \href{https://www.nndc.bnl.gov/nsr/nsrlink.jsp?1984Ca29,B}{1984Ca29}, \href{https://www.nndc.bnl.gov/nsr/nsrlink.jsp?1984CaZV,B}{1984CaZV}: \ensuremath{^{\textnormal{24}}}Mg(\ensuremath{^{\textnormal{3}}}He,p4n) E=110 MeV. In the experiments of (\href{https://www.nndc.bnl.gov/nsr/nsrlink.jsp?1983Ca01,B}{1983Ca01},}\\
\parbox[b][0.3cm]{17.7cm}{\href{https://www.nndc.bnl.gov/nsr/nsrlink.jsp?1984Ca29,B}{1984Ca29}), \ensuremath{^{\textnormal{22}}}Al recoils were stopped in a helium jet system, which was used to transport the \ensuremath{^{\textnormal{22}}}Al activity to the counting}\\
\parbox[b][0.3cm]{17.7cm}{chamber consisting of a Si \ensuremath{\Delta}E-E telescope to measure \ensuremath{\beta}-delayed protons (from \ensuremath{^{\textnormal{22}}}Al(\ensuremath{\beta}\ensuremath{^{\textnormal{+}}}p)) in coincidence at \ensuremath{\theta}\ensuremath{_{\textnormal{lab}}}=0\ensuremath{^\circ}{\textminus}70\ensuremath{^\circ}. These}\\
\parbox[b][0.3cm]{17.7cm}{experiments only investigated the \ensuremath{\beta}-delayed 2p-decay of \ensuremath{^{\textnormal{22}}}Al. The \ensuremath{\beta}-delayed \ensuremath{\alpha}-decay is not discussed. However, the authors}\\
\parbox[b][0.3cm]{17.7cm}{present Fig. 3 (\href{https://www.nndc.bnl.gov/nsr/nsrlink.jsp?1983Ca01,B}{1983Ca01}) and Fig. 1 (\href{https://www.nndc.bnl.gov/nsr/nsrlink.jsp?1984Ca29,B}{1984Ca29}), in which the decay scheme of \ensuremath{^{\textnormal{22}}}Al includes the \ensuremath{\beta}-delayed \ensuremath{\alpha}-decay branches to}\\
\parbox[b][0.3cm]{17.7cm}{the ground and first excited states of \ensuremath{^{\textnormal{18}}}Ne. No other information is provided.}\\
\vspace{12pt}
\clearpage
\vspace{0.3cm}
{\bf \small \underline{\ensuremath{^{\textnormal{22}}}Al \ensuremath{\varepsilon}\ensuremath{\alpha} decay:91.1 ms\hspace{0.2in}\href{https://www.nndc.bnl.gov/nsr/nsrlink.jsp?1997Bl03,B}{1997Bl03},\href{https://www.nndc.bnl.gov/nsr/nsrlink.jsp?2021Wu10,B}{2021Wu10} (continued)}}\\
\vspace{0.3cm}
\underline{$^{18}$Ne Levels}\\
\begin{longtable}{ccccc@{\extracolsep{\fill}}c}
\multicolumn{2}{c}{E(level)$^{{\hyperlink{NE1LEVEL0}{a}}}$}&J$^{\pi}$$^{{\hyperlink{NE1LEVEL0}{a}}}$&\multicolumn{2}{c}{T\ensuremath{_{\textnormal{1/2}}}$^{{\hyperlink{NE1LEVEL0}{a}}}$}&\\[-.2cm]
\multicolumn{2}{c}{\hrulefill}&\hrulefill&\multicolumn{2}{c}{\hrulefill}&
\endfirsthead
\multicolumn{1}{r@{}}{0}&\multicolumn{1}{@{}l}{}&\multicolumn{1}{l}{0\ensuremath{^{+}}}&\multicolumn{1}{r@{}}{1}&\multicolumn{1}{@{.}l}{66427 s}&\\
\multicolumn{1}{r@{}}{1887}&\multicolumn{1}{@{.}l}{4}&\multicolumn{1}{l}{2\ensuremath{^{+}}}&\multicolumn{1}{r@{}}{0}&\multicolumn{1}{@{.}l}{47 ps}&\\
\end{longtable}
\parbox[b][0.3cm]{17.7cm}{\makebox[1ex]{\ensuremath{^{\hypertarget{NE1LEVEL0}{a}}}} From the \ensuremath{^{\textnormal{18}}}Ne Adopted Levels.}\\
\vspace{0.5cm}
\underline{$\gamma$($^{18}$Ne)}\\
\begin{longtable}{ccccccc@{}cc@{\extracolsep{\fill}}c}
\multicolumn{2}{c}{E\ensuremath{_{\gamma}}}&\multicolumn{2}{c}{E\ensuremath{_{i}}(level)}&J\ensuremath{^{\pi}_{i}}&\multicolumn{2}{c}{E\ensuremath{_{f}}}&J\ensuremath{^{\pi}_{f}}&Comments&\\[-.2cm]
\multicolumn{2}{c}{\hrulefill}&\multicolumn{2}{c}{\hrulefill}&\hrulefill&\multicolumn{2}{c}{\hrulefill}&\hrulefill&\hrulefill&
\endfirsthead
\multicolumn{1}{r@{}}{1887}&\multicolumn{1}{@{}l}{}&\multicolumn{1}{r@{}}{1887}&\multicolumn{1}{@{.}l}{4}&\multicolumn{1}{l}{2\ensuremath{^{+}}}&\multicolumn{1}{r@{}}{0}&\multicolumn{1}{@{}l}{}&\multicolumn{1}{@{}l}{0\ensuremath{^{+}}}&\parbox[t][0.3cm]{12.975201cm}{\raggedright E\ensuremath{_{\gamma}}: Measured in (\href{https://www.nndc.bnl.gov/nsr/nsrlink.jsp?2006Ac04,B}{2006Ac04}, \href{https://www.nndc.bnl.gov/nsr/nsrlink.jsp?2021Wu10,B}{2021Wu10}). See also the \ensuremath{^{\textnormal{18}}}Ne Adopted Gammas. (\href{https://www.nndc.bnl.gov/nsr/nsrlink.jsp?1997Bl03,B}{1997Bl03}) did\vspace{0.1cm}}&\\
&&&&&&&&\parbox[t][0.3cm]{12.975201cm}{\raggedright {\ }{\ }{\ }not measure any \ensuremath{\gamma} rays.\vspace{0.1cm}}&\\
\end{longtable}
\underline{Delayed Alphas ($^{18}$Ne)}\\
\begin{longtable}{cccccccc@{\extracolsep{\fill}}c}
\multicolumn{2}{c}{E($\alpha$)}&\multicolumn{2}{c}{E($^{18}$Ne)}&\multicolumn{2}{c}{I($\alpha$)\ensuremath{^{\hyperlink{AL1DELAY0}{a}}}}&E($^{22}$Mg)&Comments&\\[-.2cm]
\multicolumn{2}{c}{\hrulefill}&\multicolumn{2}{c}{\hrulefill}&\multicolumn{2}{c}{\hrulefill}&\hrulefill&\hrulefill&
\endfirsthead
\multicolumn{1}{r@{}}{4017}&\multicolumn{1}{@{ }l}{{\it 8}}&\multicolumn{1}{r@{}}{1887}&\multicolumn{1}{@{.}l}{4}&\multicolumn{1}{r@{}}{0}&\multicolumn{1}{@{.}l}{038 {\it 17}}&\multicolumn{1}{l}{14012}&\parbox[t][0.3cm]{11.648801cm}{\raggedright E($\alpha$): Total decay energy given in the center-of-mass frame. This value is adopted\vspace{0.1cm}}&\\
&&&&&&&\parbox[t][0.3cm]{11.648801cm}{\raggedright {\ }{\ }{\ }from \ensuremath{^{\textnormal{22}}}Al Adopted Levels in (\href{https://www.nndc.bnl.gov/nsr/nsrlink.jsp?2015Ba27,B}{2015Ba27}), which is taken from (\href{https://www.nndc.bnl.gov/nsr/nsrlink.jsp?2006Ac04,B}{2006Ac04}). See\vspace{0.1cm}}&\\
&&&&&&&\parbox[t][0.3cm]{11.648801cm}{\raggedright {\ }{\ }{\ }also: E\ensuremath{_{\textnormal{c.m.}}}=3997 keV \textit{40} deduced from E\ensuremath{_{\ensuremath{\alpha}\textnormal{,lab}}}=3270 keV \textit{40} (\href{https://www.nndc.bnl.gov/nsr/nsrlink.jsp?1997Bl03,B}{1997Bl03}); and\vspace{0.1cm}}&\\
&&&&&&&\parbox[t][0.3cm]{11.648801cm}{\raggedright {\ }{\ }{\ }E\ensuremath{_{\textnormal{c.m.}}}=4030 keV \textit{10} (\href{https://www.nndc.bnl.gov/nsr/nsrlink.jsp?2021Wu10,B}{2021Wu10}). Note that the \ensuremath{\beta}-delayed \ensuremath{\alpha}-transition measured in\vspace{0.1cm}}&\\
&&&&&&&\parbox[t][0.3cm]{11.648801cm}{\raggedright {\ }{\ }{\ }(\href{https://www.nndc.bnl.gov/nsr/nsrlink.jsp?2006Ac04,B}{2006Ac04}) was superimposed on a \ensuremath{\beta}-delayed proton transition. Using \ensuremath{\gamma}-\ensuremath{\alpha}\vspace{0.1cm}}&\\
&&&&&&&\parbox[t][0.3cm]{11.648801cm}{\raggedright {\ }{\ }{\ }coincidence events, the \ensuremath{\beta}-delayed \ensuremath{\alpha}-branch was distinguished. Also, note that in\vspace{0.1cm}}&\\
&&&&&&&\parbox[t][0.3cm]{11.648801cm}{\raggedright {\ }{\ }{\ }Table 1 of (\href{https://www.nndc.bnl.gov/nsr/nsrlink.jsp?2006Ac04,B}{2006Ac04}), it is mentioned that the center-of-mass E\ensuremath{_{\ensuremath{\alpha}}} for this\vspace{0.1cm}}&\\
&&&&&&&\parbox[t][0.3cm]{11.648801cm}{\raggedright {\ }{\ }{\ }transition measured in (\href{https://www.nndc.bnl.gov/nsr/nsrlink.jsp?1997Bl03,B}{1997Bl03}) was 3997 keV \textit{49} and that this is equivalent to\vspace{0.1cm}}&\\
&&&&&&&\parbox[t][0.3cm]{11.648801cm}{\raggedright {\ }{\ }{\ }the \ensuremath{\beta}-\ensuremath{\alpha} transition measured by (\href{https://www.nndc.bnl.gov/nsr/nsrlink.jsp?1997Bl03,B}{1997Bl03}) at E=4.01 MeV \textit{5}. While (\href{https://www.nndc.bnl.gov/nsr/nsrlink.jsp?1997Bl03,B}{1997Bl03})\vspace{0.1cm}}&\\
&&&&&&&\parbox[t][0.3cm]{11.648801cm}{\raggedright {\ }{\ }{\ }mentions that the measured E\ensuremath{_{\ensuremath{\alpha}\textnormal{,lab}}}=3270 keV \textit{40} corresponds to the Q-value of\vspace{0.1cm}}&\\
&&&&&&&\parbox[t][0.3cm]{11.648801cm}{\raggedright {\ }{\ }{\ }4.01 MeV \textit{5} for an \ensuremath{\alpha} de-excitation of the \ensuremath{^{\textnormal{22}}}Mg*(IAS) towards the first excited\vspace{0.1cm}}&\\
&&&&&&&\parbox[t][0.3cm]{11.648801cm}{\raggedright {\ }{\ }{\ }state in \ensuremath{^{\textnormal{18}}}Ne. The evaluator cannot determine where the 49 keV uncertainty in\vspace{0.1cm}}&\\
&&&&&&&\parbox[t][0.3cm]{11.648801cm}{\raggedright {\ }{\ }{\ }E\ensuremath{_{\textnormal{c.m.}}}=3997 keV \textit{49} value reported by (\href{https://www.nndc.bnl.gov/nsr/nsrlink.jsp?2006Ac04,B}{2006Ac04}) for the measurement in\vspace{0.1cm}}&\\
&&&&&&&\parbox[t][0.3cm]{11.648801cm}{\raggedright {\ }{\ }{\ }(\href{https://www.nndc.bnl.gov/nsr/nsrlink.jsp?1997Bl03,B}{1997Bl03}) comes from. However, converting the E\ensuremath{_{\ensuremath{\alpha}\textnormal{,lab}}}=3270 keV \textit{40} (\href{https://www.nndc.bnl.gov/nsr/nsrlink.jsp?1997Bl03,B}{1997Bl03})\vspace{0.1cm}}&\\
&&&&&&&\parbox[t][0.3cm]{11.648801cm}{\raggedright {\ }{\ }{\ }to the center-of-mass, the evaluator deduced E\ensuremath{_{\textnormal{c.m.}}}=3997 keV \textit{40} for (\href{https://www.nndc.bnl.gov/nsr/nsrlink.jsp?1997Bl03,B}{1997Bl03}).\vspace{0.1cm}}&\\
&&&&&&&\parbox[t][0.3cm]{11.648801cm}{\raggedright I($\alpha$): Branching ratio per 100 decays of \ensuremath{^{\textnormal{22}}}Al from (\href{https://www.nndc.bnl.gov/nsr/nsrlink.jsp?2006Ac04,B}{2006Ac04}), which is\vspace{0.1cm}}&\\
&&&&&&&\parbox[t][0.3cm]{11.648801cm}{\raggedright {\ }{\ }{\ }recommended by the Adopted Levels of \ensuremath{^{\textnormal{22}}}Al (\href{https://www.nndc.bnl.gov/nsr/nsrlink.jsp?2015Ba27,B}{2015Ba27}).\vspace{0.1cm}}&\\
&&&&&&&\parbox[t][0.3cm]{11.648801cm}{\raggedright E($^{22}$Mg): 14012 keV \textit{3} (\href{https://www.nndc.bnl.gov/nsr/nsrlink.jsp?2006Ac04,B}{2006Ac04}) is the IAS of \ensuremath{^{\textnormal{22}}}Al\ensuremath{_{\textnormal{g.s}}} assuming J\ensuremath{^{\ensuremath{\pi}}}=4\ensuremath{^{\textnormal{+}}}, see the\vspace{0.1cm}}&\\
&&&&&&&\parbox[t][0.3cm]{11.648801cm}{\raggedright {\ }{\ }{\ }Adopted Levels of \ensuremath{^{\textnormal{22}}}Al in (\href{https://www.nndc.bnl.gov/nsr/nsrlink.jsp?2015Ba27,B}{2015Ba27}). See also 14046 keV \textit{5} (\href{https://www.nndc.bnl.gov/nsr/nsrlink.jsp?2021Wu10,B}{2021Wu10}) and\vspace{0.1cm}}&\\
&&&&&&&\parbox[t][0.3cm]{11.648801cm}{\raggedright {\ }{\ }{\ }14027 keV \textit{40} deduced from the reported E\ensuremath{_{\ensuremath{\alpha}\textnormal{,lab}}}=3270 keV \textit{40} (\href{https://www.nndc.bnl.gov/nsr/nsrlink.jsp?1997Bl03,B}{1997Bl03}).\vspace{0.1cm}}&\\
\end{longtable}
\parbox[b][0.3cm]{17.7cm}{\makebox[1ex]{\ensuremath{^{\hypertarget{AL1DELAY0}{a}}}} Absolute intensity per 100 decays.}\\
\vspace{0.5cm}
\clearpage
\begin{figure}[h]
\begin{center}
\includegraphics{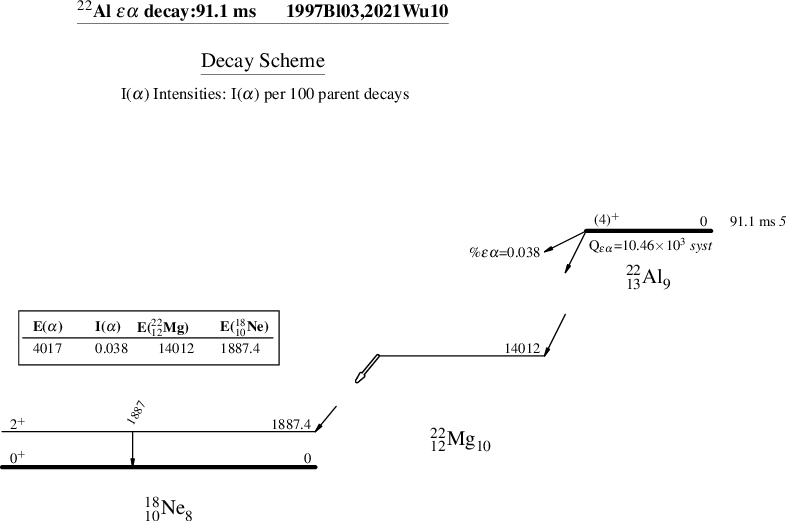}\\
\end{center}
\end{figure}
\clearpage
\subsection[\hspace{-0.2cm}\ensuremath{^{\textnormal{19}}}Na p decay]{ }
\vspace{-27pt}
\vspace{0.3cm}
\hypertarget{NA2}{{\bf \small \underline{\ensuremath{^{\textnormal{19}}}Na p decay\hspace{0.2in}\href{https://www.nndc.bnl.gov/nsr/nsrlink.jsp?2003An02,B}{2003An02},\href{https://www.nndc.bnl.gov/nsr/nsrlink.jsp?2008Pe02,B}{2008Pe02}}}}\\
\vspace{4pt}
\vspace{8pt}
\parbox[b][0.3cm]{17.7cm}{\addtolength{\parindent}{-0.2in}Parent: $^{19}$Na: E=0; J$^{\pi}$=(5/2\ensuremath{^{+}}); T$_{1/2}$$<$40 keV; Q(p)=$-$323 {\it 11}; \%p decay=100.0

}\\
\parbox[b][0.3cm]{17.7cm}{\addtolength{\parindent}{-0.2in}\ensuremath{^{19}}Na-E,J$^{\pi}$,T$_{1/2}$: From the Adopted Levels of \ensuremath{^{\textnormal{19}}}Na in ENSDF.}\\
\parbox[b][0.3cm]{17.7cm}{\addtolength{\parindent}{-0.2in}\ensuremath{^{19}}Na-Q(p): From (\href{https://www.nndc.bnl.gov/nsr/nsrlink.jsp?2021Wa16,B}{2021Wa16}).}\\
\parbox[b][0.3cm]{17.7cm}{\addtolength{\parindent}{-0.2in}\ensuremath{^{19}}Na-\%p decay: From Adopted Levels of \ensuremath{^{\textnormal{19}}}Na in the ENSDF database.}\\
\parbox[b][0.3cm]{17.7cm}{\addtolength{\parindent}{-0.2in}\href{https://www.nndc.bnl.gov/nsr/nsrlink.jsp?2003An02,B}{2003An02}, \href{https://www.nndc.bnl.gov/nsr/nsrlink.jsp?2003An28,B}{2003An28}, \href{https://www.nndc.bnl.gov/nsr/nsrlink.jsp?2004An28,B}{2004An28}: \ensuremath{^{\textnormal{1}}}H(\ensuremath{^{\textnormal{18}}}Ne,\ensuremath{^{\textnormal{19}}}Na\ensuremath{\rightarrow}p+\ensuremath{^{\textnormal{18}}}Ne), \ensuremath{^{\textnormal{1}}}H(\ensuremath{^{\textnormal{18}}}Ne,\ensuremath{^{\textnormal{19}}}Na\ensuremath{\rightarrow}2p+\ensuremath{^{\textnormal{17}}}F) E=21, 23.5, and 28 MeV; measured the}\\
\parbox[b][0.3cm]{17.7cm}{recoil protons using two segments of the Louvain-la-neuve Edinburg Detector Array (LEDA) Si array covering \ensuremath{\theta}\ensuremath{_{\textnormal{lab}}}=4.9\ensuremath{^\circ}{\textminus}11.7\ensuremath{^\circ} and}\\
\parbox[b][0.3cm]{17.7cm}{\ensuremath{\theta}\ensuremath{_{\textnormal{lab}}}=22.6\ensuremath{^\circ}{\textminus}29.9\ensuremath{^\circ}. A J\ensuremath{^{\ensuremath{\pi}}}=1/2\ensuremath{^{\textnormal{+}}} resonance was observed at E\ensuremath{_{\textnormal{c.m.}}}=1066 keV \textit{3} corresponding to \ensuremath{^{\textnormal{19}}}Na*(745 keV). The proton decay}\\
\parbox[b][0.3cm]{17.7cm}{of this state to \ensuremath{^{\textnormal{18}}}Ne\ensuremath{_{\textnormal{g.s.}}} was measured.}\\
\parbox[b][0.3cm]{17.7cm}{\addtolength{\parindent}{-0.2in}\href{https://www.nndc.bnl.gov/nsr/nsrlink.jsp?2005De15,B}{2005De15}, \href{https://www.nndc.bnl.gov/nsr/nsrlink.jsp?2006DeZU,B}{2006DeZU}: \ensuremath{^{\textnormal{1}}}H(\ensuremath{^{\textnormal{18}}}Ne,\ensuremath{^{\textnormal{19}}}Na\ensuremath{\rightarrow}\ensuremath{^{\textnormal{18}}}Ne+p), \ensuremath{^{\textnormal{1}}}H(\ensuremath{^{\textnormal{18}}}Ne,\ensuremath{^{\textnormal{19}}}Na\ensuremath{\rightarrow}2p+\ensuremath{^{\textnormal{17}}}F) E=7.2 MeV/nucleon; measured the excitation function}\\
\parbox[b][0.3cm]{17.7cm}{of \ensuremath{^{\textnormal{18}}}Ne+p quasi{\textminus}elastic scattering. The scattered protons were measured by a position sensitive \ensuremath{\Delta}E-\ensuremath{\Delta}E-E telescope covering}\\
\parbox[b][0.3cm]{17.7cm}{\ensuremath{\theta}\ensuremath{_{\textnormal{lab}}}=\ensuremath{\pm}4.5\ensuremath{^\circ} with a resolution of 30 keV \textit{10}. Two peaks at E\ensuremath{_{\textnormal{c.m.}}}\ensuremath{\approx}2400 and 3100 keV were observed on the excitation function of}\\
\parbox[b][0.3cm]{17.7cm}{p(\ensuremath{^{\textnormal{18}}}Ne,p) elastic scattering, whose shapes, energies and intensities did not match any interpretation in the context of elastic}\\
\parbox[b][0.3cm]{17.7cm}{scattering or shell model calculations for \ensuremath{^{\textnormal{19}}}Na states. An R-matrix calculation (using the ANARKI code) predicted the observed}\\
\parbox[b][0.3cm]{17.7cm}{shapes and energies of these two peaks but not the observed intensities. It was necessary to consider inelastic scattering to the}\\
\parbox[b][0.3cm]{17.7cm}{proton-unbound states of \ensuremath{^{\textnormal{18}}}Ne and the subsequent p-decay to \ensuremath{^{\textnormal{17}}}F. The two-proton decay analysis is consistent with \ensuremath{^{\textnormal{19}}}Na states at}\\
\parbox[b][0.3cm]{17.7cm}{E\ensuremath{_{\textnormal{x}}}=5585, 5809 and 5815 keV that 2p decay, via \ensuremath{^{\textnormal{18}}}Ne* intermediate states (see Fig. 8) to \ensuremath{^{\textnormal{17}}}F*(0, 495 keV).}\\
\parbox[b][0.3cm]{17.7cm}{\addtolength{\parindent}{-0.2in}\href{https://www.nndc.bnl.gov/nsr/nsrlink.jsp?2006Sk09,B}{2006Sk09}: \ensuremath{^{\textnormal{1}}}H(\ensuremath{^{\textnormal{18}}}Ne, \ensuremath{^{\textnormal{19}}}Na\ensuremath{\rightarrow}\ensuremath{^{\textnormal{18}}}Ne+p) E=56 MeV. The \ensuremath{^{\textnormal{18}}}Ne beam was purified in the TwinSol magnetic analyzer and bombarded}\\
\parbox[b][0.3cm]{17.7cm}{a thick (CH\ensuremath{_{\textnormal{2}}})\ensuremath{_{\textnormal{n}}} target. Scattered protons were measured at \ensuremath{\theta}\ensuremath{_{\textnormal{lab}}}=7.5\ensuremath{^\circ}, 22.5\ensuremath{^\circ}, and 37.5\ensuremath{^\circ} with an energy resolution of \ensuremath{\sim}30 keV. A}\\
\parbox[b][0.3cm]{17.7cm}{\ensuremath{^{\textnormal{19}}}Na state at E\ensuremath{_{\textnormal{x}}}=0.74 MeV \textit{3} was observed. No more \ensuremath{^{\textnormal{19}}}Na excited states at higher energies were observed.}\\
\parbox[b][0.3cm]{17.7cm}{\addtolength{\parindent}{-0.2in}\href{https://www.nndc.bnl.gov/nsr/nsrlink.jsp?2006AcZY,B}{2006AcZY}, \href{https://www.nndc.bnl.gov/nsr/nsrlink.jsp?2008Pe02,B}{2008Pe02}: \ensuremath{^{\textnormal{1}}}H(\ensuremath{^{\textnormal{18}}}Ne,\ensuremath{^{\textnormal{19}}}Na\ensuremath{\rightarrow}\ensuremath{^{\textnormal{18}}}Ne+p) E=66 MeV; measured the recoiling protons from the \ensuremath{^{\textnormal{19}}}Na decay using an annular}\\
\parbox[b][0.3cm]{17.7cm}{\ensuremath{\Delta}E-E Si telescope covering \ensuremath{\theta}\ensuremath{_{\textnormal{lab}}}=4.7\ensuremath{^\circ}{\textminus}20.2\ensuremath{^\circ} with an overall energy resolution of 105 keV. Elastic and inelastic scattering were}\\
\parbox[b][0.3cm]{17.7cm}{measured at seven angles. Two states of \ensuremath{^{\textnormal{19}}}Na at E\ensuremath{_{\textnormal{c.m.}}}=2.78 and 3.09 MeV were measured and J\ensuremath{^{\ensuremath{\pi}}} values of 5/2\ensuremath{^{\textnormal{+}}} and 3/2\ensuremath{^{\textnormal{+}}} were}\\
\parbox[b][0.3cm]{17.7cm}{assumed, respectively. These states proton decay to \ensuremath{^{\textnormal{18}}}Ne\ensuremath{_{\textnormal{g.s.}}}.}\\
\vspace{12pt}
\underline{$^{18}$Ne Levels}\\
\vspace{0.34cm}
\parbox[b][0.3cm]{17.7cm}{\addtolength{\parindent}{-0.254cm}\ensuremath{\Gamma}\ensuremath{_{\textnormal{p}_{\textnormal{1}}}} refers to the width of the proton transition for the \ensuremath{^{\textnormal{19}}}Na*\ensuremath{\rightarrow}p+\ensuremath{^{\textnormal{18}}}Ne decay (\href{https://www.nndc.bnl.gov/nsr/nsrlink.jsp?2005De15,B}{2005De15}).}\\
\parbox[b][0.3cm]{17.7cm}{\addtolength{\parindent}{-0.254cm}\ensuremath{\Gamma}\ensuremath{_{\textnormal{p}_{\textnormal{2}}}} refers to the width of the proton transition for the \ensuremath{^{\textnormal{18}}}Ne*\ensuremath{\rightarrow}p+\ensuremath{^{\textnormal{17}}}F decay (\href{https://www.nndc.bnl.gov/nsr/nsrlink.jsp?2005De15,B}{2005De15}).}\\
\vspace{0.34cm}
\begin{longtable}{cccccccc@{\extracolsep{\fill}}c}
\multicolumn{2}{c}{E(level)$^{}$}&J$^{\pi}$$^{}$&\multicolumn{2}{c}{\ensuremath{\Gamma} (keV)$^{{\hyperlink{NE2LEVEL1}{b}}}$}&\multicolumn{2}{c}{E\ensuremath{_{\textnormal{p}_{\textnormal{2}}}}(c.m.) (keV)$^{{\hyperlink{NE2LEVEL1}{b}}{\hyperlink{NE2LEVEL2}{c}}}$}&Comments&\\[-.2cm]
\multicolumn{2}{c}{\hrulefill}&\hrulefill&\multicolumn{2}{c}{\hrulefill}&\multicolumn{2}{c}{\hrulefill}&\hrulefill&
\endfirsthead
\multicolumn{1}{r@{}}{0}&\multicolumn{1}{@{}l}{}&\multicolumn{1}{l}{0\ensuremath{^{+}}}&\multicolumn{1}{r@{}}{1}&\multicolumn{1}{@{.}l}{66427 s}&&&\parbox[t][0.3cm]{9.7805805cm}{\raggedright E(level): Populated by the proton decay of (1) the \ensuremath{^{\textnormal{19}}}Na state at\vspace{0.1cm}}&\\
&&&&&&&\parbox[t][0.3cm]{9.7805805cm}{\raggedright {\ }{\ }{\ }E\ensuremath{_{\textnormal{x}}}=740 keV \textit{30} (\href{https://www.nndc.bnl.gov/nsr/nsrlink.jsp?2006Sk09,B}{2006Sk09}), which was also observed as a resonance\vspace{0.1cm}}&\\
&&&&&&&\parbox[t][0.3cm]{9.7805805cm}{\raggedright {\ }{\ }{\ }at E\ensuremath{_{\textnormal{c.m.}}}(p+\ensuremath{^{\textnormal{18}}}Ne)=1066 keV \textit{3} (\href{https://www.nndc.bnl.gov/nsr/nsrlink.jsp?2003An02,B}{2003An02}), and\vspace{0.1cm}}&\\
&&&&&&&\parbox[t][0.3cm]{9.7805805cm}{\raggedright {\ }{\ }{\ }E\ensuremath{_{\textnormal{c.m.}}}(p+\ensuremath{^{\textnormal{18}}}Ne)=1076 keV \textit{6} (\href{https://www.nndc.bnl.gov/nsr/nsrlink.jsp?2005De15,B}{2005De15}); (2) the \ensuremath{^{\textnormal{19}}}Na state at\vspace{0.1cm}}&\\
&&&&&&&\parbox[t][0.3cm]{9.7805805cm}{\raggedright {\ }{\ }{\ }E\ensuremath{_{\textnormal{x}}}=4371 keV \textit{10} (\href{https://www.nndc.bnl.gov/nsr/nsrlink.jsp?2005De15,B}{2005De15}); and (3) the \ensuremath{^{\textnormal{19}}}Na state at E\ensuremath{_{\textnormal{x}}}=4903\vspace{0.1cm}}&\\
&&&&&&&\parbox[t][0.3cm]{9.7805805cm}{\raggedright {\ }{\ }{\ }keV \textit{10} (\href{https://www.nndc.bnl.gov/nsr/nsrlink.jsp?2005De15,B}{2005De15}).\vspace{0.1cm}}&\\
&&&&&&&\parbox[t][0.3cm]{9.7805805cm}{\raggedright J\ensuremath{^{\pi}},T\ensuremath{_{1/2}}: From the \ensuremath{^{\textnormal{18}}}Ne Adopted Levels.\vspace{0.1cm}}&\\
\multicolumn{1}{r@{}}{1887}&\multicolumn{1}{@{.}l}{4}&\multicolumn{1}{l}{2\ensuremath{^{+}}}&\multicolumn{1}{r@{}}{0}&\multicolumn{1}{@{.}l}{47 ps}&&&\parbox[t][0.3cm]{9.7805805cm}{\raggedright E(level): From the Adopted Levels of \ensuremath{^{\textnormal{18}}}Ne. This state is populated by\vspace{0.1cm}}&\\
&&&&&&&\parbox[t][0.3cm]{9.7805805cm}{\raggedright {\ }{\ }{\ }the proton decay of (1) E\ensuremath{_{\textnormal{x}}}(\ensuremath{^{\textnormal{19}}}Na)=2459 keV \textit{32} (\href{https://www.nndc.bnl.gov/nsr/nsrlink.jsp?2008Pe02,B}{2008Pe02}); and (2)\vspace{0.1cm}}&\\
&&&&&&&\parbox[t][0.3cm]{9.7805805cm}{\raggedright {\ }{\ }{\ }E\ensuremath{_{\textnormal{x}}}(\ensuremath{^{\textnormal{19}}}Na)=2769 keV \textit{61} (\href{https://www.nndc.bnl.gov/nsr/nsrlink.jsp?2008Pe02,B}{2008Pe02}). These two \ensuremath{^{\textnormal{19}}}Na states are\vspace{0.1cm}}&\\
&&&&&&&\parbox[t][0.3cm]{9.7805805cm}{\raggedright {\ }{\ }{\ }observed by (\href{https://www.nndc.bnl.gov/nsr/nsrlink.jsp?2008Pe02,B}{2008Pe02}) at E\ensuremath{_{\textnormal{c.m.}}}(p+\ensuremath{^{\textnormal{18}}}Ne)=2.78 MeV \textit{3} and 3.09\vspace{0.1cm}}&\\
&&&&&&&\parbox[t][0.3cm]{9.7805805cm}{\raggedright {\ }{\ }{\ }MeV \textit{6}, respectively.\vspace{0.1cm}}&\\
&&&&&&&\parbox[t][0.3cm]{9.7805805cm}{\raggedright J\ensuremath{^{\pi}},T\ensuremath{_{1/2}}: From the \ensuremath{^{\textnormal{18}}}Ne Adopted Levels.\vspace{0.1cm}}&\\
\multicolumn{1}{r@{}}{4121?}&\multicolumn{1}{@{ }l}{{\it 12}}&&\multicolumn{1}{r@{}}{9}&\multicolumn{1}{@{ }l}{keV {\it +54\textminus9}}&\multicolumn{1}{r@{}}{200}&\multicolumn{1}{@{ }l}{{\it 12}}&\parbox[t][0.3cm]{9.7805805cm}{\raggedright E(level): (\href{https://www.nndc.bnl.gov/nsr/nsrlink.jsp?2005De15,B}{2005De15}) observed a transition at E\ensuremath{_{\textnormal{p}_{\textnormal{1}}}^{\textnormal{c.m.}}}=1698 keV \textit{75}\vspace{0.1cm}}&\\
&&&&&&&\parbox[t][0.3cm]{9.7805805cm}{\raggedright {\ }{\ }{\ }corresponding to the proton decay of a \ensuremath{^{\textnormal{19}}}Na* state at E\ensuremath{_{\textnormal{x}}}=5499 keV\vspace{0.1cm}}&\\
&&&&&&&\parbox[t][0.3cm]{9.7805805cm}{\raggedright {\ }{\ }{\ }\textit{76} to a \ensuremath{^{\textnormal{18}}}Ne* state at E\ensuremath{_{\textnormal{x}}}=4121 keV \textit{12}, whose width was deduced\vspace{0.1cm}}&\\
&&&&&&&\parbox[t][0.3cm]{9.7805805cm}{\raggedright {\ }{\ }{\ }as \ensuremath{\Gamma}=9 keV \textit{+54{\textminus}9}. However, the authors mentioned that no\vspace{0.1cm}}&\\
&&&&&&&\parbox[t][0.3cm]{9.7805805cm}{\raggedright {\ }{\ }{\ }correspondence could be found between the \ensuremath{^{\textnormal{18}}}Ne*(4121 keV) state\vspace{0.1cm}}&\\
&&&&&&&\parbox[t][0.3cm]{9.7805805cm}{\raggedright {\ }{\ }{\ }and any of the known \ensuremath{^{\textnormal{18}}}Ne states. The closest \ensuremath{^{\textnormal{18}}}Ne state has an\vspace{0.1cm}}&\\
&&&&&&&\parbox[t][0.3cm]{9.7805805cm}{\raggedright {\ }{\ }{\ }excitation energy that differs by about 0.4-0.5 MeV. The evaluator\vspace{0.1cm}}&\\
&&&&&&&\parbox[t][0.3cm]{9.7805805cm}{\raggedright {\ }{\ }{\ }notes that the \ensuremath{^{\textnormal{19}}}Na*(5499 keV) state does not seem to correspond to\vspace{0.1cm}}&\\
&&&&&&&\parbox[t][0.3cm]{9.7805805cm}{\raggedright {\ }{\ }{\ }any of the known \ensuremath{^{\textnormal{19}}}Na states either. Considering that the other\vspace{0.1cm}}&\\
\end{longtable}
\begin{textblock}{29}(0,27.3)
Continued on next page (footnotes at end of table)
\end{textblock}
\clearpage
\begin{longtable}{ccccccc@{\extracolsep{\fill}}c}
\\[-.4cm]
\multicolumn{8}{c}{{\bf \small \underline{\ensuremath{^{\textnormal{19}}}Na p decay\hspace{0.2in}\href{https://www.nndc.bnl.gov/nsr/nsrlink.jsp?2003An02,B}{2003An02},\href{https://www.nndc.bnl.gov/nsr/nsrlink.jsp?2008Pe02,B}{2008Pe02} (continued)}}}\\
\multicolumn{8}{c}{~}\\
\multicolumn{8}{c}{\underline{\ensuremath{^{18}}Ne Levels (continued)}}\\
\multicolumn{8}{c}{~}\\
\multicolumn{2}{c}{E(level)$^{}$}&\multicolumn{2}{c}{\ensuremath{\Gamma} (keV)$^{{\hyperlink{NE2LEVEL1}{b}}}$}&\multicolumn{2}{c}{E\ensuremath{_{\textnormal{p}_{\textnormal{2}}}}(c.m.) (keV)$^{{\hyperlink{NE2LEVEL1}{b}}{\hyperlink{NE2LEVEL2}{c}}}$}&Comments&\\[-.2cm]
\multicolumn{2}{c}{\hrulefill}&\multicolumn{2}{c}{\hrulefill}&\multicolumn{2}{c}{\hrulefill}&\hrulefill&
\endhead
&&&&&&\parbox[t][0.3cm]{10.36248cm}{\raggedright {\ }{\ }{\ }transitions observed in this study lined up with the known levels in \ensuremath{^{\textnormal{19}}}Na\vspace{0.1cm}}&\\
&&&&&&\parbox[t][0.3cm]{10.36248cm}{\raggedright {\ }{\ }{\ }and \ensuremath{^{\textnormal{18}}}Ne, the evaluator did not consider this state for the \ensuremath{^{\textnormal{18}}}Ne Adopted\vspace{0.1cm}}&\\
&&&&&&\parbox[t][0.3cm]{10.36248cm}{\raggedright {\ }{\ }{\ }Levels.\vspace{0.1cm}}&\\
&&&&&&\parbox[t][0.3cm]{10.36248cm}{\raggedright A decay to \ensuremath{^{\textnormal{17}}}F\ensuremath{_{\textnormal{g.s.}}} with a proton transition width of \ensuremath{\Gamma}\ensuremath{_{\textnormal{p}_{\textnormal{2}}}}=40 keV \textit{34} was\vspace{0.1cm}}&\\
&&&&&&\parbox[t][0.3cm]{10.36248cm}{\raggedright {\ }{\ }{\ }deduced by (\href{https://www.nndc.bnl.gov/nsr/nsrlink.jsp?2005De15,B}{2005De15}) if this state belongs to \ensuremath{^{\textnormal{18}}}Ne.\vspace{0.1cm}}&\\
\multicolumn{1}{r@{}}{4481}&\multicolumn{1}{@{ }l}{{\it 11}}&\multicolumn{1}{r@{}}{27}&\multicolumn{1}{@{ }l}{keV {\it +6\textminus9}}&\multicolumn{1}{r@{}}{560}&\multicolumn{1}{@{ }l}{{\it 11}}&\parbox[t][0.3cm]{10.36248cm}{\raggedright E(level): From (\href{https://www.nndc.bnl.gov/nsr/nsrlink.jsp?2005De15,B}{2005De15}): reported that this state may be either the\vspace{0.1cm}}&\\
&&&&&&\parbox[t][0.3cm]{10.36248cm}{\raggedright {\ }{\ }{\ }\ensuremath{^{\textnormal{18}}}Ne*(4517 keV, 1\ensuremath{^{-}}) level, or the \ensuremath{^{\textnormal{18}}}Ne*(4524 keV, 3\ensuremath{^{\textnormal{+}}}) state. Due to\vspace{0.1cm}}&\\
&&&&&&\parbox[t][0.3cm]{10.36248cm}{\raggedright {\ }{\ }{\ }uncertainty in identification of this state, we did not consider this state for\vspace{0.1cm}}&\\
&&&&&&\parbox[t][0.3cm]{10.36248cm}{\raggedright {\ }{\ }{\ }the Adopted Levels.\vspace{0.1cm}}&\\
&&&&&&\parbox[t][0.3cm]{10.36248cm}{\raggedright Decays to \ensuremath{^{\textnormal{17}}}F\ensuremath{_{\textnormal{g.s.}}} with a proton transition width of \ensuremath{\Gamma}\ensuremath{_{\textnormal{p}_{\textnormal{2}}}}=47 keV \textit{4}\vspace{0.1cm}}&\\
&&&&&&\parbox[t][0.3cm]{10.36248cm}{\raggedright {\ }{\ }{\ }(\href{https://www.nndc.bnl.gov/nsr/nsrlink.jsp?2005De15,B}{2005De15}).\vspace{0.1cm}}&\\
\multicolumn{1}{r@{}}{4590}&\multicolumn{1}{@{}l}{\ensuremath{^{{\hyperlink{NE2LEVEL0}{a}}}}}&\multicolumn{1}{r@{}}{0}&\multicolumn{1}{@{ }l}{keV {\it +37\textminus0}}&\multicolumn{1}{r@{}}{156}&\multicolumn{1}{@{ }l}{{\it 12}}&\parbox[t][0.3cm]{10.36248cm}{\raggedright Decays to the \ensuremath{^{\textnormal{17}}}F*(459 keV) state with a proton transition width of\vspace{0.1cm}}&\\
&&&&&&\parbox[t][0.3cm]{10.36248cm}{\raggedright {\ }{\ }{\ }\ensuremath{\Gamma}\ensuremath{_{\textnormal{p}_{\textnormal{2}}}}=39 keV \textit{15} (\href{https://www.nndc.bnl.gov/nsr/nsrlink.jsp?2005De15,B}{2005De15}).\vspace{0.1cm}}&\\
\multicolumn{1}{r@{}}{5117}&\multicolumn{1}{@{ }l}{{\it 11}}&\multicolumn{1}{r@{}}{33}&\multicolumn{1}{@{ }l}{keV {\it +9\textminus11}}&\multicolumn{1}{r@{}}{1196}&\multicolumn{1}{@{ }l}{{\it 11}}&\parbox[t][0.3cm]{10.36248cm}{\raggedright E(level): From (\href{https://www.nndc.bnl.gov/nsr/nsrlink.jsp?2005De15,B}{2005De15}): reported that this state may be either the\vspace{0.1cm}}&\\
&&&&&&\parbox[t][0.3cm]{10.36248cm}{\raggedright {\ }{\ }{\ }\ensuremath{^{\textnormal{18}}}Ne*(5100 keV, 2\ensuremath{^{\textnormal{+}}}) level, or the \ensuremath{^{\textnormal{18}}}Ne*(5142 keV, 3\ensuremath{^{-}}) state. Due to\vspace{0.1cm}}&\\
&&&&&&\parbox[t][0.3cm]{10.36248cm}{\raggedright {\ }{\ }{\ }uncertainty in identification of this state, we did not consider this state for\vspace{0.1cm}}&\\
&&&&&&\parbox[t][0.3cm]{10.36248cm}{\raggedright {\ }{\ }{\ }the Adopted Levels.\vspace{0.1cm}}&\\
&&&&&&\parbox[t][0.3cm]{10.36248cm}{\raggedright Decays to \ensuremath{^{\textnormal{17}}}F\ensuremath{_{\textnormal{g.s.}}} with a proton transition width of \ensuremath{\Gamma}\ensuremath{_{\textnormal{p}_{\textnormal{2}}}}=51 keV \textit{6}\vspace{0.1cm}}&\\
&&&&&&\parbox[t][0.3cm]{10.36248cm}{\raggedright {\ }{\ }{\ }(\href{https://www.nndc.bnl.gov/nsr/nsrlink.jsp?2005De15,B}{2005De15}).\vspace{0.1cm}}&\\
\end{longtable}
\parbox[b][0.3cm]{17.7cm}{\makebox[1ex]{\ensuremath{^{\hypertarget{NE2LEVEL0}{a}}}} From the Adopted Levels of \ensuremath{^{\textnormal{18}}}Ne assuming proton decay from the \ensuremath{^{\textnormal{19}}}Na*(5809 keV) state to the \ensuremath{^{\textnormal{18}}}Ne(4590 keV) as suggested}\\
\parbox[b][0.3cm]{17.7cm}{{\ }{\ }by (\href{https://www.nndc.bnl.gov/nsr/nsrlink.jsp?2005De15,B}{2005De15}: see text and Fig. 8).}\\
\parbox[b][0.3cm]{17.7cm}{\makebox[1ex]{\ensuremath{^{\hypertarget{NE2LEVEL1}{b}}}} From (\href{https://www.nndc.bnl.gov/nsr/nsrlink.jsp?2005De15,B}{2005De15}).}\\
\parbox[b][0.3cm]{17.7cm}{\makebox[1ex]{\ensuremath{^{\hypertarget{NE2LEVEL2}{c}}}} This is the center-of-mass energy of the proton emitted from the \ensuremath{^{\textnormal{18}}}Ne*\ensuremath{\rightarrow}p+\ensuremath{^{\textnormal{17}}}F\ensuremath{_{\textnormal{g.s.}}} decay (\href{https://www.nndc.bnl.gov/nsr/nsrlink.jsp?2005De15,B}{2005De15}).}\\
\vspace{0.5cm}
\underline{Protons ($^{18}$Ne)}\\
\begin{longtable}{cccccccc@{\extracolsep{\fill}}c}
\multicolumn{2}{c}{E\ensuremath{_{\textnormal{p}_{\textnormal{1}}}}(c.m.) (keV)\ensuremath{^{\hyperlink{NA2DELAY0}{a}\hyperlink{NA2DELAY1}{b}}}}&\multicolumn{2}{c}{E($^{18}$Ne)}&E\ensuremath{_{\textnormal{x}}}(\ensuremath{^{\textnormal{19}}}Na) (keV)\ensuremath{^{\hyperlink{NA2DELAY0}{a}}}&\multicolumn{2}{c}{\ensuremath{\Gamma}\ensuremath{_{\textnormal{p}_{\textnormal{1}}}} (keV)\ensuremath{^{\hyperlink{NA2DELAY0}{a}}}}&Comments&\\[-.2cm]
\multicolumn{2}{c}{\hrulefill}&\multicolumn{2}{c}{\hrulefill}&\hrulefill&\multicolumn{2}{c}{\hrulefill}&\hrulefill&
\endfirsthead
\multicolumn{1}{r@{}}{1018}&\multicolumn{1}{@{ }l}{{\it 13}}&\multicolumn{1}{r@{}}{5117}&\multicolumn{1}{@{}l}{}&\multicolumn{1}{l}{5815}&\multicolumn{1}{r@{}}{154}&\multicolumn{1}{@{ }l}{{\it 17}}&\parbox[t][0.3cm]{8.21196cm}{\raggedright E\ensuremath{_{\textnormal{x}}}(\ensuremath{^{\textnormal{19}}}Na) (keV): E\ensuremath{_{\textnormal{x}}}(\ensuremath{^{\textnormal{19}}}Na)=5815 keV \textit{17} with\vspace{0.1cm}}&\\
&&&&&&&\parbox[t][0.3cm]{8.21196cm}{\raggedright {\ }{\ }{\ }\ensuremath{\Gamma}(\ensuremath{^{\textnormal{19}}}Na)=141 keV \textit{18} (\href{https://www.nndc.bnl.gov/nsr/nsrlink.jsp?2005De15,B}{2005De15}).\vspace{0.1cm}}&\\
\multicolumn{1}{r@{}}{1424}&\multicolumn{1}{@{ }l}{{\it 30}}&\multicolumn{1}{r@{}}{4481}&\multicolumn{1}{@{}l}{}&\multicolumn{1}{l}{5585}&\multicolumn{1}{r@{}}{697}&\multicolumn{1}{@{ }l}{{\it 72}}&\parbox[t][0.3cm]{8.21196cm}{\raggedright E\ensuremath{_{\textnormal{x}}}(\ensuremath{^{\textnormal{19}}}Na) (keV): E\ensuremath{_{\textnormal{x}}}(\ensuremath{^{\textnormal{19}}}Na)=5585 keV \textit{32} with\vspace{0.1cm}}&\\
&&&&&&&\parbox[t][0.3cm]{8.21196cm}{\raggedright {\ }{\ }{\ }\ensuremath{\Gamma}(\ensuremath{^{\textnormal{19}}}Na)=695 keV \textit{72} (\href{https://www.nndc.bnl.gov/nsr/nsrlink.jsp?2005De15,B}{2005De15}).\vspace{0.1cm}}&\\
\multicolumn{1}{r@{}}{1557}&\multicolumn{1}{@{ }l}{{\it 66}}&\multicolumn{1}{r@{}}{4590}&\multicolumn{1}{@{}l}{}&\multicolumn{1}{l}{5809}&\multicolumn{1}{r@{}}{463}&\multicolumn{1}{@{ }l}{{\it 215}}&\parbox[t][0.3cm]{8.21196cm}{\raggedright E\ensuremath{_{\textnormal{x}}}(\ensuremath{^{\textnormal{19}}}Na) (keV): E\ensuremath{_{\textnormal{x}}}(\ensuremath{^{\textnormal{19}}}Na)=5809 keV \textit{76} with\vspace{0.1cm}}&\\
&&&&&&&\parbox[t][0.3cm]{8.21196cm}{\raggedright {\ }{\ }{\ }\ensuremath{\Gamma}(\ensuremath{^{\textnormal{19}}}Na)=460 keV \textit{215} (\href{https://www.nndc.bnl.gov/nsr/nsrlink.jsp?2005De15,B}{2005De15}).\vspace{0.1cm}}&\\
\multicolumn{1}{r@{}}{1698}&\multicolumn{1}{@{ }l}{{\it 75}}&\multicolumn{1}{r@{}}{4121?}&\multicolumn{1}{@{}l}{}&\multicolumn{1}{l}{5499}&\multicolumn{1}{r@{}}{541}&\multicolumn{1}{@{ }l}{{\it 178}}&\parbox[t][0.3cm]{8.21196cm}{\raggedright E\ensuremath{_{\textnormal{x}}}(\ensuremath{^{\textnormal{19}}}Na) (keV): E\ensuremath{_{\textnormal{x}}}(\ensuremath{^{\textnormal{19}}}Na)=5499 keV \textit{76} with\vspace{0.1cm}}&\\
&&&&&&&\parbox[t][0.3cm]{8.21196cm}{\raggedright {\ }{\ }{\ }\ensuremath{\Gamma}(\ensuremath{^{\textnormal{19}}}Na)=539 kV \textit{180} (\href{https://www.nndc.bnl.gov/nsr/nsrlink.jsp?2005De15,B}{2005De15}). Note that the energy\vspace{0.1cm}}&\\
&&&&&&&\parbox[t][0.3cm]{8.21196cm}{\raggedright {\ }{\ }{\ }of this \ensuremath{^{\textnormal{19}}}Na state is not in agreement within 1\ensuremath{\sigma} with any\vspace{0.1cm}}&\\
&&&&&&&\parbox[t][0.3cm]{8.21196cm}{\raggedright {\ }{\ }{\ }other \ensuremath{^{\textnormal{19}}}Na states.\vspace{0.1cm}}&\\
\end{longtable}
\parbox[b][0.3cm]{17.7cm}{\makebox[1ex]{\ensuremath{^{\hypertarget{NA2DELAY0}{a}}}} From (\href{https://www.nndc.bnl.gov/nsr/nsrlink.jsp?2005De15,B}{2005De15}).}\\
\parbox[b][0.3cm]{17.7cm}{\makebox[1ex]{\ensuremath{^{\hypertarget{NA2DELAY1}{b}}}} This is the center-of-mass energy of the proton emitted from the \ensuremath{^{\textnormal{19}}}Na*\ensuremath{\rightarrow}p+\ensuremath{^{\textnormal{18}}}Ne* decay (\href{https://www.nndc.bnl.gov/nsr/nsrlink.jsp?2005De15,B}{2005De15}).}\\
\vspace{0.5cm}
\clearpage
\begin{figure}[h]
\begin{center}
\includegraphics{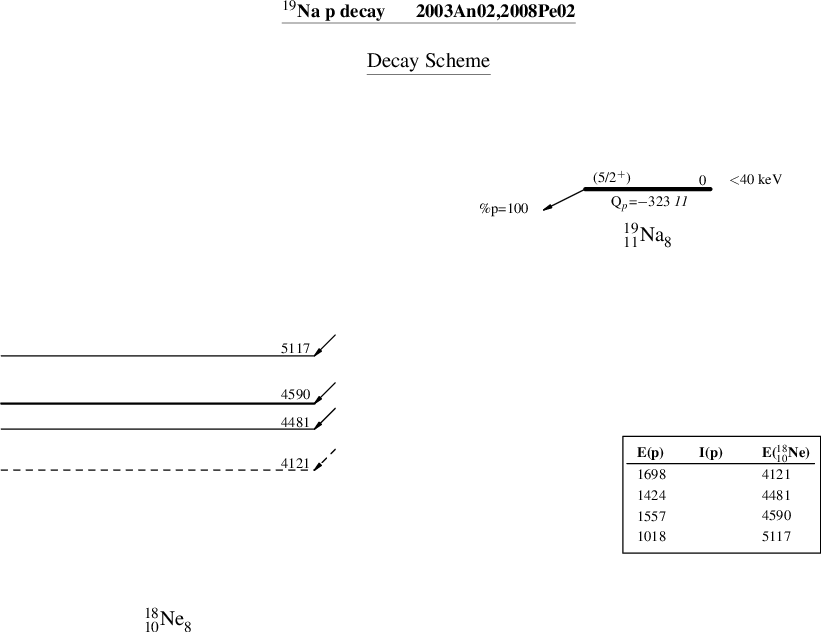}\\
\end{center}
\end{figure}
\clearpage
\subsection[\hspace{-0.2cm}\ensuremath{^{\textnormal{1}}}H(\ensuremath{^{\textnormal{17}}}F,\ensuremath{\gamma})]{ }
\vspace{-27pt}
\vspace{0.3cm}
\hypertarget{NE3}{{\bf \small \underline{\ensuremath{^{\textnormal{1}}}H(\ensuremath{^{\textnormal{17}}}F,\ensuremath{\gamma})\hspace{0.2in}\href{https://www.nndc.bnl.gov/nsr/nsrlink.jsp?1987Wi11,B}{1987Wi11},\href{https://www.nndc.bnl.gov/nsr/nsrlink.jsp?2014Al05,B}{2014Al05}}}}\\
\vspace{4pt}
\vspace{8pt}
\parbox[b][0.3cm]{17.7cm}{\addtolength{\parindent}{-0.2in}\href{https://www.nndc.bnl.gov/nsr/nsrlink.jsp?2005Fi01,B}{2005Fi01}, K. Chipps, Ph.D. Thesis, Colorado School of Mines (2008), \href{https://www.nndc.bnl.gov/nsr/nsrlink.jsp?2009Ch17,B}{2009Ch17}, \href{https://www.nndc.bnl.gov/nsr/nsrlink.jsp?2009Ch64,B}{2009Ch64}, \href{https://www.nndc.bnl.gov/nsr/nsrlink.jsp?2009Ba59,B}{2009Ba59}, \href{https://www.nndc.bnl.gov/nsr/nsrlink.jsp?2009ChZW,B}{2009ChZW}, J. R Beene \textit{et}}\\
\parbox[b][0.3cm]{17.7cm}{\textit{al}., J. Phys. G: Nucl. Part. Phys. 38 (2011) 024002: \ensuremath{^{\textnormal{1}}}H(\ensuremath{^{\textnormal{17}}}F,\ensuremath{\gamma}) E=10.83 MeV; measured the \ensuremath{\gamma}-\ensuremath{^{\textnormal{18}}}Ne* coincidences using a plastic}\\
\parbox[b][0.3cm]{17.7cm}{scintillator array surrounding the target and an ionization chamber at the focal plane of the Daresbury Recoil Separator. Measured}\\
\parbox[b][0.3cm]{17.7cm}{the resonance strength and deduced the partial \ensuremath{\gamma}-ray width of the 599.8-keV resonance using \ensuremath{\Gamma}\ensuremath{_{\textnormal{p}}}=18 keV \textit{2} (stat.) \textit{1} (sys.) from}\\
\parbox[b][0.3cm]{17.7cm}{(\href{https://www.nndc.bnl.gov/nsr/nsrlink.jsp?2000Bb04,B}{2000Bb04}). Reevaluated the \ensuremath{^{\textnormal{17}}}F(p,\ensuremath{\gamma})\ensuremath{^{\textnormal{18}}}Ne reaction rate and discussed its astrophysical impact.}\\
\vspace{0.385cm}
\parbox[b][0.3cm]{17.7cm}{\addtolength{\parindent}{-0.2in}\textit{Theory:}}\\
\parbox[b][0.3cm]{17.7cm}{\addtolength{\parindent}{-0.2in}\href{https://www.nndc.bnl.gov/nsr/nsrlink.jsp?2014Al05,B}{2014Al05}: \ensuremath{^{\textnormal{13}}}C(\ensuremath{^{\textnormal{17}}}O,\ensuremath{^{\textnormal{18}}}O) 12 MeV/nucleon; the neutron transfer measurement was performed to study \ensuremath{^{\textnormal{18}}}O, the mirror of \ensuremath{^{\textnormal{18}}}Ne. The}\\
\parbox[b][0.3cm]{17.7cm}{elastic scattering measurements were performed to extract the optical model parameters for the DWBA analysis of the neutron}\\
\parbox[b][0.3cm]{17.7cm}{transfer data. The elastic scattering angular distributions were measured at \ensuremath{\theta}\ensuremath{_{\textnormal{lab}}}=4\ensuremath{^\circ}{\textminus}25\ensuremath{^\circ} using the MDM magnetic spectrometer. The}\\
\parbox[b][0.3cm]{17.7cm}{same apparatus measured the neutron transfer data at \ensuremath{\theta}\ensuremath{_{\textnormal{lab}}}=4\ensuremath{^\circ}{\textminus}11\ensuremath{^\circ}. In a second set of measurements, a 216-MeV beam of \ensuremath{^{\textnormal{18}}}O}\\
\parbox[b][0.3cm]{17.7cm}{impinged on a \ensuremath{^{\textnormal{12}}}C enriched target and the absolute elastic scattering cross section was measured at \ensuremath{\theta}\ensuremath{_{\textnormal{lab}}}=4\ensuremath{^\circ}{\textminus}22\ensuremath{^\circ}. The Asymptotic}\\
\parbox[b][0.3cm]{17.7cm}{Normalization Coefficients were deduced from the DWBA analysis of the neutron transfer data. The ANCs were deduced for the}\\
\parbox[b][0.3cm]{17.7cm}{\ensuremath{^{\textnormal{18}}}Ne states assuming equality of the spectroscopic factors. The astrophysical S-factor and the \ensuremath{^{\textnormal{17}}}F(p,\ensuremath{\gamma}) reaction rate are calculated,}\\
\parbox[b][0.3cm]{17.7cm}{and the astrophysical implications are discussed.}\\
\vspace{0.385cm}
\parbox[b][0.3cm]{17.7cm}{\addtolength{\parindent}{-0.2in}\textit{The \ensuremath{^{18}}F(p,\ensuremath{\gamma}) Astrophysical Reaction Rate}:}\\
\parbox[b][0.3cm]{17.7cm}{\addtolength{\parindent}{-0.2in}This subsection contains experimental and theoretical studies that are focused on the determination of the \ensuremath{^{\textnormal{18}}}F(p,\ensuremath{\gamma}) astrophysical}\\
\parbox[b][0.3cm]{17.7cm}{rate. The details of the experimental studies can be found in other sections. Only a summary of the important findings with regards}\\
\parbox[b][0.3cm]{17.7cm}{to the reaction rate are mentioned here.}\\
\parbox[b][0.3cm]{17.7cm}{\addtolength{\parindent}{-0.2in}\href{https://www.nndc.bnl.gov/nsr/nsrlink.jsp?1979Wo07,B}{1979Wo07}: Compiled, evaluated available semi-empirical thermonuclear reaction data for A=16-74. Parameterized analytic formulas}\\
\parbox[b][0.3cm]{17.7cm}{are presented.}\\
\parbox[b][0.3cm]{17.7cm}{\addtolength{\parindent}{-0.2in}M. Wiescher \textit{et al}., Astrophys. J. 263 (1982) 891: The cross sections for direct proton capture transitions to two bound states in}\\
\parbox[b][0.3cm]{17.7cm}{\ensuremath{^{\textnormal{18}}}Ne at E\ensuremath{_{\textnormal{x}}}=1887 and 3616 keV were calculated using the direct capture model described by (\href{https://www.nndc.bnl.gov/nsr/nsrlink.jsp?1973Ro34,B}{1973Ro34}). The total S-factor was}\\
\parbox[b][0.3cm]{17.7cm}{computed. The resonant rate of the \ensuremath{^{\textnormal{17}}}F(p,\ensuremath{\gamma}) reaction was calculated based on the contributions of two known proton resonances}\\
\parbox[b][0.3cm]{17.7cm}{corresponding to the \ensuremath{^{\textnormal{18}}}Ne*(4.52 MeV, 1\ensuremath{^{-}}) and \ensuremath{^{\textnormal{18}}}Ne*(4.59 MeV, 0\ensuremath{^{\textnormal{+}}}) levels. The total \ensuremath{^{\textnormal{17}}}F(p,\ensuremath{\gamma}) reaction rate was calculated.}\\
\parbox[b][0.3cm]{17.7cm}{Comparison with literature reaction rate is discussed.}\\
\parbox[b][0.3cm]{17.7cm}{\addtolength{\parindent}{-0.2in}\href{https://www.nndc.bnl.gov/nsr/nsrlink.jsp?1987Wi11,B}{1987Wi11}, \href{https://www.nndc.bnl.gov/nsr/nsrlink.jsp?1988Wi08,B}{1988Wi08}: Constructed the T=1 analog states in \ensuremath{^{\textnormal{18}}}O, \ensuremath{^{\textnormal{18}}}F and \ensuremath{^{\textnormal{18}}}Ne. From these, it was evident that a J\ensuremath{^{\ensuremath{\pi}}}=3\ensuremath{^{\textnormal{+}}} proton}\\
\parbox[b][0.3cm]{17.7cm}{resonance in \ensuremath{^{\textnormal{18}}}Ne had not been observed experimentally. Calculated the excitation energy of this missing state using}\\
\parbox[b][0.3cm]{17.7cm}{Thomas-Ehrman shift calculations (and obtained E\ensuremath{_{\textnormal{x}}}=4.31 MeV); and shell model calculations, which yielded E\ensuremath{_{\textnormal{x}}}=4.33 MeV. The}\\
\parbox[b][0.3cm]{17.7cm}{authors deduced a value of E\ensuremath{_{\textnormal{x}}}(\ensuremath{^{\textnormal{18}}}Ne)=4328 keV \textit{40} (with the uncertainty coming from the nuclear structure input); estimated}\\
\parbox[b][0.3cm]{17.7cm}{\ensuremath{\Gamma}\ensuremath{\sim}\ensuremath{\Gamma}\ensuremath{_{\textnormal{p}}}=5 keV; and noted that these values lie sufficiently close to the \ensuremath{^{\textnormal{17}}}F+p threshold that the 3\ensuremath{^{\textnormal{+}}} resonance would greatly}\\
\parbox[b][0.3cm]{17.7cm}{enhance the \ensuremath{^{\textnormal{17}}}F(p,\ensuremath{\gamma}) reaction rate. Proton and \ensuremath{\gamma}-ray partial widths, resonance strength, and non-resonant direct capture}\\
\parbox[b][0.3cm]{17.7cm}{contribution were estimated for this \textit{s}-wave resonance. Recalculated the \ensuremath{^{\textnormal{17}}}F(p,\ensuremath{\gamma}) reaction rate.}\\
\parbox[b][0.3cm]{17.7cm}{\addtolength{\parindent}{-0.2in}\href{https://www.nndc.bnl.gov/nsr/nsrlink.jsp?1989Or02,B}{1989Or02}: A shell-model analysis of Coulomb energies of the A=10-55 nuclei predicted the energy of the experimentally}\\
\parbox[b][0.3cm]{17.7cm}{unobserved 3\ensuremath{^{\textnormal{+}}} state in \ensuremath{^{\textnormal{18}}}Ne to be 4.47 MeV from estimating the Coulomb energy shift between \ensuremath{^{\textnormal{18}}}O and \ensuremath{^{\textnormal{18}}}Ne mirrors using the}\\
\parbox[b][0.3cm]{17.7cm}{charge-dependent single-particle energies and two-body matrix elements.}\\
\parbox[b][0.3cm]{17.7cm}{\addtolength{\parindent}{-0.2in}\href{https://www.nndc.bnl.gov/nsr/nsrlink.jsp?1991Ga03,B}{1991Ga03}: Estimated the energy of the experimentally unobserved 3\ensuremath{^{\textnormal{+}}} state to be 4.53 MeV with \ensuremath{\Gamma}=22 keV by calculating the}\\
\parbox[b][0.3cm]{17.7cm}{wave functions for a Woods-Saxon well and a 2\textit{s}\ensuremath{_{\textnormal{1/2}}} resonance for \ensuremath{^{\textnormal{17}}}F\ensuremath{_{\textnormal{g.s.}}}+p. The authors then measured some excess counts at one}\\
\parbox[b][0.3cm]{17.7cm}{angle, and attributed these counts to the missing 3\ensuremath{^{\textnormal{+}}} state, for which they inferred E\ensuremath{_{\textnormal{x}}}=4561 keV \textit{9} (see the \ensuremath{^{\textnormal{16}}}O(\ensuremath{^{\textnormal{3}}}He,n) section) and}\\
\parbox[b][0.3cm]{17.7cm}{estimated its proton partial width to be \ensuremath{\Gamma}=25 keV. From the mirror level, they found \ensuremath{\Gamma}\ensuremath{_{\ensuremath{\gamma}}}(3\ensuremath{^{\textnormal{+}}_{\textnormal{1}}}\ensuremath{\rightarrow}2\ensuremath{^{\textnormal{+}}_{\textnormal{1}}})=25 meV \textit{16},}\\
\parbox[b][0.3cm]{17.7cm}{\ensuremath{\Gamma}\ensuremath{_{\ensuremath{\gamma}}}(3\ensuremath{^{\textnormal{+}}_{\textnormal{1}}}\ensuremath{\rightarrow}2\ensuremath{^{\textnormal{+}}_{\textnormal{2}}})=3.8 meV \textit{31}, \ensuremath{\Gamma}\ensuremath{_{\ensuremath{\gamma}}}(3\ensuremath{^{\textnormal{+}}_{\textnormal{1}}}\ensuremath{\rightarrow}4\ensuremath{^{\textnormal{+}}_{\textnormal{1}}})=0.8 meV \textit{8}. They also estimated the partial widths for the 1\ensuremath{^{-}_{\textnormal{1}}} and 0\ensuremath{^{\textnormal{+}}_{\textnormal{3}}} states. The}\\
\parbox[b][0.3cm]{17.7cm}{astrophysical S-factor at E\ensuremath{_{\textnormal{c.m.}}}=100-800 keV, and the \ensuremath{^{\textnormal{17}}}F(p,\ensuremath{\gamma}) reaction rate for T=0.1-1 GK were computed.}\\
\parbox[b][0.3cm]{17.7cm}{\addtolength{\parindent}{-0.2in}\href{https://www.nndc.bnl.gov/nsr/nsrlink.jsp?1992Ch50,B}{1992Ch50}: Provides an extensive review of the \ensuremath{^{\textnormal{17}}}F(p,\ensuremath{\gamma}) reaction rate and the structure of \ensuremath{^{\textnormal{18}}}Ne based on the available data at the}\\
\parbox[b][0.3cm]{17.7cm}{time.}\\
\parbox[b][0.3cm]{17.7cm}{\addtolength{\parindent}{-0.2in}\href{https://www.nndc.bnl.gov/nsr/nsrlink.jsp?1997Ba57,B}{1997Ba57}: Calculated the astrophysical S-factor for the \ensuremath{^{\textnormal{17}}}F(p,\ensuremath{\gamma}) reaction at E\ensuremath{_{\textnormal{c.m.}}}=50-1000 keV, and computed the REACLIB}\\
\parbox[b][0.3cm]{17.7cm}{format for the \ensuremath{^{\textnormal{17}}}F(p,\ensuremath{\gamma}) rate. Astrophysical implications are discussed.}\\
\parbox[b][0.3cm]{17.7cm}{\addtolength{\parindent}{-0.2in}\href{https://www.nndc.bnl.gov/nsr/nsrlink.jsp?1998Sh35,B}{1998Sh35}: Computed Coulomb energies for positive-parity levels in the mirror nuclei and predicted the excitation energy and width}\\
\parbox[b][0.3cm]{17.7cm}{of the astrophysically important 3\ensuremath{^{\textnormal{+}}} level at 4642 keV and \ensuremath{\Gamma}=42 keV, respectively.}\\
\parbox[b][0.3cm]{17.7cm}{\addtolength{\parindent}{-0.2in}\href{https://www.nndc.bnl.gov/nsr/nsrlink.jsp?1999Ba49,B}{1999Ba49}, \href{https://www.nndc.bnl.gov/nsr/nsrlink.jsp?2000Bb04,B}{2000Bb04}: Conclusively observed the astrophysically important 3\ensuremath{^{\textnormal{+}}} state by measuring the excitation function for the}\\
\parbox[b][0.3cm]{17.7cm}{\ensuremath{^{\textnormal{17}}}F(p,p) elastic scattering using a radioactive \ensuremath{^{\textnormal{17}}}F beam (see the \ensuremath{^{\textnormal{1}}}H(\ensuremath{^{\textnormal{17}}}F,p) section). The authors measured the excitation energy}\\
\parbox[b][0.3cm]{17.7cm}{and the width of this state. The \ensuremath{^{\textnormal{17}}}F(p,\ensuremath{\gamma}) resonant reaction rate, and the total reaction rate at T=0.1-2 GK were computed. The}\\
\parbox[b][0.3cm]{17.7cm}{properties of the resonances used in the rate calculation, the REACLIB format, and the astrophysical implications of this rate are}\\
\clearpage
\vspace{0.3cm}
{\bf \small \underline{\ensuremath{^{\textnormal{1}}}H(\ensuremath{^{\textnormal{17}}}F,\ensuremath{\gamma})\hspace{0.2in}\href{https://www.nndc.bnl.gov/nsr/nsrlink.jsp?1987Wi11,B}{1987Wi11},\href{https://www.nndc.bnl.gov/nsr/nsrlink.jsp?2014Al05,B}{2014Al05} (continued)}}\\
\vspace{0.3cm}
\parbox[b][0.3cm]{17.7cm}{presented.}\\
\parbox[b][0.3cm]{17.7cm}{\addtolength{\parindent}{-0.2in}S. T. Parete-Koon, M. Sc. Thesis, \textit{Reaction Rate of} \textit{\ensuremath{^{17}}F(p,\ensuremath{\gamma})\ensuremath{^{\textnormal{18}}}Ne and Its Implications for Nova Nucleosynthesis}, The University of}\\
\parbox[b][0.3cm]{17.7cm}{Tennessee, Knoxville (2001): The \ensuremath{^{\textnormal{17}}}F(p,\ensuremath{\gamma}) reaction rates of (\href{https://www.nndc.bnl.gov/nsr/nsrlink.jsp?1999Ba49,B}{1999Ba49}, \href{https://www.nndc.bnl.gov/nsr/nsrlink.jsp?1988Wi08,B}{1988Wi08}, \href{https://www.nndc.bnl.gov/nsr/nsrlink.jsp?1998Sh35,B}{1998Sh35}) were used to perform nova}\\
\parbox[b][0.3cm]{17.7cm}{nucleosynthesis calculations for 3 nova models, which are described and compared. The (\href{https://www.nndc.bnl.gov/nsr/nsrlink.jsp?1999Ba49,B}{1999Ba49}) rate was observed to have a}\\
\parbox[b][0.3cm]{17.7cm}{significant impact on the \ensuremath{^{\textnormal{17}}}O yields.}\\
\parbox[b][0.3cm]{17.7cm}{\addtolength{\parindent}{-0.2in}\href{https://www.nndc.bnl.gov/nsr/nsrlink.jsp?2002Il05,B}{2002Il05}: Recommended the reaction rate of (\href{https://www.nndc.bnl.gov/nsr/nsrlink.jsp?2000Bb04,B}{2000Bb04}), varied this rate by one order of magnitude up and down, and studied its}\\
\parbox[b][0.3cm]{17.7cm}{effect on nova nucleosynthesis.}\\
\parbox[b][0.3cm]{17.7cm}{\addtolength{\parindent}{-0.2in}\href{https://www.nndc.bnl.gov/nsr/nsrlink.jsp?2002SmZX,B}{2002SmZX}: Performed nova nucleosynthesis calculations. Examined the influence on nova nucleosynthesis of two reaction rates:}\\
\parbox[b][0.3cm]{17.7cm}{\ensuremath{^{\textnormal{17}}}F(p,\ensuremath{\gamma}) and \ensuremath{^{\textnormal{14}}}O(\ensuremath{\alpha},2p). Concluded that the \ensuremath{^{\textnormal{17}}}F(p,\ensuremath{\gamma}) reaction rate plays a significant role on the production of radioisotopes that}\\
\parbox[b][0.3cm]{17.7cm}{may be observable tracers of novae.}\\
\parbox[b][0.3cm]{17.7cm}{\addtolength{\parindent}{-0.2in}\href{https://www.nndc.bnl.gov/nsr/nsrlink.jsp?2003Pa54,B}{2003Pa54}: Reviewed the \ensuremath{^{\textnormal{17}}}F(p,\ensuremath{\gamma}) reaction rate, recalculated this rate for T=0.1-0.9 GK based on the information from (\href{https://www.nndc.bnl.gov/nsr/nsrlink.jsp?1988Wi08,B}{1988Wi08},}\\
\parbox[b][0.3cm]{17.7cm}{\href{https://www.nndc.bnl.gov/nsr/nsrlink.jsp?1991Ga03,B}{1991Ga03}, \href{https://www.nndc.bnl.gov/nsr/nsrlink.jsp?1998Sh35,B}{1998Sh35}, and \href{https://www.nndc.bnl.gov/nsr/nsrlink.jsp?1999Ba49,B}{1999Ba49}). Performed nova nucleosynthesis calculations for 3 different nova models and discussed the}\\
\parbox[b][0.3cm]{17.7cm}{implications of the \ensuremath{^{\textnormal{17}}}F(p,\ensuremath{\gamma}) reaction rate for nova nucleosynthesis.}\\
\parbox[b][0.3cm]{17.7cm}{\addtolength{\parindent}{-0.2in}\href{https://www.nndc.bnl.gov/nsr/nsrlink.jsp?2004De31,B}{2004De31}: Calculated the \ensuremath{^{\textnormal{17}}}F(p,\ensuremath{\gamma}) astrophysical S-factor for E\ensuremath{_{\textnormal{c.m.}}}\ensuremath{<}3 MeV using the generator coordinate method with an}\\
\parbox[b][0.3cm]{17.7cm}{extended microscopic two-cluster model.}\\
\parbox[b][0.3cm]{17.7cm}{\addtolength{\parindent}{-0.2in}\href{https://www.nndc.bnl.gov/nsr/nsrlink.jsp?2004Du02,B}{2004Du02}: Investigated the \ensuremath{^{\textnormal{17}}}F(p,\ensuremath{\gamma}) reaction rate using a two-cluster generator coordinate method. Estimated the \ensuremath{^{\textnormal{18}}}Ne levels, J\ensuremath{^{\ensuremath{\pi}}}}\\
\parbox[b][0.3cm]{17.7cm}{values, E2 and M1 reduced transition probabilities (in W.u.), and the corresponding \ensuremath{\Gamma}\ensuremath{_{\ensuremath{\gamma}}} and \ensuremath{\Gamma}\ensuremath{_{\textnormal{p}}} widths for \ensuremath{^{\textnormal{18}}}Ne resonances.}\\
\parbox[b][0.3cm]{17.7cm}{Computed the astrophysical S-factor for E\ensuremath{_{\textnormal{c.m.}}}=0.1-1.2 MeV and the \ensuremath{^{\textnormal{17}}}F(p,\ensuremath{\gamma}) reaction rate for T=0.1-2 GK.}\\
\parbox[b][0.3cm]{17.7cm}{\addtolength{\parindent}{-0.2in}\href{https://www.nndc.bnl.gov/nsr/nsrlink.jsp?2005Pa50,B}{2005Pa50}: Measured the astrophysically important 3\ensuremath{^{\textnormal{+}}} state in \ensuremath{^{\textnormal{18}}}Ne and deduced E\ensuremath{_{\textnormal{x}}}=4527 keV \textit{4} with \ensuremath{\Gamma}=17 keV \textit{4} (see the}\\
\parbox[b][0.3cm]{17.7cm}{\ensuremath{^{\textnormal{16}}}O(\ensuremath{^{\textnormal{3}}}He,n) section). Predicted that the \ensuremath{\Gamma}\ensuremath{_{\ensuremath{\gamma}}} for this state is negligible (\ensuremath{\leq}1 keV) and thus \ensuremath{\Gamma}=\ensuremath{\Gamma}\ensuremath{_{\textnormal{p}}}. The authors recommended the}\\
\parbox[b][0.3cm]{17.7cm}{\ensuremath{^{\textnormal{17}}}F(p,\ensuremath{\gamma}) reaction rate of (\href{https://www.nndc.bnl.gov/nsr/nsrlink.jsp?2000Bb04,B}{2000Bb04}).}\\
\parbox[b][0.3cm]{17.7cm}{\addtolength{\parindent}{-0.2in}\href{https://www.nndc.bnl.gov/nsr/nsrlink.jsp?2006Ch08,B}{2006Ch08}: Calculated the \ensuremath{^{\textnormal{17}}}F(p,\ensuremath{\gamma}) astrophysical S-factor at E\ensuremath{_{\textnormal{c.m.}}}\ensuremath{\leq}1 MeV for the capture reaction from both the ground state and}\\
\parbox[b][0.3cm]{17.7cm}{the first excited state in \ensuremath{^{\textnormal{17}}}F, including E1, E2, and M1 transitions. Analyzed \ensuremath{^{\textnormal{18}}}Ne level energies, widths, and spectroscopic factors}\\
\parbox[b][0.3cm]{17.7cm}{using continuum shell model. Calculated the \ensuremath{^{\textnormal{17}}}F(p,\ensuremath{\gamma}) reaction rate for T=0.1-2 GK and presented comparison with previous rates.}\\
\parbox[b][0.3cm]{17.7cm}{\addtolength{\parindent}{-0.2in}\href{https://www.nndc.bnl.gov/nsr/nsrlink.jsp?2006De47,B}{2006De47}: Calculated total and partial astrophysical S-factor at E\ensuremath{_{\textnormal{c.m.}}}=0.05-1.1 MeV for the E1, E2, and M1 multipolarities.}\\
\parbox[b][0.3cm]{17.7cm}{Calculated \ensuremath{\Gamma}\ensuremath{_{\textnormal{p}}}=21 keV and \ensuremath{\Gamma}\ensuremath{_{\ensuremath{\gamma}}}=33 meV for the astrophysically significant 3\ensuremath{^{\textnormal{+}}} state at around 4.5 MeV excitation energy.}\\
\parbox[b][0.3cm]{17.7cm}{Reviewed and discussed the microscopic, R-matrix and other models. Presented comparison with data.}\\
\parbox[b][0.3cm]{17.7cm}{\addtolength{\parindent}{-0.2in}\href{https://www.nndc.bnl.gov/nsr/nsrlink.jsp?2009Ch17,B}{2009Ch17}, \href{https://www.nndc.bnl.gov/nsr/nsrlink.jsp?2009Ch64,B}{2009Ch64}, \href{https://www.nndc.bnl.gov/nsr/nsrlink.jsp?2009ChZW,B}{2009ChZW}: Calculated the direct capture and resonant contributions to the total \ensuremath{^{\textnormal{17}}}F(p,\ensuremath{\gamma}) reaction rate at}\\
\parbox[b][0.3cm]{17.7cm}{T=0.1-0.6 GK (\href{https://www.nndc.bnl.gov/nsr/nsrlink.jsp?2009Ch64,B}{2009Ch64}) and 0.1-1 GK (\href{https://www.nndc.bnl.gov/nsr/nsrlink.jsp?2009Ch17,B}{2009Ch17}, \href{https://www.nndc.bnl.gov/nsr/nsrlink.jsp?2009ChZW,B}{2009ChZW}). Presented the E\ensuremath{_{\textnormal{c.m.}}}, J\ensuremath{^{\ensuremath{\pi}}}, \ensuremath{\Gamma}\ensuremath{_{\textnormal{p}}} and \ensuremath{\Gamma}\ensuremath{_{\ensuremath{\gamma}}} for the resonances used.}\\
\parbox[b][0.3cm]{17.7cm}{Parameterized the total reaction rate according to the REACLIB analytic form of (F. K. Thielemann, M. Arnould, and J. Truran, in}\\
\parbox[b][0.3cm]{17.7cm}{Advances in Nuclear Astrophysics, edited by E. Vangioni-Flam et al. (Editions Fronti\'{e}res, Gif-sur-Yvette, 1987), p. 525).}\\
\parbox[b][0.3cm]{17.7cm}{Presented a comparison with previous rates. Discussed the astrophysical implications of this rate for nova nucleosynthesis.}\\
\parbox[b][0.3cm]{17.7cm}{\addtolength{\parindent}{-0.2in}\href{https://www.nndc.bnl.gov/nsr/nsrlink.jsp?2010Il04,B}{2010Il04}, \href{https://www.nndc.bnl.gov/nsr/nsrlink.jsp?2010Il06,B}{2010Il06}: Analyzed and evaluated the \ensuremath{^{\textnormal{17}}}F(p,\ensuremath{\gamma}) direct capture, resonant and total reaction rate for T=0.01-10 GK.}\\
\parbox[b][0.3cm]{17.7cm}{Resonant properties for E\ensuremath{_{\textnormal{c.m.}}^{\textnormal{res}}}=596-2226 keV are discussed.}\\
\parbox[b][0.3cm]{17.7cm}{\addtolength{\parindent}{-0.2in}\href{https://www.nndc.bnl.gov/nsr/nsrlink.jsp?2014Al05,B}{2014Al05}: The astrophysical S-factor for the \ensuremath{^{\textnormal{17}}}F(p,\ensuremath{\gamma}) reaction and the asymptotic normalization coefficients were calculated for the}\\
\parbox[b][0.3cm]{17.7cm}{\ensuremath{^{\textnormal{18}}}Ne*(0, 1887, 3376, 3616 keV) states using mirror analogy of the bound states in \ensuremath{^{\textnormal{18}}}O and \ensuremath{^{\textnormal{18}}}Ne. Deduced the resonant and total}\\
\parbox[b][0.3cm]{17.7cm}{reaction rate for T=0.1-1 GK. Presented the resonance parameters used to calculate the resonance reaction rate. Discussed the}\\
\parbox[b][0.3cm]{17.7cm}{contribution of the direct capture rate to the total \ensuremath{^{\textnormal{17}}}F(p,\ensuremath{\gamma}) reaction rate, and its consequences on the production of \ensuremath{^{\textnormal{18}}}F at stellar}\\
\parbox[b][0.3cm]{17.7cm}{energies in ONe novae.}\\
\parbox[b][0.3cm]{17.7cm}{\addtolength{\parindent}{-0.2in}\href{https://www.nndc.bnl.gov/nsr/nsrlink.jsp?2017Ku27,B}{2017Ku27}: Calculated the asymptotic normalization coefficients. The direct capture cross sections together with the resulting}\\
\parbox[b][0.3cm]{17.7cm}{astrophysical S-factors for each bound state were extracted at T=0.04-0.9 GK using the RADCAP code. Deduced the resonant and}\\
\parbox[b][0.3cm]{17.7cm}{total \ensuremath{^{\textnormal{17}}}F(p,\ensuremath{\gamma}) reaction rate at the same temperature range significant for the hot CNO cycle in novae. Comparison with previous}\\
\parbox[b][0.3cm]{17.7cm}{experimental results are presented.}\\
\vspace{12pt}
\underline{$^{18}$Ne Levels}\\
\vspace{0.34cm}
\parbox[b][0.3cm]{17.7cm}{\addtolength{\parindent}{-0.254cm}(\href{https://www.nndc.bnl.gov/nsr/nsrlink.jsp?2014Al05,B}{2014Al05}) calculated the ANCs for the \ensuremath{^{\textnormal{18}}}O bound states using the \ensuremath{^{\textnormal{13}}}C(\ensuremath{^{\textnormal{17}}}O,\ensuremath{^{\textnormal{18}}}O)\ensuremath{^{\textnormal{12}}}C data; and estimated the ANCs for the}\\
\parbox[b][0.3cm]{17.7cm}{bound states of \ensuremath{^{\textnormal{18}}}Ne from analogy with the \ensuremath{^{\textnormal{18}}}O mirror levels assuming the spectroscopic factors are the same for both members of}\\
\parbox[b][0.3cm]{17.7cm}{the mirror pair.}\\
\vspace{0.34cm}
\begin{textblock}{29}(0,27.3)
Continued on next page (footnotes at end of table)
\end{textblock}
\clearpage
\vspace{0.3cm}
{\bf \small \underline{\ensuremath{^{\textnormal{1}}}H(\ensuremath{^{\textnormal{17}}}F,\ensuremath{\gamma})\hspace{0.2in}\href{https://www.nndc.bnl.gov/nsr/nsrlink.jsp?1987Wi11,B}{1987Wi11},\href{https://www.nndc.bnl.gov/nsr/nsrlink.jsp?2014Al05,B}{2014Al05} (continued)}}\\
\vspace{0.3cm}
\underline{$^{18}$Ne Levels (continued)}\\
\begin{longtable}{cccccc@{\extracolsep{\fill}}c}
\multicolumn{2}{c}{E(level)$^{{\hyperlink{NE3LEVEL0}{a}}}$}&J$^{\pi}$$^{{\hyperlink{NE3LEVEL0}{a}}}$&\multicolumn{2}{c}{S(0) (keV.b)$^{{\hyperlink{NE3LEVEL1}{b}}}$}&Comments&\\[-.2cm]
\multicolumn{2}{c}{\hrulefill}&\hrulefill&\multicolumn{2}{c}{\hrulefill}&\hrulefill&
\endfirsthead
\multicolumn{1}{r@{}}{0}&\multicolumn{1}{@{}l}{}&\multicolumn{1}{l}{0\ensuremath{^{+}}}&\multicolumn{1}{r@{}}{0}&\multicolumn{1}{@{.}l}{06 {\it 1}}&\parbox[t][0.3cm]{12.776401cm}{\raggedright ANC=C\ensuremath{^{\textnormal{2}}_{\textnormal{1}\textit{d}_{\textnormal{5/2}}}}=12.2 fm \ensuremath{^{\textnormal{$-$1}}} \textit{12} (\href{https://www.nndc.bnl.gov/nsr/nsrlink.jsp?2014Al05,B}{2014Al05}).\vspace{0.1cm}}&\\
\multicolumn{1}{r@{}}{1887}&\multicolumn{1}{@{.}l}{4}&\multicolumn{1}{l}{2\ensuremath{^{+}}}&\multicolumn{1}{r@{}}{0}&\multicolumn{1}{@{.}l}{61 {\it 11}}&\parbox[t][0.3cm]{12.776401cm}{\raggedright ANC: C\ensuremath{^{\textnormal{2}}_{\textnormal{2}\textit{s}_{\textnormal{1/2}}}}=14.9 fm \ensuremath{^{\textnormal{$-$1}}} \textit{21}; C\ensuremath{^{\textnormal{2}}_{\textnormal{1}\textit{d}_{\textnormal{5/2}}}}=2.85 fm \ensuremath{^{\textnormal{$-$1}}} \textit{32} (\href{https://www.nndc.bnl.gov/nsr/nsrlink.jsp?2014Al05,B}{2014Al05}).\vspace{0.1cm}}&\\
\multicolumn{1}{r@{}}{3376}&\multicolumn{1}{@{.}l}{4}&\multicolumn{1}{l}{4\ensuremath{^{+}}}&\multicolumn{1}{r@{}}{0}&\multicolumn{1}{@{.}l}{17 {\it 3}}&\parbox[t][0.3cm]{12.776401cm}{\raggedright ANC=C\ensuremath{^{\textnormal{2}}_{\textnormal{1}\textit{d}_{\textnormal{5/2}}}}=2.73 fm \ensuremath{^{\textnormal{$-$1}}} \textit{35} (\href{https://www.nndc.bnl.gov/nsr/nsrlink.jsp?2014Al05,B}{2014Al05}).\vspace{0.1cm}}&\\
\multicolumn{1}{r@{}}{3616}&\multicolumn{1}{@{.}l}{5}&\multicolumn{1}{l}{2\ensuremath{^{+}}}&\multicolumn{1}{r@{}}{1}&\multicolumn{1}{@{.}l}{34 {\it 24}}&\parbox[t][0.3cm]{12.776401cm}{\raggedright ANC: C\ensuremath{^{\textnormal{2}}_{\textnormal{2}\textit{s}_{\textnormal{1/2}}}}=117 fm \ensuremath{^{\textnormal{$-$1}}} \textit{20}; C\ensuremath{^{\textnormal{2}}_{\textnormal{1}\textit{d}_{\textnormal{5/2}}}}=2.46 fm \ensuremath{^{\textnormal{$-$1}}} \textit{33} (\href{https://www.nndc.bnl.gov/nsr/nsrlink.jsp?2014Al05,B}{2014Al05}).\vspace{0.1cm}}&\\
\multicolumn{1}{r@{}}{4524}&\multicolumn{1}{@{.}l}{0}&\multicolumn{1}{l}{3\ensuremath{^{+}}}&&&\parbox[t][0.3cm]{12.776401cm}{\raggedright E(level): (\href{https://www.nndc.bnl.gov/nsr/nsrlink.jsp?2000Bb04,B}{2000Bb04}) converted E\ensuremath{_{\textnormal{c.m.}}}=599.8 keV \textit{15} (stat.) \textit{20} (sys.) to E\ensuremath{_{\textnormal{x}}}=4523.7 keV \textit{29}\vspace{0.1cm}}&\\
&&&&&\parbox[t][0.3cm]{12.776401cm}{\raggedright {\ }{\ }{\ }based on an outdated \ensuremath{^{\textnormal{18}}}Ne\ensuremath{_{\textnormal{g.s.}}} mass excess of 5316.8 keV \textit{15} keV (\href{https://www.nndc.bnl.gov/nsr/nsrlink.jsp?1994Ma14,B}{1994Ma14}). The\vspace{0.1cm}}&\\
&&&&&\parbox[t][0.3cm]{12.776401cm}{\raggedright {\ }{\ }{\ }updated value is 5317.617 keV \textit{363} (\href{https://www.nndc.bnl.gov/nsr/nsrlink.jsp?2021Wa16,B}{2021Wa16}), which leads to E\ensuremath{_{\textnormal{x}}}=4522.9 keV \textit{25}.\vspace{0.1cm}}&\\
&&&&&\parbox[t][0.3cm]{12.776401cm}{\raggedright J\ensuremath{^{\pi}}: From R-matrix analysis of (\href{https://www.nndc.bnl.gov/nsr/nsrlink.jsp?2000Bb04,B}{2000Bb04}). Also (\href{https://www.nndc.bnl.gov/nsr/nsrlink.jsp?1987Wi11,B}{1987Wi11},\href{https://www.nndc.bnl.gov/nsr/nsrlink.jsp?1988Wi08,B}{1988Wi08}) deduced the same\vspace{0.1cm}}&\\
&&&&&\parbox[t][0.3cm]{12.776401cm}{\raggedright {\ }{\ }{\ }spin-parity assignment based on comparison of T=1 analog states in \ensuremath{^{\textnormal{18}}}O, \ensuremath{^{\textnormal{18}}}F, and \ensuremath{^{\textnormal{18}}}Ne.\vspace{0.1cm}}&\\
&&&&&\parbox[t][0.3cm]{12.776401cm}{\raggedright \ensuremath{\Gamma}\ensuremath{_{\textnormal{p}}}=18 keV \textit{2} (stat.) \textit{1} (sys.) (\href{https://www.nndc.bnl.gov/nsr/nsrlink.jsp?2000Bb04,B}{2000Bb04}).\vspace{0.1cm}}&\\
&&&&&\parbox[t][0.3cm]{12.776401cm}{\raggedright \ensuremath{\Gamma}\ensuremath{_{\ensuremath{\gamma}}}=56 meV \textit{24} (stat.) \textit{30} (sys.) (\href{https://www.nndc.bnl.gov/nsr/nsrlink.jsp?2009Ch17,B}{2009Ch17}, \href{https://www.nndc.bnl.gov/nsr/nsrlink.jsp?2009Ch64,B}{2009Ch64}, \href{https://www.nndc.bnl.gov/nsr/nsrlink.jsp?2009ChZW,B}{2009ChZW}: calculated from the\vspace{0.1cm}}&\\
&&&&&\parbox[t][0.3cm]{12.776401cm}{\raggedright {\ }{\ }{\ }measured \ensuremath{\omega}\ensuremath{\gamma}\ensuremath{_{\textnormal{(p,}\ensuremath{\gamma}\textnormal{)}}} and \ensuremath{\Gamma}\ensuremath{_{\textnormal{p}}}).\vspace{0.1cm}}&\\
&&&&&\parbox[t][0.3cm]{12.776401cm}{\raggedright \ensuremath{\omega}\ensuremath{\gamma}\ensuremath{_{\textnormal{(p,}\ensuremath{\gamma}\textnormal{)}}}=33 meV \textit{14} (stat.) \textit{17} (sys.) (\href{https://www.nndc.bnl.gov/nsr/nsrlink.jsp?2009Ch17,B}{2009Ch17}, \href{https://www.nndc.bnl.gov/nsr/nsrlink.jsp?2009Ch64,B}{2009Ch64}, \href{https://www.nndc.bnl.gov/nsr/nsrlink.jsp?2009ChZW,B}{2009ChZW}, \href{https://www.nndc.bnl.gov/nsr/nsrlink.jsp?2009Ba59,B}{2009Ba59}).\vspace{0.1cm}}&\\
&&&&&\parbox[t][0.3cm]{12.776401cm}{\raggedright A 2\ensuremath{\sigma} upper limit of S(E)\ensuremath{\leq}65 keV.b was deduced from (\href{https://www.nndc.bnl.gov/nsr/nsrlink.jsp?2009Ch17,B}{2009Ch17}, \href{https://www.nndc.bnl.gov/nsr/nsrlink.jsp?2009Ch64,B}{2009Ch64}) on the\vspace{0.1cm}}&\\
&&&&&\parbox[t][0.3cm]{12.776401cm}{\raggedright {\ }{\ }{\ }astrophysical S-factor of the \ensuremath{^{\textnormal{17}}}F(p,\ensuremath{\gamma}) reaction by measuring the p(\ensuremath{^{\textnormal{17}}}F,\ensuremath{\gamma}) yield off\vspace{0.1cm}}&\\
&&&&&\parbox[t][0.3cm]{12.776401cm}{\raggedright {\ }{\ }{\ }resonance at E\ensuremath{_{\textnormal{lab}}}=14.3 MeV (E\ensuremath{_{\textnormal{c.m.}}}=800 keV).\vspace{0.1cm}}&\\
\end{longtable}
\parbox[b][0.3cm]{17.7cm}{\makebox[1ex]{\ensuremath{^{\hypertarget{NE3LEVEL0}{a}}}} From the \ensuremath{^{\textnormal{18}}}Ne Adopted Levels.}\\
\parbox[b][0.3cm]{17.7cm}{\makebox[1ex]{\ensuremath{^{\hypertarget{NE3LEVEL1}{b}}}} From (\href{https://www.nndc.bnl.gov/nsr/nsrlink.jsp?2014Al05,B}{2014Al05}): the value of the total S-factor at zero energy is calculated as S\ensuremath{_{\textnormal{1$-$17}}}(0)=2.17 keV.b \textit{37}, which is 25\% lower}\\
\parbox[b][0.3cm]{17.7cm}{{\ }{\ }than S(0)=2.9 keV.b \textit{4} computed by (\href{https://www.nndc.bnl.gov/nsr/nsrlink.jsp?1991Ga03,B}{1991Ga03}).}\\
\vspace{0.5cm}
\underline{$\gamma$($^{18}$Ne)}\\
\begin{longtable}{ccccccc@{}c@{\extracolsep{\fill}}c}
\multicolumn{2}{c}{E\ensuremath{_{\gamma}}\ensuremath{^{\hyperlink{NE3GAMMA0}{a}}}}&\multicolumn{2}{c}{E\ensuremath{_{i}}(level)}&J\ensuremath{^{\pi}_{i}}&\multicolumn{2}{c}{E\ensuremath{_{f}}}&J\ensuremath{^{\pi}_{f}}&\\[-.2cm]
\multicolumn{2}{c}{\hrulefill}&\multicolumn{2}{c}{\hrulefill}&\hrulefill&\multicolumn{2}{c}{\hrulefill}&\hrulefill&
\endfirsthead
\multicolumn{1}{r@{}}{907}&\multicolumn{1}{@{.}l}{5\ensuremath{^{\hyperlink{NE3GAMMA1}{b}}}}&\multicolumn{1}{r@{}}{4524}&\multicolumn{1}{@{.}l}{0}&\multicolumn{1}{l}{3\ensuremath{^{+}}}&\multicolumn{1}{r@{}}{3616}&\multicolumn{1}{@{.}l}{5}&\multicolumn{1}{@{}l}{2\ensuremath{^{+}}}&\\
\multicolumn{1}{r@{}}{2636}&\multicolumn{1}{@{.}l}{4\ensuremath{^{\hyperlink{NE3GAMMA1}{b}}}}&\multicolumn{1}{r@{}}{4524}&\multicolumn{1}{@{.}l}{0}&\multicolumn{1}{l}{3\ensuremath{^{+}}}&\multicolumn{1}{r@{}}{1887}&\multicolumn{1}{@{.}l}{4}&\multicolumn{1}{@{}l}{2\ensuremath{^{+}}}&\\
\end{longtable}
\parbox[b][0.3cm]{17.7cm}{\makebox[1ex]{\ensuremath{^{\hypertarget{NE3GAMMA0}{a}}}} (\href{https://www.nndc.bnl.gov/nsr/nsrlink.jsp?2005Fi01,B}{2005Fi01}, \href{https://www.nndc.bnl.gov/nsr/nsrlink.jsp?2009Ch17,B}{2009Ch17}, \href{https://www.nndc.bnl.gov/nsr/nsrlink.jsp?2009Ch64,B}{2009Ch64}, \href{https://www.nndc.bnl.gov/nsr/nsrlink.jsp?2009Ba59,B}{2009Ba59}, \href{https://www.nndc.bnl.gov/nsr/nsrlink.jsp?2009ChZW,B}{2009ChZW}) did not provide any information on the \ensuremath{\gamma}-ray transition(s) measured}\\
\parbox[b][0.3cm]{17.7cm}{{\ }{\ }from the decay of the \ensuremath{^{\textnormal{18}}}Ne*(4524 keV) state. Therefore, the \ensuremath{\gamma}-ray transitions presented above are deduced by the evaluator}\\
\parbox[b][0.3cm]{17.7cm}{{\ }{\ }based on the analogy with the \ensuremath{\gamma}-ray transitions from the decay of the \ensuremath{^{\textnormal{18}}}O*(5377.8 keV, 3\ensuremath{^{\textnormal{+}}}) mirror level (\href{https://www.nndc.bnl.gov/nsr/nsrlink.jsp?1995Ti07,B}{1995Ti07}). The \ensuremath{\gamma}-ray}\\
\parbox[b][0.3cm]{17.7cm}{{\ }{\ }energies given here are calculated from the \ensuremath{^{\textnormal{18}}}Ne level energy differences corrected for the recoil energy. Since these \ensuremath{\gamma}-ray}\\
\parbox[b][0.3cm]{17.7cm}{{\ }{\ }energies are calculated, they are not included in the \ensuremath{^{\textnormal{18}}}Ne Adopted Gammas.}\\
\parbox[b][0.3cm]{17.7cm}{\makebox[1ex]{\ensuremath{^{\hypertarget{NE3GAMMA1}{b}}}} Placement of transition in the level scheme is uncertain.}\\
\vspace{0.5cm}
\clearpage
\begin{figure}[h]
\begin{center}
\includegraphics{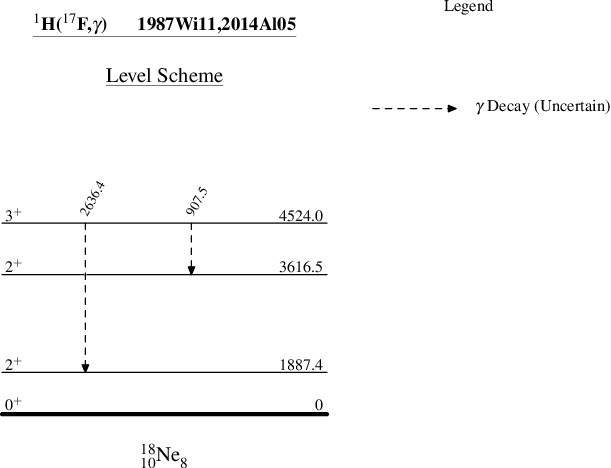}\\
\end{center}
\end{figure}
\clearpage
\subsection[\hspace{-0.2cm}\ensuremath{^{\textnormal{1}}}H(\ensuremath{^{\textnormal{17}}}F,p),(\ensuremath{^{\textnormal{17}}}F,2p),(\ensuremath{^{\textnormal{17}}}F,\ensuremath{\alpha}):res]{ }
\vspace{-27pt}
\vspace{0.3cm}
\hypertarget{NE4}{{\bf \small \underline{\ensuremath{^{\textnormal{1}}}H(\ensuremath{^{\textnormal{17}}}F,p),(\ensuremath{^{\textnormal{17}}}F,2p),(\ensuremath{^{\textnormal{17}}}F,\ensuremath{\alpha}):res\hspace{0.2in}\href{https://www.nndc.bnl.gov/nsr/nsrlink.jsp?1999Ba49,B}{1999Ba49},\href{https://www.nndc.bnl.gov/nsr/nsrlink.jsp?2020Br14,B}{2020Br14}}}}\\
\vspace{4pt}
\vspace{8pt}
\parbox[b][0.3cm]{17.7cm}{\addtolength{\parindent}{-0.2in}\href{https://www.nndc.bnl.gov/nsr/nsrlink.jsp?1999Ba49,B}{1999Ba49}, \href{https://www.nndc.bnl.gov/nsr/nsrlink.jsp?2000Bb04,B}{2000Bb04}, J. R. Beene \textit{et al}., J. Phys. G: Nucl. Part. Phys. 38 (2011) 024002: \ensuremath{^{\textnormal{1}}}H(\ensuremath{^{\textnormal{17}}}F,p) E=9-12 MeV; measured the}\\
\parbox[b][0.3cm]{17.7cm}{proton spectra and angular distributions, and p-\ensuremath{^{\textnormal{17}}}F coincidences at \ensuremath{\theta}\ensuremath{_{\textnormal{lab}}}=25\ensuremath{^\circ}{\textminus}51\ensuremath{^\circ} (\href{https://www.nndc.bnl.gov/nsr/nsrlink.jsp?1999Ba49,B}{1999Ba49}) and \ensuremath{\theta}\ensuremath{_{\textnormal{lab}}}=15\ensuremath{^\circ}{\textminus}35\ensuremath{^\circ} (\href{https://www.nndc.bnl.gov/nsr/nsrlink.jsp?2000Bb04,B}{2000Bb04}) using}\\
\parbox[b][0.3cm]{17.7cm}{the SIDAR array. Observed a previously unobserved \textit{s}-wave proton resonance with J\ensuremath{^{\ensuremath{\pi}}}=3\ensuremath{^{\textnormal{+}}} (deduced from R-matrix using the}\\
\parbox[b][0.3cm]{17.7cm}{MULTI code) at E\ensuremath{_{\textnormal{c.m.}}}=599.8 keV \textit{15} (stat.) \textit{20} (sys.) corresponding to E\ensuremath{_{\textnormal{x}}}=4523.7 keV \textit{29}. Measured \ensuremath{\Gamma}=18 keV \textit{2} (stat.) \textit{1} (sys.)}\\
\parbox[b][0.3cm]{17.7cm}{for the 3\ensuremath{^{\textnormal{+}}} resonance and discussed its J\ensuremath{^{\ensuremath{\pi}}} assignment and the implications for astrophysical processes.}\\
\parbox[b][0.3cm]{17.7cm}{\addtolength{\parindent}{-0.2in}\href{https://www.nndc.bnl.gov/nsr/nsrlink.jsp?1998HaZQ,B}{1998HaZQ}, \href{https://www.nndc.bnl.gov/nsr/nsrlink.jsp?1999Ha14,B}{1999Ha14}: \ensuremath{^{\textnormal{1}}}H(\ensuremath{^{\textnormal{17}}}F,\ensuremath{\alpha}) E\ensuremath{_{\textnormal{c.m.}}}=2.2-3.8 MeV; measured the outgoing particles$'$ energies, the \ensuremath{^{\textnormal{14}}}O-\ensuremath{\alpha} coincidences and \ensuremath{\sigma}(E)}\\
\parbox[b][0.3cm]{17.7cm}{using two position sensitive annular Si detectors covering \ensuremath{\theta}\ensuremath{_{\textnormal{lab}}}\ensuremath{\sim}13\ensuremath{^\circ}{\textminus}25\ensuremath{^\circ} for the \ensuremath{\alpha}s and \ensuremath{\theta}\ensuremath{_{\textnormal{lab}}}\ensuremath{\sim}4\ensuremath{^\circ}{\textminus}8\ensuremath{^\circ} for the \ensuremath{^{\textnormal{14}}}O ions. Observed 3}\\
\parbox[b][0.3cm]{17.7cm}{resonances at E\ensuremath{_{\textnormal{x}}}=7.60 MeV \textit{5}, 7.37 MeV \textit{6}, and 7.16 MeV \textit{15}. Assigned J\ensuremath{^{\ensuremath{\pi}}} to these resonances based on mirror levels and}\\
\parbox[b][0.3cm]{17.7cm}{Coulomb shift arguments. Deduced resonance strengths and \ensuremath{\alpha}-partial widths for these resonances. Discussed implications for}\\
\parbox[b][0.3cm]{17.7cm}{astrophysical processes.}\\
\parbox[b][0.3cm]{17.7cm}{\addtolength{\parindent}{-0.2in}\href{https://www.nndc.bnl.gov/nsr/nsrlink.jsp?2000Ga50,B}{2000Ga50}: \ensuremath{^{\textnormal{1}}}H(\ensuremath{^{\textnormal{17}}}F,p) E=33 MeV; measured recoil proton spectra and the \ensuremath{^{\textnormal{18}}}Ne excitation function using a position sensitive Si}\\
\parbox[b][0.3cm]{17.7cm}{detector covering \ensuremath{\theta}\ensuremath{_{\textnormal{c.m.}}}=160\ensuremath{^\circ}{\textminus}180\ensuremath{^\circ}. Observed three proton resonances at E\ensuremath{_{\textnormal{c.m.}}}=600, 1184 and 1231 keV, for which J\ensuremath{^{\ensuremath{\pi}}}=3\ensuremath{^{\textnormal{+}}}, 2\ensuremath{^{\textnormal{+}}},}\\
\parbox[b][0.3cm]{17.7cm}{and 3\ensuremath{^{-}} were deduced, respectively, based on an R-matrix analysis using the MULTI computer code.}\\
\parbox[b][0.3cm]{17.7cm}{\addtolength{\parindent}{-0.2in}\href{https://www.nndc.bnl.gov/nsr/nsrlink.jsp?2001Ga18,B}{2001Ga18}, \href{https://www.nndc.bnl.gov/nsr/nsrlink.jsp?2001Go01,B}{2001Go01}, \href{https://www.nndc.bnl.gov/nsr/nsrlink.jsp?2002Li66,B}{2002Li66}: \ensuremath{^{\textnormal{1}}}H(\ensuremath{^{\textnormal{17}}}F,p), \ensuremath{^{\textnormal{1}}}H(\ensuremath{^{\textnormal{17}}}F,2p) E=33 and 44 MeV; measured protons, p-p coincidences, angular}\\
\parbox[b][0.3cm]{17.7cm}{correlations, and the \ensuremath{^{\textnormal{18}}}Ne excitation function using a \ensuremath{\Delta}E-E telescope that consisted of a position sensitive Si \ensuremath{\Delta}E detector}\\
\parbox[b][0.3cm]{17.7cm}{covering \ensuremath{\theta}\ensuremath{_{\textnormal{lab}}}=\ensuremath{\pm}15\ensuremath{^\circ}, and a Si surface barrier E detector covering \ensuremath{\theta}\ensuremath{_{\textnormal{lab}}}=\ensuremath{\pm}10\ensuremath{^\circ}. The excitation function was analyzed with R-matrix}\\
\parbox[b][0.3cm]{17.7cm}{using the MULTI code. (\href{https://www.nndc.bnl.gov/nsr/nsrlink.jsp?2001Ga18,B}{2001Ga18}) deduced the properties of a \ensuremath{^{\textnormal{18}}}Ne resonance at E\ensuremath{_{\textnormal{c.m.}}}=2.22 MeV \textit{1} and considered the mode of}\\
\parbox[b][0.3cm]{17.7cm}{decay to be a simultaneous two proton decay. (\href{https://www.nndc.bnl.gov/nsr/nsrlink.jsp?2001Go01,B}{2001Go01}, \href{https://www.nndc.bnl.gov/nsr/nsrlink.jsp?2002Li66,B}{2002Li66}) deduced (using R-matrix) the properties of 4 of the \ensuremath{^{\textnormal{18}}}Ne}\\
\parbox[b][0.3cm]{17.7cm}{resonances, including the 3\ensuremath{^{\textnormal{+}}} state at E\ensuremath{_{\textnormal{c.m.}}}\ensuremath{\sim}600 keV. (\href{https://www.nndc.bnl.gov/nsr/nsrlink.jsp?2001Go01,B}{2001Go01}) estimated via R-matrix the 2p partial decay widths for the states}\\
\parbox[b][0.3cm]{17.7cm}{at 5.1 MeV and 6.15 MeV. Furthermore, (\href{https://www.nndc.bnl.gov/nsr/nsrlink.jsp?2001Go01,B}{2001Go01}) deduced the spectroscopic factors and branching ratios for 2p decay of the}\\
\parbox[b][0.3cm]{17.7cm}{6.15 MeV state assuming the diproton and the 3-body decay modes. (\href{https://www.nndc.bnl.gov/nsr/nsrlink.jsp?2001Go01,B}{2001Go01}) ruled out the 3-body decay and supported the}\\
\parbox[b][0.3cm]{17.7cm}{correlated \ensuremath{^{\textnormal{2}}}He decay from the 6.15-MeV state.}\\
\parbox[b][0.3cm]{17.7cm}{\addtolength{\parindent}{-0.2in}\href{https://www.nndc.bnl.gov/nsr/nsrlink.jsp?2000BaZV,B}{2000BaZV}, \href{https://www.nndc.bnl.gov/nsr/nsrlink.jsp?2001Bl06,B}{2001Bl06}, \href{https://www.nndc.bnl.gov/nsr/nsrlink.jsp?2003Bl11,B}{2003Bl11}: \ensuremath{^{\textnormal{1}}}H(\ensuremath{^{\textnormal{17}}}F,p), \ensuremath{^{\textnormal{1}}}H(\ensuremath{^{\textnormal{17}}}F,p\ensuremath{'}), \ensuremath{^{\textnormal{1}}}H(\ensuremath{^{\textnormal{17}}}F,\ensuremath{\alpha}) E=40.1-68.2 MeV; measured \ensuremath{^{\textnormal{14}}}O-\ensuremath{\alpha} and p-\ensuremath{^{\textnormal{17}}}F coincidences,}\\
\parbox[b][0.3cm]{17.7cm}{and \ensuremath{\sigma}(\ensuremath{\theta}) at 21 beam energies. The \ensuremath{^{\textnormal{14}}}O and \ensuremath{^{\textnormal{17}}}F recoils and the \ensuremath{\alpha}-particles were measured at \ensuremath{\theta}\ensuremath{_{\textnormal{lab}}}=3.4\ensuremath{^\circ}{\textminus}6.7\ensuremath{^\circ}, \ensuremath{\theta}\ensuremath{_{\textnormal{lab}}}=0\ensuremath{^\circ}, and}\\
\parbox[b][0.3cm]{17.7cm}{\ensuremath{\theta}\ensuremath{_{\textnormal{lab}}}=10.5\ensuremath{^\circ}{\textminus}25.8\ensuremath{^\circ}, respectively, using arrays of position sensitive Si detectors. Four proton resonances in E\ensuremath{_{\textnormal{x}}}(\ensuremath{^{\textnormal{18}}}Ne)=6-7 MeV were}\\
\parbox[b][0.3cm]{17.7cm}{observed in (\href{https://www.nndc.bnl.gov/nsr/nsrlink.jsp?2001Bl06,B}{2001Bl06}, \href{https://www.nndc.bnl.gov/nsr/nsrlink.jsp?2002BaZZ,B}{2002BaZZ}), and 13 proton resonances were observed in (\href{https://www.nndc.bnl.gov/nsr/nsrlink.jsp?2003Bl11,B}{2003Bl11}). The latter work deduced (using}\\
\parbox[b][0.3cm]{17.7cm}{R-matrix analysis) the resonance parameters for 5 resonances corresponding to E\ensuremath{_{\textnormal{x}}}(\ensuremath{^{\textnormal{18}}}Ne)=6-7 MeV, and discussed the \ensuremath{^{\textnormal{14}}}O(\ensuremath{\alpha},p)}\\
\parbox[b][0.3cm]{17.7cm}{reaction rate. (\href{https://www.nndc.bnl.gov/nsr/nsrlink.jsp?2003Bl11,B}{2003Bl11}) deduced a proton-decay branching ratio of \ensuremath{\Gamma}\ensuremath{_{\textnormal{p$'$}}}/\ensuremath{\Gamma}\ensuremath{_{\textnormal{p}}}=2.4, and \ensuremath{\Gamma}\ensuremath{_{\textnormal{tot}}}\ensuremath{\sim}58 keV for the 6.15-MeV state. The}\\
\parbox[b][0.3cm]{17.7cm}{large branch from the decay of this state to the \ensuremath{^{\textnormal{17}}}F*(495 keV) state increased the astrophysical \ensuremath{^{\textnormal{14}}}O(\ensuremath{\alpha},p)\ensuremath{^{\textnormal{17}}}F reaction rate by}\\
\parbox[b][0.3cm]{17.7cm}{factors of 3 to 60 depending on the temperature.}\\
\parbox[b][0.3cm]{17.7cm}{\addtolength{\parindent}{-0.2in}\href{https://www.nndc.bnl.gov/nsr/nsrlink.jsp?2001HaZP,B}{2001HaZP}: \ensuremath{^{\textnormal{1}}}H(\ensuremath{^{\textnormal{17}}}F,p), \ensuremath{^{\textnormal{1}}}H(\ensuremath{^{\textnormal{17}}}F,p\ensuremath{'}), \ensuremath{^{\textnormal{1}}}H(\ensuremath{^{\textnormal{17}}}F,\ensuremath{\alpha}) E=54.8-70.3 MeV; measured the excitation functions of \ensuremath{^{\textnormal{17}}}F+p, \ensuremath{\sigma}\ensuremath{_{\textnormal{(p,p}'\textnormal{)}}}, and \ensuremath{\sigma}\ensuremath{_{\textnormal{(p,}\ensuremath{\alpha}\textnormal{)}}}.}\\
\parbox[b][0.3cm]{17.7cm}{Measured p-heavy recoil coincidences using a PPAC covering \ensuremath{\theta}\ensuremath{_{\textnormal{lab}}}=1.5\ensuremath{^\circ}{\textminus}6.5\ensuremath{^\circ} (for heavy ions) and an annular position sensitive Si}\\
\parbox[b][0.3cm]{17.7cm}{\ensuremath{\Delta}E-E telescope covering \ensuremath{\theta}\ensuremath{_{\textnormal{lab}}}=7\ensuremath{^\circ}{\textminus}24.5\ensuremath{^\circ}. Six additional Si-strip detectors covered \ensuremath{\theta}\ensuremath{_{\textnormal{lab}}}=32\ensuremath{^\circ}{\textminus}56.5\ensuremath{^\circ}. Observed E\ensuremath{_{\textnormal{x}}}(\ensuremath{^{\textnormal{18}}}Ne)=7.05 MeV}\\
\parbox[b][0.3cm]{17.7cm}{and deduced a tentative J\ensuremath{^{\ensuremath{\pi}}}=4\ensuremath{^{\textnormal{+}}} based on R-matrix analysis and mirror levels arguments.}\\
\parbox[b][0.3cm]{17.7cm}{\addtolength{\parindent}{-0.2in}\href{https://www.nndc.bnl.gov/nsr/nsrlink.jsp?2001HaZQ,B}{2001HaZQ}: \ensuremath{^{\textnormal{1}}}H(\ensuremath{^{\textnormal{17}}}F,\ensuremath{^{\textnormal{17}}}F\ensuremath{'}), \ensuremath{^{\textnormal{1}}}H(\ensuremath{^{\textnormal{17}}}F,p), and \ensuremath{^{\textnormal{1}}}H(\ensuremath{^{\textnormal{17}}}F,p\ensuremath{'}) E\ensuremath{_{\textnormal{c.m.}}}=3-4 MeV; measured excitation function of \ensuremath{^{\textnormal{17}}}F+p elastic scattering at}\\
\parbox[b][0.3cm]{17.7cm}{\ensuremath{\theta}\ensuremath{_{\textnormal{lab}}}=52.5\ensuremath{^\circ}{\textminus}56.5\ensuremath{^\circ}. Deduced an upper limit of \ensuremath{\sim}5 mb/sr for the inelastic cross section of E\ensuremath{_{\textnormal{x}}}(\ensuremath{^{\textnormal{18}}}Ne)=7.05 and 7.6 MeV. Observed a}\\
\parbox[b][0.3cm]{17.7cm}{proton resonance at E\ensuremath{_{\textnormal{x}}}(\ensuremath{^{\textnormal{18}}}Ne)=7.71 MeV and deduced J\ensuremath{^{\ensuremath{\pi}}}=(1\ensuremath{^{\textnormal{+}}}, 2\ensuremath{^{-}}, 3\ensuremath{^{\textnormal{+}}}) based on the fact that this state was not populated by the}\\
\parbox[b][0.3cm]{17.7cm}{(p,\ensuremath{\alpha}) reaction. Estimated that the contribution of the inelastic scattering to the \ensuremath{^{\textnormal{14}}}O(\ensuremath{\alpha},p)\ensuremath{^{\textnormal{17}}}F reaction rate going through the \ensuremath{^{\textnormal{18}}}Ne}\\
\parbox[b][0.3cm]{17.7cm}{state at E\ensuremath{_{\textnormal{x}}}=7-7.8 MeV is less than 5\%.}\\
\parbox[b][0.3cm]{17.7cm}{\addtolength{\parindent}{-0.2in}\href{https://www.nndc.bnl.gov/nsr/nsrlink.jsp?2002BaZZ,B}{2002BaZZ}: \ensuremath{^{\textnormal{1}}}H(\ensuremath{^{\textnormal{17}}}F,p),(\ensuremath{^{\textnormal{17}}}F,\ensuremath{\alpha}) E\ensuremath{_{\textnormal{c.m.}}}\ensuremath{\approx}2-4 MeV; measured yields, deduced astrophysical reaction rates for the \ensuremath{^{\textnormal{17}}}F(p,\ensuremath{\alpha}) and}\\
\parbox[b][0.3cm]{17.7cm}{\ensuremath{^{\textnormal{14}}}O(\ensuremath{\alpha},p).}\\
\parbox[b][0.3cm]{17.7cm}{\addtolength{\parindent}{-0.2in}\href{https://www.nndc.bnl.gov/nsr/nsrlink.jsp?2002Ha15,B}{2002Ha15}, \href{https://www.nndc.bnl.gov/nsr/nsrlink.jsp?2004Ha58,B}{2004Ha58}: \ensuremath{^{\textnormal{1}}}H(\ensuremath{^{\textnormal{17}}}F,p), \ensuremath{^{\textnormal{1}}}H(\ensuremath{^{\textnormal{17}}}F,p\ensuremath{'}), \ensuremath{^{\textnormal{1}}}H(\ensuremath{^{\textnormal{17}}}F,\ensuremath{\alpha}) E=54.8-70.3 MeV; measured p-\ensuremath{^{\textnormal{17}}}F and \ensuremath{\alpha}-\ensuremath{^{\textnormal{17}}}F coincidences,}\\
\parbox[b][0.3cm]{17.7cm}{\ensuremath{\sigma}(E,\ensuremath{\theta})\ensuremath{_{\textnormal{(p,}\ensuremath{\alpha}\textnormal{)}}}, and \ensuremath{\sigma}(\ensuremath{\theta})\ensuremath{_{\textnormal{(p,p}'\textnormal{)}}} at a range of beam energies corresponding to E\ensuremath{_{\textnormal{c.m.}}}=3.07-3.94 MeV. The recoils were detected at}\\
\parbox[b][0.3cm]{17.7cm}{\ensuremath{\theta}\ensuremath{_{\textnormal{lab}}}=1.5\ensuremath{^\circ}{\textminus}6.5\ensuremath{^\circ}, and the protons and \ensuremath{\alpha}-particles were measured at \ensuremath{\theta}\ensuremath{_{\textnormal{lab}}}=7\ensuremath{^\circ}{\textminus}24.5\ensuremath{^\circ} and 32\ensuremath{^\circ}{\textminus}56.5\ensuremath{^\circ} using an array of annular position}\\
\parbox[b][0.3cm]{17.7cm}{sensitive Si detectors. Deduced an upper limit of 0.5-1 mb/sr for the \ensuremath{^{\textnormal{17}}}F(p,p\ensuremath{'}) cross section in the E\ensuremath{_{\textnormal{c.m.}}}\ensuremath{\sim}3.2 MeV. Deduced}\\
\parbox[b][0.3cm]{17.7cm}{excitation function for the elastic scattering. Extracted spectroscopic factors and resonance parameters for \ensuremath{^{\textnormal{18}}}Ne states from the}\\
\parbox[b][0.3cm]{17.7cm}{p(\ensuremath{^{\textnormal{17}}}F,\ensuremath{\alpha}) measurement. Calculated the \ensuremath{^{\textnormal{14}}}O(\ensuremath{\alpha},p) reaction rate and discussed its implication.}\\
\parbox[b][0.3cm]{17.7cm}{\addtolength{\parindent}{-0.2in}\href{https://www.nndc.bnl.gov/nsr/nsrlink.jsp?2009He16,B}{2009He16}, \href{https://www.nndc.bnl.gov/nsr/nsrlink.jsp?2010He17,B}{2010He17}: \ensuremath{^{\textnormal{1}}}H(\ensuremath{^{\textnormal{17}}}F,p), \ensuremath{^{\textnormal{1}}}H(\ensuremath{^{\textnormal{17}}}F,p\ensuremath{'}) E=44.2 MeV; measured p-\ensuremath{\gamma} coincidences from the \ensuremath{^{\textnormal{18}}}Ne\ensuremath{\rightarrow}p+\ensuremath{^{\textnormal{17}}}F*(495 keV) decay}\\
\parbox[b][0.3cm]{17.7cm}{using \ensuremath{\Delta}E-E telescopes consisting of position sensitive Si detectors covering \ensuremath{\theta}\ensuremath{_{\textnormal{lab}}}=15\ensuremath{^\circ}{\textminus}50\ensuremath{^\circ} and the Miniball array (for \ensuremath{\gamma}-ray}\\
\parbox[b][0.3cm]{17.7cm}{detection). The goal was to evaluate the contribution of the inelastic scattering to the astrophysical \ensuremath{^{\textnormal{14}}}O(\ensuremath{\alpha},p)\ensuremath{^{\textnormal{17}}}F reaction rate.}\\
\parbox[b][0.3cm]{17.7cm}{Deduced the \ensuremath{^{\textnormal{18}}}Ne resonance parameters using R-matrix analysis. Comments were made on the \ensuremath{\Gamma}\ensuremath{_{\textnormal{p}_{\textnormal{0}}}}, \ensuremath{\Gamma}\ensuremath{_{\textnormal{p}_{\textnormal{1}}}} and \ensuremath{\Gamma}\ensuremath{_{\ensuremath{\alpha}}} for the \ensuremath{^{\textnormal{18}}}Ne}\\
\parbox[b][0.3cm]{17.7cm}{state at E\ensuremath{_{\textnormal{x}}}\ensuremath{\approx}6.2 MeV. These studies found that the inelastic component would enhance the \ensuremath{^{\textnormal{14}}}O(\ensuremath{\alpha},p) reaction rate, contributing}\\
\clearpage
\vspace{0.3cm}
{\bf \small \underline{\ensuremath{^{\textnormal{1}}}H(\ensuremath{^{\textnormal{17}}}F,p),(\ensuremath{^{\textnormal{17}}}F,2p),(\ensuremath{^{\textnormal{17}}}F,\ensuremath{\alpha}):res\hspace{0.2in}\href{https://www.nndc.bnl.gov/nsr/nsrlink.jsp?1999Ba49,B}{1999Ba49},\href{https://www.nndc.bnl.gov/nsr/nsrlink.jsp?2020Br14,B}{2020Br14} (continued)}}\\
\vspace{0.3cm}
\parbox[b][0.3cm]{17.7cm}{approximately equally to the ground-state component of the reaction rate, however not to the relative degree suggested in}\\
\parbox[b][0.3cm]{17.7cm}{(\href{https://www.nndc.bnl.gov/nsr/nsrlink.jsp?2003Bl11,B}{2003Bl11}).}\\
\parbox[b][0.3cm]{17.7cm}{\addtolength{\parindent}{-0.2in}\href{https://www.nndc.bnl.gov/nsr/nsrlink.jsp?2010Ji02,B}{2010Ji02}: \ensuremath{^{\textnormal{1}}}H(\ensuremath{^{\textnormal{17}}}F,\ensuremath{\alpha}), \ensuremath{^{\textnormal{1}}}H(\ensuremath{^{\textnormal{17}}}F,p) E=55.5 MeV; measured the elastic scattering at E\ensuremath{_{\textnormal{c.m.}}}=0.4-1.7 MeV, and the (p,\ensuremath{\alpha}) channel at}\\
\parbox[b][0.3cm]{17.7cm}{E\ensuremath{_{\textnormal{c.m.}}}=2.23 MeV. Measured protons and \ensuremath{\alpha}-particles, and the excitation function for the \ensuremath{^{\textnormal{17}}}F+p elastic scattering, which was}\\
\parbox[b][0.3cm]{17.7cm}{analyzed with a multilevel R-matrix code MULTI7. A \ensuremath{\Delta}E-E telescope, which consisted of a position sensitive Si \ensuremath{\Delta}E detector}\\
\parbox[b][0.3cm]{17.7cm}{backed by a multi-guard silicon quadrant E-detector, was used at 15\ensuremath{^\circ} for elastic scattering run and at \ensuremath{\theta}\ensuremath{_{\textnormal{lab}}}=0\ensuremath{^\circ} for the (p,\ensuremath{\alpha}) run.}\\
\parbox[b][0.3cm]{17.7cm}{Deduced the proton widths and average differential cross sections (systematic uncertainty of 5.6\%). Discussed the \ensuremath{^{\textnormal{14}}}O(\ensuremath{\alpha},p) reaction}\\
\parbox[b][0.3cm]{17.7cm}{rate.}\\
\parbox[b][0.3cm]{17.7cm}{\addtolength{\parindent}{-0.2in}\href{https://www.nndc.bnl.gov/nsr/nsrlink.jsp?2010Wa18,B}{2010Wa18}: \ensuremath{^{\textnormal{1}}}H(\ensuremath{^{\textnormal{17}}}F,p) E not given; measured protons using a position sensitive \ensuremath{\Delta}E-E telescope at \ensuremath{\theta}\ensuremath{_{\textnormal{lab}}}=15\ensuremath{^\circ}. Deduced \ensuremath{^{\textnormal{18}}}Ne states}\\
\parbox[b][0.3cm]{17.7cm}{at 4.52 MeV, J\ensuremath{^{\ensuremath{\pi}}}=3\ensuremath{^{\textnormal{+}}} and 5.11 MeV, J\ensuremath{^{\ensuremath{\pi}}}=2\ensuremath{^{\textnormal{+}}} from the preliminary R-matrix analysis.}\\
\parbox[b][0.3cm]{17.7cm}{\addtolength{\parindent}{-0.2in}\href{https://www.nndc.bnl.gov/nsr/nsrlink.jsp?2010Ba21,B}{2010Ba21}: \ensuremath{^{\textnormal{1}}}H(\ensuremath{^{\textnormal{17}}}F,p), \ensuremath{^{\textnormal{1}}}H(\ensuremath{^{\textnormal{17}}}F,p\ensuremath{'}) E=52-58 MeV; measured scattered protons and p-\ensuremath{^{\textnormal{17}}}F coincidences using the SIDAR array for}\\
\parbox[b][0.3cm]{17.7cm}{protons and a gas-filled ionization counter for \ensuremath{^{\textnormal{17}}}F ions. The objective was to search for a J\ensuremath{^{\ensuremath{\pi}}}=4\ensuremath{^{\textnormal{+}}} \ensuremath{^{\textnormal{18}}}Ne resonance reported by}\\
\parbox[b][0.3cm]{17.7cm}{(\href{https://www.nndc.bnl.gov/nsr/nsrlink.jsp?2004No18,B}{2004No18}) at E\ensuremath{_{\textnormal{c.m.}}}=1.95 MeV in the \ensuremath{^{\textnormal{14}}}O+\ensuremath{\alpha} frame (\ensuremath{\sim}3.1 MeV in the time-reversed \ensuremath{^{\textnormal{17}}}F+p frame). This resonance was thought}\\
\parbox[b][0.3cm]{17.7cm}{(\href{https://www.nndc.bnl.gov/nsr/nsrlink.jsp?2004No18,B}{2004No18}) to populate the \ensuremath{^{\textnormal{17}}}F*(495 keV) state with a large decay branching ratio. (\href{https://www.nndc.bnl.gov/nsr/nsrlink.jsp?2010Ba21,B}{2010Ba21}) extracted Q-value spectra at}\\
\parbox[b][0.3cm]{17.7cm}{different beam energies to search for inelastic events. No evidence of inelastic \ensuremath{^{\textnormal{17}}}F*(495 keV)+p scattering was observed at any of}\\
\parbox[b][0.3cm]{17.7cm}{the measured incident energies. An upper limit of \ensuremath{\sim}10 mb was set on the p(\ensuremath{^{\textnormal{17}}}F,p\ensuremath{_{\textnormal{1}}}) cross section at E\ensuremath{_{\textnormal{c.m.}}}\ensuremath{\sim}3.1 MeV. Similar upper}\\
\parbox[b][0.3cm]{17.7cm}{limits were obtained at a range of bombarding energies. Upper limits on the proton decay branching ratio from this resonance to the}\\
\parbox[b][0.3cm]{17.7cm}{\ensuremath{^{\textnormal{17}}}F*(495 keV) state are tabulated as a function of the assumed width of this resonance. This decay branching ratio was constrained}\\
\parbox[b][0.3cm]{17.7cm}{to be less than 3.1\%.}\\
\parbox[b][0.3cm]{17.7cm}{\addtolength{\parindent}{-0.2in}\href{https://www.nndc.bnl.gov/nsr/nsrlink.jsp?2011He09,B}{2011He09}: \ensuremath{^{\textnormal{1}}}H(\ensuremath{^{\textnormal{17}}}F,p) E=4.22 MeV/nucleon; measured energy spectra of the recoiled protons along with protons angular}\\
\parbox[b][0.3cm]{17.7cm}{distributions using two \ensuremath{\Delta}E-E telescopes at \ensuremath{\theta}\ensuremath{_{\textnormal{lab}}}=0\ensuremath{^\circ} and 14\ensuremath{^\circ}. Observed several proton resonances in \ensuremath{^{\textnormal{18}}}Ne. Deduced their resonant}\\
\parbox[b][0.3cm]{17.7cm}{parameters using the R-matrix analysis. Identified a doublet structure around 7.10 MeV consisting of E\ensuremath{_{\textnormal{x}}}(\ensuremath{^{\textnormal{18}}}Ne)=7.05 MeV with}\\
\parbox[b][0.3cm]{17.7cm}{J\ensuremath{^{\ensuremath{\pi}}}=2\ensuremath{^{\textnormal{+}}} and E\ensuremath{_{\textnormal{x}}}(\ensuremath{^{\textnormal{18}}}Ne)=7.12 MeV with J\ensuremath{^{\ensuremath{\pi}}}=4\ensuremath{^{\textnormal{+}}}. Calculated and discussed the \ensuremath{^{\textnormal{14}}}O(\ensuremath{\alpha},p) reaction rate.}\\
\parbox[b][0.3cm]{17.7cm}{\addtolength{\parindent}{-0.2in}\href{https://www.nndc.bnl.gov/nsr/nsrlink.jsp?2012Pa02,B}{2012Pa02}: \ensuremath{^{\textnormal{1}}}H(\ensuremath{^{\textnormal{17}}}F,p) E=3.5 and 4.3 MeV/nucleon; measured p-\ensuremath{^{\textnormal{17}}}F coincidences and the proton angular distributions at}\\
\parbox[b][0.3cm]{17.7cm}{\ensuremath{\theta}\ensuremath{_{\textnormal{lab}}}=11\ensuremath{^\circ}{\textminus}65\ensuremath{^\circ} using the DINEX Si-detector array and a plastic scintillator covering \ensuremath{\theta}\ensuremath{_{\textnormal{lab}}}=0\ensuremath{^\circ}{\textminus}5\ensuremath{^\circ} to measure \ensuremath{^{\textnormal{17}}}F ions. Deduced total}\\
\parbox[b][0.3cm]{17.7cm}{reaction cross section for the quasi-elastic scattering since the inelastic excitation to the first excited state of \ensuremath{^{\textnormal{17}}}F*(495 keV) was}\\
\parbox[b][0.3cm]{17.7cm}{not resolved. Determined that the contribution of the inelastic excitation of the first excited state to the total cross section is less}\\
\parbox[b][0.3cm]{17.7cm}{than 8\%. Performed microscopic and macroscopic optical potential and DWBA analysis to probe a possible halo structure of \ensuremath{^{\textnormal{17}}}F.}\\
\parbox[b][0.3cm]{17.7cm}{\addtolength{\parindent}{-0.2in}\href{https://www.nndc.bnl.gov/nsr/nsrlink.jsp?2012LiZY,B}{2012LiZY}, L. E. Linhardt \textit{ et al}., J. Phys.: Conf. Ser. 403 (2012) 012036: \ensuremath{^{\textnormal{1}}}H(\ensuremath{^{\textnormal{17}}}F,p) and \ensuremath{^{\textnormal{1}}}H(\ensuremath{^{\textnormal{17}}}F,\ensuremath{\alpha}) 55 MeV; measured both these}\\
\parbox[b][0.3cm]{17.7cm}{reactions simultaneously; measured thick target E\ensuremath{_{\textnormal{p}}}, I\ensuremath{_{\textnormal{p}}}(\ensuremath{\theta}), E\ensuremath{_{\ensuremath{\alpha}}}, I\ensuremath{_{\ensuremath{\alpha}}}(\ensuremath{\theta}) using the \ensuremath{\Delta}E-E DSSDs of ANASEN covering the angular}\\
\parbox[b][0.3cm]{17.7cm}{range of \ensuremath{\theta}\ensuremath{_{\textnormal{lab}}}=9.5\ensuremath{^\circ}{\textminus}28\ensuremath{^\circ}; measured heavy recoils, p-recoil and \ensuremath{\alpha}-recoil coincidences using the Heavy Ion Recoil Chamber}\\
\parbox[b][0.3cm]{17.7cm}{downstream of ANASEN and at \ensuremath{\theta}\ensuremath{_{\textnormal{lab}}}=0\ensuremath{^\circ} but with \ensuremath{\theta}\ensuremath{_{\textnormal{lab}}}\ensuremath{<}1.4\ensuremath{^\circ} blocked; deduced unnormalized \ensuremath{\sigma}(E,\ensuremath{\theta}), and proton yield vs. E\ensuremath{_{\textnormal{c.m.}}};}\\
\parbox[b][0.3cm]{17.7cm}{calculated \ensuremath{\sigma}(E,\ensuremath{\theta}) using R-matrix. Statistics were too low to draw conclusive results.}\\
\parbox[b][0.3cm]{17.7cm}{\addtolength{\parindent}{-0.2in}\href{https://www.nndc.bnl.gov/nsr/nsrlink.jsp?2014Hu16,B}{2014Hu16}: \ensuremath{^{\textnormal{1}}}H(\ensuremath{^{\textnormal{17}}}F,p) E=3.6 MeV/nucleon; measured the recoil protons from elastic scattering using three \ensuremath{\Delta}E-E Si telescopes at}\\
\parbox[b][0.3cm]{17.7cm}{\ensuremath{\theta}\ensuremath{_{\textnormal{lab}}}\ensuremath{\approx}3\ensuremath{^\circ}, 10\ensuremath{^\circ} and 18\ensuremath{^\circ}. Reconstructed the excitation function for the \ensuremath{^{\textnormal{17}}}F+p elastic scattering. Analyzed the excitation function using}\\
\parbox[b][0.3cm]{17.7cm}{multi-channel R-matrix and deduced the \ensuremath{^{\textnormal{18}}}Ne level properties for 5 proton resonances observed, including a new state at}\\
\parbox[b][0.3cm]{17.7cm}{E\ensuremath{_{\textnormal{x}}}(\ensuremath{^{\textnormal{18}}}Ne)=6.85 MeV \textit{11}. Discussed a detailed comparison of the present findings with the literature results. Calculated and analyzed}\\
\parbox[b][0.3cm]{17.7cm}{the astrophysical \ensuremath{^{\textnormal{14}}}O(\ensuremath{\alpha},p) reaction rate.}\\
\parbox[b][0.3cm]{17.7cm}{\addtolength{\parindent}{-0.2in}L. E. Pratt, Ph.D. Thesis, \textit{Study of \ensuremath{^{18}}Ne using the Array for Nuclear Astrophysics and Structure with Exotic Nuclei}, Louisiana}\\
\parbox[b][0.3cm]{17.7cm}{State University and Agricultural and Mechanical College (2014): p(\ensuremath{^{\textnormal{17}}}F,p) E\ensuremath{_{\textnormal{c.m.}}}=2.053-2.985 MeV E(\ensuremath{^{\textnormal{17}}}F)=55 MeV; measured}\\
\parbox[b][0.3cm]{17.7cm}{thick target E\ensuremath{_{\textnormal{p}}} and I\ensuremath{_{\textnormal{p}}}(\ensuremath{\theta}) using the RESOLUT facility and a \ensuremath{\Delta}E-E DSSD telescope covering the angular range of \ensuremath{\theta}\ensuremath{_{\textnormal{lab}}}=8.0\ensuremath{^\circ}{\textminus}24\ensuremath{^\circ};}\\
\parbox[b][0.3cm]{17.7cm}{measured \ensuremath{^{\textnormal{17}}}F recoils and p-recoil coincidences using the Heavy Ion Recoil Chamber at \ensuremath{\theta}\ensuremath{_{\textnormal{lab}}}=0\ensuremath{^\circ} but with \ensuremath{\theta}\ensuremath{_{\textnormal{lab}}}\ensuremath{<}1.3\ensuremath{^\circ} blocked;}\\
\parbox[b][0.3cm]{17.7cm}{extracted \ensuremath{^{\textnormal{18}}}Ne resonance properties for two states using an R-matrix fit to the measured \ensuremath{\sigma}(E,\ensuremath{\theta}).}\\
\vspace{0.385cm}
\parbox[b][0.3cm]{17.7cm}{\addtolength{\parindent}{-0.2in}\textit{Reanalysis of previous data}:}\\
\parbox[b][0.3cm]{17.7cm}{\addtolength{\parindent}{-0.2in}\href{https://www.nndc.bnl.gov/nsr/nsrlink.jsp?2010HeZX,B}{2010HeZX}: \ensuremath{^{\textnormal{1}}}H(\ensuremath{^{\textnormal{17}}}F,p), \ensuremath{^{\textnormal{1}}}H(\ensuremath{^{\textnormal{17}}}F,2p) E=33 and 44 MeV. This article is not published in a peer reviewed journal. He \textit{et al}.}\\
\parbox[b][0.3cm]{17.7cm}{reanalyzed the data of the (\href{https://www.nndc.bnl.gov/nsr/nsrlink.jsp?2001Go01,B}{2001Go01}) experiment using multi-channel R-matrix. As a result, the J\ensuremath{^{\ensuremath{\pi}}}=1\ensuremath{^{-}} assignment made by}\\
\parbox[b][0.3cm]{17.7cm}{(\href{https://www.nndc.bnl.gov/nsr/nsrlink.jsp?2001Go01,B}{2001Go01}) for the E\ensuremath{_{\textnormal{x}}}(\ensuremath{^{\textnormal{18}}}Ne)=6.15 MeV was overthrown. Instead He \textit{et al}. assigned J\ensuremath{^{\ensuremath{\pi}}}=3\ensuremath{^{-}} to this state and J\ensuremath{^{\ensuremath{\pi}}}=1\ensuremath{^{-}} to another state}\\
\parbox[b][0.3cm]{17.7cm}{at E\ensuremath{_{\textnormal{x}}}(\ensuremath{^{\textnormal{18}}}Ne)=6.286 MeV. The astrophysical consequences of these new assignments regarding the \ensuremath{^{\textnormal{14}}}O(\ensuremath{\alpha},p) reaction rate were}\\
\parbox[b][0.3cm]{17.7cm}{discussed. Later on, both results of this work were overthrown by (\href{https://www.nndc.bnl.gov/nsr/nsrlink.jsp?2012Ba28,B}{2012Ba28}, \href{https://www.nndc.bnl.gov/nsr/nsrlink.jsp?2014Hu16,B}{2014Hu16}).}\\
\parbox[b][0.3cm]{17.7cm}{\addtolength{\parindent}{-0.2in}\href{https://www.nndc.bnl.gov/nsr/nsrlink.jsp?2012Ba28,B}{2012Ba28}: \ensuremath{^{\textnormal{1}}}H(\ensuremath{^{\textnormal{17}}}F,p\ensuremath{'}) E\ensuremath{_{\textnormal{c.m.}}}=2-2.6 MeV; re-analyzed \ensuremath{\sigma}(E,\ensuremath{\theta}) data from (\href{https://www.nndc.bnl.gov/nsr/nsrlink.jsp?2003Bl11,B}{2003Bl11}) corresponding to E\ensuremath{_{\textnormal{x}}}(\ensuremath{^{\textnormal{18}}}Ne)=6.15 MeV to}\\
\parbox[b][0.3cm]{17.7cm}{resolve discrepancies in the literature on the J\ensuremath{^{\ensuremath{\pi}}} assignment of the 6.15-MeV state in \ensuremath{^{\textnormal{18}}}Ne and its \ensuremath{\Gamma}\ensuremath{_{\textnormal{p$'$}}}/\ensuremath{\Gamma}\ensuremath{_{\textnormal{p}}} branching ratio.}\\
\parbox[b][0.3cm]{17.7cm}{Deduced \ensuremath{^{\textnormal{18}}}Ne proton decay branching ratios from the 6.15-MeV level to the ground and first excited state in \ensuremath{^{\textnormal{17}}}F, and p\ensuremath{_{\textnormal{0}}} and p\ensuremath{_{\textnormal{1}}}}\\
\clearpage
\vspace{0.3cm}
{\bf \small \underline{\ensuremath{^{\textnormal{1}}}H(\ensuremath{^{\textnormal{17}}}F,p),(\ensuremath{^{\textnormal{17}}}F,2p),(\ensuremath{^{\textnormal{17}}}F,\ensuremath{\alpha}):res\hspace{0.2in}\href{https://www.nndc.bnl.gov/nsr/nsrlink.jsp?1999Ba49,B}{1999Ba49},\href{https://www.nndc.bnl.gov/nsr/nsrlink.jsp?2020Br14,B}{2020Br14} (continued)}}\\
\vspace{0.3cm}
\parbox[b][0.3cm]{17.7cm}{widths using the multi-channel R-matrix code MULTI. Discussed J\ensuremath{^{\ensuremath{\pi}}} assignment of the 6.15-MeV level and the interpretation of the}\\
\parbox[b][0.3cm]{17.7cm}{\ensuremath{^{\textnormal{1}}}H(\ensuremath{^{\textnormal{17}}}F,\ensuremath{\alpha}) reaction measurements. The reanalysis of the (\href{https://www.nndc.bnl.gov/nsr/nsrlink.jsp?2003Bl11,B}{2003Bl11}) data by (\href{https://www.nndc.bnl.gov/nsr/nsrlink.jsp?2012Ba28,B}{2012Ba28}) resolved the literature discrepancies.}\\
\vspace{0.385cm}
\parbox[b][0.3cm]{17.7cm}{\addtolength{\parindent}{-0.2in}\textit{Theory}:}\\
\parbox[b][0.3cm]{17.7cm}{\addtolength{\parindent}{-0.2in}\href{https://www.nndc.bnl.gov/nsr/nsrlink.jsp?2000Fo19,B}{2000Fo19}: This work questioned some of the J\ensuremath{^{\ensuremath{\pi}}} assignments made in (\href{https://www.nndc.bnl.gov/nsr/nsrlink.jsp?1999Ha14,B}{1999Ha14}). The questions were based on the theoretical}\\
\parbox[b][0.3cm]{17.7cm}{results of (\href{https://www.nndc.bnl.gov/nsr/nsrlink.jsp?1998Sh35,B}{1998Sh35}) and on the basis of the Coulomb shifts with respect to the analog states in the mirror nucleus \ensuremath{^{\textnormal{18}}}O.}\\
\parbox[b][0.3cm]{17.7cm}{(\href{https://www.nndc.bnl.gov/nsr/nsrlink.jsp?2000Fo19,B}{2000Fo19}) concluded that the states at 7.37 and 7.60 MeV, observed in (\href{https://www.nndc.bnl.gov/nsr/nsrlink.jsp?1999Ha14,B}{1999Ha14}), cannot have J\ensuremath{^{\ensuremath{\pi}}}=4\ensuremath{^{\textnormal{+}}} assignments, and the}\\
\parbox[b][0.3cm]{17.7cm}{states at 7.16, observed in (\href{https://www.nndc.bnl.gov/nsr/nsrlink.jsp?1999Ha14,B}{1999Ha14}), and 7.37 MeV cannot have J\ensuremath{^{\ensuremath{\pi}}}=1\ensuremath{^{-}} assignments. (\href{https://www.nndc.bnl.gov/nsr/nsrlink.jsp?2000Fo19,B}{2000Fo19}) concluded that the state at 7.60}\\
\parbox[b][0.3cm]{17.7cm}{MeV could be a J\ensuremath{^{\ensuremath{\pi}}}=1\ensuremath{^{-}} state, and the 7.16-MeV state may be a J\ensuremath{^{\ensuremath{\pi}}}=4\ensuremath{^{\textnormal{+}}} state.}\\
\parbox[b][0.3cm]{17.7cm}{\addtolength{\parindent}{-0.2in}\href{https://www.nndc.bnl.gov/nsr/nsrlink.jsp?2002Gr10,B}{2002Gr10}: Calculated, using the R-matrix theory, the resonant \ensuremath{^{\textnormal{17}}}F(p,2p) production cross section on the E\ensuremath{_{\textnormal{x}}}(\ensuremath{^{\textnormal{18}}}Ne)=6.15 MeV}\\
\parbox[b][0.3cm]{17.7cm}{resonance with J\ensuremath{^{\ensuremath{\pi}}}=1\ensuremath{^{-}} and deduced an upper limit of 310 mb. Assuming identical spectroscopic factors for the \ensuremath{^{\textnormal{17}}}F\ensuremath{_{\textnormal{g.s.}}}+p and}\\
\parbox[b][0.3cm]{17.7cm}{\ensuremath{^{\textnormal{17}}}F*(495 keV)+p channels, a branching ratio of \ensuremath{\Gamma}\ensuremath{_{\textnormal{p}_{\textnormal{0}}}}/\ensuremath{\Gamma}\ensuremath{_{\textnormal{p}_{\textnormal{1}}}}\ensuremath{\sim}2 was computed. This resulted in \ensuremath{\Gamma}\ensuremath{_{\textnormal{p}_{\textnormal{0}}}}/\ensuremath{\Gamma}\ensuremath{\sim}2/3. Assuming S\ensuremath{_{\textnormal{2p}}}=S\ensuremath{_{\textnormal{1p}}} (2p}\\
\parbox[b][0.3cm]{17.7cm}{and 1p decay spectroscopic factors), upper limits were estimated for various 2p decay branching ratios. Comparisons of these results}\\
\parbox[b][0.3cm]{17.7cm}{to the experimental results of (\href{https://www.nndc.bnl.gov/nsr/nsrlink.jsp?2001Go01,B}{2001Go01}) are discussed. It was also argued that in addition to the 2p decay of the 1\ensuremath{^{-}} resonance, the}\\
\parbox[b][0.3cm]{17.7cm}{breakup of \ensuremath{^{\textnormal{17}}}F and the decay of a 2\ensuremath{^{-}} resonance at higher energy could contribute to the 2p counts observed in (\href{https://www.nndc.bnl.gov/nsr/nsrlink.jsp?2001Go01,B}{2001Go01}) because}\\
\parbox[b][0.3cm]{17.7cm}{of the use of a thick target.}\\
\parbox[b][0.3cm]{17.7cm}{\addtolength{\parindent}{-0.2in}\href{https://www.nndc.bnl.gov/nsr/nsrlink.jsp?2003Fo13,B}{2003Fo13}: Calculated excitation energies and widths for astrophysically interesting \ensuremath{^{\textnormal{18}}}Ne states at E\ensuremath{_{\textnormal{x}}}=7-8 MeV. Presented,}\\
\parbox[b][0.3cm]{17.7cm}{whenever possible, the information from mirror levels in \ensuremath{^{\textnormal{18}}}O.}\\
\parbox[b][0.3cm]{17.7cm}{\addtolength{\parindent}{-0.2in}\href{https://www.nndc.bnl.gov/nsr/nsrlink.jsp?2009Yu04,B}{2009Yu04}: \ensuremath{^{\textnormal{18}}}Ne is considered as a cluster of two protons orbiting an \ensuremath{^{\textnormal{16}}}O core. The excitation energy spectrum of \ensuremath{^{\textnormal{18}}}Ne and the}\\
\parbox[b][0.3cm]{17.7cm}{relative momentum distribution of the two protons emitted from the decay of the 6.15 MeV level of \ensuremath{^{\textnormal{18}}}Ne are calculated using the}\\
\parbox[b][0.3cm]{17.7cm}{Faddeev approach using the CD-Bonn potential for the \textit{pp} interaction. This study indicated that the probability of decay of the}\\
\parbox[b][0.3cm]{17.7cm}{6.15-MeV level via \ensuremath{^{\textnormal{2}}}He emission is half of that of the decay via three-body direct breakup involving an uncorrelated emission of}\\
\parbox[b][0.3cm]{17.7cm}{the two protons.}\\
\parbox[b][0.3cm]{17.7cm}{\addtolength{\parindent}{-0.2in}\href{https://www.nndc.bnl.gov/nsr/nsrlink.jsp?2010Ad02,B}{2010Ad02}: Investigated the structure and decay of the \ensuremath{^{\textnormal{18}}}Ne*(6135 keV) state using a microscopic cluster model involving}\\
\parbox[b][0.3cm]{17.7cm}{\ensuremath{^{\textnormal{16}}}O+p+p configurations. This description was used in the framework of the three{\textminus}cluster generator coordinate method associated}\\
\parbox[b][0.3cm]{17.7cm}{with the hyperspherical formalism. The authors analyzed the \ensuremath{^{\textnormal{18}}}Ne and \ensuremath{^{\textnormal{18}}}O low-lying states and determined the total width of the}\\
\parbox[b][0.3cm]{17.7cm}{\ensuremath{^{\textnormal{18}}}Ne*(6135 keV) level using the analytical continuation in the coupling constant method. They found \ensuremath{\Gamma}=40 keV \textit{10} keV, which is}\\
\parbox[b][0.3cm]{17.7cm}{in fair agreement with experimental value of \ensuremath{\Gamma}=50 keV \textit{5} (\href{https://www.nndc.bnl.gov/nsr/nsrlink.jsp?2001Go01,B}{2001Go01}).}\\
\parbox[b][0.3cm]{17.7cm}{\addtolength{\parindent}{-0.2in}\href{https://www.nndc.bnl.gov/nsr/nsrlink.jsp?2012Ok02,B}{2012Ok02}: Calculated asymptotic normalization coefficients and single-particle ANC as a function of the binding energy and orbital}\\
\parbox[b][0.3cm]{17.7cm}{angular momentum, separation energies for pairs of nuclei, excitation energies and widths of the first excited states. Presented}\\
\parbox[b][0.3cm]{17.7cm}{predictions of ANCs in mirror nuclei, and spectroscopic strengths in mirror pairs using the Gamow shell model and the real-energy}\\
\parbox[b][0.3cm]{17.7cm}{and complex-energy continuum-shell-model approaches. Comparison with experimental data are presented.}\\
\parbox[b][0.3cm]{17.7cm}{\addtolength{\parindent}{-0.2in}\href{https://www.nndc.bnl.gov/nsr/nsrlink.jsp?2020Br14,B}{2020Br14}: Calculated ANCs for the first two 2\ensuremath{^{\textnormal{+}}} states and the width of the 2\ensuremath{^{\textnormal{+}}_{\textnormal{3}}} states in \ensuremath{^{\textnormal{18}}}O and \ensuremath{^{\textnormal{18}}}Ne; investigated, using}\\
\parbox[b][0.3cm]{17.7cm}{R-matrix calculations, the intuitive relationship between spectroscopic factors and single-particle wave functions and their physical}\\
\parbox[b][0.3cm]{17.7cm}{counterparts, ANCs, and widths; presented comparison with experimental data. It was found that the ANC of the 2\ensuremath{^{\textnormal{+}}_{\textnormal{2}}} state in \ensuremath{^{\textnormal{18}}}Ne}\\
\parbox[b][0.3cm]{17.7cm}{deduced from the mirror state in \ensuremath{^{\textnormal{18}}}O is significantly larger than those found in previous work. Discussion on the astrophysical}\\
\parbox[b][0.3cm]{17.7cm}{implications for the \ensuremath{^{\textnormal{17}}}F(p,\ensuremath{\gamma}) reaction rate is presented.}\\
\vspace{12pt}
\underline{$^{18}$Ne Levels}\\
\begin{longtable}{ccccccc@{\extracolsep{\fill}}c}
\multicolumn{2}{c}{E(level)$^{}$}&J$^{\pi}$$^{}$&\multicolumn{2}{c}{\ensuremath{\Gamma}$^{}$}&L$^{}$&Comments&\\[-.2cm]
\multicolumn{2}{c}{\hrulefill}&\hrulefill&\multicolumn{2}{c}{\hrulefill}&\hrulefill&\hrulefill&
\endfirsthead
\multicolumn{1}{r@{}}{4522}&\multicolumn{1}{@{.}l}{9\ensuremath{^{{\hyperlink{NE4LEVEL0}{a}}}} {\it 25}}&\multicolumn{1}{l}{3\ensuremath{^{+}}}&\multicolumn{1}{r@{}}{18}&\multicolumn{1}{@{ }l}{keV {\it 2}}&\multicolumn{1}{l}{0}&\parbox[t][0.3cm]{12.2159cm}{\raggedright \ensuremath{\Gamma}\ensuremath{_{\textnormal{p}}}=18 keV \textit{2} (\href{https://www.nndc.bnl.gov/nsr/nsrlink.jsp?2010Ji02,B}{2010Ji02})\vspace{0.1cm}}&\\
&&&&&&\parbox[t][0.3cm]{12.2159cm}{\raggedright E(level): From E\ensuremath{_{\textnormal{c.m.}}}=599.8 keV \textit{25}, which is the weighted average between E\ensuremath{_{\textnormal{c.m.}}}=599.8\vspace{0.1cm}}&\\
&&&&&&\parbox[t][0.3cm]{12.2159cm}{\raggedright {\ }{\ }{\ }keV \textit{15} (stat.) \textit{20} (sys.) (\href{https://www.nndc.bnl.gov/nsr/nsrlink.jsp?1999Ba49,B}{1999Ba49}, \href{https://www.nndc.bnl.gov/nsr/nsrlink.jsp?2000Bb04,B}{2000Bb04}); E\ensuremath{_{\textnormal{c.m.}}}=0.60 MeV \textit{5} (\href{https://www.nndc.bnl.gov/nsr/nsrlink.jsp?2001Go01,B}{2001Go01}); and\vspace{0.1cm}}&\\
&&&&&&\parbox[t][0.3cm]{12.2159cm}{\raggedright {\ }{\ }{\ }E\ensuremath{_{\textnormal{c.m.}}}=598 keV \textit{25} (\href{https://www.nndc.bnl.gov/nsr/nsrlink.jsp?2010Ji02,B}{2010Ji02}). See also E\ensuremath{_{\textnormal{c.m.}}}=600 (\href{https://www.nndc.bnl.gov/nsr/nsrlink.jsp?2000Ga50,B}{2000Ga50}).\vspace{0.1cm}}&\\
&&&&&&\parbox[t][0.3cm]{12.2159cm}{\raggedright E(level): See also 4523.7 keV \textit{29} (\href{https://www.nndc.bnl.gov/nsr/nsrlink.jsp?1999Ba49,B}{1999Ba49}, \href{https://www.nndc.bnl.gov/nsr/nsrlink.jsp?2000Bb04,B}{2000Bb04}); 4.52 MeV \textit{5} (\href{https://www.nndc.bnl.gov/nsr/nsrlink.jsp?2001Go01,B}{2001Go01}); and\vspace{0.1cm}}&\\
&&&&&&\parbox[t][0.3cm]{12.2159cm}{\raggedright {\ }{\ }{\ }4520 keV \textit{25} (\href{https://www.nndc.bnl.gov/nsr/nsrlink.jsp?2010Ji02,B}{2010Ji02}, \href{https://www.nndc.bnl.gov/nsr/nsrlink.jsp?2010Wa18,B}{2010Wa18}). These are deduced with outdated Q-values (for\vspace{0.1cm}}&\\
&&&&&&\parbox[t][0.3cm]{12.2159cm}{\raggedright {\ }{\ }{\ }example, (\href{https://www.nndc.bnl.gov/nsr/nsrlink.jsp?1999Ba49,B}{1999Ba49}) deduced 4523.7 keV \textit{29} based on \ensuremath{^{\textnormal{18}}}Ne\ensuremath{_{\textnormal{g.s.}}} mass excess of 5316.8\vspace{0.1cm}}&\\
&&&&&&\parbox[t][0.3cm]{12.2159cm}{\raggedright {\ }{\ }{\ }keV \textit{15} keV from (\href{https://www.nndc.bnl.gov/nsr/nsrlink.jsp?1994Ma14,B}{1994Ma14}). The updated value is 5317.617 keV \textit{363} (\href{https://www.nndc.bnl.gov/nsr/nsrlink.jsp?2021Wa16,B}{2021Wa16})).\vspace{0.1cm}}&\\
&&&&&&\parbox[t][0.3cm]{12.2159cm}{\raggedright \ensuremath{\Gamma}: From \ensuremath{\Gamma}=18 keV \textit{2} (stat.) \textit{1} (sys.) (\href{https://www.nndc.bnl.gov/nsr/nsrlink.jsp?1999Ba49,B}{1999Ba49}, \href{https://www.nndc.bnl.gov/nsr/nsrlink.jsp?2000Bb04,B}{2000Bb04}). See also \ensuremath{\Gamma}=18 keV \textit{2}\vspace{0.1cm}}&\\
&&&&&&\parbox[t][0.3cm]{12.2159cm}{\raggedright {\ }{\ }{\ }(\href{https://www.nndc.bnl.gov/nsr/nsrlink.jsp?2001Ga18,B}{2001Ga18}, \href{https://www.nndc.bnl.gov/nsr/nsrlink.jsp?2001Go01,B}{2001Go01}, \href{https://www.nndc.bnl.gov/nsr/nsrlink.jsp?2002Li66,B}{2002Li66}); and \ensuremath{\Gamma}=18 keV confirmed by the R-matrix analysis\vspace{0.1cm}}&\\
&&&&&&\parbox[t][0.3cm]{12.2159cm}{\raggedright {\ }{\ }{\ }in (\href{https://www.nndc.bnl.gov/nsr/nsrlink.jsp?2000Ga50,B}{2000Ga50}).\vspace{0.1cm}}&\\
\end{longtable}
\begin{textblock}{29}(0,27.3)
Continued on next page (footnotes at end of table)
\end{textblock}
\clearpage
\begin{longtable}{ccccccc@{\extracolsep{\fill}}c}
\\[-.4cm]
\multicolumn{8}{c}{{\bf \small \underline{\ensuremath{^{\textnormal{1}}}H(\ensuremath{^{\textnormal{17}}}F,p),(\ensuremath{^{\textnormal{17}}}F,2p),(\ensuremath{^{\textnormal{17}}}F,\ensuremath{\alpha}):res\hspace{0.2in}\href{https://www.nndc.bnl.gov/nsr/nsrlink.jsp?1999Ba49,B}{1999Ba49},\href{https://www.nndc.bnl.gov/nsr/nsrlink.jsp?2020Br14,B}{2020Br14} (continued)}}}\\
\multicolumn{8}{c}{~}\\
\multicolumn{8}{c}{\underline{\ensuremath{^{18}}Ne Levels (continued)}}\\
\multicolumn{8}{c}{~}\\
\multicolumn{2}{c}{E(level)$^{}$}&J$^{\pi}$$^{}$&\multicolumn{2}{c}{\ensuremath{\Gamma}$^{}$}&L$^{}$&Comments&\\[-.2cm]
\multicolumn{2}{c}{\hrulefill}&\hrulefill&\multicolumn{2}{c}{\hrulefill}&\hrulefill&\hrulefill&
\endhead
&&&&&&\parbox[t][0.3cm]{12.28866cm}{\raggedright J\ensuremath{^{\pi}}: From R-matrix analyses of (\href{https://www.nndc.bnl.gov/nsr/nsrlink.jsp?1999Ba49,B}{1999Ba49}, \href{https://www.nndc.bnl.gov/nsr/nsrlink.jsp?2000Bb04,B}{2000Bb04}, \href{https://www.nndc.bnl.gov/nsr/nsrlink.jsp?2000Ga50,B}{2000Ga50}, \href{https://www.nndc.bnl.gov/nsr/nsrlink.jsp?2001Ga18,B}{2001Ga18}, \href{https://www.nndc.bnl.gov/nsr/nsrlink.jsp?2001Go01,B}{2001Go01},\vspace{0.1cm}}&\\
&&&&&&\parbox[t][0.3cm]{12.28866cm}{\raggedright {\ }{\ }{\ }\href{https://www.nndc.bnl.gov/nsr/nsrlink.jsp?2002Li66,B}{2002Li66}, \href{https://www.nndc.bnl.gov/nsr/nsrlink.jsp?2010Ji02,B}{2010Ji02}, and \href{https://www.nndc.bnl.gov/nsr/nsrlink.jsp?2010Wa18,B}{2010Wa18}).\vspace{0.1cm}}&\\
&&&&&&\parbox[t][0.3cm]{12.28866cm}{\raggedright L: From (\href{https://www.nndc.bnl.gov/nsr/nsrlink.jsp?2000Ga50,B}{2000Ga50}, \href{https://www.nndc.bnl.gov/nsr/nsrlink.jsp?2010Ji02,B}{2010Ji02}).\vspace{0.1cm}}&\\
\multicolumn{1}{r@{}}{5112}&\multicolumn{1}{@{ }l}{{\it 21}}&\multicolumn{1}{l}{2\ensuremath{^{+}}}&\multicolumn{1}{r@{}}{44}&\multicolumn{1}{@{ }l}{keV {\it 2}}&\multicolumn{1}{l}{0}&\parbox[t][0.3cm]{12.28866cm}{\raggedright \ensuremath{\Gamma}\ensuremath{_{\textnormal{p}}}=44 keV \textit{2}\vspace{0.1cm}}&\\
&&&&&&\parbox[t][0.3cm]{12.28866cm}{\raggedright E(level): Weighted average of E\ensuremath{_{\textnormal{x}}}=5113 keV \textit{30} (\href{https://www.nndc.bnl.gov/nsr/nsrlink.jsp?2010Ji02,B}{2010Ji02}: from E\ensuremath{_{\textnormal{c.m.}}}=1190 keV \textit{30});\vspace{0.1cm}}&\\
&&&&&&\parbox[t][0.3cm]{12.28866cm}{\raggedright {\ }{\ }{\ }and E\ensuremath{_{\textnormal{x}}}=5110 keV \textit{30} (\href{https://www.nndc.bnl.gov/nsr/nsrlink.jsp?2011He09,B}{2011He09}). See also E\ensuremath{_{\textnormal{x}}}=5100 keV (\href{https://www.nndc.bnl.gov/nsr/nsrlink.jsp?2009He16,B}{2009He16}); 5107 keV\vspace{0.1cm}}&\\
&&&&&&\parbox[t][0.3cm]{12.28866cm}{\raggedright {\ }{\ }{\ }(\href{https://www.nndc.bnl.gov/nsr/nsrlink.jsp?2000Ga50,B}{2000Ga50}: from E\ensuremath{_{\textnormal{c.m.}}}=1184 keV); 5103 keV (\href{https://www.nndc.bnl.gov/nsr/nsrlink.jsp?2001Go01,B}{2001Go01}: from E\ensuremath{_{\textnormal{c.m.}}}=1180 keV); and\vspace{0.1cm}}&\\
&&&&&&\parbox[t][0.3cm]{12.28866cm}{\raggedright {\ }{\ }{\ }5110 keV (\href{https://www.nndc.bnl.gov/nsr/nsrlink.jsp?2010Wa18,B}{2010Wa18}). Note that the E\ensuremath{_{\textnormal{c.m.}}} reported by (\href{https://www.nndc.bnl.gov/nsr/nsrlink.jsp?2001Go01,B}{2001Go01}) was mistyped as\vspace{0.1cm}}&\\
&&&&&&\parbox[t][0.3cm]{12.28866cm}{\raggedright {\ }{\ }{\ }1.118 MeV. The correct value is 1.18 MeV confirmed from (\href{https://www.nndc.bnl.gov/nsr/nsrlink.jsp?2002Li66,B}{2002Li66} and \href{https://www.nndc.bnl.gov/nsr/nsrlink.jsp?2010HeZX,B}{2010HeZX}).\vspace{0.1cm}}&\\
&&&&&&\parbox[t][0.3cm]{12.28866cm}{\raggedright \ensuremath{\Gamma}\ensuremath{_{\textnormal{p}}},\ensuremath{\Gamma}: Weighted average of \ensuremath{\Gamma}\ensuremath{_{\textnormal{p}}}=42 keV \textit{4} (\href{https://www.nndc.bnl.gov/nsr/nsrlink.jsp?2010Ji02,B}{2010Ji02}); and \ensuremath{\Gamma}\ensuremath{_{\textnormal{p}}}=45 keV \textit{2} (\href{https://www.nndc.bnl.gov/nsr/nsrlink.jsp?2001Ga18,B}{2001Ga18},\vspace{0.1cm}}&\\
&&&&&&\parbox[t][0.3cm]{12.28866cm}{\raggedright {\ }{\ }{\ }\href{https://www.nndc.bnl.gov/nsr/nsrlink.jsp?2001Go01,B}{2001Go01}, \href{https://www.nndc.bnl.gov/nsr/nsrlink.jsp?2002Li66,B}{2002Li66}, \href{https://www.nndc.bnl.gov/nsr/nsrlink.jsp?2009He16,B}{2009He16}, \href{https://www.nndc.bnl.gov/nsr/nsrlink.jsp?2011He09,B}{2011He09}). These were deduced from fits to elastic\vspace{0.1cm}}&\\
&&&&&&\parbox[t][0.3cm]{12.28866cm}{\raggedright {\ }{\ }{\ }scattering data that appear to have assumed the width is entirely due to L=0 proton\vspace{0.1cm}}&\\
&&&&&&\parbox[t][0.3cm]{12.28866cm}{\raggedright {\ }{\ }{\ }emission to the \ensuremath{^{\textnormal{17}}}F\ensuremath{_{\textnormal{g.s.}}}. Later on, (\href{https://www.nndc.bnl.gov/nsr/nsrlink.jsp?2020Br14,B}{2020Br14}: see Section VII.B) assigned these\vspace{0.1cm}}&\\
&&&&&&\parbox[t][0.3cm]{12.28866cm}{\raggedright {\ }{\ }{\ }measured proton widths to the 2\textit{s}\ensuremath{_{\textnormal{1/2}}} channel. Therefore, we assumed \ensuremath{\Gamma}=\ensuremath{\Gamma}\ensuremath{_{\textnormal{p}}}.\vspace{0.1cm}}&\\
&&&&&&\parbox[t][0.3cm]{12.28866cm}{\raggedright \ensuremath{\Gamma}\ensuremath{_{^{\textnormal{2}}\textnormal{He}}}=1.8\ensuremath{\times}10\ensuremath{^{\textnormal{$-$5}}} eV (\href{https://www.nndc.bnl.gov/nsr/nsrlink.jsp?2001Go01,B}{2001Go01}).\vspace{0.1cm}}&\\
&&&&&&\parbox[t][0.3cm]{12.28866cm}{\raggedright J\ensuremath{^{\pi}}: From the R-matrix analyses of (\href{https://www.nndc.bnl.gov/nsr/nsrlink.jsp?2000Ga50,B}{2000Ga50}, \href{https://www.nndc.bnl.gov/nsr/nsrlink.jsp?2001Ga18,B}{2001Ga18}, \href{https://www.nndc.bnl.gov/nsr/nsrlink.jsp?2001Go01,B}{2001Go01}, \href{https://www.nndc.bnl.gov/nsr/nsrlink.jsp?2002Li66,B}{2002Li66},\vspace{0.1cm}}&\\
&&&&&&\parbox[t][0.3cm]{12.28866cm}{\raggedright {\ }{\ }{\ }\href{https://www.nndc.bnl.gov/nsr/nsrlink.jsp?2009He16,B}{2009He16}, \href{https://www.nndc.bnl.gov/nsr/nsrlink.jsp?2010Ji02,B}{2010Ji02}, \href{https://www.nndc.bnl.gov/nsr/nsrlink.jsp?2010Wa18,B}{2010Wa18}, and \href{https://www.nndc.bnl.gov/nsr/nsrlink.jsp?2011He09,B}{2011He09}).\vspace{0.1cm}}&\\
&&&&&&\parbox[t][0.3cm]{12.28866cm}{\raggedright L: From (\href{https://www.nndc.bnl.gov/nsr/nsrlink.jsp?2010Ji02,B}{2010Ji02}, \href{https://www.nndc.bnl.gov/nsr/nsrlink.jsp?2011He09,B}{2011He09}).\vspace{0.1cm}}&\\
\multicolumn{1}{r@{}}{5154}&\multicolumn{1}{@{}l}{\ensuremath{^{{\hyperlink{NE4LEVEL0}{a}}}}}&\multicolumn{1}{l}{3\ensuremath{^{-}}}&&&&\parbox[t][0.3cm]{12.28866cm}{\raggedright E(level): From E\ensuremath{_{\textnormal{c.m.}}}=1231 keV (\href{https://www.nndc.bnl.gov/nsr/nsrlink.jsp?2000Ga50,B}{2000Ga50}).\vspace{0.1cm}}&\\
&&&&&&\parbox[t][0.3cm]{12.28866cm}{\raggedright J\ensuremath{^{\pi}}: From (\href{https://www.nndc.bnl.gov/nsr/nsrlink.jsp?2000Ga50,B}{2000Ga50}) based on R-matrix analysis.\vspace{0.1cm}}&\\
\multicolumn{1}{r@{}}{6135}&\multicolumn{1}{@{ }l}{{\it 1}}&\multicolumn{1}{l}{1\ensuremath{^{-}}}&\multicolumn{1}{r@{}}{53}&\multicolumn{1}{@{.}l}{0 keV {\it 18}}&\multicolumn{1}{l}{1}&\parbox[t][0.3cm]{12.28866cm}{\raggedright \ensuremath{\Gamma}\ensuremath{\alpha}=6.9 eV \textit{20}\vspace{0.1cm}}&\\
&&&&&&\parbox[t][0.3cm]{12.28866cm}{\raggedright \ensuremath{\Gamma}\ensuremath{_{\textnormal{p0}}}/\ensuremath{\Gamma}=0.70 \textit{4} (\href{https://www.nndc.bnl.gov/nsr/nsrlink.jsp?2012Ba28,B}{2012Ba28}); \ensuremath{\Gamma}\ensuremath{_{\textnormal{p1}}}/\ensuremath{\Gamma}=0.30 \textit{2} (\href{https://www.nndc.bnl.gov/nsr/nsrlink.jsp?2012Ba28,B}{2012Ba28})\vspace{0.1cm}}&\\
&&&&&&\parbox[t][0.3cm]{12.28866cm}{\raggedright \ensuremath{\Gamma}\ensuremath{_{\textnormal{p1}}}/\ensuremath{\Gamma}\ensuremath{_{\textnormal{p0}}}=0.42 \textit{3} (\href{https://www.nndc.bnl.gov/nsr/nsrlink.jsp?2012Ba28,B}{2012Ba28})\vspace{0.1cm}}&\\
&&&&&&\parbox[t][0.3cm]{12.28866cm}{\raggedright E(level): Weighted average of 6.14 MeV \textit{1} (\href{https://www.nndc.bnl.gov/nsr/nsrlink.jsp?2001Go01,B}{2001Go01}: from E\ensuremath{_{\textnormal{c.m.}}}=2.22 MeV \textit{1}); 6.18\vspace{0.1cm}}&\\
&&&&&&\parbox[t][0.3cm]{12.28866cm}{\raggedright {\ }{\ }{\ }MeV \textit{6} (\href{https://www.nndc.bnl.gov/nsr/nsrlink.jsp?2009He16,B}{2009He16}, \href{https://www.nndc.bnl.gov/nsr/nsrlink.jsp?2010He17,B}{2010He17}: from E\ensuremath{_{\textnormal{c.m.}}}=2.26 MeV \textit{6}); 6135 keV \textit{1} (\href{https://www.nndc.bnl.gov/nsr/nsrlink.jsp?2012Ba28,B}{2012Ba28}: from\vspace{0.1cm}}&\\
&&&&&&\parbox[t][0.3cm]{12.28866cm}{\raggedright {\ }{\ }{\ }E\ensuremath{_{\textnormal{c.m.}}}=2212 keV \textit{1}); 6150 keV \textit{30} (\href{https://www.nndc.bnl.gov/nsr/nsrlink.jsp?2014Hu16,B}{2014Hu16}); and 6142 keV \textit{5} (stat.) \textit{8} (sys.) from (L.\vspace{0.1cm}}&\\
&&&&&&\parbox[t][0.3cm]{12.28866cm}{\raggedright {\ }{\ }{\ }E. Pratt (Ph.D. Thesis, 2014)).\vspace{0.1cm}}&\\
&&&&&&\parbox[t][0.3cm]{12.28866cm}{\raggedright E(level): See also 6150 keV (\href{https://www.nndc.bnl.gov/nsr/nsrlink.jsp?2001Bl06,B}{2001Bl06}, \href{https://www.nndc.bnl.gov/nsr/nsrlink.jsp?2001Ga18,B}{2001Ga18}); 6.15 MeV (\href{https://www.nndc.bnl.gov/nsr/nsrlink.jsp?2002Ha15,B}{2002Ha15} and \href{https://www.nndc.bnl.gov/nsr/nsrlink.jsp?2010Ji02,B}{2010Ji02}:\vspace{0.1cm}}&\\
&&&&&&\parbox[t][0.3cm]{12.28866cm}{\raggedright {\ }{\ }{\ }from E\ensuremath{_{\textnormal{c.m.}}}(\ensuremath{^{\textnormal{14}}}O+\ensuremath{\alpha})=1.04 MeV and E\ensuremath{_{\textnormal{c.m.}}}(p+\ensuremath{^{\textnormal{17}}}F)=2.23 MeV); 6137 keV (\href{https://www.nndc.bnl.gov/nsr/nsrlink.jsp?2003Bl11,B}{2003Bl11});\vspace{0.1cm}}&\\
&&&&&&\parbox[t][0.3cm]{12.28866cm}{\raggedright {\ }{\ }{\ }and 6120 keV (\href{https://www.nndc.bnl.gov/nsr/nsrlink.jsp?2010HeZX,B}{2010HeZX}: considered this state to be the same as the 6150 keV state\vspace{0.1cm}}&\\
&&&&&&\parbox[t][0.3cm]{12.28866cm}{\raggedright {\ }{\ }{\ }observed in (\href{https://www.nndc.bnl.gov/nsr/nsrlink.jsp?1996Ha26,B}{1996Ha26})).\vspace{0.1cm}}&\\
&&&&&&\parbox[t][0.3cm]{12.28866cm}{\raggedright E(level): This state is visible in the excitation function of \ensuremath{^{\textnormal{1}}}H(\ensuremath{^{\textnormal{17}}}F,p\ensuremath{_{\textnormal{1}}}\ensuremath{\gamma}) in (\href{https://www.nndc.bnl.gov/nsr/nsrlink.jsp?2009He16,B}{2009He16},\vspace{0.1cm}}&\\
&&&&&&\parbox[t][0.3cm]{12.28866cm}{\raggedright {\ }{\ }{\ }\href{https://www.nndc.bnl.gov/nsr/nsrlink.jsp?2010He17,B}{2010He17}), when the spectrum is conditioned exclusively in coincidence with the \ensuremath{\gamma}-rays\vspace{0.1cm}}&\\
&&&&&&\parbox[t][0.3cm]{12.28866cm}{\raggedright {\ }{\ }{\ }from the decay of \ensuremath{^{\textnormal{17}}}F*(495 keV).\vspace{0.1cm}}&\\
&&&&&&\parbox[t][0.3cm]{12.28866cm}{\raggedright \ensuremath{\Gamma}: Weighted average of \ensuremath{\Gamma}=50 keV \textit{5} (\href{https://www.nndc.bnl.gov/nsr/nsrlink.jsp?2001Ga18,B}{2001Ga18}, \href{https://www.nndc.bnl.gov/nsr/nsrlink.jsp?2001Go01,B}{2001Go01}, \href{https://www.nndc.bnl.gov/nsr/nsrlink.jsp?2002Li66,B}{2002Li66}); \ensuremath{\Gamma}=53.7 keV \textit{20}\vspace{0.1cm}}&\\
&&&&&&\parbox[t][0.3cm]{12.28866cm}{\raggedright {\ }{\ }{\ }(\href{https://www.nndc.bnl.gov/nsr/nsrlink.jsp?2012Ba28,B}{2012Ba28}: sum of \ensuremath{\Gamma}\ensuremath{_{\textnormal{p}_{\textnormal{0}}}}=37.8 keV \textit{19} and \ensuremath{\Gamma}\ensuremath{_{\textnormal{p}_{\textnormal{1}}}}=15.9 keV \textit{7} from their best R-matrix\vspace{0.1cm}}&\\
&&&&&&\parbox[t][0.3cm]{12.28866cm}{\raggedright {\ }{\ }{\ }fit. (\href{https://www.nndc.bnl.gov/nsr/nsrlink.jsp?2014Hu16,B}{2014Hu16}) mistakenly reported this sum as \ensuremath{\Gamma}=53.7 keV \textit{26}); \ensuremath{\Gamma}=50 keV \textit{15}\vspace{0.1cm}}&\\
&&&&&&\parbox[t][0.3cm]{12.28866cm}{\raggedright {\ }{\ }{\ }(\href{https://www.nndc.bnl.gov/nsr/nsrlink.jsp?2014Hu16,B}{2014Hu16}); and \ensuremath{\Gamma}=45 keV \textit{12} (L. E. Pratt, Ph.D. Thesis, 2014).\vspace{0.1cm}}&\\
&&&&&&\parbox[t][0.3cm]{12.28866cm}{\raggedright \ensuremath{\Gamma}: See also 58 keV \textit{2} (\href{https://www.nndc.bnl.gov/nsr/nsrlink.jsp?2003Bl11,B}{2003Bl11}: this value is excluded because (\href{https://www.nndc.bnl.gov/nsr/nsrlink.jsp?2012Ba28,B}{2012Ba28}) reanalyzed\vspace{0.1cm}}&\\
&&&&&&\parbox[t][0.3cm]{12.28866cm}{\raggedright {\ }{\ }{\ }these data); \ensuremath{\Gamma}(res)\ensuremath{\approx}120 keV \textit{60} (\href{https://www.nndc.bnl.gov/nsr/nsrlink.jsp?2009He16,B}{2009He16}: the experimental resolution is poor (80\vspace{0.1cm}}&\\
&&&&&&\parbox[t][0.3cm]{12.28866cm}{\raggedright {\ }{\ }{\ }keV) and the resonance width is not obtained from the observed width); and \ensuremath{\Gamma}= 40 keV\vspace{0.1cm}}&\\
&&&&&&\parbox[t][0.3cm]{12.28866cm}{\raggedright {\ }{\ }{\ }\textit{10} calculated by (\href{https://www.nndc.bnl.gov/nsr/nsrlink.jsp?2010Ad02,B}{2010Ad02}).\vspace{0.1cm}}&\\
&&&&&&\parbox[t][0.3cm]{12.28866cm}{\raggedright \ensuremath{\Gamma}\ensuremath{_{\ensuremath{\alpha}}}: Weighted average (with external errors) of \ensuremath{\Gamma}\ensuremath{_{\ensuremath{\alpha}}}=8 eV \textit{2} (\href{https://www.nndc.bnl.gov/nsr/nsrlink.jsp?2003Bl11,B}{2003Bl11}) and \ensuremath{\Gamma}\ensuremath{_{\ensuremath{\alpha}}}=3.2 eV\vspace{0.1cm}}&\\
&&&&&&\parbox[t][0.3cm]{12.28866cm}{\raggedright {\ }{\ }{\ }\textit{+50{\textminus}20} (\href{https://www.nndc.bnl.gov/nsr/nsrlink.jsp?2002Ha15,B}{2002Ha15}). See also the theoretical result of \ensuremath{\Gamma}\ensuremath{_{\ensuremath{\alpha}}}=3.9 eV \textit{10} and S\ensuremath{_{\ensuremath{\alpha}}}=0.23\vspace{0.1cm}}&\\
&&&&&&\parbox[t][0.3cm]{12.28866cm}{\raggedright {\ }{\ }{\ }(spectroscopic factor) from (\href{https://www.nndc.bnl.gov/nsr/nsrlink.jsp?2012Fo12,B}{2012Fo12}). These values are calculated for \ensuremath{^{\textnormal{18}}}Ne*(6135\vspace{0.1cm}}&\\
&&&&&&\parbox[t][0.3cm]{12.28866cm}{\raggedright {\ }{\ }{\ }keV, 1\ensuremath{^{-}}) state from the mirror level at \ensuremath{^{\textnormal{18}}}O*(6.20 MeV, 1\ensuremath{^{-}}) assuming equal\vspace{0.1cm}}&\\
&&&&&&\parbox[t][0.3cm]{12.28866cm}{\raggedright {\ }{\ }{\ }spectroscopic factors for mirror states. See also (\href{https://www.nndc.bnl.gov/nsr/nsrlink.jsp?2012Ok02,B}{2012Ok02}) for the theoretical ANC\vspace{0.1cm}}&\\
&&&&&&\parbox[t][0.3cm]{12.28866cm}{\raggedright {\ }{\ }{\ }values.\vspace{0.1cm}}&\\
&&&&&&\parbox[t][0.3cm]{12.28866cm}{\raggedright \ensuremath{\Gamma}\ensuremath{_{\textnormal{p}_{\textnormal{0}}}}: See also \ensuremath{\Gamma}\ensuremath{_{\textnormal{p}_{\textnormal{0}}}}\ensuremath{>}15 keV at 90\% confidence level (L. E. Pratt, Ph.D. Thesis, 2014).\vspace{0.1cm}}&\\
&&&&&&\parbox[t][0.3cm]{12.28866cm}{\raggedright J\ensuremath{^{\pi}}: From the R-matrix analyses of (\href{https://www.nndc.bnl.gov/nsr/nsrlink.jsp?2001Ga18,B}{2001Ga18}, \href{https://www.nndc.bnl.gov/nsr/nsrlink.jsp?2001Go01,B}{2001Go01}, \href{https://www.nndc.bnl.gov/nsr/nsrlink.jsp?2002Li66,B}{2002Li66}, \href{https://www.nndc.bnl.gov/nsr/nsrlink.jsp?2003Bl11,B}{2003Bl11},\vspace{0.1cm}}&\\
&&&&&&\parbox[t][0.3cm]{12.28866cm}{\raggedright {\ }{\ }{\ }\href{https://www.nndc.bnl.gov/nsr/nsrlink.jsp?2009He16,B}{2009He16}, \href{https://www.nndc.bnl.gov/nsr/nsrlink.jsp?2010He17,B}{2010He17}, \href{https://www.nndc.bnl.gov/nsr/nsrlink.jsp?2010Ji02,B}{2010Ji02}, \href{https://www.nndc.bnl.gov/nsr/nsrlink.jsp?2012Ba28,B}{2012Ba28}, \href{https://www.nndc.bnl.gov/nsr/nsrlink.jsp?2014Hu16,B}{2014Hu16}, and L. E. Pratt Ph.D. Thesis\vspace{0.1cm}}&\\
&&&&&&\parbox[t][0.3cm]{12.28866cm}{\raggedright {\ }{\ }{\ }2014), and comparison of excitation function measured by (\href{https://www.nndc.bnl.gov/nsr/nsrlink.jsp?2002Ha15,B}{2002Ha15}) with R-matrix\vspace{0.1cm}}&\\
&&&&&&\parbox[t][0.3cm]{12.28866cm}{\raggedright {\ }{\ }{\ }calculations for J\ensuremath{^{\ensuremath{\pi}}}=1\ensuremath{^{-}}.\vspace{0.1cm}}&\\
&&&&&&\parbox[t][0.3cm]{12.28866cm}{\raggedright J\ensuremath{^{\pi}}: (\href{https://www.nndc.bnl.gov/nsr/nsrlink.jsp?2014Hu16,B}{2014Hu16}) specifically ruled out the unnatural-parity J\ensuremath{^{\ensuremath{\pi}}}=2\ensuremath{^{-}} assignment made by\vspace{0.1cm}}&\\
&&&&&&\parbox[t][0.3cm]{12.28866cm}{\raggedright {\ }{\ }{\ }(\href{https://www.nndc.bnl.gov/nsr/nsrlink.jsp?2010HeZX,B}{2010HeZX}, unpublished) because such an assignment would be unlikely based on the\vspace{0.1cm}}&\\
&&&&&&\parbox[t][0.3cm]{12.28866cm}{\raggedright {\ }{\ }{\ }discussions of the 2p-emission from this state (\href{https://www.nndc.bnl.gov/nsr/nsrlink.jsp?2001Go01,B}{2001Go01}, \href{https://www.nndc.bnl.gov/nsr/nsrlink.jsp?2008Ra12,B}{2008Ra12}). (\href{https://www.nndc.bnl.gov/nsr/nsrlink.jsp?2014Hu16,B}{2014Hu16}) also\vspace{0.1cm}}&\\
\end{longtable}
\begin{textblock}{29}(0,27.3)
Continued on next page (footnotes at end of table)
\end{textblock}
\clearpage
\begin{longtable}{ccccccc@{\extracolsep{\fill}}c}
\\[-.4cm]
\multicolumn{8}{c}{{\bf \small \underline{\ensuremath{^{\textnormal{1}}}H(\ensuremath{^{\textnormal{17}}}F,p),(\ensuremath{^{\textnormal{17}}}F,2p),(\ensuremath{^{\textnormal{17}}}F,\ensuremath{\alpha}):res\hspace{0.2in}\href{https://www.nndc.bnl.gov/nsr/nsrlink.jsp?1999Ba49,B}{1999Ba49},\href{https://www.nndc.bnl.gov/nsr/nsrlink.jsp?2020Br14,B}{2020Br14} (continued)}}}\\
\multicolumn{8}{c}{~}\\
\multicolumn{8}{c}{\underline{\ensuremath{^{18}}Ne Levels (continued)}}\\
\multicolumn{8}{c}{~}\\
\multicolumn{2}{c}{E(level)$^{}$}&J$^{\pi}$$^{}$&\multicolumn{2}{c}{\ensuremath{\Gamma}$^{}$}&L$^{}$&Comments&\\[-.2cm]
\multicolumn{2}{c}{\hrulefill}&\hrulefill&\multicolumn{2}{c}{\hrulefill}&\hrulefill&\hrulefill&
\endhead
&&&&&&\parbox[t][0.3cm]{10.904081cm}{\raggedright {\ }{\ }{\ }ruled out the J\ensuremath{^{\ensuremath{\pi}}}=3\ensuremath{^{-}} assignment made by (\href{https://www.nndc.bnl.gov/nsr/nsrlink.jsp?2010HeZX,B}{2010HeZX}) due to the large\vspace{0.1cm}}&\\
&&&&&&\parbox[t][0.3cm]{10.904081cm}{\raggedright {\ }{\ }{\ }inelastic branch observed for this state in (\href{https://www.nndc.bnl.gov/nsr/nsrlink.jsp?2014Hu16,B}{2014Hu16}: see Fig. 3(c)).\vspace{0.1cm}}&\\
&&&&&&\parbox[t][0.3cm]{10.904081cm}{\raggedright (\href{https://www.nndc.bnl.gov/nsr/nsrlink.jsp?2014Hu16,B}{2014Hu16}) considered this state to have a 4p-2h configuration, where h (holes)\vspace{0.1cm}}&\\
&&&&&&\parbox[t][0.3cm]{10.904081cm}{\raggedright {\ }{\ }{\ }are in 1\textit{p}\ensuremath{_{\textnormal{3/2}}} and p (particles) are in 2\textit{s}\ensuremath{_{\textnormal{1/2}}} or 1\textit{d}\ensuremath{_{\textnormal{3/2}}} orbitals.\vspace{0.1cm}}&\\
&&&&&&\parbox[t][0.3cm]{10.904081cm}{\raggedright L: From (\href{https://www.nndc.bnl.gov/nsr/nsrlink.jsp?2014Hu16,B}{2014Hu16}).\vspace{0.1cm}}&\\
&&&&&&\parbox[t][0.3cm]{10.904081cm}{\raggedright \ensuremath{\omega}\ensuremath{\gamma}\ensuremath{_{\textnormal{(p,p}'\textnormal{)}}}=22 keV \textit{7} (\href{https://www.nndc.bnl.gov/nsr/nsrlink.jsp?2009He16,B}{2009He16}). An enhanced p\ensuremath{_{\textnormal{1}}} branch is preferred by\vspace{0.1cm}}&\\
&&&&&&\parbox[t][0.3cm]{10.904081cm}{\raggedright {\ }{\ }{\ }(\href{https://www.nndc.bnl.gov/nsr/nsrlink.jsp?2009He16,B}{2009He16}).\vspace{0.1cm}}&\\
&&&&&&\parbox[t][0.3cm]{10.904081cm}{\raggedright (\href{https://www.nndc.bnl.gov/nsr/nsrlink.jsp?2002Ha15,B}{2002Ha15}): \ensuremath{\omega}\ensuremath{\gamma}\ensuremath{_{\textnormal{(p,}\ensuremath{\alpha}\textnormal{)}}}=0.8 eV \textit{+12{\textminus}5} and \ensuremath{\Theta}\ensuremath{^{\textnormal{2}}_{\ensuremath{\alpha}}}=0.19 \textit{+30{\textminus}10}, which is labeled\vspace{0.1cm}}&\\
&&&&&&\parbox[t][0.3cm]{10.904081cm}{\raggedright {\ }{\ }{\ }as the spectroscopic factor in (\href{https://www.nndc.bnl.gov/nsr/nsrlink.jsp?2002Ha15,B}{2002Ha15}).\vspace{0.1cm}}&\\
&&&&&&\parbox[t][0.3cm]{10.904081cm}{\raggedright (\href{https://www.nndc.bnl.gov/nsr/nsrlink.jsp?2003Bl11,B}{2003Bl11}) reported that the decay mode of this state is via proton decay to the\vspace{0.1cm}}&\\
&&&&&&\parbox[t][0.3cm]{10.904081cm}{\raggedright {\ }{\ }{\ }\ensuremath{^{\textnormal{17}}}F*(495 keV, 1/2\ensuremath{^{\textnormal{+}}}) state. (\href{https://www.nndc.bnl.gov/nsr/nsrlink.jsp?2001Ga18,B}{2001Ga18}, \href{https://www.nndc.bnl.gov/nsr/nsrlink.jsp?2001Go01,B}{2001Go01}, \href{https://www.nndc.bnl.gov/nsr/nsrlink.jsp?2002Li66,B}{2002Li66}) reported a 2p\vspace{0.1cm}}&\\
&&&&&&\parbox[t][0.3cm]{10.904081cm}{\raggedright {\ }{\ }{\ }decay to \ensuremath{^{\textnormal{16}}}O\ensuremath{_{\textnormal{g.s.}}}(0\ensuremath{^{+}}) with \ensuremath{\Gamma}\ensuremath{_{^{\textnormal{2}}\textnormal{He}}}=21 eV \textit{3} (\href{https://www.nndc.bnl.gov/nsr/nsrlink.jsp?2001Go01,B}{2001Go01}) for decay via\vspace{0.1cm}}&\\
&&&&&&\parbox[t][0.3cm]{10.904081cm}{\raggedright {\ }{\ }{\ }emission of a diproton and with \ensuremath{\Gamma}\ensuremath{_{\textnormal{2p}}}=57 eV \textit{6} (\href{https://www.nndc.bnl.gov/nsr/nsrlink.jsp?2001Go01,B}{2001Go01}) assuming a\vspace{0.1cm}}&\\
&&&&&&\parbox[t][0.3cm]{10.904081cm}{\raggedright {\ }{\ }{\ }3-body decay. Because of limited angular coverage in these experiments, the\vspace{0.1cm}}&\\
&&&&&&\parbox[t][0.3cm]{10.904081cm}{\raggedright {\ }{\ }{\ }data did not differentiate between simultaneous decay, diproton (\ensuremath{^{\textnormal{2}}}He)\vspace{0.1cm}}&\\
&&&&&&\parbox[t][0.3cm]{10.904081cm}{\raggedright {\ }{\ }{\ }emission, or direct three-body decay. However, (\href{https://www.nndc.bnl.gov/nsr/nsrlink.jsp?2001Ga18,B}{2001Ga18}) suggests that the\vspace{0.1cm}}&\\
&&&&&&\parbox[t][0.3cm]{10.904081cm}{\raggedright {\ }{\ }{\ }mode of decay is most likely via simultaneous two proton emission, and\vspace{0.1cm}}&\\
&&&&&&\parbox[t][0.3cm]{10.904081cm}{\raggedright {\ }{\ }{\ }(\href{https://www.nndc.bnl.gov/nsr/nsrlink.jsp?2001Go01,B}{2001Go01}) finds a 2p decay branching ratio of 4.2\ensuremath{\times}10\ensuremath{^{\textnormal{$-$4}}} for the \ensuremath{^{\textnormal{2}}}He\vspace{0.1cm}}&\\
&&&&&&\parbox[t][0.3cm]{10.904081cm}{\raggedright {\ }{\ }{\ }emission mode and 1.1\ensuremath{\times}10\ensuremath{^{\textnormal{$-$3}}} assuming a three-body decay. (\href{https://www.nndc.bnl.gov/nsr/nsrlink.jsp?2001Go01,B}{2001Go01}) rules\vspace{0.1cm}}&\\
&&&&&&\parbox[t][0.3cm]{10.904081cm}{\raggedright {\ }{\ }{\ }out the 3-body decay and supports a decay mode via the emission of \ensuremath{^{\textnormal{2}}}He.\vspace{0.1cm}}&\\
&&&&&&\parbox[t][0.3cm]{10.904081cm}{\raggedright {\ }{\ }{\ }(\href{https://www.nndc.bnl.gov/nsr/nsrlink.jsp?2001Go01,B}{2001Go01}) estimated an spectroscopic factor of 0.35 for the 2p decay of this\vspace{0.1cm}}&\\
&&&&&&\parbox[t][0.3cm]{10.904081cm}{\raggedright {\ }{\ }{\ }state and calculated and discussed the spectroscopic factors obtained if the\vspace{0.1cm}}&\\
&&&&&&\parbox[t][0.3cm]{10.904081cm}{\raggedright {\ }{\ }{\ }decay is assumed to be of 3-body form.\vspace{0.1cm}}&\\
\multicolumn{1}{r@{}}{6280}&\multicolumn{1}{@{}l}{\ensuremath{^{{\hyperlink{NE4LEVEL0}{a}}}} {\it 30}}&\multicolumn{1}{l}{3\ensuremath{^{-}}}&\multicolumn{1}{r@{}}{20}&\multicolumn{1}{@{ }l}{keV {\it 15}}&\multicolumn{1}{l}{1}&\parbox[t][0.3cm]{10.904081cm}{\raggedright E(level): From (\href{https://www.nndc.bnl.gov/nsr/nsrlink.jsp?2014Hu16,B}{2014Hu16}: corresponds to E\ensuremath{_{\textnormal{c.m.}}}=2.36 MeV. The resonance\vspace{0.1cm}}&\\
&&&&&&\parbox[t][0.3cm]{10.904081cm}{\raggedright {\ }{\ }{\ }parameters of this state used in the R-matrix fit of (\href{https://www.nndc.bnl.gov/nsr/nsrlink.jsp?2014Hu16,B}{2014Hu16}) were taken\vspace{0.1cm}}&\\
&&&&&&\parbox[t][0.3cm]{10.904081cm}{\raggedright {\ }{\ }{\ }from (\href{https://www.nndc.bnl.gov/nsr/nsrlink.jsp?1996Ha26,B}{1996Ha26}), which resulted in a good fit in (\href{https://www.nndc.bnl.gov/nsr/nsrlink.jsp?2014Hu16,B}{2014Hu16})). See alo\vspace{0.1cm}}&\\
&&&&&&\parbox[t][0.3cm]{10.904081cm}{\raggedright {\ }{\ }{\ }E\ensuremath{_{\textnormal{x}}}=6.29 MeV (\href{https://www.nndc.bnl.gov/nsr/nsrlink.jsp?2002Ha15,B}{2002Ha15}: from E\ensuremath{_{\textnormal{c.m.}}}=2.37 MeV); 6240 keV (\href{https://www.nndc.bnl.gov/nsr/nsrlink.jsp?2010HeZX,B}{2010HeZX});\vspace{0.1cm}}&\\
&&&&&&\parbox[t][0.3cm]{10.904081cm}{\raggedright {\ }{\ }{\ }and \ensuremath{\approx}6310 keV (\href{https://www.nndc.bnl.gov/nsr/nsrlink.jsp?2003Bl11,B}{2003Bl11}).\vspace{0.1cm}}&\\
&&&&&&\parbox[t][0.3cm]{10.904081cm}{\raggedright J\ensuremath{^{\pi}},\ensuremath{\Gamma},L: From an R-matrix analysis by (\href{https://www.nndc.bnl.gov/nsr/nsrlink.jsp?2014Hu16,B}{2014Hu16}).\vspace{0.1cm}}&\\
&&&&&&\parbox[t][0.3cm]{10.904081cm}{\raggedright \ensuremath{\Gamma}\ensuremath{_{\textnormal{p}_{\textnormal{0}}}}\ensuremath{<}4 keV from (L. E. Pratt, Ph.D. Thesis, 2014) at 90\% confidence level.\vspace{0.1cm}}&\\
&&&&&&\parbox[t][0.3cm]{10.904081cm}{\raggedright J\ensuremath{^{\pi}}: See also (\href{https://www.nndc.bnl.gov/nsr/nsrlink.jsp?2002Ha15,B}{2002Ha15}), where a comparison was made for the measured\vspace{0.1cm}}&\\
&&&&&&\parbox[t][0.3cm]{10.904081cm}{\raggedright {\ }{\ }{\ }excitation function and the R-matrix calculations using J\ensuremath{^{\ensuremath{\pi}}}=3\ensuremath{^{-}}.\vspace{0.1cm}}&\\
&&&&&&\parbox[t][0.3cm]{10.904081cm}{\raggedright \ensuremath{\omega}\ensuremath{\gamma}\ensuremath{_{\textnormal{(p,}\ensuremath{\alpha}\textnormal{)}}}=0.2 eV (\href{https://www.nndc.bnl.gov/nsr/nsrlink.jsp?2002Ha15,B}{2002Ha15}).\vspace{0.1cm}}&\\
\multicolumn{1}{r@{}}{6341}&\multicolumn{1}{@{ }l}{{\it 10}}&\multicolumn{1}{l}{2\ensuremath{^{-}}}&\multicolumn{1}{r@{}}{10}&\multicolumn{1}{@{ }l}{keV {\it 5}}&\multicolumn{1}{l}{1}&\parbox[t][0.3cm]{10.904081cm}{\raggedright E(level): Weighted average of 6.34 MeV \textit{1} (\href{https://www.nndc.bnl.gov/nsr/nsrlink.jsp?2001Ga18,B}{2001Ga18}, \href{https://www.nndc.bnl.gov/nsr/nsrlink.jsp?2001Go01,B}{2001Go01}, \href{https://www.nndc.bnl.gov/nsr/nsrlink.jsp?2002Li66,B}{2002Li66}:\vspace{0.1cm}}&\\
&&&&&&\parbox[t][0.3cm]{10.904081cm}{\raggedright {\ }{\ }{\ }from E\ensuremath{_{\textnormal{c.m.}}}=2.42 MeV \textit{1}); and 6350 keV \textit{30} (\href{https://www.nndc.bnl.gov/nsr/nsrlink.jsp?2014Hu16,B}{2014Hu16}). See also E\ensuremath{_{\textnormal{x}}}=6.35\vspace{0.1cm}}&\\
&&&&&&\parbox[t][0.3cm]{10.904081cm}{\raggedright {\ }{\ }{\ }MeV (\href{https://www.nndc.bnl.gov/nsr/nsrlink.jsp?2002Ha15,B}{2002Ha15}: from E\ensuremath{_{\textnormal{c.m.}}}=2.43 MeV); \ensuremath{\approx}6310 keV (\href{https://www.nndc.bnl.gov/nsr/nsrlink.jsp?2003Bl11,B}{2003Bl11}); 6320 keV\vspace{0.1cm}}&\\
&&&&&&\parbox[t][0.3cm]{10.904081cm}{\raggedright {\ }{\ }{\ }(\href{https://www.nndc.bnl.gov/nsr/nsrlink.jsp?2010HeZX,B}{2010HeZX}); and 6373 keV \textit{8} (stat.) \textit{8} (sys.) (L. E. Pratt, Ph.D. Thesis, 2014:\vspace{0.1cm}}&\\
&&&&&&\parbox[t][0.3cm]{10.904081cm}{\raggedright {\ }{\ }{\ }this state was a 2\ensuremath{^{-}} state, whose J\ensuremath{^{\ensuremath{\pi}}}, energy and width were deduced from a\vspace{0.1cm}}&\\
&&&&&&\parbox[t][0.3cm]{10.904081cm}{\raggedright {\ }{\ }{\ }phenomenological R-matrix calculation. The state was unresolved from the\vspace{0.1cm}}&\\
&&&&&&\parbox[t][0.3cm]{10.904081cm}{\raggedright {\ }{\ }{\ }nearby 3\ensuremath{^{-}} state at 6280 keV \textit{30}. Pratt set an upper limit on the proton partial\vspace{0.1cm}}&\\
&&&&&&\parbox[t][0.3cm]{10.904081cm}{\raggedright {\ }{\ }{\ }width of the 3\ensuremath{^{-}} state to the \ensuremath{^{\textnormal{17}}}F\ensuremath{_{\textnormal{g.s.}}} of \ensuremath{\Gamma}\ensuremath{_{\textnormal{p}}}\ensuremath{<}4 keV at 90\% confidence level).\vspace{0.1cm}}&\\
&&&&&&\parbox[t][0.3cm]{10.904081cm}{\raggedright \ensuremath{\Gamma},J\ensuremath{^{\pi}},L: From the R-matrix analysis of (\href{https://www.nndc.bnl.gov/nsr/nsrlink.jsp?2014Hu16,B}{2014Hu16}).\vspace{0.1cm}}&\\
&&&&&&\parbox[t][0.3cm]{10.904081cm}{\raggedright \ensuremath{\Gamma}: See also \ensuremath{\Gamma}\ensuremath{\approx}50 keV (\href{https://www.nndc.bnl.gov/nsr/nsrlink.jsp?2001Go01,B}{2001Go01}: however, the R-matrix fit does not\vspace{0.1cm}}&\\
&&&&&&\parbox[t][0.3cm]{10.904081cm}{\raggedright {\ }{\ }{\ }describe all of the observed features of the data in the vicinity of this state);\vspace{0.1cm}}&\\
&&&&&&\parbox[t][0.3cm]{10.904081cm}{\raggedright {\ }{\ }{\ }and \ensuremath{\Gamma}=87 keV \textit{15} deduced from an R-matrix analysis for the 6373 keV\vspace{0.1cm}}&\\
&&&&&&\parbox[t][0.3cm]{10.904081cm}{\raggedright {\ }{\ }{\ }unresolved doublet observed in L. E. Pratt Ph.D. Thesis (2014).\vspace{0.1cm}}&\\
&&&&&&\parbox[t][0.3cm]{10.904081cm}{\raggedright J\ensuremath{^{\pi}}: See also (\href{https://www.nndc.bnl.gov/nsr/nsrlink.jsp?2002Ha15,B}{2002Ha15}: comparison of excitation function with R-matrix\vspace{0.1cm}}&\\
&&&&&&\parbox[t][0.3cm]{10.904081cm}{\raggedright {\ }{\ }{\ }calculations) and (\href{https://www.nndc.bnl.gov/nsr/nsrlink.jsp?2010HeZX,B}{2010HeZX}: unpublished, R-matrix analysis).\vspace{0.1cm}}&\\
\multicolumn{1}{r@{}}{6.85\ensuremath{\times10^{3}}}&\multicolumn{1}{@{ }l}{{\it 11}}&\multicolumn{1}{l}{(0\ensuremath{^{+}},0\ensuremath{^{-}})}&\multicolumn{1}{r@{}}{50}&\multicolumn{1}{@{ }l}{keV {\it 30}}&\multicolumn{1}{l}{(2,3)}&\parbox[t][0.3cm]{10.904081cm}{\raggedright \ensuremath{\Gamma}\ensuremath{\alpha}=149 eV (\href{https://www.nndc.bnl.gov/nsr/nsrlink.jsp?2014Hu16,B}{2014Hu16})\vspace{0.1cm}}&\\
&&&&&&\parbox[t][0.3cm]{10.904081cm}{\raggedright E(level): From (\href{https://www.nndc.bnl.gov/nsr/nsrlink.jsp?2014Hu16,B}{2014Hu16}). The energy uncertainty of this state observed in\vspace{0.1cm}}&\\
&&&&&&\parbox[t][0.3cm]{10.904081cm}{\raggedright {\ }{\ }{\ }(\href{https://www.nndc.bnl.gov/nsr/nsrlink.jsp?2014Hu16,B}{2014Hu16}) is sometimes mentioned as 100 keV in the text but in the abstract\vspace{0.1cm}}&\\
&&&&&&\parbox[t][0.3cm]{10.904081cm}{\raggedright {\ }{\ }{\ }and tables, it is reported as 110 keV.\vspace{0.1cm}}&\\
&&&&&&\parbox[t][0.3cm]{10.904081cm}{\raggedright E(level): Based on all the evidence discussed in (\href{https://www.nndc.bnl.gov/nsr/nsrlink.jsp?2014Hu16,B}{2014Hu16}), they concluded\vspace{0.1cm}}&\\
&&&&&&\parbox[t][0.3cm]{10.904081cm}{\raggedright {\ }{\ }{\ }that very likely a new state around 6.8 MeV exists in \ensuremath{^{\textnormal{18}}}Ne. In addition to\vspace{0.1cm}}&\\
&&&&&&\parbox[t][0.3cm]{10.904081cm}{\raggedright {\ }{\ }{\ }those arguments, (\href{https://www.nndc.bnl.gov/nsr/nsrlink.jsp?2011He09,B}{2011He09}) also argued that their R-matrix fit to the\vspace{0.1cm}}&\\
\end{longtable}
\begin{textblock}{29}(0,27.3)
Continued on next page (footnotes at end of table)
\end{textblock}
\clearpage
\begin{longtable}{ccccccc@{\extracolsep{\fill}}c}
\\[-.4cm]
\multicolumn{8}{c}{{\bf \small \underline{\ensuremath{^{\textnormal{1}}}H(\ensuremath{^{\textnormal{17}}}F,p),(\ensuremath{^{\textnormal{17}}}F,2p),(\ensuremath{^{\textnormal{17}}}F,\ensuremath{\alpha}):res\hspace{0.2in}\href{https://www.nndc.bnl.gov/nsr/nsrlink.jsp?1999Ba49,B}{1999Ba49},\href{https://www.nndc.bnl.gov/nsr/nsrlink.jsp?2020Br14,B}{2020Br14} (continued)}}}\\
\multicolumn{8}{c}{~}\\
\multicolumn{8}{c}{\underline{\ensuremath{^{18}}Ne Levels (continued)}}\\
\multicolumn{8}{c}{~}\\
\multicolumn{2}{c}{E(level)$^{}$}&J$^{\pi}$$^{}$&\multicolumn{2}{c}{\ensuremath{\Gamma}$^{}$}&L$^{}$&Comments&\\[-.2cm]
\multicolumn{2}{c}{\hrulefill}&\hrulefill&\multicolumn{2}{c}{\hrulefill}&\hrulefill&\hrulefill&
\endhead
&&&&&&\parbox[t][0.3cm]{11.625261cm}{\raggedright {\ }{\ }{\ }measured \ensuremath{^{\textnormal{18}}}Ne excitation function improved when considering a doublet near 7.1\vspace{0.1cm}}&\\
&&&&&&\parbox[t][0.3cm]{11.625261cm}{\raggedright {\ }{\ }{\ }MeV. Their best fit with a doublet resulted in two states at 6.97 MeV and 7.09\vspace{0.1cm}}&\\
&&&&&&\parbox[t][0.3cm]{11.625261cm}{\raggedright {\ }{\ }{\ }MeV with J\ensuremath{^{\ensuremath{\pi}}}=2\ensuremath{^{\textnormal{+}}} (with L=2) and J\ensuremath{^{\ensuremath{\pi}}}=4\ensuremath{^{\textnormal{+}}} (with L=2), respectively. However, they\vspace{0.1cm}}&\\
&&&&&&\parbox[t][0.3cm]{11.625261cm}{\raggedright {\ }{\ }{\ }did not report an uncertainty for these excitation energies. (\href{https://www.nndc.bnl.gov/nsr/nsrlink.jsp?2011He09,B}{2011He09}) mention in\vspace{0.1cm}}&\\
&&&&&&\parbox[t][0.3cm]{11.625261cm}{\raggedright {\ }{\ }{\ }the text that the uncertainty in excitation energies were estimated to be 30 keV but\vspace{0.1cm}}&\\
&&&&&&\parbox[t][0.3cm]{11.625261cm}{\raggedright {\ }{\ }{\ }it is not clear if this would still be the case for this doublet. (\href{https://www.nndc.bnl.gov/nsr/nsrlink.jsp?2011He09,B}{2011He09}) did not\vspace{0.1cm}}&\\
&&&&&&\parbox[t][0.3cm]{11.625261cm}{\raggedright {\ }{\ }{\ }suggest the energies of the doublet as their final result. Instead, they took the\vspace{0.1cm}}&\\
&&&&&&\parbox[t][0.3cm]{11.625261cm}{\raggedright {\ }{\ }{\ }7.05-MeV and 7.12-MeV for this doublet from (\href{https://www.nndc.bnl.gov/nsr/nsrlink.jsp?1996Ha26,B}{1996Ha26}).\vspace{0.1cm}}&\\
&&&&&&\parbox[t][0.3cm]{11.625261cm}{\raggedright \ensuremath{\Gamma}: From (\href{https://www.nndc.bnl.gov/nsr/nsrlink.jsp?2014Hu16,B}{2014Hu16}).\vspace{0.1cm}}&\\
&&&&&&\parbox[t][0.3cm]{11.625261cm}{\raggedright J\ensuremath{^{\pi}}: From R-matrix analysis of (\href{https://www.nndc.bnl.gov/nsr/nsrlink.jsp?2014Hu16,B}{2014Hu16}): J\ensuremath{^{\ensuremath{\pi}}}=(0\ensuremath{^{+}},0\ensuremath{^{-}}). (\href{https://www.nndc.bnl.gov/nsr/nsrlink.jsp?2014Hu16,B}{2014Hu16}) prefers the\vspace{0.1cm}}&\\
&&&&&&\parbox[t][0.3cm]{11.625261cm}{\raggedright {\ }{\ }{\ }J\ensuremath{^{\ensuremath{\pi}}}=0\ensuremath{^{\textnormal{+}}} assignment for this state based on the evidence of this state potentially being\vspace{0.1cm}}&\\
&&&&&&\parbox[t][0.3cm]{11.625261cm}{\raggedright {\ }{\ }{\ }populated in the direct \ensuremath{\alpha}(\ensuremath{^{\textnormal{14}}}O,p) reaction in (\href{https://www.nndc.bnl.gov/nsr/nsrlink.jsp?2004No14,B}{2004No14}, \href{https://www.nndc.bnl.gov/nsr/nsrlink.jsp?2004No18,B}{2004No18}).\vspace{0.1cm}}&\\
&&&&&&\parbox[t][0.3cm]{11.625261cm}{\raggedright L: From (\href{https://www.nndc.bnl.gov/nsr/nsrlink.jsp?2014Hu16,B}{2014Hu16}): L=2 for J\ensuremath{^{\ensuremath{\pi}}}=0\ensuremath{^{\textnormal{+}}}, and L=3 for J\ensuremath{^{\ensuremath{\pi}}}=0\ensuremath{^{-}}, which is less preferred by\vspace{0.1cm}}&\\
&&&&&&\parbox[t][0.3cm]{11.625261cm}{\raggedright {\ }{\ }{\ }(\href{https://www.nndc.bnl.gov/nsr/nsrlink.jsp?2014Hu16,B}{2014Hu16}).\vspace{0.1cm}}&\\
&&&&&&\parbox[t][0.3cm]{11.625261cm}{\raggedright \ensuremath{\omega}\ensuremath{\gamma}\ensuremath{_{\textnormal{(}\ensuremath{\alpha}\textnormal{,p)}}}=149 eV assuming J\ensuremath{^{\ensuremath{\pi}}}=0\ensuremath{^{\textnormal{+}}} and C\ensuremath{^{\textnormal{2}}}S=0.01 from (\href{https://www.nndc.bnl.gov/nsr/nsrlink.jsp?2014Hu16,B}{2014Hu16}). These values\vspace{0.1cm}}&\\
&&&&&&\parbox[t][0.3cm]{11.625261cm}{\raggedright {\ }{\ }{\ }together with \ensuremath{\Gamma}\ensuremath{_{\ensuremath{\alpha}}}=149 eV are calculated by (\href{https://www.nndc.bnl.gov/nsr/nsrlink.jsp?2014Hu16,B}{2014Hu16}).\vspace{0.1cm}}&\\
\multicolumn{1}{r@{}}{7054}&\multicolumn{1}{@{ }l}{{\it 28}}&\multicolumn{1}{l}{4\ensuremath{^{+}}}&\multicolumn{1}{r@{}}{94}&\multicolumn{1}{@{ }l}{keV {\it 20}}&\multicolumn{1}{l}{2}&\parbox[t][0.3cm]{11.625261cm}{\raggedright \ensuremath{\Gamma}\ensuremath{_{\textnormal{p}}}=94 eV \textit{20}\vspace{0.1cm}}&\\
&&&&&&\parbox[t][0.3cm]{11.625261cm}{\raggedright \ensuremath{\Gamma}\ensuremath{\alpha}=40 eV \textit{14} (\href{https://www.nndc.bnl.gov/nsr/nsrlink.jsp?2002Ha15,B}{2002Ha15})\vspace{0.1cm}}&\\
&&&&&&\parbox[t][0.3cm]{11.625261cm}{\raggedright E(level): Weighted average of 7.16 MeV \textit{15} (\href{https://www.nndc.bnl.gov/nsr/nsrlink.jsp?1999Ha14,B}{1999Ha14}: associated with E\ensuremath{_{\textnormal{c.m.}}}=3.24\vspace{0.1cm}}&\\
&&&&&&\parbox[t][0.3cm]{11.625261cm}{\raggedright {\ }{\ }{\ }MeV); 7.05 MeV \textit{10} (\href{https://www.nndc.bnl.gov/nsr/nsrlink.jsp?2002Ha15,B}{2002Ha15}: from E\ensuremath{_{\textnormal{c.m.}}}=3.13 MeV \textit{10}); and 7.05 MeV \textit{3}\vspace{0.1cm}}&\\
&&&&&&\parbox[t][0.3cm]{11.625261cm}{\raggedright {\ }{\ }{\ }(\href{https://www.nndc.bnl.gov/nsr/nsrlink.jsp?2014Hu16,B}{2014Hu16}: associated with E\ensuremath{_{\textnormal{c.m.}}}=3.13 MeV). See also \ensuremath{\approx}7070 keV (\href{https://www.nndc.bnl.gov/nsr/nsrlink.jsp?2001Bl06,B}{2001Bl06});\vspace{0.1cm}}&\\
&&&&&&\parbox[t][0.3cm]{11.625261cm}{\raggedright {\ }{\ }{\ }7050 keV (\href{https://www.nndc.bnl.gov/nsr/nsrlink.jsp?2001HaZP,B}{2001HaZP}); 7050 (\href{https://www.nndc.bnl.gov/nsr/nsrlink.jsp?2001HaZQ,B}{2001HaZQ}); and 7092 keV (\href{https://www.nndc.bnl.gov/nsr/nsrlink.jsp?2003Bl11,B}{2003Bl11}).\vspace{0.1cm}}&\\
&&&&&&\parbox[t][0.3cm]{11.625261cm}{\raggedright \ensuremath{\Gamma}: Weighted average of \ensuremath{\Gamma}\ensuremath{_{\textnormal{p}}}=90 keV \textit{40} (\href{https://www.nndc.bnl.gov/nsr/nsrlink.jsp?2002Ha15,B}{2002Ha15}: Assuming \ensuremath{\Gamma}=\ensuremath{\Gamma}\ensuremath{_{\textnormal{p}}}) and \ensuremath{\Gamma}=95\vspace{0.1cm}}&\\
&&&&&&\parbox[t][0.3cm]{11.625261cm}{\raggedright {\ }{\ }{\ }keV \textit{20} (\href{https://www.nndc.bnl.gov/nsr/nsrlink.jsp?2014Hu16,B}{2014Hu16}). Evaluator assumed \ensuremath{\Gamma}\ensuremath{\approx}\ensuremath{\Gamma}\ensuremath{_{\textnormal{p}}}.\vspace{0.1cm}}&\\
&&&&&&\parbox[t][0.3cm]{11.625261cm}{\raggedright \ensuremath{\Gamma}\ensuremath{_{\textnormal{p$'$}}}\ensuremath{<}1 keV (\href{https://www.nndc.bnl.gov/nsr/nsrlink.jsp?2002Ha15,B}{2002Ha15}).\vspace{0.1cm}}&\\
&&&&&&\parbox[t][0.3cm]{11.625261cm}{\raggedright J\ensuremath{^{\pi}}: From (\href{https://www.nndc.bnl.gov/nsr/nsrlink.jsp?2001HaZQ,B}{2001HaZQ}: based on angular momentum selection rules); (\href{https://www.nndc.bnl.gov/nsr/nsrlink.jsp?2002Ha15,B}{2002Ha15}:\vspace{0.1cm}}&\\
&&&&&&\parbox[t][0.3cm]{11.625261cm}{\raggedright {\ }{\ }{\ }comparison of measured excitation function with R-matrix analysis); (\href{https://www.nndc.bnl.gov/nsr/nsrlink.jsp?2011He09,B}{2011He09}:\vspace{0.1cm}}&\\
&&&&&&\parbox[t][0.3cm]{11.625261cm}{\raggedright {\ }{\ }{\ }R-matrix analysis with L=2, see text); and (\href{https://www.nndc.bnl.gov/nsr/nsrlink.jsp?2014Hu16,B}{2014Hu16}: R-matrix analysis with\vspace{0.1cm}}&\\
&&&&&&\parbox[t][0.3cm]{11.625261cm}{\raggedright {\ }{\ }{\ }L=2). See also J\ensuremath{^{\ensuremath{\pi}}}=(1\ensuremath{^{-}}) (\href{https://www.nndc.bnl.gov/nsr/nsrlink.jsp?1999Ha14,B}{1999Ha14}: mirror level and Coulomb shift analyses). This\vspace{0.1cm}}&\\
&&&&&&\parbox[t][0.3cm]{11.625261cm}{\raggedright {\ }{\ }{\ }assignment was ruled out by (\href{https://www.nndc.bnl.gov/nsr/nsrlink.jsp?2000Fo19,B}{2000Fo19}), who suggested J\ensuremath{^{\ensuremath{\pi}}}=(4\ensuremath{^{\textnormal{+}}}) based on the\vspace{0.1cm}}&\\
&&&&&&\parbox[t][0.3cm]{11.625261cm}{\raggedright {\ }{\ }{\ }theoretical calculations in (\href{https://www.nndc.bnl.gov/nsr/nsrlink.jsp?1998Sh35,B}{1998Sh35}) and the Coulomb shift between mirror levels.\vspace{0.1cm}}&\\
&&&&&&\parbox[t][0.3cm]{11.625261cm}{\raggedright L: From (\href{https://www.nndc.bnl.gov/nsr/nsrlink.jsp?2011He09,B}{2011He09}: L=2 for J\ensuremath{^{\ensuremath{\pi}}}=4\ensuremath{^{\textnormal{+}}}, see text); and (\href{https://www.nndc.bnl.gov/nsr/nsrlink.jsp?2014Hu16,B}{2014Hu16}).\vspace{0.1cm}}&\\
&&&&&&\parbox[t][0.3cm]{11.625261cm}{\raggedright Proton spectroscopic factor: \ensuremath{\Theta}\ensuremath{^{\textnormal{2}}_{\textnormal{p}}}=0.25 (\href{https://www.nndc.bnl.gov/nsr/nsrlink.jsp?2002Ha15,B}{2002Ha15}) for J\ensuremath{^{\ensuremath{\pi}}}=4\ensuremath{^{\textnormal{+}}}.\vspace{0.1cm}}&\\
&&&&&&\parbox[t][0.3cm]{11.625261cm}{\raggedright The \ensuremath{\alpha}-particle spectroscopic factor: \ensuremath{\Theta}\ensuremath{^{\textnormal{2}}_{\ensuremath{\alpha}}}=0.50 \textit{17} (\href{https://www.nndc.bnl.gov/nsr/nsrlink.jsp?2002Ha15,B}{2002Ha15}: for J\ensuremath{^{\ensuremath{\pi}}}=4\ensuremath{^{\textnormal{+}}}).\vspace{0.1cm}}&\\
&&&&&&\parbox[t][0.3cm]{11.625261cm}{\raggedright \ensuremath{\omega}\ensuremath{\gamma}\ensuremath{_{\textnormal{(p,}\ensuremath{\alpha}\textnormal{)}}}=29 eV \textit{10} (\href{https://www.nndc.bnl.gov/nsr/nsrlink.jsp?2002Ha15,B}{2002Ha15}: for J\ensuremath{^{\ensuremath{\pi}}}=4\ensuremath{^{\textnormal{+}}}). Calculations performed by (\href{https://www.nndc.bnl.gov/nsr/nsrlink.jsp?2012Fo29,B}{2012Fo29})\vspace{0.1cm}}&\\
&&&&&&\parbox[t][0.3cm]{11.625261cm}{\raggedright {\ }{\ }{\ }suggest that the resonance strength may be weaker.\vspace{0.1cm}}&\\
&&&&&&\parbox[t][0.3cm]{11.625261cm}{\raggedright (\href{https://www.nndc.bnl.gov/nsr/nsrlink.jsp?2010Ba21,B}{2010Ba21}) set an upper limit of \ensuremath{\sim}10 mb for the p(\ensuremath{^{\textnormal{17}}}F,p\ensuremath{_{\textnormal{1}}}) cross section populating\vspace{0.1cm}}&\\
&&&&&&\parbox[t][0.3cm]{11.625261cm}{\raggedright {\ }{\ }{\ }this state. (\href{https://www.nndc.bnl.gov/nsr/nsrlink.jsp?2010Ba21,B}{2010Ba21}) set upper limits on the \ensuremath{\Gamma}\ensuremath{_{\textnormal{p}_{\textnormal{1}}}}/\ensuremath{\Gamma}\ensuremath{_{\textnormal{p}_{\textnormal{0}}}} as a function of assumed\vspace{0.1cm}}&\\
&&&&&&\parbox[t][0.3cm]{11.625261cm}{\raggedright {\ }{\ }{\ }width and estimated that this branching ratio is less than 3.1\%.\vspace{0.1cm}}&\\
\multicolumn{1}{r@{}}{7.40\ensuremath{\times10^{3}}}&\multicolumn{1}{@{}l}{\ensuremath{^{{\hyperlink{NE4LEVEL0}{a}}}} {\it 6}}&\multicolumn{1}{l}{(2\ensuremath{^{+}})}&\multicolumn{1}{r@{}}{70}&\multicolumn{1}{@{ }l}{keV {\it 60}}&\multicolumn{1}{l}{2}&\parbox[t][0.3cm]{11.625261cm}{\raggedright \ensuremath{\Gamma}\ensuremath{\alpha}=40 eV \textit{30} (\href{https://www.nndc.bnl.gov/nsr/nsrlink.jsp?2002Ha15,B}{2002Ha15})\vspace{0.1cm}}&\\
&&&&&&\parbox[t][0.3cm]{11.625261cm}{\raggedright E(level): From (\href{https://www.nndc.bnl.gov/nsr/nsrlink.jsp?2002Ha15,B}{2002Ha15}) using E\ensuremath{_{\textnormal{c.m.}}}=3.48 MeV \textit{6}. Note that (\href{https://www.nndc.bnl.gov/nsr/nsrlink.jsp?1999Ha14,B}{1999Ha14}) reported\vspace{0.1cm}}&\\
&&&&&&\parbox[t][0.3cm]{11.625261cm}{\raggedright {\ }{\ }{\ }E\ensuremath{_{\textnormal{x}}}=7.37 MeV \textit{6} associated with E\ensuremath{_{\textnormal{c.m.}}}=3.48 MeV. However, this excitation energy\vspace{0.1cm}}&\\
&&&&&&\parbox[t][0.3cm]{11.625261cm}{\raggedright {\ }{\ }{\ }appears to be computed wrongly. See also E\ensuremath{_{\textnormal{x}}}=7330 keV (\href{https://www.nndc.bnl.gov/nsr/nsrlink.jsp?2011He09,B}{2011He09}); \ensuremath{\approx}7350 keV\vspace{0.1cm}}&\\
&&&&&&\parbox[t][0.3cm]{11.625261cm}{\raggedright {\ }{\ }{\ }(\href{https://www.nndc.bnl.gov/nsr/nsrlink.jsp?2001Bl06,B}{2001Bl06}); and 7323 keV (\href{https://www.nndc.bnl.gov/nsr/nsrlink.jsp?2003Bl11,B}{2003Bl11}).\vspace{0.1cm}}&\\
&&&&&&\parbox[t][0.3cm]{11.625261cm}{\raggedright \ensuremath{\Gamma}: From (\href{https://www.nndc.bnl.gov/nsr/nsrlink.jsp?2002Ha15,B}{2002Ha15}).\vspace{0.1cm}}&\\
&&&&&&\parbox[t][0.3cm]{11.625261cm}{\raggedright \ensuremath{\Gamma}\ensuremath{_{\ensuremath{\alpha}}}=40 eV \textit{30} (\href{https://www.nndc.bnl.gov/nsr/nsrlink.jsp?2002Ha15,B}{2002Ha15}: for J\ensuremath{^{\ensuremath{\pi}}}=2\ensuremath{^{\textnormal{+}}}). See also (\href{https://www.nndc.bnl.gov/nsr/nsrlink.jsp?1999Ha14,B}{1999Ha14}: \ensuremath{\Gamma}\ensuremath{_{\ensuremath{\alpha}}}=0.08 keV for\vspace{0.1cm}}&\\
&&&&&&\parbox[t][0.3cm]{11.625261cm}{\raggedright {\ }{\ }{\ }J\ensuremath{^{\ensuremath{\pi}}}=1\ensuremath{^{-}} and \ensuremath{\Gamma}\ensuremath{_{\ensuremath{\alpha}}}=0.03 keV for J\ensuremath{^{\ensuremath{\pi}}}=4\ensuremath{^{-}}).\vspace{0.1cm}}&\\
&&&&&&\parbox[t][0.3cm]{11.625261cm}{\raggedright J\ensuremath{^{\pi}}: From (\href{https://www.nndc.bnl.gov/nsr/nsrlink.jsp?2011He09,B}{2011He09}) based on R-matrix analysis with L=2 (see Fig. 7b). See also\vspace{0.1cm}}&\\
&&&&&&\parbox[t][0.3cm]{11.625261cm}{\raggedright {\ }{\ }{\ }(\href{https://www.nndc.bnl.gov/nsr/nsrlink.jsp?2002Ha15,B}{2002Ha15}), who makes no J\ensuremath{^{\ensuremath{\pi}}} assignment for this state other than it must be\vspace{0.1cm}}&\\
&&&&&&\parbox[t][0.3cm]{11.625261cm}{\raggedright {\ }{\ }{\ }natural parity, but by inspection of the mirror levels and Coulomb shift analysis,\vspace{0.1cm}}&\\
&&&&&&\parbox[t][0.3cm]{11.625261cm}{\raggedright {\ }{\ }{\ }(\href{https://www.nndc.bnl.gov/nsr/nsrlink.jsp?2002Ha15,B}{2002Ha15}) speculates it may be a 2\ensuremath{^{\textnormal{+}}} assignment; and J\ensuremath{^{\ensuremath{\pi}}}=1\ensuremath{^{-}} (\href{https://www.nndc.bnl.gov/nsr/nsrlink.jsp?1999Ha14,B}{1999Ha14}) based on\vspace{0.1cm}}&\\
&&&&&&\parbox[t][0.3cm]{11.625261cm}{\raggedright {\ }{\ }{\ }mirror level and Coulomb shift analyses. J\ensuremath{^{\ensuremath{\pi}}}=1\ensuremath{^{-}} was ruled out by (\href{https://www.nndc.bnl.gov/nsr/nsrlink.jsp?2000Fo19,B}{2000Fo19}) based\vspace{0.1cm}}&\\
&&&&&&\parbox[t][0.3cm]{11.625261cm}{\raggedright {\ }{\ }{\ }on the calculations of (\href{https://www.nndc.bnl.gov/nsr/nsrlink.jsp?1998Sh35,B}{1998Sh35}) and mirror level arguments. For the same\vspace{0.1cm}}&\\
\end{longtable}
\begin{textblock}{29}(0,27.3)
Continued on next page (footnotes at end of table)
\end{textblock}
\clearpage
\begin{longtable}{ccccccc@{\extracolsep{\fill}}c}
\\[-.4cm]
\multicolumn{8}{c}{{\bf \small \underline{\ensuremath{^{\textnormal{1}}}H(\ensuremath{^{\textnormal{17}}}F,p),(\ensuremath{^{\textnormal{17}}}F,2p),(\ensuremath{^{\textnormal{17}}}F,\ensuremath{\alpha}):res\hspace{0.2in}\href{https://www.nndc.bnl.gov/nsr/nsrlink.jsp?1999Ba49,B}{1999Ba49},\href{https://www.nndc.bnl.gov/nsr/nsrlink.jsp?2020Br14,B}{2020Br14} (continued)}}}\\
\multicolumn{8}{c}{~}\\
\multicolumn{8}{c}{\underline{\ensuremath{^{18}}Ne Levels (continued)}}\\
\multicolumn{8}{c}{~}\\
\multicolumn{2}{c}{E(level)$^{}$}&J$^{\pi}$$^{}$&\multicolumn{2}{c}{\ensuremath{\Gamma}$^{}$}&L$^{}$&Comments&\\[-.2cm]
\multicolumn{2}{c}{\hrulefill}&\hrulefill&\multicolumn{2}{c}{\hrulefill}&\hrulefill&\hrulefill&
\endhead
&&&&&&\parbox[t][0.3cm]{11.49258cm}{\raggedright {\ }{\ }{\ }reasons, (\href{https://www.nndc.bnl.gov/nsr/nsrlink.jsp?2000Fo19,B}{2000Fo19}) also ruled out the J\ensuremath{^{\ensuremath{\pi}}}=4\ensuremath{^{\textnormal{+}}} assignment made also by\vspace{0.1cm}}&\\
&&&&&&\parbox[t][0.3cm]{11.49258cm}{\raggedright {\ }{\ }{\ }(\href{https://www.nndc.bnl.gov/nsr/nsrlink.jsp?1999Ha14,B}{1999Ha14}).\vspace{0.1cm}}&\\
&&&&&&\parbox[t][0.3cm]{11.49258cm}{\raggedright L: From (\href{https://www.nndc.bnl.gov/nsr/nsrlink.jsp?2011He09,B}{2011He09}).\vspace{0.1cm}}&\\
&&&&&&\parbox[t][0.3cm]{11.49258cm}{\raggedright \ensuremath{\omega}\ensuremath{\gamma}\ensuremath{_{\textnormal{(p,}\ensuremath{\alpha}\textnormal{)}}}=18 eV \textit{14} (\href{https://www.nndc.bnl.gov/nsr/nsrlink.jsp?2002Ha15,B}{2002Ha15}: J\ensuremath{^{\ensuremath{\pi}}}=2\ensuremath{^{\textnormal{+}}}). See also \ensuremath{\omega}\ensuremath{\gamma}\ensuremath{_{\textnormal{(p,}\ensuremath{\alpha}\textnormal{)}}}=23 eV \textit{12} (\href{https://www.nndc.bnl.gov/nsr/nsrlink.jsp?1999Ha14,B}{1999Ha14}:\vspace{0.1cm}}&\\
&&&&&&\parbox[t][0.3cm]{11.49258cm}{\raggedright {\ }{\ }{\ }J\ensuremath{^{\ensuremath{\pi}}}=1\ensuremath{^{-}}).\vspace{0.1cm}}&\\
&&&&&&\parbox[t][0.3cm]{11.49258cm}{\raggedright \ensuremath{\Theta}\ensuremath{^{\textnormal{2}}_{\ensuremath{\alpha}}}=0.004 \textit{3} (\href{https://www.nndc.bnl.gov/nsr/nsrlink.jsp?2002Ha15,B}{2002Ha15}: J\ensuremath{^{\ensuremath{\pi}}}=2\ensuremath{^{\textnormal{+}}}). \ensuremath{\Theta}\ensuremath{^{\textnormal{2}}_{\ensuremath{\alpha}}} is labeled as the spectroscopic factor in\vspace{0.1cm}}&\\
&&&&&&\parbox[t][0.3cm]{11.49258cm}{\raggedright {\ }{\ }{\ }(\href{https://www.nndc.bnl.gov/nsr/nsrlink.jsp?2002Ha15,B}{2002Ha15}). See also \ensuremath{\Theta}\ensuremath{^{\textnormal{2}}_{\ensuremath{\alpha}}}=0.002 for J\ensuremath{^{\ensuremath{\pi}}}=1\ensuremath{^{-}} assignment and \ensuremath{\Theta}\ensuremath{^{\textnormal{2}}_{\ensuremath{\alpha}}}=0.09 for J\ensuremath{^{\ensuremath{\pi}}}=4\ensuremath{^{\textnormal{+}}}\vspace{0.1cm}}&\\
&&&&&&\parbox[t][0.3cm]{11.49258cm}{\raggedright {\ }{\ }{\ }assignment made by (\href{https://www.nndc.bnl.gov/nsr/nsrlink.jsp?1999Ha14,B}{1999Ha14}).\vspace{0.1cm}}&\\
\multicolumn{1}{r@{}}{7.61\ensuremath{\times10^{3}}}&\multicolumn{1}{@{}l}{\ensuremath{^{{\hyperlink{NE4LEVEL0}{a}}}} {\it 5}}&\multicolumn{1}{l}{(1\ensuremath{^{-}})}&\multicolumn{1}{r@{}}{75}&\multicolumn{1}{@{ }l}{keV {\it 20}}&\multicolumn{1}{l}{(1)}&\parbox[t][0.3cm]{11.49258cm}{\raggedright \ensuremath{\Gamma}\ensuremath{\alpha}=1.2 keV (\href{https://www.nndc.bnl.gov/nsr/nsrlink.jsp?1999Ha14,B}{1999Ha14}); \ensuremath{\Gamma}\ensuremath{\alpha}=1000 eV \textit{120} (\href{https://www.nndc.bnl.gov/nsr/nsrlink.jsp?2002Ha15,B}{2002Ha15})\vspace{0.1cm}}&\\
&&&&&&\parbox[t][0.3cm]{11.49258cm}{\raggedright \ensuremath{\Gamma}\ensuremath{_{\textnormal{p}}}=72 keV \textit{20} (\href{https://www.nndc.bnl.gov/nsr/nsrlink.jsp?2002Ha15,B}{2002Ha15})\vspace{0.1cm}}&\\
&&&&&&\parbox[t][0.3cm]{11.49258cm}{\raggedright E(level): From (\href{https://www.nndc.bnl.gov/nsr/nsrlink.jsp?2002Ha15,B}{2002Ha15}) using E\ensuremath{_{\textnormal{c.m.}}}=3.69 MeV \textit{5}. Note that (\href{https://www.nndc.bnl.gov/nsr/nsrlink.jsp?1999Ha14,B}{1999Ha14})\vspace{0.1cm}}&\\
&&&&&&\parbox[t][0.3cm]{11.49258cm}{\raggedright {\ }{\ }{\ }reported E\ensuremath{_{\textnormal{x}}}=7.60 MeV \textit{5} associated with E\ensuremath{_{\textnormal{c.m.}}}=3.69 MeV. However, this\vspace{0.1cm}}&\\
&&&&&&\parbox[t][0.3cm]{11.49258cm}{\raggedright {\ }{\ }{\ }excitation energy appears to be computed wrongly. See also E\ensuremath{_{\textnormal{x}}}\ensuremath{\approx}7600 keV\vspace{0.1cm}}&\\
&&&&&&\parbox[t][0.3cm]{11.49258cm}{\raggedright {\ }{\ }{\ }(\href{https://www.nndc.bnl.gov/nsr/nsrlink.jsp?2001Bl06,B}{2001Bl06}), 7584 keV (\href{https://www.nndc.bnl.gov/nsr/nsrlink.jsp?2003Bl11,B}{2003Bl11}), 7600 keV (\href{https://www.nndc.bnl.gov/nsr/nsrlink.jsp?2001HaZP,B}{2001HaZP}), and (\ensuremath{\approx}7500) keV\vspace{0.1cm}}&\\
&&&&&&\parbox[t][0.3cm]{11.49258cm}{\raggedright {\ }{\ }{\ }(\href{https://www.nndc.bnl.gov/nsr/nsrlink.jsp?2011He09,B}{2011He09}). (\href{https://www.nndc.bnl.gov/nsr/nsrlink.jsp?2011He09,B}{2011He09}) describes this state as a ``groove-like'' structure, due to\vspace{0.1cm}}&\\
&&&&&&\parbox[t][0.3cm]{11.49258cm}{\raggedright {\ }{\ }{\ }the interference with nearby resonances observed in the \ensuremath{^{\textnormal{17}}}F+p excitation function\vspace{0.1cm}}&\\
&&&&&&\parbox[t][0.3cm]{11.49258cm}{\raggedright {\ }{\ }{\ }at E\ensuremath{_{\textnormal{c.m.}}}\ensuremath{\sim}3.6 MeV. Due to poor statistics in (\href{https://www.nndc.bnl.gov/nsr/nsrlink.jsp?2011He09,B}{2011He09}), they found it difficult to\vspace{0.1cm}}&\\
&&&&&&\parbox[t][0.3cm]{11.49258cm}{\raggedright {\ }{\ }{\ }fit this feature.\vspace{0.1cm}}&\\
&&&&&&\parbox[t][0.3cm]{11.49258cm}{\raggedright \ensuremath{\Gamma}: From \ensuremath{\Gamma}=\ensuremath{\Gamma}\ensuremath{_{\textnormal{p}}}+\ensuremath{\Gamma}\ensuremath{_{\ensuremath{\alpha}}}+\ensuremath{\Gamma}\ensuremath{_{\textnormal{p$'$}}}, where each partial width is from (\href{https://www.nndc.bnl.gov/nsr/nsrlink.jsp?2002Ha15,B}{2002Ha15}) listed\vspace{0.1cm}}&\\
&&&&&&\parbox[t][0.3cm]{11.49258cm}{\raggedright {\ }{\ }{\ }above; and \ensuremath{\Gamma}\ensuremath{_{\textnormal{p$'$}}}\ensuremath{<}2 keV (\href{https://www.nndc.bnl.gov/nsr/nsrlink.jsp?2002Ha15,B}{2002Ha15}).\vspace{0.1cm}}&\\
&&&&&&\parbox[t][0.3cm]{11.49258cm}{\raggedright J\ensuremath{^{\pi}}: From (\href{https://www.nndc.bnl.gov/nsr/nsrlink.jsp?1999Ha14,B}{1999Ha14}, \href{https://www.nndc.bnl.gov/nsr/nsrlink.jsp?2002Ha15,B}{2002Ha15}: based on mirror levels and Coulomb shift\vspace{0.1cm}}&\\
&&&&&&\parbox[t][0.3cm]{11.49258cm}{\raggedright {\ }{\ }{\ }arguments) and (\href{https://www.nndc.bnl.gov/nsr/nsrlink.jsp?2011He09,B}{2011He09}: R-matrix analysis). Note that (\href{https://www.nndc.bnl.gov/nsr/nsrlink.jsp?1999Ha14,B}{1999Ha14}) also\vspace{0.1cm}}&\\
&&&&&&\parbox[t][0.3cm]{11.49258cm}{\raggedright {\ }{\ }{\ }assigned J\ensuremath{^{\ensuremath{\pi}}}=(2\ensuremath{^{\textnormal{+}}}, 3\ensuremath{^{-}}) to this state. However, (\href{https://www.nndc.bnl.gov/nsr/nsrlink.jsp?2000Fo19,B}{2000Fo19}) suggested J\ensuremath{^{\ensuremath{\pi}}}=(1\ensuremath{^{-}}) for\vspace{0.1cm}}&\\
&&&&&&\parbox[t][0.3cm]{11.49258cm}{\raggedright {\ }{\ }{\ }this state based on the theoretical calculations of (\href{https://www.nndc.bnl.gov/nsr/nsrlink.jsp?1998Sh35,B}{1998Sh35}) and mirror levels\vspace{0.1cm}}&\\
&&&&&&\parbox[t][0.3cm]{11.49258cm}{\raggedright {\ }{\ }{\ }arguments.\vspace{0.1cm}}&\\
&&&&&&\parbox[t][0.3cm]{11.49258cm}{\raggedright L: From (\href{https://www.nndc.bnl.gov/nsr/nsrlink.jsp?2011He09,B}{2011He09}).\vspace{0.1cm}}&\\
&&&&&&\parbox[t][0.3cm]{11.49258cm}{\raggedright \ensuremath{\omega}\ensuremath{\gamma}\ensuremath{_{\textnormal{(p,}\ensuremath{\alpha}\textnormal{)}}}=271 eV \textit{30} is the weighted average of 300 eV \textit{40} (\href{https://www.nndc.bnl.gov/nsr/nsrlink.jsp?1999Ha14,B}{1999Ha14}: for\vspace{0.1cm}}&\\
&&&&&&\parbox[t][0.3cm]{11.49258cm}{\raggedright {\ }{\ }{\ }J\ensuremath{^{\ensuremath{\pi}}}=(1\ensuremath{^{-}})), and \ensuremath{\omega}\ensuremath{\gamma}\ensuremath{_{\textnormal{(p,}\ensuremath{\alpha}\textnormal{)}}}=255 eV \textit{30} (\href{https://www.nndc.bnl.gov/nsr/nsrlink.jsp?2002Ha15,B}{2002Ha15}: for J\ensuremath{^{\ensuremath{\pi}}}=(1\ensuremath{^{-}})).\vspace{0.1cm}}&\\
&&&&&&\parbox[t][0.3cm]{11.49258cm}{\raggedright \ensuremath{\Theta}\ensuremath{^{\textnormal{2}}_{\ensuremath{\alpha}}}=0.015 (\href{https://www.nndc.bnl.gov/nsr/nsrlink.jsp?1999Ha14,B}{1999Ha14}: for J\ensuremath{^{\ensuremath{\pi}}}=(1\ensuremath{^{-}})) and \ensuremath{\Theta}\ensuremath{^{\textnormal{2}}_{\ensuremath{\alpha}}}=0.0130 \textit{15} (\href{https://www.nndc.bnl.gov/nsr/nsrlink.jsp?2002Ha15,B}{2002Ha15}: for J\ensuremath{^{\ensuremath{\pi}}}=(1\ensuremath{^{-}}),\vspace{0.1cm}}&\\
&&&&&&\parbox[t][0.3cm]{11.49258cm}{\raggedright {\ }{\ }{\ }this is labeled as the spectroscopic factor).\vspace{0.1cm}}&\\
&&&&&&\parbox[t][0.3cm]{11.49258cm}{\raggedright \ensuremath{\Theta}\ensuremath{^{\textnormal{2}}_{\textnormal{p}}}=0.034 (\href{https://www.nndc.bnl.gov/nsr/nsrlink.jsp?2002Ha15,B}{2002Ha15}: for J\ensuremath{^{\ensuremath{\pi}}}=(1\ensuremath{^{-}})).\vspace{0.1cm}}&\\
&&&&&&\parbox[t][0.3cm]{11.49258cm}{\raggedright (\href{https://www.nndc.bnl.gov/nsr/nsrlink.jsp?2002Ha15,B}{2002Ha15}) also considered J\ensuremath{^{\ensuremath{\pi}}}=(2\ensuremath{^{\textnormal{+}}} and 3\ensuremath{^{-}}) assignments for this state. As a result,\vspace{0.1cm}}&\\
&&&&&&\parbox[t][0.3cm]{11.49258cm}{\raggedright {\ }{\ }{\ }(\href{https://www.nndc.bnl.gov/nsr/nsrlink.jsp?2002Ha15,B}{2002Ha15}) estimated \ensuremath{\Gamma}\ensuremath{_{\ensuremath{\alpha}}}=610 eV \textit{70} and extracted a spectroscopic factor of\vspace{0.1cm}}&\\
&&&&&&\parbox[t][0.3cm]{11.49258cm}{\raggedright {\ }{\ }{\ }\ensuremath{\Theta}\ensuremath{^{\textnormal{2}}_{\ensuremath{\alpha}}}=0.0200 \textit{25} assuming J\ensuremath{^{\ensuremath{\pi}}}=(2\ensuremath{^{\textnormal{+}}}), and \ensuremath{\Gamma}\ensuremath{_{\ensuremath{\alpha}}}=440 eV \textit{50} and \ensuremath{\Theta}\ensuremath{^{\textnormal{2}}_{\ensuremath{\alpha}}}=0.070 \textit{8}\vspace{0.1cm}}&\\
&&&&&&\parbox[t][0.3cm]{11.49258cm}{\raggedright {\ }{\ }{\ }assuming J\ensuremath{^{\ensuremath{\pi}}}=(3\ensuremath{^{-}}). These J\ensuremath{^{\ensuremath{\pi}}} assignments are not favored.\vspace{0.1cm}}&\\
&&&&&&\parbox[t][0.3cm]{11.49258cm}{\raggedright (\href{https://www.nndc.bnl.gov/nsr/nsrlink.jsp?1999Ha14,B}{1999Ha14}) estimated \ensuremath{\Gamma}\ensuremath{_{\ensuremath{\alpha}}}=0.7 keV and \ensuremath{\Theta}\ensuremath{^{\textnormal{2}}_{\ensuremath{\alpha}}}=0.026 for J\ensuremath{^{\ensuremath{\pi}}}=2\ensuremath{^{\textnormal{+}}}; and \ensuremath{\Gamma}\ensuremath{_{\ensuremath{\alpha}}}=0.5 keV\vspace{0.1cm}}&\\
&&&&&&\parbox[t][0.3cm]{11.49258cm}{\raggedright {\ }{\ }{\ }and \ensuremath{\Theta}\ensuremath{^{\textnormal{2}}_{\ensuremath{\alpha}}}=0.09 for J\ensuremath{^{\ensuremath{\pi}}}=3\ensuremath{^{-}}. These J\ensuremath{^{\ensuremath{\pi}}} assignments are not favored.\vspace{0.1cm}}&\\
\multicolumn{1}{r@{}}{7710}&\multicolumn{1}{@{ }l}{{\it 26}}&\multicolumn{1}{l}{2\ensuremath{^{-}}}&\multicolumn{1}{r@{}}{70}&\multicolumn{1}{@{ }l}{keV {\it 25}}&\multicolumn{1}{l}{3}&\parbox[t][0.3cm]{11.49258cm}{\raggedright \ensuremath{\Gamma}\ensuremath{_{\textnormal{p}}}=59 keV \textit{25} (\href{https://www.nndc.bnl.gov/nsr/nsrlink.jsp?2002Ha15,B}{2002Ha15})\vspace{0.1cm}}&\\
&&&&&&\parbox[t][0.3cm]{11.49258cm}{\raggedright E(level): Weighted average of 7.71 MeV \textit{5} (\href{https://www.nndc.bnl.gov/nsr/nsrlink.jsp?2002Ha15,B}{2002Ha15}: from E\ensuremath{_{\textnormal{c.m.}}}=3.79 MeV \textit{5});\vspace{0.1cm}}&\\
&&&&&&\parbox[t][0.3cm]{11.49258cm}{\raggedright {\ }{\ }{\ }and 7710 keV \textit{30} (\href{https://www.nndc.bnl.gov/nsr/nsrlink.jsp?2011He09,B}{2011He09}). See also E\ensuremath{_{\textnormal{x}}}=7710 keV (\href{https://www.nndc.bnl.gov/nsr/nsrlink.jsp?2001HaZP,B}{2001HaZP}).\vspace{0.1cm}}&\\
&&&&&&\parbox[t][0.3cm]{11.49258cm}{\raggedright \ensuremath{\Gamma}: From (\href{https://www.nndc.bnl.gov/nsr/nsrlink.jsp?2002Ha15,B}{2002Ha15}); however, they report \ensuremath{\Gamma}=70 keV \textit{30}. The evaluator revised\vspace{0.1cm}}&\\
&&&&&&\parbox[t][0.3cm]{11.49258cm}{\raggedright {\ }{\ }{\ }the uncertainty to be consistent with their reported partial widths: \ensuremath{\Gamma}=\ensuremath{\Gamma}\ensuremath{_{\textnormal{p}}}+\ensuremath{\Gamma}\ensuremath{_{\textnormal{p$'$}}}=70\vspace{0.1cm}}&\\
&&&&&&\parbox[t][0.3cm]{11.49258cm}{\raggedright {\ }{\ }{\ }keV \textit{25}.\vspace{0.1cm}}&\\
&&&&&&\parbox[t][0.3cm]{11.49258cm}{\raggedright \ensuremath{\Gamma}\ensuremath{_{\textnormal{p$'$}}}=11 keV \textit{5} (\href{https://www.nndc.bnl.gov/nsr/nsrlink.jsp?2002Ha15,B}{2002Ha15}).\vspace{0.1cm}}&\\
&&&&&&\parbox[t][0.3cm]{11.49258cm}{\raggedright J\ensuremath{^{\pi}}: From (\href{https://www.nndc.bnl.gov/nsr/nsrlink.jsp?2001HaZP,B}{2001HaZP}, \href{https://www.nndc.bnl.gov/nsr/nsrlink.jsp?2002Ha15,B}{2002Ha15}: assignment is based on mirror levels and\vspace{0.1cm}}&\\
&&&&&&\parbox[t][0.3cm]{11.49258cm}{\raggedright {\ }{\ }{\ }Coulomb shift arguments) and (\href{https://www.nndc.bnl.gov/nsr/nsrlink.jsp?2011He09,B}{2011He09}: R-matrix analysis).\vspace{0.1cm}}&\\
&&&&&&\parbox[t][0.3cm]{11.49258cm}{\raggedright J\ensuremath{^{\pi}}: See also J\ensuremath{^{\ensuremath{\pi}}}=(1\ensuremath{^{\textnormal{+}}}, 2\ensuremath{^{-}}, 3\ensuremath{^{\textnormal{+}}}) from the R-matrix analysis of (\href{https://www.nndc.bnl.gov/nsr/nsrlink.jsp?2001HaZP,B}{2001HaZP}) and\vspace{0.1cm}}&\\
&&&&&&\parbox[t][0.3cm]{11.49258cm}{\raggedright {\ }{\ }{\ }J\ensuremath{^{\ensuremath{\pi}}}=(3\ensuremath{^{-}}, 2\ensuremath{^{-}}, 1\ensuremath{^{-}}) from R-matrix analysis of (\href{https://www.nndc.bnl.gov/nsr/nsrlink.jsp?2011He09,B}{2011He09}).\vspace{0.1cm}}&\\
&&&&&&\parbox[t][0.3cm]{11.49258cm}{\raggedright L: From (\href{https://www.nndc.bnl.gov/nsr/nsrlink.jsp?2011He09,B}{2011He09}).\vspace{0.1cm}}&\\
\end{longtable}
\parbox[b][0.3cm]{17.7cm}{\makebox[1ex]{\ensuremath{^{\hypertarget{NE4LEVEL0}{a}}}} The level energy is deduced from E\ensuremath{_{\textnormal{x}}}(\ensuremath{^{\textnormal{18}}}Ne)=E\ensuremath{_{\textnormal{c.m.}}}+q, where E\ensuremath{_{\textnormal{c.m.}}} is the resonance energy, and q is the Q-value of the \ensuremath{^{\textnormal{17}}}F+p}\\
\parbox[b][0.3cm]{17.7cm}{{\ }{\ }resonant elastic scattering. To calculate the Q-value, the masses of \ensuremath{^{\textnormal{17}}}F, \ensuremath{^{\textnormal{18}}}Ne and p are taken from (\href{https://www.nndc.bnl.gov/nsr/nsrlink.jsp?2021Wa16,B}{2021Wa16}: AME-2020).}\\
\vspace{0.5cm}
\clearpage
\subsection[\hspace{-0.2cm}\ensuremath{^{\textnormal{1}}}H(\ensuremath{^{\textnormal{18}}}Ne,p\ensuremath{'})]{ }
\vspace{-27pt}
\vspace{0.3cm}
\hypertarget{NE5}{{\bf \small \underline{\ensuremath{^{\textnormal{1}}}H(\ensuremath{^{\textnormal{18}}}Ne,p\ensuremath{'})\hspace{0.2in}\href{https://www.nndc.bnl.gov/nsr/nsrlink.jsp?1999Ri05,B}{1999Ri05},\href{https://www.nndc.bnl.gov/nsr/nsrlink.jsp?2019Mi21,B}{2019Mi21}}}}\\
\vspace{4pt}
\vspace{8pt}
\parbox[b][0.3cm]{17.7cm}{\addtolength{\parindent}{-0.2in}\href{https://www.nndc.bnl.gov/nsr/nsrlink.jsp?1999Ri05,B}{1999Ri05}: \ensuremath{^{\textnormal{1}}}H(\ensuremath{^{\textnormal{18}}}Ne,p\ensuremath{'}) E=30 MeV/nucleon; measured inelastic proton scattering for the 0\ensuremath{^{\textnormal{+}}_{\textnormal{g.s.}}} \ensuremath{\rightarrow} 2\ensuremath{^{\textnormal{+}}_{\textnormal{1}}} transition in inverse}\\
\parbox[b][0.3cm]{17.7cm}{kinematics. Measured proton spectra, \ensuremath{\sigma}\ensuremath{_{\textnormal{inelastic}}}(\ensuremath{\theta}), and angular distributions (\ensuremath{\theta}\ensuremath{_{\textnormal{lab}}}=65\ensuremath{^\circ}{\textminus}80\ensuremath{^\circ}) for protons scattered from the}\\
\parbox[b][0.3cm]{17.7cm}{\ensuremath{^{\textnormal{18}}}Ne\ensuremath{_{\textnormal{g.s.}}} and \ensuremath{^{\textnormal{18}}}Ne*(1.89 MeV, 2\ensuremath{^{\textnormal{+}}_{\textnormal{1}}}). Deduced hadronic mirror symmetry, using folding model calculations, for quadrupole}\\
\parbox[b][0.3cm]{17.7cm}{transitions in the \textit{s}-\textit{d} shell nuclei. The theoretical results reproduced the experimental data very well, particularly for \ensuremath{\theta}\ensuremath{_{\textnormal{c.m.}}}\ensuremath{<}60\ensuremath{^\circ}.}\\
\parbox[b][0.3cm]{17.7cm}{Comparison with n+\ensuremath{^{\textnormal{18}}}O scattering are discussed.}\\
\vspace{0.385cm}
\parbox[b][0.3cm]{17.7cm}{\addtolength{\parindent}{-0.2in}\textit{Theory:}}\\
\parbox[b][0.3cm]{17.7cm}{\addtolength{\parindent}{-0.2in}\href{https://www.nndc.bnl.gov/nsr/nsrlink.jsp?2005Gu03,B}{2005Gu03}: \ensuremath{^{\textnormal{1}}}H(\ensuremath{^{\textnormal{18}}}Ne,p); investigated the experimental data obtained via low energy (E\ensuremath{<}100 MeV/nucleon) proton scattering on}\\
\parbox[b][0.3cm]{17.7cm}{\ensuremath{^{\textnormal{18}}}Ne (\href{https://www.nndc.bnl.gov/nsr/nsrlink.jsp?1999Ri05,B}{1999Ri05}) and \ensuremath{^{\textnormal{18}}}O (\href{https://www.nndc.bnl.gov/nsr/nsrlink.jsp?1974Es02,B}{1974Es02}) using folding model approach via two different effective NN interactions. Calculated \ensuremath{\sigma}(E,\ensuremath{\theta})}\\
\parbox[b][0.3cm]{17.7cm}{using folding model approach. Comparison of effective interactions are discussed. Recalculated the quadrupole deformations of \ensuremath{^{\textnormal{18}}}O}\\
\parbox[b][0.3cm]{17.7cm}{and \ensuremath{^{\textnormal{18}}}Ne using the experimental B(E2) values from (\href{https://www.nndc.bnl.gov/nsr/nsrlink.jsp?1987Ra01,B}{1987Ra01}). The results are \ensuremath{\delta}=0.33078 and \ensuremath{\delta}=0.59284 for \ensuremath{^{\textnormal{18}}}O and \ensuremath{^{\textnormal{18}}}Ne,}\\
\parbox[b][0.3cm]{17.7cm}{respectively.}\\
\parbox[b][0.3cm]{17.7cm}{\addtolength{\parindent}{-0.2in}\href{https://www.nndc.bnl.gov/nsr/nsrlink.jsp?2014Ja05,B}{2014Ja05}: \ensuremath{^{\textnormal{18}}}Ne(p,p\ensuremath{'}); used the coupled-channel representation of the Gamow shell model to describe the low energy (E\ensuremath{_{\textnormal{c.m.}}}=0.5-3}\\
\parbox[b][0.3cm]{17.7cm}{MeV) experimental data obtained via elastic and inelastic scattering of protons on \ensuremath{^{\textnormal{18}}}Ne at \ensuremath{\theta}\ensuremath{_{\textnormal{c.m.}}}=105\ensuremath{^\circ}{\textminus}180\ensuremath{^\circ} (\href{https://www.nndc.bnl.gov/nsr/nsrlink.jsp?2006Sk09,B}{2006Sk09}, \href{https://www.nndc.bnl.gov/nsr/nsrlink.jsp?2003An02,B}{2003An02},}\\
\parbox[b][0.3cm]{17.7cm}{and \href{https://www.nndc.bnl.gov/nsr/nsrlink.jsp?2005De15,B}{2005De15}). Deduced the theoretical low-lying excited states of \ensuremath{^{\textnormal{18}}}Ne and \ensuremath{^{\textnormal{19}}}Na, the two proton separation energy for \ensuremath{^{\textnormal{18}}}Ne, the}\\
\parbox[b][0.3cm]{17.7cm}{p+\ensuremath{^{\textnormal{18}}}Ne excitation function, and the elastic/inelastic differential cross sections in the \ensuremath{^{\textnormal{18}}}Ne(p,p\ensuremath{'}) reaction, calculated in (\textit{sd}-\textit{p}) and}\\
\parbox[b][0.3cm]{17.7cm}{(\textit{sd}) model spaces, for E\ensuremath{_{\textnormal{c.m.}}}=0.1-5 MeV. Comparison with experimental data are discussed.}\\
\parbox[b][0.3cm]{17.7cm}{\addtolength{\parindent}{-0.2in}\href{https://www.nndc.bnl.gov/nsr/nsrlink.jsp?2019Mi21,B}{2019Mi21}: \ensuremath{^{\textnormal{18}}}Ne(p,p); Calculated proton radial densities, binding energy, Coulomb energies of the ground and first few excited}\\
\parbox[b][0.3cm]{17.7cm}{states, and energies and widths of the low-lying states of the isotones of \ensuremath{^{\textnormal{16}}}O, including \ensuremath{^{\textnormal{18}}}Ne, using Gamow shell model with the}\\
\parbox[b][0.3cm]{17.7cm}{A-dependent EFT interaction. Calculated differential \ensuremath{\sigma}(E) for \ensuremath{^{\textnormal{18}}}Ne(p,p) at E\ensuremath{_{\textnormal{c.m.}}}=0.5-2.5 MeV and \ensuremath{\theta}\ensuremath{_{\textnormal{c.m.}}}=105\ensuremath{^\circ}{\textminus}180\ensuremath{^\circ} using Gamow}\\
\parbox[b][0.3cm]{17.7cm}{shell model and resonating group method with the A-dependent EFT interaction. Comparison with experimental data are discussed.}\\
\parbox[b][0.3cm]{17.7cm}{The calculated excitation functions of the \ensuremath{^{\textnormal{18}}}Ne(p,p) reaction compare well with experimental data (\href{https://www.nndc.bnl.gov/nsr/nsrlink.jsp?2003An02,B}{2003An02}, \href{https://www.nndc.bnl.gov/nsr/nsrlink.jsp?2005De15,B}{2005De15}) for all}\\
\parbox[b][0.3cm]{17.7cm}{considered angles.}\\
\vspace{12pt}
\underline{$^{18}$Ne Levels}\\
\begin{longtable}{cccc@{\extracolsep{\fill}}c}
\multicolumn{2}{c}{E(level)$^{{\hyperlink{NE5LEVEL0}{a}}}$}&J$^{\pi}$$^{{\hyperlink{NE5LEVEL0}{a}}}$&Comments&\\[-.2cm]
\multicolumn{2}{c}{\hrulefill}&\hrulefill&\hrulefill&
\endfirsthead
\multicolumn{1}{r@{}}{0}&\multicolumn{1}{@{}l}{}&\multicolumn{1}{l}{0\ensuremath{^{+}}}&&\\
\multicolumn{1}{r@{}}{1.89\ensuremath{\times10^{3}}}&\multicolumn{1}{@{}l}{}&\multicolumn{1}{l}{2\ensuremath{^{+}}}&\parbox[t][0.3cm]{15.102401cm}{\raggedright (\href{https://www.nndc.bnl.gov/nsr/nsrlink.jsp?2005Gu03,B}{2005Gu03}) calculated a quadrupole deformation parameter of \ensuremath{\delta}=0.59284 for this state.\vspace{0.1cm}}&\\
&&&\parbox[t][0.3cm]{15.102401cm}{\raggedright The neutron and proton multipole matrix element for the empirical quadrupole densities are M\ensuremath{_{\textnormal{n}}}=6.18 fm\ensuremath{^{\textnormal{2}}} and\vspace{0.1cm}}&\\
&&&\parbox[t][0.3cm]{15.102401cm}{\raggedright {\ }{\ }{\ }M\ensuremath{_{\textnormal{p}}}=3.00 fm\ensuremath{^{\textnormal{2}}}, with M\ensuremath{_{\textnormal{n}}}/M\ensuremath{_{\textnormal{p}}}=2.06 in the convention in which B(E2; 0\ensuremath{^{\textnormal{+}}_{\textnormal{g.s.}}}\ensuremath{\rightarrow}2\ensuremath{^{\textnormal{+}}_{\textnormal{1}}})=5M\ensuremath{^{\textnormal{2}}_{\textnormal{p}}}=45 fm\ensuremath{^{\textnormal{4}}} (\href{https://www.nndc.bnl.gov/nsr/nsrlink.jsp?1999Ri05,B}{1999Ri05}).\vspace{0.1cm}}&\\
&&&\parbox[t][0.3cm]{15.102401cm}{\raggedright {\ }{\ }{\ }The empirical densities have M\ensuremath{_{\textnormal{p}}}/M\ensuremath{_{\textnormal{n}}}\ensuremath{>}Z/N for \ensuremath{^{\textnormal{18}}}Ne and M\ensuremath{_{\textnormal{n}}}/M\ensuremath{_{\textnormal{p}}}\ensuremath{>}N/Z for \ensuremath{^{\textnormal{18}}}O (\href{https://www.nndc.bnl.gov/nsr/nsrlink.jsp?1999Ri05,B}{1999Ri05}). The result M\ensuremath{_{\textnormal{p}}}/M\ensuremath{_{\textnormal{n}}}\vspace{0.1cm}}&\\
&&&\parbox[t][0.3cm]{15.102401cm}{\raggedright {\ }{\ }{\ }\ensuremath{>} Z/N is expected for low-lying quadrupole transitions in a closed neutron shell nucleus like \ensuremath{^{\textnormal{18}}}Ne\vspace{0.1cm}}&\\
&&&\parbox[t][0.3cm]{15.102401cm}{\raggedright {\ }{\ }{\ }(\href{https://www.nndc.bnl.gov/nsr/nsrlink.jsp?1999Ri05,B}{1999Ri05}).\vspace{0.1cm}}&\\
\end{longtable}
\parbox[b][0.3cm]{17.7cm}{\makebox[1ex]{\ensuremath{^{\hypertarget{NE5LEVEL0}{a}}}} From (\href{https://www.nndc.bnl.gov/nsr/nsrlink.jsp?1999Ri05,B}{1999Ri05}).}\\
\vspace{0.5cm}
\clearpage
\subsection[\hspace{-0.2cm}\ensuremath{^{\textnormal{2}}}H(\ensuremath{^{\textnormal{17}}}F,n)]{ }
\vspace{-27pt}
\vspace{0.3cm}
\hypertarget{NE6}{{\bf \small \underline{\ensuremath{^{\textnormal{2}}}H(\ensuremath{^{\textnormal{17}}}F,n)\hspace{0.2in}\href{https://www.nndc.bnl.gov/nsr/nsrlink.jsp?2017Ku27,B}{2017Ku27}}}}\\
\vspace{4pt}
\vspace{8pt}
\parbox[b][0.3cm]{17.7cm}{\addtolength{\parindent}{-0.2in}\href{https://www.nndc.bnl.gov/nsr/nsrlink.jsp?2017Ku27,B}{2017Ku27}: \ensuremath{^{\textnormal{2}}}H(\ensuremath{^{\textnormal{17}}}F,n)\ensuremath{^{\textnormal{18}}}Ne*(p)\ensuremath{^{\textnormal{17}}}F E=95.5 MeV; measured the n-\ensuremath{^{\textnormal{18}}}Ne*(\ensuremath{\rightarrow}p+\ensuremath{^{\textnormal{17}}}F+\ensuremath{\gamma}) coincidences using the RESONEUT detector}\\
\parbox[b][0.3cm]{17.7cm}{array at \ensuremath{\theta}\ensuremath{_{\textnormal{lab}}}=90\ensuremath{^\circ} covering \ensuremath{\theta}\ensuremath{_{\textnormal{c.m.}}}=3\ensuremath{^\circ}{\textminus}10\ensuremath{^\circ} (to measure neutrons with E=100-600 keV), an annular position sensitive silicon \ensuremath{\Delta}E-E}\\
\parbox[b][0.3cm]{17.7cm}{telescope covering \ensuremath{\theta}\ensuremath{_{\textnormal{lab}}}=8\ensuremath{^\circ}{\textminus}21\ensuremath{^\circ} (for protons), a cylindrical nine-fold segments position sensitive gas ionization chamber (for \ensuremath{^{\textnormal{18}}}Ne*}\\
\parbox[b][0.3cm]{17.7cm}{and \ensuremath{^{\textnormal{17}}}F ions), and a barrel-shaped array surrounding the target consisting of 20 position sensitive NaI(Tl) detectors (for \ensuremath{\gamma} rays).}\\
\parbox[b][0.3cm]{17.7cm}{Deduced the \ensuremath{^{\textnormal{18}}}Ne excitation energies from a kinematic reconstruction and invariant mass analysis of the p+\ensuremath{^{\textnormal{17}}}F+\ensuremath{\gamma} events.}\\
\parbox[b][0.3cm]{17.7cm}{\addtolength{\parindent}{-0.2in}Total cross sections of 15.3 mb \textit{3} (stat.) \textit{12} (sys.) and 2.30 mb \textit{35} (stat.) \textit{29} (sys.) were extracted for the prominent \ensuremath{^{\textnormal{18}}}Ne states}\\
\parbox[b][0.3cm]{17.7cm}{populated at E\ensuremath{_{\textnormal{c.m.}}}=0.60 and 1.165 MeV, respectively. The neutron angular distributions were deduced from coupled reaction}\\
\parbox[b][0.3cm]{17.7cm}{channel calculations using the FRESCO code. The spectroscopic factors and Asymptotic Normalization Coefficients were deduced}\\
\parbox[b][0.3cm]{17.7cm}{for \ensuremath{^{\textnormal{18}}}Ne states. The astrophysical S-factors were deduced using the RADCAP code. The total S-factor for the \ensuremath{^{\textnormal{17}}}F(p,\ensuremath{\gamma}) reaction}\\
\parbox[b][0.3cm]{17.7cm}{was deduced as S(0)=2.44 keV.b \textit{97}.}\\
\parbox[b][0.3cm]{17.7cm}{\addtolength{\parindent}{-0.2in}D. W. Bardayan {et al}., AIP Conf. Proc. 2160 (2019) 070010: \ensuremath{^{\textnormal{2}}}H(\ensuremath{^{\textnormal{17}}}F,n) E=57.2 MeV; measured neutrons using the VANDLE and}\\
\parbox[b][0.3cm]{17.7cm}{UMDSA arrays covering the angular ranges of \ensuremath{\theta}\ensuremath{_{\textnormal{lab}}}=40\ensuremath{^\circ}{\textminus}170\ensuremath{^\circ} and \ensuremath{\theta}\ensuremath{_{\textnormal{lab}}}=18\ensuremath{^\circ}{\textminus}120\ensuremath{^\circ}, respectively. The \ensuremath{^{\textnormal{18}}}Ne recoils were detected in}\\
\parbox[b][0.3cm]{17.7cm}{a plastic scintillator downstream the target. This detector provided the reference signal for measuring the neutron time-of-flight.}\\
\parbox[b][0.3cm]{17.7cm}{Analysis and results were not discussed.}\\
\vspace{12pt}
\underline{$^{18}$Ne Levels}\\
\vspace{0.34cm}
\parbox[b][0.3cm]{17.7cm}{\addtolength{\parindent}{-0.254cm}The 2\textit{s}\ensuremath{_{\textnormal{1/2}}} and 1\textit{d}\ensuremath{_{\textnormal{5/2}}} ANCs were not independently determined in (\href{https://www.nndc.bnl.gov/nsr/nsrlink.jsp?2017Ku27,B}{2017Ku27}). The ratio from the mirror nucleus was assumed. See}\\
\parbox[b][0.3cm]{17.7cm}{also (\href{https://www.nndc.bnl.gov/nsr/nsrlink.jsp?2020Br14,B}{2020Br14}) for the theoretical ANCs.}\\
\vspace{0.34cm}
\begin{longtable}{ccccc@{\extracolsep{\fill}}c}
\multicolumn{2}{c}{E(level)$^{{\hyperlink{NE6LEVEL0}{a}}}$}&J$^{\pi}$$^{{\hyperlink{NE6LEVEL2}{c}}}$&L$^{{\hyperlink{NE6LEVEL2}{c}}}$&Comments&\\[-.2cm]
\multicolumn{2}{c}{\hrulefill}&\hrulefill&\hrulefill&\hrulefill&
\endfirsthead
\multicolumn{1}{r@{}}{0}&\multicolumn{1}{@{}l}{}&&&\parbox[t][0.3cm]{14.07324cm}{\raggedright E(level): From (D. W. Bardayan {et al}., AIP Conf. Proc. 2160 (2019) 070010).\vspace{0.1cm}}&\\
\multicolumn{1}{r@{}}{1888}&\multicolumn{1}{@{}l}{\ensuremath{^{{\hyperlink{NE6LEVEL1}{b}}}}}&\multicolumn{1}{l}{2\ensuremath{^{+}}}&\multicolumn{1}{l}{0,2$^{{\hyperlink{NE6LEVEL3}{d}}}$}&\parbox[t][0.3cm]{14.07324cm}{\raggedright E(level): See also 1887 keV (D. W. Bardayan {et al}., AIP Conf. Proc. 2160 (2019) 070010).\vspace{0.1cm}}&\\
&&&&\parbox[t][0.3cm]{14.07324cm}{\raggedright C\ensuremath{^{\textnormal{2}}}S: Based on the measured differential cross section of d\ensuremath{\sigma}/d\ensuremath{\Omega}\ensuremath{_{\textnormal{c.m.}}}=13.2 mb/sr \textit{63} (stat.) \textit{17} (sys.) at\vspace{0.1cm}}&\\
&&&&\parbox[t][0.3cm]{14.07324cm}{\raggedright {\ }{\ }{\ }\ensuremath{\theta}\ensuremath{_{\textnormal{c.m.}}}=11\ensuremath{^\circ}: C\ensuremath{^{\textnormal{2}}}S(2\textit{s}\ensuremath{_{\textnormal{1/2}}})=0.22 \textit{11} (stat.) \textit{3} (sys.) for \textit{l}=0, and C\ensuremath{^{\textnormal{2}}}S(1\textit{d}\ensuremath{_{\textnormal{5/2}}})=0.88 \textit{42} (stat.) \textit{11} (sys.) for\vspace{0.1cm}}&\\
&&&&\parbox[t][0.3cm]{14.07324cm}{\raggedright {\ }{\ }{\ }\textit{l}=2 contributions assuming the same mixing between \textit{l}=0 and \textit{l}=2 as that in the mirror reaction\vspace{0.1cm}}&\\
&&&&\parbox[t][0.3cm]{14.07324cm}{\raggedright {\ }{\ }{\ }(\href{https://www.nndc.bnl.gov/nsr/nsrlink.jsp?2017Ku27,B}{2017Ku27}).\vspace{0.1cm}}&\\
&&&&\parbox[t][0.3cm]{14.07324cm}{\raggedright (\href{https://www.nndc.bnl.gov/nsr/nsrlink.jsp?2017Ku27,B}{2017Ku27}): ANC(2\textit{s}\ensuremath{_{\textnormal{1/2}}})=22 fm\ensuremath{^{\textnormal{$-$1}}} \textit{10} (stat.) \textit{3} (sys.) assuming pure \textit{l}=0 (from 2\textit{s}\ensuremath{_{\textnormal{1/2}}}) and with no \textit{l}\vspace{0.1cm}}&\\
&&&&\parbox[t][0.3cm]{14.07324cm}{\raggedright {\ }{\ }{\ }mixing. Assuming the same mixing between \textit{l}=0 (2\textit{s}\ensuremath{_{\textnormal{1/2}}}) and \textit{l}=2 (1\textit{d}\ensuremath{_{\textnormal{5/2}}}) as that in the mirror\vspace{0.1cm}}&\\
&&&&\parbox[t][0.3cm]{14.07324cm}{\raggedright {\ }{\ }{\ }reaction: ANC(2\textit{s}\ensuremath{_{\textnormal{1/2}}})=16.0 fm\ensuremath{^{\textnormal{$-$1}}} \textit{77} (stat.) \textit{21} (sys.) for \textit{l}=0 and ANC(1\textit{d}\ensuremath{_{\textnormal{5/2}}})=2.6 fm\ensuremath{^{\textnormal{$-$1}}} \textit{12} (stat.) \textit{3}\vspace{0.1cm}}&\\
&&&&\parbox[t][0.3cm]{14.07324cm}{\raggedright {\ }{\ }{\ }(sys.) for \textit{l}=2.\vspace{0.1cm}}&\\
\multicolumn{1}{r@{}}{3376}&\multicolumn{1}{@{}l}{\ensuremath{^{{\hyperlink{NE6LEVEL1}{b}}}}}&\multicolumn{1}{l}{4\ensuremath{^{+}}}&\multicolumn{1}{l}{2$^{{\hyperlink{NE6LEVEL3}{d}}}$}&\parbox[t][0.3cm]{14.07324cm}{\raggedright C\ensuremath{^{\textnormal{2}}}S: C\ensuremath{^{\textnormal{2}}}S(1\textit{d}\ensuremath{_{\textnormal{5/2}}})=1.42 \textit{50} (stat.) \textit{20} (sys.) (\href{https://www.nndc.bnl.gov/nsr/nsrlink.jsp?2017Ku27,B}{2017Ku27}) based on d\ensuremath{\sigma}/d\ensuremath{\Omega}\ensuremath{_{\textnormal{c.m.}}}=16.7 mb/sr \textit{56} (stat.) \textit{23}\vspace{0.1cm}}&\\
&&&&\parbox[t][0.3cm]{14.07324cm}{\raggedright {\ }{\ }{\ }(sys.) at \ensuremath{\theta}\ensuremath{_{\textnormal{c.m.}}}=8\ensuremath{^\circ}.\vspace{0.1cm}}&\\
&&&&\parbox[t][0.3cm]{14.07324cm}{\raggedright ANC(1\textit{d}\ensuremath{_{\textnormal{5/2}}})=2.8 fm\ensuremath{^{\textnormal{$-$1}}} \textit{9} (stat.) \textit{4} (sys.) (\href{https://www.nndc.bnl.gov/nsr/nsrlink.jsp?2017Ku27,B}{2017Ku27}).\vspace{0.1cm}}&\\
\multicolumn{1}{r@{}}{3576}&\multicolumn{1}{@{}l}{}&&&\parbox[t][0.3cm]{14.07324cm}{\raggedright E(level): Possibly populated but unresolved from the 3376-, 3576-, and 3616-keV states (see D. W.\vspace{0.1cm}}&\\
&&&&\parbox[t][0.3cm]{14.07324cm}{\raggedright {\ }{\ }{\ }Bardayan {et al}., AIP Conf. Proc. 2160 (2019) 070010).\vspace{0.1cm}}&\\
\multicolumn{1}{r@{}}{3616}&\multicolumn{1}{@{}l}{\ensuremath{^{{\hyperlink{NE6LEVEL1}{b}}}}}&\multicolumn{1}{l}{2\ensuremath{^{+}}}&\multicolumn{1}{l}{0,2$^{{\hyperlink{NE6LEVEL3}{d}}}$}&\parbox[t][0.3cm]{14.07324cm}{\raggedright E(level): (D. W. Bardayan {et al}., AIP Conf. Proc. 2160 (2019) 070010) observed a large peak, which\vspace{0.1cm}}&\\
&&&&\parbox[t][0.3cm]{14.07324cm}{\raggedright {\ }{\ }{\ }was attributed to the unresolved states at 3376 keV, 3576 keV and 3616 keV (see Fig. 4).\vspace{0.1cm}}&\\
&&&&\parbox[t][0.3cm]{14.07324cm}{\raggedright C\ensuremath{^{\textnormal{2}}}S: C\ensuremath{^{\textnormal{2}}}S(2\textit{s}\ensuremath{_{\textnormal{1/2}}})=0.41 \textit{14} (stat.) \textit{6} (sys.) for \textit{l}=0, and C\ensuremath{^{\textnormal{2}}}S(1\textit{d}\ensuremath{_{\textnormal{5/2}}})=0.76 \textit{26} (stat.) \textit{11} (sys.) for \textit{l}=2\vspace{0.1cm}}&\\
&&&&\parbox[t][0.3cm]{14.07324cm}{\raggedright {\ }{\ }{\ }contributions (based on d\ensuremath{\sigma}/d\ensuremath{\Omega}\ensuremath{_{\textnormal{c.m.}}}=25.9 mb/sr \textit{90} (stat.) \textit{39} (sys.) at \ensuremath{\theta}\ensuremath{_{\textnormal{c.m.}}}=8\ensuremath{^\circ}) assuming the same\vspace{0.1cm}}&\\
&&&&\parbox[t][0.3cm]{14.07324cm}{\raggedright {\ }{\ }{\ }mixing between \textit{l}=0 and \textit{l}=2 as that in the mirror reaction (\href{https://www.nndc.bnl.gov/nsr/nsrlink.jsp?2017Ku27,B}{2017Ku27}).\vspace{0.1cm}}&\\
&&&&\parbox[t][0.3cm]{14.07324cm}{\raggedright ANC(2\textit{s}\ensuremath{_{\textnormal{1/2}}})=188 fm\ensuremath{^{\textnormal{$-$1}}} \textit{66} (stat.) \textit{28} (sys.) (\href{https://www.nndc.bnl.gov/nsr/nsrlink.jsp?2017Ku27,B}{2017Ku27}): assuming pure \textit{l}=0 (from 2\textit{s}\ensuremath{_{\textnormal{1/2}}}) and with no \textit{l}\vspace{0.1cm}}&\\
&&&&\parbox[t][0.3cm]{14.07324cm}{\raggedright {\ }{\ }{\ }mixing. Aassuming the same mixing between \textit{l}=0 and \textit{l}=2 as that in the mirror reaction (\href{https://www.nndc.bnl.gov/nsr/nsrlink.jsp?2017Ku27,B}{2017Ku27}):\vspace{0.1cm}}&\\
&&&&\parbox[t][0.3cm]{14.07324cm}{\raggedright {\ }{\ }{\ }ANC(2\textit{s}\ensuremath{_{\textnormal{1/2}}})=148 fm\ensuremath{^{\textnormal{$-$1}}} \textit{52} (stat.) \textit{22} (sys.) for \textit{l}=0 and ANC(1\textit{d}\ensuremath{_{\textnormal{5/2}}})=3.1 fm\ensuremath{^{\textnormal{$-$1}}} \textit{11} (stat.) \textit{5} (sys.) for\vspace{0.1cm}}&\\
&&&&\parbox[t][0.3cm]{14.07324cm}{\raggedright {\ }{\ }{\ }\textit{l}=2.\vspace{0.1cm}}&\\
\end{longtable}
\begin{textblock}{29}(0,27.3)
Continued on next page (footnotes at end of table)
\end{textblock}
\clearpage
\begin{longtable}{ccccccccc@{\extracolsep{\fill}}c}
\\[-.4cm]
\multicolumn{10}{c}{{\bf \small \underline{\ensuremath{^{\textnormal{2}}}H(\ensuremath{^{\textnormal{17}}}F,n)\hspace{0.2in}\href{https://www.nndc.bnl.gov/nsr/nsrlink.jsp?2017Ku27,B}{2017Ku27} (continued)}}}\\
\multicolumn{10}{c}{~}\\
\multicolumn{10}{c}{\underline{\ensuremath{^{18}}Ne Levels (continued)}}\\
\multicolumn{10}{c}{~}\\
\multicolumn{2}{c}{E(level)$^{{\hyperlink{NE6LEVEL0}{a}}}$}&J$^{\pi}$$^{{\hyperlink{NE6LEVEL2}{c}}}$&\multicolumn{2}{c}{\ensuremath{\Gamma}\ensuremath{_{\textnormal{p}}} (keV)$^{}$}&L$^{{\hyperlink{NE6LEVEL2}{c}}}$&\multicolumn{2}{c}{C\ensuremath{^{\textnormal{2}}}S$^{{\hyperlink{NE6LEVEL4}{e}}}$}&Comments&\\[-.2cm]
\multicolumn{2}{c}{\hrulefill}&\hrulefill&\multicolumn{2}{c}{\hrulefill}&\hrulefill&\multicolumn{2}{c}{\hrulefill}&\hrulefill&
\endhead
\multicolumn{1}{r@{}}{4523}&\multicolumn{1}{@{}l}{}&\multicolumn{1}{l}{3\ensuremath{^{+}}}&\multicolumn{1}{r@{}}{14}&\multicolumn{1}{@{.}l}{2 keV {\it 11}}&\multicolumn{1}{l}{0}&\multicolumn{1}{r@{}}{0}&\multicolumn{1}{@{.}l}{79 {\it 6}}&\parbox[t][0.3cm]{10.4356cm}{\raggedright E(level): The corresponding E\ensuremath{_{\textnormal{c.m.}}}=599.8 keV was mentioned in\vspace{0.1cm}}&\\
&&&&&&&&\parbox[t][0.3cm]{10.4356cm}{\raggedright {\ }{\ }{\ }(\href{https://www.nndc.bnl.gov/nsr/nsrlink.jsp?2017Ku27,B}{2017Ku27}). See also 4561 keV (D. W. Bardayan {et al}., AIP Conf. Proc.\vspace{0.1cm}}&\\
&&&&&&&&\parbox[t][0.3cm]{10.4356cm}{\raggedright {\ }{\ }{\ }2160 (2019) 070010).\vspace{0.1cm}}&\\
&&&&&&&&\parbox[t][0.3cm]{10.4356cm}{\raggedright \ensuremath{\Gamma}\ensuremath{_{\textnormal{p}}}=14.2 keV \textit{3} (stat.) \textit{11} (sys.) (\href{https://www.nndc.bnl.gov/nsr/nsrlink.jsp?2017Ku27,B}{2017Ku27}).\vspace{0.1cm}}&\\
&&&&&&&&\parbox[t][0.3cm]{10.4356cm}{\raggedright C\ensuremath{^{\textnormal{2}}}S: Weighted average of C\ensuremath{^{\textnormal{2}}}S=0.78 \textit{2} (stat.) \textit{6} (sys.) (\href{https://www.nndc.bnl.gov/nsr/nsrlink.jsp?2017Ku27,B}{2017Ku27}: from the\vspace{0.1cm}}&\\
&&&&&&&&\parbox[t][0.3cm]{10.4356cm}{\raggedright {\ }{\ }{\ }p-\ensuremath{^{\textnormal{17}}}F coincidence events using \ensuremath{\sigma}=15.3 mb \textit{3} (stat.) \textit{12} (sys.)); and\vspace{0.1cm}}&\\
&&&&&&&&\parbox[t][0.3cm]{10.4356cm}{\raggedright {\ }{\ }{\ }C\ensuremath{^{\textnormal{2}}}S=0.84 \textit{10} (stat.) \textit{10} (sys.) (\href{https://www.nndc.bnl.gov/nsr/nsrlink.jsp?2017Ku27,B}{2017Ku27}: from the reconstructed \ensuremath{^{\textnormal{18}}}Ne\vspace{0.1cm}}&\\
&&&&&&&&\parbox[t][0.3cm]{10.4356cm}{\raggedright {\ }{\ }{\ }excitation energy spectrum based on the neutron time-of-flight:\vspace{0.1cm}}&\\
&&&&&&&&\parbox[t][0.3cm]{10.4356cm}{\raggedright {\ }{\ }{\ }d\ensuremath{\sigma}/d\ensuremath{\Omega}\ensuremath{_{\textnormal{c.m.}}}=41 mb/sr \textit{5} (stat.) \textit{5} (sys.) at \ensuremath{\theta}\ensuremath{_{\textnormal{c.m.}}}=6\ensuremath{^\circ} using the coupled\vspace{0.1cm}}&\\
&&&&&&&&\parbox[t][0.3cm]{10.4356cm}{\raggedright {\ }{\ }{\ }reaction channels calculations (see text) and assuming J\ensuremath{^{\ensuremath{\pi}}}=3\ensuremath{^{\textnormal{+}}}).\vspace{0.1cm}}&\\
\multicolumn{1}{r@{}}{5.09\ensuremath{\times10^{3}}}&\multicolumn{1}{@{}l}{}&\multicolumn{1}{l}{2\ensuremath{^{+}}}&\multicolumn{1}{r@{}}{50}&\multicolumn{1}{@{ }l}{keV {\it 11}}&&&&\parbox[t][0.3cm]{10.4356cm}{\raggedright E(level): The corresponding E\ensuremath{_{\textnormal{c.m.}}}=1.165 keV was mentioned in (\href{https://www.nndc.bnl.gov/nsr/nsrlink.jsp?2017Ku27,B}{2017Ku27}).\vspace{0.1cm}}&\\
&&&&&&&&\parbox[t][0.3cm]{10.4356cm}{\raggedright \ensuremath{\Gamma}\ensuremath{_{\textnormal{p}}}=50 keV \textit{8} (stat.) \textit{8} (sys.) (\href{https://www.nndc.bnl.gov/nsr/nsrlink.jsp?2017Ku27,B}{2017Ku27}).\vspace{0.1cm}}&\\
&&&&&&&&\parbox[t][0.3cm]{10.4356cm}{\raggedright J\ensuremath{^{\pi}}: (\href{https://www.nndc.bnl.gov/nsr/nsrlink.jsp?2017Ku27,B}{2017Ku27}) excludes J\ensuremath{^{\ensuremath{\pi}}}=3\ensuremath{^{-}} because it would lead to a much smaller\vspace{0.1cm}}&\\
&&&&&&&&\parbox[t][0.3cm]{10.4356cm}{\raggedright {\ }{\ }{\ }spectroscopic factor and a cross section too small to be observed in this\vspace{0.1cm}}&\\
&&&&&&&&\parbox[t][0.3cm]{10.4356cm}{\raggedright {\ }{\ }{\ }experiment.\vspace{0.1cm}}&\\
&&&&&&&&\parbox[t][0.3cm]{10.4356cm}{\raggedright C\ensuremath{^{\textnormal{2}}}S=0.20 \textit{+3{\textminus}2} (stat.) \textit{+3{\textminus}2} (sys.) (\href{https://www.nndc.bnl.gov/nsr/nsrlink.jsp?2017Ku27,B}{2017Ku27}) from the p-\ensuremath{^{\textnormal{17}}}F coincidences.\vspace{0.1cm}}&\\
&&&&&&&&\parbox[t][0.3cm]{10.4356cm}{\raggedright \ensuremath{\sigma}=2.30 mb \textit{35} (stat.) \textit{29} (sys.) (\href{https://www.nndc.bnl.gov/nsr/nsrlink.jsp?2017Ku27,B}{2017Ku27}).\vspace{0.1cm}}&\\
\end{longtable}
\parbox[b][0.3cm]{17.7cm}{\makebox[1ex]{\ensuremath{^{\hypertarget{NE6LEVEL0}{a}}}} From (\href{https://www.nndc.bnl.gov/nsr/nsrlink.jsp?2017Ku27,B}{2017Ku27}), unless noted otherwise.}\\
\parbox[b][0.3cm]{17.7cm}{\makebox[1ex]{\ensuremath{^{\hypertarget{NE6LEVEL1}{b}}}} This state was observed in (\href{https://www.nndc.bnl.gov/nsr/nsrlink.jsp?2017Ku27,B}{2017Ku27}) from the neutron time-of-flight spectrum when a gate on the \ensuremath{\gamma}-ray coincident events was}\\
\parbox[b][0.3cm]{17.7cm}{{\ }{\ }applied.}\\
\parbox[b][0.3cm]{17.7cm}{\makebox[1ex]{\ensuremath{^{\hypertarget{NE6LEVEL2}{c}}}} Deduced (1) for the proton unbound states (E\ensuremath{_{\textnormal{x}}}(\ensuremath{^{\textnormal{18}}}Ne)\ensuremath{>}4 MeV) from the coupled reactions channel calculations using FRESCO}\\
\parbox[b][0.3cm]{17.7cm}{{\ }{\ }(\href{https://www.nndc.bnl.gov/nsr/nsrlink.jsp?2017Ku27,B}{2017Ku27}), and (2) for the proton bound states (E\ensuremath{_{\textnormal{x}}}(\ensuremath{^{\textnormal{18}}}Ne)\ensuremath{<}4 MeV) from the Monte Carlo simulation of the correlation between}\\
\parbox[b][0.3cm]{17.7cm}{{\ }{\ }the energies of the detected charged particle reaction products and the center-of-mass neutron emission angles (from the deuteron}\\
\parbox[b][0.3cm]{17.7cm}{{\ }{\ }breakup). As a result, the simulated spectra of p+\ensuremath{^{\textnormal{17}}}F energy distributions, assuming various angular momentum transfers, were}\\
\parbox[b][0.3cm]{17.7cm}{{\ }{\ }used to fit the experimental spectra to find the best fit, and thus the best \textit{l} value (\href{https://www.nndc.bnl.gov/nsr/nsrlink.jsp?2017Ku27,B}{2017Ku27}).}\\
\parbox[b][0.3cm]{17.7cm}{\makebox[1ex]{\ensuremath{^{\hypertarget{NE6LEVEL3}{d}}}} (\href{https://www.nndc.bnl.gov/nsr/nsrlink.jsp?2017Ku27,B}{2017Ku27}) does not present the Monte Carlo fits, from which the angular momenta are deduced for the transfer reaction}\\
\parbox[b][0.3cm]{17.7cm}{{\ }{\ }populating this state.}\\
\parbox[b][0.3cm]{17.7cm}{\makebox[1ex]{\ensuremath{^{\hypertarget{NE6LEVEL4}{e}}}} An upper limit of C\ensuremath{^{\textnormal{2}}}S\ensuremath{\leq}0.1 was estimated for the spectroscopic factors of potential additional low-lying resonances in \ensuremath{^{\textnormal{18}}}Ne}\\
\parbox[b][0.3cm]{17.7cm}{{\ }{\ }(\href{https://www.nndc.bnl.gov/nsr/nsrlink.jsp?2017Ku27,B}{2017Ku27}).}\\
\vspace{0.5cm}
\clearpage
\subsection[\hspace{-0.2cm}\ensuremath{^{\textnormal{4}}}He(\ensuremath{^{\textnormal{14}}}O,P),(\ensuremath{^{\textnormal{14}}}O,2p):res]{ }
\vspace{-27pt}
\vspace{0.3cm}
\hypertarget{NE7}{{\bf \small \underline{\ensuremath{^{\textnormal{4}}}He(\ensuremath{^{\textnormal{14}}}O,P),(\ensuremath{^{\textnormal{14}}}O,2p):res\hspace{0.2in}\href{https://www.nndc.bnl.gov/nsr/nsrlink.jsp?1987Wi11,B}{1987Wi11},\href{https://www.nndc.bnl.gov/nsr/nsrlink.jsp?2018Na26,B}{2018Na26}}}}\\
\vspace{4pt}
\vspace{8pt}
\parbox[b][0.3cm]{17.7cm}{\addtolength{\parindent}{-0.2in}\href{https://www.nndc.bnl.gov/nsr/nsrlink.jsp?2004No14,B}{2004No14}, \href{https://www.nndc.bnl.gov/nsr/nsrlink.jsp?2004No18,B}{2004No18}, \href{https://www.nndc.bnl.gov/nsr/nsrlink.jsp?2006Ku17,B}{2006Ku17}: \ensuremath{^{\textnormal{4}}}He(\ensuremath{^{\textnormal{14}}}O,p) E=43 MeV; measured protons by a \ensuremath{\Delta}E-E telescope at \ensuremath{\theta}\ensuremath{_{\textnormal{lab}}}=0\ensuremath{^\circ}; measured E\ensuremath{_{\textnormal{p}}} and}\\
\parbox[b][0.3cm]{17.7cm}{\ensuremath{\sigma}(E) for \ensuremath{^{\textnormal{14}}}O(\ensuremath{\alpha},p) at E\ensuremath{_{\textnormal{c.m.}}}(\ensuremath{^{\textnormal{14}}}O+\ensuremath{\alpha})=0.8-3.8 MeV (E\ensuremath{_{\textnormal{c.m.}}}(\ensuremath{^{\textnormal{17}}}F+p)=2{\textminus}4.8 MeV) using thick target yield technique; observed}\\
\parbox[b][0.3cm]{17.7cm}{enhancements in the cross section near E\ensuremath{_{\textnormal{x}}}(\ensuremath{^{\textnormal{18}}}Ne)=6.1, 7.1, and 7.4 MeV; observed for the first time a weak transition at}\\
\parbox[b][0.3cm]{17.7cm}{E\ensuremath{_{\textnormal{c.m.}}}(\ensuremath{^{\textnormal{14}}}O+\ensuremath{\alpha})=1.5 MeV; discussed astrophysical implications and the \ensuremath{^{\textnormal{14}}}O(\ensuremath{\alpha},p) reaction rate.}\\
\parbox[b][0.3cm]{17.7cm}{\addtolength{\parindent}{-0.2in}\href{https://www.nndc.bnl.gov/nsr/nsrlink.jsp?2007Fu09,B}{2007Fu09}, \href{https://www.nndc.bnl.gov/nsr/nsrlink.jsp?2008FuZZ,B}{2008FuZZ}: \ensuremath{^{\textnormal{4}}}He(\ensuremath{^{\textnormal{14}}}O,X)\ensuremath{^{\textnormal{16}}}O E=66 MeV; used the MARS recoil separator and measured protons emitted from the decay}\\
\parbox[b][0.3cm]{17.7cm}{of \ensuremath{^{\textnormal{18}}}Ne states using an array of 4 segmented Silicon detectors followed by a Silicon telescope at \ensuremath{\theta}\ensuremath{_{\textnormal{lab}}}=0\ensuremath{^\circ}; energy resolution was}\\
\parbox[b][0.3cm]{17.7cm}{200 keV (FWHM); measured E\ensuremath{_{\textnormal{p}}}, \ensuremath{\theta}\ensuremath{_{\textnormal{p}}}, p-p-coincidences; thick target technique; deduced \ensuremath{^{\textnormal{18}}}Ne excitation function and decay cross}\\
\parbox[b][0.3cm]{17.7cm}{sections; constructed Dalitz plots for the (\ensuremath{\alpha},2p) events assuming a sequential decay via \ensuremath{^{\textnormal{18}}}Ne* \ensuremath{\rightarrow} p\ensuremath{_{\textnormal{1}}}+\ensuremath{^{\textnormal{17}}}F* \ensuremath{\rightarrow} p\ensuremath{_{\textnormal{1}}}+\ensuremath{^{\textnormal{16}}}O\ensuremath{_{\textnormal{g.s.}}}+p\ensuremath{_{\textnormal{2}}};}\\
\parbox[b][0.3cm]{17.7cm}{dominant intensity of 2p events corresponded to the known \ensuremath{^{\textnormal{17}}}F*(3.10, 3.86, 5.22 MeV); observed diproton decay of \ensuremath{^{\textnormal{18}}}Ne*(8.45}\\
\parbox[b][0.3cm]{17.7cm}{MeV); calculated relative energy distribution for the two protons from the decay of the 8.45 MeV state using the Faddeev approach;}\\
\parbox[b][0.3cm]{17.7cm}{observed 4 new states of \ensuremath{^{\textnormal{18}}}Ne at 10.12, 10.66, 11.29, 11.8 MeV; discussed astrophysical implications for the \ensuremath{^{\textnormal{14}}}O(\ensuremath{\alpha},p) reaction}\\
\parbox[b][0.3cm]{17.7cm}{rate.}\\
\parbox[b][0.3cm]{17.7cm}{\addtolength{\parindent}{-0.2in}\href{https://www.nndc.bnl.gov/nsr/nsrlink.jsp?2015Ki07,B}{2015Ki07}: \ensuremath{^{\textnormal{4}}}He(\ensuremath{^{\textnormal{14}}}O,p\ensuremath{_{\textnormal{0,1,2,3}}}) E=33.8 MeV; measured single protons and p-p coincidences at E\ensuremath{_{\textnormal{c.m.}}}=2.1-5.3 MeV using a set of}\\
\parbox[b][0.3cm]{17.7cm}{position sensitive \ensuremath{\Delta}E-E Si telescopes that covered \ensuremath{\theta}\ensuremath{_{\textnormal{lab}}}={\textminus}12\ensuremath{^\circ}{\textminus}78\ensuremath{^\circ}; data show much less statistics for the democratic 2p decay than}\\
\parbox[b][0.3cm]{17.7cm}{for the sequential decay so the 2p events were excluded from the analysis but were considered in error estimations; it was generally}\\
\parbox[b][0.3cm]{17.7cm}{possible to distinguish the \ensuremath{^{\textnormal{14}}}O(\ensuremath{\alpha},p\ensuremath{_{\textnormal{0,1,2,3}}}) groups; deduced excitation function of \ensuremath{^{\textnormal{18}}}Ne; analyzed the excitation function using}\\
\parbox[b][0.3cm]{17.7cm}{SAMMY R-matrix code; performed shell model calculations to deduce spins and parities for the observed states above 8.1 MeV;}\\
\parbox[b][0.3cm]{17.7cm}{identified \ensuremath{^{\textnormal{18}}}Ne resonances significant to the \ensuremath{^{\textnormal{14}}}O(\ensuremath{\alpha},p) astrophysical reaction rate; discussed astrophysical implications.}\\
\parbox[b][0.3cm]{17.7cm}{\addtolength{\parindent}{-0.2in}\href{https://www.nndc.bnl.gov/nsr/nsrlink.jsp?2023Pa44,B}{2023Pa44}: \ensuremath{^{\textnormal{4}}}He(\ensuremath{^{\textnormal{14}}}O,p) E=3.845 MeV/nucleon; measured \ensuremath{\sigma}(E) for the \ensuremath{\alpha}(\ensuremath{^{\textnormal{14}}}O,p)\ensuremath{^{\textnormal{17}}}F reaction around E\ensuremath{_{\textnormal{c.m.}}}=7 MeV using the}\\
\parbox[b][0.3cm]{17.7cm}{upgraded TexAT active target chamber (TexAT\_v2) together with the MARS beamline. This study was carried out to commission}\\
\parbox[b][0.3cm]{17.7cm}{TexAT\_v2. Analysis is in progress. The authors indicated that the direct cross section measurement of the \ensuremath{\alpha}(\ensuremath{^{\textnormal{14}}}O,p)\ensuremath{^{\textnormal{17}}}F reaction is}\\
\parbox[b][0.3cm]{17.7cm}{proposed to be performed at the Center for Nuclear Study, RIKEN Nishina Center using the TexAT detector.}\\
\vspace{0.385cm}
\parbox[b][0.3cm]{17.7cm}{\addtolength{\parindent}{-0.2in}\textit{Theory:}}\\
\parbox[b][0.3cm]{17.7cm}{\addtolength{\parindent}{-0.2in}\href{https://www.nndc.bnl.gov/nsr/nsrlink.jsp?2018Na26,B}{2018Na26}: Calculated bound and resonant 0\ensuremath{^{\textnormal{+}}} levels, decay widths, monopole transition strengths, and Coulomb shifts for the mirror}\\
\parbox[b][0.3cm]{17.7cm}{cluster \ensuremath{\alpha}+\ensuremath{^{\textnormal{14}}}O system using the orthogonality condition model; results interpreted in terms of the extension of the Thomas-Ehrman}\\
\parbox[b][0.3cm]{17.7cm}{shift (TES); comparison with experimental values; proposed combination of the cluster TES and the monopole transitions as}\\
\parbox[b][0.3cm]{17.7cm}{experimental probe for cluster structure in mirror systems.}\\
\vspace{0.385cm}
\parbox[b][0.3cm]{17.7cm}{\addtolength{\parindent}{-0.2in}\textit{The Astrophysical \ensuremath{^{14}}O(\ensuremath{\alpha},p) Reaction Rate}:}\\
\parbox[b][0.3cm]{17.7cm}{\addtolength{\parindent}{-0.2in}This subsection contains experimental and theoretical studies that are focused on the determination of the \ensuremath{^{\textnormal{14}}}O(\ensuremath{\alpha},p) astrophysical}\\
\parbox[b][0.3cm]{17.7cm}{rate. The details of the experimental studies can be found in other sections.}\\
\parbox[b][0.3cm]{17.7cm}{\addtolength{\parindent}{-0.2in}R. K. Wallace and S. E. Woosley, Astrophys. J. Suppl. Ser. 45 (1981) 389: Calculated the \ensuremath{^{\textnormal{14}}}O(\ensuremath{\alpha},p) reaction rate based on the}\\
\parbox[b][0.3cm]{17.7cm}{resonant contribution from two \ensuremath{\alpha}-unbound levels at E\ensuremath{_{\textnormal{x}}}=6.3 and 6.35 MeV in \ensuremath{^{\textnormal{18}}}Ne; approximated the Hauser-Feshbach rate;}\\
\parbox[b][0.3cm]{17.7cm}{deduced an analytical expression for the reaction rate.}\\
\parbox[b][0.3cm]{17.7cm}{\addtolength{\parindent}{-0.2in}\href{https://www.nndc.bnl.gov/nsr/nsrlink.jsp?1987Wi11,B}{1987Wi11}: Calculated the resonant and non-resonant contributions to the \ensuremath{^{\textnormal{14}}}O(\ensuremath{\alpha},p)\ensuremath{^{\textnormal{17}}}F and \ensuremath{^{\textnormal{14}}}O(\ensuremath{\alpha},\ensuremath{\gamma})\ensuremath{^{\textnormal{18}}}Ne reaction rates at 0.1-10}\\
\parbox[b][0.3cm]{17.7cm}{GK; calculated depletion rate ratios, S-factor, reaction \ensuremath{\sigma}, resonance energies, widths, spectroscopic factors, and resonance strengths.}\\
\parbox[b][0.3cm]{17.7cm}{\addtolength{\parindent}{-0.2in}\href{https://www.nndc.bnl.gov/nsr/nsrlink.jsp?1988Ca26,B}{1988Ca26}: Calculated an analytical expression for this reaction rate, which was used to obtain the rate from 0.001-10 GK.}\\
\parbox[b][0.3cm]{17.7cm}{\addtolength{\parindent}{-0.2in}\href{https://www.nndc.bnl.gov/nsr/nsrlink.jsp?1988Fu02,B}{1988Fu02}, \href{https://www.nndc.bnl.gov/nsr/nsrlink.jsp?1989Fu01,B}{1989Fu01}: Calculated the \ensuremath{^{\textnormal{14}}}O(\ensuremath{\alpha},p) astrophysical reaction rate based on microscopic coupled-channel generator}\\
\parbox[b][0.3cm]{17.7cm}{coordinate method; calculated reaction \ensuremath{\sigma} and the astrophysical S-factor vs. E for E\ensuremath{_{\textnormal{c.m.}}}=0.4-4 MeV; deduced \ensuremath{^{\textnormal{18}}}Ne resonance}\\
\parbox[b][0.3cm]{17.7cm}{properties; pointed out the important contributions to the \ensuremath{^{\textnormal{14}}}O(\ensuremath{\alpha},p) reaction rate by the direct capture process for temperatures}\\
\parbox[b][0.3cm]{17.7cm}{T\ensuremath{\leq}0.3 GK; comparison with the results of (\href{https://www.nndc.bnl.gov/nsr/nsrlink.jsp?1987Wi11,B}{1987Wi11}). (\href{https://www.nndc.bnl.gov/nsr/nsrlink.jsp?1989Fu01,B}{1989Fu01}) included inelastic scattering of \ensuremath{^{\textnormal{14}}}O+\ensuremath{\alpha} in this calculation and}\\
\parbox[b][0.3cm]{17.7cm}{recalculated the S-factor.}\\
\parbox[b][0.3cm]{17.7cm}{\addtolength{\parindent}{-0.2in}\href{https://www.nndc.bnl.gov/nsr/nsrlink.jsp?1992Ch50,B}{1992Ch50}: Provides a review of the \ensuremath{^{\textnormal{14}}}O(\ensuremath{\alpha},p) reaction rate and the structure of \ensuremath{^{\textnormal{18}}}Ne based on the available data at the time.}\\
\parbox[b][0.3cm]{17.7cm}{\addtolength{\parindent}{-0.2in}\href{https://www.nndc.bnl.gov/nsr/nsrlink.jsp?1996Ha26,B}{1996Ha26}: Calculated astrophysical S-factor vs. E and the \ensuremath{^{\textnormal{14}}}O(\ensuremath{\alpha},p) astrophysical reaction rate considering E\ensuremath{_{\textnormal{c.m.}}}\ensuremath{<}2.5 MeV and for}\\
\parbox[b][0.3cm]{17.7cm}{0.1-1 GK; deduced the \ensuremath{^{\textnormal{18}}}Ne resonance energies, J\ensuremath{^{\ensuremath{\pi}}}, C\ensuremath{^{\textnormal{2}}}S\ensuremath{_{\textnormal{p}}}, C\ensuremath{^{\textnormal{2}}}S\ensuremath{_{\ensuremath{\alpha}}}, \ensuremath{\Gamma}\ensuremath{_{\textnormal{p}}}, \ensuremath{\Gamma}\ensuremath{_{\ensuremath{\alpha}}}, \ensuremath{\Gamma}, and \ensuremath{\omega}\ensuremath{\gamma}\ensuremath{_{\textnormal{(}\ensuremath{\alpha}\textnormal{,p)}}} for resonances with E\ensuremath{_{\textnormal{c.m.}}}\ensuremath{<}2.5}\\
\parbox[b][0.3cm]{17.7cm}{MeV; discussed the astrophysical implications.}\\
\parbox[b][0.3cm]{17.7cm}{\addtolength{\parindent}{-0.2in}\href{https://www.nndc.bnl.gov/nsr/nsrlink.jsp?1997Ba57,B}{1997Ba57}: Calculated the \ensuremath{^{\textnormal{14}}}O(\ensuremath{\alpha},p) and \ensuremath{^{\textnormal{17}}}F(p,\ensuremath{\gamma}) reaction rates for 0.1-1 GK based on the experimental data of (\href{https://www.nndc.bnl.gov/nsr/nsrlink.jsp?1991Ga03,B}{1991Ga03},}\\
\parbox[b][0.3cm]{17.7cm}{\href{https://www.nndc.bnl.gov/nsr/nsrlink.jsp?1996Ha26,B}{1996Ha26}); calculated the S-factor for the \ensuremath{^{\textnormal{17}}}F(p,\ensuremath{\gamma}) reaction for E\ensuremath{_{\textnormal{c.m.}}}=\ensuremath{\sim}30-1000 keV; deduced the REACLIB analytical}\\
\parbox[b][0.3cm]{17.7cm}{expressions and parameters for these rates.}\\
\parbox[b][0.3cm]{17.7cm}{\addtolength{\parindent}{-0.2in}\href{https://www.nndc.bnl.gov/nsr/nsrlink.jsp?2002Ha15,B}{2002Ha15}: Analyzed data and deduced \ensuremath{^{\textnormal{18}}}Ne resonance energies, J\ensuremath{^{\ensuremath{\pi}}}, \ensuremath{\Gamma}\ensuremath{_{\textnormal{p}}}, \ensuremath{\Gamma}\ensuremath{_{\ensuremath{\alpha}}}, \ensuremath{\Gamma}\ensuremath{_{\textnormal{p$'$}}}, \ensuremath{\Gamma} and \ensuremath{\omega}\ensuremath{\gamma}\ensuremath{_{\textnormal{(}\ensuremath{\alpha}\textnormal{,p)}}} for 7 states with}\\
\parbox[b][0.3cm]{17.7cm}{E\ensuremath{_{\textnormal{x}}}=6.15-7.71 MeV; used these to calculate the \ensuremath{^{\textnormal{14}}}O(\ensuremath{\alpha},p) resonant reaction rate for 0.5-3 GK; comparison with the rate of}\\
\parbox[b][0.3cm]{17.7cm}{(\href{https://www.nndc.bnl.gov/nsr/nsrlink.jsp?1996Ha26,B}{1996Ha26}) is presented.}\\
\clearpage
\vspace{0.3cm}
{\bf \small \underline{\ensuremath{^{\textnormal{4}}}He(\ensuremath{^{\textnormal{14}}}O,P),(\ensuremath{^{\textnormal{14}}}O,2p):res\hspace{0.2in}\href{https://www.nndc.bnl.gov/nsr/nsrlink.jsp?1987Wi11,B}{1987Wi11},\href{https://www.nndc.bnl.gov/nsr/nsrlink.jsp?2018Na26,B}{2018Na26} (continued)}}\\
\vspace{0.3cm}
\parbox[b][0.3cm]{17.7cm}{\addtolength{\parindent}{-0.2in}\href{https://www.nndc.bnl.gov/nsr/nsrlink.jsp?2002Il05,B}{2002Il05}: Recommended the rate of (\href{https://www.nndc.bnl.gov/nsr/nsrlink.jsp?1988Ca26,B}{1988Ca26}).}\\
\parbox[b][0.3cm]{17.7cm}{\addtolength{\parindent}{-0.2in}Luc L. Dessieux, Jr., M. Sc. Thesis, \textit{2p or Not 2p: The \ensuremath{^{14}}O(\ensuremath{\alpha},2p) \ensuremath{^{\textnormal{16}}}O Reaction Rate and Its Implications On Nova and X-ray}}\\
\parbox[b][0.3cm]{17.7cm}{\textit{Burst Nucleosynthesis}, The University of Tennessee, Knoxville (2002), and \href{https://www.nndc.bnl.gov/nsr/nsrlink.jsp?2004Sm08,B}{2004Sm08}: Evaluated the \ensuremath{^{\textnormal{14}}}O(\ensuremath{\alpha},p) and \ensuremath{^{\textnormal{14}}}O(\ensuremath{\alpha},2p)}\\
\parbox[b][0.3cm]{17.7cm}{reaction rates at temperatures of interest for novae and type I X-ray bursts and discussed their astrophysical implications.}\\
\parbox[b][0.3cm]{17.7cm}{\addtolength{\parindent}{-0.2in}\href{https://www.nndc.bnl.gov/nsr/nsrlink.jsp?2003Bl11,B}{2003Bl11}: Calculated the \ensuremath{^{\textnormal{14}}}O(\ensuremath{\alpha},p) reaction rate for 0.1-1.5 GK incorporating the branch to the first excited state in \ensuremath{^{\textnormal{17}}}F; discussed}\\
\parbox[b][0.3cm]{17.7cm}{the reaction rate on the basis of their experimental results; comparison with previous rates is presented.}\\
\parbox[b][0.3cm]{17.7cm}{\addtolength{\parindent}{-0.2in}\href{https://www.nndc.bnl.gov/nsr/nsrlink.jsp?2004No14,B}{2004No14}, \href{https://www.nndc.bnl.gov/nsr/nsrlink.jsp?2004No18,B}{2004No18}, \href{https://www.nndc.bnl.gov/nsr/nsrlink.jsp?2006Ku17,B}{2006Ku17}: Discussed the \ensuremath{^{\textnormal{14}}}O(\ensuremath{\alpha},p) reaction rate based on the results of their measurements (see the first}\\
\parbox[b][0.3cm]{17.7cm}{subsection of this dataset).}\\
\parbox[b][0.3cm]{17.7cm}{\addtolength{\parindent}{-0.2in}\href{https://www.nndc.bnl.gov/nsr/nsrlink.jsp?2007Fu09,B}{2007Fu09}, \href{https://www.nndc.bnl.gov/nsr/nsrlink.jsp?2008FuZZ,B}{2008FuZZ}: Based on the measured properties of the \ensuremath{^{\textnormal{14}}}O+\ensuremath{\alpha} interaction, the authors disputed the assumptions made in}\\
\parbox[b][0.3cm]{17.7cm}{(\href{https://www.nndc.bnl.gov/nsr/nsrlink.jsp?2004No18,B}{2004No18}, \href{https://www.nndc.bnl.gov/nsr/nsrlink.jsp?2006Ku17,B}{2006Ku17}) for the interpretation of the \ensuremath{^{\textnormal{14}}}O(\ensuremath{\alpha}, p) rate at astrophysical energies. On the contrary to what was stated in}\\
\parbox[b][0.3cm]{17.7cm}{(\href{https://www.nndc.bnl.gov/nsr/nsrlink.jsp?2006Ku17,B}{2006Ku17}) that the \ensuremath{^{\textnormal{14}}}O(\ensuremath{\alpha},p) reaction dominates in the production of low energy protons, (\href{https://www.nndc.bnl.gov/nsr/nsrlink.jsp?2007Fu09,B}{2007Fu09}, \href{https://www.nndc.bnl.gov/nsr/nsrlink.jsp?2008FuZZ,B}{2008FuZZ}) concluded that}\\
\parbox[b][0.3cm]{17.7cm}{the \ensuremath{\alpha}+\ensuremath{^{\textnormal{14}}}O interaction results in populating high energy \ensuremath{^{\textnormal{18}}}Ne resonances (E\ensuremath{_{\textnormal{x}}}=7-12 MeV) with energies above the astrophysically}\\
\parbox[b][0.3cm]{17.7cm}{significant energy region. The protons from the sequential decays (via proton-unbound \ensuremath{^{\textnormal{17}}}F levels) of these states, therefore, create a}\\
\parbox[b][0.3cm]{17.7cm}{large background in the astrophysically important energy region, which would make it harder to distinguish \ensuremath{^{\textnormal{18}}}Ne proton decay}\\
\parbox[b][0.3cm]{17.7cm}{paths.}\\
\parbox[b][0.3cm]{17.7cm}{\addtolength{\parindent}{-0.2in}\href{https://www.nndc.bnl.gov/nsr/nsrlink.jsp?2010Ha15,B}{2010Ha15}: Discussed the \ensuremath{^{\textnormal{14}}}O(\ensuremath{\alpha},p) reaction rate based on the results of their measurement (see \ensuremath{^{\textnormal{4}}}He(\ensuremath{^{\textnormal{14}}}O,\ensuremath{\alpha}) section).}\\
\parbox[b][0.3cm]{17.7cm}{\addtolength{\parindent}{-0.2in}\href{https://www.nndc.bnl.gov/nsr/nsrlink.jsp?2011He09,B}{2011He09}: Calculated the total resonant \ensuremath{^{\textnormal{14}}}O(\ensuremath{\alpha},p) reaction rate at T=0.5-3 GK. Presented the resonance properties (E\ensuremath{_{\textnormal{x}}}, J\ensuremath{^{\ensuremath{\pi}}} and}\\
\parbox[b][0.3cm]{17.7cm}{\ensuremath{\omega}\ensuremath{\gamma}\ensuremath{_{\textnormal{(}\ensuremath{\alpha}\textnormal{,p)}}}) for the resonances used. Comparison with previous rates is presented.}\\
\parbox[b][0.3cm]{17.7cm}{\addtolength{\parindent}{-0.2in}\href{https://www.nndc.bnl.gov/nsr/nsrlink.jsp?2012Al11,B}{2012Al11}: Calculated the resonant \ensuremath{^{\textnormal{14}}}O(\ensuremath{\alpha},p) reaction rate for 0.1-5 GK using 8 \ensuremath{^{\textnormal{18}}}Ne resonances with E\ensuremath{_{\textnormal{x}}}=5.1-8.1 MeV; presented}\\
\parbox[b][0.3cm]{17.7cm}{the recommended resonance energy, J\ensuremath{^{\ensuremath{\pi}}}, \ensuremath{\Gamma}\ensuremath{_{\ensuremath{\alpha}}}, \ensuremath{\Gamma}\ensuremath{_{\textnormal{p}}}, \ensuremath{\Gamma}, \ensuremath{\omega}\ensuremath{\gamma}\ensuremath{_{\textnormal{(}\ensuremath{\alpha}\textnormal{,p)}}} for these resonances; comparison with previous rates; discussed the}\\
\parbox[b][0.3cm]{17.7cm}{2p decay of the 6.15-MeV \ensuremath{^{\textnormal{18}}}Ne state and its effect on the reaction rate; discussed the astrophysical implications.}\\
\parbox[b][0.3cm]{17.7cm}{\addtolength{\parindent}{-0.2in}\href{https://www.nndc.bnl.gov/nsr/nsrlink.jsp?2014Hu16,B}{2014Hu16}: Recommended the resonance energy, J\ensuremath{^{\ensuremath{\pi}}}, \ensuremath{\Gamma}\ensuremath{_{\ensuremath{\alpha}}}, \ensuremath{\Gamma}\ensuremath{_{\textnormal{p}}}, \ensuremath{\Gamma}\ensuremath{_{\textnormal{p$'$}}}, \ensuremath{\Gamma}, and \ensuremath{\omega}\ensuremath{\gamma}\ensuremath{_{\textnormal{(}\ensuremath{\alpha}\textnormal{,p)}}} for 8 of the \ensuremath{^{\textnormal{18}}}Ne states with E\ensuremath{_{\textnormal{x}}}=5.153-8.11}\\
\parbox[b][0.3cm]{17.7cm}{MeV; calculated the resonance \ensuremath{^{\textnormal{14}}}O(\ensuremath{\alpha},p) reaction rate for 0.25-3.0 GK using these resonances; recommended the direct capture rate}\\
\parbox[b][0.3cm]{17.7cm}{calculated by (\href{https://www.nndc.bnl.gov/nsr/nsrlink.jsp?1988Fu02,B}{1988Fu02}); discussed the interferences between the direct and resonant captures for the \ensuremath{^{\textnormal{18}}}Ne*(6.15 MeV) state;}\\
\parbox[b][0.3cm]{17.7cm}{deduced an analytical expression for the reaction rate; discussed the astrophysical implications; comparison with previous reaction}\\
\parbox[b][0.3cm]{17.7cm}{rates is presented.}\\
\parbox[b][0.3cm]{17.7cm}{\addtolength{\parindent}{-0.2in}\href{https://www.nndc.bnl.gov/nsr/nsrlink.jsp?2015Ki07,B}{2015Ki07}: Deduced the resonance energies, \ensuremath{\Gamma}\ensuremath{_{\ensuremath{\alpha}}}, \ensuremath{\Gamma}\ensuremath{_{\textnormal{p}}}, \ensuremath{\omega}\ensuremath{\gamma}\ensuremath{_{\textnormal{(}\ensuremath{\alpha}\textnormal{,p)}}}, J\ensuremath{^{\ensuremath{\pi}}} for \ensuremath{^{\textnormal{18}}}Ne resonances corresponding to the states with}\\
\parbox[b][0.3cm]{17.7cm}{E\ensuremath{_{\textnormal{x}}}=7.35-8.10 MeV; calculated the \ensuremath{^{\textnormal{14}}}O(\ensuremath{\alpha},p) resonant reaction rate for 0.5-4 GK; comparison with previous rates is presented.}\\
\vspace{12pt}
\underline{$^{18}$Ne Levels}\\
\vspace{0.34cm}
\parbox[b][0.3cm]{17.7cm}{\addtolength{\parindent}{-0.254cm}(\href{https://www.nndc.bnl.gov/nsr/nsrlink.jsp?2007Fu09,B}{2007Fu09}, \href{https://www.nndc.bnl.gov/nsr/nsrlink.jsp?2008FuZZ,B}{2008FuZZ}) reported that the sequential 2p decays (\ensuremath{^{\textnormal{18}}}Ne*\ensuremath{\rightarrow}(\ensuremath{^{\textnormal{17}}}F*\ensuremath{\rightarrow}\ensuremath{^{\textnormal{16}}}O\ensuremath{_{\textnormal{g.s.}}}+p)+p) were dominant in the high energy}\\
\parbox[b][0.3cm]{17.7cm}{region (E\ensuremath{_{\textnormal{x}}}(\ensuremath{^{\textnormal{18}}}Ne)\ensuremath{>}8 MeV).}\\
\vspace{0.34cm}
\begin{longtable}{cccccc@{\extracolsep{\fill}}c}
\multicolumn{2}{c}{E(level)$^{}$}&J$^{\pi}$$^{{\hyperlink{NE7LEVEL4}{e}}}$&\multicolumn{2}{c}{\ensuremath{\Gamma}$^{{\hyperlink{NE7LEVEL2}{c}}}$}&Comments&\\[-.2cm]
\multicolumn{2}{c}{\hrulefill}&\hrulefill&\multicolumn{2}{c}{\hrulefill}&\hrulefill&
\endfirsthead
\multicolumn{1}{r@{}}{6150}&\multicolumn{1}{@{}l}{}&&&&\parbox[t][0.3cm]{12.12584cm}{\raggedright E(level): From (\href{https://www.nndc.bnl.gov/nsr/nsrlink.jsp?2004No14,B}{2004No14}, \href{https://www.nndc.bnl.gov/nsr/nsrlink.jsp?2004No18,B}{2004No18}, \href{https://www.nndc.bnl.gov/nsr/nsrlink.jsp?2006Ku17,B}{2006Ku17}).\vspace{0.1cm}}&\\
\multicolumn{1}{r@{}}{6290}&\multicolumn{1}{@{}l}{}&&&&\parbox[t][0.3cm]{12.12584cm}{\raggedright E(level): From (\href{https://www.nndc.bnl.gov/nsr/nsrlink.jsp?2004No18,B}{2004No18}, \href{https://www.nndc.bnl.gov/nsr/nsrlink.jsp?2006Ku17,B}{2006Ku17}).\vspace{0.1cm}}&\\
\multicolumn{1}{r@{}}{7050}&\multicolumn{1}{@{}l}{\ensuremath{^{{\hyperlink{NE7LEVEL0}{a}}}}}&&&&\parbox[t][0.3cm]{12.12584cm}{\raggedright (\href{https://www.nndc.bnl.gov/nsr/nsrlink.jsp?2004No14,B}{2004No14}, \href{https://www.nndc.bnl.gov/nsr/nsrlink.jsp?2004No18,B}{2004No18}, \href{https://www.nndc.bnl.gov/nsr/nsrlink.jsp?2006Ku17,B}{2006Ku17}): evidence from the direct measurement of \ensuremath{^{\textnormal{14}}}O(\ensuremath{\alpha},p)\vspace{0.1cm}}&\\
&&&&&\parbox[t][0.3cm]{12.12584cm}{\raggedright {\ }{\ }{\ }supports proton decay to \ensuremath{^{\textnormal{17}}}F*(495 keV).\vspace{0.1cm}}&\\
\multicolumn{1}{r@{}}{7120}&\multicolumn{1}{@{}l}{\ensuremath{^{{\hyperlink{NE7LEVEL0}{a}}}}}&&&&\parbox[t][0.3cm]{12.12584cm}{\raggedright (\href{https://www.nndc.bnl.gov/nsr/nsrlink.jsp?2004No14,B}{2004No14}, \href{https://www.nndc.bnl.gov/nsr/nsrlink.jsp?2004No18,B}{2004No18}, \href{https://www.nndc.bnl.gov/nsr/nsrlink.jsp?2006Ku17,B}{2006Ku17}): evidence from the direct measurement of \ensuremath{^{\textnormal{14}}}O(\ensuremath{\alpha},p)\vspace{0.1cm}}&\\
&&&&&\parbox[t][0.3cm]{12.12584cm}{\raggedright {\ }{\ }{\ }supports proton decay to \ensuremath{^{\textnormal{17}}}F*(495 keV) but no such evidence is observed by\vspace{0.1cm}}&\\
&&&&&\parbox[t][0.3cm]{12.12584cm}{\raggedright {\ }{\ }{\ }(\href{https://www.nndc.bnl.gov/nsr/nsrlink.jsp?2010Ba21,B}{2010Ba21}).\vspace{0.1cm}}&\\
\multicolumn{1}{r@{}}{7.35\ensuremath{\times10^{3}}}&\multicolumn{1}{@{}l}{\ensuremath{^{{\hyperlink{NE7LEVEL1}{b}}}} {\it 3}}&\multicolumn{1}{l}{(1\ensuremath{^{-}})}&\multicolumn{1}{r@{}}{390}&\multicolumn{1}{@{ }l}{keV {\it 40}}&\parbox[t][0.3cm]{12.12584cm}{\raggedright \ensuremath{\Gamma}\ensuremath{\alpha}=3.1 keV \textit{2}; \ensuremath{\Gamma}\ensuremath{_{\textnormal{p}}}=387 keV \textit{40} (\href{https://www.nndc.bnl.gov/nsr/nsrlink.jsp?2015Ki07,B}{2015Ki07})\vspace{0.1cm}}&\\
&&&&&\parbox[t][0.3cm]{12.12584cm}{\raggedright E(level): See also 7350 keV (\href{https://www.nndc.bnl.gov/nsr/nsrlink.jsp?2004No18,B}{2004No18}, \href{https://www.nndc.bnl.gov/nsr/nsrlink.jsp?2006Ku17,B}{2006Ku17}) and 7370 keV (\href{https://www.nndc.bnl.gov/nsr/nsrlink.jsp?2004No14,B}{2004No14}). Note\vspace{0.1cm}}&\\
&&&&&\parbox[t][0.3cm]{12.12584cm}{\raggedright {\ }{\ }{\ }that (\href{https://www.nndc.bnl.gov/nsr/nsrlink.jsp?2015Ki07,B}{2015Ki07}) had difficulty identifying this state due to poor statistics and claimed\vspace{0.1cm}}&\\
&&&&&\parbox[t][0.3cm]{12.12584cm}{\raggedright {\ }{\ }{\ }that the data for this state were featureless. However, their fit to the nearby states\vspace{0.1cm}}&\\
&&&&&\parbox[t][0.3cm]{12.12584cm}{\raggedright {\ }{\ }{\ }depended on the existence of this state and on its J\ensuremath{^{\ensuremath{\pi}}} assignment.\vspace{0.1cm}}&\\
&&&&&\parbox[t][0.3cm]{12.12584cm}{\raggedright E\ensuremath{_{\textnormal{c.m.}}}(\ensuremath{^{\textnormal{14}}}O+\ensuremath{\alpha})=2.25 MeV (\href{https://www.nndc.bnl.gov/nsr/nsrlink.jsp?2015Ki07,B}{2015Ki07}).\vspace{0.1cm}}&\\
&&&&&\parbox[t][0.3cm]{12.12584cm}{\raggedright \ensuremath{\omega}\ensuremath{\gamma}\ensuremath{_{\textnormal{(}\ensuremath{\alpha}\textnormal{,p)}}}=9.32 keV (\href{https://www.nndc.bnl.gov/nsr/nsrlink.jsp?2015Ki07,B}{2015Ki07}).\vspace{0.1cm}}&\\
&&&&&\parbox[t][0.3cm]{12.12584cm}{\raggedright The evaluator notes that the statistics for this state is very poor (only a few counts),\vspace{0.1cm}}&\\
&&&&&\parbox[t][0.3cm]{12.12584cm}{\raggedright {\ }{\ }{\ }which makes these results unreliable. Therefore, these results are excluded from the\vspace{0.1cm}}&\\
&&&&&\parbox[t][0.3cm]{12.12584cm}{\raggedright {\ }{\ }{\ }\ensuremath{^{\textnormal{18}}}Ne Adopted Levels.\vspace{0.1cm}}&\\
\end{longtable}
\begin{textblock}{29}(0,27.3)
Continued on next page (footnotes at end of table)
\end{textblock}
\clearpage
\begin{longtable}{cccccc@{\extracolsep{\fill}}c}
\\[-.4cm]
\multicolumn{7}{c}{{\bf \small \underline{\ensuremath{^{\textnormal{4}}}He(\ensuremath{^{\textnormal{14}}}O,P),(\ensuremath{^{\textnormal{14}}}O,2p):res\hspace{0.2in}\href{https://www.nndc.bnl.gov/nsr/nsrlink.jsp?1987Wi11,B}{1987Wi11},\href{https://www.nndc.bnl.gov/nsr/nsrlink.jsp?2018Na26,B}{2018Na26} (continued)}}}\\
\multicolumn{7}{c}{~}\\
\multicolumn{7}{c}{\underline{\ensuremath{^{18}}Ne Levels (continued)}}\\
\multicolumn{7}{c}{~}\\
\multicolumn{2}{c}{E(level)$^{}$}&J$^{\pi}$$^{{\hyperlink{NE7LEVEL4}{e}}}$&\multicolumn{2}{c}{\ensuremath{\Gamma}$^{{\hyperlink{NE7LEVEL2}{c}}}$}&Comments&\\[-.2cm]
\multicolumn{2}{c}{\hrulefill}&\hrulefill&\multicolumn{2}{c}{\hrulefill}&\hrulefill&
\endhead
\multicolumn{1}{r@{}}{7.58\ensuremath{\times10^{3}}}&\multicolumn{1}{@{}l}{\ensuremath{^{{\hyperlink{NE7LEVEL1}{b}}}} {\it 2}}&\multicolumn{1}{l}{(0\ensuremath{^{+}})}&\multicolumn{1}{r@{}}{104}&\multicolumn{1}{@{ }l}{keV {\it 26}}&\parbox[t][0.3cm]{11.54804cm}{\raggedright \ensuremath{\Gamma}\ensuremath{\alpha}=1.5 keV \textit{5}; \ensuremath{\Gamma}\ensuremath{_{\textnormal{p}}}=102 keV \textit{26} (\href{https://www.nndc.bnl.gov/nsr/nsrlink.jsp?2015Ki07,B}{2015Ki07})\vspace{0.1cm}}&\\
&&&&&\parbox[t][0.3cm]{11.54804cm}{\raggedright E\ensuremath{_{\textnormal{c.m.}}}(\ensuremath{^{\textnormal{14}}}O+\ensuremath{\alpha})=2.46 MeV (\href{https://www.nndc.bnl.gov/nsr/nsrlink.jsp?2015Ki07,B}{2015Ki07}). Evaluator notes that this state is relatively\vspace{0.1cm}}&\\
&&&&&\parbox[t][0.3cm]{11.54804cm}{\raggedright {\ }{\ }{\ }featureless and suffers from poor statistics.\vspace{0.1cm}}&\\
&&&&&\parbox[t][0.3cm]{11.54804cm}{\raggedright E(level): See also 7600 (\href{https://www.nndc.bnl.gov/nsr/nsrlink.jsp?2004No14,B}{2004No14}); 7660 keV (\href{https://www.nndc.bnl.gov/nsr/nsrlink.jsp?2007Fu09,B}{2007Fu09}, \href{https://www.nndc.bnl.gov/nsr/nsrlink.jsp?2008FuZZ,B}{2008FuZZ}); and 7620\vspace{0.1cm}}&\\
&&&&&\parbox[t][0.3cm]{11.54804cm}{\raggedright {\ }{\ }{\ }keV (\href{https://www.nndc.bnl.gov/nsr/nsrlink.jsp?2004No18,B}{2004No18}, \href{https://www.nndc.bnl.gov/nsr/nsrlink.jsp?2006Ku17,B}{2006Ku17}).\vspace{0.1cm}}&\\
&&&&&\parbox[t][0.3cm]{11.54804cm}{\raggedright \ensuremath{\Gamma}: See also \ensuremath{\Gamma}(FWHM)=0.31 MeV from (\href{https://www.nndc.bnl.gov/nsr/nsrlink.jsp?2008FuZZ,B}{2008FuZZ}).\vspace{0.1cm}}&\\
&&&&&\parbox[t][0.3cm]{11.54804cm}{\raggedright \ensuremath{\omega}\ensuremath{\gamma}\ensuremath{_{\textnormal{(}\ensuremath{\alpha}\textnormal{,p)}}}=1.52 keV (\href{https://www.nndc.bnl.gov/nsr/nsrlink.jsp?2015Ki07,B}{2015Ki07}).\vspace{0.1cm}}&\\
&&&&&\parbox[t][0.3cm]{11.54804cm}{\raggedright (\href{https://www.nndc.bnl.gov/nsr/nsrlink.jsp?2015Ki07,B}{2015Ki07}): this state could be a candidate for the mirror state of the \ensuremath{^{\textnormal{18}}}O(7.796\vspace{0.1cm}}&\\
&&&&&\parbox[t][0.3cm]{11.54804cm}{\raggedright {\ }{\ }{\ }MeV, 0\ensuremath{^{\textnormal{+}}}) state, whose nature is proposed to be 6-particle-4-hole (6p-4h)\vspace{0.1cm}}&\\
&&&&&\parbox[t][0.3cm]{11.54804cm}{\raggedright {\ }{\ }{\ }(\href{https://www.nndc.bnl.gov/nsr/nsrlink.jsp?2011Fo12,B}{2011Fo12}).\vspace{0.1cm}}&\\
&&&&&\parbox[t][0.3cm]{11.54804cm}{\raggedright \ensuremath{\sigma}=0.11 mb (\href{https://www.nndc.bnl.gov/nsr/nsrlink.jsp?2008FuZZ,B}{2008FuZZ}: deduced for 2p decay for the 7660-keV level).\vspace{0.1cm}}&\\
\multicolumn{1}{r@{}}{7.72\ensuremath{\times10^{3}}?}&\multicolumn{1}{@{}l}{\ensuremath{^{{\hyperlink{NE7LEVEL1}{b}}}} {\it 2}}&\multicolumn{1}{l}{(2\ensuremath{^{+}},3\ensuremath{^{-}})}&\multicolumn{1}{r@{}}{281}&\multicolumn{1}{@{ }l}{keV {\it 61}}&\parbox[t][0.3cm]{11.54804cm}{\raggedright \ensuremath{\Gamma}\ensuremath{\alpha}=1.9 keV \textit{3}; \ensuremath{\Gamma}\ensuremath{_{\textnormal{p}}}=279 keV \textit{61} (\href{https://www.nndc.bnl.gov/nsr/nsrlink.jsp?2015Ki07,B}{2015Ki07})\vspace{0.1cm}}&\\
&&&&&\parbox[t][0.3cm]{11.54804cm}{\raggedright E(level): Evaluator notes that this state is too wide to be consistent with the\vspace{0.1cm}}&\\
&&&&&\parbox[t][0.3cm]{11.54804cm}{\raggedright {\ }{\ }{\ }7.7-MeV state observed by others (see the \ensuremath{^{\textnormal{18}}}Ne Adopted Levels). This state, as\vspace{0.1cm}}&\\
&&&&&\parbox[t][0.3cm]{11.54804cm}{\raggedright {\ }{\ }{\ }well as the 8.1-MeV state from (\href{https://www.nndc.bnl.gov/nsr/nsrlink.jsp?2015Ki07,B}{2015Ki07}) may be part of unresolved group of\vspace{0.1cm}}&\\
&&&&&\parbox[t][0.3cm]{11.54804cm}{\raggedright {\ }{\ }{\ }states. This is why these two states and their properties are reported here as\vspace{0.1cm}}&\\
&&&&&\parbox[t][0.3cm]{11.54804cm}{\raggedright {\ }{\ }{\ }tentative.\vspace{0.1cm}}&\\
&&&&&\parbox[t][0.3cm]{11.54804cm}{\raggedright E\ensuremath{_{\textnormal{c.m.}}}(\ensuremath{^{\textnormal{14}}}O+\ensuremath{\alpha})=2.61 MeV (\href{https://www.nndc.bnl.gov/nsr/nsrlink.jsp?2015Ki07,B}{2015Ki07}).\vspace{0.1cm}}&\\
&&&&&\parbox[t][0.3cm]{11.54804cm}{\raggedright \ensuremath{\omega}\ensuremath{\gamma}\ensuremath{_{\textnormal{(}\ensuremath{\alpha}\textnormal{,p)}}}=5.38 keV (\href{https://www.nndc.bnl.gov/nsr/nsrlink.jsp?2015Ki07,B}{2015Ki07}).\vspace{0.1cm}}&\\
\multicolumn{1}{r@{}}{7950}&\multicolumn{1}{@{}l}{}&&&&\parbox[t][0.3cm]{11.54804cm}{\raggedright E(level): From (\href{https://www.nndc.bnl.gov/nsr/nsrlink.jsp?2004No18,B}{2004No18}, \href{https://www.nndc.bnl.gov/nsr/nsrlink.jsp?2006Ku17,B}{2006Ku17}).\vspace{0.1cm}}&\\
\multicolumn{1}{r@{}}{8.10\ensuremath{\times10^{3}}?}&\multicolumn{1}{@{}l}{\ensuremath{^{{\hyperlink{NE7LEVEL1}{b}}}} {\it 10}}&\multicolumn{1}{l}{(0\ensuremath{^{+}})}&\multicolumn{1}{r@{}}{338}&\multicolumn{1}{@{ }l}{keV {\it 38}}&\parbox[t][0.3cm]{11.54804cm}{\raggedright \ensuremath{\Gamma}\ensuremath{\alpha}=40.4 keV \textit{49}; \ensuremath{\Gamma}\ensuremath{_{\textnormal{p}}}=298 keV \textit{38} (\href{https://www.nndc.bnl.gov/nsr/nsrlink.jsp?2015Ki07,B}{2015Ki07})\vspace{0.1cm}}&\\
&&&&&\parbox[t][0.3cm]{11.54804cm}{\raggedright E(level): See also 8100 keV (\href{https://www.nndc.bnl.gov/nsr/nsrlink.jsp?2004No18,B}{2004No18}, \href{https://www.nndc.bnl.gov/nsr/nsrlink.jsp?2006Ku17,B}{2006Ku17}). Evaluator notes that this level\vspace{0.1cm}}&\\
&&&&&\parbox[t][0.3cm]{11.54804cm}{\raggedright {\ }{\ }{\ }is most likely an unresolved state, which is comprised of more than one level.\vspace{0.1cm}}&\\
&&&&&\parbox[t][0.3cm]{11.54804cm}{\raggedright {\ }{\ }{\ }(\href{https://www.nndc.bnl.gov/nsr/nsrlink.jsp?1981Ne09,B}{1981Ne09}) observed a triplet at E\ensuremath{_{\textnormal{x}}}\ensuremath{\sim}8 MeV via \ensuremath{^{\textnormal{16}}}O(\ensuremath{^{\textnormal{3}}}He,n). Those authors did\vspace{0.1cm}}&\\
&&&&&\parbox[t][0.3cm]{11.54804cm}{\raggedright {\ }{\ }{\ }not specifically mention which states were part of the triplet. The evaluator\vspace{0.1cm}}&\\
&&&&&\parbox[t][0.3cm]{11.54804cm}{\raggedright {\ }{\ }{\ }estimated that the member states are the 7712-, 7915-, and 8100-keV levels as\vspace{0.1cm}}&\\
&&&&&\parbox[t][0.3cm]{11.54804cm}{\raggedright {\ }{\ }{\ }reported by (\href{https://www.nndc.bnl.gov/nsr/nsrlink.jsp?1981Ne09,B}{1981Ne09}), see the discussion in the \ensuremath{^{\textnormal{16}}}O(\ensuremath{^{\textnormal{3}}}He,n) dataset. There is yet\vspace{0.1cm}}&\\
&&&&&\parbox[t][0.3cm]{11.54804cm}{\raggedright {\ }{\ }{\ }another state at 7949 keV as reported by (\href{https://www.nndc.bnl.gov/nsr/nsrlink.jsp?1981Ne09,B}{1981Ne09}) but this state was not\vspace{0.1cm}}&\\
&&&&&\parbox[t][0.3cm]{11.54804cm}{\raggedright {\ }{\ }{\ }populated via \ensuremath{^{\textnormal{16}}}O(\ensuremath{^{\textnormal{3}}}He,n). Instead, \ensuremath{^{\textnormal{20}}}Ne(p,t) was used to populate it, and thus it\vspace{0.1cm}}&\\
&&&&&\parbox[t][0.3cm]{11.54804cm}{\raggedright {\ }{\ }{\ }is not part of this triplet. The three states constructing the triplet mentioned above\vspace{0.1cm}}&\\
&&&&&\parbox[t][0.3cm]{11.54804cm}{\raggedright {\ }{\ }{\ }were all assigned widths of up to \ensuremath{\sim}50 keV by (\href{https://www.nndc.bnl.gov/nsr/nsrlink.jsp?1981Ne09,B}{1981Ne09}). The 7949-keV state is\vspace{0.1cm}}&\\
&&&&&\parbox[t][0.3cm]{11.54804cm}{\raggedright {\ }{\ }{\ }presumed to be comparable in width but slightly wider. The 8.1-MeV state\vspace{0.1cm}}&\\
&&&&&\parbox[t][0.3cm]{11.54804cm}{\raggedright {\ }{\ }{\ }reported by (\href{https://www.nndc.bnl.gov/nsr/nsrlink.jsp?2015Ki07,B}{2015Ki07}) is about 4 times as wide as each of the above mentioned\vspace{0.1cm}}&\\
&&&&&\parbox[t][0.3cm]{11.54804cm}{\raggedright {\ }{\ }{\ }states and is somewhat featureless (see Fig. 11 in (\href{https://www.nndc.bnl.gov/nsr/nsrlink.jsp?2015Ki07,B}{2015Ki07})). Therefore, it is\vspace{0.1cm}}&\\
&&&&&\parbox[t][0.3cm]{11.54804cm}{\raggedright {\ }{\ }{\ }conceivable to think that the 8.1-MeV state could be an unresolved peak, which is\vspace{0.1cm}}&\\
&&&&&\parbox[t][0.3cm]{11.54804cm}{\raggedright {\ }{\ }{\ }consisted of more than one peak.\vspace{0.1cm}}&\\
&&&&&\parbox[t][0.3cm]{11.54804cm}{\raggedright E\ensuremath{_{\textnormal{c.m.}}}(\ensuremath{^{\textnormal{14}}}O+\ensuremath{\alpha})=2.99 MeV (\href{https://www.nndc.bnl.gov/nsr/nsrlink.jsp?2015Ki07,B}{2015Ki07}).\vspace{0.1cm}}&\\
&&&&&\parbox[t][0.3cm]{11.54804cm}{\raggedright J\ensuremath{^{\pi}}: The 0\ensuremath{^{\textnormal{+}}} assignment in (\href{https://www.nndc.bnl.gov/nsr/nsrlink.jsp?2015Ki07,B}{2015Ki07}) yielded the best \ensuremath{\chi}\ensuremath{^{\textnormal{2}}} for fitting both the\vspace{0.1cm}}&\\
&&&&&\parbox[t][0.3cm]{11.54804cm}{\raggedright {\ }{\ }{\ }E\ensuremath{_{\textnormal{x}}}\ensuremath{<}8.2 MeV and E\ensuremath{_{\textnormal{x}}}\ensuremath{>}8.2 MeV regions.\vspace{0.1cm}}&\\
&&&&&\parbox[t][0.3cm]{11.54804cm}{\raggedright \ensuremath{\omega}\ensuremath{\gamma}\ensuremath{_{\textnormal{(}\ensuremath{\alpha}\textnormal{,p)}}}=35.0 keV (\href{https://www.nndc.bnl.gov/nsr/nsrlink.jsp?2015Ki07,B}{2015Ki07}) assuming a single state.\vspace{0.1cm}}&\\
\multicolumn{1}{r@{}}{8300}&\multicolumn{1}{@{}l}{}&&&&\parbox[t][0.3cm]{11.54804cm}{\raggedright E(level): From (\href{https://www.nndc.bnl.gov/nsr/nsrlink.jsp?2004No18,B}{2004No18}, \href{https://www.nndc.bnl.gov/nsr/nsrlink.jsp?2006Ku17,B}{2006Ku17}).\vspace{0.1cm}}&\\
\multicolumn{1}{r@{}}{8.50\ensuremath{\times10^{3}}}&\multicolumn{1}{@{}l}{\ensuremath{^{{\hyperlink{NE7LEVEL1}{b}}}} {\it 10}}&\multicolumn{1}{l}{(1\ensuremath{^{-}},2\ensuremath{^{+}})}&\multicolumn{1}{r@{}}{630}&\multicolumn{1}{@{ }l}{keV {\it 64}}&\parbox[t][0.3cm]{11.54804cm}{\raggedright \ensuremath{\Gamma}\ensuremath{\alpha}=84.5 keV \textit{210}; \ensuremath{\Gamma}\ensuremath{_{\textnormal{p}}}=546 keV \textit{60} (\href{https://www.nndc.bnl.gov/nsr/nsrlink.jsp?2015Ki07,B}{2015Ki07})\vspace{0.1cm}}&\\
&&&&&\parbox[t][0.3cm]{11.54804cm}{\raggedright E(level): See also 8450 keV (\href{https://www.nndc.bnl.gov/nsr/nsrlink.jsp?2007Fu09,B}{2007Fu09}, \href{https://www.nndc.bnl.gov/nsr/nsrlink.jsp?2008FuZZ,B}{2008FuZZ}) with \ensuremath{\Gamma}(FWHM)=0.47 MeV\vspace{0.1cm}}&\\
&&&&&\parbox[t][0.3cm]{11.54804cm}{\raggedright {\ }{\ }{\ }(\href{https://www.nndc.bnl.gov/nsr/nsrlink.jsp?2008FuZZ,B}{2008FuZZ}).\vspace{0.1cm}}&\\
&&&&&\parbox[t][0.3cm]{11.54804cm}{\raggedright E\ensuremath{_{\textnormal{c.m.}}}(\ensuremath{^{\textnormal{14}}}O+\ensuremath{\alpha})=3.39 MeV (\href{https://www.nndc.bnl.gov/nsr/nsrlink.jsp?2015Ki07,B}{2015Ki07}).\vspace{0.1cm}}&\\
&&&&&\parbox[t][0.3cm]{11.54804cm}{\raggedright \ensuremath{\sigma}=0.73 mb (\href{https://www.nndc.bnl.gov/nsr/nsrlink.jsp?2008FuZZ,B}{2008FuZZ}) deduced for 2p decay of the state at 8450 keV.\vspace{0.1cm}}&\\
&&&&&\parbox[t][0.3cm]{11.54804cm}{\raggedright (\href{https://www.nndc.bnl.gov/nsr/nsrlink.jsp?2007Fu09,B}{2007Fu09}, \href{https://www.nndc.bnl.gov/nsr/nsrlink.jsp?2008FuZZ,B}{2008FuZZ}): evidence suggests that the state at 8450 keV decays to\vspace{0.1cm}}&\\
&&&&&\parbox[t][0.3cm]{11.54804cm}{\raggedright {\ }{\ }{\ }\ensuremath{^{\textnormal{16}}}O\ensuremath{_{\textnormal{g.s.}}} via \ensuremath{^{\textnormal{2}}}He emission. This is supported by a predicted decay spectrum using\vspace{0.1cm}}&\\
&&&&&\parbox[t][0.3cm]{11.54804cm}{\raggedright {\ }{\ }{\ }the Faddeev approach. The angular momenta of L=0,1,2,3 were assumed for the\vspace{0.1cm}}&\\
&&&&&\parbox[t][0.3cm]{11.54804cm}{\raggedright {\ }{\ }{\ }\ensuremath{^{\textnormal{2}}}He relative to the \ensuremath{^{\textnormal{16}}}O core when calculating the Faddeev model. The\vspace{0.1cm}}&\\
&&&&&\parbox[t][0.3cm]{11.54804cm}{\raggedright {\ }{\ }{\ }experimentally measured relative energy between the two protons emitted from the\vspace{0.1cm}}&\\
&&&&&\parbox[t][0.3cm]{11.54804cm}{\raggedright {\ }{\ }{\ }level at 8.45 MeV in \ensuremath{^{\textnormal{18}}}Ne was most consistent with L=3 (see Fig. 6 of\vspace{0.1cm}}&\\
&&&&&\parbox[t][0.3cm]{11.54804cm}{\raggedright {\ }{\ }{\ }(\href{https://www.nndc.bnl.gov/nsr/nsrlink.jsp?2008FuZZ,B}{2008FuZZ})).\vspace{0.1cm}}&\\
\end{longtable}
\begin{textblock}{29}(0,27.3)
Continued on next page (footnotes at end of table)
\end{textblock}
\clearpage
\begin{longtable}{cccccc@{\extracolsep{\fill}}c}
\\[-.4cm]
\multicolumn{7}{c}{{\bf \small \underline{\ensuremath{^{\textnormal{4}}}He(\ensuremath{^{\textnormal{14}}}O,P),(\ensuremath{^{\textnormal{14}}}O,2p):res\hspace{0.2in}\href{https://www.nndc.bnl.gov/nsr/nsrlink.jsp?1987Wi11,B}{1987Wi11},\href{https://www.nndc.bnl.gov/nsr/nsrlink.jsp?2018Na26,B}{2018Na26} (continued)}}}\\
\multicolumn{7}{c}{~}\\
\multicolumn{7}{c}{\underline{\ensuremath{^{18}}Ne Levels (continued)}}\\
\multicolumn{7}{c}{~}\\
\multicolumn{2}{c}{E(level)$^{}$}&J$^{\pi}$$^{{\hyperlink{NE7LEVEL4}{e}}}$&\multicolumn{2}{c}{\ensuremath{\Gamma}$^{{\hyperlink{NE7LEVEL2}{c}}}$}&Comments&\\[-.2cm]
\multicolumn{2}{c}{\hrulefill}&\hrulefill&\multicolumn{2}{c}{\hrulefill}&\hrulefill&
\endhead
\multicolumn{1}{r@{}}{9.14\ensuremath{\times10^{3}}}&\multicolumn{1}{@{}l}{\ensuremath{^{{\hyperlink{NE7LEVEL1}{b}}}} {\it 10}}&\multicolumn{1}{l}{3\ensuremath{^{-}}}&\multicolumn{1}{r@{}}{494}&\multicolumn{1}{@{ }l}{keV {\it 59}}&\parbox[t][0.3cm]{11.09864cm}{\raggedright \ensuremath{\Gamma}\ensuremath{\alpha}=27.0 keV \textit{60}; \ensuremath{\Gamma}\ensuremath{_{\textnormal{p}}}=467.5 keV \textit{590} (\href{https://www.nndc.bnl.gov/nsr/nsrlink.jsp?2015Ki07,B}{2015Ki07})\vspace{0.1cm}}&\\
&&&&&\parbox[t][0.3cm]{11.09864cm}{\raggedright E\ensuremath{_{\textnormal{c.m.}}}(\ensuremath{^{\textnormal{14}}}O+\ensuremath{\alpha})=4.03 MeV (\href{https://www.nndc.bnl.gov/nsr/nsrlink.jsp?2015Ki07,B}{2015Ki07}). This study pairs the 9.14 MeV \textit{10} state\vspace{0.1cm}}&\\
&&&&&\parbox[t][0.3cm]{11.09864cm}{\raggedright {\ }{\ }{\ }with the 9.2-MeV state observed in the (p,t) measurements.\vspace{0.1cm}}&\\
&&&&&\parbox[t][0.3cm]{11.09864cm}{\raggedright E(level): A state was observed in (\href{https://www.nndc.bnl.gov/nsr/nsrlink.jsp?2007Fu09,B}{2007Fu09}) with E\ensuremath{_{\textnormal{x}}}=9.4 MeV, whose width\vspace{0.1cm}}&\\
&&&&&\parbox[t][0.3cm]{11.09864cm}{\raggedright {\ }{\ }{\ }was deduced as \ensuremath{\Gamma}(FWHM)=0.42 MeV (\href{https://www.nndc.bnl.gov/nsr/nsrlink.jsp?2008FuZZ,B}{2008FuZZ}). (\href{https://www.nndc.bnl.gov/nsr/nsrlink.jsp?2007Fu09,B}{2007Fu09}) suggests that\vspace{0.1cm}}&\\
&&&&&\parbox[t][0.3cm]{11.09864cm}{\raggedright {\ }{\ }{\ }this state is likely to be the 9.2-MeV state observed in the (p,t) measurements.\vspace{0.1cm}}&\\
&&&&&\parbox[t][0.3cm]{11.09864cm}{\raggedright {\ }{\ }{\ }Evidence suggests that the 9.4-MeV state proton decays to \ensuremath{^{\textnormal{17}}}F*(3.1 MeV) and\vspace{0.1cm}}&\\
&&&&&\parbox[t][0.3cm]{11.09864cm}{\raggedright {\ }{\ }{\ }to \ensuremath{^{\textnormal{17}}}F*(3.86 MeV), see Fig. 5 of (\href{https://www.nndc.bnl.gov/nsr/nsrlink.jsp?2008FuZZ,B}{2008FuZZ}). \ensuremath{\sigma}=9.92 mb was deduced in\vspace{0.1cm}}&\\
&&&&&\parbox[t][0.3cm]{11.09864cm}{\raggedright {\ }{\ }{\ }(\href{https://www.nndc.bnl.gov/nsr/nsrlink.jsp?2008FuZZ,B}{2008FuZZ}) for the 2p decay of the 9.4-MeV level.\vspace{0.1cm}}&\\
\multicolumn{1}{r@{}}{9.60\ensuremath{\times10^{3}}}&\multicolumn{1}{@{}l}{\ensuremath{^{{\hyperlink{NE7LEVEL1}{b}}}} {\it 9}}&\multicolumn{1}{l}{(1\ensuremath{^{-}},0\ensuremath{^{+}})}&\multicolumn{1}{r@{}}{628}&\multicolumn{1}{@{ }l}{MeV {\it 104}}&\parbox[t][0.3cm]{11.09864cm}{\raggedright \ensuremath{\Gamma}\ensuremath{\alpha}=620 keV \textit{104}; \ensuremath{\Gamma}\ensuremath{_{\textnormal{p}}}=7.8 keV \textit{66} (\href{https://www.nndc.bnl.gov/nsr/nsrlink.jsp?2015Ki07,B}{2015Ki07})\vspace{0.1cm}}&\\
&&&&&\parbox[t][0.3cm]{11.09864cm}{\raggedright E\ensuremath{_{\textnormal{c.m.}}}(\ensuremath{^{\textnormal{14}}}O+\ensuremath{\alpha})=4.49 MeV (\href{https://www.nndc.bnl.gov/nsr/nsrlink.jsp?2015Ki07,B}{2015Ki07}).\vspace{0.1cm}}&\\
&&&&&\parbox[t][0.3cm]{11.09864cm}{\raggedright (\href{https://www.nndc.bnl.gov/nsr/nsrlink.jsp?2015Ki07,B}{2015Ki07}): this state may be a superposition of two or more states.\vspace{0.1cm}}&\\
&&&&&\parbox[t][0.3cm]{11.09864cm}{\raggedright E(level): (\href{https://www.nndc.bnl.gov/nsr/nsrlink.jsp?2015Ki07,B}{2015Ki07}) claims that this state was populated, for the first time, in\vspace{0.1cm}}&\\
&&&&&\parbox[t][0.3cm]{11.09864cm}{\raggedright {\ }{\ }{\ }their experiment. Note that (\href{https://www.nndc.bnl.gov/nsr/nsrlink.jsp?1996Ha26,B}{1996Ha26}) observed a state at 9580 keV \textit{20} using\vspace{0.1cm}}&\\
&&&&&\parbox[t][0.3cm]{11.09864cm}{\raggedright {\ }{\ }{\ }the \ensuremath{^{\textnormal{12}}}C(\ensuremath{^{\textnormal{12}}}C,\ensuremath{^{\textnormal{6}}}He) reaction, but the J\ensuremath{^{\ensuremath{\pi}}} assignment was not deduced by\vspace{0.1cm}}&\\
&&&&&\parbox[t][0.3cm]{11.09864cm}{\raggedright {\ }{\ }{\ }(\href{https://www.nndc.bnl.gov/nsr/nsrlink.jsp?1996Ha26,B}{1996Ha26}).\vspace{0.1cm}}&\\
\multicolumn{1}{r@{}}{10.07\ensuremath{\times10^{3}}}&\multicolumn{1}{@{}l}{\ensuremath{^{{\hyperlink{NE7LEVEL1}{b}}}} {\it 8}}&\multicolumn{1}{l}{(2\ensuremath{^{+}},1\ensuremath{^{-}})}&\multicolumn{1}{r@{}}{92}&\multicolumn{1}{@{ }l}{keV {\it 18}}&\parbox[t][0.3cm]{11.09864cm}{\raggedright \ensuremath{\Gamma}\ensuremath{\alpha}=1.5 keV \textit{8}; \ensuremath{\Gamma}\ensuremath{_{\textnormal{p}}}=90 keV \textit{18} (\href{https://www.nndc.bnl.gov/nsr/nsrlink.jsp?2015Ki07,B}{2015Ki07})\vspace{0.1cm}}&\\
&&&&&\parbox[t][0.3cm]{11.09864cm}{\raggedright E(level): This state was observed for the first time in (\href{https://www.nndc.bnl.gov/nsr/nsrlink.jsp?2015Ki07,B}{2015Ki07}). See also 10.12\vspace{0.1cm}}&\\
&&&&&\parbox[t][0.3cm]{11.09864cm}{\raggedright {\ }{\ }{\ }MeV (\href{https://www.nndc.bnl.gov/nsr/nsrlink.jsp?2007Fu09,B}{2007Fu09}, \href{https://www.nndc.bnl.gov/nsr/nsrlink.jsp?2008FuZZ,B}{2008FuZZ}: \ensuremath{\Gamma}(FWHM)=0.54 MeV) observed for the first\vspace{0.1cm}}&\\
&&&&&\parbox[t][0.3cm]{11.09864cm}{\raggedright {\ }{\ }{\ }time in (\href{https://www.nndc.bnl.gov/nsr/nsrlink.jsp?2007Fu09,B}{2007Fu09}).\vspace{0.1cm}}&\\
&&&&&\parbox[t][0.3cm]{11.09864cm}{\raggedright E\ensuremath{_{\textnormal{c.m.}}}(\ensuremath{^{\textnormal{14}}}O+\ensuremath{\alpha})=4.96 MeV (\href{https://www.nndc.bnl.gov/nsr/nsrlink.jsp?2015Ki07,B}{2015Ki07}).\vspace{0.1cm}}&\\
&&&&&\parbox[t][0.3cm]{11.09864cm}{\raggedright J\ensuremath{^{\pi}}: The R-matrix analysis of (\href{https://www.nndc.bnl.gov/nsr/nsrlink.jsp?2015Ki07,B}{2015Ki07}) with J\ensuremath{^{\ensuremath{\pi}}}=2\ensuremath{^{\textnormal{+}}} yields a slightly better \ensuremath{\chi}\ensuremath{^{\textnormal{2}}}\vspace{0.1cm}}&\\
&&&&&\parbox[t][0.3cm]{11.09864cm}{\raggedright {\ }{\ }{\ }for the fit.\vspace{0.1cm}}&\\
&&&&&\parbox[t][0.3cm]{11.09864cm}{\raggedright \ensuremath{\sigma}=29.2 mb (\href{https://www.nndc.bnl.gov/nsr/nsrlink.jsp?2008FuZZ,B}{2008FuZZ}) for the 2p decay of the state at 10.12 MeV.\vspace{0.1cm}}&\\
&&&&&\parbox[t][0.3cm]{11.09864cm}{\raggedright (\href{https://www.nndc.bnl.gov/nsr/nsrlink.jsp?2007Fu09,B}{2007Fu09}, \href{https://www.nndc.bnl.gov/nsr/nsrlink.jsp?2008FuZZ,B}{2008FuZZ}): evidence suggests that the state at 10.12 MeV proton\vspace{0.1cm}}&\\
&&&&&\parbox[t][0.3cm]{11.09864cm}{\raggedright {\ }{\ }{\ }decays to \ensuremath{^{\textnormal{17}}}F*(3.1 MeV) and to \ensuremath{^{\textnormal{17}}}F*(3.86 MeV), see Fig. 5 of (\href{https://www.nndc.bnl.gov/nsr/nsrlink.jsp?2008FuZZ,B}{2008FuZZ}).\vspace{0.1cm}}&\\
\multicolumn{1}{r@{}}{10660}&\multicolumn{1}{@{}l}{\ensuremath{^{{\hyperlink{NE7LEVEL3}{d}}}}}&&\multicolumn{1}{r@{}}{0}&\multicolumn{1}{@{.}l}{73 MeV}&\parbox[t][0.3cm]{11.09864cm}{\raggedright \ensuremath{\Gamma}: From \ensuremath{\Gamma}(FWHM) obtained by (\href{https://www.nndc.bnl.gov/nsr/nsrlink.jsp?2008FuZZ,B}{2008FuZZ}).\vspace{0.1cm}}&\\
&&&&&\parbox[t][0.3cm]{11.09864cm}{\raggedright \ensuremath{\sigma}=25.9 mb (\href{https://www.nndc.bnl.gov/nsr/nsrlink.jsp?2008FuZZ,B}{2008FuZZ}) for the 2p decay of this state.\vspace{0.1cm}}&\\
&&&&&\parbox[t][0.3cm]{11.09864cm}{\raggedright (\href{https://www.nndc.bnl.gov/nsr/nsrlink.jsp?2007Fu09,B}{2007Fu09}, \href{https://www.nndc.bnl.gov/nsr/nsrlink.jsp?2008FuZZ,B}{2008FuZZ}): evidence suggests that this state proton decays to\vspace{0.1cm}}&\\
&&&&&\parbox[t][0.3cm]{11.09864cm}{\raggedright {\ }{\ }{\ }\ensuremath{^{\textnormal{17}}}F*(3.1 MeV) and to \ensuremath{^{\textnormal{17}}}F*(3.86 MeV) with a weaker branch, see Fig. 5 of\vspace{0.1cm}}&\\
&&&&&\parbox[t][0.3cm]{11.09864cm}{\raggedright {\ }{\ }{\ }(\href{https://www.nndc.bnl.gov/nsr/nsrlink.jsp?2008FuZZ,B}{2008FuZZ}).\vspace{0.1cm}}&\\
\multicolumn{1}{r@{}}{11290}&\multicolumn{1}{@{}l}{\ensuremath{^{{\hyperlink{NE7LEVEL3}{d}}}}}&&\multicolumn{1}{r@{}}{0}&\multicolumn{1}{@{.}l}{66 MeV}&\parbox[t][0.3cm]{11.09864cm}{\raggedright \ensuremath{\Gamma}: From \ensuremath{\Gamma}(FWHM) obtained by (\href{https://www.nndc.bnl.gov/nsr/nsrlink.jsp?2008FuZZ,B}{2008FuZZ}).\vspace{0.1cm}}&\\
&&&&&\parbox[t][0.3cm]{11.09864cm}{\raggedright \ensuremath{\sigma}=49.9 mb (\href{https://www.nndc.bnl.gov/nsr/nsrlink.jsp?2008FuZZ,B}{2008FuZZ}) for the 2p decay of this state.\vspace{0.1cm}}&\\
&&&&&\parbox[t][0.3cm]{11.09864cm}{\raggedright (\href{https://www.nndc.bnl.gov/nsr/nsrlink.jsp?2007Fu09,B}{2007Fu09}, \href{https://www.nndc.bnl.gov/nsr/nsrlink.jsp?2008FuZZ,B}{2008FuZZ}): evidence suggests that this state proton decays to\vspace{0.1cm}}&\\
&&&&&\parbox[t][0.3cm]{11.09864cm}{\raggedright {\ }{\ }{\ }\ensuremath{^{\textnormal{17}}}F*(3.1 MeV), \ensuremath{^{\textnormal{17}}}F*(3.86 MeV) and \ensuremath{^{\textnormal{17}}}F*(5.22 MeV), see Fig. 5 of\vspace{0.1cm}}&\\
&&&&&\parbox[t][0.3cm]{11.09864cm}{\raggedright {\ }{\ }{\ }(\href{https://www.nndc.bnl.gov/nsr/nsrlink.jsp?2008FuZZ,B}{2008FuZZ}).\vspace{0.1cm}}&\\
\multicolumn{1}{r@{}}{11800}&\multicolumn{1}{@{}l}{\ensuremath{^{{\hyperlink{NE7LEVEL3}{d}}}}}&&\multicolumn{1}{r@{}}{0}&\multicolumn{1}{@{.}l}{52 MeV}&\parbox[t][0.3cm]{11.09864cm}{\raggedright \ensuremath{\Gamma}: From \ensuremath{\Gamma}(FWHM) obtained by (\href{https://www.nndc.bnl.gov/nsr/nsrlink.jsp?2008FuZZ,B}{2008FuZZ}).\vspace{0.1cm}}&\\
&&&&&\parbox[t][0.3cm]{11.09864cm}{\raggedright \ensuremath{\sigma}=42.4 mb (\href{https://www.nndc.bnl.gov/nsr/nsrlink.jsp?2008FuZZ,B}{2008FuZZ}) for the 2p decay of this level.\vspace{0.1cm}}&\\
\end{longtable}
\parbox[b][0.3cm]{17.7cm}{\makebox[1ex]{\ensuremath{^{\hypertarget{NE7LEVEL0}{a}}}} (\href{https://www.nndc.bnl.gov/nsr/nsrlink.jsp?2004No14,B}{2004No14}, \href{https://www.nndc.bnl.gov/nsr/nsrlink.jsp?2004No18,B}{2004No18}): these studies observed a weak, previously unobserved transition at E\ensuremath{_{\textnormal{c.m.}}}(\ensuremath{^{\textnormal{14}}}O+\ensuremath{\alpha})=1.5 MeV,}\\
\parbox[b][0.3cm]{17.7cm}{{\ }{\ }corresponding to E\ensuremath{_{\textnormal{c.m.}}}(\ensuremath{^{\textnormal{17}}}F+p)=2.7 MeV (\href{https://www.nndc.bnl.gov/nsr/nsrlink.jsp?2006Ku17,B}{2006Ku17}). Since the \ensuremath{^{\textnormal{17}}}F+p elastic scattering measurement of (\href{https://www.nndc.bnl.gov/nsr/nsrlink.jsp?2001Bl06,B}{2001Bl06}) did not}\\
\parbox[b][0.3cm]{17.7cm}{{\ }{\ }reveal any \ensuremath{^{\textnormal{18}}}Ne state with a large proton width, and because the \ensuremath{^{\textnormal{18}}}Ne states in this energy region (near the \ensuremath{\alpha}-threshold) cannot}\\
\parbox[b][0.3cm]{17.7cm}{{\ }{\ }have a large \ensuremath{\alpha}-width, (\href{https://www.nndc.bnl.gov/nsr/nsrlink.jsp?2006Ku17,B}{2006Ku17}) ruled out the possibility that the observed transition at E\ensuremath{_{\textnormal{c.m.}}}(\ensuremath{^{\textnormal{14}}}O+\ensuremath{\alpha})=1.5 MeV could be a}\\
\parbox[b][0.3cm]{17.7cm}{{\ }{\ }\ensuremath{^{\textnormal{18}}}Ne resonance. Instead, (\href{https://www.nndc.bnl.gov/nsr/nsrlink.jsp?2004No14,B}{2004No14}, \href{https://www.nndc.bnl.gov/nsr/nsrlink.jsp?2006Ku17,B}{2006Ku17}) considered this transition to be an evidence supporting the proton decay of the}\\
\parbox[b][0.3cm]{17.7cm}{{\ }{\ }\ensuremath{^{\textnormal{18}}}Ne state around 7.1 MeV to the \ensuremath{^{\textnormal{17}}}F*(495 keV), which enhanced the deduced \ensuremath{^{\textnormal{14}}}O(\ensuremath{\alpha},p) reaction rate by 50\% around 2 GK}\\
\parbox[b][0.3cm]{17.7cm}{{\ }{\ }(\href{https://www.nndc.bnl.gov/nsr/nsrlink.jsp?2004No14,B}{2004No14}). However, (\href{https://www.nndc.bnl.gov/nsr/nsrlink.jsp?2010Ba21,B}{2010Ba21}) did not see any evidence of inelastic \ensuremath{^{\textnormal{17}}}F+p scattering in this energy region, and the decay}\\
\parbox[b][0.3cm]{17.7cm}{{\ }{\ }branching ratio of \ensuremath{\Gamma}\ensuremath{_{\textnormal{p$'$}}}/\ensuremath{\Gamma}\ensuremath{_{\textnormal{p}}} was constrained to be less than 0.03. (\href{https://www.nndc.bnl.gov/nsr/nsrlink.jsp?2014Hu16,B}{2014Hu16}) observed a new state at 6.85 MeV, which may be}\\
\parbox[b][0.3cm]{17.7cm}{{\ }{\ }what (\href{https://www.nndc.bnl.gov/nsr/nsrlink.jsp?2004No14,B}{2004No14}, \href{https://www.nndc.bnl.gov/nsr/nsrlink.jsp?2004No18,B}{2004No18}) had observed at E\ensuremath{_{\textnormal{c.m.}}}(\ensuremath{^{\textnormal{14}}}O+\ensuremath{\alpha})=1.5 MeV.}\\
\parbox[b][0.3cm]{17.7cm}{\makebox[1ex]{\ensuremath{^{\hypertarget{NE7LEVEL1}{b}}}} From (\href{https://www.nndc.bnl.gov/nsr/nsrlink.jsp?2015Ki07,B}{2015Ki07}).}\\
\parbox[b][0.3cm]{17.7cm}{\makebox[1ex]{\ensuremath{^{\hypertarget{NE7LEVEL2}{c}}}} \ensuremath{\Gamma}=\ensuremath{\Gamma}\ensuremath{_{\ensuremath{\alpha}}}+\ensuremath{\Gamma}\ensuremath{_{\textnormal{p}}} deduced from the R-matrix analysis of (\href{https://www.nndc.bnl.gov/nsr/nsrlink.jsp?2015Ki07,B}{2015Ki07}) unless noted otherwise.}\\
\parbox[b][0.3cm]{17.7cm}{\makebox[1ex]{\ensuremath{^{\hypertarget{NE7LEVEL3}{d}}}} This state was observed for the first time in (\href{https://www.nndc.bnl.gov/nsr/nsrlink.jsp?2007Fu09,B}{2007Fu09}).}\\
\parbox[b][0.3cm]{17.7cm}{\makebox[1ex]{\ensuremath{^{\hypertarget{NE7LEVEL4}{e}}}} From the R-matrix analysis of (\href{https://www.nndc.bnl.gov/nsr/nsrlink.jsp?2015Ki07,B}{2015Ki07}).}\\
\vspace{0.5cm}
\clearpage
\subsection[\hspace{-0.2cm}\ensuremath{^{\textnormal{4}}}He(\ensuremath{^{\textnormal{14}}}O,\ensuremath{\alpha}):res]{ }
\vspace{-27pt}
\vspace{0.3cm}
\hypertarget{NE8}{{\bf \small \underline{\ensuremath{^{\textnormal{4}}}He(\ensuremath{^{\textnormal{14}}}O,\ensuremath{\alpha}):res\hspace{0.2in}\href{https://www.nndc.bnl.gov/nsr/nsrlink.jsp?2008Fu07,B}{2008Fu07},\href{https://www.nndc.bnl.gov/nsr/nsrlink.jsp?2022Ba39,B}{2022Ba39}}}}\\
\vspace{4pt}
\vspace{8pt}
\parbox[b][0.3cm]{17.7cm}{\addtolength{\parindent}{-0.2in}\href{https://www.nndc.bnl.gov/nsr/nsrlink.jsp?2008Fu07,B}{2008Fu07}: \ensuremath{^{\textnormal{4}}}He(\ensuremath{^{\textnormal{14}}}O,\ensuremath{\alpha}) E=33, 42, and 57 MeV; measured the \ensuremath{^{\textnormal{14}}}O+\ensuremath{\alpha} elastic scattering excitation function using the MARS recoil}\\
\parbox[b][0.3cm]{17.7cm}{spectrometer. Measured the reaction products using 4 large position sensitive Si detectors. Energy resolution was \ensuremath{\sim}40 keV in the}\\
\parbox[b][0.3cm]{17.7cm}{center-of-mass frame. Measured angular distributions of the \ensuremath{\alpha}-particles. The measured \ensuremath{^{\textnormal{14}}}O+\ensuremath{\alpha} excitation function was analyzed}\\
\parbox[b][0.3cm]{17.7cm}{using R-matrix and shows \ensuremath{\alpha}-cluster structure in \ensuremath{^{\textnormal{18}}}Ne. Numerous \ensuremath{^{\textnormal{18}}}Ne resonances with E\ensuremath{_{\textnormal{c.m.}}}=3.6-11.4 MeV were measured.}\\
\parbox[b][0.3cm]{17.7cm}{Authors concluded that there is a large increase in the radius of the \ensuremath{\alpha}-cluster states in \ensuremath{^{\textnormal{18}}}Ne. Comparisons between mirror levels in}\\
\parbox[b][0.3cm]{17.7cm}{\ensuremath{^{\textnormal{18}}}O and \ensuremath{^{\textnormal{18}}}Ne are discussed.}\\
\parbox[b][0.3cm]{17.7cm}{\addtolength{\parindent}{-0.2in}\href{https://www.nndc.bnl.gov/nsr/nsrlink.jsp?2010Ha15,B}{2010Ha15}: \ensuremath{^{\textnormal{4}}}He(\ensuremath{^{\textnormal{14}}}O,\ensuremath{\alpha}) E=24 and 35 MeV; measured the excitation function of the \ensuremath{^{\textnormal{14}}}O(\ensuremath{\alpha},\ensuremath{\alpha}) resonant elastic scattering and \ensuremath{\sigma}(\ensuremath{\theta})}\\
\parbox[b][0.3cm]{17.7cm}{(in inverse kinematics) using the thick target method. Energy range: E\ensuremath{_{\textnormal{c.m.}}}=2.1\ensuremath{\sim}8.0 MeV. The \ensuremath{\alpha}-particles were measured using a}\\
\parbox[b][0.3cm]{17.7cm}{set of \ensuremath{\Delta}E-E telescopes consisting of position sensitive Si detectors and Si surface barrier detectors covering \ensuremath{\theta}\ensuremath{_{\textnormal{lab}}}=0\ensuremath{^\circ}{\textminus}70\ensuremath{^\circ}. Only the}\\
\parbox[b][0.3cm]{17.7cm}{data collected by the telescope at 0\ensuremath{^\circ} were used in this study. Energy resolution was 40 keV. A new state was observed at}\\
\parbox[b][0.3cm]{17.7cm}{E\ensuremath{_{\textnormal{c.m.}}}=6900 keV. Deduced the \ensuremath{^{\textnormal{18}}}Ne resonance parameters for 3 resonances using an R-matrix analysis. The results are in}\\
\parbox[b][0.3cm]{17.7cm}{agreement with the previous results obtained by (\href{https://www.nndc.bnl.gov/nsr/nsrlink.jsp?2008Fu07,B}{2008Fu07}).}\\
\parbox[b][0.3cm]{17.7cm}{\addtolength{\parindent}{-0.2in}\href{https://www.nndc.bnl.gov/nsr/nsrlink.jsp?2022Ba39,B}{2022Ba39}: \ensuremath{^{\textnormal{4}}}He(\ensuremath{^{\textnormal{14}}}O,\ensuremath{\alpha}) E=61.8 MeV; used the TexAT active target filled with \ensuremath{^{\textnormal{4}}}He(96\%)+CO\ensuremath{_{\textnormal{2}}}(4\%) gas mixture. The light reaction}\\
\parbox[b][0.3cm]{17.7cm}{products were identified via \ensuremath{\Delta}E-E telescopes. Deduced the \ensuremath{^{\textnormal{14}}}O+\ensuremath{\alpha} resonant elastic scattering excitation function and the associated}\\
\parbox[b][0.3cm]{17.7cm}{angular distributions. A multi-channel R-matrix analysis using the MINRMATRIX code was performed, which included the data of}\\
\parbox[b][0.3cm]{17.7cm}{(\href{https://www.nndc.bnl.gov/nsr/nsrlink.jsp?2008Fu07,B}{2008Fu07}) experiment with an assigned error of 30\%. The dimensionless reduced \ensuremath{\alpha} and proton widths (\ensuremath{\theta}\ensuremath{^{\textnormal{2}}_{\ensuremath{\alpha}}} and \ensuremath{\theta}\ensuremath{^{\textnormal{2}}_{\textnormal{(p}_{\textnormal{0}}\textnormal{+p}_{\textnormal{1}}\textnormal{)}}}), and}\\
\parbox[b][0.3cm]{17.7cm}{J\ensuremath{^{\ensuremath{\pi}}} were deduced from R-matrix. Discussed mirror levels and performed shell model calculations to deduce spectroscopic factors.}\\
\vspace{0.385cm}
\parbox[b][0.3cm]{17.7cm}{\addtolength{\parindent}{-0.2in}\textit{Theory}:}\\
\parbox[b][0.3cm]{17.7cm}{\addtolength{\parindent}{-0.2in}A. Volya, M. Barbui, V. Z. Goldberg, and G. V. Rogachev, Commun. Phys. 5 (2022) 322: This theoretical article discusses}\\
\parbox[b][0.3cm]{17.7cm}{superradiance in terms of continuum coupling. Analyzed the data of (\href{https://www.nndc.bnl.gov/nsr/nsrlink.jsp?2022Ba39,B}{2022Ba39}) for \ensuremath{^{\textnormal{18}}}Ne and (\href{https://www.nndc.bnl.gov/nsr/nsrlink.jsp?2014Av04,B}{2014Av04}) for \ensuremath{^{\textnormal{18}}}O levels with a}\\
\parbox[b][0.3cm]{17.7cm}{substantial reduced \ensuremath{\alpha} width; found that among the state populated, only one 0\ensuremath{^{\textnormal{+}}} state with E\ensuremath{_{\textnormal{x}}}=9.8 MeV \textit{3} in \ensuremath{^{\textnormal{18}}}Ne and E\ensuremath{_{\textnormal{x}}}=9.9}\\
\parbox[b][0.3cm]{17.7cm}{MeV \textit{1} in \ensuremath{^{\textnormal{18}}}O could be explained by the superradiance phenomenon.}\\
\vspace{12pt}
\underline{$^{18}$Ne Levels}\\
\vspace{0.34cm}
\parbox[b][0.3cm]{17.7cm}{\addtolength{\parindent}{-0.254cm}(\href{https://www.nndc.bnl.gov/nsr/nsrlink.jsp?2022Ba39,B}{2022Ba39}): \ensuremath{\theta}\ensuremath{_{\textnormal{p}}^{\textnormal{2}}}=(\ensuremath{\gamma}\ensuremath{_{\textnormal{p}_{\textnormal{0}}}^{\textnormal{2}}}+\ensuremath{\gamma}\ensuremath{_{\textnormal{p}_{\textnormal{1}}}^{\textnormal{2}}})/\ensuremath{\gamma}\ensuremath{_{\textnormal{SP}}^{\textnormal{2}}}, where \ensuremath{\gamma}\ensuremath{_{\textnormal{p}_{\textnormal{0}}}^{\textnormal{2}}} and \ensuremath{\gamma}\ensuremath{_{\textnormal{p}_{\textnormal{1}}}^{\textnormal{2}}} are proton reduced partial widths for the channels populating the}\\
\parbox[b][0.3cm]{17.7cm}{ground and the first excited states of \ensuremath{^{\textnormal{17}}}F, respectively. \ensuremath{\theta}\ensuremath{_{\textnormal{p}}^{\textnormal{2}}} is the proton dimensionless reduced width.}\\
\parbox[b][0.3cm]{17.7cm}{\addtolength{\parindent}{-0.254cm}\ensuremath{\Gamma}, \ensuremath{\Gamma}\ensuremath{_{\ensuremath{\alpha}}}, \ensuremath{\Gamma}\ensuremath{_{\textnormal{p}}}, \ensuremath{\theta}\ensuremath{_{\ensuremath{\alpha}}^{\textnormal{2}}} and \ensuremath{\theta}\ensuremath{_{\textnormal{p}}^{\textnormal{2}}} are from R-matrix fits. The last two values are only provided by (\href{https://www.nndc.bnl.gov/nsr/nsrlink.jsp?2022Ba39,B}{2022Ba39}), while the widths are}\\
\parbox[b][0.3cm]{17.7cm}{provided for a few states in (\href{https://www.nndc.bnl.gov/nsr/nsrlink.jsp?2008Fu07,B}{2008Fu07}, \href{https://www.nndc.bnl.gov/nsr/nsrlink.jsp?2010Ha15,B}{2010Ha15}) and for all states measured by (\href{https://www.nndc.bnl.gov/nsr/nsrlink.jsp?2022Ba39,B}{2022Ba39}).}\\
\vspace{0.34cm}

\begin{textblock}{29}(0,27.3)
Continued on next page (footnotes at end of table)
\end{textblock}
\clearpage
%
\begin{textblock}{29}(0,27.3)
Continued on next page (footnotes at end of table)
\end{textblock}
\clearpage
%
\begin{textblock}{29}(0,27.3)
Continued on next page (footnotes at end of table)
\end{textblock}
\clearpage
%
\parbox[b][0.3cm]{17.7cm}{\makebox[1ex]{\ensuremath{^{\hypertarget{NE8LEVEL0}{a}}}} Level energies from (\href{https://www.nndc.bnl.gov/nsr/nsrlink.jsp?2008Fu07,B}{2008Fu07}) and (\href{https://www.nndc.bnl.gov/nsr/nsrlink.jsp?2010Ha15,B}{2010Ha15}) that are reported here are recalculated using E\ensuremath{_{\textnormal{x}}}(\ensuremath{^{\textnormal{18}}}Ne)=E\ensuremath{_{\textnormal{c.m.}}}+S\ensuremath{_{\ensuremath{\alpha}}}(\ensuremath{^{\textnormal{18}}}Ne), where}\\
\parbox[b][0.3cm]{17.7cm}{{\ }{\ }S\ensuremath{_{\ensuremath{\alpha}}} is the \ensuremath{\alpha}-separation energy for \ensuremath{^{\textnormal{18}}}Ne, and E\ensuremath{_{\textnormal{c.m.}}} values are from (\href{https://www.nndc.bnl.gov/nsr/nsrlink.jsp?2008Fu07,B}{2008Fu07}) and/or (\href{https://www.nndc.bnl.gov/nsr/nsrlink.jsp?2010Ha15,B}{2010Ha15}). To obtain S\ensuremath{_{\ensuremath{\alpha}}}, the masses of}\\
\parbox[b][0.3cm]{17.7cm}{{\ }{\ }\ensuremath{^{\textnormal{14}}}O, \ensuremath{^{\textnormal{18}}}Ne and \ensuremath{^{\textnormal{4}}}He are taken from (\href{https://www.nndc.bnl.gov/nsr/nsrlink.jsp?2021Wa16,B}{2021Wa16}: AME-2020).}\\
\parbox[b][0.3cm]{17.7cm}{\makebox[1ex]{\ensuremath{^{\hypertarget{NE8LEVEL1}{b}}}} The data for the states located near the low- or high-energy edge of the excitation function measured by (\href{https://www.nndc.bnl.gov/nsr/nsrlink.jsp?2022Ba39,B}{2022Ba39}) are}\\
\parbox[b][0.3cm]{17.7cm}{{\ }{\ }incomplete. Also, due to the high thresholds in some detectors used in the experiment of (\href{https://www.nndc.bnl.gov/nsr/nsrlink.jsp?2022Ba39,B}{2022Ba39}), the angular distributions in}\\
\parbox[b][0.3cm]{17.7cm}{{\ }{\ }the excitation energy range from 7 to 8 MeV are incomplete. Therefore, (\href{https://www.nndc.bnl.gov/nsr/nsrlink.jsp?2022Ba39,B}{2022Ba39}) reported these states in parentheses and}\\
\parbox[b][0.3cm]{17.7cm}{{\ }{\ }presumably considered them as tentative.}\\
\parbox[b][0.3cm]{17.7cm}{\makebox[1ex]{\ensuremath{^{\hypertarget{NE8LEVEL2}{c}}}} The lower limit spin estimations given by (\href{https://www.nndc.bnl.gov/nsr/nsrlink.jsp?2008Fu07,B}{2008Fu07}) for the high energy \ensuremath{^{\textnormal{18}}}Ne resonances are based on the assumption that the}\\
\parbox[b][0.3cm]{17.7cm}{{\ }{\ }reduced width for the proton decay channel with the most favorable penetrability is \ensuremath{\sim}10\% of the reduced width for the \ensuremath{\alpha}-decay}\\
\parbox[b][0.3cm]{17.7cm}{{\ }{\ }channel. If this ratio changes to 15\%, the estimated spin values from (\href{https://www.nndc.bnl.gov/nsr/nsrlink.jsp?2008Fu07,B}{2008Fu07}) should be increased by at least one unit.}\\
\parbox[b][0.3cm]{17.7cm}{\makebox[1ex]{\ensuremath{^{\hypertarget{NE8LEVEL3}{d}}}} This level has a pronounced \ensuremath{\alpha}-cluster structure reported by (\href{https://www.nndc.bnl.gov/nsr/nsrlink.jsp?2022Ba39,B}{2022Ba39}), which is demonstrated with the large (\ensuremath{>}0.1)}\\
\parbox[b][0.3cm]{17.7cm}{{\ }{\ }dimensionless reduced width, \ensuremath{\theta}\ensuremath{_{\ensuremath{\alpha}}^{\textnormal{2}}}.}\\
\parbox[b][0.3cm]{17.7cm}{\makebox[1ex]{\ensuremath{^{\hypertarget{NE8LEVEL4}{e}}}} (\href{https://www.nndc.bnl.gov/nsr/nsrlink.jsp?2022Ba39,B}{2022Ba39}): \ensuremath{\theta}\ensuremath{_{\ensuremath{\alpha}}^{\textnormal{2}}}=\ensuremath{\gamma}\ensuremath{_{\ensuremath{\alpha}}^{\textnormal{2}}}/\ensuremath{\gamma}\ensuremath{_{\textnormal{SP}}^{\textnormal{2}}}, where \ensuremath{\gamma}\ensuremath{_{\ensuremath{\alpha}}^{\textnormal{2}}} is the \ensuremath{\alpha}-reduced partial width, and \ensuremath{\gamma}\ensuremath{_{\textnormal{SP}}^{\textnormal{2}}}=\ensuremath{\hslash}\ensuremath{^{\textnormal{2}}}/\ensuremath{\mu}R\ensuremath{^{\textnormal{2}}} is the single-particle limit}\\
\parbox[b][0.3cm]{17.7cm}{{\ }{\ }calculated at channel radius R=5.2 fm. \ensuremath{\theta}\ensuremath{_{\ensuremath{\alpha}}^{\textnormal{2}}} is the \ensuremath{\alpha}-dimensionless reduced width.}\\
\vspace{0.5cm}
\clearpage
\subsection[\hspace{-0.2cm}\ensuremath{^{\textnormal{9}}}Be(\ensuremath{^{\textnormal{17}}}Ne,\ensuremath{^{\textnormal{18}}}Ne)]{ }
\vspace{-27pt}
\vspace{0.3cm}
\hypertarget{NE9}{{\bf \small \underline{\ensuremath{^{\textnormal{9}}}Be(\ensuremath{^{\textnormal{17}}}Ne,\ensuremath{^{\textnormal{18}}}Ne)\hspace{0.2in}\href{https://www.nndc.bnl.gov/nsr/nsrlink.jsp?2019Ch16,B}{2019Ch16}}}}\\
\vspace{4pt}
\vspace{8pt}
\parbox[b][0.3cm]{17.7cm}{\addtolength{\parindent}{-0.2in}\href{https://www.nndc.bnl.gov/nsr/nsrlink.jsp?2019Ch16,B}{2019Ch16}: \ensuremath{^{\textnormal{9}}}Be(\ensuremath{^{\textnormal{17}}}Ne,X) E=58.2 MeV; measured the reaction products using the HiRA High-Resolution position sensitive \ensuremath{\Delta}E-E}\\
\parbox[b][0.3cm]{17.7cm}{telescope array covering \ensuremath{\phi}\ensuremath{_{\textnormal{lab,zenith}}}=2\ensuremath{^\circ} to 13.9\ensuremath{^\circ}. The target was surrounded by the \ensuremath{\gamma}-ray CAESAR CAESium iodide ARray}\\
\parbox[b][0.3cm]{17.7cm}{covering \ensuremath{\theta}\ensuremath{_{\textnormal{lab,polar}}}=57.5\ensuremath{^\circ}{\textminus}122.4\ensuremath{^\circ} with a complete azimuthal coverage. Measured \ensuremath{\gamma}-particle, p-\ensuremath{^{\textnormal{17}}}F, and \ensuremath{\alpha}-\ensuremath{^{\textnormal{14}}}O coincidences. Used}\\
\parbox[b][0.3cm]{17.7cm}{invariant mass spectroscopy and deduced the \ensuremath{^{\textnormal{18}}}Ne invariant mass excitation spectrum. Performed shell model calculations using}\\
\parbox[b][0.3cm]{17.7cm}{OXBASH to obtain theoretical spectroscopic factors. Deduced branching ratios and \ensuremath{\sigma} for \ensuremath{^{\textnormal{18}}}Ne states.}\\
\vspace{12pt}
\underline{$^{18}$Ne Levels}\\
\vspace{0.34cm}
\parbox[b][0.3cm]{17.7cm}{\addtolength{\parindent}{-0.254cm}A systematic uncertainty of \ensuremath{\pm}15\% is added to the cross sections of the following states according to the discussions found in}\\
\parbox[b][0.3cm]{17.7cm}{(\href{https://www.nndc.bnl.gov/nsr/nsrlink.jsp?2019Ch16,B}{2019Ch16}).}\\
\vspace{0.34cm}
\begin{longtable}{cccccc@{\extracolsep{\fill}}c}
\multicolumn{2}{c}{E(level)$^{{\hyperlink{NE9LEVEL6}{g}}}$}&J$^{\pi}$$^{{\hyperlink{NE9LEVEL6}{g}}{\hyperlink{NE9LEVEL7}{h}}}$&\multicolumn{2}{c}{\ensuremath{\Gamma}\ensuremath{_{\textnormal{p}_{\textnormal{1}}}}/\ensuremath{\Gamma}$^{{\hyperlink{NE9LEVEL6}{g}}{\hyperlink{NE9LEVEL8}{i}}}$}&Comments&\\[-.2cm]
\multicolumn{2}{c}{\hrulefill}&\hrulefill&\multicolumn{2}{c}{\hrulefill}&\hrulefill&
\endfirsthead
\multicolumn{1}{r@{}}{4514}&\multicolumn{1}{@{}l}{\ensuremath{^{{\hyperlink{NE9LEVEL1}{b}}}} {\it 4}}&\multicolumn{1}{l}{1\ensuremath{^{-}}}&\multicolumn{1}{r@{}}{$<$0}&\multicolumn{1}{@{.}l}{125\ensuremath{^{{\hyperlink{NE9LEVEL9}{j}}}}}&\parbox[t][0.3cm]{12.2756405cm}{\raggedright \ensuremath{\Gamma}\ensuremath{_{\textnormal{p}_{\textnormal{1}}}}/\ensuremath{\Gamma}: (\href{https://www.nndc.bnl.gov/nsr/nsrlink.jsp?2019Ch16,B}{2019Ch16}) mentions that there is a possibility for this result to be overestimated\vspace{0.1cm}}&\\
&&&&&\parbox[t][0.3cm]{12.2756405cm}{\raggedright {\ }{\ }{\ }due to other sources of background that may have been unaccounted for, or from the\vspace{0.1cm}}&\\
&&&&&\parbox[t][0.3cm]{12.2756405cm}{\raggedright {\ }{\ }{\ }overlap of some of the nearby states. Even though (\href{https://www.nndc.bnl.gov/nsr/nsrlink.jsp?2019Ch16,B}{2019Ch16}) deduced an upper limit\vspace{0.1cm}}&\\
&&&&&\parbox[t][0.3cm]{12.2756405cm}{\raggedright {\ }{\ }{\ }of 12.5\% for the branching ratio of the decay of this state to \ensuremath{^{\textnormal{17}}}F*(495 keV), the authors\vspace{0.1cm}}&\\
&&&&&\parbox[t][0.3cm]{12.2756405cm}{\raggedright {\ }{\ }{\ }expect the actual value to be extremely small as it is energetically more favorable for\vspace{0.1cm}}&\\
&&&&&\parbox[t][0.3cm]{12.2756405cm}{\raggedright {\ }{\ }{\ }this state to proton decay to \ensuremath{^{\textnormal{17}}}F\ensuremath{_{\textnormal{g.s.}}} rather than to \ensuremath{^{\textnormal{17}}}F*(495 keV) level. A shell-model\vspace{0.1cm}}&\\
&&&&&\parbox[t][0.3cm]{12.2756405cm}{\raggedright {\ }{\ }{\ }estimation of \ensuremath{\Gamma}\ensuremath{_{\textnormal{p}_{\textnormal{1}}}}/\ensuremath{\Gamma}\ensuremath{_{\textnormal{tot}}}=1.32\ensuremath{\times}10\ensuremath{^{\textnormal{$-$6}}} by (\href{https://www.nndc.bnl.gov/nsr/nsrlink.jsp?2019Ch16,B}{2019Ch16}) confirms this hypothesis.\vspace{0.1cm}}&\\
&&&&&\parbox[t][0.3cm]{12.2756405cm}{\raggedright \ensuremath{\sigma}\ensuremath{_{\textnormal{peak}}}=133 \ensuremath{\mu}b \textit{8} (stat.) \textit{20} (sys.) (\href{https://www.nndc.bnl.gov/nsr/nsrlink.jsp?2019Ch16,B}{2019Ch16}).\vspace{0.1cm}}&\\
&&&&&\parbox[t][0.3cm]{12.2756405cm}{\raggedright C\ensuremath{^{\textnormal{2}}}S(\textit{d}\ensuremath{_{\textnormal{3/2}}})=0.015 and C\ensuremath{^{\textnormal{2}}}S(\textit{s}\ensuremath{_{\textnormal{1/2}}})=0.365 (\href{https://www.nndc.bnl.gov/nsr/nsrlink.jsp?2019Ch16,B}{2019Ch16}: both from shell model calculations).\vspace{0.1cm}}&\\
&&&&&\parbox[t][0.3cm]{12.2756405cm}{\raggedright {\ }{\ }{\ }However, the \textit{s}-wave transfer should be suppressed due to momentum mismatch. Note\vspace{0.1cm}}&\\
&&&&&\parbox[t][0.3cm]{12.2756405cm}{\raggedright {\ }{\ }{\ }that the authors state that \textit{either the effect of the momentum mismatch is not as large as}\vspace{0.1cm}}&\\
&&&&&\parbox[t][0.3cm]{12.2756405cm}{\raggedright {\ }{\ }{\ }\textit{expected, or these shell model predictions are in error}.\vspace{0.1cm}}&\\
\multicolumn{1}{r@{}}{4594}&\multicolumn{1}{@{}l}{\ensuremath{^{{\hyperlink{NE9LEVEL0}{a}}}} {\it 12}}&\multicolumn{1}{l}{0\ensuremath{^{+}}}&\multicolumn{1}{r@{}}{$>$0}&\multicolumn{1}{@{.}l}{16\ensuremath{^{{\hyperlink{NE9LEVEL9}{j}}}}}&\parbox[t][0.3cm]{12.2756405cm}{\raggedright \ensuremath{\Gamma}\ensuremath{_{\textnormal{p}_{\textnormal{1}}}}/\ensuremath{\Gamma}\ensuremath{_{\textnormal{tot}}}=0.036 (\href{https://www.nndc.bnl.gov/nsr/nsrlink.jsp?2019Ch16,B}{2019Ch16}: from a shell model calculation).\vspace{0.1cm}}&\\
&&&&&\parbox[t][0.3cm]{12.2756405cm}{\raggedright \ensuremath{\sigma}\ensuremath{_{\textnormal{peak}}}=11 \ensuremath{\mu}b \textit{3} (stat.) \textit{2} (sys.) (\href{https://www.nndc.bnl.gov/nsr/nsrlink.jsp?2019Ch16,B}{2019Ch16}). In the text, this value was reported as\vspace{0.1cm}}&\\
&&&&&\parbox[t][0.3cm]{12.2756405cm}{\raggedright {\ }{\ }{\ }\ensuremath{\sigma}\ensuremath{_{\textnormal{peak}}}=13 \ensuremath{\mu}b \textit{3} (stat.) \textit{2} (sys.). The authors also mention that this cross section is only\vspace{0.1cm}}&\\
&&&&&\parbox[t][0.3cm]{12.2756405cm}{\raggedright {\ }{\ }{\ }for the observed branch from the decay to \ensuremath{^{\textnormal{17}}}F(495 keV). They estimate that the total\vspace{0.1cm}}&\\
&&&&&\parbox[t][0.3cm]{12.2756405cm}{\raggedright {\ }{\ }{\ }cross section for this state must be less than 81 \ensuremath{\mu}b. Considering the \ensuremath{\pm}15\% systematic\vspace{0.1cm}}&\\
&&&&&\parbox[t][0.3cm]{12.2756405cm}{\raggedright {\ }{\ }{\ }uncertainty, this upper limit should be \ensuremath{\sigma}\ensuremath{_{\textnormal{tot}}}\ensuremath{<}69 \ensuremath{\mu}b.\vspace{0.1cm}}&\\
&&&&&\parbox[t][0.3cm]{12.2756405cm}{\raggedright C\ensuremath{^{\textnormal{2}}}S(1\textit{p}\ensuremath{_{\textnormal{1/2}}})=0.66 (\href{https://www.nndc.bnl.gov/nsr/nsrlink.jsp?2019Ch16,B}{2019Ch16}: from shell model calculations). However, the larger\vspace{0.1cm}}&\\
&&&&&\parbox[t][0.3cm]{12.2756405cm}{\raggedright {\ }{\ }{\ }momentum mismtach for \textit{p}-wave capture should suppress this yield relative to those for\vspace{0.1cm}}&\\
&&&&&\parbox[t][0.3cm]{12.2756405cm}{\raggedright {\ }{\ }{\ }\textit{d}-wave capture.\vspace{0.1cm}}&\\
\multicolumn{1}{r@{}}{5135}&\multicolumn{1}{@{}l}{\ensuremath{^{{\hyperlink{NE9LEVEL1}{b}}}} {\it 7}}&\multicolumn{1}{l}{3\ensuremath{^{-}}}&\multicolumn{1}{r@{}}{$<$0}&\multicolumn{1}{@{.}l}{009\ensuremath{^{{\hyperlink{NE9LEVEL9}{j}}}}}&\parbox[t][0.3cm]{12.2756405cm}{\raggedright E(level): The uncertainty in the excitation energy of this state deduced from the invariant\vspace{0.1cm}}&\\
&&&&&\parbox[t][0.3cm]{12.2756405cm}{\raggedright {\ }{\ }{\ }mass of p+\ensuremath{^{\textnormal{17}}}F is reported as 2 keV in Table IV of (\href{https://www.nndc.bnl.gov/nsr/nsrlink.jsp?2019Ch16,B}{2019Ch16}). Furthermore,\vspace{0.1cm}}&\\
&&&&&\parbox[t][0.3cm]{12.2756405cm}{\raggedright {\ }{\ }{\ }(\href{https://www.nndc.bnl.gov/nsr/nsrlink.jsp?2019Ch16,B}{2019Ch16}) reports a \ensuremath{\pm}6.6 keV systematic uncertainty (see text) for this level. So, the\vspace{0.1cm}}&\\
&&&&&\parbox[t][0.3cm]{12.2756405cm}{\raggedright {\ }{\ }{\ }evaluator increased the uncertainty to 7 keV by adding the statistical and systematic\vspace{0.1cm}}&\\
&&&&&\parbox[t][0.3cm]{12.2756405cm}{\raggedright {\ }{\ }{\ }uncertainties in quadrature such that E\ensuremath{_{\textnormal{x}}}=5135 keV \textit{2} (stat.) \textit{7} (sys.).\vspace{0.1cm}}&\\
&&&&&\parbox[t][0.3cm]{12.2756405cm}{\raggedright J\ensuremath{^{\pi}}: J\ensuremath{^{\ensuremath{\pi}}}=2\ensuremath{^{\textnormal{+}}} is ruled out because it is not expected to be populated by the \ensuremath{^{\textnormal{9}}}Be(\ensuremath{^{\textnormal{17}}}Ne,\ensuremath{^{\textnormal{18}}}Ne)\vspace{0.1cm}}&\\
&&&&&\parbox[t][0.3cm]{12.2756405cm}{\raggedright {\ }{\ }{\ }reaction.\vspace{0.1cm}}&\\
&&&&&\parbox[t][0.3cm]{12.2756405cm}{\raggedright \ensuremath{\Gamma}\ensuremath{_{\textnormal{p}_{\textnormal{1}}}}/\ensuremath{\Gamma}\ensuremath{_{\textnormal{tot}}}=3.6\ensuremath{\times}10\ensuremath{^{\textnormal{$-$4}}} (\href{https://www.nndc.bnl.gov/nsr/nsrlink.jsp?2019Ch16,B}{2019Ch16}: from a shell model calculation).\vspace{0.1cm}}&\\
&&&&&\parbox[t][0.3cm]{12.2756405cm}{\raggedright \ensuremath{\sigma}\ensuremath{_{\textnormal{peak}}}=1206 \ensuremath{\mu}b \textit{20} (stat.) \textit{181} (sys.) (\href{https://www.nndc.bnl.gov/nsr/nsrlink.jsp?2019Ch16,B}{2019Ch16}).\vspace{0.1cm}}&\\
&&&&&\parbox[t][0.3cm]{12.2756405cm}{\raggedright C\ensuremath{^{\textnormal{2}}}S(1\textit{d}\ensuremath{_{\textnormal{5/2}}})=0.65 (\href{https://www.nndc.bnl.gov/nsr/nsrlink.jsp?2019Ch16,B}{2019Ch16}: from shell model calculations).\vspace{0.1cm}}&\\
\multicolumn{1}{r@{}}{5457}&\multicolumn{1}{@{}l}{\ensuremath{^{{\hyperlink{NE9LEVEL1}{b}}}} {\it 8}}&\multicolumn{1}{l}{2\ensuremath{^{-}}}&\multicolumn{1}{r@{}}{$<$0}&\multicolumn{1}{@{.}l}{19\ensuremath{^{{\hyperlink{NE9LEVEL9}{j}}}}}&\parbox[t][0.3cm]{12.2756405cm}{\raggedright \ensuremath{\Gamma}\ensuremath{_{\textnormal{p}_{\textnormal{1}}}}/\ensuremath{\Gamma}\ensuremath{_{\textnormal{tot}}}=0.0022 (\href{https://www.nndc.bnl.gov/nsr/nsrlink.jsp?2019Ch16,B}{2019Ch16}: from a shell model calculation).\vspace{0.1cm}}&\\
&&&&&\parbox[t][0.3cm]{12.2756405cm}{\raggedright \ensuremath{\sigma}\ensuremath{_{\textnormal{peak}}}=186 \ensuremath{\mu}b \textit{13} (stat.) \textit{28} (sys.) (\href{https://www.nndc.bnl.gov/nsr/nsrlink.jsp?2019Ch16,B}{2019Ch16}).\vspace{0.1cm}}&\\
&&&&&\parbox[t][0.3cm]{12.2756405cm}{\raggedright C\ensuremath{^{\textnormal{2}}}S(1\textit{d}\ensuremath{_{\textnormal{5/2}}})=0.23 and C\ensuremath{^{\textnormal{2}}}S(\textit{d}\ensuremath{_{\textnormal{3/2}}})=0.12 (\href{https://www.nndc.bnl.gov/nsr/nsrlink.jsp?2019Ch16,B}{2019Ch16}: both from shell model calculations).\vspace{0.1cm}}&\\
\multicolumn{1}{r@{}}{6150}&\multicolumn{1}{@{}l}{\ensuremath{^{{\hyperlink{NE9LEVEL1}{b}}}}}&\multicolumn{1}{l}{1\ensuremath{^{-}}}&\multicolumn{1}{r@{}}{0}&\multicolumn{1}{@{.}l}{65\ensuremath{^{{\hyperlink{NE9LEVEL9}{j}}}}}&\parbox[t][0.3cm]{12.2756405cm}{\raggedright E(level): This energy was fixed to 6150 keV from (\href{https://www.nndc.bnl.gov/nsr/nsrlink.jsp?1995Ti07,B}{1995Ti07}).\vspace{0.1cm}}&\\
&&&&&\parbox[t][0.3cm]{12.2756405cm}{\raggedright \ensuremath{\Gamma}\ensuremath{_{\textnormal{p}_{\textnormal{1}}}}/\ensuremath{\Gamma}: Fixed to \ensuremath{\Gamma}\ensuremath{_{\textnormal{p}_{\textnormal{1}}}}/\ensuremath{\Gamma}=0.65 from (\href{https://www.nndc.bnl.gov/nsr/nsrlink.jsp?2003Bl11,B}{2003Bl11}).\vspace{0.1cm}}&\\
&&&&&\parbox[t][0.3cm]{12.2756405cm}{\raggedright \ensuremath{\sigma}\ensuremath{_{\textnormal{peak}}}\ensuremath{<}46 \ensuremath{\mu}b: note that this upper limit is reported as 54 \ensuremath{\mu}b at 2\ensuremath{\sigma} level (\href{https://www.nndc.bnl.gov/nsr/nsrlink.jsp?2019Ch16,B}{2019Ch16}).\vspace{0.1cm}}&\\
&&&&&\parbox[t][0.3cm]{12.2756405cm}{\raggedright {\ }{\ }{\ }Considering the 15\% systematic uncertainty on this value suggested by (\href{https://www.nndc.bnl.gov/nsr/nsrlink.jsp?2019Ch16,B}{2019Ch16}), the\vspace{0.1cm}}&\\
&&&&&\parbox[t][0.3cm]{12.2756405cm}{\raggedright {\ }{\ }{\ }upper limit is altered by the evaluator to 46 \ensuremath{\mu}b.\vspace{0.1cm}}&\\
\multicolumn{1}{r@{}}{$\approx$6.3\ensuremath{\times10^{3}}}&\multicolumn{1}{@{}l}{\ensuremath{^{{\hyperlink{NE9LEVEL1}{b}}}}}&\multicolumn{1}{l}{(2\ensuremath{^{-}},3\ensuremath{^{-}})}&\multicolumn{1}{r@{}}{$<$0}&\multicolumn{1}{@{.}l}{12\ensuremath{^{{\hyperlink{NE9LEVEL9}{j}}}}}&\parbox[t][0.3cm]{12.2756405cm}{\raggedright E(level): Likely a doublet consisting of E\ensuremath{_{\textnormal{x}}}=6279 keV \textit{36} and E\ensuremath{_{\textnormal{x}}}=6369 keV \textit{36}\vspace{0.1cm}}&\\
&&&&&\parbox[t][0.3cm]{12.2756405cm}{\raggedright {\ }{\ }{\ }(\href{https://www.nndc.bnl.gov/nsr/nsrlink.jsp?2019Ch16,B}{2019Ch16}).\vspace{0.1cm}}&\\
\end{longtable}
\begin{textblock}{29}(0,27.3)
Continued on next page (footnotes at end of table)
\end{textblock}
\clearpage
\begin{longtable}{ccccc@{\extracolsep{\fill}}c}
\\[-.4cm]
\multicolumn{6}{c}{{\bf \small \underline{\ensuremath{^{\textnormal{9}}}Be(\ensuremath{^{\textnormal{17}}}Ne,\ensuremath{^{\textnormal{18}}}Ne)\hspace{0.2in}\href{https://www.nndc.bnl.gov/nsr/nsrlink.jsp?2019Ch16,B}{2019Ch16} (continued)}}}\\
\multicolumn{6}{c}{~}\\
\multicolumn{6}{c}{\underline{\ensuremath{^{18}}Ne Levels (continued)}}\\
\multicolumn{6}{c}{~}\\
\multicolumn{2}{c}{E(level)$^{{\hyperlink{NE9LEVEL6}{g}}}$}&\multicolumn{2}{c}{\ensuremath{\Gamma}$^{{\hyperlink{NE9LEVEL6}{g}}}$}&Comments&\\[-.2cm]
\multicolumn{2}{c}{\hrulefill}&\multicolumn{2}{c}{\hrulefill}&\hrulefill&
\endhead
&&&&\parbox[t][0.3cm]{12.9101305cm}{\raggedright \ensuremath{\Gamma}\ensuremath{_{\textnormal{p}_{\textnormal{1}}}}/\ensuremath{\Gamma}: The reported upper limit branching ratio is for the pair of the two members of the\vspace{0.1cm}}&\\
&&&&\parbox[t][0.3cm]{12.9101305cm}{\raggedright {\ }{\ }{\ }6.3-MeV unresolved doublet (see above).\vspace{0.1cm}}&\\
&&&&\parbox[t][0.3cm]{12.9101305cm}{\raggedright \ensuremath{\sigma}\ensuremath{_{\textnormal{peak}}}=354 \ensuremath{\mu}b \textit{17} (stat.) \textit{53} (sys.) (\href{https://www.nndc.bnl.gov/nsr/nsrlink.jsp?2019Ch16,B}{2019Ch16}): this is the total cross section for the likely\vspace{0.1cm}}&\\
&&&&\parbox[t][0.3cm]{12.9101305cm}{\raggedright {\ }{\ }{\ }unresolved doublet states.\vspace{0.1cm}}&\\
\multicolumn{1}{r@{}}{9111}&\multicolumn{1}{@{}l}{\ensuremath{^{{\hyperlink{NE9LEVEL2}{c}}{\hyperlink{NE9LEVEL4}{e}}}} {\it 25}}&\multicolumn{1}{r@{}}{$<$60}&\multicolumn{1}{@{}l}{\ensuremath{^{{\hyperlink{NE9LEVEL5}{f}}}} keV}&\parbox[t][0.3cm]{12.9101305cm}{\raggedright E(level): (\href{https://www.nndc.bnl.gov/nsr/nsrlink.jsp?2019Ch16,B}{2019Ch16}) argues that this state is not the same as the 9.2 MeV with \ensuremath{\Gamma}=300 keV\vspace{0.1cm}}&\\
&&&&\parbox[t][0.3cm]{12.9101305cm}{\raggedright {\ }{\ }{\ }level observed in (\href{https://www.nndc.bnl.gov/nsr/nsrlink.jsp?2008Fu07,B}{2008Fu07}), and that the 9.111-MeV state observed in (\href{https://www.nndc.bnl.gov/nsr/nsrlink.jsp?2019Ch16,B}{2019Ch16}) does\vspace{0.1cm}}&\\
&&&&\parbox[t][0.3cm]{12.9101305cm}{\raggedright {\ }{\ }{\ }not have a strong \ensuremath{\alpha}-cluster structure contribution to be excited via \ensuremath{\alpha} scattering.\vspace{0.1cm}}&\\
&&&&\parbox[t][0.3cm]{12.9101305cm}{\raggedright \ensuremath{\sigma}\ensuremath{_{\textnormal{peak}}}=52 \ensuremath{\mu}b \textit{5} (stat.) \textit{8} (sys.) (\href{https://www.nndc.bnl.gov/nsr/nsrlink.jsp?2019Ch16,B}{2019Ch16}).\vspace{0.1cm}}&\\
\multicolumn{1}{r@{}}{11584}&\multicolumn{1}{@{}l}{\ensuremath{^{{\hyperlink{NE9LEVEL2}{c}}{\hyperlink{NE9LEVEL4}{e}}}} {\it 64}}&\multicolumn{1}{r@{}}{$<$650}&\multicolumn{1}{@{}l}{\ensuremath{^{{\hyperlink{NE9LEVEL5}{f}}}} keV}&\parbox[t][0.3cm]{12.9101305cm}{\raggedright \ensuremath{\sigma}\ensuremath{_{\textnormal{peak}}}\ensuremath{\sim}18 \ensuremath{\mu}b \textit{3} (sys.) (\href{https://www.nndc.bnl.gov/nsr/nsrlink.jsp?2019Ch16,B}{2019Ch16}).\vspace{0.1cm}}&\\
\multicolumn{1}{r@{}}{16794}&\multicolumn{1}{@{}l}{\ensuremath{^{{\hyperlink{NE9LEVEL3}{d}}{\hyperlink{NE9LEVEL4}{e}}}} {\it 29}}&\multicolumn{1}{r@{}}{328}&\multicolumn{1}{@{ }l}{keV {\it 68}}&\parbox[t][0.3cm]{12.9101305cm}{\raggedright T=2 (\href{https://www.nndc.bnl.gov/nsr/nsrlink.jsp?2019Ch16,B}{2019Ch16})\vspace{0.1cm}}&\\
&&&&\parbox[t][0.3cm]{12.9101305cm}{\raggedright \ensuremath{\sigma}\ensuremath{_{\textnormal{peak}}}=182 \ensuremath{\mu}b \textit{11} (stat.) \textit{27} (sys.) (\href{https://www.nndc.bnl.gov/nsr/nsrlink.jsp?2019Ch16,B}{2019Ch16}).\vspace{0.1cm}}&\\
&&&&\parbox[t][0.3cm]{12.9101305cm}{\raggedright (\href{https://www.nndc.bnl.gov/nsr/nsrlink.jsp?2019Ch16,B}{2019Ch16}): this state appears to have an exotic decay involving isospin symmetry breaking \ensuremath{\alpha}\vspace{0.1cm}}&\\
&&&&\parbox[t][0.3cm]{12.9101305cm}{\raggedright {\ }{\ }{\ }and proton decay branches: (1) p+\ensuremath{^{\textnormal{17}}}F*(11.192 MeV, 1/2\ensuremath{^{-}}, T=3/2) \ensuremath{\rightarrow} \ensuremath{\alpha}+\ensuremath{^{\textnormal{13}}}N*(2.365 MeV,\vspace{0.1cm}}&\\
&&&&\parbox[t][0.3cm]{12.9101305cm}{\raggedright {\ }{\ }{\ }1/2\ensuremath{^{\textnormal{+}}}) \ensuremath{\rightarrow} p+\ensuremath{^{\textnormal{12}}}C\ensuremath{_{\textnormal{g.s.}}}, and (2) p+\ensuremath{^{\textnormal{17}}}F*(11.192 MeV, 1/2\ensuremath{^{-}}, T=3/2) \ensuremath{\rightarrow} p+\ensuremath{^{\textnormal{16}}}O*(9.585 MeV, 1\ensuremath{^{-}})\vspace{0.1cm}}&\\
&&&&\parbox[t][0.3cm]{12.9101305cm}{\raggedright {\ }{\ }{\ }\ensuremath{\rightarrow} \ensuremath{\alpha}+\ensuremath{^{\textnormal{12}}}C\ensuremath{_{\textnormal{g.s.}}}. An additional smaller branch to the \ensuremath{^{\textnormal{16}}}O*(2\ensuremath{^{\textnormal{+}}_{\textnormal{2}}}) state, which then decays to the\vspace{0.1cm}}&\\
&&&&\parbox[t][0.3cm]{12.9101305cm}{\raggedright {\ }{\ }{\ }\ensuremath{^{\textnormal{12}}}C\ensuremath{_{\textnormal{g.s.}}} via \ensuremath{\alpha}-emission cannot be ruled out. The branching ratio of the decay of the isobaric\vspace{0.1cm}}&\\
&&&&\parbox[t][0.3cm]{12.9101305cm}{\raggedright {\ }{\ }{\ }analog state in \ensuremath{^{\textnormal{17}}}F*(11.192 MeV), which is part of the exotic decay of this state, is\vspace{0.1cm}}&\\
&&&&\parbox[t][0.3cm]{12.9101305cm}{\raggedright {\ }{\ }{\ }estimated to be \ensuremath{\Gamma}\ensuremath{_{\ensuremath{\alpha}}}/\ensuremath{\Gamma}\ensuremath{_{\textnormal{p}}}=65\% \textit{9} (\href{https://www.nndc.bnl.gov/nsr/nsrlink.jsp?2019Ch16,B}{2019Ch16}).\vspace{0.1cm}}&\\
&&&&\parbox[t][0.3cm]{12.9101305cm}{\raggedright (\href{https://www.nndc.bnl.gov/nsr/nsrlink.jsp?2019Ch16,B}{2019Ch16}): using the isobaric multiplet mass equation for the J\ensuremath{^{\ensuremath{\pi}}}=2\ensuremath{^{-}_{\textnormal{1}}} and 3\ensuremath{^{-}_{\textnormal{1}}} states in A=18\vspace{0.1cm}}&\\
&&&&\parbox[t][0.3cm]{12.9101305cm}{\raggedright {\ }{\ }{\ }(T=2 multiplet), it appears that this state lines up (with a larger than expected deviation of\vspace{0.1cm}}&\\
&&&&\parbox[t][0.3cm]{12.9101305cm}{\raggedright {\ }{\ }{\ }140 keV \textit{34}) with the J\ensuremath{^{\ensuremath{\pi}}}=3\ensuremath{^{-}} analog states in \ensuremath{^{\textnormal{18}}}N, \ensuremath{^{\textnormal{18}}}O, and \ensuremath{^{\textnormal{18}}}Na. However, the intrinsic\vspace{0.1cm}}&\\
&&&&\parbox[t][0.3cm]{12.9101305cm}{\raggedright {\ }{\ }{\ }width of this state (328 keV \textit{68}) casts doubt on the J\ensuremath{^{\ensuremath{\pi}}}=3\ensuremath{^{-}} assumption because in that case,\vspace{0.1cm}}&\\
&&&&\parbox[t][0.3cm]{12.9101305cm}{\raggedright {\ }{\ }{\ }its analog state in \ensuremath{^{\textnormal{18}}}Na*(0.83 MeV, 3\ensuremath{^{-}}) has a much narrower width of \ensuremath{\Gamma}=42 keV \textit{10} (see\vspace{0.1cm}}&\\
&&&&\parbox[t][0.3cm]{12.9101305cm}{\raggedright {\ }{\ }{\ }Adopted Levels of \ensuremath{^{\textnormal{18}}}Na in ENSDF). The state with a comparable width in \ensuremath{^{\textnormal{18}}}Na is the\vspace{0.1cm}}&\\
&&&&\parbox[t][0.3cm]{12.9101305cm}{\raggedright {\ }{\ }{\ }\ensuremath{^{\textnormal{18}}}Na*(0.59 MeV, 0\ensuremath{^{-}}) state, whose J\ensuremath{^{\ensuremath{\pi}}} assignment is not of interest and it occurs at 240 keV\vspace{0.1cm}}&\\
&&&&\parbox[t][0.3cm]{12.9101305cm}{\raggedright {\ }{\ }{\ }lower energy than the \ensuremath{^{\textnormal{18}}}Na*(0.83 MeV, 3\ensuremath{^{-}}) state. So, (\href{https://www.nndc.bnl.gov/nsr/nsrlink.jsp?2019Ch16,B}{2019Ch16}) reported that it is possible\vspace{0.1cm}}&\\
&&&&\parbox[t][0.3cm]{12.9101305cm}{\raggedright {\ }{\ }{\ }that the \ensuremath{^{\textnormal{18}}}Ne state reported here is a multiplet of a number of unresolved states of \ensuremath{^{\textnormal{18}}}Ne.\vspace{0.1cm}}&\\
\end{longtable}
\parbox[b][0.3cm]{17.7cm}{\makebox[1ex]{\ensuremath{^{\hypertarget{NE9LEVEL0}{a}}}} Decays to \ensuremath{^{\textnormal{17}}}F*(495 keV)+p.}\\
\parbox[b][0.3cm]{17.7cm}{\makebox[1ex]{\ensuremath{^{\hypertarget{NE9LEVEL1}{b}}}} Decays to \ensuremath{^{\textnormal{17}}}F\ensuremath{_{\textnormal{g.s.}}}+p.}\\
\parbox[b][0.3cm]{17.7cm}{\makebox[1ex]{\ensuremath{^{\hypertarget{NE9LEVEL2}{c}}}} Decays to \ensuremath{^{\textnormal{14}}}O\ensuremath{_{\textnormal{g.s.}}}+\ensuremath{\alpha}.}\\
\parbox[b][0.3cm]{17.7cm}{\makebox[1ex]{\ensuremath{^{\hypertarget{NE9LEVEL3}{d}}}} Observed in the \ensuremath{^{\textnormal{12}}}C+\ensuremath{\alpha}+2p decay channel.}\\
\parbox[b][0.3cm]{17.7cm}{\makebox[1ex]{\ensuremath{^{\hypertarget{NE9LEVEL4}{e}}}} Due to low efficiency and poor resolution for observing this state in the p+\ensuremath{^{\textnormal{17}}}F decay channel, the sensitivity to detect evidence}\\
\parbox[b][0.3cm]{17.7cm}{{\ }{\ }for the decay of this state via p+\ensuremath{^{\textnormal{17}}}F is significantly reduced (\href{https://www.nndc.bnl.gov/nsr/nsrlink.jsp?2019Ch16,B}{2019Ch16}).}\\
\parbox[b][0.3cm]{17.7cm}{\makebox[1ex]{\ensuremath{^{\hypertarget{NE9LEVEL5}{f}}}} At 1\ensuremath{\sigma} level.}\\
\parbox[b][0.3cm]{17.7cm}{\makebox[1ex]{\ensuremath{^{\hypertarget{NE9LEVEL6}{g}}}} From (\href{https://www.nndc.bnl.gov/nsr/nsrlink.jsp?2019Ch16,B}{2019Ch16}).}\\
\parbox[b][0.3cm]{17.7cm}{\makebox[1ex]{\ensuremath{^{\hypertarget{NE9LEVEL7}{h}}}} From comparison with \ensuremath{^{\textnormal{18}}}O mirror levels, comparison with prior measurements, and by considering the angular momentum}\\
\parbox[b][0.3cm]{17.7cm}{{\ }{\ }selection rules for the \ensuremath{^{\textnormal{9}}}Be+\ensuremath{^{\textnormal{17}}}Ne reaction given that the momentum mismatch (due to fast beam$'$s higher energy) favors neutron}\\
\parbox[b][0.3cm]{17.7cm}{{\ }{\ }capture to \textit{d}\ensuremath{_{\textnormal{3/2}}} or \textit{d}\ensuremath{_{\textnormal{5/2}}} shells.}\\
\parbox[b][0.3cm]{17.7cm}{\makebox[1ex]{\ensuremath{^{\hypertarget{NE9LEVEL8}{i}}}} Branching ratio for the decay of \ensuremath{^{\textnormal{18}}}Ne* state via \ensuremath{^{\textnormal{17}}}F*(495 keV)+p channel. (\href{https://www.nndc.bnl.gov/nsr/nsrlink.jsp?2019Ch16,B}{2019Ch16}) reports that these results \textit{are probably an}}\\
\parbox[b][0.3cm]{17.7cm}{{\ }{\ }\textit{overestimation of these excited state branches as other} [not accounted for] \textit{sources of background are present}.}\\
\parbox[b][0.3cm]{17.7cm}{\makebox[1ex]{\ensuremath{^{\hypertarget{NE9LEVEL9}{j}}}} At 2\ensuremath{\sigma} level.}\\
\vspace{0.5cm}
\clearpage
\subsection[\hspace{-0.2cm}\ensuremath{^{\textnormal{9}}}Be(\ensuremath{^{\textnormal{18}}}Ne,\ensuremath{^{\textnormal{18}}}Ne\ensuremath{'})]{ }
\vspace{-27pt}
\vspace{0.3cm}
\hypertarget{NE10}{{\bf \small \underline{\ensuremath{^{\textnormal{9}}}Be(\ensuremath{^{\textnormal{18}}}Ne,\ensuremath{^{\textnormal{18}}}Ne\ensuremath{'})\hspace{0.2in}\href{https://www.nndc.bnl.gov/nsr/nsrlink.jsp?2001ZeZZ,B}{2001ZeZZ}}}}\\
\vspace{4pt}
\vspace{8pt}
\parbox[b][0.3cm]{17.7cm}{\addtolength{\parindent}{-0.2in}\href{https://www.nndc.bnl.gov/nsr/nsrlink.jsp?2001ZeZZ,B}{2001ZeZZ}: \ensuremath{^{\textnormal{9}}}Be(\ensuremath{^{\textnormal{18}}}Ne,\ensuremath{^{\textnormal{18}}}Ne\ensuremath{\rightarrow}p+\ensuremath{^{\textnormal{17}}}F), \ensuremath{^{\textnormal{9}}}Be(\ensuremath{^{\textnormal{18}}}Ne,\ensuremath{^{\textnormal{18}}}Ne\ensuremath{\rightarrow}2p+\ensuremath{^{\textnormal{16}}}O) E\ensuremath{\approx}30-40 MeV/nucleon; measured p-\ensuremath{^{\textnormal{17}}}F coincidences. The 2p}\\
\parbox[b][0.3cm]{17.7cm}{emission from \ensuremath{^{\textnormal{18}}}Ne* was measured via detecting triple p-p-\ensuremath{^{\textnormal{16}}}O coincidences measured by the MUST detector assembly and the}\\
\parbox[b][0.3cm]{17.7cm}{SPEG spectrometer. Deduced the \ensuremath{^{\textnormal{18}}}Ne invariant mass and the excitation function of \ensuremath{^{\textnormal{18}}}Ne. The measured excitation energy and}\\
\parbox[b][0.3cm]{17.7cm}{angular distributions of protons and decay fragments agree well with the predictions of a break-up model calculation based on}\\
\parbox[b][0.3cm]{17.7cm}{solving the time-dependent Schr\"{o}dinger equation. The incomplete reconstruction of the missing mass prevented detection of any}\\
\parbox[b][0.3cm]{17.7cm}{\ensuremath{^{\textnormal{2}}}He emission from the \ensuremath{^{\textnormal{18}}}Ne*(6.15-MeV) state. So this study can neither confirm or negate the results of (\href{https://www.nndc.bnl.gov/nsr/nsrlink.jsp?2001Go01,B}{2001Go01}) regarding to}\\
\parbox[b][0.3cm]{17.7cm}{the diproton decay of the 6.15-MeV state in \ensuremath{^{\textnormal{18}}}Ne.}\\
\vspace{12pt}
\underline{$^{18}$Ne Levels}\\
\begin{longtable}{ccccc@{\extracolsep{\fill}}c}
\multicolumn{2}{c}{E(level)$^{{\hyperlink{NE10LEVEL0}{a}}}$}&\multicolumn{2}{c}{\ensuremath{\Gamma} (MeV)$^{{\hyperlink{NE10LEVEL0}{a}}}$}&Comments&\\[-.2cm]
\multicolumn{2}{c}{\hrulefill}&\multicolumn{2}{c}{\hrulefill}&\hrulefill&
\endfirsthead
\multicolumn{1}{r@{}}{4.91\ensuremath{\times10^{3}}}&\multicolumn{1}{@{ }l}{{\it 1}}&\multicolumn{1}{r@{}}{0}&\multicolumn{1}{@{.}l}{74 MeV {\it 2}}&\parbox[t][0.3cm]{13.296241cm}{\raggedright E(level): From reconstruction of the \ensuremath{^{\textnormal{18}}}Ne missing mass via detecting the proton-\ensuremath{^{\textnormal{17}}}F coincidences.\vspace{0.1cm}}&\\
&&&&\parbox[t][0.3cm]{13.296241cm}{\raggedright (\href{https://www.nndc.bnl.gov/nsr/nsrlink.jsp?2001ZeZZ,B}{2001ZeZZ}) reports that this state corresponds to a mixture of the \ensuremath{^{\textnormal{18}}}Ne excited states located\vspace{0.1cm}}&\\
&&&&\parbox[t][0.3cm]{13.296241cm}{\raggedright {\ }{\ }{\ }between 4.52 MeV and 5.45 MeV. A simulation carried out assuming that the observed state is\vspace{0.1cm}}&\\
&&&&\parbox[t][0.3cm]{13.296241cm}{\raggedright {\ }{\ }{\ }solely from the proton decay of the resonant state at 5.11 MeV gave rise to a peak of\vspace{0.1cm}}&\\
&&&&\parbox[t][0.3cm]{13.296241cm}{\raggedright {\ }{\ }{\ }approximately 300 keV wide. The broad width suggests multiple unresolved states are\vspace{0.1cm}}&\\
&&&&\parbox[t][0.3cm]{13.296241cm}{\raggedright {\ }{\ }{\ }populated. Therefore, this result is not included in the Adopted Levels.\vspace{0.1cm}}&\\
\end{longtable}
\parbox[b][0.3cm]{17.7cm}{\makebox[1ex]{\ensuremath{^{\hypertarget{NE10LEVEL0}{a}}}} From (\href{https://www.nndc.bnl.gov/nsr/nsrlink.jsp?2001ZeZZ,B}{2001ZeZZ}).}\\
\vspace{0.5cm}
\clearpage
\subsection[\hspace{-0.2cm}\ensuremath{^{\textnormal{9}}}Be(\ensuremath{^{\textnormal{20}}}Mg,X)]{ }
\vspace{-27pt}
\vspace{0.3cm}
\hypertarget{NE11}{{\bf \small \underline{\ensuremath{^{\textnormal{9}}}Be(\ensuremath{^{\textnormal{20}}}Mg,X)\hspace{0.2in}\href{https://www.nndc.bnl.gov/nsr/nsrlink.jsp?2004Ze05,B}{2004Ze05},\href{https://www.nndc.bnl.gov/nsr/nsrlink.jsp?2010Mu12,B}{2010Mu12}}}}\\
\vspace{4pt}
\vspace{8pt}
\parbox[b][0.3cm]{17.7cm}{\addtolength{\parindent}{-0.2in}\href{https://www.nndc.bnl.gov/nsr/nsrlink.jsp?2004Ze05,B}{2004Ze05}: \ensuremath{^{\textnormal{9}}}Be(\ensuremath{^{\textnormal{20}}}Mg,\ensuremath{^{\textnormal{19}}}Na) E=43 MeV/nucleon; studied breakup of \ensuremath{^{\textnormal{19}}}Na into p+\ensuremath{^{\textnormal{18}}}Ne; measured the energy and scattering angle}\\
\parbox[b][0.3cm]{17.7cm}{of the recoiling protons using the SPEG spectrometer and the MUST detector array covering \ensuremath{\theta}\ensuremath{_{\textnormal{lab}}}=2\ensuremath{^\circ}{\textminus}25\ensuremath{^\circ}. The \ensuremath{^{\textnormal{18}}}Ne decay}\\
\parbox[b][0.3cm]{17.7cm}{particles were measured using the focal plane detector of the SPEC spectrograph located at 0\ensuremath{^\circ} covering \ensuremath{\theta}\ensuremath{_{\textnormal{lab}}}=\ensuremath{\pm}2\ensuremath{^\circ} in both}\\
\parbox[b][0.3cm]{17.7cm}{horizontal and vertical directions. Deduced the invariant mass of \ensuremath{^{\textnormal{18}}}Ne+p events originated from the decay of an unresolved \ensuremath{^{\textnormal{19}}}Na}\\
\parbox[b][0.3cm]{17.7cm}{state at 0.16 MeV \textit{11}. The experimental resolution was 250 keV \textit{50}.}\\
\parbox[b][0.3cm]{17.7cm}{\addtolength{\parindent}{-0.2in}\href{https://www.nndc.bnl.gov/nsr/nsrlink.jsp?2010Mu12,B}{2010Mu12}: \ensuremath{^{\textnormal{9}}}Be(\ensuremath{^{\textnormal{20}}}Mg,X) E=450 MeV/nucleon; investigated the 2p decay of \ensuremath{^{\textnormal{20}}}Mg and 1p decay of \ensuremath{^{\textnormal{19}}}Na. The trajectories of}\\
\parbox[b][0.3cm]{17.7cm}{protons and the respective heavy ion particles, originating from the in-flight decay of their parent state, were measured using the}\\
\parbox[b][0.3cm]{17.7cm}{Projectile Fragment Separator at GSI together with a tracking technique utilizing microstrip detectors. Reconstructed the angular}\\
\parbox[b][0.3cm]{17.7cm}{correlations of the decay fragments. No momenta were measured. Energies and widths of the parent states were deduced using}\\
\parbox[b][0.3cm]{17.7cm}{GEANT simulations. From \ensuremath{^{\textnormal{18}}}Ne+2p events measured in coincidence, the (p\ensuremath{_{\textnormal{1}}}-\ensuremath{^{\textnormal{18}}}Ne)-(p\ensuremath{_{\textnormal{2}}}-\ensuremath{^{\textnormal{18}}}Ne) angular correlations were deduced.}\\
\vspace{12pt}
\underline{$^{18}$Ne Levels}\\
\begin{longtable}{ccc@{\extracolsep{\fill}}c}
\multicolumn{2}{c}{E(level)$^{}$}&Comments&\\[-.2cm]
\multicolumn{2}{c}{\hrulefill}&\hrulefill&
\endfirsthead
\multicolumn{1}{r@{}}{0}&\multicolumn{1}{@{}l}{}&\parbox[t][0.3cm]{16.24712cm}{\raggedright E(level): The \ensuremath{^{\textnormal{18}}}Ne\ensuremath{_{\textnormal{g.s.}}} is populated from the 1p decay of \ensuremath{^{\textnormal{19}}}Na\ensuremath{_{\textnormal{g.s.}}} (\href{https://www.nndc.bnl.gov/nsr/nsrlink.jsp?2010Mu12,B}{2010Mu12}), and from the in-flight proton decay of\vspace{0.1cm}}&\\
&&\parbox[t][0.3cm]{16.24712cm}{\raggedright {\ }{\ }{\ }a state observed at 0.16 MeV \textit{11}, which is the unresolved contribution of \ensuremath{^{\textnormal{19}}}Na\ensuremath{_{\textnormal{g.s.}}} and \ensuremath{^{\textnormal{19}}}Na*(120 keV) (\href{https://www.nndc.bnl.gov/nsr/nsrlink.jsp?2004Ze05,B}{2004Ze05}).\vspace{0.1cm}}&\\
\end{longtable}
\clearpage
\subsection[\hspace{-0.2cm}\ensuremath{^{\textnormal{9}}}Be(\ensuremath{^{\textnormal{34}}}Ar,\ensuremath{^{\textnormal{18}}}Ne\ensuremath{\gamma})]{ }
\vspace{-27pt}
\vspace{0.3cm}
\hypertarget{NE12}{{\bf \small \underline{\ensuremath{^{\textnormal{9}}}Be(\ensuremath{^{\textnormal{34}}}Ar,\ensuremath{^{\textnormal{18}}}Ne\ensuremath{\gamma})\hspace{0.2in}\href{https://www.nndc.bnl.gov/nsr/nsrlink.jsp?2006Ob03,B}{2006Ob03}}}}\\
\vspace{4pt}
\vspace{8pt}
\parbox[b][0.3cm]{17.7cm}{\addtolength{\parindent}{-0.2in}\href{https://www.nndc.bnl.gov/nsr/nsrlink.jsp?2006Ob03,B}{2006Ob03}: \ensuremath{^{\textnormal{9}}}Be(\ensuremath{^{\textnormal{34}}}Ar,\ensuremath{^{\textnormal{18}}}Ne\ensuremath{\gamma}) E\ensuremath{\approx}94-110 MeV/nucleon; measured E\ensuremath{_{\ensuremath{\gamma}}}, I\ensuremath{_{\ensuremath{\gamma}}}, particle-\ensuremath{\gamma} coincidences, energy loss, and time-of-flight}\\
\parbox[b][0.3cm]{17.7cm}{of charged particles using the S800 spectrometer$'$s focal plane detection system. The beam was a cocktail of \ensuremath{^{\textnormal{24}}}Mg, \ensuremath{^{\textnormal{25}}}Al, \ensuremath{^{\textnormal{26}}}Si, and}\\
\parbox[b][0.3cm]{17.7cm}{\ensuremath{^{\textnormal{34}}}Ar ions. Deduced the relative population of excited states of the \textit{sd}-shell nuclei produced by fragmentation. Discussed}\\
\parbox[b][0.3cm]{17.7cm}{nondissipative population. Performed shell model calculations using the Oxbash code. Deduced reaction mechanisms.}\\
\vspace{12pt}
\underline{$^{18}$Ne Levels}\\
\vspace{0.34cm}
\parbox[b][0.3cm]{17.7cm}{\addtolength{\parindent}{-0.254cm}The relative populations (combining the statistical feeding with the nondissipative population) of \ensuremath{^{\textnormal{18}}}Ne\ensuremath{_{\textnormal{g.s.}}} from the interaction of}\\
\parbox[b][0.3cm]{17.7cm}{\ensuremath{^{\textnormal{9}}}Be with \ensuremath{^{\textnormal{24}}}Mg, \ensuremath{^{\textnormal{25}}}Al, \ensuremath{^{\textnormal{26}}}Si, and \ensuremath{^{\textnormal{34}}}Ar beams at E\ensuremath{_{\textnormal{lab}}}=94, 102, 109, and 110 MeV/nucleon, respectively, were deduced as P\ensuremath{_{\textnormal{k}}}=0.8,}\\
\parbox[b][0.3cm]{17.7cm}{0.3, 0, and 0.2, respectively (\href{https://www.nndc.bnl.gov/nsr/nsrlink.jsp?2006Ob03,B}{2006Ob03}).}\\
\parbox[b][0.3cm]{17.7cm}{\addtolength{\parindent}{-0.254cm}The weights of the single-particle component for the interaction of \ensuremath{^{\textnormal{9}}}Be with \ensuremath{^{\textnormal{24}}}Mg, \ensuremath{^{\textnormal{25}}}Al, \ensuremath{^{\textnormal{26}}}Si, and \ensuremath{^{\textnormal{34}}}Ar beams at E\ensuremath{_{\textnormal{lab}}}=94, 102,}\\
\parbox[b][0.3cm]{17.7cm}{109, and 110 MeV/nucleon, respectively, were deduced as \ensuremath{\alpha}=0.0 \textit{8}, 0.0 \textit{9}, 0.4 \textit{6}, and 0.0 \textit{8}, respectively (\href{https://www.nndc.bnl.gov/nsr/nsrlink.jsp?2006Ob03,B}{2006Ob03}).}\\
\vspace{0.34cm}
\begin{longtable}{ccccc@{\extracolsep{\fill}}c}
\multicolumn{2}{c}{E(level)$^{{\hyperlink{NE12LEVEL0}{a}}}$}&J$^{\pi}$$^{{\hyperlink{NE12LEVEL0}{a}}}$&\multicolumn{2}{c}{P\ensuremath{_{\textnormal{k}}} (\%)$^{{\hyperlink{NE12LEVEL1}{b}}}$}&\\[-.2cm]
\multicolumn{2}{c}{\hrulefill}&\hrulefill&\multicolumn{2}{c}{\hrulefill}&
\endfirsthead
\multicolumn{1}{r@{}}{0}&\multicolumn{1}{@{}l}{}&\multicolumn{1}{l}{0\ensuremath{^{+}}}&\multicolumn{1}{r@{}}{20}&\multicolumn{1}{@{}l}{}&\\
\multicolumn{1}{r@{}}{1887}&\multicolumn{1}{@{.}l}{4}&\multicolumn{1}{l}{2\ensuremath{^{+}}}&\multicolumn{1}{r@{}}{40}&\multicolumn{1}{@{}l}{}&\\
\multicolumn{1}{r@{}}{3376}&\multicolumn{1}{@{.}l}{4}&\multicolumn{1}{l}{4\ensuremath{^{+}}}&\multicolumn{1}{r@{}}{26}&\multicolumn{1}{@{}l}{}&\\
\multicolumn{1}{r@{}}{3576}&\multicolumn{1}{@{.}l}{3}&\multicolumn{1}{l}{0\ensuremath{^{+}}}&\multicolumn{1}{r@{}}{4}&\multicolumn{1}{@{}l}{}&\\
\multicolumn{1}{r@{}}{3616}&\multicolumn{1}{@{.}l}{5}&\multicolumn{1}{l}{2\ensuremath{^{+}}}&\multicolumn{1}{r@{}}{10}&\multicolumn{1}{@{}l}{}&\\
\end{longtable}
\parbox[b][0.3cm]{17.7cm}{\makebox[1ex]{\ensuremath{^{\hypertarget{NE12LEVEL0}{a}}}} From the \ensuremath{^{\textnormal{18}}}Ne Adopted Levels.}\\
\parbox[b][0.3cm]{17.7cm}{\makebox[1ex]{\ensuremath{^{\hypertarget{NE12LEVEL1}{b}}}} The relative population of \ensuremath{^{\textnormal{18}}}Ne in a particular state obtained from (\href{https://www.nndc.bnl.gov/nsr/nsrlink.jsp?2006Ob03,B}{2006Ob03}). \ensuremath{^{\textnormal{18}}}Ne is populated from interaction of \ensuremath{^{\textnormal{34}}}Ar at}\\
\parbox[b][0.3cm]{17.7cm}{{\ }{\ }110 MeV/nucleon with a \ensuremath{^{\textnormal{9}}}Be target. The given values are estimated by the evaluator based on Fig. 4 in (\href{https://www.nndc.bnl.gov/nsr/nsrlink.jsp?2006Ob03,B}{2006Ob03}).}\\
\vspace{0.5cm}
\underline{$\gamma$($^{18}$Ne)}\\
\begin{longtable}{ccccccc@{}c@{\extracolsep{\fill}}c}
\multicolumn{2}{c}{E\ensuremath{_{\gamma}}\ensuremath{^{\hyperlink{NE12GAMMA0}{a}}}}&\multicolumn{2}{c}{E\ensuremath{_{i}}(level)}&J\ensuremath{^{\pi}_{i}}&\multicolumn{2}{c}{E\ensuremath{_{f}}}&J\ensuremath{^{\pi}_{f}}&\\[-.2cm]
\multicolumn{2}{c}{\hrulefill}&\multicolumn{2}{c}{\hrulefill}&\hrulefill&\multicolumn{2}{c}{\hrulefill}&\hrulefill&
\endfirsthead
\multicolumn{1}{r@{}}{1488}&\multicolumn{1}{@{.}l}{9}&\multicolumn{1}{r@{}}{3376}&\multicolumn{1}{@{.}l}{4}&\multicolumn{1}{l}{4\ensuremath{^{+}}}&\multicolumn{1}{r@{}}{1887}&\multicolumn{1}{@{.}l}{4}&\multicolumn{1}{@{}l}{2\ensuremath{^{+}}}&\\
\multicolumn{1}{r@{}}{1689}&\multicolumn{1}{@{}l}{}&\multicolumn{1}{r@{}}{3576}&\multicolumn{1}{@{.}l}{3}&\multicolumn{1}{l}{0\ensuremath{^{+}}}&\multicolumn{1}{r@{}}{1887}&\multicolumn{1}{@{.}l}{4}&\multicolumn{1}{@{}l}{2\ensuremath{^{+}}}&\\
\multicolumn{1}{r@{}}{1729}&\multicolumn{1}{@{.}l}{2}&\multicolumn{1}{r@{}}{3616}&\multicolumn{1}{@{.}l}{5}&\multicolumn{1}{l}{2\ensuremath{^{+}}}&\multicolumn{1}{r@{}}{1887}&\multicolumn{1}{@{.}l}{4}&\multicolumn{1}{@{}l}{2\ensuremath{^{+}}}&\\
\multicolumn{1}{r@{}}{1887}&\multicolumn{1}{@{.}l}{3}&\multicolumn{1}{r@{}}{1887}&\multicolumn{1}{@{.}l}{4}&\multicolumn{1}{l}{2\ensuremath{^{+}}}&\multicolumn{1}{r@{}}{0}&\multicolumn{1}{@{}l}{}&\multicolumn{1}{@{}l}{0\ensuremath{^{+}}}&\\
\end{longtable}
\parbox[b][0.3cm]{17.7cm}{\makebox[1ex]{\ensuremath{^{\hypertarget{NE12GAMMA0}{a}}}} From the \ensuremath{^{\textnormal{18}}}Ne Adopted Gammas. Note that no information is provided by (\href{https://www.nndc.bnl.gov/nsr/nsrlink.jsp?2006Ob03,B}{2006Ob03}) regarding the measured \ensuremath{^{\textnormal{18}}}Ne \ensuremath{\gamma}-ray}\\
\parbox[b][0.3cm]{17.7cm}{{\ }{\ }transitions, their energies and intensities. The authors only mention that the highest \ensuremath{\gamma}-ray energy observed for all populated}\\
\parbox[b][0.3cm]{17.7cm}{{\ }{\ }transitions (not specific to \ensuremath{^{\textnormal{18}}}Ne) was 2.5 MeV.}\\
\vspace{0.5cm}
\clearpage
\begin{figure}[h]
\begin{center}
\includegraphics{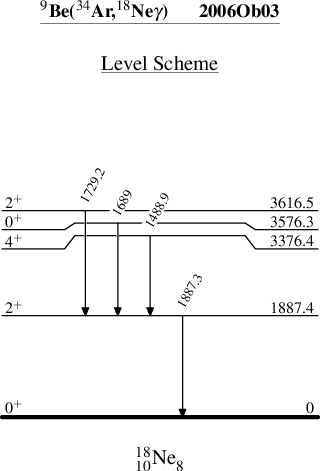}\\
\end{center}
\end{figure}
\clearpage
\subsection[\hspace{-0.2cm}\ensuremath{^{\textnormal{12}}}C(\ensuremath{^{\textnormal{12}}}C,\ensuremath{^{\textnormal{6}}}He)]{ }
\vspace{-27pt}
\vspace{0.3cm}
\hypertarget{NE13}{{\bf \small \underline{\ensuremath{^{\textnormal{12}}}C(\ensuremath{^{\textnormal{12}}}C,\ensuremath{^{\textnormal{6}}}He)\hspace{0.2in}\href{https://www.nndc.bnl.gov/nsr/nsrlink.jsp?1992HaZZ,B}{1992HaZZ},\href{https://www.nndc.bnl.gov/nsr/nsrlink.jsp?1996Ha26,B}{1996Ha26}}}}\\
\vspace{4pt}
\vspace{8pt}
\parbox[b][0.3cm]{17.7cm}{\addtolength{\parindent}{-0.2in}\href{https://www.nndc.bnl.gov/nsr/nsrlink.jsp?1992HaZZ,B}{1992HaZZ}, \href{https://www.nndc.bnl.gov/nsr/nsrlink.jsp?1996Ha26,B}{1996Ha26}: \ensuremath{^{\textnormal{12}}}C(\ensuremath{^{\textnormal{12}}}C,\ensuremath{^{\textnormal{6}}}He) E=80 MeV; measured \ensuremath{^{\textnormal{6}}}He ejectiles at the focal plane of an Enge split-pole spectrograph at}\\
\parbox[b][0.3cm]{17.7cm}{\ensuremath{\theta}\ensuremath{_{\textnormal{lab}}}=1\ensuremath{^\circ}, 2\ensuremath{^\circ}, 4\ensuremath{^\circ}, 6\ensuremath{^\circ}, 7\ensuremath{^\circ}, and 10\ensuremath{^\circ}. Energy resolution was \ensuremath{\Delta}E\ensuremath{\sim}70 keV (FWHM) at all angles except at \ensuremath{\theta}\ensuremath{_{\textnormal{lab}}}=1\ensuremath{^\circ}, where the}\\
\parbox[b][0.3cm]{17.7cm}{resolution was \ensuremath{\sim}100 keV (FWHM). Deduced \ensuremath{^{\textnormal{18}}}Ne excited states, differential cross sections, and \ensuremath{^{\textnormal{6}}}He angular distributions.}\\
\parbox[b][0.3cm]{17.7cm}{(d\ensuremath{\sigma}/d\ensuremath{\Omega})\ensuremath{_{\textnormal{max}}}\ensuremath{\sim}1 \ensuremath{\mu}b/sr was reported. Performed Hauser-Feshbach calculations using the STATIS code to compute the compound}\\
\parbox[b][0.3cm]{17.7cm}{nucleus cross sections for the \ensuremath{^{\textnormal{12}}}C(\ensuremath{^{\textnormal{12}}}C,\ensuremath{^{\textnormal{6}}}He) reaction. Most angular distributions did not present clear features that could be used to}\\
\parbox[b][0.3cm]{17.7cm}{deduce reliable J\ensuremath{^{\ensuremath{\pi}}} values.}\\
\vspace{12pt}
\underline{$^{18}$Ne Levels}\\
\begin{longtable}{cccc@{\extracolsep{\fill}}c}
\multicolumn{2}{c}{E(level)$^{{\hyperlink{NE13LEVEL0}{a}}}$}&J$^{\pi}$$^{{\hyperlink{NE13LEVEL7}{h}}}$&Comments&\\[-.2cm]
\multicolumn{2}{c}{\hrulefill}&\hrulefill&\hrulefill&
\endfirsthead
\multicolumn{1}{r@{}}{0}&\multicolumn{1}{@{}l}{\ensuremath{^{{\hyperlink{NE13LEVEL1}{b}}{\hyperlink{NE13LEVEL3}{d}}}}}&\multicolumn{1}{l}{(0\ensuremath{^{+}})\ensuremath{^{{\hyperlink{NE13LEVEL8}{i}}}}}&&\\
\multicolumn{1}{r@{}}{1.89\ensuremath{\times10^{3}}}&\multicolumn{1}{@{}l}{\ensuremath{^{{\hyperlink{NE13LEVEL1}{b}}{\hyperlink{NE13LEVEL3}{d}}}}}&\multicolumn{1}{l}{(2\ensuremath{^{+}})\ensuremath{^{{\hyperlink{NE13LEVEL8}{i}}}}}&&\\
\multicolumn{1}{r@{}}{3.38\ensuremath{\times10^{3}}}&\multicolumn{1}{@{}l}{\ensuremath{^{{\hyperlink{NE13LEVEL1}{b}}{\hyperlink{NE13LEVEL3}{d}}}}}&&&\\
\multicolumn{1}{r@{}}{3.58\ensuremath{\times10^{3}}}&\multicolumn{1}{@{}l}{\ensuremath{^{{\hyperlink{NE13LEVEL1}{b}}{\hyperlink{NE13LEVEL2}{c}}}}}&&&\\
\multicolumn{1}{r@{}}{3.62\ensuremath{\times10^{3}}}&\multicolumn{1}{@{}l}{\ensuremath{^{{\hyperlink{NE13LEVEL1}{b}}{\hyperlink{NE13LEVEL2}{c}}}}}&&&\\
\multicolumn{1}{r@{}}{4.52\ensuremath{\times10^{3}}}&\multicolumn{1}{@{}l}{\ensuremath{^{{\hyperlink{NE13LEVEL4}{e}}}}}&&&\\
\multicolumn{1}{r@{}}{4.56\ensuremath{\times10^{3}}}&\multicolumn{1}{@{}l}{\ensuremath{^{{\hyperlink{NE13LEVEL4}{e}}}}}&&\parbox[t][0.3cm]{14.2314205cm}{\raggedright E(level): This energy must have come from the previous \ensuremath{^{\textnormal{18}}}Ne evaluation (\href{https://www.nndc.bnl.gov/nsr/nsrlink.jsp?1995Ti07,B}{1995Ti07}) and originally\vspace{0.1cm}}&\\
&&&\parbox[t][0.3cm]{14.2314205cm}{\raggedright {\ }{\ }{\ }reported by (\href{https://www.nndc.bnl.gov/nsr/nsrlink.jsp?1991Ga03,B}{1991Ga03}). Note that this state was observed at E\ensuremath{_{\textnormal{beam}}}=10.9 MeV and only at one angle\vspace{0.1cm}}&\\
&&&\parbox[t][0.3cm]{14.2314205cm}{\raggedright {\ }{\ }{\ }of \ensuremath{\theta}\ensuremath{_{\textnormal{lab}}}=124.7\ensuremath{^\circ}. (\href{https://www.nndc.bnl.gov/nsr/nsrlink.jsp?2005Pa50,B}{2005Pa50}) saw no evidence for this level and suggested that it is likely that the\vspace{0.1cm}}&\\
&&&\parbox[t][0.3cm]{14.2314205cm}{\raggedright {\ }{\ }{\ }experiment of (\href{https://www.nndc.bnl.gov/nsr/nsrlink.jsp?1991Ga03,B}{1991Ga03}), whose resolution was a factor of 2 poorer, was unable to resolve the states\vspace{0.1cm}}&\\
&&&\parbox[t][0.3cm]{14.2314205cm}{\raggedright {\ }{\ }{\ }observed in (\href{https://www.nndc.bnl.gov/nsr/nsrlink.jsp?2005Pa50,B}{2005Pa50}) at 4519 keV and 4527 keV.\vspace{0.1cm}}&\\
\multicolumn{1}{r@{}}{4.59\ensuremath{\times10^{3}}}&\multicolumn{1}{@{}l}{\ensuremath{^{{\hyperlink{NE13LEVEL4}{e}}}}}&&&\\
\multicolumn{1}{r@{}}{5.11\ensuremath{\times10^{3}}}&\multicolumn{1}{@{}l}{\ensuremath{^{{\hyperlink{NE13LEVEL5}{f}}}}}&&&\\
\multicolumn{1}{r@{}}{5.15\ensuremath{\times10^{3}}}&\multicolumn{1}{@{}l}{\ensuremath{^{{\hyperlink{NE13LEVEL5}{f}}}}}&&&\\
\multicolumn{1}{r@{}}{5.45\ensuremath{\times10^{3}}}&\multicolumn{1}{@{}l}{\ensuremath{^{{\hyperlink{NE13LEVEL3}{d}}}}}&\multicolumn{1}{l}{(2\ensuremath{^{-}})\ensuremath{^{{\hyperlink{NE13LEVEL8}{i}}}}}&\parbox[t][0.3cm]{14.2314205cm}{\raggedright J\ensuremath{^{\pi}}: The angular distribution of this state populated by the \ensuremath{^{\textnormal{12}}}C(\ensuremath{^{\textnormal{12}}}C,\ensuremath{^{\textnormal{6}}}He) reaction has the characteristics\vspace{0.1cm}}&\\
&&&\parbox[t][0.3cm]{14.2314205cm}{\raggedright {\ }{\ }{\ }of an unnatural-parity state, with very weak population at forward angles peaking at backward angles.\vspace{0.1cm}}&\\
\multicolumn{1}{r@{}}{6.15\ensuremath{\times10^{3}}}&\multicolumn{1}{@{ }l}{{\it 2}}&\multicolumn{1}{l}{(1\ensuremath{^{-}})}&&\\
\multicolumn{1}{r@{}}{6.30\ensuremath{\times10^{3}}}&\multicolumn{1}{@{}l}{\ensuremath{^{{\hyperlink{NE13LEVEL3}{d}}{\hyperlink{NE13LEVEL6}{g}}}}}&&&\\
\multicolumn{1}{r@{}}{7.12\ensuremath{\times10^{3}}}&\multicolumn{1}{@{ }l}{{\it 2}}&&&\\
\multicolumn{1}{r@{}}{7.35\ensuremath{\times10^{3}}}&\multicolumn{1}{@{ }l}{{\it 2}}&\multicolumn{1}{l}{(1\ensuremath{^{-}})}&&\\
\multicolumn{1}{r@{}}{7.62\ensuremath{\times10^{3}}}&\multicolumn{1}{@{ }l}{{\it 2}}&&&\\
\multicolumn{1}{r@{}}{7.73\ensuremath{\times10^{3}}}&\multicolumn{1}{@{ }l}{{\it 2}}&&&\\
\multicolumn{1}{r@{}}{7.94\ensuremath{\times10^{3}}}&\multicolumn{1}{@{ }l}{{\it 2}}&&&\\
\multicolumn{1}{r@{}}{8.11\ensuremath{\times10^{3}}}&\multicolumn{1}{@{}l}{\ensuremath{^{{\hyperlink{NE13LEVEL3}{d}}}}}&&&\\
\multicolumn{1}{r@{}}{8.30\ensuremath{\times10^{3}}}&\multicolumn{1}{@{ }l}{{\it 2}}&&&\\
\multicolumn{1}{r@{}}{8.45\ensuremath{\times10^{3}}?}&\multicolumn{1}{@{ }l}{{\it 3}}&&\parbox[t][0.3cm]{14.2314205cm}{\raggedright E(level): This tentative state, if exists, has poor statistics and is not even labeled on the spectrum shown\vspace{0.1cm}}&\\
&&&\parbox[t][0.3cm]{14.2314205cm}{\raggedright {\ }{\ }{\ }in Fig. 6 of (\href{https://www.nndc.bnl.gov/nsr/nsrlink.jsp?1996Ha26,B}{1996Ha26}). Therefore, it was not considered for the Adopted Levels.\vspace{0.1cm}}&\\
\multicolumn{1}{r@{}}{8.55\ensuremath{\times10^{3}}}&\multicolumn{1}{@{ }l}{{\it 3}}&&&\\
\multicolumn{1}{r@{}}{8.94\ensuremath{\times10^{3}}}&\multicolumn{1}{@{ }l}{{\it 2}}&&&\\
\multicolumn{1}{r@{}}{9.18\ensuremath{\times10^{3}}}&\multicolumn{1}{@{ }l}{{\it 2}}&&&\\
\multicolumn{1}{r@{}}{9.58\ensuremath{\times10^{3}}}&\multicolumn{1}{@{ }l}{{\it 2}}&&&\\
\end{longtable}
\parbox[b][0.3cm]{17.7cm}{\makebox[1ex]{\ensuremath{^{\hypertarget{NE13LEVEL0}{a}}}} From (\href{https://www.nndc.bnl.gov/nsr/nsrlink.jsp?1996Ha26,B}{1996Ha26}: weighted average values of the excitation energies measured at \ensuremath{\theta}\ensuremath{_{\textnormal{lab}}}=2\ensuremath{^\circ}, 4\ensuremath{^\circ}, 6\ensuremath{^\circ}, 7\ensuremath{^\circ}, and 10\ensuremath{^\circ}).}\\
\parbox[b][0.3cm]{17.7cm}{\makebox[1ex]{\ensuremath{^{\hypertarget{NE13LEVEL1}{b}}}} From Figs. 6 and 7 of (\href{https://www.nndc.bnl.gov/nsr/nsrlink.jsp?1996Ha26,B}{1996Ha26}).}\\
\parbox[b][0.3cm]{17.7cm}{\makebox[1ex]{\ensuremath{^{\hypertarget{NE13LEVEL2}{c}}}} The 3.58 MeV and 3.62 MeV doublet could not be resolved (see Fig. 6 of (\href{https://www.nndc.bnl.gov/nsr/nsrlink.jsp?1996Ha26,B}{1996Ha26})).}\\
\parbox[b][0.3cm]{17.7cm}{\makebox[1ex]{\ensuremath{^{\hypertarget{NE13LEVEL3}{d}}}} E\ensuremath{_{\textnormal{x}}} used as a calibration point in (\href{https://www.nndc.bnl.gov/nsr/nsrlink.jsp?1996Ha26,B}{1996Ha26}).}\\
\parbox[b][0.3cm]{17.7cm}{\makebox[1ex]{\ensuremath{^{\hypertarget{NE13LEVEL4}{e}}}} The 4.52-4.56-4.59-MeV triplet states were unresolved (see Fig. 6 of (\href{https://www.nndc.bnl.gov/nsr/nsrlink.jsp?1996Ha26,B}{1996Ha26})).}\\
\parbox[b][0.3cm]{17.7cm}{\makebox[1ex]{\ensuremath{^{\hypertarget{NE13LEVEL5}{f}}}} The 5.11 MeV and 5.15 MeV doublet could not be resolved (see Fig. 6 of (\href{https://www.nndc.bnl.gov/nsr/nsrlink.jsp?1996Ha26,B}{1996Ha26})).}\\
\parbox[b][0.3cm]{17.7cm}{\makebox[1ex]{\ensuremath{^{\hypertarget{NE13LEVEL6}{g}}}} The 6.30 MeV and 6.35 MeV doublet could not be resolved (see Table II of (\href{https://www.nndc.bnl.gov/nsr/nsrlink.jsp?1996Ha26,B}{1996Ha26})). The calibration favored the 6.30 MeV}\\
\parbox[b][0.3cm]{17.7cm}{{\ }{\ }state, and thus this value was used as a calibration point.}\\
\parbox[b][0.3cm]{17.7cm}{\makebox[1ex]{\ensuremath{^{\hypertarget{NE13LEVEL7}{h}}}} From (\href{https://www.nndc.bnl.gov/nsr/nsrlink.jsp?1996Ha26,B}{1996Ha26}) based on Hauser-Feshbach statistical-model calculations (using the STATIS computer code) compared to the}\\
\begin{textblock}{29}(0,27.3)
Continued on next page (footnotes at end of table)
\end{textblock}
\clearpage
\vspace*{-0.5cm}
{\bf \small \underline{\ensuremath{^{\textnormal{12}}}C(\ensuremath{^{\textnormal{12}}}C,\ensuremath{^{\textnormal{6}}}He)\hspace{0.2in}\href{https://www.nndc.bnl.gov/nsr/nsrlink.jsp?1992HaZZ,B}{1992HaZZ},\href{https://www.nndc.bnl.gov/nsr/nsrlink.jsp?1996Ha26,B}{1996Ha26} (continued)}}\\
\vspace{0.3cm}
\underline{$^{18}$Ne Levels (continued)}\\
\vspace{0.3cm}
\parbox[b][0.3cm]{17.7cm}{{\ }{\ }experimental angular distributions for the \ensuremath{^{\textnormal{18}}}Ne states populated in the \ensuremath{^{\textnormal{12}}}C(\ensuremath{^{\textnormal{12}}}C,\ensuremath{^{\textnormal{6}}}He) reaction at E\ensuremath{_{\textnormal{lab}}}=80 MeV.}\\
\parbox[b][0.3cm]{17.7cm}{\makebox[1ex]{\ensuremath{^{\hypertarget{NE13LEVEL8}{i}}}} From Fig. 8 of (\href{https://www.nndc.bnl.gov/nsr/nsrlink.jsp?1996Ha26,B}{1996Ha26}).}\\
\vspace{0.5cm}
\clearpage
\subsection[\hspace{-0.2cm}\ensuremath{^{\textnormal{14}}}N(\ensuremath{^{\textnormal{17}}}F,\ensuremath{^{\textnormal{18}}}Ne)]{ }
\vspace{-27pt}
\vspace{0.3cm}
\hypertarget{NE14}{{\bf \small \underline{\ensuremath{^{\textnormal{14}}}N(\ensuremath{^{\textnormal{17}}}F,\ensuremath{^{\textnormal{18}}}Ne)\hspace{0.2in}\href{https://www.nndc.bnl.gov/nsr/nsrlink.jsp?2003Bl12,B}{2003Bl12},\href{https://www.nndc.bnl.gov/nsr/nsrlink.jsp?2004Bl21,B}{2004Bl21}}}}\\
\vspace{4pt}
\vspace{8pt}
\parbox[b][0.3cm]{17.7cm}{\addtolength{\parindent}{-0.2in}\href{https://www.nndc.bnl.gov/nsr/nsrlink.jsp?2003Bl12,B}{2003Bl12}, \href{https://www.nndc.bnl.gov/nsr/nsrlink.jsp?2004Bl21,B}{2004Bl21}: \ensuremath{^{\textnormal{14}}}N(\ensuremath{^{\textnormal{17}}}F,\ensuremath{^{\textnormal{17}}}F), \ensuremath{^{\textnormal{14}}}N(\ensuremath{^{\textnormal{17}}}F,\ensuremath{^{\textnormal{18}}}Ne) E=170 MeV; measured the reaction products by a pair of position sensitive}\\
\parbox[b][0.3cm]{17.7cm}{\ensuremath{\Delta}E-E silicon telescopes covering \ensuremath{\theta}\ensuremath{_{\textnormal{lab}}}=2\ensuremath{^\circ}{\textminus}9\ensuremath{^\circ}. Measured \ensuremath{\gamma}-\ensuremath{^{\textnormal{18}}}Ne coincidences using the CLARION array that consisted of 11}\\
\parbox[b][0.3cm]{17.7cm}{segmented clover Ge-detectors. Deduced the differential cross sections of the \ensuremath{^{\textnormal{14}}}N(\ensuremath{^{\textnormal{17}}}F,\ensuremath{^{\textnormal{18}}}Ne)\ensuremath{^{\textnormal{13}}}C reaction populating \ensuremath{^{\textnormal{18}}}Ne states by}\\
\parbox[b][0.3cm]{17.7cm}{gating charged particles on individual \ensuremath{\gamma}-ray transitions. The J\ensuremath{^{\ensuremath{\pi}}} values for strongest transitions were deduced using DWBA}\\
\parbox[b][0.3cm]{17.7cm}{analysis, which is discussed in detail.}\\
\vspace{12pt}
\underline{$^{18}$Ne Levels}\\
\begin{longtable}{ccccc@{\extracolsep{\fill}}c}
\multicolumn{2}{c}{E(level)$^{{\hyperlink{NE14LEVEL0}{a}}}$}&J$^{\pi}$$^{{\hyperlink{NE14LEVEL0}{a}}}$&L$^{}$&Comments&\\[-.2cm]
\multicolumn{2}{c}{\hrulefill}&\hrulefill&\hrulefill&\hrulefill&
\endfirsthead
\multicolumn{1}{r@{}}{0}&\multicolumn{1}{@{}l}{}&\multicolumn{1}{l}{0\ensuremath{^{+}}}&&&\\
\multicolumn{1}{r@{}}{1887}&\multicolumn{1}{@{.}l}{4}&\multicolumn{1}{l}{2\ensuremath{^{+}}}&&&\\
\multicolumn{1}{r@{}}{3376}&\multicolumn{1}{@{.}l}{4}&\multicolumn{1}{l}{4\ensuremath{^{+}}}&\multicolumn{1}{l}{2}&\parbox[t][0.3cm]{14.505521cm}{\raggedright E(level),J\ensuremath{^{\pi}},L: From (\href{https://www.nndc.bnl.gov/nsr/nsrlink.jsp?2004Bl21,B}{2004Bl21}). L and J are determined from a DWBA analysis (using a \textit{d}\ensuremath{_{\textnormal{5/2}}} proton\vspace{0.1cm}}&\\
&&&&\parbox[t][0.3cm]{14.505521cm}{\raggedright {\ }{\ }{\ }transfer to populate the 3376 keV state in \ensuremath{^{\textnormal{18}}}Ne), and by comparison with the single particle neutron\vspace{0.1cm}}&\\
&&&&\parbox[t][0.3cm]{14.505521cm}{\raggedright {\ }{\ }{\ }amplitude in \ensuremath{^{\textnormal{18}}}O. The spectroscopic amplitudes for the 4\ensuremath{^{\textnormal{+}}} state are reported in (\href{https://www.nndc.bnl.gov/nsr/nsrlink.jsp?2004Bl21,B}{2004Bl21}) to be about\vspace{0.1cm}}&\\
&&&&\parbox[t][0.3cm]{14.505521cm}{\raggedright {\ }{\ }{\ }30\% larger than expected.\vspace{0.1cm}}&\\
\multicolumn{1}{r@{}}{3616}&\multicolumn{1}{@{.}l}{5}&\multicolumn{1}{l}{2\ensuremath{^{+}}}&&&\\
\end{longtable}
\parbox[b][0.3cm]{17.7cm}{\makebox[1ex]{\ensuremath{^{\hypertarget{NE14LEVEL0}{a}}}} From the Adopted Levels of \ensuremath{^{\textnormal{18}}}Ne unless otherwise noted.}\\
\vspace{0.5cm}
\underline{$\gamma$($^{18}$Ne)}\\
\begin{longtable}{ccccccc@{}c@{\extracolsep{\fill}}c}
\multicolumn{2}{c}{E\ensuremath{_{\gamma}}\ensuremath{^{\hyperlink{NE14GAMMA0}{a}}}}&\multicolumn{2}{c}{E\ensuremath{_{i}}(level)}&J\ensuremath{^{\pi}_{i}}&\multicolumn{2}{c}{E\ensuremath{_{f}}}&J\ensuremath{^{\pi}_{f}}&\\[-.2cm]
\multicolumn{2}{c}{\hrulefill}&\multicolumn{2}{c}{\hrulefill}&\hrulefill&\multicolumn{2}{c}{\hrulefill}&\hrulefill&
\endfirsthead
\multicolumn{1}{r@{}}{1488}&\multicolumn{1}{@{.}l}{9}&\multicolumn{1}{r@{}}{3376}&\multicolumn{1}{@{.}l}{4}&\multicolumn{1}{l}{4\ensuremath{^{+}}}&\multicolumn{1}{r@{}}{1887}&\multicolumn{1}{@{.}l}{4}&\multicolumn{1}{@{}l}{2\ensuremath{^{+}}}&\\
\multicolumn{1}{r@{}}{1729}&\multicolumn{1}{@{.}l}{2}&\multicolumn{1}{r@{}}{3616}&\multicolumn{1}{@{.}l}{5}&\multicolumn{1}{l}{2\ensuremath{^{+}}}&\multicolumn{1}{r@{}}{1887}&\multicolumn{1}{@{.}l}{4}&\multicolumn{1}{@{}l}{2\ensuremath{^{+}}}&\\
\multicolumn{1}{r@{}}{1887}&\multicolumn{1}{@{.}l}{3}&\multicolumn{1}{r@{}}{1887}&\multicolumn{1}{@{.}l}{4}&\multicolumn{1}{l}{2\ensuremath{^{+}}}&\multicolumn{1}{r@{}}{0}&\multicolumn{1}{@{}l}{}&\multicolumn{1}{@{}l}{0\ensuremath{^{+}}}&\\
\end{longtable}
\parbox[b][0.3cm]{17.7cm}{\makebox[1ex]{\ensuremath{^{\hypertarget{NE14GAMMA0}{a}}}} The \ensuremath{\gamma}-rays were observed in (\href{https://www.nndc.bnl.gov/nsr/nsrlink.jsp?2003Bl12,B}{2003Bl12}: see Fig. 1); however, the authors did not present the \ensuremath{\gamma}-ray energies or the excitation}\\
\parbox[b][0.3cm]{17.7cm}{{\ }{\ }energies of the states involved in those transitions. The \ensuremath{\gamma}-ray energies are therefore taken from the Adopted Gammas of \ensuremath{^{\textnormal{18}}}Ne.}\\
\vspace{0.5cm}
\clearpage
\begin{figure}[h]
\begin{center}
\includegraphics{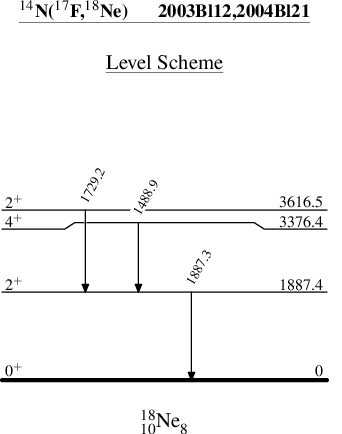}\\
\end{center}
\end{figure}
\clearpage
\subsection[\hspace{-0.2cm}\ensuremath{^{\textnormal{16}}}O(\ensuremath{^{\textnormal{3}}}He,n)]{ }
\vspace{-27pt}
\vspace{0.3cm}
\hypertarget{NE15}{{\bf \small \underline{\ensuremath{^{\textnormal{16}}}O(\ensuremath{^{\textnormal{3}}}He,n)\hspace{0.2in}\href{https://www.nndc.bnl.gov/nsr/nsrlink.jsp?1953Ku08,B}{1953Ku08},\href{https://www.nndc.bnl.gov/nsr/nsrlink.jsp?2012Fo29,B}{2012Fo29}}}}\\
\vspace{4pt}
\vspace{8pt}
\parbox[b][0.3cm]{17.7cm}{\addtolength{\parindent}{-0.2in}\href{https://www.nndc.bnl.gov/nsr/nsrlink.jsp?1953Ku08,B}{1953Ku08}: \ensuremath{^{\textnormal{16}}}O(\ensuremath{^{\textnormal{3}}}He,n)\ensuremath{^{\textnormal{18}}}Ne(\ensuremath{\beta}\ensuremath{^{\textnormal{+}}})\ensuremath{^{\textnormal{18}}}F, \ensuremath{^{\textnormal{16}}}O(\ensuremath{^{\textnormal{3}}}He,p) E=21 MeV; measured the \ensuremath{^{\textnormal{18}}}F radiative decay and evaluated the relative cross}\\
\parbox[b][0.3cm]{17.7cm}{sections of the \ensuremath{^{\textnormal{3}}}He+\ensuremath{^{\textnormal{nat}}}O and \ensuremath{^{\textnormal{3}}}He+\ensuremath{^{\textnormal{nat}}}Ni reactions.}\\
\parbox[b][0.3cm]{17.7cm}{\addtolength{\parindent}{-0.2in}\href{https://www.nndc.bnl.gov/nsr/nsrlink.jsp?1960Aj01,B}{1960Aj01}: \ensuremath{^{\textnormal{16}}}O(\ensuremath{^{\textnormal{3}}}He,n); measured the mass of \ensuremath{^{\textnormal{18}}}Ne.}\\
\parbox[b][0.3cm]{17.7cm}{\addtolength{\parindent}{-0.2in}\href{https://www.nndc.bnl.gov/nsr/nsrlink.jsp?1960Aj03,B}{1960Aj03}: \ensuremath{^{\textnormal{16}}}O(\ensuremath{^{\textnormal{3}}}He,n) E=5.51 MeV; measured E\ensuremath{_{\textnormal{n}}} using emulsions plates mounted at \ensuremath{\theta}\ensuremath{_{\textnormal{lab}}}=0\ensuremath{^\circ}, 15\ensuremath{^\circ}, 30\ensuremath{^\circ}, 45\ensuremath{^\circ}, 90\ensuremath{^\circ}, and 135\ensuremath{^\circ};}\\
\parbox[b][0.3cm]{17.7cm}{populated \ensuremath{^{\textnormal{18}}}Ne\ensuremath{_{\textnormal{g.s.}}}; measured \ensuremath{\sigma}(\ensuremath{\theta}) for \ensuremath{^{\textnormal{18}}}Ne\ensuremath{_{\textnormal{g.s.}}}; deduced Q\ensuremath{_{\textnormal{g.s.}}}={\textminus}3.19 MeV \textit{4}; deduced mass(\ensuremath{^{\textnormal{18}}}Ne)=18.00570 amu \textit{4} and mass}\\
\parbox[b][0.3cm]{17.7cm}{excess(\ensuremath{^{\textnormal{18}}}Ne)=10.64 MeV \textit{4}; deduced d\ensuremath{\sigma}/d\ensuremath{\Omega}(\ensuremath{\theta}=0\ensuremath{^\circ})=1.6 mb/sr \textit{3} for the \ensuremath{^{\textnormal{16}}}O(\ensuremath{^{\textnormal{3}}}He,n)\ensuremath{^{\textnormal{18}}}Ne\ensuremath{_{\textnormal{g.s.}}} reaction. No evidence for the}\\
\parbox[b][0.3cm]{17.7cm}{existence of the \ensuremath{^{\textnormal{18}}}Ne*(114 keV \textit{15}) proposed by (\href{https://www.nndc.bnl.gov/nsr/nsrlink.jsp?1959Du81,B}{1959Du81}) was found.}\\
\parbox[b][0.3cm]{17.7cm}{\addtolength{\parindent}{-0.2in}\href{https://www.nndc.bnl.gov/nsr/nsrlink.jsp?1961Du02,B}{1961Du02}: \ensuremath{^{\textnormal{16}}}O(\ensuremath{^{\textnormal{3}}}He,n) E=4.5 and 5.6 MeV; measured the excitation energy spectrum of the \ensuremath{^{\textnormal{16}}}O(\ensuremath{^{\textnormal{3}}}He,n) reaction, and the yield of}\\
\parbox[b][0.3cm]{17.7cm}{fast neutrons as well as the ratio of slow-to-fast neutrons as a function of the incident energy; measured the ground state threshold}\\
\parbox[b][0.3cm]{17.7cm}{energy of 3811 keV \textit{15} for the \ensuremath{^{\textnormal{16}}}O(\ensuremath{^{\textnormal{3}}}He,n) reaction (published in (\href{https://www.nndc.bnl.gov/nsr/nsrlink.jsp?1959Du81,B}{1959Du81})); interpreted the observed break in the excitation}\\
\parbox[b][0.3cm]{17.7cm}{curve at 3.95 MeV as a resonance in the compound nucleus (\ensuremath{^{\textnormal{19}}}Ne); deduced m=18.011446 amu \textit{14} and \ensuremath{\Delta}M=10.658 MeV \textit{13} for}\\
\parbox[b][0.3cm]{17.7cm}{the atomic mass and mass excess of \ensuremath{^{\textnormal{18}}}Ne, respectively; deduced Q\ensuremath{_{\textnormal{g.s.}}}={\textminus}3206 keV \textit{13}; deduced the end-point positron energy of}\\
\parbox[b][0.3cm]{17.7cm}{3423 keV \textit{2} for the decay of \ensuremath{^{\textnormal{18}}}Ne\ensuremath{_{\textnormal{g.s.}}} to \ensuremath{^{\textnormal{18}}}F\ensuremath{_{\textnormal{g.s.}}}. No excited state of \ensuremath{^{\textnormal{18}}}Ne was observed and the authors ruled out the E\ensuremath{_{\textnormal{x}}}=114 keV}\\
\parbox[b][0.3cm]{17.7cm}{\textit{15} state proposed by (\href{https://www.nndc.bnl.gov/nsr/nsrlink.jsp?1959Du81,B}{1959Du81}).}\\
\parbox[b][0.3cm]{17.7cm}{\addtolength{\parindent}{-0.2in}\href{https://www.nndc.bnl.gov/nsr/nsrlink.jsp?1961Ga01,B}{1961Ga01}: \ensuremath{^{\textnormal{16}}}O(\ensuremath{^{\textnormal{3}}}He,n) E=5.6 and 6.09 MeV; measured neutrons using a pulsed beam time-of-flight neutron spectrometer;}\\
\parbox[b][0.3cm]{17.7cm}{measured \ensuremath{\sigma}(\ensuremath{\theta}) of n\ensuremath{_{\textnormal{0}}} group at \ensuremath{\theta}\ensuremath{_{\textnormal{c.m.}}}=5\ensuremath{^\circ}{\textminus}175\ensuremath{^\circ}. Angular distributions were observed to be forward peaked. The neutron angular}\\
\parbox[b][0.3cm]{17.7cm}{distribution data were fitted with double-stripping theoretical cross sections based on the zero order spherical Bessel functions}\\
\parbox[b][0.3cm]{17.7cm}{formalism of (\href{https://www.nndc.bnl.gov/nsr/nsrlink.jsp?1960Ne21,B}{1960Ne21}). Good agreement was obtained with the theoretical prediction for \textit{l}=0 double stripping. This led to the}\\
\parbox[b][0.3cm]{17.7cm}{J\ensuremath{^{\ensuremath{\pi}}}=0\ensuremath{^{\textnormal{+}}} assignment for the \ensuremath{^{\textnormal{18}}}Ne\ensuremath{_{\textnormal{g.s.}}}; however a knock-out mechanism could not be excluded.}\\
\parbox[b][0.3cm]{17.7cm}{\addtolength{\parindent}{-0.2in}\href{https://www.nndc.bnl.gov/nsr/nsrlink.jsp?1961To03,B}{1961To03}: \ensuremath{^{\textnormal{16}}}O(\ensuremath{^{\textnormal{3}}}He,n) E\ensuremath{\leq}10 MeV; measured the \ensuremath{^{\textnormal{18}}}Ne excitation energy spectra for E(\ensuremath{^{\textnormal{3}}}He)=2-5.7 MeV using two long BF\ensuremath{_{\textnormal{3}}}}\\
\parbox[b][0.3cm]{17.7cm}{counters at \ensuremath{\theta}\ensuremath{_{\textnormal{lab}}}=0\ensuremath{^\circ} and 90\ensuremath{^\circ}; measured the time-of-flight spectrum of the \ensuremath{^{\textnormal{16}}}O(\ensuremath{^{\textnormal{3}}}He,n) reaction at \ensuremath{\theta}\ensuremath{_{\textnormal{lab}}}=20\ensuremath{^\circ}, 25\ensuremath{^\circ}, and 35\ensuremath{^\circ}; deduced}\\
\parbox[b][0.3cm]{17.7cm}{Q\ensuremath{_{\textnormal{g.s.}}}={\textminus}3199 keV \textit{6} and ground state threshold energy of 3802 keV \textit{7} for the \ensuremath{^{\textnormal{16}}}O(\ensuremath{^{\textnormal{3}}}He,n) reaction; deduced \ensuremath{\Delta}M=10.649 MeV \textit{8}}\\
\parbox[b][0.3cm]{17.7cm}{and an end point energy of 3439 keV \textit{12} for the \ensuremath{^{\textnormal{18}}}Ne\ensuremath{_{\textnormal{g.s.}}}\ensuremath{\rightarrow}\ensuremath{^{\textnormal{18}}}F*(1042 keV)+\ensuremath{\beta}\ensuremath{^{\textnormal{+}}}; confirmed the existence of no excited state below}\\
\parbox[b][0.3cm]{17.7cm}{E\ensuremath{_{\textnormal{x}}}=1.5 MeV in \ensuremath{^{\textnormal{18}}}Ne; deduced 3 excited levels in \ensuremath{^{\textnormal{18}}}Ne at E\ensuremath{_{\textnormal{x}}}=1880 keV \textit{10}, 3362 keV \textit{11}, and 3608 keV \textit{12} from the measured}\\
\parbox[b][0.3cm]{17.7cm}{ratio of slow-to-fast neutrons at incident energies of 3.5-10 MeV; deduced Q-values of {\textminus}5079 keV \textit{8}, {\textminus}6561 keV \textit{9}, and {\textminus}6807 keV}\\
\parbox[b][0.3cm]{17.7cm}{\textit{10} for the three measured excited states, respectively.}\\
\parbox[b][0.3cm]{17.7cm}{\addtolength{\parindent}{-0.2in}\href{https://www.nndc.bnl.gov/nsr/nsrlink.jsp?1962Ma61,B}{1962Ma61}: \ensuremath{^{\textnormal{16}}}O(\ensuremath{^{\textnormal{3}}}He,n) E=1-31 MeV; measured the induced radioactivity from the activation by calibrated end-window}\\
\parbox[b][0.3cm]{17.7cm}{proportional counters and NaI scintillators; measured the combined cross section of the \ensuremath{^{\textnormal{16}}}O(\ensuremath{^{\textnormal{3}}}He,p) and \ensuremath{^{\textnormal{16}}}O(\ensuremath{^{\textnormal{3}}}He,n) reactions; the}\\
\parbox[b][0.3cm]{17.7cm}{summed cross section was highest at 7.7 MeV incident energy and was measured to be \ensuremath{\sigma}=400 mb for the \ensuremath{^{\textnormal{16}}}O(\ensuremath{^{\textnormal{3}}}He,n)\ensuremath{^{\textnormal{18}}}Ne\ensuremath{_{\textnormal{g.s.}}}}\\
\parbox[b][0.3cm]{17.7cm}{reaction. The authors deduced Q\ensuremath{_{\textnormal{g.s.}}}={\textminus}3000 keV.}\\
\parbox[b][0.3cm]{17.7cm}{\addtolength{\parindent}{-0.2in}\href{https://www.nndc.bnl.gov/nsr/nsrlink.jsp?1964Br13,B}{1964Br13}: \ensuremath{^{\textnormal{16}}}O(\ensuremath{^{\textnormal{3}}}He,n) E=25.4 MeV; measured neutrons with E\ensuremath{_{\textnormal{n}}}=0-25 MeV at \ensuremath{\theta}\ensuremath{_{\textnormal{lab}}}=0\ensuremath{^\circ} using a liquid hydrogen bubble chamber at}\\
\parbox[b][0.3cm]{17.7cm}{26 K; measured \ensuremath{\sigma}(E\ensuremath{_{\textnormal{n}}},\ensuremath{\theta}) and deduced d\ensuremath{\sigma}/d\ensuremath{\Omega}(\ensuremath{\theta}\ensuremath{_{\textnormal{lab}}}=0\ensuremath{^\circ})=2.0 mb/sr \textit{4}; observed 5 levels of \ensuremath{^{\textnormal{18}}}Ne at E\ensuremath{_{\textnormal{x}}}=0, 1.88 MeV, 3.36 MeV,}\\
\parbox[b][0.3cm]{17.7cm}{3.61 MeV, and either one or a group of levels at 5.2 MeV \textit{3}; deduced upper limits of 15.8 MeV and 16.4 MeV on the neutron}\\
\parbox[b][0.3cm]{17.7cm}{energies from the \ensuremath{^{\textnormal{16}}}O(\ensuremath{^{\textnormal{3}}}He,n2p) and \ensuremath{^{\textnormal{16}}}O(\ensuremath{^{\textnormal{3}}}He,np) reactions, respectively. These reactions produced a continuum in the neutron}\\
\parbox[b][0.3cm]{17.7cm}{spectrum.}\\
\parbox[b][0.3cm]{17.7cm}{\addtolength{\parindent}{-0.2in}\href{https://www.nndc.bnl.gov/nsr/nsrlink.jsp?1965Br42,B}{1965Br42}: \ensuremath{^{\textnormal{16}}}O(\ensuremath{^{\textnormal{3}}}He,n) E=4.9-31 MeV; measured products of various \ensuremath{^{\textnormal{3}}}He induced reactions; measured the combined cross section}\\
\parbox[b][0.3cm]{17.7cm}{of \ensuremath{^{\textnormal{16}}}O(\ensuremath{^{\textnormal{3}}}He,p) and \ensuremath{^{\textnormal{16}}}O(\ensuremath{^{\textnormal{3}}}He,n) reactions; estimated that the cross section of the latter reaction populating \ensuremath{^{\textnormal{18}}}Ne\ensuremath{_{\textnormal{g.s.}}} is \ensuremath{\sigma}=20 mb.}\\
\parbox[b][0.3cm]{17.7cm}{\addtolength{\parindent}{-0.2in}\href{https://www.nndc.bnl.gov/nsr/nsrlink.jsp?1966Kr05,B}{1966Kr05}: \ensuremath{^{\textnormal{16}}}O(\ensuremath{^{\textnormal{3}}}He,n) E=11 MeV; measured neutrons with a double scatter time-of-flight spectrometer with 1 MeV resolution;}\\
\parbox[b][0.3cm]{17.7cm}{measured \ensuremath{\sigma}(E\ensuremath{_{\textnormal{n}}},\ensuremath{\theta}) at 8 angles between \ensuremath{\theta}\ensuremath{_{\textnormal{lab}}}=6\ensuremath{^\circ}{\textminus}50\ensuremath{^\circ} and deduced (d\ensuremath{\sigma}/d\ensuremath{\Omega})\ensuremath{^{\textnormal{max}}_{\textnormal{c.m.}}}=21 mb/sr \textit{4} at \ensuremath{\theta}\ensuremath{_{\textnormal{c.m.}}}\ensuremath{\sim}15\ensuremath{^\circ}; populated \ensuremath{^{\textnormal{18}}}Ne\ensuremath{_{\textnormal{g.s.}}}}\\
\parbox[b][0.3cm]{17.7cm}{and \ensuremath{^{\textnormal{18}}}Ne*(1.88 MeV). Using the plane wave double stripping theory of (\href{https://www.nndc.bnl.gov/nsr/nsrlink.jsp?1960Ne21,B}{1960Ne21}), the authors deduced L=0 and L=2, and}\\
\parbox[b][0.3cm]{17.7cm}{J\ensuremath{^{\ensuremath{\pi}}}=0\ensuremath{^{\textnormal{+}}} and J\ensuremath{^{\ensuremath{\pi}}}=2\ensuremath{^{\textnormal{+}}} for these states, respectively.}\\
\parbox[b][0.3cm]{17.7cm}{\addtolength{\parindent}{-0.2in}\href{https://www.nndc.bnl.gov/nsr/nsrlink.jsp?1967Mc03,B}{1967Mc03}: \ensuremath{^{\textnormal{16}}}O(\ensuremath{^{\textnormal{3}}}He,n) E=4.9, 5.2, and 5.6 MeV; measured neutrons using a time-of-flight neutron spectrometer that consisted of a}\\
\parbox[b][0.3cm]{17.7cm}{NE-213 liquid scintillator coupled to a PMT and long counters placed at \ensuremath{\theta}\ensuremath{_{\textnormal{lab}}}=90\ensuremath{^\circ} and 120\ensuremath{^\circ}; measured \ensuremath{\sigma}(E;E\ensuremath{_{\textnormal{n}}},\ensuremath{\theta}) for}\\
\parbox[b][0.3cm]{17.7cm}{\ensuremath{^{\textnormal{16}}}O(\ensuremath{^{\textnormal{3}}}He,n\ensuremath{_{\textnormal{0}}}); performed a plane wave Born approximation calculation and deduced L=0 for \ensuremath{^{\textnormal{18}}}Ne\ensuremath{_{\textnormal{g.s.}}}.}\\
\parbox[b][0.3cm]{17.7cm}{\addtolength{\parindent}{-0.2in}\href{https://www.nndc.bnl.gov/nsr/nsrlink.jsp?1968Sh09,B}{1968Sh09}: \ensuremath{^{\textnormal{16}}}O(\ensuremath{^{\textnormal{3}}}He,n) E=8.5, 9, 9.5, and 10 MeV; measured neutrons using a time-of-flight spectrometer that consisted of a Pilot}\\
\parbox[b][0.3cm]{17.7cm}{B scintillator coupled to a PMT; populated \ensuremath{^{\textnormal{18}}}Ne levels at E\ensuremath{_{\textnormal{x}}}=0, 1.88, 3.36, 3.61, and 4.55 MeV; measured \ensuremath{\sigma}(E\ensuremath{_{\textnormal{n}}},\ensuremath{\theta}) for \ensuremath{^{\textnormal{18}}}Ne\ensuremath{_{\textnormal{g.s.}}} at}\\
\parbox[b][0.3cm]{17.7cm}{10 MeV. Energy resolution: 70-300 keV.}\\
\parbox[b][0.3cm]{17.7cm}{\addtolength{\parindent}{-0.2in}\href{https://www.nndc.bnl.gov/nsr/nsrlink.jsp?1968To09,B}{1968To09}, J. H. Towle and G. J. Wall, Conf. on Low and medium energy nuclear physics, A. E. R. E., Harwell, March 1968:}\\
\parbox[b][0.3cm]{17.7cm}{\ensuremath{^{\textnormal{16}}}O(\ensuremath{^{\textnormal{3}}}He,n) E=9.15-10.55 MeV; measured neutrons using a neutron time-of-flight spectrometer that consisted of a liquid NE-213}\\
\parbox[b][0.3cm]{17.7cm}{scintillator coupled to 2 PMTs; measured \ensuremath{\sigma}(E;E\ensuremath{_{\textnormal{n}}},\ensuremath{\theta}) at \ensuremath{\theta}\ensuremath{_{\textnormal{c.m.}}}=0\ensuremath{^\circ}{\textminus}150\ensuremath{^\circ}; measured \ensuremath{^{\textnormal{18}}}Ne excitation energy spectrum at \ensuremath{\theta}\ensuremath{_{\textnormal{lab}}}=20\ensuremath{^\circ};}\\
\parbox[b][0.3cm]{17.7cm}{deduced \ensuremath{^{\textnormal{18}}}Ne level energies, L, J, and \ensuremath{\pi} for the \ensuremath{^{\textnormal{18}}}Ne(0, 1.88, 3.36, 3.61, 4.59 MeV) states from a double stripping DWBA}\\
\parbox[b][0.3cm]{17.7cm}{analysis.}\\
\clearpage
\vspace{0.3cm}
{\bf \small \underline{\ensuremath{^{\textnormal{16}}}O(\ensuremath{^{\textnormal{3}}}He,n)\hspace{0.2in}\href{https://www.nndc.bnl.gov/nsr/nsrlink.jsp?1953Ku08,B}{1953Ku08},\href{https://www.nndc.bnl.gov/nsr/nsrlink.jsp?2012Fo29,B}{2012Fo29} (continued)}}\\
\vspace{0.3cm}
\parbox[b][0.3cm]{17.7cm}{\addtolength{\parindent}{-0.2in}\href{https://www.nndc.bnl.gov/nsr/nsrlink.jsp?1970Ad02,B}{1970Ad02}: \ensuremath{^{\textnormal{16}}}O(\ensuremath{^{\textnormal{3}}}He,n) E=8.84-13.5 MeV; measured neutrons using a time-of-flight spectrometer consisting of a Pilot B plastic}\\
\parbox[b][0.3cm]{17.7cm}{scintillator placed at \ensuremath{\theta}\ensuremath{_{\textnormal{lab}}}=0\ensuremath{^\circ}{\textminus}150\ensuremath{^\circ}; energy resolution was \ensuremath{<}50 keV for 1 MeV neutrons; measured \ensuremath{\sigma}(E;E\ensuremath{_{\textnormal{n}}},\ensuremath{\theta}) at E=9, 9.5, 10.5,}\\
\parbox[b][0.3cm]{17.7cm}{11.5, and 12.5 MeV; deduced \ensuremath{^{\textnormal{18}}}Ne levels, L and J\ensuremath{^{\ensuremath{\pi}}} for the states up to E\ensuremath{_{\textnormal{x}}}=5.14 MeV. Comparisons with (\ensuremath{^{\textnormal{3}}}He,p) and (t,p)}\\
\parbox[b][0.3cm]{17.7cm}{reactions are discussed.}\\
\parbox[b][0.3cm]{17.7cm}{\addtolength{\parindent}{-0.2in}\href{https://www.nndc.bnl.gov/nsr/nsrlink.jsp?1971NeZR,B}{1971NeZR}, \href{https://www.nndc.bnl.gov/nsr/nsrlink.jsp?1974Ne04,B}{1974Ne04}: \ensuremath{^{\textnormal{16}}}O(\ensuremath{^{\textnormal{3}}}He,n) E=10-20 MeV; measured neutrons using a time-of-flight spectrometer; measured \ensuremath{\sigma}(E\ensuremath{_{\textnormal{n}}},\ensuremath{\theta});}\\
\parbox[b][0.3cm]{17.7cm}{deduced \ensuremath{^{\textnormal{18}}}Ne level energies up to E\ensuremath{_{\textnormal{x}}}=8.5 MeV, L and J\ensuremath{^{\ensuremath{\pi}}} for the \ensuremath{^{\textnormal{18}}}Ne*(4513, 4587 keV) states; discussed mirror states and}\\
\parbox[b][0.3cm]{17.7cm}{calculated their Coulomb shift; discussed two-nucleon configurations for the 0\ensuremath{^{\textnormal{+}}_{\textnormal{3}}} state in \ensuremath{^{\textnormal{18}}}Ne; models that assign most of the \textit{s}\ensuremath{_{\textnormal{1/2}}}}\\
\parbox[b][0.3cm]{17.7cm}{strength to the \ensuremath{^{\textnormal{18}}}Ne(0\ensuremath{^{\textnormal{+}}_{\textnormal{3}}}) state (\href{https://www.nndc.bnl.gov/nsr/nsrlink.jsp?1969Be94,B}{1969Be94}, \href{https://www.nndc.bnl.gov/nsr/nsrlink.jsp?1970El23,B}{1970El23}) rather than to the \ensuremath{^{\textnormal{18}}}Ne(0\ensuremath{^{\textnormal{+}}_{\textnormal{2}}}) level (\href{https://www.nndc.bnl.gov/nsr/nsrlink.jsp?1965En02,B}{1965En02}, \href{https://www.nndc.bnl.gov/nsr/nsrlink.jsp?1969Zu03,B}{1969Zu03}, \href{https://www.nndc.bnl.gov/nsr/nsrlink.jsp?1972Ka01,B}{1972Ka01}) are}\\
\parbox[b][0.3cm]{17.7cm}{preferred.}\\
\parbox[b][0.3cm]{17.7cm}{\addtolength{\parindent}{-0.2in}\href{https://www.nndc.bnl.gov/nsr/nsrlink.jsp?1974PeZO,B}{1974PeZO}, \href{https://www.nndc.bnl.gov/nsr/nsrlink.jsp?1975Pe11,B}{1975Pe11}: \ensuremath{^{\textnormal{16}}}O(\ensuremath{^{\textnormal{3}}}He,n) E=18.3 MeV; measured neutrons using a neutron time-of-flight spectrometer at \ensuremath{\theta}\ensuremath{_{\textnormal{lab}}}=0\ensuremath{^\circ}{\textminus}45\ensuremath{^\circ}}\\
\parbox[b][0.3cm]{17.7cm}{with a resolution of 120 keV (FWHM); measured \ensuremath{\sigma}(E\ensuremath{_{\textnormal{n}}},\ensuremath{\theta}); deduced \ensuremath{^{\textnormal{18}}}Ne level energies, L, and J\ensuremath{^{\ensuremath{\pi}}} values for the ground and first}\\
\parbox[b][0.3cm]{17.7cm}{excited states using a zero-range DWBA analysis via DWUCK4.}\\
\parbox[b][0.3cm]{17.7cm}{\addtolength{\parindent}{-0.2in}\href{https://www.nndc.bnl.gov/nsr/nsrlink.jsp?1977Ev01,B}{1977Ev01}: \ensuremath{^{\textnormal{16}}}O(\ensuremath{^{\textnormal{3}}}He,n) E=15, 18, and 21 MeV; measured neutrons using the Munich neutron time-of-flight spectrometer with six}\\
\parbox[b][0.3cm]{17.7cm}{counters at \ensuremath{\theta}\ensuremath{_{\textnormal{lab}}}=0\ensuremath{^\circ}{\textminus}40\ensuremath{^\circ} and two NE-213 scintillators at \ensuremath{\theta}\ensuremath{_{\textnormal{lab}}}=5\ensuremath{^\circ} and 15\ensuremath{^\circ}; measured \ensuremath{\sigma}(E\ensuremath{_{\textnormal{n}}},\ensuremath{\theta}); deduced \ensuremath{^{\textnormal{18}}}Ne level energies for the}\\
\parbox[b][0.3cm]{17.7cm}{states with E\ensuremath{_{\textnormal{x}}}\ensuremath{\leq}8070 keV; deduced J\ensuremath{^{\ensuremath{\pi}}} values using a zero-range DWBA analysis via DWUCK2; comparison with shell model}\\
\parbox[b][0.3cm]{17.7cm}{calculations and the two-proton amplitudes used in the DWUCK analysis are presented.}\\
\parbox[b][0.3cm]{17.7cm}{\addtolength{\parindent}{-0.2in}\href{https://www.nndc.bnl.gov/nsr/nsrlink.jsp?1981Ne09,B}{1981Ne09}: \ensuremath{^{\textnormal{16}}}O(\ensuremath{^{\textnormal{3}}}He,n) E=10-22 MeV; measured neutrons using a time-of-flight spectrometer that consisted of a plastic scintillator;}\\
\parbox[b][0.3cm]{17.7cm}{measured \ensuremath{\sigma}(E\ensuremath{_{\textnormal{n}}},\ensuremath{\theta}) for \ensuremath{\theta}\ensuremath{_{\textnormal{c.m.}}}=0\ensuremath{^\circ}{\textminus}150\ensuremath{^\circ}; deduced \ensuremath{^{\textnormal{18}}}Ne level energies, widths, and J\ensuremath{^{\ensuremath{\pi}}} values (using a DWBA analysis with the code}\\
\parbox[b][0.3cm]{17.7cm}{JULIE) for states with E\ensuremath{_{\textnormal{x}}}\ensuremath{\leq}8100 keV; calculated two-particle Coulomb energy shifts for A=18, T=1 states and found that the}\\
\parbox[b][0.3cm]{17.7cm}{difference between the excitation energy of the 4.59-MeV 0\ensuremath{^{\textnormal{+}}} state and the analog state in \ensuremath{^{\textnormal{18}}}O could be accounted for by a double}\\
\parbox[b][0.3cm]{17.7cm}{Thomas-Ehrman shift, which gives strong evidence for the predominantly \textit{s}\ensuremath{_{\textnormal{1/2}}} character of these states.}\\
\parbox[b][0.3cm]{17.7cm}{\addtolength{\parindent}{-0.2in}\href{https://www.nndc.bnl.gov/nsr/nsrlink.jsp?1989GaZW,B}{1989GaZW}, \href{https://www.nndc.bnl.gov/nsr/nsrlink.jsp?1990GaZR,B}{1990GaZR}, \href{https://www.nndc.bnl.gov/nsr/nsrlink.jsp?1990GaZW,B}{1990GaZW}, \href{https://www.nndc.bnl.gov/nsr/nsrlink.jsp?1991Ga03,B}{1991Ga03}: \ensuremath{^{\textnormal{16}}}O(\ensuremath{^{\textnormal{3}}}He,n) E=9.5-10.5, 11, and 12 MeV; measured neutrons using a}\\
\parbox[b][0.3cm]{17.7cm}{time-of-flight spectrometer consisting of plastic scintillators; measured \ensuremath{\sigma}(E\ensuremath{_{\textnormal{n}}},\ensuremath{\theta}) at \ensuremath{\theta}\ensuremath{_{\textnormal{lab}}}=0\ensuremath{^\circ}, 42\ensuremath{^\circ}, and 126\ensuremath{^\circ} (E\ensuremath{_{\textnormal{beam}}}=9.8-11 MeV); at}\\
\parbox[b][0.3cm]{17.7cm}{\ensuremath{\theta}\ensuremath{_{\textnormal{lab}}}=0\ensuremath{^\circ} and 35\ensuremath{^\circ} (E\ensuremath{_{\textnormal{beam}}}=10.5, 11 and 12 MeV); and at \ensuremath{\theta}\ensuremath{_{\textnormal{lab}}}=0\ensuremath{^\circ} and 124.7\ensuremath{^\circ} (E\ensuremath{_{\textnormal{beam}}}=10.9 MeV); observed evidence at}\\
\parbox[b][0.3cm]{17.7cm}{\ensuremath{\theta}\ensuremath{_{\textnormal{lab}}}=124.7\ensuremath{^\circ} for a previously unresolved level in the form of excess counts that could not be attributed to background or any source}\\
\parbox[b][0.3cm]{17.7cm}{other than \ensuremath{^{\textnormal{18}}}Ne. Using the \ensuremath{^{\textnormal{18}}}Ne mass excess of 5319 keV \textit{5} (\href{https://www.nndc.bnl.gov/nsr/nsrlink.jsp?1987Aj02,B}{1987Aj02}), an excitation energy of 4561 keV \textit{9} was inferred for this}\\
\parbox[b][0.3cm]{17.7cm}{missing 3\ensuremath{^{\textnormal{+}}} state by (\href{https://www.nndc.bnl.gov/nsr/nsrlink.jsp?1991Ga03,B}{1991Ga03}). Deduced \ensuremath{^{\textnormal{18}}}Ne level energies, Q-values and J\ensuremath{^{\ensuremath{\pi}}} values for the \ensuremath{^{\textnormal{18}}}Ne*(0, 1887, 3376, 3576, 3616,}\\
\parbox[b][0.3cm]{17.7cm}{4520, 4561, and 4589 keV) states; deduced widths for the \ensuremath{^{\textnormal{18}}}Ne*(4520, 4561, and 4589 keV) states; computed the \ensuremath{^{\textnormal{17}}}F(p,\ensuremath{\gamma})}\\
\parbox[b][0.3cm]{17.7cm}{resonance properties, reaction rate, and the astrophysical S-factor; discussed the astrophysical implications.}\\
\parbox[b][0.3cm]{17.7cm}{\addtolength{\parindent}{-0.2in}\href{https://www.nndc.bnl.gov/nsr/nsrlink.jsp?1994Ma14,B}{1994Ma14}: \ensuremath{^{\textnormal{16}}}O(\ensuremath{^{\textnormal{3}}}He,n) E=7.31 MeV; measured the neutrons corresponding to the \ensuremath{^{\textnormal{18}}}Ne*(0, 1.89, 3.38 MeV) states using a}\\
\parbox[b][0.3cm]{17.7cm}{time-of-flight spectrometer consisting of two liquid scintillators at \ensuremath{\theta}\ensuremath{_{\textnormal{lab}}}=0\ensuremath{^\circ} and 8.75\ensuremath{^\circ}; deduced \ensuremath{^{\textnormal{18}}}Ne level energies and and}\\
\parbox[b][0.3cm]{17.7cm}{Q-values for the states at E\ensuremath{_{\textnormal{x}}}=0, 1.89 MeV, and 3.38 MeV; deduced mass excesses for the \ensuremath{^{\textnormal{18}}}Ne*(0, 1.89 MeV) states.}\\
\parbox[b][0.3cm]{17.7cm}{\addtolength{\parindent}{-0.2in}\href{https://www.nndc.bnl.gov/nsr/nsrlink.jsp?1996Ha26,B}{1996Ha26}: \ensuremath{^{\textnormal{16}}}O(\ensuremath{^{\textnormal{3}}}He,n) E=10.9-14.5 MeV; measured neutrons using a time-of-flight spectrometer consisting of 3 liquid scintillators,}\\
\parbox[b][0.3cm]{17.7cm}{one fixed at \ensuremath{\theta}\ensuremath{_{\textnormal{lab}}}=0\ensuremath{^\circ}, the other two at \ensuremath{\theta}\ensuremath{_{\textnormal{lab}}}=11\ensuremath{^\circ}, 23\ensuremath{^\circ}, 34\ensuremath{^\circ}, 47\ensuremath{^\circ}, 64\ensuremath{^\circ}, and 79\ensuremath{^\circ}; measured \ensuremath{\sigma}(\ensuremath{\theta}) at E\ensuremath{_{\textnormal{lab}}}=14.5 MeV; deduced \ensuremath{^{\textnormal{18}}}Ne}\\
\parbox[b][0.3cm]{17.7cm}{level energies, \ensuremath{\Gamma}, and J\ensuremath{^{\ensuremath{\pi}}} values for the E\ensuremath{_{\textnormal{x}}}\ensuremath{\leq}8110 keV states. The J\ensuremath{^{\ensuremath{\pi}}} values were deduced using a combination of penetrability}\\
\parbox[b][0.3cm]{17.7cm}{considerations (i.e., generally a larger width is expected for a state that decays via a proton carrying a lower orbital angular}\\
\parbox[b][0.3cm]{17.7cm}{momentum than a state emitting a proton with a higher angular momentum) as well as zero-range DWBA analysis using}\\
\parbox[b][0.3cm]{17.7cm}{DWUCK4. The authors calculated the Coulomb shifts and widths for all states of \ensuremath{^{\textnormal{18}}}Ne up to E\ensuremath{_{\textnormal{x}}}=7.6 MeV; calculated the \ensuremath{^{\textnormal{14}}}O(\ensuremath{\alpha},p)}\\
\parbox[b][0.3cm]{17.7cm}{reation rate and the astrophysical S-factor and discussed the astrophysical implications.}\\
\parbox[b][0.3cm]{17.7cm}{\addtolength{\parindent}{-0.2in}\href{https://www.nndc.bnl.gov/nsr/nsrlink.jsp?2005Pa50,B}{2005Pa50}: \ensuremath{^{\textnormal{16}}}O(\ensuremath{^{\textnormal{3}}}He,n) E=9.9 MeV at \ensuremath{\theta}\ensuremath{_{\textnormal{beam}}}=0\ensuremath{^\circ}, 10.1 MeV at \ensuremath{\theta}=30\ensuremath{^\circ}, and 10.4 MeV at \ensuremath{\theta}=60\ensuremath{^\circ}; measured neutron energy spectra}\\
\parbox[b][0.3cm]{17.7cm}{using an array of two plastic and one NE-213 liquid scintillators; measured \ensuremath{\sigma}(\ensuremath{\theta}); energy resolutions: 10, 12, and 16 keV for the}\\
\parbox[b][0.3cm]{17.7cm}{aforementioned beam energies, respectively. The goal was to resolve the discrepancy between (\href{https://www.nndc.bnl.gov/nsr/nsrlink.jsp?1991Ga03,B}{1991Ga03}) and (\href{https://www.nndc.bnl.gov/nsr/nsrlink.jsp?2000Bb04,B}{2000Bb04})}\\
\parbox[b][0.3cm]{17.7cm}{regarding the excitation energy and width of the \ensuremath{^{\textnormal{18}}}Ne*(3\ensuremath{^{\textnormal{+}}_{\textnormal{1}}}) state. Deduced \ensuremath{^{\textnormal{18}}}Ne level properties for the \ensuremath{^{\textnormal{18}}}Ne*(0, 1887, 3376,}\\
\parbox[b][0.3cm]{17.7cm}{3576, 3616, 4519, 4527, and 4590 keV) states; performed Hauser-Feshbach calculations to compare with the experimental}\\
\parbox[b][0.3cm]{17.7cm}{differential cross sections for all the observed levels.}\\
\parbox[b][0.3cm]{17.7cm}{\addtolength{\parindent}{-0.2in}\href{https://www.nndc.bnl.gov/nsr/nsrlink.jsp?2010AlZZ,B}{2010AlZZ}, \href{https://www.nndc.bnl.gov/nsr/nsrlink.jsp?2010Ta17,B}{2010Ta17}, \href{https://www.nndc.bnl.gov/nsr/nsrlink.jsp?2012Al11,B}{2012Al11}: \ensuremath{^{\textnormal{16}}}O(\ensuremath{^{\textnormal{3}}}He,n) E=15 MeV; measured E(particle) for the charged-particles from the decay of \ensuremath{^{\textnormal{18}}}Ne*}\\
\parbox[b][0.3cm]{17.7cm}{resonances at E\ensuremath{_{\textnormal{x}}}=5.1-8.09 MeV using the LESA array that consisted of 4 Si-pad detectors covering \ensuremath{\theta}\ensuremath{_{\textnormal{lab}}}=90\ensuremath{^\circ}{\textminus}150\ensuremath{^\circ}; measured}\\
\parbox[b][0.3cm]{17.7cm}{neutrons time-of-flight using 16 liquid scintillators covering \ensuremath{\theta}\ensuremath{_{\textnormal{lab}}}=11\ensuremath{^\circ} to 39\ensuremath{^\circ}; measured I\ensuremath{_{\textnormal{p/}\ensuremath{\alpha}}}(\ensuremath{\theta}), E\ensuremath{_{\textnormal{n}}}, I\ensuremath{_{\textnormal{n}}}(\ensuremath{\theta}), p-n coincidences and}\\
\parbox[b][0.3cm]{17.7cm}{\ensuremath{\alpha}-n coincidences; considered only the p, p$'$, \ensuremath{^{\textnormal{2}}}He and \ensuremath{\alpha}-decay from the populated \ensuremath{^{\textnormal{18}}}Ne*(5.1, 6.15, 6.30, 7.06, 7.95, and 8.09 MeV)}\\
\parbox[b][0.3cm]{17.7cm}{levels to the \ensuremath{^{\textnormal{17}}}F(0, 495 keV) states and to \ensuremath{^{\textnormal{16}}}O\ensuremath{_{\textnormal{g.s.}}}; performed a complete, event-by-event kinematics reconstruction to deduce}\\
\parbox[b][0.3cm]{17.7cm}{\ensuremath{^{\textnormal{18}}}Ne states; measured angular distributions of the protons from the \ensuremath{^{\textnormal{18}}}Ne*(p\ensuremath{_{\textnormal{0}}}+\ensuremath{^{\textnormal{17}}}F\ensuremath{_{\textnormal{g.s.}}}) decay channels. (\href{https://www.nndc.bnl.gov/nsr/nsrlink.jsp?2012Al11,B}{2012Al11}) deduced level}\\
\parbox[b][0.3cm]{17.7cm}{decay branching ratios for p\ensuremath{_{\textnormal{0}}} (to \ensuremath{^{\textnormal{17}}}F\ensuremath{_{\textnormal{g.s.}}}), p$'$(to \ensuremath{^{\textnormal{17}}}F(495 keV)), 2p and \ensuremath{\alpha} decay modes (to \ensuremath{^{\textnormal{16}}}O\ensuremath{_{\textnormal{g.s.}}}); discussed the astrophysical}\\
\parbox[b][0.3cm]{17.7cm}{reaction rate for \ensuremath{^{\textnormal{14}}}O(\ensuremath{\alpha},p); comparison with literature results are given. Some of the branching ratios deduced by (\href{https://www.nndc.bnl.gov/nsr/nsrlink.jsp?2012Al11,B}{2012Al11}) are}\\
\clearpage
\vspace{0.3cm}
{\bf \small \underline{\ensuremath{^{\textnormal{16}}}O(\ensuremath{^{\textnormal{3}}}He,n)\hspace{0.2in}\href{https://www.nndc.bnl.gov/nsr/nsrlink.jsp?1953Ku08,B}{1953Ku08},\href{https://www.nndc.bnl.gov/nsr/nsrlink.jsp?2012Fo29,B}{2012Fo29} (continued)}}\\
\vspace{0.3cm}
\parbox[b][0.3cm]{17.7cm}{disputed in the literature. See for example (\href{https://www.nndc.bnl.gov/nsr/nsrlink.jsp?2012Fo29,B}{2012Fo29}, \href{https://www.nndc.bnl.gov/nsr/nsrlink.jsp?2019Ch16,B}{2019Ch16}, and \href{https://www.nndc.bnl.gov/nsr/nsrlink.jsp?2020Br14,B}{2020Br14}).}\\
\vspace{0.385cm}
\parbox[b][0.3cm]{17.7cm}{\addtolength{\parindent}{-0.2in}\textit{The \ensuremath{^{16}}O(\ensuremath{^{\textnormal{3}}}He,n) Studies with Relevant Information on the \ensuremath{^{\textnormal{18}}}Ne(\ensuremath{\beta}\ensuremath{^{\textnormal{+}}})\ensuremath{^{\textnormal{18}}}F Decay}:}\\
\parbox[b][0.3cm]{17.7cm}{\addtolength{\parindent}{-0.2in}\href{https://www.nndc.bnl.gov/nsr/nsrlink.jsp?1959Du81,B}{1959Du81}, \href{https://www.nndc.bnl.gov/nsr/nsrlink.jsp?1959Du83,B}{1959Du83}: \ensuremath{^{\textnormal{16}}}O(\ensuremath{^{}}He,n); measured the yield of slow-to-fast neutrons; deduced the excitation curve of \ensuremath{^{\textnormal{18}}}Ne as a function}\\
\parbox[b][0.3cm]{17.7cm}{of bombarding energy; observed an upward break in the excitation curve at 3.95 MeV, which was interpreted as the threshold for a}\\
\parbox[b][0.3cm]{17.7cm}{possible new excited state in \ensuremath{^{\textnormal{18}}}Ne at 114 keV \textit{15}; measured the threshold energy for the \ensuremath{^{\textnormal{16}}}O(\ensuremath{^{\textnormal{3}}}He,n) reaction to be 3.811 MeV \textit{15};}\\
\parbox[b][0.3cm]{17.7cm}{deduced the half-life of \ensuremath{^{\textnormal{18}}}Ne\ensuremath{_{\textnormal{g.s.}}} as T\ensuremath{_{\textnormal{1/2}}}=1.25 s \textit{20}, see (\href{https://www.nndc.bnl.gov/nsr/nsrlink.jsp?1961Bu05,B}{1961Bu05}). The state at E\ensuremath{_{\textnormal{x}}}=114 keV was later voided (see above).}\\
\parbox[b][0.3cm]{17.7cm}{\addtolength{\parindent}{-0.2in}\href{https://www.nndc.bnl.gov/nsr/nsrlink.jsp?1960Bu03,B}{1960Bu03}: \ensuremath{^{\textnormal{16}}}O(\ensuremath{^{}}He,n); measured the decay of \ensuremath{^{\textnormal{18}}}Ne\ensuremath{_{\textnormal{g.s.}}}(\ensuremath{\beta}\ensuremath{^{\textnormal{+}}})\ensuremath{^{\textnormal{18}}}F\ensuremath{_{\textnormal{g.s.}}}. This study was the first to indicate that \ensuremath{^{\textnormal{18}}}Ne decays by}\\
\parbox[b][0.3cm]{17.7cm}{positron emission to the 1.04- and 1.70-MeV levels of \ensuremath{^{\textnormal{18}}}F.}\\
\parbox[b][0.3cm]{17.7cm}{\addtolength{\parindent}{-0.2in}\href{https://www.nndc.bnl.gov/nsr/nsrlink.jsp?1961Ec02,B}{1961Ec02}: \ensuremath{^{\textnormal{16}}}O(\ensuremath{^{\textnormal{3}}}He,n) E=5.2 MeV; measured the \ensuremath{\gamma}-rays from the \ensuremath{^{\textnormal{18}}}Ne\ensuremath{_{\textnormal{g.s.}}}(\ensuremath{\beta}\ensuremath{^{\textnormal{+}}})\ensuremath{^{\textnormal{18}}}F*\ensuremath{\rightarrow}\ensuremath{\gamma}+\ensuremath{^{\textnormal{18}}}F\ensuremath{_{\textnormal{g.s.}}} decay at \ensuremath{\theta}\ensuremath{_{\textnormal{lab}}}=90\ensuremath{^\circ} using a}\\
\parbox[b][0.3cm]{17.7cm}{lead-shielded NaI(Tl) detector. The beam was switched on and off every 2, 10, and 50 seconds using a shutter downstream the}\\
\parbox[b][0.3cm]{17.7cm}{accelerator. This study only observed the superallowed decay branch, and its \ensuremath{\beta}-delayed \ensuremath{\gamma}-ray energy was measured as E\ensuremath{_{\ensuremath{\gamma}}}=1035}\\
\parbox[b][0.3cm]{17.7cm}{keV \textit{10}. Its intensity decreased with the lifetime of \ensuremath{^{\textnormal{18}}}Ne\ensuremath{_{\textnormal{g.s.}}}, which was deduced as \ensuremath{\tau}=2.5 s \textit{6}.}\\
\parbox[b][0.3cm]{17.7cm}{\addtolength{\parindent}{-0.2in}\href{https://www.nndc.bnl.gov/nsr/nsrlink.jsp?1961Bu05,B}{1961Bu05}: \ensuremath{^{\textnormal{16}}}O(\ensuremath{^{\textnormal{3}}}He,n) E=5.2 MeV; measured E\ensuremath{_{\ensuremath{\gamma}}}=1041 keV \textit{5} for the superallowed \ensuremath{\beta}-delayed \ensuremath{\gamma}-ray transition using a NaI(Tl)}\\
\parbox[b][0.3cm]{17.7cm}{crystal. The \ensuremath{\gamma}-ray energy was measured relative to the \ensuremath{^{\textnormal{22}}}Na \ensuremath{\gamma}-ray at E\ensuremath{_{\ensuremath{\gamma}}}=1273.6 keV \textit{16}. Deduced T\ensuremath{_{\textnormal{1/2}}}(\ensuremath{^{\textnormal{18}}}Ne\ensuremath{_{\textnormal{g.s.}}})= 1.46 s \textit{7} by}\\
\parbox[b][0.3cm]{17.7cm}{observing the annihilation radiation due to \ensuremath{^{\textnormal{18}}}Ne positron decay; measured branching ratios of 93\% \textit{2} and 7\% \textit{2} for the}\\
\parbox[b][0.3cm]{17.7cm}{\ensuremath{^{\textnormal{18}}}Ne\ensuremath{_{\textnormal{g.s.}}}(\ensuremath{\beta})\ensuremath{^{\textnormal{18}}}F\ensuremath{_{\textnormal{g.s.}}} and \ensuremath{^{\textnormal{18}}}Ne\ensuremath{_{\textnormal{g.s.}}}(\ensuremath{\beta})\ensuremath{^{\textnormal{18}}}F*(1041 keV) decay branches, respectively, by a comparison of the intensities of the}\\
\parbox[b][0.3cm]{17.7cm}{photopeaks of the annihilation radiation and that of the 1.04-MeV \ensuremath{\gamma}-ray. Using the results of (\href{https://www.nndc.bnl.gov/nsr/nsrlink.jsp?1959Du81,B}{1959Du81}), the authors deduced the}\\
\parbox[b][0.3cm]{17.7cm}{end point energy of 3423 keV \textit{13} for the positrons from the \ensuremath{^{\textnormal{18}}}Ne\ensuremath{_{\textnormal{g.s.}}}(\ensuremath{\beta}\ensuremath{^{\textnormal{+}}})\ensuremath{^{\textnormal{18}}}F\ensuremath{_{\textnormal{g.s.}}} decay.}\\
\parbox[b][0.3cm]{17.7cm}{\addtolength{\parindent}{-0.2in}\href{https://www.nndc.bnl.gov/nsr/nsrlink.jsp?1963Fr10,B}{1963Fr10}, \href{https://www.nndc.bnl.gov/nsr/nsrlink.jsp?1965Fr09,B}{1965Fr09}: \ensuremath{^{\textnormal{16}}}O(\ensuremath{^{\textnormal{3}}}He,n) E=5.2 MeV; measured the positrons from the \ensuremath{^{\textnormal{18}}}Ne\ensuremath{_{\textnormal{g.s.}}}(\ensuremath{\beta}\ensuremath{^{\textnormal{+}}})\ensuremath{^{\textnormal{18}}}F\ensuremath{_{\textnormal{g.s.}}} decay by focusing them onto a}\\
\parbox[b][0.3cm]{17.7cm}{bell-type Geiger-Muller counter using a Siegbahn-Sl\"{a}tis intermediate-image spectrometer (with a resolution of 4\%). The activated}\\
\parbox[b][0.3cm]{17.7cm}{target was moved outside the spectrometer and the \ensuremath{\beta}-delayed \ensuremath{\gamma}-rays were measured using a NaI crystal with a resolution of 8\% at}\\
\parbox[b][0.3cm]{17.7cm}{661 keV. Deduced T\ensuremath{_{\textnormal{1/2}}}(\ensuremath{^{\textnormal{18}}}Ne\ensuremath{_{\textnormal{g.s.}}})=1.47 s \textit{10}. Measured E\ensuremath{_{\ensuremath{\gamma}}}=1035 keV \textit{20} for the superallowed \ensuremath{\beta}-delayed \ensuremath{\gamma}-ray transition. The}\\
\parbox[b][0.3cm]{17.7cm}{branching ratios and the positron end-point energies were measured to be 91\% \textit{3} and 3416 keV \textit{9}, and 9\% \textit{3} and 2373 keV \textit{11} for}\\
\parbox[b][0.3cm]{17.7cm}{the \ensuremath{^{\textnormal{18}}}Ne\ensuremath{_{\textnormal{g.s.}}}(\ensuremath{\beta})\ensuremath{^{\textnormal{18}}}F\ensuremath{_{\textnormal{g.s.}}} and superallowed decay branches, respectively. Deduced log \textit{ft} values, and \ensuremath{^{\textnormal{18}}}Ne mass excess of 10.651}\\
\parbox[b][0.3cm]{17.7cm}{MeV \textit{10}.}\\
\parbox[b][0.3cm]{17.7cm}{\addtolength{\parindent}{-0.2in}\href{https://www.nndc.bnl.gov/nsr/nsrlink.jsp?1967Mi02,B}{1967Mi02}: \ensuremath{^{\textnormal{16}}}O(\ensuremath{^{\textnormal{3}}}He,n\ensuremath{_{\textnormal{0}}}) E=10-12 MeV; the \ensuremath{^{\textnormal{18}}}Ne\ensuremath{_{\textnormal{g.s.}}} was populated as a contaminant in various spectra due to the oxygen}\\
\parbox[b][0.3cm]{17.7cm}{contamination in the targets.}\\
\parbox[b][0.3cm]{17.7cm}{\addtolength{\parindent}{-0.2in}\href{https://www.nndc.bnl.gov/nsr/nsrlink.jsp?1968Go05,B}{1968Go05}: \ensuremath{^{\textnormal{16}}}O(\ensuremath{^{\textnormal{3}}}He,n) E=5.6 MeV; populated the \ensuremath{^{\textnormal{18}}}Ne\ensuremath{_{\textnormal{g.s.}}} and measured the \ensuremath{\beta}-delayed \ensuremath{\gamma}-rays from the decay of}\\
\parbox[b][0.3cm]{17.7cm}{\ensuremath{^{\textnormal{18}}}Ne\ensuremath{_{\textnormal{g.s.}}}(\ensuremath{\beta})\ensuremath{^{\textnormal{18}}}F*(\ensuremath{\gamma})\ensuremath{^{\textnormal{18}}}F using a Ge(Li) detector. Measured E\ensuremath{_{\ensuremath{\gamma}}}=1043 keV \textit{1} for the \ensuremath{\gamma}-ray transition following the superallowed}\\
\parbox[b][0.3cm]{17.7cm}{decay branch. Deduced the branching ratios and log \textit{ft} values for the \ensuremath{^{\textnormal{18}}}Ne decay to the \ensuremath{^{\textnormal{18}}}F*(1.08, 1.70, 2.10 MeV) states. The}\\
\parbox[b][0.3cm]{17.7cm}{deduced branching ratios are reported as \ensuremath{<}0.7\%, \ensuremath{<}0.9\%, and \ensuremath{<}1.5\%, respectively.}\\
\parbox[b][0.3cm]{17.7cm}{\addtolength{\parindent}{-0.2in}\href{https://www.nndc.bnl.gov/nsr/nsrlink.jsp?1970Al11,B}{1970Al11}: \ensuremath{^{\textnormal{16}}}O(\ensuremath{^{\textnormal{3}}}He,n) E=5.5{\textminus}MeV; measured \ensuremath{\beta}-rays from the \ensuremath{^{\textnormal{18}}}Ne\ensuremath{_{\textnormal{g.s.}}}(\ensuremath{\beta})\ensuremath{^{\textnormal{18}}}F decay using a plastic scintillator with beam cycles}\\
\parbox[b][0.3cm]{17.7cm}{of 4 s beam on, and 102 s beam off. Deduced T\ensuremath{_{\textnormal{1/2}}} for the \ensuremath{^{\textnormal{18}}}Ne\ensuremath{_{\textnormal{g.s.}}} in 3 separate measurements and obtained T\ensuremath{_{\textnormal{1/2}}}=1.701 s \textit{32},}\\
\parbox[b][0.3cm]{17.7cm}{1.650 s \textit{33}, and 1.658 s \textit{25}; deduced the recommended half-life of 1.67 s \textit{2}; deduced log {ft}, and a branching ratio of 92.4\% \textit{17} for}\\
\parbox[b][0.3cm]{17.7cm}{the \ensuremath{^{\textnormal{18}}}Ne\ensuremath{_{\textnormal{g.s.}}}(\ensuremath{\beta})\ensuremath{^{\textnormal{18}}}F\ensuremath{_{\textnormal{g.s.}}} decay branch.}\\
\parbox[b][0.3cm]{17.7cm}{\addtolength{\parindent}{-0.2in}\href{https://www.nndc.bnl.gov/nsr/nsrlink.jsp?1970As06,B}{1970As06}: \ensuremath{^{\textnormal{16}}}O(\ensuremath{^{\textnormal{3}}}He,n) E=10-11 MeV; measured the positron decay of \ensuremath{^{\textnormal{18}}}Ne\ensuremath{_{\textnormal{g.s.}}}: measured I\ensuremath{_{\ensuremath{\gamma}}} for the \ensuremath{\beta}-delayed \ensuremath{\gamma}-ray transitions}\\
\parbox[b][0.3cm]{17.7cm}{from the \ensuremath{^{\textnormal{18}}}F*(1042, 1700 keV) states using a Ge(Li) detector with an energy resolution of 3 keV at 1.33 MeV; deduced T\ensuremath{_{\textnormal{1/2}}}=1.69}\\
\parbox[b][0.3cm]{17.7cm}{s \textit{4} for \ensuremath{^{\textnormal{18}}}Ne\ensuremath{_{\textnormal{g.s.}}}; deduced I\ensuremath{_{\ensuremath{\beta}}}, branching ratios and log \textit{ft} for the \ensuremath{\beta}\ensuremath{^{\textnormal{+}}} decay branches to \ensuremath{^{\textnormal{18}}}F*(1042, 1700 keV). The resulting}\\
\parbox[b][0.3cm]{17.7cm}{branching ratios are 92.5\% \textit{2} for the ground state branch, 7.3\% \textit{2} for the superallowed branch, and 0.17\% \textit{5} for the}\\
\parbox[b][0.3cm]{17.7cm}{\ensuremath{^{\textnormal{18}}}Ne\ensuremath{_{\textnormal{g.s.}}}(\ensuremath{\beta}\ensuremath{^{\textnormal{+}}})\ensuremath{^{\textnormal{18}}}F*(1700 keV) branch.}\\
\parbox[b][0.3cm]{17.7cm}{\addtolength{\parindent}{-0.2in}\href{https://www.nndc.bnl.gov/nsr/nsrlink.jsp?1972Ha58,B}{1972Ha58}: \ensuremath{^{\textnormal{16}}}O(\ensuremath{^{\textnormal{3}}}He,n) E=12 MeV; measured the \ensuremath{\beta}-delayed \ensuremath{\gamma}-rays from the \ensuremath{^{\textnormal{18}}}Ne\ensuremath{_{\textnormal{g.s.}}}(\ensuremath{\beta}\ensuremath{^{\textnormal{+}}})\ensuremath{^{\textnormal{18}}}F* decay using a Ge(Li) detector.}\\
\parbox[b][0.3cm]{17.7cm}{Deduced T\ensuremath{_{\textnormal{1/2}}}=1655 ms \textit{25} for \ensuremath{^{\textnormal{18}}}Ne\ensuremath{_{\textnormal{g.s.}}}; obtained a branching ratio of 7.65\% \textit{26} for the superallowed decay branch; and deduced}\\
\parbox[b][0.3cm]{17.7cm}{log \textit{ft}. The half-life deduced in (\href{https://www.nndc.bnl.gov/nsr/nsrlink.jsp?1975Ha21,B}{1975Ha21}) supersedes the preliminary result of (\href{https://www.nndc.bnl.gov/nsr/nsrlink.jsp?1972Ha58,B}{1972Ha58}).}\\
\parbox[b][0.3cm]{17.7cm}{\addtolength{\parindent}{-0.2in}\href{https://www.nndc.bnl.gov/nsr/nsrlink.jsp?1975Al27,B}{1975Al27}: \ensuremath{^{\textnormal{16}}}O(\ensuremath{^{\textnormal{3}}}He,n) E=5.7 MeV; measured the \ensuremath{\beta}-decay of \ensuremath{^{\textnormal{18}}}Ne\ensuremath{_{\textnormal{g.s.}}} using two NE102 plastic scintillators. The half-life of}\\
\parbox[b][0.3cm]{17.7cm}{\ensuremath{^{\textnormal{18}}}Ne\ensuremath{_{\textnormal{g.s.}}} was deduced as T\ensuremath{_{\textnormal{1/2}}}=1.669 s \textit{4} by multi-scaling at 0.1 s/channel for 300-350 channels. The authors deduced A=18}\\
\parbox[b][0.3cm]{17.7cm}{\ensuremath{\beta}-decay mirror asymmetry of \ensuremath{\delta}\ensuremath{_{\textnormal{exp}}}={\textminus}0.86\% \textit{80}, which is the asymmetry in the \ensuremath{\beta}-decay \textit{ft} values, i.e., \ensuremath{\delta}=[log(\textit{ft}\ensuremath{^{\textnormal{+}}})/log(\textit{ft}\ensuremath{^{-}})]{\textminus}1.}\\
\parbox[b][0.3cm]{17.7cm}{\addtolength{\parindent}{-0.2in}\href{https://www.nndc.bnl.gov/nsr/nsrlink.jsp?1975Ha21,B}{1975Ha21}: \ensuremath{^{\textnormal{16}}}O(\ensuremath{^{\textnormal{3}}}He,n) E=12 MeV; measured the \ensuremath{\beta}-delayed \ensuremath{\gamma}-rays from the decay of \ensuremath{^{\textnormal{18}}}Ne using a Ge(Li) detector. Measured}\\
\parbox[b][0.3cm]{17.7cm}{E\ensuremath{_{\ensuremath{\gamma}}}=659.4 keV \textit{10}, 1041.3 keV \textit{10}, and 1699.6 keV \textit{20} with relative intensities (2.1\% \textit{3}, 100\%, and 0.71\% \textit{17}, respectively);}\\
\parbox[b][0.3cm]{17.7cm}{measured absolute branching ratios of 92.11\% \textit{21}, 7.66\% \textit{21}, and 0.23\% \textit{3}, respectively, for the decay to the \ensuremath{^{\textnormal{18}}}F*(0, 1042, 1700}\\
\parbox[b][0.3cm]{17.7cm}{keV) states. Deduced E\ensuremath{_{\textnormal{max}}}(\ensuremath{\beta})=2383.0 keV \textit{47}, and T\ensuremath{_{\textnormal{1/2}}}=1687 s \textit{9} for \ensuremath{^{\textnormal{18}}}Ne\ensuremath{_{\textnormal{g.s.}}} extracted by comparing the resultant decay curve}\\
\parbox[b][0.3cm]{17.7cm}{with a single component exponential using least squares method; deduced log \textit{ft} values for each decay branch; determined the}\\
\parbox[b][0.3cm]{17.7cm}{corrected \textit{Ft} value.}\\
\clearpage
\vspace{0.3cm}
{\bf \small \underline{\ensuremath{^{\textnormal{16}}}O(\ensuremath{^{\textnormal{3}}}He,n)\hspace{0.2in}\href{https://www.nndc.bnl.gov/nsr/nsrlink.jsp?1953Ku08,B}{1953Ku08},\href{https://www.nndc.bnl.gov/nsr/nsrlink.jsp?2012Fo29,B}{2012Fo29} (continued)}}\\
\vspace{0.3cm}
\parbox[b][0.3cm]{17.7cm}{\addtolength{\parindent}{-0.2in}\href{https://www.nndc.bnl.gov/nsr/nsrlink.jsp?1981Ad01,B}{1981Ad01}: \ensuremath{^{\textnormal{16}}}O(\ensuremath{^{\textnormal{3}}}He,n) E=12 MeV; measured the energy and intensity of the \ensuremath{\beta}-delayed \ensuremath{\gamma}-rays from the \ensuremath{^{\textnormal{18}}}Ne(\ensuremath{\beta})\ensuremath{^{\textnormal{18}}}F* decay using}\\
\parbox[b][0.3cm]{17.7cm}{a Ge(Li) detector. Data collection started 0.1 seconds after each beam bombardment ended and continued for 1.7 seconds. Observed}\\
\parbox[b][0.3cm]{17.7cm}{\ensuremath{^{\textnormal{18}}}F \ensuremath{\gamma}-rays at 659-, 1042-, 1081-, and 1700-keV. Measured the branching ratio of the \ensuremath{^{\textnormal{18}}}Ne\ensuremath{_{\textnormal{g.s.}}}(\ensuremath{\beta})\ensuremath{^{\textnormal{18}}}F*(1081 keV) decay for the}\\
\parbox[b][0.3cm]{17.7cm}{first time. Deduced I\ensuremath{_{\ensuremath{\gamma}}}=100.0, 1.71\ensuremath{\times}10\ensuremath{^{\textnormal{$-$2}}} \textit{41}, and 2.47 \textit{5} for the transitions to the 1042-, 1081-, and 1700-keV levels in \ensuremath{^{\textnormal{18}}}F,}\\
\parbox[b][0.3cm]{17.7cm}{respectively. Deduced \textit{Ft} value for the \ensuremath{^{\textnormal{18}}}Ne(g.s., 0\ensuremath{^{\textnormal{+}}})\ensuremath{\rightarrow}\ensuremath{^{\textnormal{18}}}F(1081, 0\ensuremath{^{-}}) branch; deduced the strength of the parity non-conserving}\\
\parbox[b][0.3cm]{17.7cm}{\ensuremath{\pi}-exchange NN interaction.}\\
\parbox[b][0.3cm]{17.7cm}{\addtolength{\parindent}{-0.2in}\href{https://www.nndc.bnl.gov/nsr/nsrlink.jsp?1982He04,B}{1982He04}: \ensuremath{^{\textnormal{16}}}O(\ensuremath{^{\textnormal{3}}}He,n) E=15 and 18 MeV; measured E\ensuremath{_{\ensuremath{\gamma}}} and I\ensuremath{_{\ensuremath{\gamma}}} for the \ensuremath{\beta}-delayed \ensuremath{\gamma}-rays from the decay of \ensuremath{^{\textnormal{18}}}F* states using a}\\
\parbox[b][0.3cm]{17.7cm}{Ge(Li) detector with a NaI anti-Compton shield. The \ensuremath{^{\textnormal{18}}}F \ensuremath{\gamma}-rays of 659-, 1042-, 1081-, and 1700-keV were observed. A \ensuremath{\gamma}-ray}\\
\parbox[b][0.3cm]{17.7cm}{with E\ensuremath{_{\ensuremath{\gamma}}}=1164 keV was observed, which could be the \ensuremath{^{\textnormal{18}}}F*(2100 keV, 2\ensuremath{^{-}})\ensuremath{\rightarrow}\ensuremath{^{\textnormal{18}}}F*(937 keV, 3\ensuremath{^{\textnormal{+}}})+\ensuremath{\gamma} transition but the authors were}\\
\parbox[b][0.3cm]{17.7cm}{doubtful that the 2100-keV state could be fed by the \ensuremath{\beta}\ensuremath{^{\textnormal{+}}} decay of \ensuremath{^{\textnormal{18}}}Ne (see also (\href{https://www.nndc.bnl.gov/nsr/nsrlink.jsp?1968Go05,B}{1968Go05})). Deduced \textit{Ft} and the strength of the}\\
\parbox[b][0.3cm]{17.7cm}{parity non-conserving \ensuremath{\pi}-exchange NN interaction. Deduced \ensuremath{^{\textnormal{18}}}F levels, and absolute branching ratios for the \ensuremath{\beta}-decay branches to}\\
\parbox[b][0.3cm]{17.7cm}{the \ensuremath{^{\textnormal{18}}}F*(1042, 1081, 1700 keV) states. The results are 7.70\% \textit{21}, 2.14\ensuremath{\times}10\ensuremath{^{\textnormal{$-$3}}}\% \textit{26}, and 0.183\% \textit{6}, respectively.}\\
\parbox[b][0.3cm]{17.7cm}{\addtolength{\parindent}{-0.2in}\href{https://www.nndc.bnl.gov/nsr/nsrlink.jsp?1982DaZZ,B}{1982DaZZ}, \href{https://www.nndc.bnl.gov/nsr/nsrlink.jsp?1983Ad03,B}{1983Ad03}: \ensuremath{^{\textnormal{16}}}O(\ensuremath{^{\textnormal{3}}}He,n) E=12 MeV; measured E\ensuremath{_{\ensuremath{\gamma}}} and I\ensuremath{_{\ensuremath{\gamma}}} of the \ensuremath{\beta}-delayed \ensuremath{\gamma}-rays corresponding to the decays from}\\
\parbox[b][0.3cm]{17.7cm}{the \ensuremath{^{\textnormal{18}}}F*(1042, 1081, and 1700 keV) states using a shielded Ge(Li) detector. Deduced absolute branching ratios of 92.11\% \textit{21},}\\
\parbox[b][0.3cm]{17.7cm}{7.70\% \textit{21}, 2.07\ensuremath{\times}10\ensuremath{^{\textnormal{$-$3}}}\% \textit{28}, and 0.188 \textit{6} for the \ensuremath{^{\textnormal{18}}}Ne\ensuremath{_{\textnormal{g.s.}}}(\ensuremath{\beta}\ensuremath{^{\textnormal{+}}})\ensuremath{^{\textnormal{18}}}F*(0, 1042, 1081, 1700 keV) decay branches, respectively. Deduced}\\
\parbox[b][0.3cm]{17.7cm}{\textit{Ft} values. Discussed the \ensuremath{\pi}-exchange contribution to the parity non-conserving NN force and weak pion coupling constant.}\\
\parbox[b][0.3cm]{17.7cm}{Comparison to shell model calculations are provided.}\\
\parbox[b][0.3cm]{17.7cm}{\addtolength{\parindent}{-0.2in}\href{https://www.nndc.bnl.gov/nsr/nsrlink.jsp?2002Vo11,B}{2002Vo11}: \ensuremath{^{\textnormal{16}}}O(\ensuremath{^{\textnormal{3}}}He,n) E=10 MeV; measured E\ensuremath{_{\ensuremath{\gamma}}} and \ensuremath{\beta}-\ensuremath{\gamma} coincidences using 14 Si(Li) and 2 HPGe-detectors; deduced \ensuremath{\beta}-\ensuremath{\nu}}\\
\parbox[b][0.3cm]{17.7cm}{angular correlation coefficient (\ensuremath{\alpha}=+1.06 \textit{19}) and an upper limit for the presence of a scalar interaction.}\\
\vspace{0.385cm}
\parbox[b][0.3cm]{17.7cm}{\addtolength{\parindent}{-0.2in}\textit{Theory}:}\\
\parbox[b][0.3cm]{17.7cm}{\addtolength{\parindent}{-0.2in}\href{https://www.nndc.bnl.gov/nsr/nsrlink.jsp?1964He06,B}{1964He06}: X(\ensuremath{^{\textnormal{3}}}He,n) E=20 MeV; discussed the two-nucleon stripping reactions with a particular reference to the (\ensuremath{^{\textnormal{3}}}He,n) reaction.}\\
\parbox[b][0.3cm]{17.7cm}{Three models (plane wave and distorted wave Born approximations, and a simple diffraction model) are studied and compared.}\\
\parbox[b][0.3cm]{17.7cm}{DWBA is used to calculate absolute differential cross sections to various final states for the (\ensuremath{^{\textnormal{3}}}He,n) reaction on \ensuremath{^{\textnormal{12}}}C, \ensuremath{^{\textnormal{16}}}O, Ni, and}\\
\parbox[b][0.3cm]{17.7cm}{Sn targets at 20 MeV incident \ensuremath{^{\textnormal{3}}}He ions. Comparison with experimental data is made where available and an agreement is found.}\\
\parbox[b][0.3cm]{17.7cm}{To further such comparisons, summed cross sections were computed to several low-lying states of the final nucleus. Spectroscopic}\\
\parbox[b][0.3cm]{17.7cm}{weights are obtained for pure and mixed configurations of single-particle wave functions.}\\
\parbox[b][0.3cm]{17.7cm}{\addtolength{\parindent}{-0.2in}\href{https://www.nndc.bnl.gov/nsr/nsrlink.jsp?2012Fo29,B}{2012Fo29}: \ensuremath{^{\textnormal{16}}}O(\ensuremath{^{\textnormal{3}}}He,n); reviewed and compared the available experimental and theoretical properties of the \ensuremath{^{\textnormal{18}}}Ne*(7.06 MeV, 4\ensuremath{^{\textnormal{+}}})}\\
\parbox[b][0.3cm]{17.7cm}{state; calculated an upper limit of 2\ensuremath{\times}10\ensuremath{^{\textnormal{$-$4}}} for the p\ensuremath{_{\textnormal{1}}}/p\ensuremath{_{\textnormal{0}}} decay branching ratio for the state mentioned above. Therefore, the author}\\
\parbox[b][0.3cm]{17.7cm}{disputed the p\ensuremath{_{\textnormal{1}}}/p\ensuremath{_{\textnormal{0}}} decay branching ratio for this state deduced by (\href{https://www.nndc.bnl.gov/nsr/nsrlink.jsp?2012Al11,B}{2012Al11}); discussed the problems with the p\ensuremath{_{\textnormal{1}}} branching ratio}\\
\parbox[b][0.3cm]{17.7cm}{obtained by (\href{https://www.nndc.bnl.gov/nsr/nsrlink.jsp?2012Al11,B}{2012Al11}) and concluded that the reported (by \href{https://www.nndc.bnl.gov/nsr/nsrlink.jsp?2012Al11,B}{2012Al11}) p\ensuremath{_{\textnormal{1}}} decay must be from a nearby state {\textminus} perhaps at E\ensuremath{_{\textnormal{x}}}=7.37}\\
\parbox[b][0.3cm]{17.7cm}{MeV.}\\
\vspace{0.385cm}
\parbox[b][0.3cm]{17.7cm}{\addtolength{\parindent}{-0.2in}\textit{Others}:}\\
\parbox[b][0.3cm]{17.7cm}{\addtolength{\parindent}{-0.2in}\href{https://www.nndc.bnl.gov/nsr/nsrlink.jsp?1977Fi13,B}{1977Fi13}: \ensuremath{^{\textnormal{16}}}O(\ensuremath{^{\textnormal{3}}}He,p), \ensuremath{^{\textnormal{16}}}O(\ensuremath{^{\textnormal{3}}}He,n) E=14-41 MeV; measured \ensuremath{\sigma}(E) for production of radio isotope \ensuremath{^{\textnormal{18}}}F used for gamma}\\
\parbox[b][0.3cm]{17.7cm}{scintigraphy for imaging bones and bony lesions. Cross sections of these reactions are given as a function of incident energy.}\\
\parbox[b][0.3cm]{17.7cm}{\addtolength{\parindent}{-0.2in}\href{https://www.nndc.bnl.gov/nsr/nsrlink.jsp?1991Gu05,B}{1991Gu05}: \ensuremath{^{\textnormal{16}}}O(\ensuremath{^{\textnormal{3}}}He,n) E=36 MeV; reviewed the available data (at the time) on the production of \ensuremath{^{\textnormal{18}}}F\ensuremath{_{\textnormal{g.s.}}}; included a discussion of}\\
\parbox[b][0.3cm]{17.7cm}{and recommendations for production techniques in the context of proton emission tomography. This work presented the \ensuremath{^{\textnormal{18}}}F}\\
\parbox[b][0.3cm]{17.7cm}{production yield per 2h irradiation at E=36 MeV from (\href{https://www.nndc.bnl.gov/nsr/nsrlink.jsp?1983Kn11,B}{1983Kn11} and \href{https://www.nndc.bnl.gov/nsr/nsrlink.jsp?1977Fi13,B}{1977Fi13}).}\\
\vspace{12pt}
\underline{$^{18}$Ne Levels}\\
\vspace{0.34cm}
\parbox[b][0.3cm]{17.7cm}{\addtolength{\parindent}{-0.254cm}The neutron angular distribution data of (\href{https://www.nndc.bnl.gov/nsr/nsrlink.jsp?1968To09,B}{1968To09}) show very similar patterns at different beam energies, which are indicative of}\\
\parbox[b][0.3cm]{17.7cm}{direct reaction mechanism, except for the broad resonance at 3.36 MeV.}\\
\parbox[b][0.3cm]{17.7cm}{\addtolength{\parindent}{-0.254cm}(d\ensuremath{\sigma}/d\ensuremath{\Omega})\ensuremath{_{\textnormal{tot.,lab}}}=84 mb/sr for 5 MeV\ensuremath{<}E\ensuremath{_{\textnormal{lab}}}\ensuremath{<}35 MeV and at \ensuremath{\theta}\ensuremath{_{\textnormal{lab}}}=0\ensuremath{^\circ} (\href{https://www.nndc.bnl.gov/nsr/nsrlink.jsp?1964Br13,B}{1964Br13}: see Table 2).}\\
\vspace{0.34cm}
\begin{longtable}{ccccccccc@{\extracolsep{\fill}}c}
\multicolumn{2}{c}{E(level)$^{}$}&J$^{\pi}$$^{{\hyperlink{NE15LEVEL0}{a}}}$&\multicolumn{2}{c}{T\ensuremath{_{\textnormal{1/2}}} or \ensuremath{\Gamma}$^{}$}&L$^{{\hyperlink{NE15LEVEL0}{a}}{\hyperlink{NE15LEVEL1}{b}}{\hyperlink{NE15LEVEL2}{c}}}$&\multicolumn{2}{c}{Relative C\ensuremath{^{\textnormal{2}}}S$^{{\hyperlink{NE15LEVEL3}{d}}}$}&Comments&\\[-.2cm]
\multicolumn{2}{c}{\hrulefill}&\hrulefill&\multicolumn{2}{c}{\hrulefill}&\hrulefill&\multicolumn{2}{c}{\hrulefill}&\hrulefill&
\endfirsthead
\multicolumn{1}{r@{}}{0}&\multicolumn{1}{@{}l}{}&\multicolumn{1}{l}{0\ensuremath{^{+}}}&\multicolumn{1}{r@{}}{1}&\multicolumn{1}{@{.}l}{671 s {\it 6}}&\multicolumn{1}{l}{0}&\multicolumn{1}{r@{}}{1}&\multicolumn{1}{@{}l}{}&\parbox[t][0.3cm]{9.877061cm}{\raggedright \%\ensuremath{\varepsilon}+\%\ensuremath{\beta}\ensuremath{^{\textnormal{+}}}=100\vspace{0.1cm}}&\\
&&&&&&&&\parbox[t][0.3cm]{9.877061cm}{\raggedright T=1 (\href{https://www.nndc.bnl.gov/nsr/nsrlink.jsp?1961Bu05,B}{1961Bu05},\href{https://www.nndc.bnl.gov/nsr/nsrlink.jsp?1963Fr10,B}{1963Fr10},\href{https://www.nndc.bnl.gov/nsr/nsrlink.jsp?1975Ha21,B}{1975Ha21},\href{https://www.nndc.bnl.gov/nsr/nsrlink.jsp?1981Ne09,B}{1981Ne09},\href{https://www.nndc.bnl.gov/nsr/nsrlink.jsp?1981Ad01,B}{1981Ad01})\vspace{0.1cm}}&\\
&&&&&&&&\parbox[t][0.3cm]{9.877061cm}{\raggedright T\ensuremath{_{\textnormal{z}}}={\textminus}1 (\href{https://www.nndc.bnl.gov/nsr/nsrlink.jsp?1968Gi09,B}{1968Gi09}, \href{https://www.nndc.bnl.gov/nsr/nsrlink.jsp?1975Ha21,B}{1975Ha21}).\vspace{0.1cm}}&\\
&&&&&&&&\parbox[t][0.3cm]{9.877061cm}{\raggedright E(level): From (\href{https://www.nndc.bnl.gov/nsr/nsrlink.jsp?1959Du81,B}{1959Du81}, \href{https://www.nndc.bnl.gov/nsr/nsrlink.jsp?1960Aj03,B}{1960Aj03}, \href{https://www.nndc.bnl.gov/nsr/nsrlink.jsp?1961Bu05,B}{1961Bu05}, \href{https://www.nndc.bnl.gov/nsr/nsrlink.jsp?1961Du02,B}{1961Du02}, \href{https://www.nndc.bnl.gov/nsr/nsrlink.jsp?1961Ga01,B}{1961Ga01},\vspace{0.1cm}}&\\
&&&&&&&&\parbox[t][0.3cm]{9.877061cm}{\raggedright {\ }{\ }{\ }\href{https://www.nndc.bnl.gov/nsr/nsrlink.jsp?1961To03,B}{1961To03}, \href{https://www.nndc.bnl.gov/nsr/nsrlink.jsp?1962Ma61,B}{1962Ma61}, \href{https://www.nndc.bnl.gov/nsr/nsrlink.jsp?1964Br13,B}{1964Br13} (see Fig. 10 and the text), \href{https://www.nndc.bnl.gov/nsr/nsrlink.jsp?1965Br42,B}{1965Br42},\vspace{0.1cm}}&\\
\end{longtable}
\begin{textblock}{29}(0,27.3)
Continued on next page (footnotes at end of table)
\end{textblock}
\clearpage
\begin{longtable}{ccccccc@{\extracolsep{\fill}}c}
\\[-.4cm]
\multicolumn{8}{c}{{\bf \small \underline{\ensuremath{^{\textnormal{16}}}O(\ensuremath{^{\textnormal{3}}}He,n)\hspace{0.2in}\href{https://www.nndc.bnl.gov/nsr/nsrlink.jsp?1953Ku08,B}{1953Ku08},\href{https://www.nndc.bnl.gov/nsr/nsrlink.jsp?2012Fo29,B}{2012Fo29} (continued)}}}\\
\multicolumn{8}{c}{~}\\
\multicolumn{8}{c}{\underline{\ensuremath{^{18}}Ne Levels (continued)}}\\
\multicolumn{8}{c}{~}\\
\multicolumn{2}{c}{E(level)$^{}$}&J$^{\pi}$$^{{\hyperlink{NE15LEVEL0}{a}}}$&L$^{{\hyperlink{NE15LEVEL0}{a}}{\hyperlink{NE15LEVEL1}{b}}{\hyperlink{NE15LEVEL2}{c}}}$&\multicolumn{2}{c}{Relative C\ensuremath{^{\textnormal{2}}}S$^{{\hyperlink{NE15LEVEL3}{d}}}$}&Comments&\\[-.2cm]
\multicolumn{2}{c}{\hrulefill}&\hrulefill&\hrulefill&\multicolumn{2}{c}{\hrulefill}&\hrulefill&
\endhead
&&&&&&\parbox[t][0.3cm]{11.383441cm}{\raggedright {\ }{\ }{\ }\href{https://www.nndc.bnl.gov/nsr/nsrlink.jsp?1966Kr05,B}{1966Kr05}, \href{https://www.nndc.bnl.gov/nsr/nsrlink.jsp?1967Mc03,B}{1967Mc03}, \href{https://www.nndc.bnl.gov/nsr/nsrlink.jsp?1967Mi02,B}{1967Mi02}, \href{https://www.nndc.bnl.gov/nsr/nsrlink.jsp?1968Sh09,B}{1968Sh09}, \href{https://www.nndc.bnl.gov/nsr/nsrlink.jsp?1968To09,B}{1968To09}, \href{https://www.nndc.bnl.gov/nsr/nsrlink.jsp?1970Ad02,B}{1970Ad02}, \href{https://www.nndc.bnl.gov/nsr/nsrlink.jsp?1975Pe11,B}{1975Pe11},\vspace{0.1cm}}&\\
&&&&&&\parbox[t][0.3cm]{11.383441cm}{\raggedright {\ }{\ }{\ }\href{https://www.nndc.bnl.gov/nsr/nsrlink.jsp?1977Ev01,B}{1977Ev01}, \href{https://www.nndc.bnl.gov/nsr/nsrlink.jsp?1981Ne09,B}{1981Ne09}, \href{https://www.nndc.bnl.gov/nsr/nsrlink.jsp?1991Ga03,B}{1991Ga03} (used as a calibration point), \href{https://www.nndc.bnl.gov/nsr/nsrlink.jsp?1994Ma14,B}{1994Ma14},\vspace{0.1cm}}&\\
&&&&&&\parbox[t][0.3cm]{11.383441cm}{\raggedright {\ }{\ }{\ }\href{https://www.nndc.bnl.gov/nsr/nsrlink.jsp?1996Ha26,B}{1996Ha26}, \href{https://www.nndc.bnl.gov/nsr/nsrlink.jsp?2005Pa50,B}{2005Pa50}, \href{https://www.nndc.bnl.gov/nsr/nsrlink.jsp?2010AlZZ,B}{2010AlZZ}, \href{https://www.nndc.bnl.gov/nsr/nsrlink.jsp?2010Ta17,B}{2010Ta17}, and \href{https://www.nndc.bnl.gov/nsr/nsrlink.jsp?2012Al11,B}{2012Al11}).\vspace{0.1cm}}&\\
&&&&&&\parbox[t][0.3cm]{11.383441cm}{\raggedright A tentative state was observed by (\href{https://www.nndc.bnl.gov/nsr/nsrlink.jsp?1959Du81,B}{1959Du81}) at 114 keV \textit{15}, the existence of\vspace{0.1cm}}&\\
&&&&&&\parbox[t][0.3cm]{11.383441cm}{\raggedright {\ }{\ }{\ }which was ruled out by (\href{https://www.nndc.bnl.gov/nsr/nsrlink.jsp?1960Aj03,B}{1960Aj03}, \href{https://www.nndc.bnl.gov/nsr/nsrlink.jsp?1961Du02,B}{1961Du02}).\vspace{0.1cm}}&\\
&&&&&&\parbox[t][0.3cm]{11.383441cm}{\raggedright T\ensuremath{_{1/2}}: Weighted average (with external errors) of 1.46 s \textit{7} (\href{https://www.nndc.bnl.gov/nsr/nsrlink.jsp?1961Bu05,B}{1961Bu05}); 1.73 s \textit{42}\vspace{0.1cm}}&\\
&&&&&&\parbox[t][0.3cm]{11.383441cm}{\raggedright {\ }{\ }{\ }(\href{https://www.nndc.bnl.gov/nsr/nsrlink.jsp?1961Ec02,B}{1961Ec02}: from \ensuremath{\tau}=2.5 s \textit{6}); 1.47 s \textit{10} (\href{https://www.nndc.bnl.gov/nsr/nsrlink.jsp?1963Fr10,B}{1963Fr10}, \href{https://www.nndc.bnl.gov/nsr/nsrlink.jsp?1965Fr09,B}{1965Fr09}); 1.69 s \textit{4}\vspace{0.1cm}}&\\
&&&&&&\parbox[t][0.3cm]{11.383441cm}{\raggedright {\ }{\ }{\ }(\href{https://www.nndc.bnl.gov/nsr/nsrlink.jsp?1970As06,B}{1970As06}); 1.67 s {2} (\href{https://www.nndc.bnl.gov/nsr/nsrlink.jsp?1970Al11,B}{1970Al11}); 1.669 s \textit{4} (\href{https://www.nndc.bnl.gov/nsr/nsrlink.jsp?1975Al27,B}{1975Al27}); and 1.687 s \textit{9}\vspace{0.1cm}}&\\
&&&&&&\parbox[t][0.3cm]{11.383441cm}{\raggedright {\ }{\ }{\ }(\href{https://www.nndc.bnl.gov/nsr/nsrlink.jsp?1975Ha21,B}{1975Ha21}). Note that the 1.25 s \textit{20} (\href{https://www.nndc.bnl.gov/nsr/nsrlink.jsp?1959Du81,B}{1959Du81}, see \href{https://www.nndc.bnl.gov/nsr/nsrlink.jsp?1961Bu05,B}{1961Bu05}) value is marked\vspace{0.1cm}}&\\
&&&&&&\parbox[t][0.3cm]{11.383441cm}{\raggedright {\ }{\ }{\ }as an outlier by Chauvenet$'$s criterion, and thus is excluded.\vspace{0.1cm}}&\\
&&&&&&\parbox[t][0.3cm]{11.383441cm}{\raggedright T\ensuremath{_{1/2}}: If the 1.46 s \textit{7} (\href{https://www.nndc.bnl.gov/nsr/nsrlink.jsp?1961Bu05,B}{1961Bu05}) and 1.47 s \textit{10} (\href{https://www.nndc.bnl.gov/nsr/nsrlink.jsp?1963Fr10,B}{1963Fr10}, \href{https://www.nndc.bnl.gov/nsr/nsrlink.jsp?1965Fr09,B}{1965Fr09}) values are\vspace{0.1cm}}&\\
&&&&&&\parbox[t][0.3cm]{11.383441cm}{\raggedright {\ }{\ }{\ }excluded from the weighted average, the \ensuremath{\chi}\ensuremath{^{\textnormal{2}}}/\ensuremath{\nu} improves from 2.8 to 0.89, and the\vspace{0.1cm}}&\\
&&&&&&\parbox[t][0.3cm]{11.383441cm}{\raggedright {\ }{\ }{\ }weighted average is 1.672 s \textit{4}.\vspace{0.1cm}}&\\
&&&&&&\parbox[t][0.3cm]{11.383441cm}{\raggedright J\ensuremath{^{\pi}},L: From the plane wave Born approximation calculations of (\href{https://www.nndc.bnl.gov/nsr/nsrlink.jsp?1961Ga01,B}{1961Ga01},\vspace{0.1cm}}&\\
&&&&&&\parbox[t][0.3cm]{11.383441cm}{\raggedright {\ }{\ }{\ }\href{https://www.nndc.bnl.gov/nsr/nsrlink.jsp?1966Kr05,B}{1966Kr05}, and \href{https://www.nndc.bnl.gov/nsr/nsrlink.jsp?1967Mc03,B}{1967Mc03}: see Fig. 13, at E(\ensuremath{^{\textnormal{3}}}He)=5.6 MeV); and the zero-range\vspace{0.1cm}}&\\
&&&&&&\parbox[t][0.3cm]{11.383441cm}{\raggedright {\ }{\ }{\ }DWBA analyses of (\href{https://www.nndc.bnl.gov/nsr/nsrlink.jsp?1968Sh09,B}{1968Sh09}, \href{https://www.nndc.bnl.gov/nsr/nsrlink.jsp?1968To09,B}{1968To09}, \href{https://www.nndc.bnl.gov/nsr/nsrlink.jsp?1970Ad02,B}{1970Ad02}, \href{https://www.nndc.bnl.gov/nsr/nsrlink.jsp?1975Pe11,B}{1975Pe11}, \href{https://www.nndc.bnl.gov/nsr/nsrlink.jsp?1977Ev01,B}{1977Ev01},\vspace{0.1cm}}&\\
&&&&&&\parbox[t][0.3cm]{11.383441cm}{\raggedright {\ }{\ }{\ }\href{https://www.nndc.bnl.gov/nsr/nsrlink.jsp?1981Ne09,B}{1981Ne09}, and \href{https://www.nndc.bnl.gov/nsr/nsrlink.jsp?1996Ha26,B}{1996Ha26}), all of which assign L=0 and J\ensuremath{^{\ensuremath{\pi}}}=0\ensuremath{^{\textnormal{+}}}. Note that the\vspace{0.1cm}}&\\
&&&&&&\parbox[t][0.3cm]{11.383441cm}{\raggedright {\ }{\ }{\ }DWBA fit of (\href{https://www.nndc.bnl.gov/nsr/nsrlink.jsp?1977Ev01,B}{1977Ev01}) and the measured neutron angular distributions data for\vspace{0.1cm}}&\\
&&&&&&\parbox[t][0.3cm]{11.383441cm}{\raggedright {\ }{\ }{\ }this state diverge at \ensuremath{\theta}\ensuremath{_{\textnormal{c.m.}}}\ensuremath{>}20\ensuremath{^\circ}.\vspace{0.1cm}}&\\
&&&&&&\parbox[t][0.3cm]{11.383441cm}{\raggedright Mass of \ensuremath{^{\textnormal{18}}}Ne: 18.00570 amu \textit{4} (\href{https://www.nndc.bnl.gov/nsr/nsrlink.jsp?1960Aj03,B}{1960Aj03}); and 18.011446 amu \textit{14} (\href{https://www.nndc.bnl.gov/nsr/nsrlink.jsp?1961Du02,B}{1961Du02}:\vspace{0.1cm}}&\\
&&&&&&\parbox[t][0.3cm]{11.383441cm}{\raggedright {\ }{\ }{\ }based on \ensuremath{^{\textnormal{16}}}O mass as the standard).\vspace{0.1cm}}&\\
&&&&&&\parbox[t][0.3cm]{11.383441cm}{\raggedright Mass excess of \ensuremath{^{\textnormal{18}}}Ne: 10.64 MeV \textit{4} (\href{https://www.nndc.bnl.gov/nsr/nsrlink.jsp?1960Aj03,B}{1960Aj03}); 10658 keV \textit{13} (\href{https://www.nndc.bnl.gov/nsr/nsrlink.jsp?1961Du02,B}{1961Du02}: based\vspace{0.1cm}}&\\
&&&&&&\parbox[t][0.3cm]{11.383441cm}{\raggedright {\ }{\ }{\ }on \ensuremath{^{\textnormal{16}}}O mass as the standard); 10649 keV \textit{8} (\href{https://www.nndc.bnl.gov/nsr/nsrlink.jsp?1961To03,B}{1961To03}); 10651 keV \textit{10}\vspace{0.1cm}}&\\
&&&&&&\parbox[t][0.3cm]{11.383441cm}{\raggedright {\ }{\ }{\ }(\href{https://www.nndc.bnl.gov/nsr/nsrlink.jsp?1963Fr10,B}{1963Fr10}); and 5316.8 keV \textit{15} (\href{https://www.nndc.bnl.gov/nsr/nsrlink.jsp?1994Ma14,B}{1994Ma14}: inferred by measuring the energy\vspace{0.1cm}}&\\
&&&&&&\parbox[t][0.3cm]{11.383441cm}{\raggedright {\ }{\ }{\ }loss of \ensuremath{^{\textnormal{3}}}He ions through the target at E\ensuremath{_{\textnormal{lab}}}(\ensuremath{^{\textnormal{3}}}He)=7.31 MeV and by using\vspace{0.1cm}}&\\
&&&&&&\parbox[t][0.3cm]{11.383441cm}{\raggedright {\ }{\ }{\ }E\ensuremath{_{\textnormal{x}}}(\ensuremath{^{\textnormal{18}}}Ne)=1887.3 keV \textit{2} from (\href{https://www.nndc.bnl.gov/nsr/nsrlink.jsp?1987Aj02,B}{1987Aj02}). The energy loss was determined by\vspace{0.1cm}}&\\
&&&&&&\parbox[t][0.3cm]{11.383441cm}{\raggedright {\ }{\ }{\ }measuring the \ensuremath{^{\textnormal{16}}}O(\ensuremath{\alpha},\ensuremath{\gamma}) excitation function over the narrow resonance at\vspace{0.1cm}}&\\
&&&&&&\parbox[t][0.3cm]{11.383441cm}{\raggedright {\ }{\ }{\ }E\ensuremath{_{\ensuremath{\alpha}}}=6928 keV \textit{4} (\href{https://www.nndc.bnl.gov/nsr/nsrlink.jsp?1987Aj02,B}{1987Aj02})).\vspace{0.1cm}}&\\
&&&&&&\parbox[t][0.3cm]{11.383441cm}{\raggedright Q(\ensuremath{^{\textnormal{16}}}He,n)\ensuremath{^{\textnormal{18}}}Ne\ensuremath{_{\textnormal{g.s.}}}: {\textminus}3190 keV \textit{4} (\href{https://www.nndc.bnl.gov/nsr/nsrlink.jsp?1960Aj03,B}{1960Aj03}); {\textminus}3206 keV \textit{13} (\href{https://www.nndc.bnl.gov/nsr/nsrlink.jsp?1961Du02,B}{1961Du02}); {\textminus}3199\vspace{0.1cm}}&\\
&&&&&&\parbox[t][0.3cm]{11.383441cm}{\raggedright {\ }{\ }{\ }keV \textit{6} (\href{https://www.nndc.bnl.gov/nsr/nsrlink.jsp?1961To03,B}{1961To03}: weighted average between Q={\textminus}3199 keV \textit{20} measured from the\vspace{0.1cm}}&\\
&&&&&&\parbox[t][0.3cm]{11.383441cm}{\raggedright {\ }{\ }{\ }neutron time-of-flight spectrum and Q={\textminus}3199 keV \textit{6} measured using a calibrated\vspace{0.1cm}}&\\
&&&&&&\parbox[t][0.3cm]{11.383441cm}{\raggedright {\ }{\ }{\ }NMR to accurately obtain the beam energy); {\textminus}3000 (\href{https://www.nndc.bnl.gov/nsr/nsrlink.jsp?1962Ma61,B}{1962Ma61}); {\textminus}3196 keV\vspace{0.1cm}}&\\
&&&&&&\parbox[t][0.3cm]{11.383441cm}{\raggedright {\ }{\ }{\ }(\href{https://www.nndc.bnl.gov/nsr/nsrlink.jsp?1991Ga03,B}{1991Ga03}); and {\textminus}3194.0 keV \textit{15} (\href{https://www.nndc.bnl.gov/nsr/nsrlink.jsp?1994Ma14,B}{1994Ma14}).\vspace{0.1cm}}&\\
&&&&&&\parbox[t][0.3cm]{11.383441cm}{\raggedright Threshold energy(\ensuremath{^{\textnormal{16}}}O(\ensuremath{^{\textnormal{3}}}He,n)\ensuremath{^{\textnormal{18}}}Ne\ensuremath{_{\textnormal{g.s.}}}): 3811 keV \textit{15} (\href{https://www.nndc.bnl.gov/nsr/nsrlink.jsp?1961Du02,B}{1961Du02}); and 3802 keV \textit{7}\vspace{0.1cm}}&\\
&&&&&&\parbox[t][0.3cm]{11.383441cm}{\raggedright {\ }{\ }{\ }(\href{https://www.nndc.bnl.gov/nsr/nsrlink.jsp?1961To03,B}{1961To03}).\vspace{0.1cm}}&\\
&&&&&&\parbox[t][0.3cm]{11.383441cm}{\raggedright d\ensuremath{\sigma}/d\ensuremath{\Omega}(\ensuremath{\theta}=0\ensuremath{^\circ})=1.6 mb/sr \textit{3} for the \ensuremath{^{\textnormal{16}}}O(\ensuremath{^{\textnormal{3}}}He,n)\ensuremath{^{\textnormal{18}}}Ne\ensuremath{_{\textnormal{g.s.}}} reaction at 5.5 MeV\vspace{0.1cm}}&\\
&&&&&&\parbox[t][0.3cm]{11.383441cm}{\raggedright {\ }{\ }{\ }(\href{https://www.nndc.bnl.gov/nsr/nsrlink.jsp?1960Aj03,B}{1960Aj03}).\vspace{0.1cm}}&\\
&&&&&&\parbox[t][0.3cm]{11.383441cm}{\raggedright For the \ensuremath{^{\textnormal{16}}}O(\ensuremath{^{\textnormal{3}}}He,n)\ensuremath{^{\textnormal{18}}}Ne\ensuremath{_{\textnormal{g.s.}}} reaction: \ensuremath{\sigma}=400 mb at E(\ensuremath{^{\textnormal{3}}}He)=7.7 MeV (\href{https://www.nndc.bnl.gov/nsr/nsrlink.jsp?1962Ma61,B}{1962Ma61});\vspace{0.1cm}}&\\
&&&&&&\parbox[t][0.3cm]{11.383441cm}{\raggedright {\ }{\ }{\ }and \ensuremath{\sigma}=20 mb (\href{https://www.nndc.bnl.gov/nsr/nsrlink.jsp?1965Br42,B}{1965Br42}).\vspace{0.1cm}}&\\
&&&&&&\parbox[t][0.3cm]{11.383441cm}{\raggedright d\ensuremath{\sigma}/d\ensuremath{\Omega}(\ensuremath{\theta}\ensuremath{_{\textnormal{lab}}}=0\ensuremath{^\circ})=2.0 mb/sr \textit{4} at E=25.4 MeV (\href{https://www.nndc.bnl.gov/nsr/nsrlink.jsp?1964Br13,B}{1964Br13}).\vspace{0.1cm}}&\\
&&&&&&\parbox[t][0.3cm]{11.383441cm}{\raggedright d\ensuremath{\sigma}/d\ensuremath{\Omega}\ensuremath{^{\textnormal{max}}_{\textnormal{c.m.}}}=21 mb/sr \textit{4} at \ensuremath{\theta}\ensuremath{_{\textnormal{c.m.}}}\ensuremath{\sim}15\ensuremath{^\circ} at 2.54 MeV (\href{https://www.nndc.bnl.gov/nsr/nsrlink.jsp?1966Kr05,B}{1966Kr05}).\vspace{0.1cm}}&\\
&&&&&&\parbox[t][0.3cm]{11.383441cm}{\raggedright d\ensuremath{\sigma}/d\ensuremath{\Omega}\ensuremath{_{\textnormal{c.m.}}}(\ensuremath{\theta}\ensuremath{_{\textnormal{c.m.}}}=0\ensuremath{^\circ})=3.65 mb/sr at E=15 MeV (\href{https://www.nndc.bnl.gov/nsr/nsrlink.jsp?1977Ev01,B}{1977Ev01}).\vspace{0.1cm}}&\\
&&&&&&\parbox[t][0.3cm]{11.383441cm}{\raggedright d\ensuremath{\sigma}/d\ensuremath{\Omega}\ensuremath{_{\textnormal{c.m.}}}(\ensuremath{\theta}\ensuremath{_{\textnormal{c.m.}}}=0\ensuremath{^\circ})=3.69 mb/sr at E=18 MeV (\href{https://www.nndc.bnl.gov/nsr/nsrlink.jsp?1977Ev01,B}{1977Ev01}).\vspace{0.1cm}}&\\
&&&&&&\parbox[t][0.3cm]{11.383441cm}{\raggedright d\ensuremath{\sigma}/d\ensuremath{\Omega}\ensuremath{_{\textnormal{c.m.}}}(\ensuremath{\theta}\ensuremath{_{\textnormal{c.m.}}}=0\ensuremath{^\circ})=2.40 mb/sr at E=21 MeV (\href{https://www.nndc.bnl.gov/nsr/nsrlink.jsp?1977Ev01,B}{1977Ev01}).\vspace{0.1cm}}&\\
&&&&&&\parbox[t][0.3cm]{11.383441cm}{\raggedright (\href{https://www.nndc.bnl.gov/nsr/nsrlink.jsp?1967Mc03,B}{1967Mc03}): performed plane wave Born approximation calculations to fit the\vspace{0.1cm}}&\\
&&&&&&\parbox[t][0.3cm]{11.383441cm}{\raggedright {\ }{\ }{\ }\ensuremath{\sigma}(E\ensuremath{_{\textnormal{n}}},\ensuremath{\theta}) measured at 3 beam energies. The cross sections obtained at backward\vspace{0.1cm}}&\\
&&&&&&\parbox[t][0.3cm]{11.383441cm}{\raggedright {\ }{\ }{\ }angles were inconsistent with those obtained by (\href{https://www.nndc.bnl.gov/nsr/nsrlink.jsp?1961Ga01,B}{1961Ga01}).\vspace{0.1cm}}&\\
\multicolumn{1}{r@{}}{1886}&\multicolumn{1}{@{.}l}{5 {\it 13}}&\multicolumn{1}{l}{2\ensuremath{^{+}}}&\multicolumn{1}{l}{2}&\multicolumn{1}{r@{}}{0}&\multicolumn{1}{@{.}l}{74}&\parbox[t][0.3cm]{11.383441cm}{\raggedright T=1 (\href{https://www.nndc.bnl.gov/nsr/nsrlink.jsp?1981Ne09,B}{1981Ne09})\vspace{0.1cm}}&\\
&&&&&&\parbox[t][0.3cm]{11.383441cm}{\raggedright E(level): Weighted average of 1880 keV \textit{10} (\href{https://www.nndc.bnl.gov/nsr/nsrlink.jsp?1961To03,B}{1961To03}); 2.0 MeV \textit{1} (\href{https://www.nndc.bnl.gov/nsr/nsrlink.jsp?1966Kr05,B}{1966Kr05});\vspace{0.1cm}}&\\
\end{longtable}
\begin{textblock}{29}(0,27.3)
Continued on next page (footnotes at end of table)
\end{textblock}
\clearpage
\begin{longtable}{ccccccccc@{\extracolsep{\fill}}c}
\\[-.4cm]
\multicolumn{10}{c}{{\bf \small \underline{\ensuremath{^{\textnormal{16}}}O(\ensuremath{^{\textnormal{3}}}He,n)\hspace{0.2in}\href{https://www.nndc.bnl.gov/nsr/nsrlink.jsp?1953Ku08,B}{1953Ku08},\href{https://www.nndc.bnl.gov/nsr/nsrlink.jsp?2012Fo29,B}{2012Fo29} (continued)}}}\\
\multicolumn{10}{c}{~}\\
\multicolumn{10}{c}{\underline{\ensuremath{^{18}}Ne Levels (continued)}}\\
\multicolumn{10}{c}{~}\\
\multicolumn{2}{c}{E(level)$^{}$}&J$^{\pi}$$^{{\hyperlink{NE15LEVEL0}{a}}}$&\multicolumn{2}{c}{T\ensuremath{_{\textnormal{1/2}}} or \ensuremath{\Gamma}$^{}$}&L$^{{\hyperlink{NE15LEVEL0}{a}}{\hyperlink{NE15LEVEL1}{b}}{\hyperlink{NE15LEVEL2}{c}}}$&\multicolumn{2}{c}{Relative C\ensuremath{^{\textnormal{2}}}S$^{{\hyperlink{NE15LEVEL3}{d}}}$}&Comments&\\[-.2cm]
\multicolumn{2}{c}{\hrulefill}&\hrulefill&\multicolumn{2}{c}{\hrulefill}&\hrulefill&\multicolumn{2}{c}{\hrulefill}&\hrulefill&
\endhead
&&&&&&&&\parbox[t][0.3cm]{9.589391cm}{\raggedright {\ }{\ }{\ }1886 keV \textit{7} (\href{https://www.nndc.bnl.gov/nsr/nsrlink.jsp?1977Ev01,B}{1977Ev01}); 1886.5 keV \textit{15} (\href{https://www.nndc.bnl.gov/nsr/nsrlink.jsp?1994Ma14,B}{1994Ma14}: from\vspace{0.1cm}}&\\
&&&&&&&&\parbox[t][0.3cm]{9.589391cm}{\raggedright {\ }{\ }{\ }\ensuremath{\Delta}(\ensuremath{^{\textnormal{18}}}Ne*(2\ensuremath{^{\textnormal{+}}_{\textnormal{1}}}))=7204.1 keV \textit{15}); and 1887 keV \textit{3} (\href{https://www.nndc.bnl.gov/nsr/nsrlink.jsp?2005Pa50,B}{2005Pa50}). Note\vspace{0.1cm}}&\\
&&&&&&&&\parbox[t][0.3cm]{9.589391cm}{\raggedright {\ }{\ }{\ }that \ensuremath{\Delta}\ensuremath{_{\textnormal{ex}}}=\ensuremath{\Delta}\ensuremath{_{\textnormal{g.s.}}}+E\ensuremath{_{\textnormal{x}}}, where \ensuremath{\Delta}\ensuremath{_{\textnormal{ex}}} is the mass excess for an excited\vspace{0.1cm}}&\\
&&&&&&&&\parbox[t][0.3cm]{9.589391cm}{\raggedright {\ }{\ }{\ }state at E\ensuremath{_{\textnormal{x}}} and \ensuremath{\Delta}\ensuremath{_{\textnormal{g.s.}}} is the ground state mass excess that is\vspace{0.1cm}}&\\
&&&&&&&&\parbox[t][0.3cm]{9.589391cm}{\raggedright {\ }{\ }{\ }\ensuremath{\Delta}\ensuremath{_{\textnormal{g.s.}}}(\ensuremath{^{\textnormal{18}}}Ne)=5317.617 keV \textit{363} (\href{https://www.nndc.bnl.gov/nsr/nsrlink.jsp?2021Wa16,B}{2021Wa16}). Note that the E\ensuremath{_{\textnormal{x}}}=2.0\vspace{0.1cm}}&\\
&&&&&&&&\parbox[t][0.3cm]{9.589391cm}{\raggedright {\ }{\ }{\ }MeV \textit{1} was considered an outlier by the Chauvenet criterion, and\vspace{0.1cm}}&\\
&&&&&&&&\parbox[t][0.3cm]{9.589391cm}{\raggedright {\ }{\ }{\ }thus excluded.\vspace{0.1cm}}&\\
&&&&&&&&\parbox[t][0.3cm]{9.589391cm}{\raggedright E(level): See also 1880 keV (\href{https://www.nndc.bnl.gov/nsr/nsrlink.jsp?1964Br13,B}{1964Br13}, \href{https://www.nndc.bnl.gov/nsr/nsrlink.jsp?1968Sh09,B}{1968Sh09}, \href{https://www.nndc.bnl.gov/nsr/nsrlink.jsp?1968To09,B}{1968To09}); 1.89\vspace{0.1cm}}&\\
&&&&&&&&\parbox[t][0.3cm]{9.589391cm}{\raggedright {\ }{\ }{\ }MeV (\href{https://www.nndc.bnl.gov/nsr/nsrlink.jsp?1970Ad02,B}{1970Ad02}, \href{https://www.nndc.bnl.gov/nsr/nsrlink.jsp?1975Pe11,B}{1975Pe11}, \href{https://www.nndc.bnl.gov/nsr/nsrlink.jsp?1994Ma14,B}{1994Ma14}, \href{https://www.nndc.bnl.gov/nsr/nsrlink.jsp?1996Ha26,B}{1996Ha26}, \href{https://www.nndc.bnl.gov/nsr/nsrlink.jsp?2010AlZZ,B}{2010AlZZ},\vspace{0.1cm}}&\\
&&&&&&&&\parbox[t][0.3cm]{9.589391cm}{\raggedright {\ }{\ }{\ }\href{https://www.nndc.bnl.gov/nsr/nsrlink.jsp?2010Ta17,B}{2010Ta17}, \href{https://www.nndc.bnl.gov/nsr/nsrlink.jsp?2012Al11,B}{2012Al11}); 1887.3 keV (\href{https://www.nndc.bnl.gov/nsr/nsrlink.jsp?1981Ne09,B}{1981Ne09}: this state was used\vspace{0.1cm}}&\\
&&&&&&&&\parbox[t][0.3cm]{9.589391cm}{\raggedright {\ }{\ }{\ }as a calibration point, whose energy was taken from (\href{https://www.nndc.bnl.gov/nsr/nsrlink.jsp?1969Ro08,B}{1969Ro08}));\vspace{0.1cm}}&\\
&&&&&&&&\parbox[t][0.3cm]{9.589391cm}{\raggedright {\ }{\ }{\ }1887 keV (\href{https://www.nndc.bnl.gov/nsr/nsrlink.jsp?1991Ga03,B}{1991Ga03}: this state was used as a calibration point).\vspace{0.1cm}}&\\
&&&&&&&&\parbox[t][0.3cm]{9.589391cm}{\raggedright J\ensuremath{^{\pi}},L: From plane wave Born approximation calculations of\vspace{0.1cm}}&\\
&&&&&&&&\parbox[t][0.3cm]{9.589391cm}{\raggedright {\ }{\ }{\ }(\href{https://www.nndc.bnl.gov/nsr/nsrlink.jsp?1966Kr05,B}{1966Kr05}) using the (\href{https://www.nndc.bnl.gov/nsr/nsrlink.jsp?1960Ne21,B}{1960Ne21}) formalism; and the zero-range\vspace{0.1cm}}&\\
&&&&&&&&\parbox[t][0.3cm]{9.589391cm}{\raggedright {\ }{\ }{\ }DWBA analyses of (\href{https://www.nndc.bnl.gov/nsr/nsrlink.jsp?1968To09,B}{1968To09}, \href{https://www.nndc.bnl.gov/nsr/nsrlink.jsp?1970Ad02,B}{1970Ad02}, \href{https://www.nndc.bnl.gov/nsr/nsrlink.jsp?1975Pe11,B}{1975Pe11}, \href{https://www.nndc.bnl.gov/nsr/nsrlink.jsp?1977Ev01,B}{1977Ev01},\vspace{0.1cm}}&\\
&&&&&&&&\parbox[t][0.3cm]{9.589391cm}{\raggedright {\ }{\ }{\ }\href{https://www.nndc.bnl.gov/nsr/nsrlink.jsp?1981Ne09,B}{1981Ne09}, \href{https://www.nndc.bnl.gov/nsr/nsrlink.jsp?1996Ha26,B}{1996Ha26}). All of of these studies assigned J\ensuremath{^{\ensuremath{\pi}}}=2\ensuremath{^{\textnormal{+}}} and\vspace{0.1cm}}&\\
&&&&&&&&\parbox[t][0.3cm]{9.589391cm}{\raggedright {\ }{\ }{\ }L=2 to this state. Note that the DWBA curves of (\href{https://www.nndc.bnl.gov/nsr/nsrlink.jsp?1970Ad02,B}{1970Ad02}:\vspace{0.1cm}}&\\
&&&&&&&&\parbox[t][0.3cm]{9.589391cm}{\raggedright {\ }{\ }{\ }J\ensuremath{^{\ensuremath{\pi}}}=2\ensuremath{^{\textnormal{+}}}) fail to reproduce the deep minimum observed at forward\vspace{0.1cm}}&\\
&&&&&&&&\parbox[t][0.3cm]{9.589391cm}{\raggedright {\ }{\ }{\ }angles, which is most pronounced around 10 MeV. At 9.5 MeV, the\vspace{0.1cm}}&\\
&&&&&&&&\parbox[t][0.3cm]{9.589391cm}{\raggedright {\ }{\ }{\ }neutron angular distribution data were indicative of a resonance-like\vspace{0.1cm}}&\\
&&&&&&&&\parbox[t][0.3cm]{9.589391cm}{\raggedright {\ }{\ }{\ }structure due to the compound nuclear reaction contributions. The\vspace{0.1cm}}&\\
&&&&&&&&\parbox[t][0.3cm]{9.589391cm}{\raggedright {\ }{\ }{\ }neutron angular distribution for this state measured by (\href{https://www.nndc.bnl.gov/nsr/nsrlink.jsp?1970Ad02,B}{1970Ad02})\vspace{0.1cm}}&\\
&&&&&&&&\parbox[t][0.3cm]{9.589391cm}{\raggedright {\ }{\ }{\ }disagrees with that of (\href{https://www.nndc.bnl.gov/nsr/nsrlink.jsp?1966Kr05,B}{1966Kr05}) in magnitude and shape.\vspace{0.1cm}}&\\
&&&&&&&&\parbox[t][0.3cm]{9.589391cm}{\raggedright Mass excess for the 2\ensuremath{^{\textnormal{+}}_{\textnormal{1}}} state=7204.1 keV \textit{15} (\href{https://www.nndc.bnl.gov/nsr/nsrlink.jsp?1994Ma14,B}{1994Ma14}).\vspace{0.1cm}}&\\
&&&&&&&&\parbox[t][0.3cm]{9.589391cm}{\raggedright Q-value={\textminus}5079 keV \textit{8} (\href{https://www.nndc.bnl.gov/nsr/nsrlink.jsp?1961To03,B}{1961To03}); and {\textminus}5083 keV (\href{https://www.nndc.bnl.gov/nsr/nsrlink.jsp?1991Ga03,B}{1991Ga03}).\vspace{0.1cm}}&\\
&&&&&&&&\parbox[t][0.3cm]{9.589391cm}{\raggedright Threshold energy=6038 keV \textit{9} (\href{https://www.nndc.bnl.gov/nsr/nsrlink.jsp?1961To03,B}{1961To03}).\vspace{0.1cm}}&\\
&&&&&&&&\parbox[t][0.3cm]{9.589391cm}{\raggedright d\ensuremath{\sigma}/d\ensuremath{\Omega}\ensuremath{_{\textnormal{c.m.}}}=11 mb/sr \textit{2} at \ensuremath{\theta}\ensuremath{_{\textnormal{c.m.}}}\ensuremath{\sim}15\ensuremath{^\circ} and E\ensuremath{_{\textnormal{lab}}}=11 MeV (\href{https://www.nndc.bnl.gov/nsr/nsrlink.jsp?1966Kr05,B}{1966Kr05}).\vspace{0.1cm}}&\\
&&&&&&&&\parbox[t][0.3cm]{9.589391cm}{\raggedright \ensuremath{\sigma}\ensuremath{_{\textnormal{rel}}}=1.8 \textit{2} at E\ensuremath{_{\textnormal{lab}}}=18.3 MeV (\href{https://www.nndc.bnl.gov/nsr/nsrlink.jsp?1975Pe11,B}{1975Pe11}). This cross section is\vspace{0.1cm}}&\\
&&&&&&&&\parbox[t][0.3cm]{9.589391cm}{\raggedright {\ }{\ }{\ }reported relative to that of the \ensuremath{^{\textnormal{16}}}O(\ensuremath{^{\textnormal{3}}}He,n)\ensuremath{^{\textnormal{18}}}Ne\ensuremath{_{\textnormal{g.s.}}}. The same\vspace{0.1cm}}&\\
&&&&&&&&\parbox[t][0.3cm]{9.589391cm}{\raggedright {\ }{\ }{\ }relative cross section calculated by (\href{https://www.nndc.bnl.gov/nsr/nsrlink.jsp?1975Pe11,B}{1975Pe11}) using shell model is\vspace{0.1cm}}&\\
&&&&&&&&\parbox[t][0.3cm]{9.589391cm}{\raggedright {\ }{\ }{\ }0.76.\vspace{0.1cm}}&\\
&&&&&&&&\parbox[t][0.3cm]{9.589391cm}{\raggedright d\ensuremath{\sigma}/d\ensuremath{\Omega}\ensuremath{_{\textnormal{c.m.}}}(\ensuremath{\theta}\ensuremath{_{\textnormal{c.m.}}}=0\ensuremath{^\circ})=4.55 mb/sr at E=15 MeV (\href{https://www.nndc.bnl.gov/nsr/nsrlink.jsp?1977Ev01,B}{1977Ev01}).\vspace{0.1cm}}&\\
&&&&&&&&\parbox[t][0.3cm]{9.589391cm}{\raggedright d\ensuremath{\sigma}/d\ensuremath{\Omega}\ensuremath{_{\textnormal{c.m.}}}(\ensuremath{\theta}\ensuremath{_{\textnormal{c.m.}}}=0\ensuremath{^\circ})=6.80 mb/sr at E=18 MeV (\href{https://www.nndc.bnl.gov/nsr/nsrlink.jsp?1977Ev01,B}{1977Ev01}).\vspace{0.1cm}}&\\
&&&&&&&&\parbox[t][0.3cm]{9.589391cm}{\raggedright d\ensuremath{\sigma}/d\ensuremath{\Omega}\ensuremath{_{\textnormal{c.m.}}}(\ensuremath{\theta}\ensuremath{_{\textnormal{c.m.}}}=0\ensuremath{^\circ})=6.15 mb/sr at E=21 MeV (\href{https://www.nndc.bnl.gov/nsr/nsrlink.jsp?1977Ev01,B}{1977Ev01}).\vspace{0.1cm}}&\\
\multicolumn{1}{r@{}}{3375}&\multicolumn{1}{@{ }l}{{\it 3}}&\multicolumn{1}{l}{(4\ensuremath{^{+}})}&\multicolumn{1}{r@{}}{$<$30}&\multicolumn{1}{@{ }l}{keV}&\multicolumn{1}{l}{(4)}&\multicolumn{1}{r@{}}{1}&\multicolumn{1}{@{.}l}{54}&\parbox[t][0.3cm]{9.589391cm}{\raggedright E(level): Weighted average of 3362 keV \textit{11} (\href{https://www.nndc.bnl.gov/nsr/nsrlink.jsp?1961To03,B}{1961To03}); 3375 keV\vspace{0.1cm}}&\\
&&&&&&&&\parbox[t][0.3cm]{9.589391cm}{\raggedright {\ }{\ }{\ }\textit{15} (\href{https://www.nndc.bnl.gov/nsr/nsrlink.jsp?1970Ad02,B}{1970Ad02}); 3374 keV \textit{7} (\href{https://www.nndc.bnl.gov/nsr/nsrlink.jsp?1977Ev01,B}{1977Ev01}); and 3376 keV \textit{3}\vspace{0.1cm}}&\\
&&&&&&&&\parbox[t][0.3cm]{9.589391cm}{\raggedright {\ }{\ }{\ }(\href{https://www.nndc.bnl.gov/nsr/nsrlink.jsp?2005Pa50,B}{2005Pa50}). See also 3360 keV (\href{https://www.nndc.bnl.gov/nsr/nsrlink.jsp?1964Br13,B}{1964Br13}, \href{https://www.nndc.bnl.gov/nsr/nsrlink.jsp?1968Sh09,B}{1968Sh09}, \href{https://www.nndc.bnl.gov/nsr/nsrlink.jsp?1968To09,B}{1968To09});\vspace{0.1cm}}&\\
&&&&&&&&\parbox[t][0.3cm]{9.589391cm}{\raggedright {\ }{\ }{\ }3376.2 keV (\href{https://www.nndc.bnl.gov/nsr/nsrlink.jsp?1981Ne09,B}{1981Ne09}: this state was used as a calibration point,\vspace{0.1cm}}&\\
&&&&&&&&\parbox[t][0.3cm]{9.589391cm}{\raggedright {\ }{\ }{\ }whose energy was taken from (\href{https://www.nndc.bnl.gov/nsr/nsrlink.jsp?1969Ro08,B}{1969Ro08})); 3376 keV (\href{https://www.nndc.bnl.gov/nsr/nsrlink.jsp?1991Ga03,B}{1991Ga03}:\vspace{0.1cm}}&\\
&&&&&&&&\parbox[t][0.3cm]{9.589391cm}{\raggedright {\ }{\ }{\ }this state was used as a calibration point); and 3380 keV\vspace{0.1cm}}&\\
&&&&&&&&\parbox[t][0.3cm]{9.589391cm}{\raggedright {\ }{\ }{\ }(\href{https://www.nndc.bnl.gov/nsr/nsrlink.jsp?1994Ma14,B}{1994Ma14}).\vspace{0.1cm}}&\\
&&&&&&&&\parbox[t][0.3cm]{9.589391cm}{\raggedright (\href{https://www.nndc.bnl.gov/nsr/nsrlink.jsp?1968Sh09,B}{1968Sh09}, M. H. Shapiro et al., Bull. Amer. Phys. Soc. 13 (1968)\vspace{0.1cm}}&\\
&&&&&&&&\parbox[t][0.3cm]{9.589391cm}{\raggedright {\ }{\ }{\ }698) suggested that the 3.36 MeV level in \ensuremath{^{\textnormal{18}}}Ne is a closely spaced\vspace{0.1cm}}&\\
&&&&&&&&\parbox[t][0.3cm]{9.589391cm}{\raggedright {\ }{\ }{\ }doublet, see (\href{https://www.nndc.bnl.gov/nsr/nsrlink.jsp?1968Gi09,B}{1968Gi09}).\vspace{0.1cm}}&\\
&&&&&&&&\parbox[t][0.3cm]{9.589391cm}{\raggedright \ensuremath{\Gamma}: From (\href{https://www.nndc.bnl.gov/nsr/nsrlink.jsp?1970Ad02,B}{1970Ad02}).\vspace{0.1cm}}&\\
&&&&&&&&\parbox[t][0.3cm]{9.589391cm}{\raggedright J\ensuremath{^{\pi}},L: From the zero-range DWBA analyses of (\href{https://www.nndc.bnl.gov/nsr/nsrlink.jsp?1968To09,B}{1968To09}: J\ensuremath{^{\ensuremath{\pi}}}=(4\ensuremath{^{\textnormal{+}}})\vspace{0.1cm}}&\\
&&&&&&&&\parbox[t][0.3cm]{9.589391cm}{\raggedright {\ }{\ }{\ }and L=(4)); (\href{https://www.nndc.bnl.gov/nsr/nsrlink.jsp?1970Ad02,B}{1970Ad02}: J\ensuremath{^{\ensuremath{\pi}}}=(4\ensuremath{^{\textnormal{+}}})); and (\href{https://www.nndc.bnl.gov/nsr/nsrlink.jsp?1977Ev01,B}{1977Ev01}: J\ensuremath{^{\ensuremath{\pi}}}=4\ensuremath{^{\textnormal{+}}} but the\vspace{0.1cm}}&\\
&&&&&&&&\parbox[t][0.3cm]{9.589391cm}{\raggedright {\ }{\ }{\ }agreement between the DWBA fits and the data at E(\ensuremath{^{\textnormal{3}}}He)=15 MeV\vspace{0.1cm}}&\\
&&&&&&&&\parbox[t][0.3cm]{9.589391cm}{\raggedright {\ }{\ }{\ }and 18 MeV is not good). See also a similar assignment from\vspace{0.1cm}}&\\
&&&&&&&&\parbox[t][0.3cm]{9.589391cm}{\raggedright {\ }{\ }{\ }(\href{https://www.nndc.bnl.gov/nsr/nsrlink.jsp?1991Ga03,B}{1991Ga03}: J\ensuremath{^{\ensuremath{\pi}}}=4\ensuremath{^{\textnormal{+}}} from the analysis of mirror states in \ensuremath{^{\textnormal{18}}}Ne and\vspace{0.1cm}}&\\
&&&&&&&&\parbox[t][0.3cm]{9.589391cm}{\raggedright {\ }{\ }{\ }\ensuremath{^{\textnormal{18}}}O).\vspace{0.1cm}}&\\
&&&&&&&&\parbox[t][0.3cm]{9.589391cm}{\raggedright Note that (\href{https://www.nndc.bnl.gov/nsr/nsrlink.jsp?1968To09,B}{1968To09}) suggested that this state may be an indication\vspace{0.1cm}}&\\
&&&&&&&&\parbox[t][0.3cm]{9.589391cm}{\raggedright {\ }{\ }{\ }of high spin modes of excitation in \ensuremath{^{\textnormal{19}}}Ne compound nucleus in the\vspace{0.1cm}}&\\
\end{longtable}
\begin{textblock}{29}(0,27.3)
Continued on next page (footnotes at end of table)
\end{textblock}
\clearpage
\begin{longtable}{ccccccccc@{\extracolsep{\fill}}c}
\\[-.4cm]
\multicolumn{10}{c}{{\bf \small \underline{\ensuremath{^{\textnormal{16}}}O(\ensuremath{^{\textnormal{3}}}He,n)\hspace{0.2in}\href{https://www.nndc.bnl.gov/nsr/nsrlink.jsp?1953Ku08,B}{1953Ku08},\href{https://www.nndc.bnl.gov/nsr/nsrlink.jsp?2012Fo29,B}{2012Fo29} (continued)}}}\\
\multicolumn{10}{c}{~}\\
\multicolumn{10}{c}{\underline{\ensuremath{^{18}}Ne Levels (continued)}}\\
\multicolumn{10}{c}{~}\\
\multicolumn{2}{c}{E(level)$^{}$}&J$^{\pi}$$^{{\hyperlink{NE15LEVEL0}{a}}}$&\multicolumn{2}{c}{T\ensuremath{_{\textnormal{1/2}}} or \ensuremath{\Gamma}$^{}$}&L$^{{\hyperlink{NE15LEVEL0}{a}}{\hyperlink{NE15LEVEL1}{b}}{\hyperlink{NE15LEVEL2}{c}}}$&\multicolumn{2}{c}{Relative C\ensuremath{^{\textnormal{2}}}S$^{{\hyperlink{NE15LEVEL3}{d}}}$}&Comments&\\[-.2cm]
\multicolumn{2}{c}{\hrulefill}&\hrulefill&\multicolumn{2}{c}{\hrulefill}&\hrulefill&\multicolumn{2}{c}{\hrulefill}&\hrulefill&
\endhead
&&&&&&&&\parbox[t][0.3cm]{9.607421cm}{\raggedright {\ }{\ }{\ }region of 16.5 MeV because the neutron angular distributions for\vspace{0.1cm}}&\\
&&&&&&&&\parbox[t][0.3cm]{9.607421cm}{\raggedright {\ }{\ }{\ }this state were not consistent with direct reaction mechanism (see\vspace{0.1cm}}&\\
&&&&&&&&\parbox[t][0.3cm]{9.607421cm}{\raggedright {\ }{\ }{\ }also (\href{https://www.nndc.bnl.gov/nsr/nsrlink.jsp?1970Ad02,B}{1970Ad02}), E(\ensuremath{^{\textnormal{3}}}He)=9.5 MeV), and thus DWBA analysis\vspace{0.1cm}}&\\
&&&&&&&&\parbox[t][0.3cm]{9.607421cm}{\raggedright {\ }{\ }{\ }failed to describe the data well. The authors pointed out that the\vspace{0.1cm}}&\\
&&&&&&&&\parbox[t][0.3cm]{9.607421cm}{\raggedright {\ }{\ }{\ }DWBA analysis is further complicated due to the potential presence\vspace{0.1cm}}&\\
&&&&&&&&\parbox[t][0.3cm]{9.607421cm}{\raggedright {\ }{\ }{\ }of an unresolved 0\ensuremath{^{\textnormal{+}}} state in the vicinity of this state in their data,\vspace{0.1cm}}&\\
&&&&&&&&\parbox[t][0.3cm]{9.607421cm}{\raggedright {\ }{\ }{\ }which may explain their observed high cross section for \ensuremath{\theta}\ensuremath{_{\textnormal{c.m.}}}\ensuremath{<}30\ensuremath{^\circ}.\vspace{0.1cm}}&\\
&&&&&&&&\parbox[t][0.3cm]{9.607421cm}{\raggedright Q-value={\textminus}6561 keV \textit{9} (\href{https://www.nndc.bnl.gov/nsr/nsrlink.jsp?1961To03,B}{1961To03}); {\textminus}6572 keV (\href{https://www.nndc.bnl.gov/nsr/nsrlink.jsp?1991Ga03,B}{1991Ga03}); and {\textminus}6568\vspace{0.1cm}}&\\
&&&&&&&&\parbox[t][0.3cm]{9.607421cm}{\raggedright {\ }{\ }{\ }keV \textit{2} (\href{https://www.nndc.bnl.gov/nsr/nsrlink.jsp?1994Ma14,B}{1994Ma14}: using the Q\ensuremath{_{\textnormal{g.s.}}} and E\ensuremath{_{\textnormal{x}}}=3376.2 keV \textit{4} from\vspace{0.1cm}}&\\
&&&&&&&&\parbox[t][0.3cm]{9.607421cm}{\raggedright {\ }{\ }{\ }(\href{https://www.nndc.bnl.gov/nsr/nsrlink.jsp?1987Aj02,B}{1987Aj02})).\vspace{0.1cm}}&\\
&&&&&&&&\parbox[t][0.3cm]{9.607421cm}{\raggedright Threshold energy=7800 keV \textit{10} (\href{https://www.nndc.bnl.gov/nsr/nsrlink.jsp?1961To03,B}{1961To03}).\vspace{0.1cm}}&\\
&&&&&&&&\parbox[t][0.3cm]{9.607421cm}{\raggedright d\ensuremath{\sigma}/d\ensuremath{\Omega}\ensuremath{_{\textnormal{c.m.}}}(\ensuremath{\theta}\ensuremath{_{\textnormal{c.m.}}}=0\ensuremath{^\circ})=3.30 mb/sr at E=15 MeV (\href{https://www.nndc.bnl.gov/nsr/nsrlink.jsp?1977Ev01,B}{1977Ev01}).\vspace{0.1cm}}&\\
&&&&&&&&\parbox[t][0.3cm]{9.607421cm}{\raggedright d\ensuremath{\sigma}/d\ensuremath{\Omega}\ensuremath{_{\textnormal{c.m.}}}(\ensuremath{\theta}\ensuremath{_{\textnormal{c.m.}}}=0\ensuremath{^\circ})=2.62 mb/sr at E=18 MeV (\href{https://www.nndc.bnl.gov/nsr/nsrlink.jsp?1977Ev01,B}{1977Ev01}).\vspace{0.1cm}}&\\
&&&&&&&&\parbox[t][0.3cm]{9.607421cm}{\raggedright d\ensuremath{\sigma}/d\ensuremath{\Omega}\ensuremath{_{\textnormal{c.m.}}}(\ensuremath{\theta}\ensuremath{_{\textnormal{c.m.}}}=0\ensuremath{^\circ})=3.50 mb/sr at E=21 MeV (\href{https://www.nndc.bnl.gov/nsr/nsrlink.jsp?1977Ev01,B}{1977Ev01}).\vspace{0.1cm}}&\\
\multicolumn{1}{r@{}}{3576}&\multicolumn{1}{@{ }l}{{\it 3}}&\multicolumn{1}{l}{(0\ensuremath{^{+}})}&&&\multicolumn{1}{l}{(0)}&\multicolumn{1}{r@{}}{$<$0}&\multicolumn{1}{@{.}l}{6}&\parbox[t][0.3cm]{9.607421cm}{\raggedright E(level): Weighted average of 3564 keV \textit{20} (\href{https://www.nndc.bnl.gov/nsr/nsrlink.jsp?1970Ad02,B}{1970Ad02}); and 3576\vspace{0.1cm}}&\\
&&&&&&&&\parbox[t][0.3cm]{9.607421cm}{\raggedright {\ }{\ }{\ }keV \textit{3} (\href{https://www.nndc.bnl.gov/nsr/nsrlink.jsp?2005Pa50,B}{2005Pa50}). See also 3576 keV (\href{https://www.nndc.bnl.gov/nsr/nsrlink.jsp?1991Ga03,B}{1991Ga03}: this state was\vspace{0.1cm}}&\\
&&&&&&&&\parbox[t][0.3cm]{9.607421cm}{\raggedright {\ }{\ }{\ }used as a calibration point); and 3.5 MeV (\href{https://www.nndc.bnl.gov/nsr/nsrlink.jsp?2010AlZZ,B}{2010AlZZ}, \href{https://www.nndc.bnl.gov/nsr/nsrlink.jsp?2010Ta17,B}{2010Ta17},\vspace{0.1cm}}&\\
&&&&&&&&\parbox[t][0.3cm]{9.607421cm}{\raggedright {\ }{\ }{\ }\href{https://www.nndc.bnl.gov/nsr/nsrlink.jsp?2012Al11,B}{2012Al11}).\vspace{0.1cm}}&\\
&&&&&&&&\parbox[t][0.3cm]{9.607421cm}{\raggedright J\ensuremath{^{\pi}}: From (\href{https://www.nndc.bnl.gov/nsr/nsrlink.jsp?1970Ad02,B}{1970Ad02}: recommended J\ensuremath{^{\ensuremath{\pi}}}=0\ensuremath{^{\textnormal{+}}} from the isobaric analog\vspace{0.1cm}}&\\
&&&&&&&&\parbox[t][0.3cm]{9.607421cm}{\raggedright {\ }{\ }{\ }levels analysis and supported by L=2 DWBA fit to the 3.564+3.610\vspace{0.1cm}}&\\
&&&&&&&&\parbox[t][0.3cm]{9.607421cm}{\raggedright {\ }{\ }{\ }MeV unresolved states, see below). See also (\href{https://www.nndc.bnl.gov/nsr/nsrlink.jsp?1991Ga03,B}{1991Ga03}: from the\vspace{0.1cm}}&\\
&&&&&&&&\parbox[t][0.3cm]{9.607421cm}{\raggedright {\ }{\ }{\ }analysis of mirror states in \ensuremath{^{\textnormal{18}}}Ne and \ensuremath{^{\textnormal{18}}}O).\vspace{0.1cm}}&\\
&&&&&&&&\parbox[t][0.3cm]{9.607421cm}{\raggedright L: (\href{https://www.nndc.bnl.gov/nsr/nsrlink.jsp?1970Ad02,B}{1970Ad02}): a zero-range DWBA fit with L=2 was fitted to the\vspace{0.1cm}}&\\
&&&&&&&&\parbox[t][0.3cm]{9.607421cm}{\raggedright {\ }{\ }{\ }combined cross sections of the unresolved 3564- and 3610-keV\vspace{0.1cm}}&\\
&&&&&&&&\parbox[t][0.3cm]{9.607421cm}{\raggedright {\ }{\ }{\ }doublet measured at E=9, 9.5, 10.5 and 11.5 MeV, which implied\vspace{0.1cm}}&\\
&&&&&&&&\parbox[t][0.3cm]{9.607421cm}{\raggedright {\ }{\ }{\ }that one member of the doublet (presumably a 0\ensuremath{^{\textnormal{+}}} state with L=0)\vspace{0.1cm}}&\\
&&&&&&&&\parbox[t][0.3cm]{9.607421cm}{\raggedright {\ }{\ }{\ }was weakly populated, as expected by comparison with the \ensuremath{^{\textnormal{16}}}O(t,\vspace{0.1cm}}&\\
&&&&&&&&\parbox[t][0.3cm]{9.607421cm}{\raggedright {\ }{\ }{\ }p) reaction.\vspace{0.1cm}}&\\
&&&&&&&&\parbox[t][0.3cm]{9.607421cm}{\raggedright Q-value={\textminus}6772 keV (\href{https://www.nndc.bnl.gov/nsr/nsrlink.jsp?1991Ga03,B}{1991Ga03}).\vspace{0.1cm}}&\\
&&&&&&&&\parbox[t][0.3cm]{9.607421cm}{\raggedright (\href{https://www.nndc.bnl.gov/nsr/nsrlink.jsp?1977Ev01,B}{1977Ev01}) predicted that (d\ensuremath{\sigma}/d\ensuremath{\Omega})\ensuremath{_{\textnormal{c.m.}}}\ensuremath{<}300 \ensuremath{\mu}b/sr at \ensuremath{\theta}\ensuremath{_{\textnormal{c.m.}}}=0\ensuremath{^\circ} (at\vspace{0.1cm}}&\\
&&&&&&&&\parbox[t][0.3cm]{9.607421cm}{\raggedright {\ }{\ }{\ }E(\ensuremath{^{\textnormal{3}}}He)=15, 18, and 21 MeV) for the 0\ensuremath{^{\textnormal{+}}} state at 3576 keV, which\vspace{0.1cm}}&\\
&&&&&&&&\parbox[t][0.3cm]{9.607421cm}{\raggedright {\ }{\ }{\ }remained unobserved in (\href{https://www.nndc.bnl.gov/nsr/nsrlink.jsp?1977Ev01,B}{1977Ev01}). The cross section value is\vspace{0.1cm}}&\\
&&&&&&&&\parbox[t][0.3cm]{9.607421cm}{\raggedright {\ }{\ }{\ }deduced compared to (d\ensuremath{\sigma}/d\ensuremath{\Omega})\ensuremath{_{\textnormal{c.m.}}}=3.6 mb/sr for the ground-state\vspace{0.1cm}}&\\
&&&&&&&&\parbox[t][0.3cm]{9.607421cm}{\raggedright {\ }{\ }{\ }transition at E(\ensuremath{^{\textnormal{3}}}He)=18 MeV.\vspace{0.1cm}}&\\
&&&&&&&&\parbox[t][0.3cm]{9.607421cm}{\raggedright (\href{https://www.nndc.bnl.gov/nsr/nsrlink.jsp?1970El23,B}{1970El23}, \href{https://www.nndc.bnl.gov/nsr/nsrlink.jsp?1973Mc06,B}{1973Mc06}) predicted that this state may have a pure\vspace{0.1cm}}&\\
&&&&&&&&\parbox[t][0.3cm]{9.607421cm}{\raggedright {\ }{\ }{\ }2p-0h configuration.\vspace{0.1cm}}&\\
\multicolumn{1}{r@{}}{3615}&\multicolumn{1}{@{ }l}{{\it 3}}&\multicolumn{1}{l}{2\ensuremath{^{+}}}&&&\multicolumn{1}{l}{2}&\multicolumn{1}{r@{}}{0}&\multicolumn{1}{@{.}l}{22}&\parbox[t][0.3cm]{9.607421cm}{\raggedright E(level): Weighted average of 3608 keV \textit{12} (\href{https://www.nndc.bnl.gov/nsr/nsrlink.jsp?1961To03,B}{1961To03}); 3610 keV \textit{15}\vspace{0.1cm}}&\\
&&&&&&&&\parbox[t][0.3cm]{9.607421cm}{\raggedright {\ }{\ }{\ }(\href{https://www.nndc.bnl.gov/nsr/nsrlink.jsp?1970Ad02,B}{1970Ad02}); 3603 keV \textit{15} (\href{https://www.nndc.bnl.gov/nsr/nsrlink.jsp?1977Ev01,B}{1977Ev01}); and 3616 keV \textit{3}\vspace{0.1cm}}&\\
&&&&&&&&\parbox[t][0.3cm]{9.607421cm}{\raggedright {\ }{\ }{\ }(\href{https://www.nndc.bnl.gov/nsr/nsrlink.jsp?2005Pa50,B}{2005Pa50}). See also 3610 keV (\href{https://www.nndc.bnl.gov/nsr/nsrlink.jsp?1964Br13,B}{1964Br13}, \href{https://www.nndc.bnl.gov/nsr/nsrlink.jsp?1968Sh09,B}{1968Sh09}, \href{https://www.nndc.bnl.gov/nsr/nsrlink.jsp?1968To09,B}{1968To09});\vspace{0.1cm}}&\\
&&&&&&&&\parbox[t][0.3cm]{9.607421cm}{\raggedright {\ }{\ }{\ }3616.4 keV (\href{https://www.nndc.bnl.gov/nsr/nsrlink.jsp?1981Ne09,B}{1981Ne09}: the state was used as calibration and its\vspace{0.1cm}}&\\
&&&&&&&&\parbox[t][0.3cm]{9.607421cm}{\raggedright {\ }{\ }{\ }energy was taken from (\href{https://www.nndc.bnl.gov/nsr/nsrlink.jsp?1969Ro08,B}{1969Ro08})); and 3616 keV (\href{https://www.nndc.bnl.gov/nsr/nsrlink.jsp?1991Ga03,B}{1991Ga03}: this\vspace{0.1cm}}&\\
&&&&&&&&\parbox[t][0.3cm]{9.607421cm}{\raggedright {\ }{\ }{\ }state was used as a calibration point).\vspace{0.1cm}}&\\
&&&&&&&&\parbox[t][0.3cm]{9.607421cm}{\raggedright J\ensuremath{^{\pi}},L: From the zero-range DWBA analyses of (\href{https://www.nndc.bnl.gov/nsr/nsrlink.jsp?1968To09,B}{1968To09}: J\ensuremath{^{\ensuremath{\pi}}}=2\ensuremath{^{\textnormal{+}}} and\vspace{0.1cm}}&\\
&&&&&&&&\parbox[t][0.3cm]{9.607421cm}{\raggedright {\ }{\ }{\ }L=2, see Fig. 11) and (\href{https://www.nndc.bnl.gov/nsr/nsrlink.jsp?1977Ev01,B}{1977Ev01}: J\ensuremath{^{\ensuremath{\pi}}}=2\ensuremath{^{\textnormal{+}}}, see Fig. 4 and\vspace{0.1cm}}&\\
&&&&&&&&\parbox[t][0.3cm]{9.607421cm}{\raggedright {\ }{\ }{\ }E(\ensuremath{^{\textnormal{3}}}He)=15 MeV). See also a similar assignment from (\href{https://www.nndc.bnl.gov/nsr/nsrlink.jsp?1991Ga03,B}{1991Ga03}:\vspace{0.1cm}}&\\
&&&&&&&&\parbox[t][0.3cm]{9.607421cm}{\raggedright {\ }{\ }{\ }J\ensuremath{^{\ensuremath{\pi}}}=2\ensuremath{^{\textnormal{+}}} from the analysis of mirror states in \ensuremath{^{\textnormal{18}}}Ne and \ensuremath{^{\textnormal{18}}}O).\vspace{0.1cm}}&\\
&&&&&&&&\parbox[t][0.3cm]{9.607421cm}{\raggedright Q-value={\textminus}6807 keV \textit{10} (\href{https://www.nndc.bnl.gov/nsr/nsrlink.jsp?1961To03,B}{1961To03}); and {\textminus}6812 keV (\href{https://www.nndc.bnl.gov/nsr/nsrlink.jsp?1991Ga03,B}{1991Ga03}).\vspace{0.1cm}}&\\
&&&&&&&&\parbox[t][0.3cm]{9.607421cm}{\raggedright Threshold energy=8092 keV \textit{11} (\href{https://www.nndc.bnl.gov/nsr/nsrlink.jsp?1961To03,B}{1961To03}).\vspace{0.1cm}}&\\
&&&&&&&&\parbox[t][0.3cm]{9.607421cm}{\raggedright d\ensuremath{\sigma}/d\ensuremath{\Omega}\ensuremath{_{\textnormal{c.m.}}}(\ensuremath{\theta}\ensuremath{_{\textnormal{c.m.}}}=0\ensuremath{^\circ})=0.85 mb/sr at E=15 MeV (\href{https://www.nndc.bnl.gov/nsr/nsrlink.jsp?1977Ev01,B}{1977Ev01}).\vspace{0.1cm}}&\\
&&&&&&&&\parbox[t][0.3cm]{9.607421cm}{\raggedright d\ensuremath{\sigma}/d\ensuremath{\Omega}\ensuremath{_{\textnormal{c.m.}}}(\ensuremath{\theta}\ensuremath{_{\textnormal{c.m.}}}=0\ensuremath{^\circ})=0.92 mb/sr at E=18 MeV (\href{https://www.nndc.bnl.gov/nsr/nsrlink.jsp?1977Ev01,B}{1977Ev01}).\vspace{0.1cm}}&\\
&&&&&&&&\parbox[t][0.3cm]{9.607421cm}{\raggedright d\ensuremath{\sigma}/d\ensuremath{\Omega}\ensuremath{_{\textnormal{c.m.}}}(\ensuremath{\theta}\ensuremath{_{\textnormal{c.m.}}}=0\ensuremath{^\circ})=0.55 mb/sr at E=21 MeV (\href{https://www.nndc.bnl.gov/nsr/nsrlink.jsp?1977Ev01,B}{1977Ev01}).\vspace{0.1cm}}&\\
\multicolumn{1}{r@{}}{4518}&\multicolumn{1}{@{ }l}{{\it 4}}&\multicolumn{1}{l}{1\ensuremath{^{-}}}&\multicolumn{1}{r@{}}{9}&\multicolumn{1}{@{ }l}{keV {\it 6}}&\multicolumn{1}{l}{1}&&&\parbox[t][0.3cm]{9.607421cm}{\raggedright E(level): Weighted average of 4505 keV \textit{15} (\href{https://www.nndc.bnl.gov/nsr/nsrlink.jsp?1970Ad02,B}{1970Ad02}); 4520 keV \textit{1}\vspace{0.1cm}}&\\
&&&&&&&&\parbox[t][0.3cm]{9.607421cm}{\raggedright {\ }{\ }{\ }(stat.) \textit{7} (sys.) (\href{https://www.nndc.bnl.gov/nsr/nsrlink.jsp?1991Ga03,B}{1991Ga03}: this excitation energy was computed\vspace{0.1cm}}&\\
&&&&&&&&\parbox[t][0.3cm]{9.607421cm}{\raggedright {\ }{\ }{\ }using a \ensuremath{^{\textnormal{18}}}Ne mass excess of 5319 keV \textit{5} (\href{https://www.nndc.bnl.gov/nsr/nsrlink.jsp?1987Aj02,B}{1987Aj02}). The current\vspace{0.1cm}}&\\
&&&&&&&&\parbox[t][0.3cm]{9.607421cm}{\raggedright {\ }{\ }{\ }\ensuremath{^{\textnormal{18}}}Ne mass excess is 5317.617 keV \textit{363} (\href{https://www.nndc.bnl.gov/nsr/nsrlink.jsp?2021Wa16,B}{2021Wa16}), which is\vspace{0.1cm}}&\\
&&&&&&&&\parbox[t][0.3cm]{9.607421cm}{\raggedright {\ }{\ }{\ }consistent with the one used by the authors, so the evaluator did not\vspace{0.1cm}}&\\
\end{longtable}
\begin{textblock}{29}(0,27.3)
Continued on next page (footnotes at end of table)
\end{textblock}
\clearpage
\begin{longtable}{ccccccccc@{\extracolsep{\fill}}c}
\\[-.4cm]
\multicolumn{10}{c}{{\bf \small \underline{\ensuremath{^{\textnormal{16}}}O(\ensuremath{^{\textnormal{3}}}He,n)\hspace{0.2in}\href{https://www.nndc.bnl.gov/nsr/nsrlink.jsp?1953Ku08,B}{1953Ku08},\href{https://www.nndc.bnl.gov/nsr/nsrlink.jsp?2012Fo29,B}{2012Fo29} (continued)}}}\\
\multicolumn{10}{c}{~}\\
\multicolumn{10}{c}{\underline{\ensuremath{^{18}}Ne Levels (continued)}}\\
\multicolumn{10}{c}{~}\\
\multicolumn{2}{c}{E(level)$^{}$}&J$^{\pi}$$^{{\hyperlink{NE15LEVEL0}{a}}}$&\multicolumn{2}{c}{T\ensuremath{_{\textnormal{1/2}}} or \ensuremath{\Gamma}$^{}$}&L$^{{\hyperlink{NE15LEVEL0}{a}}{\hyperlink{NE15LEVEL1}{b}}{\hyperlink{NE15LEVEL2}{c}}}$&\multicolumn{2}{c}{Relative C\ensuremath{^{\textnormal{2}}}S$^{{\hyperlink{NE15LEVEL3}{d}}}$}&Comments&\\[-.2cm]
\multicolumn{2}{c}{\hrulefill}&\hrulefill&\multicolumn{2}{c}{\hrulefill}&\hrulefill&\multicolumn{2}{c}{\hrulefill}&\hrulefill&
\endhead
&&&&&&&&\parbox[t][0.3cm]{9.877061cm}{\raggedright {\ }{\ }{\ }re-scale this excitation energy); and 4519 keV \textit{1} (stat.) \textit{4} (sys.)\vspace{0.1cm}}&\\
&&&&&&&&\parbox[t][0.3cm]{9.877061cm}{\raggedright {\ }{\ }{\ }(\href{https://www.nndc.bnl.gov/nsr/nsrlink.jsp?2005Pa50,B}{2005Pa50}).\vspace{0.1cm}}&\\
&&&&&&&&\parbox[t][0.3cm]{9.877061cm}{\raggedright E(level): See also 4513 keV \textit{13} (\href{https://www.nndc.bnl.gov/nsr/nsrlink.jsp?1974Ne04,B}{1974Ne04}: this state is an unresolved\vspace{0.1cm}}&\\
&&&&&&&&\parbox[t][0.3cm]{9.877061cm}{\raggedright {\ }{\ }{\ }doublet. The width of the peak was found to increase as \ensuremath{\theta}\ensuremath{_{\textnormal{lab}}} was\vspace{0.1cm}}&\\
&&&&&&&&\parbox[t][0.3cm]{9.877061cm}{\raggedright {\ }{\ }{\ }changed from 20\ensuremath{^\circ} to 30\ensuremath{^\circ}, and the excitation energy appeared to\vspace{0.1cm}}&\\
&&&&&&&&\parbox[t][0.3cm]{9.877061cm}{\raggedright {\ }{\ }{\ }decrease by about 30 keV. This indicated that at forward angles the\vspace{0.1cm}}&\\
&&&&&&&&\parbox[t][0.3cm]{9.877061cm}{\raggedright {\ }{\ }{\ }bulk of the strength was due to the higher-lying member of the\vspace{0.1cm}}&\\
&&&&&&&&\parbox[t][0.3cm]{9.877061cm}{\raggedright {\ }{\ }{\ }doublet; at larger angles, where the observed cross section was small,\vspace{0.1cm}}&\\
&&&&&&&&\parbox[t][0.3cm]{9.877061cm}{\raggedright {\ }{\ }{\ }neither of the states dominated); 4537 keV \textit{10} (\href{https://www.nndc.bnl.gov/nsr/nsrlink.jsp?1977Ev01,B}{1977Ev01}: this state is\vspace{0.1cm}}&\\
&&&&&&&&\parbox[t][0.3cm]{9.877061cm}{\raggedright {\ }{\ }{\ }an unresolved doublet, whose reported energy is the mean energy of\vspace{0.1cm}}&\\
&&&&&&&&\parbox[t][0.3cm]{9.877061cm}{\raggedright {\ }{\ }{\ }the doublet); 4513 keV \textit{13} (\href{https://www.nndc.bnl.gov/nsr/nsrlink.jsp?1981Ne09,B}{1981Ne09}: this state is a member of a\vspace{0.1cm}}&\\
&&&&&&&&\parbox[t][0.3cm]{9.877061cm}{\raggedright {\ }{\ }{\ }doublet. At \ensuremath{\theta}\ensuremath{_{\textnormal{c.m.}}}=0\ensuremath{^\circ}, the population of this doublet is dominated by\vspace{0.1cm}}&\\
&&&&&&&&\parbox[t][0.3cm]{9.877061cm}{\raggedright {\ }{\ }{\ }the higher-lying member of the doublet and that at large angles the\vspace{0.1cm}}&\\
&&&&&&&&\parbox[t][0.3cm]{9.877061cm}{\raggedright {\ }{\ }{\ }two contributions become comparable); and 4.5 MeV (\href{https://www.nndc.bnl.gov/nsr/nsrlink.jsp?2010AlZZ,B}{2010AlZZ},\vspace{0.1cm}}&\\
&&&&&&&&\parbox[t][0.3cm]{9.877061cm}{\raggedright {\ }{\ }{\ }\href{https://www.nndc.bnl.gov/nsr/nsrlink.jsp?2010Ta17,B}{2010Ta17}, \href{https://www.nndc.bnl.gov/nsr/nsrlink.jsp?2012Al11,B}{2012Al11}).\vspace{0.1cm}}&\\
&&&&&&&&\parbox[t][0.3cm]{9.877061cm}{\raggedright (\href{https://www.nndc.bnl.gov/nsr/nsrlink.jsp?1970Ad02,B}{1970Ad02}, E\ensuremath{_{\textnormal{x}}}=4505 keV \textit{15}): the authors could not eliminate the\vspace{0.1cm}}&\\
&&&&&&&&\parbox[t][0.3cm]{9.877061cm}{\raggedright {\ }{\ }{\ }\ensuremath{^{\textnormal{17}}}O(\ensuremath{^{\textnormal{3}}}He,n) reaction as the source for populating this state. However,\vspace{0.1cm}}&\\
&&&&&&&&\parbox[t][0.3cm]{9.877061cm}{\raggedright {\ }{\ }{\ }this would imply an unreasonable cross section of 500 mb/sr at\vspace{0.1cm}}&\\
&&&&&&&&\parbox[t][0.3cm]{9.877061cm}{\raggedright {\ }{\ }{\ }\ensuremath{\theta}\ensuremath{_{\textnormal{c.m.}}}=0\ensuremath{^\circ} and E(\ensuremath{^{\textnormal{3}}}He)=12.5 MeV. So, the authors concluded that this\vspace{0.1cm}}&\\
&&&&&&&&\parbox[t][0.3cm]{9.877061cm}{\raggedright {\ }{\ }{\ }state belonged to \ensuremath{^{\textnormal{18}}}Ne. This study also pointed out that this state may\vspace{0.1cm}}&\\
&&&&&&&&\parbox[t][0.3cm]{9.877061cm}{\raggedright {\ }{\ }{\ }be the mirror of the \ensuremath{^{\textnormal{18}}}O*(4.455 MeV, 1\ensuremath{^{-}}) state, in which case the\vspace{0.1cm}}&\\
&&&&&&&&\parbox[t][0.3cm]{9.877061cm}{\raggedright {\ }{\ }{\ }large displacement between this state and its mirror would be due to a\vspace{0.1cm}}&\\
&&&&&&&&\parbox[t][0.3cm]{9.877061cm}{\raggedright {\ }{\ }{\ }double Thomas-Ehrman shift.\vspace{0.1cm}}&\\
&&&&&&&&\parbox[t][0.3cm]{9.877061cm}{\raggedright \ensuremath{\Gamma}: From (\href{https://www.nndc.bnl.gov/nsr/nsrlink.jsp?1991Ga03,B}{1991Ga03}). See also \ensuremath{\Gamma}\ensuremath{\leq}40 keV (\href{https://www.nndc.bnl.gov/nsr/nsrlink.jsp?1970Ad02,B}{1970Ad02}: deduced from\vspace{0.1cm}}&\\
&&&&&&&&\parbox[t][0.3cm]{9.877061cm}{\raggedright {\ }{\ }{\ }folding the line shape of the corresponding neutron group with a\vspace{0.1cm}}&\\
&&&&&&&&\parbox[t][0.3cm]{9.877061cm}{\raggedright {\ }{\ }{\ }negligible intrinsic width into a Breit-Wigner curve).\vspace{0.1cm}}&\\
&&&&&&&&\parbox[t][0.3cm]{9.877061cm}{\raggedright \ensuremath{\Gamma}: See also (\href{https://www.nndc.bnl.gov/nsr/nsrlink.jsp?1991Ga03,B}{1991Ga03}): deduced \ensuremath{\Gamma}\ensuremath{_{\textnormal{p}}}\ensuremath{<}0.1 keV for this state using the\vspace{0.1cm}}&\\
&&&&&&&&\parbox[t][0.3cm]{9.877061cm}{\raggedright {\ }{\ }{\ }measured proton spectroscopic factor of the mirror state in \ensuremath{^{\textnormal{18}}}O from\vspace{0.1cm}}&\\
&&&&&&&&\parbox[t][0.3cm]{9.877061cm}{\raggedright {\ }{\ }{\ }(\href{https://www.nndc.bnl.gov/nsr/nsrlink.jsp?1976Li01,B}{1976Li01}) and the single-particle width deduced from optical model\vspace{0.1cm}}&\\
&&&&&&&&\parbox[t][0.3cm]{9.877061cm}{\raggedright {\ }{\ }{\ }calculations of (\href{https://www.nndc.bnl.gov/nsr/nsrlink.jsp?1991Ga03,B}{1991Ga03}). They assumed a negligible branch for the\vspace{0.1cm}}&\\
&&&&&&&&\parbox[t][0.3cm]{9.877061cm}{\raggedright {\ }{\ }{\ }\ensuremath{^{\textnormal{18}}}Ne*(4518 keV)\ensuremath{\rightarrow}\ensuremath{^{\textnormal{17}}}F*(495 keV)+p decay such that \ensuremath{\Gamma}\ensuremath{\approx}\ensuremath{\Gamma}\ensuremath{_{\textnormal{p}_{\textnormal{0}}}}.\vspace{0.1cm}}&\\
&&&&&&&&\parbox[t][0.3cm]{9.877061cm}{\raggedright {\ }{\ }{\ }(\href{https://www.nndc.bnl.gov/nsr/nsrlink.jsp?1991Ga03,B}{1991Ga03}) used \ensuremath{\Gamma}\ensuremath{_{\textnormal{p}}}\ensuremath{<}0.1 keV (as opposed to their experimentally\vspace{0.1cm}}&\\
&&&&&&&&\parbox[t][0.3cm]{9.877061cm}{\raggedright {\ }{\ }{\ }determined value of \ensuremath{\Gamma}\ensuremath{_{\textnormal{tot}}}=9 keV \textit{6}) to calculate the \ensuremath{^{\textnormal{17}}}F(p,\ensuremath{\gamma}) reaction\vspace{0.1cm}}&\\
&&&&&&&&\parbox[t][0.3cm]{9.877061cm}{\raggedright {\ }{\ }{\ }rate. They argued that the observed width of \ensuremath{\Gamma}=9 keV \textit{6} is consistent\vspace{0.1cm}}&\\
&&&&&&&&\parbox[t][0.3cm]{9.877061cm}{\raggedright {\ }{\ }{\ }with zero [at 2\ensuremath{\sigma}], and therefore deduced \ensuremath{\Gamma}\ensuremath{_{\textnormal{p}}}\ensuremath{<}0.1 keV from the mirror\vspace{0.1cm}}&\\
&&&&&&&&\parbox[t][0.3cm]{9.877061cm}{\raggedright {\ }{\ }{\ }level.\vspace{0.1cm}}&\\
&&&&&&&&\parbox[t][0.3cm]{9.877061cm}{\raggedright J\ensuremath{^{\pi}},L: From the zero-range DWBA analysis of (\href{https://www.nndc.bnl.gov/nsr/nsrlink.jsp?1977Ev01,B}{1977Ev01}) using\vspace{0.1cm}}&\\
&&&&&&&&\parbox[t][0.3cm]{9.877061cm}{\raggedright {\ }{\ }{\ }DWUCK2 with L=1. The authors assigned J\ensuremath{^{\ensuremath{\pi}}}=0\ensuremath{^{\textnormal{+}}}, 1\ensuremath{^{-}} to this state but\vspace{0.1cm}}&\\
&&&&&&&&\parbox[t][0.3cm]{9.877061cm}{\raggedright {\ }{\ }{\ }only present the L=1 fits (see Fig. 4). The agreements between\vspace{0.1cm}}&\\
&&&&&&&&\parbox[t][0.3cm]{9.877061cm}{\raggedright {\ }{\ }{\ }DWBA fits and the neutron angular distribution data at E(\ensuremath{^{\textnormal{3}}}He)=15\vspace{0.1cm}}&\\
&&&&&&&&\parbox[t][0.3cm]{9.877061cm}{\raggedright {\ }{\ }{\ }MeV and 18 MeV are very good for L=1. L=0 fits are not presented;\vspace{0.1cm}}&\\
&&&&&&&&\parbox[t][0.3cm]{9.877061cm}{\raggedright {\ }{\ }{\ }however, the authors reported J\ensuremath{^{\ensuremath{\pi}}}=0\ensuremath{^{\textnormal{+}}} and 1\ensuremath{^{-}} because this state was\vspace{0.1cm}}&\\
&&&&&&&&\parbox[t][0.3cm]{9.877061cm}{\raggedright {\ }{\ }{\ }expected to be an unresolved doublet. The evaluator adopted J\ensuremath{^{\ensuremath{\pi}}}=1\ensuremath{^{-}}\vspace{0.1cm}}&\\
&&&&&&&&\parbox[t][0.3cm]{9.877061cm}{\raggedright {\ }{\ }{\ }with L=1 from these data and the other supporting evidence mentioned\vspace{0.1cm}}&\\
&&&&&&&&\parbox[t][0.3cm]{9.877061cm}{\raggedright {\ }{\ }{\ }below.\vspace{0.1cm}}&\\
&&&&&&&&\parbox[t][0.3cm]{9.877061cm}{\raggedright J\ensuremath{^{\pi}},L: See also (\href{https://www.nndc.bnl.gov/nsr/nsrlink.jsp?1970Ad02,B}{1970Ad02}: J\ensuremath{^{\ensuremath{\pi}}}=(0\ensuremath{^{\textnormal{+}}}, 1\ensuremath{^{-}}) deduced from the analog levels\vspace{0.1cm}}&\\
&&&&&&&&\parbox[t][0.3cm]{9.877061cm}{\raggedright {\ }{\ }{\ }analysis); (\href{https://www.nndc.bnl.gov/nsr/nsrlink.jsp?1981Ne09,B}{1981Ne09}: J\ensuremath{^{\ensuremath{\pi}}}=(0\ensuremath{^{\textnormal{+}}}, 1\ensuremath{^{-}}) from DWBA analysis with L=0\vspace{0.1cm}}&\\
&&&&&&&&\parbox[t][0.3cm]{9.877061cm}{\raggedright {\ }{\ }{\ }and L=1); (\href{https://www.nndc.bnl.gov/nsr/nsrlink.jsp?1991Ga03,B}{1991Ga03}: J\ensuremath{^{\ensuremath{\pi}}}=1\ensuremath{^{-}} from the analysis of mirror states in\vspace{0.1cm}}&\\
&&&&&&&&\parbox[t][0.3cm]{9.877061cm}{\raggedright {\ }{\ }{\ }\ensuremath{^{\textnormal{18}}}Ne and \ensuremath{^{\textnormal{18}}}O); and (\href{https://www.nndc.bnl.gov/nsr/nsrlink.jsp?2005Pa50,B}{2005Pa50}: J=1 from comparison of the measured\vspace{0.1cm}}&\\
&&&&&&&&\parbox[t][0.3cm]{9.877061cm}{\raggedright {\ }{\ }{\ }differential cross sections to those predicted by the Hauser-Feshbach\vspace{0.1cm}}&\\
&&&&&&&&\parbox[t][0.3cm]{9.877061cm}{\raggedright {\ }{\ }{\ }theory, which is sensitive to J but not to \ensuremath{\pi} changes).\vspace{0.1cm}}&\\
&&&&&&&&\parbox[t][0.3cm]{9.877061cm}{\raggedright (\href{https://www.nndc.bnl.gov/nsr/nsrlink.jsp?2005Pa50,B}{2005Pa50}): Figs. 4 and 5 show measured and Hauser-Feshbach\vspace{0.1cm}}&\\
&&&&&&&&\parbox[t][0.3cm]{9.877061cm}{\raggedright {\ }{\ }{\ }differential cross sections of the observed states in graphical formats\vspace{0.1cm}}&\\
&&&&&&&&\parbox[t][0.3cm]{9.877061cm}{\raggedright {\ }{\ }{\ }per beam energy.\vspace{0.1cm}}&\\
&&&&&&&&\parbox[t][0.3cm]{9.877061cm}{\raggedright Q-value={\textminus}7716 keV \textit{1} (stat.) \textit{5} (sys.) (\href{https://www.nndc.bnl.gov/nsr/nsrlink.jsp?1991Ga03,B}{1991Ga03}).\vspace{0.1cm}}&\\
\end{longtable}
\begin{textblock}{29}(0,27.3)
Continued on next page (footnotes at end of table)
\end{textblock}
\clearpage
\begin{longtable}{ccccccc@{\extracolsep{\fill}}c}
\\[-.4cm]
\multicolumn{8}{c}{{\bf \small \underline{\ensuremath{^{\textnormal{16}}}O(\ensuremath{^{\textnormal{3}}}He,n)\hspace{0.2in}\href{https://www.nndc.bnl.gov/nsr/nsrlink.jsp?1953Ku08,B}{1953Ku08},\href{https://www.nndc.bnl.gov/nsr/nsrlink.jsp?2012Fo29,B}{2012Fo29} (continued)}}}\\
\multicolumn{8}{c}{~}\\
\multicolumn{8}{c}{\underline{\ensuremath{^{18}}Ne Levels (continued)}}\\
\multicolumn{8}{c}{~}\\
\multicolumn{2}{c}{E(level)$^{}$}&J$^{\pi}$$^{{\hyperlink{NE15LEVEL0}{a}}}$&\multicolumn{2}{c}{T\ensuremath{_{\textnormal{1/2}}} or \ensuremath{\Gamma}$^{}$}&L$^{{\hyperlink{NE15LEVEL0}{a}}{\hyperlink{NE15LEVEL1}{b}}{\hyperlink{NE15LEVEL2}{c}}}$&Comments&\\[-.2cm]
\multicolumn{2}{c}{\hrulefill}&\hrulefill&\multicolumn{2}{c}{\hrulefill}&\hrulefill&\hrulefill&
\endhead
\multicolumn{1}{r@{}}{4527}&\multicolumn{1}{@{ }l}{{\it 4}}&\multicolumn{1}{l}{3\ensuremath{^{+}}}&\multicolumn{1}{r@{}}{17}&\multicolumn{1}{@{ }l}{keV {\it 4}}&&\parbox[t][0.3cm]{11.747241cm}{\raggedright E(level): From E\ensuremath{_{\textnormal{x}}}=4527 keV \textit{1} (stat.) \textit{4} (sys.) (\href{https://www.nndc.bnl.gov/nsr/nsrlink.jsp?2005Pa50,B}{2005Pa50}). See also 4537 keV \textit{10}\vspace{0.1cm}}&\\
&&&&&&\parbox[t][0.3cm]{11.747241cm}{\raggedright {\ }{\ }{\ }(\href{https://www.nndc.bnl.gov/nsr/nsrlink.jsp?1977Ev01,B}{1977Ev01}: this state is identified as an unresolved doublet, whose reported energy\vspace{0.1cm}}&\\
&&&&&&\parbox[t][0.3cm]{11.747241cm}{\raggedright {\ }{\ }{\ }is the mean energy of the doublet); and 4561 keV \textit{6} (stat.) \textit{7} (sys.) (\href{https://www.nndc.bnl.gov/nsr/nsrlink.jsp?1991Ga03,B}{1991Ga03}: this\vspace{0.1cm}}&\\
&&&&&&\parbox[t][0.3cm]{11.747241cm}{\raggedright {\ }{\ }{\ }excitation energy was computed using a \ensuremath{^{\textnormal{18}}}Ne mass excess of 5319 keV \textit{5}\vspace{0.1cm}}&\\
&&&&&&\parbox[t][0.3cm]{11.747241cm}{\raggedright {\ }{\ }{\ }(\href{https://www.nndc.bnl.gov/nsr/nsrlink.jsp?1987Aj02,B}{1987Aj02}). The current \ensuremath{^{\textnormal{18}}}Ne mass excess is 5317.617 keV \textit{363} (\href{https://www.nndc.bnl.gov/nsr/nsrlink.jsp?2021Wa16,B}{2021Wa16}),\vspace{0.1cm}}&\\
&&&&&&\parbox[t][0.3cm]{11.747241cm}{\raggedright {\ }{\ }{\ }which is consistent with the one used by the authors, so the evaluator did not\vspace{0.1cm}}&\\
&&&&&&\parbox[t][0.3cm]{11.747241cm}{\raggedright {\ }{\ }{\ }re-scale this excitation energy. Also note that this state was only observed at\vspace{0.1cm}}&\\
&&&&&&\parbox[t][0.3cm]{11.747241cm}{\raggedright {\ }{\ }{\ }E\ensuremath{_{\textnormal{beam}}}=10.9 MeV and at \ensuremath{\theta}\ensuremath{_{\textnormal{lab}}}=124.7\ensuremath{^\circ}). (\href{https://www.nndc.bnl.gov/nsr/nsrlink.jsp?2005Pa50,B}{2005Pa50}) saw no evidence for this level\vspace{0.1cm}}&\\
&&&&&&\parbox[t][0.3cm]{11.747241cm}{\raggedright {\ }{\ }{\ }and suggested that it is likely that the experiment of (\href{https://www.nndc.bnl.gov/nsr/nsrlink.jsp?1991Ga03,B}{1991Ga03}), whose resolution\vspace{0.1cm}}&\\
&&&&&&\parbox[t][0.3cm]{11.747241cm}{\raggedright {\ }{\ }{\ }was a factor of 2 poorer, was unable to resolve the states observed in (\href{https://www.nndc.bnl.gov/nsr/nsrlink.jsp?2005Pa50,B}{2005Pa50}) at\vspace{0.1cm}}&\\
&&&&&&\parbox[t][0.3cm]{11.747241cm}{\raggedright {\ }{\ }{\ }4519 keV and 4527 keV.\vspace{0.1cm}}&\\
&&&&&&\parbox[t][0.3cm]{11.747241cm}{\raggedright \ensuremath{\Gamma}: From (\href{https://www.nndc.bnl.gov/nsr/nsrlink.jsp?2005Pa50,B}{2005Pa50}). The error on the width is dominated by the systematic\vspace{0.1cm}}&\\
&&&&&&\parbox[t][0.3cm]{11.747241cm}{\raggedright {\ }{\ }{\ }uncertainty (\href{https://www.nndc.bnl.gov/nsr/nsrlink.jsp?2005Pa50,B}{2005Pa50}). See also \ensuremath{\Gamma}=25 keV (\href{https://www.nndc.bnl.gov/nsr/nsrlink.jsp?1991Ga03,B}{1991Ga03}: not measured but estimated\vspace{0.1cm}}&\\
&&&&&&\parbox[t][0.3cm]{11.747241cm}{\raggedright {\ }{\ }{\ }from a Woods-Saxon calculation for E\ensuremath{_{\textnormal{x}}}=4561 keV \textit{9}).\vspace{0.1cm}}&\\
&&&&&&\parbox[t][0.3cm]{11.747241cm}{\raggedright (\href{https://www.nndc.bnl.gov/nsr/nsrlink.jsp?1991Ga03,B}{1991Ga03}) recommended \ensuremath{\Gamma}\ensuremath{_{\ensuremath{\gamma}}}(3\ensuremath{^{\textnormal{+}}_{\textnormal{1}}}\ensuremath{\rightarrow}2\ensuremath{^{\textnormal{+}}_{\textnormal{1}}})=25 meV \textit{16}, \ensuremath{\Gamma}\ensuremath{_{\ensuremath{\gamma}}}(3\ensuremath{^{\textnormal{+}}_{\textnormal{1}}}\ensuremath{\rightarrow}2\ensuremath{^{\textnormal{+}}_{\textnormal{2}}})=3.8 meV \textit{31},\vspace{0.1cm}}&\\
&&&&&&\parbox[t][0.3cm]{11.747241cm}{\raggedright {\ }{\ }{\ }and \ensuremath{\Gamma}\ensuremath{_{\ensuremath{\gamma}}}(3\ensuremath{^{\textnormal{+}}_{\textnormal{1}}}\ensuremath{\rightarrow}4\ensuremath{^{\textnormal{+}}_{\textnormal{1}}})=0.8 meV \textit{8}. Under the assumption that the \ensuremath{^{\textnormal{18}}}Ne B(M1)\vspace{0.1cm}}&\\
&&&&&&\parbox[t][0.3cm]{11.747241cm}{\raggedright {\ }{\ }{\ }transitions are identical to those in \ensuremath{^{\textnormal{18}}}O, (\href{https://www.nndc.bnl.gov/nsr/nsrlink.jsp?1991Ga03,B}{1991Ga03}) deduced the lower-value \ensuremath{\gamma}-ray\vspace{0.1cm}}&\\
&&&&&&\parbox[t][0.3cm]{11.747241cm}{\raggedright {\ }{\ }{\ }partial widths mentioned above. The higher values were deduced by assuming that\vspace{0.1cm}}&\\
&&&&&&\parbox[t][0.3cm]{11.747241cm}{\raggedright {\ }{\ }{\ }the actual B(M1) transitions could be twice as large as the shell-model predictions\vspace{0.1cm}}&\\
&&&&&&\parbox[t][0.3cm]{11.747241cm}{\raggedright {\ }{\ }{\ }obtained from a private communication between the authors and B. A. Brown.\vspace{0.1cm}}&\\
&&&&&&\parbox[t][0.3cm]{11.747241cm}{\raggedright J\ensuremath{^{\pi}},L: From (\href{https://www.nndc.bnl.gov/nsr/nsrlink.jsp?2005Pa50,B}{2005Pa50}: J=3 from comparing the measured differential cross sections\vspace{0.1cm}}&\\
&&&&&&\parbox[t][0.3cm]{11.747241cm}{\raggedright {\ }{\ }{\ }to the Hauser-Feshbach predicted ones); and (\href{https://www.nndc.bnl.gov/nsr/nsrlink.jsp?1991Ga03,B}{1991Ga03}: J\ensuremath{^{\ensuremath{\pi}}}=3\ensuremath{^{\textnormal{+}}} from the analysis of\vspace{0.1cm}}&\\
&&&&&&\parbox[t][0.3cm]{11.747241cm}{\raggedright {\ }{\ }{\ }mirror states in \ensuremath{^{\textnormal{18}}}Ne and \ensuremath{^{\textnormal{18}}}O).\vspace{0.1cm}}&\\
&&&&&&\parbox[t][0.3cm]{11.747241cm}{\raggedright Q-value={\textminus}7757 keV \textit{6} (stat.) \textit{5} (sys.) (\href{https://www.nndc.bnl.gov/nsr/nsrlink.jsp?1991Ga03,B}{1991Ga03}: assuming E\ensuremath{_{\textnormal{x}}}=4561 keV \textit{9}).\vspace{0.1cm}}&\\
\multicolumn{1}{r@{}}{4589}&\multicolumn{1}{@{}l}{\ensuremath{^{{\hyperlink{NE15LEVEL4}{e}}}} {\it 4}}&\multicolumn{1}{l}{0\ensuremath{^{+}}}&\multicolumn{1}{r@{}}{4}&\multicolumn{1}{@{ }l}{keV {\it 4}}&\multicolumn{1}{l}{0}&\parbox[t][0.3cm]{11.747241cm}{\raggedright T=1 (\href{https://www.nndc.bnl.gov/nsr/nsrlink.jsp?1974Ne04,B}{1974Ne04},\href{https://www.nndc.bnl.gov/nsr/nsrlink.jsp?1981Ne09,B}{1981Ne09})\vspace{0.1cm}}&\\
&&&&&&\parbox[t][0.3cm]{11.747241cm}{\raggedright E(level): Weighted average of 4590 keV \textit{30} (\href{https://www.nndc.bnl.gov/nsr/nsrlink.jsp?1968To09,B}{1968To09}); 4571 keV \textit{15} (\href{https://www.nndc.bnl.gov/nsr/nsrlink.jsp?1970Ad02,B}{1970Ad02});\vspace{0.1cm}}&\\
&&&&&&\parbox[t][0.3cm]{11.747241cm}{\raggedright {\ }{\ }{\ }4589 keV \textit{1} (stat.) \textit{7} (sys.) (\href{https://www.nndc.bnl.gov/nsr/nsrlink.jsp?1991Ga03,B}{1991Ga03}: this excitation energy was computed using a\vspace{0.1cm}}&\\
&&&&&&\parbox[t][0.3cm]{11.747241cm}{\raggedright {\ }{\ }{\ }\ensuremath{^{\textnormal{18}}}Ne mass excess of 5319 keV \textit{5} (\href{https://www.nndc.bnl.gov/nsr/nsrlink.jsp?1987Aj02,B}{1987Aj02}). The current \ensuremath{^{\textnormal{18}}}Ne mass excess is\vspace{0.1cm}}&\\
&&&&&&\parbox[t][0.3cm]{11.747241cm}{\raggedright {\ }{\ }{\ }5317.617 keV \textit{363} (\href{https://www.nndc.bnl.gov/nsr/nsrlink.jsp?2021Wa16,B}{2021Wa16}), which is consistent with the one used by the\vspace{0.1cm}}&\\
&&&&&&\parbox[t][0.3cm]{11.747241cm}{\raggedright {\ }{\ }{\ }authors, so the evaluator did not re-scale this excitation energy); and 4590 keV \textit{1}\vspace{0.1cm}}&\\
&&&&&&\parbox[t][0.3cm]{11.747241cm}{\raggedright {\ }{\ }{\ }(stat.) \textit{4} (sys.) (\href{https://www.nndc.bnl.gov/nsr/nsrlink.jsp?2005Pa50,B}{2005Pa50}).\vspace{0.1cm}}&\\
&&&&&&\parbox[t][0.3cm]{11.747241cm}{\raggedright E(level): See also 4550 keV (\href{https://www.nndc.bnl.gov/nsr/nsrlink.jsp?1968Sh09,B}{1968Sh09}); 4587 keV \textit{13} (\href{https://www.nndc.bnl.gov/nsr/nsrlink.jsp?1974Ne04,B}{1974Ne04}: this state is\vspace{0.1cm}}&\\
&&&&&&\parbox[t][0.3cm]{11.747241cm}{\raggedright {\ }{\ }{\ }identified as an unresolved doublet. The width of the peak was found to increase as\vspace{0.1cm}}&\\
&&&&&&\parbox[t][0.3cm]{11.747241cm}{\raggedright {\ }{\ }{\ }\ensuremath{\theta}\ensuremath{_{\textnormal{lab}}} was changed from 20\ensuremath{^\circ} to 30\ensuremath{^\circ}, and the excitation energy appeared to decrease\vspace{0.1cm}}&\\
&&&&&&\parbox[t][0.3cm]{11.747241cm}{\raggedright {\ }{\ }{\ }by about 30 keV. This indicated that at forward angles the bulk of the strength was\vspace{0.1cm}}&\\
&&&&&&\parbox[t][0.3cm]{11.747241cm}{\raggedright {\ }{\ }{\ }due to the higher-lying member of the doublet; at larger angles, where the observed\vspace{0.1cm}}&\\
&&&&&&\parbox[t][0.3cm]{11.747241cm}{\raggedright {\ }{\ }{\ }cross section was small, neither of the states dominated); and 4587 keV \textit{13}\vspace{0.1cm}}&\\
&&&&&&\parbox[t][0.3cm]{11.747241cm}{\raggedright {\ }{\ }{\ }(\href{https://www.nndc.bnl.gov/nsr/nsrlink.jsp?1981Ne09,B}{1981Ne09}: this state is a member of a doublet. At \ensuremath{\theta}\ensuremath{_{\textnormal{c.m.}}}=0\ensuremath{^\circ}, the population of this\vspace{0.1cm}}&\\
&&&&&&\parbox[t][0.3cm]{11.747241cm}{\raggedright {\ }{\ }{\ }doublet is dominated by the higher-lying member of the doublet and at large angles\vspace{0.1cm}}&\\
&&&&&&\parbox[t][0.3cm]{11.747241cm}{\raggedright {\ }{\ }{\ }the two contributions become comparable).\vspace{0.1cm}}&\\
&&&&&&\parbox[t][0.3cm]{11.747241cm}{\raggedright (\href{https://www.nndc.bnl.gov/nsr/nsrlink.jsp?1970Ad02,B}{1970Ad02}): the \ensuremath{^{\textnormal{17}}}O(\ensuremath{^{\textnormal{3}}}He,n) reaction could not be eliminated as the source for the\vspace{0.1cm}}&\\
&&&&&&\parbox[t][0.3cm]{11.747241cm}{\raggedright {\ }{\ }{\ }production of this state; however, this would imply an unreasonably large cross\vspace{0.1cm}}&\\
&&&&&&\parbox[t][0.3cm]{11.747241cm}{\raggedright {\ }{\ }{\ }section of 500 mb/sr at \ensuremath{\theta}\ensuremath{_{\textnormal{c.m.}}}=0\ensuremath{^\circ} and E(\ensuremath{^{\textnormal{3}}}He)=12.5 MeV. So, the authors concluded\vspace{0.1cm}}&\\
&&&&&&\parbox[t][0.3cm]{11.747241cm}{\raggedright {\ }{\ }{\ }that this state belonged to \ensuremath{^{\textnormal{18}}}Ne. They also mentioned that this state may be the\vspace{0.1cm}}&\\
&&&&&&\parbox[t][0.3cm]{11.747241cm}{\raggedright {\ }{\ }{\ }mirror to the \ensuremath{^{\textnormal{18}}}O*(5336 keV,0\ensuremath{^{\textnormal{+}}}) state, in which case the large shift is due to the\vspace{0.1cm}}&\\
&&&&&&\parbox[t][0.3cm]{11.747241cm}{\raggedright {\ }{\ }{\ }double Thomas-Ehrman shift.\vspace{0.1cm}}&\\
&&&&&&\parbox[t][0.3cm]{11.747241cm}{\raggedright \ensuremath{\Gamma}: From \ensuremath{\Gamma}=4 keV \textit{4} measured by (\href{https://www.nndc.bnl.gov/nsr/nsrlink.jsp?1991Ga03,B}{1991Ga03}). See also \ensuremath{\Gamma}\ensuremath{\leq}130 keV (\href{https://www.nndc.bnl.gov/nsr/nsrlink.jsp?1968To09,B}{1968To09}: the\vspace{0.1cm}}&\\
&&&&&&\parbox[t][0.3cm]{11.747241cm}{\raggedright {\ }{\ }{\ }width of the corresponding neutron group); \ensuremath{\Gamma}\ensuremath{\leq}40 keV (\href{https://www.nndc.bnl.gov/nsr/nsrlink.jsp?1970Ad02,B}{1970Ad02}: deduced by\vspace{0.1cm}}&\\
&&&&&&\parbox[t][0.3cm]{11.747241cm}{\raggedright {\ }{\ }{\ }folding the line shape of the corresponding neutron group with a negligible intrinsic\vspace{0.1cm}}&\\
&&&&&&\parbox[t][0.3cm]{11.747241cm}{\raggedright {\ }{\ }{\ }width into a Breit-Wigner curve); and \ensuremath{\Gamma}\ensuremath{_{\textnormal{p}}}=1 keV (\href{https://www.nndc.bnl.gov/nsr/nsrlink.jsp?1991Ga03,B}{1991Ga03}). This value was\vspace{0.1cm}}&\\
&&&&&&\parbox[t][0.3cm]{11.747241cm}{\raggedright {\ }{\ }{\ }obtained for the state populated at E\ensuremath{_{\textnormal{x}}}=4589 keV. Its \ensuremath{\Gamma}\ensuremath{_{\textnormal{p}}}=1 keV width was deduced\vspace{0.1cm}}&\\
&&&&&&\parbox[t][0.3cm]{11.747241cm}{\raggedright {\ }{\ }{\ }using the experimentally determined proton spectroscopic factor of the mirror level\vspace{0.1cm}}&\\
&&&&&&\parbox[t][0.3cm]{11.747241cm}{\raggedright {\ }{\ }{\ }in \ensuremath{^{\textnormal{18}}}O (\href{https://www.nndc.bnl.gov/nsr/nsrlink.jsp?1976Li01,B}{1976Li01}) and the single-particle reduced width, which was determined\vspace{0.1cm}}&\\
\end{longtable}
\begin{textblock}{29}(0,27.3)
Continued on next page (footnotes at end of table)
\end{textblock}
\clearpage
\begin{longtable}{ccccccc@{\extracolsep{\fill}}c}
\\[-.4cm]
\multicolumn{8}{c}{{\bf \small \underline{\ensuremath{^{\textnormal{16}}}O(\ensuremath{^{\textnormal{3}}}He,n)\hspace{0.2in}\href{https://www.nndc.bnl.gov/nsr/nsrlink.jsp?1953Ku08,B}{1953Ku08},\href{https://www.nndc.bnl.gov/nsr/nsrlink.jsp?2012Fo29,B}{2012Fo29} (continued)}}}\\
\multicolumn{8}{c}{~}\\
\multicolumn{8}{c}{\underline{\ensuremath{^{18}}Ne Levels (continued)}}\\
\multicolumn{8}{c}{~}\\
\multicolumn{2}{c}{E(level)$^{}$}&J$^{\pi}$$^{{\hyperlink{NE15LEVEL0}{a}}}$&\multicolumn{2}{c}{T\ensuremath{_{\textnormal{1/2}}} or \ensuremath{\Gamma}$^{}$}&L$^{{\hyperlink{NE15LEVEL0}{a}}{\hyperlink{NE15LEVEL1}{b}}{\hyperlink{NE15LEVEL2}{c}}}$&Comments&\\[-.2cm]
\multicolumn{2}{c}{\hrulefill}&\hrulefill&\multicolumn{2}{c}{\hrulefill}&\hrulefill&\hrulefill&
\endhead
&&&&&&\parbox[t][0.3cm]{11.48525cm}{\raggedright {\ }{\ }{\ }from an optical model calculation by (\href{https://www.nndc.bnl.gov/nsr/nsrlink.jsp?1991Ga03,B}{1991Ga03}). \ensuremath{\Gamma}\ensuremath{_{\textnormal{p}}}=1 keV was used by the\vspace{0.1cm}}&\\
&&&&&&\parbox[t][0.3cm]{11.48525cm}{\raggedright {\ }{\ }{\ }authors to deduce the \ensuremath{^{\textnormal{17}}}F(p,\ensuremath{\gamma}) reaction rate. It was assumed that \ensuremath{\Gamma}\ensuremath{\approx}\ensuremath{\Gamma}\ensuremath{_{\textnormal{p}}}, where\vspace{0.1cm}}&\\
&&&&&&\parbox[t][0.3cm]{11.48525cm}{\raggedright {\ }{\ }{\ }\ensuremath{\Gamma}\ensuremath{_{\textnormal{p}}}=1 keV is consistent with the observed width of \ensuremath{\Gamma}=4 keV \textit{4}. (\href{https://www.nndc.bnl.gov/nsr/nsrlink.jsp?1991Ga03,B}{1991Ga03})\vspace{0.1cm}}&\\
&&&&&&\parbox[t][0.3cm]{11.48525cm}{\raggedright {\ }{\ }{\ }assumed that this level proton decays to the \ensuremath{^{\textnormal{17}}}F\ensuremath{_{\textnormal{g.s.}}}, which implies that \ensuremath{\Gamma}\ensuremath{\approx}\ensuremath{\Gamma}\ensuremath{_{\textnormal{p}_{\textnormal{0}}}}.\vspace{0.1cm}}&\\
&&&&&&\parbox[t][0.3cm]{11.48525cm}{\raggedright J\ensuremath{^{\pi}}: From the DWBA analyses of (1) (\href{https://www.nndc.bnl.gov/nsr/nsrlink.jsp?1974Ne04,B}{1974Ne04}): these data are not presented but\vspace{0.1cm}}&\\
&&&&&&\parbox[t][0.3cm]{11.48525cm}{\raggedright {\ }{\ }{\ }the authors mention that the DWBA analysis favors L=0 for the higher energy\vspace{0.1cm}}&\\
&&&&&&\parbox[t][0.3cm]{11.48525cm}{\raggedright {\ }{\ }{\ }state in the 4.5-MeV doublet (this state). This is confirmed with the presented\vspace{0.1cm}}&\\
&&&&&&\parbox[t][0.3cm]{11.48525cm}{\raggedright {\ }{\ }{\ }DWBA analysis of this state using their \ensuremath{^{\textnormal{20}}}Ne(p,t) data. (2) (\href{https://www.nndc.bnl.gov/nsr/nsrlink.jsp?1981Ne09,B}{1981Ne09}): J\ensuremath{^{\ensuremath{\pi}}}=(0\ensuremath{^{\textnormal{+}}},\vspace{0.1cm}}&\\
&&&&&&\parbox[t][0.3cm]{11.48525cm}{\raggedright {\ }{\ }{\ }1\ensuremath{^{-}}) and L=(0,1). (3) (\href{https://www.nndc.bnl.gov/nsr/nsrlink.jsp?1991Ga03,B}{1991Ga03}): J\ensuremath{^{\ensuremath{\pi}}}=0\ensuremath{^{\textnormal{+}}} from mirror state analysis. (4)\vspace{0.1cm}}&\\
&&&&&&\parbox[t][0.3cm]{11.48525cm}{\raggedright {\ }{\ }{\ }(\href{https://www.nndc.bnl.gov/nsr/nsrlink.jsp?2005Pa50,B}{2005Pa50}): deduced J=0 by comparing the measured differential cross sections to\vspace{0.1cm}}&\\
&&&&&&\parbox[t][0.3cm]{11.48525cm}{\raggedright {\ }{\ }{\ }the Hauser-Feshbach predicted cross sections. The evaluator adopted J\ensuremath{^{\ensuremath{\pi}}}=0\ensuremath{^{\textnormal{+}}} and\vspace{0.1cm}}&\\
&&&&&&\parbox[t][0.3cm]{11.48525cm}{\raggedright {\ }{\ }{\ }L=0 because the 1\ensuremath{^{-}} state in this energy range is established to be at E\ensuremath{_{\textnormal{x}}}=4519 keV\vspace{0.1cm}}&\\
&&&&&&\parbox[t][0.3cm]{11.48525cm}{\raggedright {\ }{\ }{\ }as reported by (\href{https://www.nndc.bnl.gov/nsr/nsrlink.jsp?2005Pa50,B}{2005Pa50}).\vspace{0.1cm}}&\\
&&&&&&\parbox[t][0.3cm]{11.48525cm}{\raggedright (\href{https://www.nndc.bnl.gov/nsr/nsrlink.jsp?2005Pa50,B}{2005Pa50}) presents the measured and Hauser-Feshbach differential cross sections of\vspace{0.1cm}}&\\
&&&&&&\parbox[t][0.3cm]{11.48525cm}{\raggedright {\ }{\ }{\ }the observed states in graphical formats per beam energy (see Figs. 4 and 5).\vspace{0.1cm}}&\\
&&&&&&\parbox[t][0.3cm]{11.48525cm}{\raggedright Q-value={\textminus}7785 keV \textit{1} (stat.) \textit{5} (sys.) (\href{https://www.nndc.bnl.gov/nsr/nsrlink.jsp?1991Ga03,B}{1991Ga03}).\vspace{0.1cm}}&\\
&&&&&&\parbox[t][0.3cm]{11.48525cm}{\raggedright (\href{https://www.nndc.bnl.gov/nsr/nsrlink.jsp?1977Ev01,B}{1977Ev01}): the mean energy of the 4.5-MeV doublet shows that at E(\ensuremath{^{\textnormal{3}}}He)=18\vspace{0.1cm}}&\\
&&&&&&\parbox[t][0.3cm]{11.48525cm}{\raggedright {\ }{\ }{\ }MeV the 0\ensuremath{^{\textnormal{+}}} and 1\ensuremath{^{-}} states are almost equally populated implying a cross section of\vspace{0.1cm}}&\\
&&&&&&\parbox[t][0.3cm]{11.48525cm}{\raggedright {\ }{\ }{\ }(d\ensuremath{\sigma}/d\ensuremath{\Omega})\ensuremath{_{\textnormal{c.m.}}}=1 mb/sr at \ensuremath{\theta}\ensuremath{_{\textnormal{c.m.}}}=0\ensuremath{^\circ} for the 0\ensuremath{^{\textnormal{+}}} member of the doublet.\vspace{0.1cm}}&\\
&&&&&&\parbox[t][0.3cm]{11.48525cm}{\raggedright (\href{https://www.nndc.bnl.gov/nsr/nsrlink.jsp?1981Ne09,B}{1981Ne09}): predicted a predominantly \textit{s}\ensuremath{_{\textnormal{1/2}}^{\textnormal{2}}} character for this state.\vspace{0.1cm}}&\\
\multicolumn{1}{r@{}}{5098}&\multicolumn{1}{@{ }l}{{\it 14}}&\multicolumn{1}{l}{2\ensuremath{^{+}}}&\multicolumn{1}{r@{}}{50}&\multicolumn{1}{@{ }l}{keV {\it 10}}&\multicolumn{1}{l}{(2,3)}&\parbox[t][0.3cm]{11.48525cm}{\raggedright E(level): Weighted average (with external errors) of 5106 keV \textit{8} (\href{https://www.nndc.bnl.gov/nsr/nsrlink.jsp?1996Ha26,B}{1996Ha26}); and\vspace{0.1cm}}&\\
&&&&&&\parbox[t][0.3cm]{11.48525cm}{\raggedright {\ }{\ }{\ }5075 keV \textit{13} (\href{https://www.nndc.bnl.gov/nsr/nsrlink.jsp?1974Ne04,B}{1974Ne04}, \href{https://www.nndc.bnl.gov/nsr/nsrlink.jsp?1981Ne09,B}{1981Ne09}: this state is a member of a doublet. The\vspace{0.1cm}}&\\
&&&&&&\parbox[t][0.3cm]{11.48525cm}{\raggedright {\ }{\ }{\ }population of this doublet is dominated by the lower-lying member (this state) at\vspace{0.1cm}}&\\
&&&&&&\parbox[t][0.3cm]{11.48525cm}{\raggedright {\ }{\ }{\ }small angles).\vspace{0.1cm}}&\\
&&&&&&\parbox[t][0.3cm]{11.48525cm}{\raggedright E(level): See also 5.2 MeV \textit{3} (\href{https://www.nndc.bnl.gov/nsr/nsrlink.jsp?1964Br13,B}{1964Br13}: indicated that this state may be formed by\vspace{0.1cm}}&\\
&&&&&&\parbox[t][0.3cm]{11.48525cm}{\raggedright {\ }{\ }{\ }a group of unresolved neutrons); 5.14 MeV {2} (\href{https://www.nndc.bnl.gov/nsr/nsrlink.jsp?1970Ad02,B}{1970Ad02}: may be an unresolved\vspace{0.1cm}}&\\
&&&&&&\parbox[t][0.3cm]{11.48525cm}{\raggedright {\ }{\ }{\ }doublet); 5.15 MeV (\href{https://www.nndc.bnl.gov/nsr/nsrlink.jsp?2010AlZZ,B}{2010AlZZ}, \href{https://www.nndc.bnl.gov/nsr/nsrlink.jsp?2010Ta17,B}{2010Ta17}); and 5.10 MeV \textit{10} (\href{https://www.nndc.bnl.gov/nsr/nsrlink.jsp?2012Al11,B}{2012Al11}: an\vspace{0.1cm}}&\\
&&&&&&\parbox[t][0.3cm]{11.48525cm}{\raggedright {\ }{\ }{\ }unresolved doublet, see Table II. Note that Fig. 6 and Table I both report this\vspace{0.1cm}}&\\
&&&&&&\parbox[t][0.3cm]{11.48525cm}{\raggedright {\ }{\ }{\ }excitation energy as 5.15 MeV).\vspace{0.1cm}}&\\
&&&&&&\parbox[t][0.3cm]{11.48525cm}{\raggedright \ensuremath{\Gamma}: From (\href{https://www.nndc.bnl.gov/nsr/nsrlink.jsp?1996Ha26,B}{1996Ha26}). See also \ensuremath{\Gamma}=100 keV \textit{40} (\href{https://www.nndc.bnl.gov/nsr/nsrlink.jsp?1970Ad02,B}{1970Ad02}: deduced by folding the\vspace{0.1cm}}&\\
&&&&&&\parbox[t][0.3cm]{11.48525cm}{\raggedright {\ }{\ }{\ }line shape of the corresponding neutron group with a negligible intrinsic width\vspace{0.1cm}}&\\
&&&&&&\parbox[t][0.3cm]{11.48525cm}{\raggedright {\ }{\ }{\ }into a Breit-Wigner curve. The state observed was an unresolved doublet so this\vspace{0.1cm}}&\\
&&&&&&\parbox[t][0.3cm]{11.48525cm}{\raggedright {\ }{\ }{\ }measurement is excluded); and \ensuremath{\Gamma}\ensuremath{\leq}60 keV (\href{https://www.nndc.bnl.gov/nsr/nsrlink.jsp?1974Ne04,B}{1974Ne04}, \href{https://www.nndc.bnl.gov/nsr/nsrlink.jsp?1981Ne09,B}{1981Ne09}).\vspace{0.1cm}}&\\
&&&&&&\parbox[t][0.3cm]{11.48525cm}{\raggedright (\href{https://www.nndc.bnl.gov/nsr/nsrlink.jsp?2012Al11,B}{2012Al11}) reported \ensuremath{\Gamma}\ensuremath{_{\textnormal{p}_{\textnormal{0}}}}/\ensuremath{\Gamma}=89\% \textit{3}, \ensuremath{\Gamma}\ensuremath{_{\textnormal{p}_{\textnormal{1}}}}/\ensuremath{\Gamma}=11\% \textit{4}, and \ensuremath{\Gamma}\ensuremath{_{\textnormal{p}_{\textnormal{1}}}}/\ensuremath{\Gamma}\ensuremath{_{\textnormal{p}_{\textnormal{0}}}}=0.124.\vspace{0.1cm}}&\\
&&&&&&\parbox[t][0.3cm]{11.48525cm}{\raggedright {\ }{\ }{\ }However, (\href{https://www.nndc.bnl.gov/nsr/nsrlink.jsp?2020Br14,B}{2020Br14}) argues these branching ratios are for a combination of the\vspace{0.1cm}}&\\
&&&&&&\parbox[t][0.3cm]{11.48525cm}{\raggedright {\ }{\ }{\ }5.10- and 5.15-MeV states, which remained unresolved in (\href{https://www.nndc.bnl.gov/nsr/nsrlink.jsp?2012Al11,B}{2012Al11}). If the\vspace{0.1cm}}&\\
&&&&&&\parbox[t][0.3cm]{11.48525cm}{\raggedright {\ }{\ }{\ }\ensuremath{^{\textnormal{18}}}Ne*(5097 keV, 2\ensuremath{^{\textnormal{+}}}) state were to decay to \ensuremath{^{\textnormal{17}}}F*(495 keV, 1/2\ensuremath{^{\textnormal{+}}}), the proton\vspace{0.1cm}}&\\
&&&&&&\parbox[t][0.3cm]{11.48525cm}{\raggedright {\ }{\ }{\ }must carry an angular momentum of L=2. Considering the Coulomb barrier\vspace{0.1cm}}&\\
&&&&&&\parbox[t][0.3cm]{11.48525cm}{\raggedright {\ }{\ }{\ }associated with this decay, (\href{https://www.nndc.bnl.gov/nsr/nsrlink.jsp?2020Br14,B}{2020Br14}) suggested that the \ensuremath{\Gamma}\ensuremath{_{\textnormal{p}_{\textnormal{1}}}} branch observed by\vspace{0.1cm}}&\\
&&&&&&\parbox[t][0.3cm]{11.48525cm}{\raggedright {\ }{\ }{\ }(\href{https://www.nndc.bnl.gov/nsr/nsrlink.jsp?2012Al11,B}{2012Al11}) cannot involve the \ensuremath{^{\textnormal{18}}}Ne*(5097 keV, 2\ensuremath{^{\textnormal{+}}}) state. See also (\href{https://www.nndc.bnl.gov/nsr/nsrlink.jsp?2019Ch16,B}{2019Ch16}),\vspace{0.1cm}}&\\
&&&&&&\parbox[t][0.3cm]{11.48525cm}{\raggedright {\ }{\ }{\ }which discussed the \ensuremath{\Gamma}\ensuremath{_{\textnormal{p}_{\textnormal{1}}}}/\ensuremath{\Gamma}=11\% \textit{4} result of (\href{https://www.nndc.bnl.gov/nsr/nsrlink.jsp?2012Al11,B}{2012Al11}) that is in disagreement\vspace{0.1cm}}&\\
&&&&&&\parbox[t][0.3cm]{11.48525cm}{\raggedright {\ }{\ }{\ }with \ensuremath{\Gamma}\ensuremath{_{\textnormal{p}_{\textnormal{1}}}}/\ensuremath{\Gamma}\ensuremath{<}0.009 deduced in (\href{https://www.nndc.bnl.gov/nsr/nsrlink.jsp?2019Ch16,B}{2019Ch16}).\vspace{0.1cm}}&\\
&&&&&&\parbox[t][0.3cm]{11.48525cm}{\raggedright J\ensuremath{^{\pi}},L: From the DWBA analyses of (\href{https://www.nndc.bnl.gov/nsr/nsrlink.jsp?1970Ad02,B}{1970Ad02}: J\ensuremath{^{\ensuremath{\pi}}}=2\ensuremath{^{\textnormal{+}}}, 3\ensuremath{^{-}} and L=2,3); (\href{https://www.nndc.bnl.gov/nsr/nsrlink.jsp?1981Ne09,B}{1981Ne09}:\vspace{0.1cm}}&\\
&&&&&&\parbox[t][0.3cm]{11.48525cm}{\raggedright {\ }{\ }{\ }J\ensuremath{^{\ensuremath{\pi}}}=(2\ensuremath{^{\textnormal{+}}}, 3\ensuremath{^{-}}) and L=2,3); and (\href{https://www.nndc.bnl.gov/nsr/nsrlink.jsp?1996Ha26,B}{1996Ha26}: J\ensuremath{^{\ensuremath{\pi}}}=2\ensuremath{^{\textnormal{+}}} from a combination of Coulomb\vspace{0.1cm}}&\\
&&&&&&\parbox[t][0.3cm]{11.48525cm}{\raggedright {\ }{\ }{\ }barrier penetrability considerations, level widths, mirror levels analysis, and\vspace{0.1cm}}&\\
&&&&&&\parbox[t][0.3cm]{11.48525cm}{\raggedright {\ }{\ }{\ }Woods-Saxon calculations of the Coulomb energy shifts).\vspace{0.1cm}}&\\
&&&&&&\parbox[t][0.3cm]{11.48525cm}{\raggedright (\href{https://www.nndc.bnl.gov/nsr/nsrlink.jsp?2012Al11,B}{2012Al11}): a\ensuremath{_{\textnormal{0}}}=1506.90, a\ensuremath{_{\textnormal{2}}}={\textminus}819.45, a\ensuremath{_{\textnormal{4}}}={\textminus}547.42, and a\ensuremath{_{\textnormal{6}}}={\textminus}802.13, where a\ensuremath{_{\textnormal{i}}} are the\vspace{0.1cm}}&\\
&&&&&&\parbox[t][0.3cm]{11.48525cm}{\raggedright {\ }{\ }{\ }Legendre polynomial fit coefficients for the measured proton$'$s (from the decay of\vspace{0.1cm}}&\\
&&&&&&\parbox[t][0.3cm]{11.48525cm}{\raggedright {\ }{\ }{\ }\ensuremath{^{\textnormal{18}}}Ne state) angular distribution.\vspace{0.1cm}}&\\
\multicolumn{1}{r@{}}{5142}&\multicolumn{1}{@{ }l}{{\it 8}}&\multicolumn{1}{l}{3\ensuremath{^{-}}}&\multicolumn{1}{r@{}}{$\leq$20}&\multicolumn{1}{@{ }l}{keV}&\multicolumn{1}{l}{(2,3)}&\parbox[t][0.3cm]{11.48525cm}{\raggedright E(level): Weighted average (with external errors) of 5130 keV \textit{10} (\href{https://www.nndc.bnl.gov/nsr/nsrlink.jsp?1977Ev01,B}{1977Ev01}); 5135\vspace{0.1cm}}&\\
&&&&&&\parbox[t][0.3cm]{11.48525cm}{\raggedright {\ }{\ }{\ }keV \textit{12} (\href{https://www.nndc.bnl.gov/nsr/nsrlink.jsp?1974Ne04,B}{1974Ne04}, \href{https://www.nndc.bnl.gov/nsr/nsrlink.jsp?1981Ne09,B}{1981Ne09}: this state is a member of a doublet. The population\vspace{0.1cm}}&\\
&&&&&&\parbox[t][0.3cm]{11.48525cm}{\raggedright {\ }{\ }{\ }of this doublet is dominated by the higher-lying member (this state) at small\vspace{0.1cm}}&\\
&&&&&&\parbox[t][0.3cm]{11.48525cm}{\raggedright {\ }{\ }{\ }angles); and 5153 keV \textit{8} (\href{https://www.nndc.bnl.gov/nsr/nsrlink.jsp?1996Ha26,B}{1996Ha26}).\vspace{0.1cm}}&\\
\end{longtable}
\begin{textblock}{29}(0,27.3)
Continued on next page (footnotes at end of table)
\end{textblock}
\clearpage
\begin{longtable}{ccccccc@{\extracolsep{\fill}}c}
\\[-.4cm]
\multicolumn{8}{c}{{\bf \small \underline{\ensuremath{^{\textnormal{16}}}O(\ensuremath{^{\textnormal{3}}}He,n)\hspace{0.2in}\href{https://www.nndc.bnl.gov/nsr/nsrlink.jsp?1953Ku08,B}{1953Ku08},\href{https://www.nndc.bnl.gov/nsr/nsrlink.jsp?2012Fo29,B}{2012Fo29} (continued)}}}\\
\multicolumn{8}{c}{~}\\
\multicolumn{8}{c}{\underline{\ensuremath{^{18}}Ne Levels (continued)}}\\
\multicolumn{8}{c}{~}\\
\multicolumn{2}{c}{E(level)$^{}$}&J$^{\pi}$$^{{\hyperlink{NE15LEVEL0}{a}}}$&\multicolumn{2}{c}{T\ensuremath{_{\textnormal{1/2}}} or \ensuremath{\Gamma}$^{}$}&L$^{{\hyperlink{NE15LEVEL0}{a}}{\hyperlink{NE15LEVEL1}{b}}{\hyperlink{NE15LEVEL2}{c}}}$&Comments&\\[-.2cm]
\multicolumn{2}{c}{\hrulefill}&\hrulefill&\multicolumn{2}{c}{\hrulefill}&\hrulefill&\hrulefill&
\endhead
&&&&&&\parbox[t][0.3cm]{11.25413cm}{\raggedright E(level): See also 5.2 MeV \textit{3} (\href{https://www.nndc.bnl.gov/nsr/nsrlink.jsp?1964Br13,B}{1964Br13}: possibly a group of unresolved levels);\vspace{0.1cm}}&\\
&&&&&&\parbox[t][0.3cm]{11.25413cm}{\raggedright {\ }{\ }{\ }and 5.14 MeV \textit{20} (\href{https://www.nndc.bnl.gov/nsr/nsrlink.jsp?1970Ad02,B}{1970Ad02}: this state is an unresolved doublet).\vspace{0.1cm}}&\\
&&&&&&\parbox[t][0.3cm]{11.25413cm}{\raggedright (\href{https://www.nndc.bnl.gov/nsr/nsrlink.jsp?1977Ev01,B}{1977Ev01}): the measured energy of 5130 keV \textit{10} corresponds to an unresolved\vspace{0.1cm}}&\\
&&&&&&\parbox[t][0.3cm]{11.25413cm}{\raggedright {\ }{\ }{\ }doublet, whose reported energy is the mean energy of the members of the\vspace{0.1cm}}&\\
&&&&&&\parbox[t][0.3cm]{11.25413cm}{\raggedright {\ }{\ }{\ }doublet. However, this energy was observed to be more or less constant as a\vspace{0.1cm}}&\\
&&&&&&\parbox[t][0.3cm]{11.25413cm}{\raggedright {\ }{\ }{\ }function of incident energy, and it showed a small shift with angle. These were\vspace{0.1cm}}&\\
&&&&&&\parbox[t][0.3cm]{11.25413cm}{\raggedright {\ }{\ }{\ }indicative that most likely only one member of the doublet (5.14-MeV state) was\vspace{0.1cm}}&\\
&&&&&&\parbox[t][0.3cm]{11.25413cm}{\raggedright {\ }{\ }{\ }populated in their experiment.\vspace{0.1cm}}&\\
&&&&&&\parbox[t][0.3cm]{11.25413cm}{\raggedright \ensuremath{\Gamma}: From (\href{https://www.nndc.bnl.gov/nsr/nsrlink.jsp?1996Ha26,B}{1996Ha26}). See also \ensuremath{\Gamma}\ensuremath{\leq}60 keV (\href{https://www.nndc.bnl.gov/nsr/nsrlink.jsp?1974Ne04,B}{1974Ne04}, \href{https://www.nndc.bnl.gov/nsr/nsrlink.jsp?1981Ne09,B}{1981Ne09}); \ensuremath{\Gamma}=100 keV\vspace{0.1cm}}&\\
&&&&&&\parbox[t][0.3cm]{11.25413cm}{\raggedright {\ }{\ }{\ }\textit{40} (\href{https://www.nndc.bnl.gov/nsr/nsrlink.jsp?1970Ad02,B}{1970Ad02}: deduced for a state that is unresolved from another nearby state).\vspace{0.1cm}}&\\
&&&&&&\parbox[t][0.3cm]{11.25413cm}{\raggedright J\ensuremath{^{\pi}},L: From (\href{https://www.nndc.bnl.gov/nsr/nsrlink.jsp?1977Ev01,B}{1977Ev01}: J\ensuremath{^{\ensuremath{\pi}}}=3\ensuremath{^{-}} from the zero-range DWBA analysis. L is not\vspace{0.1cm}}&\\
&&&&&&\parbox[t][0.3cm]{11.25413cm}{\raggedright {\ }{\ }{\ }given but the evaluator assigned L=3 from the selection rules. The agreement\vspace{0.1cm}}&\\
&&&&&&\parbox[t][0.3cm]{11.25413cm}{\raggedright {\ }{\ }{\ }between the DWBA fit and the neutron angular distribution data at E(\ensuremath{^{\textnormal{3}}}He)=15\vspace{0.1cm}}&\\
&&&&&&\parbox[t][0.3cm]{11.25413cm}{\raggedright {\ }{\ }{\ }MeV and 18 MeV is very good); (\href{https://www.nndc.bnl.gov/nsr/nsrlink.jsp?1981Ne09,B}{1981Ne09}: J\ensuremath{^{\ensuremath{\pi}}}=(2\ensuremath{^{\textnormal{+}}}, 3\ensuremath{^{-}}) from DWBA analysis\vspace{0.1cm}}&\\
&&&&&&\parbox[t][0.3cm]{11.25413cm}{\raggedright {\ }{\ }{\ }with L=2,3); and (\href{https://www.nndc.bnl.gov/nsr/nsrlink.jsp?1996Ha26,B}{1996Ha26}: J\ensuremath{^{\ensuremath{\pi}}}=3\ensuremath{^{-}} from a combination of penetrability\vspace{0.1cm}}&\\
&&&&&&\parbox[t][0.3cm]{11.25413cm}{\raggedright {\ }{\ }{\ }considerations, level widths, mirror levels analysis, and Woods-Saxon\vspace{0.1cm}}&\\
&&&&&&\parbox[t][0.3cm]{11.25413cm}{\raggedright {\ }{\ }{\ }calculations of the Coulomb energy shifts).\vspace{0.1cm}}&\\
\multicolumn{1}{r@{}}{5454}&\multicolumn{1}{@{ }l}{{\it 8}}&\multicolumn{1}{l}{(2\ensuremath{^{-}})}&\multicolumn{1}{r@{}}{$\leq$20}&\multicolumn{1}{@{ }l}{keV}&&\parbox[t][0.3cm]{11.25413cm}{\raggedright E(level),\ensuremath{\Gamma}: From (\href{https://www.nndc.bnl.gov/nsr/nsrlink.jsp?1996Ha26,B}{1996Ha26}).\vspace{0.1cm}}&\\
&&&&&&\parbox[t][0.3cm]{11.25413cm}{\raggedright J\ensuremath{^{\pi}},L: From (\href{https://www.nndc.bnl.gov/nsr/nsrlink.jsp?1996Ha26,B}{1996Ha26}: based on a combination of penetrability considerations,\vspace{0.1cm}}&\\
&&&&&&\parbox[t][0.3cm]{11.25413cm}{\raggedright {\ }{\ }{\ }level widths, mirror levels analysis, and Woods-Saxon calculations of the\vspace{0.1cm}}&\\
&&&&&&\parbox[t][0.3cm]{11.25413cm}{\raggedright {\ }{\ }{\ }Coulomb energy shifts. The un-natural spin-parity assignment was assigned\vspace{0.1cm}}&\\
&&&&&&\parbox[t][0.3cm]{11.25413cm}{\raggedright {\ }{\ }{\ }because the neutron angular distribution was not forward peaked).\vspace{0.1cm}}&\\
\multicolumn{1}{r@{}}{6150}&\multicolumn{1}{@{}l}{\ensuremath{^{{\hyperlink{NE15LEVEL5}{f}}}} {\it 10}}&\multicolumn{1}{l}{(1\ensuremath{^{-}})}&\multicolumn{1}{r@{}}{$\leq$40}&\multicolumn{1}{@{ }l}{keV}&\multicolumn{1}{l}{(1)}&\parbox[t][0.3cm]{11.25413cm}{\raggedright \ensuremath{\Gamma}\ensuremath{_{\textnormal{p0}}}/\ensuremath{\Gamma}\ensuremath{\leq}0.52\vspace{0.1cm}}&\\
&&&&&&\parbox[t][0.3cm]{11.25413cm}{\raggedright \ensuremath{\Gamma}\ensuremath{_{\textnormal{p1}}}/\ensuremath{\Gamma}\ensuremath{\leq}0.21 (\href{https://www.nndc.bnl.gov/nsr/nsrlink.jsp?2012Al11,B}{2012Al11})\vspace{0.1cm}}&\\
&&&&&&\parbox[t][0.3cm]{11.25413cm}{\raggedright E(level): Weighted average of 6150 keV \textit{10} (\href{https://www.nndc.bnl.gov/nsr/nsrlink.jsp?1996Ha26,B}{1996Ha26}) and 6.15 MeV \textit{8}\vspace{0.1cm}}&\\
&&&&&&\parbox[t][0.3cm]{11.25413cm}{\raggedright {\ }{\ }{\ }(\href{https://www.nndc.bnl.gov/nsr/nsrlink.jsp?2012Al11,B}{2012Al11}: see Table II). See also 6.1 MeV (\href{https://www.nndc.bnl.gov/nsr/nsrlink.jsp?2010AlZZ,B}{2010AlZZ}, \href{https://www.nndc.bnl.gov/nsr/nsrlink.jsp?2010Ta17,B}{2010Ta17}).\vspace{0.1cm}}&\\
&&&&&&\parbox[t][0.3cm]{11.25413cm}{\raggedright \ensuremath{\Gamma}: From (\href{https://www.nndc.bnl.gov/nsr/nsrlink.jsp?1996Ha26,B}{1996Ha26}).\vspace{0.1cm}}&\\
&&&&&&\parbox[t][0.3cm]{11.25413cm}{\raggedright (\href{https://www.nndc.bnl.gov/nsr/nsrlink.jsp?2012Al11,B}{2012Al11}) deduced \ensuremath{\Gamma}\ensuremath{_{^{\textnormal{2}}\textnormal{He}}}/\ensuremath{\Gamma}\ensuremath{\leq}0.27. The authors reported an excess of counts in\vspace{0.1cm}}&\\
&&&&&&\parbox[t][0.3cm]{11.25413cm}{\raggedright {\ }{\ }{\ }the energy region corresponding to the 2p-decay channel for this state. This is\vspace{0.1cm}}&\\
&&&&&&\parbox[t][0.3cm]{11.25413cm}{\raggedright {\ }{\ }{\ }not a direct evidence for 2p decay since (\href{https://www.nndc.bnl.gov/nsr/nsrlink.jsp?2012Al11,B}{2012Al11}) did not detect the two\vspace{0.1cm}}&\\
&&&&&&\parbox[t][0.3cm]{11.25413cm}{\raggedright {\ }{\ }{\ }individual protons from the decay of this state.\vspace{0.1cm}}&\\
&&&&&&\parbox[t][0.3cm]{11.25413cm}{\raggedright (\href{https://www.nndc.bnl.gov/nsr/nsrlink.jsp?2012Al11,B}{2012Al11}) reported \ensuremath{\Gamma}\ensuremath{_{\ensuremath{\alpha}}}\ensuremath{\approx}2 eV from a private communication of the authors with\vspace{0.1cm}}&\\
&&&&&&\parbox[t][0.3cm]{11.25413cm}{\raggedright {\ }{\ }{\ }G. V. Rogachev. This width was determined from the ANC deduced for the\vspace{0.1cm}}&\\
&&&&&&\parbox[t][0.3cm]{11.25413cm}{\raggedright {\ }{\ }{\ }mirror level in \ensuremath{^{\textnormal{18}}}O (see also (\href{https://www.nndc.bnl.gov/nsr/nsrlink.jsp?2009Jo07,B}{2009Jo07})). With this and the other branching\vspace{0.1cm}}&\\
&&&&&&\parbox[t][0.3cm]{11.25413cm}{\raggedright {\ }{\ }{\ }ratios deduced by (\href{https://www.nndc.bnl.gov/nsr/nsrlink.jsp?2012Al11,B}{2012Al11}), the evaluator determined \ensuremath{\Gamma}\ensuremath{_{\textnormal{p}_{\textnormal{0}}}}/\ensuremath{\Gamma}\ensuremath{\leq}0.52. This\vspace{0.1cm}}&\\
&&&&&&\parbox[t][0.3cm]{11.25413cm}{\raggedright {\ }{\ }{\ }value implies \ensuremath{\Gamma}\ensuremath{_{\textnormal{p}_{\textnormal{1}}}}/\ensuremath{\Gamma}\ensuremath{_{\textnormal{p}_{\textnormal{0}}}}\ensuremath{\leq}0.40, which disagrees with \ensuremath{\Gamma}\ensuremath{_{\textnormal{p}_{\textnormal{1}}}}/\ensuremath{\Gamma}\ensuremath{_{\textnormal{p}_{\textnormal{0}}}}\ensuremath{\leq}0.27 reported in\vspace{0.1cm}}&\\
&&&&&&\parbox[t][0.3cm]{11.25413cm}{\raggedright {\ }{\ }{\ }Table IV of (\href{https://www.nndc.bnl.gov/nsr/nsrlink.jsp?2012Al11,B}{2012Al11}). The evaluator assumed that the latter value is erroneous\vspace{0.1cm}}&\\
&&&&&&\parbox[t][0.3cm]{11.25413cm}{\raggedright {\ }{\ }{\ }based on \ensuremath{\Gamma}\ensuremath{_{\textnormal{p}_{\textnormal{1}}}}=15.9 keV \textit{7} and \ensuremath{\Gamma}\ensuremath{_{\textnormal{p}_{\textnormal{0}}}}=37.8 keV \textit{19} deduced by (\href{https://www.nndc.bnl.gov/nsr/nsrlink.jsp?2012Ba28,B}{2012Ba28}: from\vspace{0.1cm}}&\\
&&&&&&\parbox[t][0.3cm]{11.25413cm}{\raggedright {\ }{\ }{\ }\ensuremath{^{\textnormal{17}}}F(p,p\ensuremath{'}) data of (\href{https://www.nndc.bnl.gov/nsr/nsrlink.jsp?2003Bl11,B}{2003Bl11})).\vspace{0.1cm}}&\\
&&&&&&\parbox[t][0.3cm]{11.25413cm}{\raggedright (\href{https://www.nndc.bnl.gov/nsr/nsrlink.jsp?2012Al11,B}{2012Al11}): upper limits for branching ratios are estimated at 95\% confidence\vspace{0.1cm}}&\\
&&&&&&\parbox[t][0.3cm]{11.25413cm}{\raggedright {\ }{\ }{\ }level using a Bayesian approach assuming that the net excess of counts were\vspace{0.1cm}}&\\
&&&&&&\parbox[t][0.3cm]{11.25413cm}{\raggedright {\ }{\ }{\ }from real decay events.\vspace{0.1cm}}&\\
&&&&&&\parbox[t][0.3cm]{11.25413cm}{\raggedright J\ensuremath{^{\pi}},L: From (\href{https://www.nndc.bnl.gov/nsr/nsrlink.jsp?1996Ha26,B}{1996Ha26}: J\ensuremath{^{\ensuremath{\pi}}}=(1\ensuremath{^{-}}) from a combination of zero-range DWBA\vspace{0.1cm}}&\\
&&&&&&\parbox[t][0.3cm]{11.25413cm}{\raggedright {\ }{\ }{\ }analysis, penetrability considerations, level widths, mirror levels analysis, and\vspace{0.1cm}}&\\
&&&&&&\parbox[t][0.3cm]{11.25413cm}{\raggedright {\ }{\ }{\ }Woods-Saxon calculations of the Coulomb energy shifts. The DWBA fits with\vspace{0.1cm}}&\\
&&&&&&\parbox[t][0.3cm]{11.25413cm}{\raggedright {\ }{\ }{\ }L=0, L=1, and L=2 are presented; however, these fits do not describe the\vspace{0.1cm}}&\\
&&&&&&\parbox[t][0.3cm]{11.25413cm}{\raggedright {\ }{\ }{\ }measured neutron angular distribution data well). The evaluator selected a\vspace{0.1cm}}&\\
&&&&&&\parbox[t][0.3cm]{11.25413cm}{\raggedright {\ }{\ }{\ }tentative L=(1) as the best \textit{l}-value for this case.\vspace{0.1cm}}&\\
&&&&&&\parbox[t][0.3cm]{11.25413cm}{\raggedright (\href{https://www.nndc.bnl.gov/nsr/nsrlink.jsp?2012Al11,B}{2012Al11}): a\ensuremath{_{\textnormal{0}}}=194.80, a\ensuremath{_{\textnormal{2}}}={\textminus}43.05, a\ensuremath{_{\textnormal{4}}}={\textminus}22.06, and a\ensuremath{_{\textnormal{6}}}={\textminus}94.73, where a\ensuremath{_{\textnormal{i}}} are the\vspace{0.1cm}}&\\
&&&&&&\parbox[t][0.3cm]{11.25413cm}{\raggedright {\ }{\ }{\ }Legendre polynomial fit coefficients for the measured proton$'$s (from proton\vspace{0.1cm}}&\\
&&&&&&\parbox[t][0.3cm]{11.25413cm}{\raggedright {\ }{\ }{\ }decay of this state) angular distribution.\vspace{0.1cm}}&\\
\multicolumn{1}{r@{}}{6299}&\multicolumn{1}{@{ }l}{{\it 9}}&\multicolumn{1}{l}{(3\ensuremath{^{-}})}&\multicolumn{1}{r@{}}{180}&\multicolumn{1}{@{ }l}{keV {\it 60}}&&\parbox[t][0.3cm]{11.25413cm}{\raggedright \ensuremath{\Gamma}\ensuremath{_{\textnormal{p0}}}/\ensuremath{\Gamma}=1.00 \textit{2} (\href{https://www.nndc.bnl.gov/nsr/nsrlink.jsp?2012Al11,B}{2012Al11})\vspace{0.1cm}}&\\
&&&&&&\parbox[t][0.3cm]{11.25413cm}{\raggedright E(level): Weighted average of 6291 keV \textit{30} (\href{https://www.nndc.bnl.gov/nsr/nsrlink.jsp?1974Ne04,B}{1974Ne04}, \href{https://www.nndc.bnl.gov/nsr/nsrlink.jsp?1981Ne09,B}{1981Ne09}); 6300 keV \textit{10}\vspace{0.1cm}}&\\
&&&&&&\parbox[t][0.3cm]{11.25413cm}{\raggedright {\ }{\ }{\ }(\href{https://www.nndc.bnl.gov/nsr/nsrlink.jsp?1996Ha26,B}{1996Ha26}); and 6.30 MeV \textit{5} (\href{https://www.nndc.bnl.gov/nsr/nsrlink.jsp?2012Al11,B}{2012Al11}: Table II). See also 6.3 MeV\vspace{0.1cm}}&\\
\end{longtable}
\begin{textblock}{29}(0,27.3)
Continued on next page (footnotes at end of table)
\end{textblock}
\clearpage
\begin{longtable}{ccccccc@{\extracolsep{\fill}}c}
\\[-.4cm]
\multicolumn{8}{c}{{\bf \small \underline{\ensuremath{^{\textnormal{16}}}O(\ensuremath{^{\textnormal{3}}}He,n)\hspace{0.2in}\href{https://www.nndc.bnl.gov/nsr/nsrlink.jsp?1953Ku08,B}{1953Ku08},\href{https://www.nndc.bnl.gov/nsr/nsrlink.jsp?2012Fo29,B}{2012Fo29} (continued)}}}\\
\multicolumn{8}{c}{~}\\
\multicolumn{8}{c}{\underline{\ensuremath{^{18}}Ne Levels (continued)}}\\
\multicolumn{8}{c}{~}\\
\multicolumn{2}{c}{E(level)$^{}$}&J$^{\pi}$$^{{\hyperlink{NE15LEVEL0}{a}}}$&\multicolumn{2}{c}{T\ensuremath{_{\textnormal{1/2}}} or \ensuremath{\Gamma}$^{}$}&L$^{{\hyperlink{NE15LEVEL0}{a}}{\hyperlink{NE15LEVEL1}{b}}{\hyperlink{NE15LEVEL2}{c}}}$&Comments&\\[-.2cm]
\multicolumn{2}{c}{\hrulefill}&\hrulefill&\multicolumn{2}{c}{\hrulefill}&\hrulefill&\hrulefill&
\endhead
&&&&&&\parbox[t][0.3cm]{10.729831cm}{\raggedright {\ }{\ }{\ }(\href{https://www.nndc.bnl.gov/nsr/nsrlink.jsp?2010AlZZ,B}{2010AlZZ}, \href{https://www.nndc.bnl.gov/nsr/nsrlink.jsp?2010Ta17,B}{2010Ta17}).\vspace{0.1cm}}&\\
&&&&&&\parbox[t][0.3cm]{10.729831cm}{\raggedright \ensuremath{\Gamma}: From (\href{https://www.nndc.bnl.gov/nsr/nsrlink.jsp?1974Ne04,B}{1974Ne04}, \href{https://www.nndc.bnl.gov/nsr/nsrlink.jsp?1981Ne09,B}{1981Ne09}).\vspace{0.1cm}}&\\
&&&&&&\parbox[t][0.3cm]{10.729831cm}{\raggedright J\ensuremath{^{\pi}}: From (\href{https://www.nndc.bnl.gov/nsr/nsrlink.jsp?1996Ha26,B}{1996Ha26}: J\ensuremath{^{\ensuremath{\pi}}}=(3\ensuremath{^{-}}) deduced from a combination of penetrability\vspace{0.1cm}}&\\
&&&&&&\parbox[t][0.3cm]{10.729831cm}{\raggedright {\ }{\ }{\ }considerations, level widths, mirror levels analysis, and Woods-Saxon\vspace{0.1cm}}&\\
&&&&&&\parbox[t][0.3cm]{10.729831cm}{\raggedright {\ }{\ }{\ }calculations of the Coulomb energy shifts).\vspace{0.1cm}}&\\
&&&&&&\parbox[t][0.3cm]{10.729831cm}{\raggedright (\href{https://www.nndc.bnl.gov/nsr/nsrlink.jsp?2012Al11,B}{2012Al11}): a\ensuremath{_{\textnormal{0}}}=532.94, a\ensuremath{_{\textnormal{2}}}={\textminus}193.96, a\ensuremath{_{\textnormal{4}}}=111.55, and a\ensuremath{_{\textnormal{6}}}=0, where a\ensuremath{_{\textnormal{i}}} are the\vspace{0.1cm}}&\\
&&&&&&\parbox[t][0.3cm]{10.729831cm}{\raggedright {\ }{\ }{\ }Legendre polynomial fit coefficients for the measured proton$'$s (from proton\vspace{0.1cm}}&\\
&&&&&&\parbox[t][0.3cm]{10.729831cm}{\raggedright {\ }{\ }{\ }decay of this state) angular distribution.\vspace{0.1cm}}&\\
\multicolumn{1}{r@{}}{6350}&\multicolumn{1}{@{ }l}{{\it 10}}&\multicolumn{1}{l}{(2\ensuremath{^{-}})}&&&&\parbox[t][0.3cm]{10.729831cm}{\raggedright E(level): From (\href{https://www.nndc.bnl.gov/nsr/nsrlink.jsp?1996Ha26,B}{1996Ha26}).\vspace{0.1cm}}&\\
&&&&&&\parbox[t][0.3cm]{10.729831cm}{\raggedright J\ensuremath{^{\pi}}: From (\href{https://www.nndc.bnl.gov/nsr/nsrlink.jsp?1996Ha26,B}{1996Ha26}: J\ensuremath{^{\ensuremath{\pi}}}=(2\ensuremath{^{-}}), deduced from a combination of penetrability\vspace{0.1cm}}&\\
&&&&&&\parbox[t][0.3cm]{10.729831cm}{\raggedright {\ }{\ }{\ }considerations, level widths, mirror levels analysis, and Woods-Saxon\vspace{0.1cm}}&\\
&&&&&&\parbox[t][0.3cm]{10.729831cm}{\raggedright {\ }{\ }{\ }calculations of the Coulomb energy shifts).\vspace{0.1cm}}&\\
\multicolumn{1}{r@{}}{7050?}&\multicolumn{1}{@{ }l}{{\it 30}}&\multicolumn{1}{l}{(1\ensuremath{^{-}},2\ensuremath{^{+}},4\ensuremath{^{+}})}&\multicolumn{1}{r@{}}{$\leq$120}&\multicolumn{1}{@{ }l}{keV}&\multicolumn{1}{l}{1,2}&\parbox[t][0.3cm]{10.729831cm}{\raggedright E(level): From (\href{https://www.nndc.bnl.gov/nsr/nsrlink.jsp?1996Ha26,B}{1996Ha26}): assuming a doublet. This state is the lower energy\vspace{0.1cm}}&\\
&&&&&&\parbox[t][0.3cm]{10.729831cm}{\raggedright {\ }{\ }{\ }member of a doublet and was reported as a tentative state, see below. See\vspace{0.1cm}}&\\
&&&&&&\parbox[t][0.3cm]{10.729831cm}{\raggedright {\ }{\ }{\ }also 7062 keV \textit{12} (\href{https://www.nndc.bnl.gov/nsr/nsrlink.jsp?1974Ne04,B}{1974Ne04}, \href{https://www.nndc.bnl.gov/nsr/nsrlink.jsp?1981Ne09,B}{1981Ne09}); 7051 keV \textit{18} (\href{https://www.nndc.bnl.gov/nsr/nsrlink.jsp?1977Ev01,B}{1977Ev01}); 7.0\vspace{0.1cm}}&\\
&&&&&&\parbox[t][0.3cm]{10.729831cm}{\raggedright {\ }{\ }{\ }MeV (\href{https://www.nndc.bnl.gov/nsr/nsrlink.jsp?2010AlZZ,B}{2010AlZZ}, \href{https://www.nndc.bnl.gov/nsr/nsrlink.jsp?2010Ta17,B}{2010Ta17}); and 7.06 MeV \textit{4} (\href{https://www.nndc.bnl.gov/nsr/nsrlink.jsp?2012Al11,B}{2012Al11}: Table II). These\vspace{0.1cm}}&\\
&&&&&&\parbox[t][0.3cm]{10.729831cm}{\raggedright {\ }{\ }{\ }states are not considered because they were all fitted using a single peak,\vspace{0.1cm}}&\\
&&&&&&\parbox[t][0.3cm]{10.729831cm}{\raggedright {\ }{\ }{\ }and thus constitute an unresolved doublet.\vspace{0.1cm}}&\\
&&&&&&\parbox[t][0.3cm]{10.729831cm}{\raggedright E(level): (\href{https://www.nndc.bnl.gov/nsr/nsrlink.jsp?1996Ha26,B}{1996Ha26}): this level may be a member of a close lying doublet.\vspace{0.1cm}}&\\
&&&&&&\parbox[t][0.3cm]{10.729831cm}{\raggedright {\ }{\ }{\ }Considering a single peak, (\href{https://www.nndc.bnl.gov/nsr/nsrlink.jsp?1996Ha26,B}{1996Ha26}) deduced E\ensuremath{_{\textnormal{x}}}=7070 keV \textit{10} and\vspace{0.1cm}}&\\
&&&&&&\parbox[t][0.3cm]{10.729831cm}{\raggedright {\ }{\ }{\ }\ensuremath{\Gamma}=200 keV \textit{40} for this state. However, the fit to the spectrum considering\vspace{0.1cm}}&\\
&&&&&&\parbox[t][0.3cm]{10.729831cm}{\raggedright {\ }{\ }{\ }two states (a doublet) resulted in a smaller \ensuremath{\chi}\ensuremath{^{\textnormal{2}}}, and was thus a better fit. As\vspace{0.1cm}}&\\
&&&&&&\parbox[t][0.3cm]{10.729831cm}{\raggedright {\ }{\ }{\ }a result, two tentative states were deduced at (7050 keV \textit{30}) and (7120 keV\vspace{0.1cm}}&\\
&&&&&&\parbox[t][0.3cm]{10.729831cm}{\raggedright {\ }{\ }{\ }\textit{30}), both of which with \ensuremath{\Gamma}\ensuremath{\leq}120 keV.\vspace{0.1cm}}&\\
&&&&&&\parbox[t][0.3cm]{10.729831cm}{\raggedright \ensuremath{\Gamma}: From (\href{https://www.nndc.bnl.gov/nsr/nsrlink.jsp?1996Ha26,B}{1996Ha26}) considering a doublet. See also \ensuremath{\Gamma}=180 keV \textit{50}\vspace{0.1cm}}&\\
&&&&&&\parbox[t][0.3cm]{10.729831cm}{\raggedright {\ }{\ }{\ }(\href{https://www.nndc.bnl.gov/nsr/nsrlink.jsp?1974Ne04,B}{1974Ne04}, \href{https://www.nndc.bnl.gov/nsr/nsrlink.jsp?1981Ne09,B}{1981Ne09}); and \ensuremath{\Gamma}=200 keV \textit{40} (\href{https://www.nndc.bnl.gov/nsr/nsrlink.jsp?1996Ha26,B}{1996Ha26}) considering a\vspace{0.1cm}}&\\
&&&&&&\parbox[t][0.3cm]{10.729831cm}{\raggedright {\ }{\ }{\ }single state.\vspace{0.1cm}}&\\
&&&&&&\parbox[t][0.3cm]{10.729831cm}{\raggedright \ensuremath{\Gamma}\ensuremath{_{\ensuremath{\alpha}}}=22.6 eV \textit{32} and \ensuremath{\Gamma}\ensuremath{_{\textnormal{p}}}=64 keV \textit{13} are calculated by (\href{https://www.nndc.bnl.gov/nsr/nsrlink.jsp?2003Fo13,B}{2003Fo13}) assuming\vspace{0.1cm}}&\\
&&&&&&\parbox[t][0.3cm]{10.729831cm}{\raggedright {\ }{\ }{\ }mirror symmetry of spectroscopic factors and J\ensuremath{^{\ensuremath{\pi}}}=4\ensuremath{^{\textnormal{+}}}.\vspace{0.1cm}}&\\
&&&&&&\parbox[t][0.3cm]{10.729831cm}{\raggedright \ensuremath{\Gamma}\ensuremath{_{\textnormal{p}_{\textnormal{1}}}}\ensuremath{<}14 eV and \ensuremath{\Gamma}\ensuremath{_{\textnormal{p}_{\textnormal{1}}}}/\ensuremath{\Gamma}\ensuremath{_{\textnormal{p}_{\textnormal{0}}}}\ensuremath{<}2\ensuremath{\times}10\ensuremath{^{\textnormal{$-$4}}} are calculated by (\href{https://www.nndc.bnl.gov/nsr/nsrlink.jsp?2012Fo29,B}{2012Fo29}). This work\vspace{0.1cm}}&\\
&&&&&&\parbox[t][0.3cm]{10.729831cm}{\raggedright {\ }{\ }{\ }disputed the validity and overthrew the following experimental branching\vspace{0.1cm}}&\\
&&&&&&\parbox[t][0.3cm]{10.729831cm}{\raggedright {\ }{\ }{\ }ratios deduced by (\href{https://www.nndc.bnl.gov/nsr/nsrlink.jsp?2012Al11,B}{2012Al11}) for this state: \ensuremath{\Gamma}\ensuremath{_{\textnormal{p}_{\textnormal{0}}}}/\ensuremath{\Gamma}=0.83 \textit{3}, \ensuremath{\Gamma}\ensuremath{_{\textnormal{p}_{\textnormal{1}}}}/\ensuremath{\Gamma}=0.16 \textit{7},\vspace{0.1cm}}&\\
&&&&&&\parbox[t][0.3cm]{10.729831cm}{\raggedright {\ }{\ }{\ }\ensuremath{\Gamma}\ensuremath{_{\ensuremath{\alpha}}}/\ensuremath{\Gamma}\ensuremath{\leq}0.01 (see Fig. 7), \ensuremath{\Gamma}\ensuremath{_{\textnormal{p$'$}}}/\ensuremath{\Gamma}\ensuremath{_{\textnormal{p}}}=0.19, and \ensuremath{\Gamma}\ensuremath{_{\ensuremath{\alpha}}}/\ensuremath{\Gamma}\ensuremath{_{\textnormal{p}}}\ensuremath{\leq}0.01 (\href{https://www.nndc.bnl.gov/nsr/nsrlink.jsp?2012Al11,B}{2012Al11}).\vspace{0.1cm}}&\\
&&&&&&\parbox[t][0.3cm]{10.729831cm}{\raggedright J\ensuremath{^{\pi}},L: From (\href{https://www.nndc.bnl.gov/nsr/nsrlink.jsp?1977Ev01,B}{1977Ev01}: J\ensuremath{^{\ensuremath{\pi}}}=2\ensuremath{^{\textnormal{+}}} from the zero-range DWBA analysis. L is not\vspace{0.1cm}}&\\
&&&&&&\parbox[t][0.3cm]{10.729831cm}{\raggedright {\ }{\ }{\ }given but the evaluator assigned L=2 from this DWBA analysis. The\vspace{0.1cm}}&\\
&&&&&&\parbox[t][0.3cm]{10.729831cm}{\raggedright {\ }{\ }{\ }agreement between DWBA fit and the neutron angular distribution data at 15\vspace{0.1cm}}&\\
&&&&&&\parbox[t][0.3cm]{10.729831cm}{\raggedright {\ }{\ }{\ }and 18 MeV incident energies is very good); (\href{https://www.nndc.bnl.gov/nsr/nsrlink.jsp?1981Ne09,B}{1981Ne09}: J\ensuremath{^{\ensuremath{\pi}}}=(1\ensuremath{^{-}}, 2\ensuremath{^{\textnormal{+}}}) from\vspace{0.1cm}}&\\
&&&&&&\parbox[t][0.3cm]{10.729831cm}{\raggedright {\ }{\ }{\ }DWBA analysis with L=1,2. L=2 (and thus J\ensuremath{^{\ensuremath{\pi}}}=2\ensuremath{^{\textnormal{+}}}) was preferred based on\vspace{0.1cm}}&\\
&&&&&&\parbox[t][0.3cm]{10.729831cm}{\raggedright {\ }{\ }{\ }the reaction mechanism); (\href{https://www.nndc.bnl.gov/nsr/nsrlink.jsp?1996Ha26,B}{1996Ha26}: (J\ensuremath{^{\ensuremath{\pi}}}=4\ensuremath{^{\textnormal{+}}}) from a combination of\vspace{0.1cm}}&\\
&&&&&&\parbox[t][0.3cm]{10.729831cm}{\raggedright {\ }{\ }{\ }penetrability considerations, level widths, mirror levels analysis, and\vspace{0.1cm}}&\\
&&&&&&\parbox[t][0.3cm]{10.729831cm}{\raggedright {\ }{\ }{\ }Woods-Saxon calculations of the Coulomb energy shifts); and (\href{https://www.nndc.bnl.gov/nsr/nsrlink.jsp?2000Fo19,B}{2000Fo19}:\vspace{0.1cm}}&\\
&&&&&&\parbox[t][0.3cm]{10.729831cm}{\raggedright {\ }{\ }{\ }J\ensuremath{^{\ensuremath{\pi}}}=4\ensuremath{^{\textnormal{+}}} predicted from shell model calculations).\vspace{0.1cm}}&\\
&&&&&&\parbox[t][0.3cm]{10.729831cm}{\raggedright (\href{https://www.nndc.bnl.gov/nsr/nsrlink.jsp?2012Al11,B}{2012Al11}): a\ensuremath{_{\textnormal{0}}}=1554.26, a\ensuremath{_{\textnormal{2}}}={\textminus}109.72, a\ensuremath{_{\textnormal{4}}}={\textminus}1230.54, and a\ensuremath{_{\textnormal{6}}}={\textminus}589.35, where a\ensuremath{_{\textnormal{i}}}\vspace{0.1cm}}&\\
&&&&&&\parbox[t][0.3cm]{10.729831cm}{\raggedright {\ }{\ }{\ }are the Legendre polynomial fit coefficients for the measured proton$'$s (from\vspace{0.1cm}}&\\
&&&&&&\parbox[t][0.3cm]{10.729831cm}{\raggedright {\ }{\ }{\ }proton decay of this state) angular distribution.\vspace{0.1cm}}&\\
&&&&&&\parbox[t][0.3cm]{10.729831cm}{\raggedright (\href{https://www.nndc.bnl.gov/nsr/nsrlink.jsp?1981Ne09,B}{1981Ne09}) suggests that this state may contain substantial \textit{d}\ensuremath{_{\textnormal{3/2}}} strength.\vspace{0.1cm}}&\\
&&&&&&\parbox[t][0.3cm]{10.729831cm}{\raggedright (\href{https://www.nndc.bnl.gov/nsr/nsrlink.jsp?2012Fo29,B}{2012Fo29}) reports this state to be of four-particle two-hole (4p-2h)\vspace{0.1cm}}&\\
&&&&&&\parbox[t][0.3cm]{10.729831cm}{\raggedright {\ }{\ }{\ }configuration, i.e., (\textit{sd})\ensuremath{^{\textnormal{4}}}(1\textit{p})\ensuremath{^{\textnormal{$-$2}}}, and calculated \ensuremath{\Gamma}\ensuremath{_{\textnormal{sp}}}(\textit{l}=4)=0.68 keV for the\vspace{0.1cm}}&\\
&&&&&&\parbox[t][0.3cm]{10.729831cm}{\raggedright {\ }{\ }{\ }proton decay of \ensuremath{^{\textnormal{18}}}Ne*(7.06 MeV, 4\ensuremath{^{\textnormal{+}}})\ensuremath{\rightarrow}p+\ensuremath{^{\textnormal{17}}}F*(495 keV, 1/2\ensuremath{^{\textnormal{+}}}). (\href{https://www.nndc.bnl.gov/nsr/nsrlink.jsp?2012Fo29,B}{2012Fo29})\vspace{0.1cm}}&\\
&&&&&&\parbox[t][0.3cm]{10.729831cm}{\raggedright {\ }{\ }{\ }assumed C\ensuremath{^{\textnormal{2}}}S=0.01-0.02 for this \ensuremath{^{\textnormal{18}}}Ne state and calculated the \ensuremath{\Gamma}\ensuremath{_{\textnormal{p}_{\textnormal{1}}}} indicated\vspace{0.1cm}}&\\
&&&&&&\parbox[t][0.3cm]{10.729831cm}{\raggedright {\ }{\ }{\ }above, which yields a \ensuremath{\Gamma}\ensuremath{_{\textnormal{p}_{\textnormal{1}}}}/\ensuremath{\Gamma}\ensuremath{_{\textnormal{p}_{\textnormal{0}}}} branching ratio of less than 2\ensuremath{\times}10\ensuremath{^{\textnormal{$-$4}}}. This\vspace{0.1cm}}&\\
&&&&&&\parbox[t][0.3cm]{10.729831cm}{\raggedright {\ }{\ }{\ }value disagrees with the \ensuremath{\Gamma}\ensuremath{_{\textnormal{p}_{\textnormal{1}}}}/\ensuremath{\Gamma}\ensuremath{_{\textnormal{p}_{\textnormal{0}}}}=0.19 deduced by (\href{https://www.nndc.bnl.gov/nsr/nsrlink.jsp?2012Al11,B}{2012Al11}).\vspace{0.1cm}}&\\
\multicolumn{1}{r@{}}{7120?}&\multicolumn{1}{@{}l}{\ensuremath{^{{\hyperlink{NE15LEVEL5}{f}}}} {\it 30}}&&\multicolumn{1}{r@{}}{$\leq$120}&\multicolumn{1}{@{ }l}{keV}&&\parbox[t][0.3cm]{10.729831cm}{\raggedright E(level): From (\href{https://www.nndc.bnl.gov/nsr/nsrlink.jsp?1996Ha26,B}{1996Ha26}): this level is a member of a close lying doublet.\vspace{0.1cm}}&\\
&&&&&&\parbox[t][0.3cm]{10.729831cm}{\raggedright {\ }{\ }{\ }(\href{https://www.nndc.bnl.gov/nsr/nsrlink.jsp?1996Ha26,B}{1996Ha26}) considered a single peak in this region and deduced E\ensuremath{_{\textnormal{x}}}=7070\vspace{0.1cm}}&\\
&&&&&&\parbox[t][0.3cm]{10.729831cm}{\raggedright {\ }{\ }{\ }keV \textit{10} and \ensuremath{\Gamma}=200 keV \textit{40}. However, the fit with two states (a doublet) led\vspace{0.1cm}}&\\
\end{longtable}
\begin{textblock}{29}(0,27.3)
Continued on next page (footnotes at end of table)
\end{textblock}
\clearpage
\begin{longtable}{ccccccc@{\extracolsep{\fill}}c}
\\[-.4cm]
\multicolumn{8}{c}{{\bf \small \underline{\ensuremath{^{\textnormal{16}}}O(\ensuremath{^{\textnormal{3}}}He,n)\hspace{0.2in}\href{https://www.nndc.bnl.gov/nsr/nsrlink.jsp?1953Ku08,B}{1953Ku08},\href{https://www.nndc.bnl.gov/nsr/nsrlink.jsp?2012Fo29,B}{2012Fo29} (continued)}}}\\
\multicolumn{8}{c}{~}\\
\multicolumn{8}{c}{\underline{\ensuremath{^{18}}Ne Levels (continued)}}\\
\multicolumn{8}{c}{~}\\
\multicolumn{2}{c}{E(level)$^{}$}&J$^{\pi}$$^{{\hyperlink{NE15LEVEL0}{a}}}$&\multicolumn{2}{c}{T\ensuremath{_{\textnormal{1/2}}} or \ensuremath{\Gamma}$^{}$}&L$^{{\hyperlink{NE15LEVEL0}{a}}{\hyperlink{NE15LEVEL1}{b}}{\hyperlink{NE15LEVEL2}{c}}}$&Comments&\\[-.2cm]
\multicolumn{2}{c}{\hrulefill}&\hrulefill&\multicolumn{2}{c}{\hrulefill}&\hrulefill&\hrulefill&
\endhead
&&&&&&\parbox[t][0.3cm]{10.83469cm}{\raggedright {\ }{\ }{\ }to a smaller \ensuremath{\chi}\ensuremath{^{\textnormal{2}}}, and was thus a better fit. Consequently, (\href{https://www.nndc.bnl.gov/nsr/nsrlink.jsp?1996Ha26,B}{1996Ha26})\vspace{0.1cm}}&\\
&&&&&&\parbox[t][0.3cm]{10.83469cm}{\raggedright {\ }{\ }{\ }obtained two tentative states: E\ensuremath{_{\textnormal{x}}}=(7050 keV \textit{30}) and (7120 keV \textit{30}) both\vspace{0.1cm}}&\\
&&&&&&\parbox[t][0.3cm]{10.83469cm}{\raggedright {\ }{\ }{\ }with \ensuremath{\Gamma}\ensuremath{\leq}120 keV.\vspace{0.1cm}}&\\
&&&&&&\parbox[t][0.3cm]{10.83469cm}{\raggedright E(level): The combination of the \ensuremath{^{\textnormal{12}}}C(\ensuremath{^{\textnormal{12}}}C,\ensuremath{^{\textnormal{6}}}He)\ensuremath{^{\textnormal{18}}}Ne data from (\href{https://www.nndc.bnl.gov/nsr/nsrlink.jsp?1996Ha26,B}{1996Ha26})\vspace{0.1cm}}&\\
&&&&&&\parbox[t][0.3cm]{10.83469cm}{\raggedright {\ }{\ }{\ }together with the evidence reported previously is consistent with two levels at\vspace{0.1cm}}&\\
&&&&&&\parbox[t][0.3cm]{10.83469cm}{\raggedright {\ }{\ }{\ }E\ensuremath{_{\textnormal{x}}}=7.05 MeV and E\ensuremath{_{\textnormal{x}}}=7.12 MeV.\vspace{0.1cm}}&\\
&&&&&&\parbox[t][0.3cm]{10.83469cm}{\raggedright \ensuremath{\Gamma}: From (\href{https://www.nndc.bnl.gov/nsr/nsrlink.jsp?1996Ha26,B}{1996Ha26}) considering a doublet.\vspace{0.1cm}}&\\
\multicolumn{1}{r@{}}{7350}&\multicolumn{1}{@{}l}{\ensuremath{^{{\hyperlink{NE15LEVEL5}{f}}}} {\it 18}}&\multicolumn{1}{l}{(1\ensuremath{^{-}})}&\multicolumn{1}{r@{}}{$\leq$50}&\multicolumn{1}{@{ }l}{keV}&\multicolumn{1}{l}{(1)}&\parbox[t][0.3cm]{10.83469cm}{\raggedright E(level),\ensuremath{\Gamma}: From (\href{https://www.nndc.bnl.gov/nsr/nsrlink.jsp?1996Ha26,B}{1996Ha26}).\vspace{0.1cm}}&\\
&&&&&&\parbox[t][0.3cm]{10.83469cm}{\raggedright J\ensuremath{^{\pi}},L: (\href{https://www.nndc.bnl.gov/nsr/nsrlink.jsp?1996Ha26,B}{1996Ha26}: J\ensuremath{^{\ensuremath{\pi}}}=(0\ensuremath{^{\textnormal{+}}}, 1\ensuremath{^{-}}, 2\ensuremath{^{\textnormal{+}}})): from a zero-range DWBA analysis with\vspace{0.1cm}}&\\
&&&&&&\parbox[t][0.3cm]{10.83469cm}{\raggedright {\ }{\ }{\ }L=(0,1,2). However, from a combination of Coulomb penetrability\vspace{0.1cm}}&\\
&&&&&&\parbox[t][0.3cm]{10.83469cm}{\raggedright {\ }{\ }{\ }considerations, level widths, mirror levels analysis, and Woods-Saxon\vspace{0.1cm}}&\\
&&&&&&\parbox[t][0.3cm]{10.83469cm}{\raggedright {\ }{\ }{\ }calculations of the Coulomb energy shifts, (\href{https://www.nndc.bnl.gov/nsr/nsrlink.jsp?1996Ha26,B}{1996Ha26}) assigned J\ensuremath{^{\ensuremath{\pi}}}=(1\ensuremath{^{-}}) to\vspace{0.1cm}}&\\
&&&&&&\parbox[t][0.3cm]{10.83469cm}{\raggedright {\ }{\ }{\ }this state.\vspace{0.1cm}}&\\
\multicolumn{1}{r@{}}{7718}&\multicolumn{1}{@{ }l}{{\it 9}}&&\multicolumn{1}{r@{}}{$\leq$30}&\multicolumn{1}{@{ }l}{keV}&&\parbox[t][0.3cm]{10.83469cm}{\raggedright E(level): Weighted average of 7712 keV \textit{20} (\href{https://www.nndc.bnl.gov/nsr/nsrlink.jsp?1974Ne04,B}{1974Ne04}, \href{https://www.nndc.bnl.gov/nsr/nsrlink.jsp?1981Ne09,B}{1981Ne09}); and 7720\vspace{0.1cm}}&\\
&&&&&&\parbox[t][0.3cm]{10.83469cm}{\raggedright {\ }{\ }{\ }keV \textit{10} (\href{https://www.nndc.bnl.gov/nsr/nsrlink.jsp?1996Ha26,B}{1996Ha26}). (\href{https://www.nndc.bnl.gov/nsr/nsrlink.jsp?1981Ne09,B}{1981Ne09}) observed a triplet of states near 8 MeV.\vspace{0.1cm}}&\\
&&&&&&\parbox[t][0.3cm]{10.83469cm}{\raggedright {\ }{\ }{\ }These states were unresolved at 14.5 MeV incident energy (see the last peak\vspace{0.1cm}}&\\
&&&&&&\parbox[t][0.3cm]{10.83469cm}{\raggedright {\ }{\ }{\ }in Fig. 5). However, at 16.1 MeV incident energy, these 3 states are more or\vspace{0.1cm}}&\\
&&&&&&\parbox[t][0.3cm]{10.83469cm}{\raggedright {\ }{\ }{\ }less resolved at \ensuremath{\theta}\ensuremath{_{\textnormal{lab}}}=5\ensuremath{^\circ} (see Fig. 6) and very well separated at \ensuremath{\theta}\ensuremath{_{\textnormal{lab}}}=25\ensuremath{^\circ}.\vspace{0.1cm}}&\\
&&&&&&\parbox[t][0.3cm]{10.83469cm}{\raggedright {\ }{\ }{\ }(\href{https://www.nndc.bnl.gov/nsr/nsrlink.jsp?1981Ne09,B}{1981Ne09}) does not report which states construct the triplet and only\vspace{0.1cm}}&\\
&&&&&&\parbox[t][0.3cm]{10.83469cm}{\raggedright {\ }{\ }{\ }mentions ``an 8-MeV triplet''. The neutron energy spectra presented in\vspace{0.1cm}}&\\
&&&&&&\parbox[t][0.3cm]{10.83469cm}{\raggedright {\ }{\ }{\ }(\href{https://www.nndc.bnl.gov/nsr/nsrlink.jsp?1981Ne09,B}{1981Ne09}) is in logarithmic scale, which makes it harder to reliably\vspace{0.1cm}}&\\
&&&&&&\parbox[t][0.3cm]{10.83469cm}{\raggedright {\ }{\ }{\ }determine which states are included in the triplet. Evaluator notes that the\vspace{0.1cm}}&\\
&&&&&&\parbox[t][0.3cm]{10.83469cm}{\raggedright {\ }{\ }{\ }7062-keV state is a strongly populated, broad peak in (\href{https://www.nndc.bnl.gov/nsr/nsrlink.jsp?1981Ne09,B}{1981Ne09}: see Fig. 8).\vspace{0.1cm}}&\\
&&&&&&\parbox[t][0.3cm]{10.83469cm}{\raggedright {\ }{\ }{\ }Based on that, we estimated that the 3 states constructing the triplet are most\vspace{0.1cm}}&\\
&&&&&&\parbox[t][0.3cm]{10.83469cm}{\raggedright {\ }{\ }{\ }likely the 7712-keV, 7915-keV, and 8100-keV states, all of which were\vspace{0.1cm}}&\\
&&&&&&\parbox[t][0.3cm]{10.83469cm}{\raggedright {\ }{\ }{\ }estimated to have \ensuremath{\Gamma}\ensuremath{\leq}50 keV. Note that the 7949-keV state observed in\vspace{0.1cm}}&\\
&&&&&&\parbox[t][0.3cm]{10.83469cm}{\raggedright {\ }{\ }{\ }(\href{https://www.nndc.bnl.gov/nsr/nsrlink.jsp?1981Ne09,B}{1981Ne09}) via \ensuremath{^{\textnormal{20}}}Ne(p,t) measurement is not part of this triplet because this\vspace{0.1cm}}&\\
&&&&&&\parbox[t][0.3cm]{10.83469cm}{\raggedright {\ }{\ }{\ }latter state was not populated via the \ensuremath{^{\textnormal{16}}}O(\ensuremath{^{\textnormal{3}}}He,n) reaction, so no neutron\vspace{0.1cm}}&\\
&&&&&&\parbox[t][0.3cm]{10.83469cm}{\raggedright {\ }{\ }{\ }peak in the triplet can be associated with that state. Moreover, no evidence of\vspace{0.1cm}}&\\
&&&&&&\parbox[t][0.3cm]{10.83469cm}{\raggedright {\ }{\ }{\ }the state at E\ensuremath{_{\textnormal{x}}}=8294 keV (see the \ensuremath{^{\textnormal{18}}}Ne Adopted Levels) is mentioned in\vspace{0.1cm}}&\\
&&&&&&\parbox[t][0.3cm]{10.83469cm}{\raggedright {\ }{\ }{\ }(\href{https://www.nndc.bnl.gov/nsr/nsrlink.jsp?1981Ne09,B}{1981Ne09}). Therefore, the \ensuremath{^{\textnormal{18}}}Ne*(7718 keV) level recommended here is\vspace{0.1cm}}&\\
&&&&&&\parbox[t][0.3cm]{10.83469cm}{\raggedright {\ }{\ }{\ }most likely one of the members of this triplet.\vspace{0.1cm}}&\\
&&&&&&\parbox[t][0.3cm]{10.83469cm}{\raggedright \ensuremath{\Gamma}: From (\href{https://www.nndc.bnl.gov/nsr/nsrlink.jsp?1996Ha26,B}{1996Ha26}). See also \ensuremath{\Gamma}\ensuremath{\leq}50 keV (\href{https://www.nndc.bnl.gov/nsr/nsrlink.jsp?1974Ne04,B}{1974Ne04}, \href{https://www.nndc.bnl.gov/nsr/nsrlink.jsp?1981Ne09,B}{1981Ne09}).\vspace{0.1cm}}&\\
\multicolumn{1}{r@{}}{7924}&\multicolumn{1}{@{ }l}{{\it 11}}&\multicolumn{1}{l}{(1\ensuremath{^{-}},2\ensuremath{^{+}})}&\multicolumn{1}{r@{}}{40}&\multicolumn{1}{@{ }l}{keV {\it 10}}&\multicolumn{1}{l}{(1,2)}&\parbox[t][0.3cm]{10.83469cm}{\raggedright E(level): Weighted average (with external errors) of 7915 keV \textit{12} (\href{https://www.nndc.bnl.gov/nsr/nsrlink.jsp?1974Ne04,B}{1974Ne04},\vspace{0.1cm}}&\\
&&&&&&\parbox[t][0.3cm]{10.83469cm}{\raggedright {\ }{\ }{\ }\href{https://www.nndc.bnl.gov/nsr/nsrlink.jsp?1981Ne09,B}{1981Ne09}); 7903 keV \textit{15} (\href{https://www.nndc.bnl.gov/nsr/nsrlink.jsp?1977Ev01,B}{1977Ev01}); and 7940 keV \textit{10} (\href{https://www.nndc.bnl.gov/nsr/nsrlink.jsp?1996Ha26,B}{1996Ha26}).\vspace{0.1cm}}&\\
&&&&&&\parbox[t][0.3cm]{10.83469cm}{\raggedright {\ }{\ }{\ }Evaluator notes that in this region, a close lying doublet is expected. See also\vspace{0.1cm}}&\\
&&&&&&\parbox[t][0.3cm]{10.83469cm}{\raggedright {\ }{\ }{\ }7.9 MeV (\href{https://www.nndc.bnl.gov/nsr/nsrlink.jsp?2010AlZZ,B}{2010AlZZ}, \href{https://www.nndc.bnl.gov/nsr/nsrlink.jsp?2010Ta17,B}{2010Ta17}), and 7950 keV \textit{30} (\href{https://www.nndc.bnl.gov/nsr/nsrlink.jsp?2012Al11,B}{2012Al11}: Table II).\vspace{0.1cm}}&\\
&&&&&&\parbox[t][0.3cm]{10.83469cm}{\raggedright {\ }{\ }{\ }This state was excluded since the poor neutron time-of-flight resolution of\vspace{0.1cm}}&\\
&&&&&&\parbox[t][0.3cm]{10.83469cm}{\raggedright {\ }{\ }{\ }(\href{https://www.nndc.bnl.gov/nsr/nsrlink.jsp?2012Al11,B}{2012Al11}) was not sufficient enough to separate the doublet, and a wide\vspace{0.1cm}}&\\
&&&&&&\parbox[t][0.3cm]{10.83469cm}{\raggedright {\ }{\ }{\ }neutron peak is observed, where this doublet is expected.\vspace{0.1cm}}&\\
&&&&&&\parbox[t][0.3cm]{10.83469cm}{\raggedright E(level): This state is most likely another member of the 8-MeV triplet\vspace{0.1cm}}&\\
&&&&&&\parbox[t][0.3cm]{10.83469cm}{\raggedright {\ }{\ }{\ }observed in (\href{https://www.nndc.bnl.gov/nsr/nsrlink.jsp?1981Ne09,B}{1981Ne09}) discussed above.\vspace{0.1cm}}&\\
&&&&&&\parbox[t][0.3cm]{10.83469cm}{\raggedright \ensuremath{\Gamma}: From (\href{https://www.nndc.bnl.gov/nsr/nsrlink.jsp?1996Ha26,B}{1996Ha26}). See also \ensuremath{\Gamma}\ensuremath{\leq}50 keV (\href{https://www.nndc.bnl.gov/nsr/nsrlink.jsp?1974Ne04,B}{1974Ne04}, \href{https://www.nndc.bnl.gov/nsr/nsrlink.jsp?1981Ne09,B}{1981Ne09}).\vspace{0.1cm}}&\\
&&&&&&\parbox[t][0.3cm]{10.83469cm}{\raggedright J\ensuremath{^{\pi}},L: From (\href{https://www.nndc.bnl.gov/nsr/nsrlink.jsp?1981Ne09,B}{1981Ne09}: J\ensuremath{^{\ensuremath{\pi}}}=(1\ensuremath{^{-}},2\ensuremath{^{\textnormal{+}}}) based on comparison of the shape of\vspace{0.1cm}}&\\
&&&&&&\parbox[t][0.3cm]{10.83469cm}{\raggedright {\ }{\ }{\ }neutron angular distribution data with the DWBA curves obtained with\vspace{0.1cm}}&\\
&&&&&&\parbox[t][0.3cm]{10.83469cm}{\raggedright {\ }{\ }{\ }L=1,2). See also (\href{https://www.nndc.bnl.gov/nsr/nsrlink.jsp?1996Ha26,B}{1996Ha26}): J\ensuremath{^{\ensuremath{\pi}}}=(1\ensuremath{^{-}},2\ensuremath{^{\textnormal{+}}},3\ensuremath{^{-}}) from zero-range DWBA analysis\vspace{0.1cm}}&\\
&&&&&&\parbox[t][0.3cm]{10.83469cm}{\raggedright {\ }{\ }{\ }with L=1,2,3, respectively. These DWBA fits do not seem to describe the\vspace{0.1cm}}&\\
&&&&&&\parbox[t][0.3cm]{10.83469cm}{\raggedright {\ }{\ }{\ }measured neutron angular distribution data well.\vspace{0.1cm}}&\\
&&&&&&\parbox[t][0.3cm]{10.83469cm}{\raggedright (\href{https://www.nndc.bnl.gov/nsr/nsrlink.jsp?2012Al11,B}{2012Al11}) deduced branching ratios of \ensuremath{\Gamma}\ensuremath{_{\textnormal{p}_{\textnormal{0}}}}/\ensuremath{\Gamma}=64\% 8, \ensuremath{\Gamma}\ensuremath{_{\textnormal{p}_{\textnormal{1}}}}/\ensuremath{\Gamma}=16\% 10, and\vspace{0.1cm}}&\\
&&&&&&\parbox[t][0.3cm]{10.83469cm}{\raggedright {\ }{\ }{\ }\ensuremath{\Gamma}\ensuremath{_{\ensuremath{\alpha}}}/\ensuremath{\Gamma}=20\% 9 for this state. However, there is a close doublet expected in\vspace{0.1cm}}&\\
&&&&&&\parbox[t][0.3cm]{10.83469cm}{\raggedright {\ }{\ }{\ }this region. But (\href{https://www.nndc.bnl.gov/nsr/nsrlink.jsp?2012Al11,B}{2012Al11}) appears to have ignored this. They used \ensuremath{\Gamma}=55\vspace{0.1cm}}&\\
&&&&&&\parbox[t][0.3cm]{10.83469cm}{\raggedright {\ }{\ }{\ }keV \textit{20} keV and deduced \ensuremath{\Gamma}\ensuremath{_{\textnormal{p$'$}}}/\ensuremath{\Gamma}\ensuremath{_{\textnormal{p}}}=0.25 and \ensuremath{\Gamma}\ensuremath{_{\ensuremath{\alpha}}}/\ensuremath{\Gamma}\ensuremath{_{\textnormal{p}}}=0.31 in addition to\vspace{0.1cm}}&\\
&&&&&&\parbox[t][0.3cm]{10.83469cm}{\raggedright {\ }{\ }{\ }branching ratios mentioned before. The \ensuremath{\Gamma}\ensuremath{_{\textnormal{tot}}} used must have been the\vspace{0.1cm}}&\\
&&&&&&\parbox[t][0.3cm]{10.83469cm}{\raggedright {\ }{\ }{\ }unweighted average (with an inflated uncertainty by 5 keV) of \ensuremath{\Gamma}=40 keV \textit{10}\vspace{0.1cm}}&\\
&&&&&&\parbox[t][0.3cm]{10.83469cm}{\raggedright {\ }{\ }{\ }from (\href{https://www.nndc.bnl.gov/nsr/nsrlink.jsp?1996Ha26,B}{1996Ha26}: (\ensuremath{^{\textnormal{3}}}He,n) data) and \ensuremath{\Gamma}=70 keV \textit{20} from (\href{https://www.nndc.bnl.gov/nsr/nsrlink.jsp?1996Ha26,B}{1996Ha26}: (p,t)\vspace{0.1cm}}&\\
\end{longtable}
\begin{textblock}{29}(0,27.3)
Continued on next page (footnotes at end of table)
\end{textblock}
\clearpage
\begin{longtable}{ccccc@{\extracolsep{\fill}}c}
\\[-.4cm]
\multicolumn{6}{c}{{\bf \small \underline{\ensuremath{^{\textnormal{16}}}O(\ensuremath{^{\textnormal{3}}}He,n)\hspace{0.2in}\href{https://www.nndc.bnl.gov/nsr/nsrlink.jsp?1953Ku08,B}{1953Ku08},\href{https://www.nndc.bnl.gov/nsr/nsrlink.jsp?2012Fo29,B}{2012Fo29} (continued)}}}\\
\multicolumn{6}{c}{~}\\
\multicolumn{6}{c}{\underline{\ensuremath{^{18}}Ne Levels (continued)}}\\
\multicolumn{6}{c}{~}\\
\multicolumn{2}{c}{E(level)$^{}$}&\multicolumn{2}{c}{T\ensuremath{_{\textnormal{1/2}}} or \ensuremath{\Gamma}$^{}$}&Comments&\\[-.2cm]
\multicolumn{2}{c}{\hrulefill}&\multicolumn{2}{c}{\hrulefill}&\hrulefill&
\endhead
&&&&\parbox[t][0.3cm]{14.005811cm}{\raggedright {\ }{\ }{\ }data using the K600 spectrograph). However, the state observed in the (p,t) measurement with a\vspace{0.1cm}}&\\
&&&&\parbox[t][0.3cm]{14.005811cm}{\raggedright {\ }{\ }{\ }larger width is most likely the higher energy member of the doublet. So, these results are unreliable.\vspace{0.1cm}}&\\
&&&&\parbox[t][0.3cm]{14.005811cm}{\raggedright {\ }{\ }{\ }The evaluator notes that (\href{https://www.nndc.bnl.gov/nsr/nsrlink.jsp?1996Ha26,B}{1996Ha26}) also did not consider the doublet as separate states when\vspace{0.1cm}}&\\
&&&&\parbox[t][0.3cm]{14.005811cm}{\raggedright {\ }{\ }{\ }presenting their results in tabular format (presumably because the deduced energies were\vspace{0.1cm}}&\\
&&&&\parbox[t][0.3cm]{14.005811cm}{\raggedright {\ }{\ }{\ }indistinguishable) but they seem to acknowledge the doublet in their discussions (see section V.E in\vspace{0.1cm}}&\\
&&&&\parbox[t][0.3cm]{14.005811cm}{\raggedright {\ }{\ }{\ }that article).\vspace{0.1cm}}&\\
&&&&\parbox[t][0.3cm]{14.005811cm}{\raggedright (\href{https://www.nndc.bnl.gov/nsr/nsrlink.jsp?2012Al11,B}{2012Al11}) observed, for the first time, the \ensuremath{\alpha}-decay from the level measured at 7950 keV \textit{30} with a\vspace{0.1cm}}&\\
&&&&\parbox[t][0.3cm]{14.005811cm}{\raggedright {\ }{\ }{\ }significant branching ratio. However, as mentioned above, it is not clear which of the members of\vspace{0.1cm}}&\\
&&&&\parbox[t][0.3cm]{14.005811cm}{\raggedright {\ }{\ }{\ }this doublet is responsible for this.\vspace{0.1cm}}&\\
&&&&\parbox[t][0.3cm]{14.005811cm}{\raggedright (\href{https://www.nndc.bnl.gov/nsr/nsrlink.jsp?2012Al11,B}{2012Al11}): a\ensuremath{_{\textnormal{0}}}=1116.96, a\ensuremath{_{\textnormal{2}}}={\textminus}18.64, a\ensuremath{_{\textnormal{4}}}={\textminus}428.29, and a\ensuremath{_{\textnormal{6}}}={\textminus}495.91, where a\ensuremath{_{\textnormal{i}}} are the Legendre\vspace{0.1cm}}&\\
&&&&\parbox[t][0.3cm]{14.005811cm}{\raggedright {\ }{\ }{\ }polynomial fit coefficients for the measured angular distribution of protons from the decay of this\vspace{0.1cm}}&\\
&&&&\parbox[t][0.3cm]{14.005811cm}{\raggedright {\ }{\ }{\ }state. Evaluator notes that these results may be unreliable due to reasons mentioned above.\vspace{0.1cm}}&\\
\multicolumn{1}{r@{}}{8098}&\multicolumn{1}{@{ }l}{{\it 9}}&\multicolumn{1}{r@{}}{$\leq$30}&\multicolumn{1}{@{ }l}{keV}&\parbox[t][0.3cm]{14.005811cm}{\raggedright \ensuremath{\Gamma}\ensuremath{_{\textnormal{p0}}}/\ensuremath{\Gamma}=0.65 \textit{12} (\href{https://www.nndc.bnl.gov/nsr/nsrlink.jsp?2012Al11,B}{2012Al11})\vspace{0.1cm}}&\\
&&&&\parbox[t][0.3cm]{14.005811cm}{\raggedright \ensuremath{\Gamma}\ensuremath{_{\textnormal{p1}}}/\ensuremath{\Gamma}=0.14 \textit{12} (\href{https://www.nndc.bnl.gov/nsr/nsrlink.jsp?2012Al11,B}{2012Al11})\vspace{0.1cm}}&\\
&&&&\parbox[t][0.3cm]{14.005811cm}{\raggedright \ensuremath{\Gamma}\ensuremath{\alpha}/\ensuremath{\Gamma}=0.21 \textit{13} (\href{https://www.nndc.bnl.gov/nsr/nsrlink.jsp?2012Al11,B}{2012Al11})\vspace{0.1cm}}&\\
&&&&\parbox[t][0.3cm]{14.005811cm}{\raggedright \ensuremath{\Gamma}\ensuremath{\alpha}/\ensuremath{\Gamma}\ensuremath{_{\textnormal{p}}}=0.32 (\href{https://www.nndc.bnl.gov/nsr/nsrlink.jsp?2012Al11,B}{2012Al11})\vspace{0.1cm}}&\\
&&&&\parbox[t][0.3cm]{14.005811cm}{\raggedright E(level): Weighted average (with external errors) of 8100 keV \textit{14} (\href{https://www.nndc.bnl.gov/nsr/nsrlink.jsp?1974Ne04,B}{1974Ne04}, \href{https://www.nndc.bnl.gov/nsr/nsrlink.jsp?1981Ne09,B}{1981Ne09}); 8070 keV\vspace{0.1cm}}&\\
&&&&\parbox[t][0.3cm]{14.005811cm}{\raggedright {\ }{\ }{\ }\textit{15} (\href{https://www.nndc.bnl.gov/nsr/nsrlink.jsp?1977Ev01,B}{1977Ev01}); 8110 keV \textit{10} (\href{https://www.nndc.bnl.gov/nsr/nsrlink.jsp?1996Ha26,B}{1996Ha26}); and 8090 keV \textit{30} (\href{https://www.nndc.bnl.gov/nsr/nsrlink.jsp?2012Al11,B}{2012Al11}: Table II). See also 8.1 MeV\vspace{0.1cm}}&\\
&&&&\parbox[t][0.3cm]{14.005811cm}{\raggedright {\ }{\ }{\ }(\href{https://www.nndc.bnl.gov/nsr/nsrlink.jsp?2010AlZZ,B}{2010AlZZ}, \href{https://www.nndc.bnl.gov/nsr/nsrlink.jsp?2010Ta17,B}{2010Ta17}). Evaluator notes that this state is most likely a member of the 8-MeV triplet\vspace{0.1cm}}&\\
&&&&\parbox[t][0.3cm]{14.005811cm}{\raggedright {\ }{\ }{\ }reported by (\href{https://www.nndc.bnl.gov/nsr/nsrlink.jsp?1981Ne09,B}{1981Ne09}) based on their \ensuremath{^{\textnormal{16}}}O(\ensuremath{^{\textnormal{3}}}He,n) data (see Figs. 6 and 7).\vspace{0.1cm}}&\\
&&&&\parbox[t][0.3cm]{14.005811cm}{\raggedright \ensuremath{\Gamma}: From (\href{https://www.nndc.bnl.gov/nsr/nsrlink.jsp?1996Ha26,B}{1996Ha26}). See also \ensuremath{\Gamma}\ensuremath{\leq}50 keV (\href{https://www.nndc.bnl.gov/nsr/nsrlink.jsp?1974Ne04,B}{1974Ne04}, \href{https://www.nndc.bnl.gov/nsr/nsrlink.jsp?1981Ne09,B}{1981Ne09}).\vspace{0.1cm}}&\\
&&&&\parbox[t][0.3cm]{14.005811cm}{\raggedright \ensuremath{\Gamma}\ensuremath{_{\textnormal{p$'$}}}/\ensuremath{\Gamma}\ensuremath{_{\textnormal{p}}}=0.22 (\href{https://www.nndc.bnl.gov/nsr/nsrlink.jsp?2012Al11,B}{2012Al11}).\vspace{0.1cm}}&\\
&&&&\parbox[t][0.3cm]{14.005811cm}{\raggedright (\href{https://www.nndc.bnl.gov/nsr/nsrlink.jsp?2012Al11,B}{2012Al11}): a\ensuremath{_{\textnormal{0}}}=642.28, a\ensuremath{_{\textnormal{2}}}={\textminus}47.26, a\ensuremath{_{\textnormal{4}}}={\textminus}276.58, and a\ensuremath{_{\textnormal{6}}}={\textminus}299.72, where a\ensuremath{_{\textnormal{i}}} are the Legendre polynomial\vspace{0.1cm}}&\\
&&&&\parbox[t][0.3cm]{14.005811cm}{\raggedright {\ }{\ }{\ }fit coefficients for the measured proton$'$s (from proton decay of this state) angular distribution.\vspace{0.1cm}}&\\
\multicolumn{1}{r@{}}{8500}&\multicolumn{1}{@{ }l}{{\it 30}}&\multicolumn{1}{r@{}}{$\leq$120}&\multicolumn{1}{@{ }l}{keV}&\parbox[t][0.3cm]{14.005811cm}{\raggedright E(level),\ensuremath{\Gamma}: From (\href{https://www.nndc.bnl.gov/nsr/nsrlink.jsp?1974Ne04,B}{1974Ne04}).\vspace{0.1cm}}&\\
\end{longtable}
\parbox[b][0.3cm]{17.7cm}{\makebox[1ex]{\ensuremath{^{\hypertarget{NE15LEVEL0}{a}}}} (\href{https://www.nndc.bnl.gov/nsr/nsrlink.jsp?1968Sh09,B}{1968Sh09}): the double stripping selection rules for the (\ensuremath{^{\textnormal{3}}}He,n) reactions require J\ensuremath{_{\textnormal{f}}}=L and \ensuremath{\pi}\ensuremath{_{\textnormal{f}}} =({\textminus}1)\ensuremath{^{\textnormal{L}}} when a target nucleus with}\\
\parbox[b][0.3cm]{17.7cm}{{\ }{\ }J\ensuremath{^{\ensuremath{\pi}}}=0\ensuremath{^{\textnormal{+}}} is bombarded and a direct, one-step diproton in a relative \textit{s}=0 state is transferred. Here, J\ensuremath{_{\textnormal{f}}} and \ensuremath{\pi}\ensuremath{_{\textnormal{f}}} are the spin and parity}\\
\parbox[b][0.3cm]{17.7cm}{{\ }{\ }of the final nucleus and L is the transferred orbital angular momentum.}\\
\parbox[b][0.3cm]{17.7cm}{\makebox[1ex]{\ensuremath{^{\hypertarget{NE15LEVEL1}{b}}}} (\href{https://www.nndc.bnl.gov/nsr/nsrlink.jsp?1970Ad02,B}{1970Ad02}): the L=2 transitions in the DWBA calculations do not reproduce the forward angle dip in the data but the agreement}\\
\parbox[b][0.3cm]{17.7cm}{{\ }{\ }improves with the increase of the bombarding energy, as it does for the L=4 transitions. Note that although there is evidence that}\\
\parbox[b][0.3cm]{17.7cm}{{\ }{\ }compound nuclear effects are not completely negligible in the energy region studied in this work, the simplified DWBA}\\
\parbox[b][0.3cm]{17.7cm}{{\ }{\ }calculations appear to be in rough quantitative agreement with the measured angular distributions.}\\
\parbox[b][0.3cm]{17.7cm}{\makebox[1ex]{\ensuremath{^{\hypertarget{NE15LEVEL2}{c}}}} The angular momentum transfer is not given by (\href{https://www.nndc.bnl.gov/nsr/nsrlink.jsp?1977Ev01,B}{1977Ev01}) except that of the state at 4537 keV. So, L is deduced by the}\\
\parbox[b][0.3cm]{17.7cm}{{\ }{\ }evaluator based on the angular momentum selection rules for the \ensuremath{^{\textnormal{16}}}O(\ensuremath{^{\textnormal{3}}}He,n) reaction for each state with reported DWBA}\\
\parbox[b][0.3cm]{17.7cm}{{\ }{\ }analysis from (\href{https://www.nndc.bnl.gov/nsr/nsrlink.jsp?1977Ev01,B}{1977Ev01}: see Fig. 4).}\\
\parbox[b][0.3cm]{17.7cm}{\makebox[1ex]{\ensuremath{^{\hypertarget{NE15LEVEL3}{d}}}} From (\href{https://www.nndc.bnl.gov/nsr/nsrlink.jsp?1970Ad02,B}{1970Ad02}): the relative spectroscopic factors for the \ensuremath{^{\textnormal{16}}}O(\ensuremath{^{\textnormal{3}}}He,n) reaction are obtained at various incident energies between}\\
\parbox[b][0.3cm]{17.7cm}{{\ }{\ }9-12.5 MeV and for two sets of optical potential models (see Tables 5 and 6). (\href{https://www.nndc.bnl.gov/nsr/nsrlink.jsp?1970Ad02,B}{1970Ad02}) has arbitrarily set the spectroscopic}\\
\parbox[b][0.3cm]{17.7cm}{{\ }{\ }factor equal to unity for the transition to the \ensuremath{^{\textnormal{18}}}Ne\ensuremath{_{\textnormal{g.s.}}} measured at E(\ensuremath{^{\textnormal{3}}}He)=11.5 MeV. The values given here are the relative}\\
\parbox[b][0.3cm]{17.7cm}{{\ }{\ }spectroscopic factors obtained from a consistent analysis of the \ensuremath{^{\textnormal{16}}}O(t,p), \ensuremath{^{\textnormal{16}}}O(\ensuremath{^{\textnormal{3}}}He,p), and \ensuremath{^{\textnormal{16}}}O(\ensuremath{^{\textnormal{3}}}He,n) reactions and for optical}\\
\parbox[b][0.3cm]{17.7cm}{{\ }{\ }potential model A.}\\
\parbox[b][0.3cm]{17.7cm}{\makebox[1ex]{\ensuremath{^{\hypertarget{NE15LEVEL4}{e}}}} This state was first observed in (\href{https://www.nndc.bnl.gov/nsr/nsrlink.jsp?1968To09,B}{1968To09}).}\\
\parbox[b][0.3cm]{17.7cm}{\makebox[1ex]{\ensuremath{^{\hypertarget{NE15LEVEL5}{f}}}} This state was first observed in (\href{https://www.nndc.bnl.gov/nsr/nsrlink.jsp?1996Ha26,B}{1996Ha26}).}\\
\vspace{0.5cm}
\clearpage
\subsection[\hspace{-0.2cm}\ensuremath{^{\textnormal{16}}}O(\ensuremath{^{\textnormal{3}}}He,n\ensuremath{\gamma})]{ }
\vspace{-27pt}
\vspace{0.3cm}
\hypertarget{NE16}{{\bf \small \underline{\ensuremath{^{\textnormal{16}}}O(\ensuremath{^{\textnormal{3}}}He,n\ensuremath{\gamma})\hspace{0.2in}\href{https://www.nndc.bnl.gov/nsr/nsrlink.jsp?1968Gi09,B}{1968Gi09},\href{https://www.nndc.bnl.gov/nsr/nsrlink.jsp?2003Ta13,B}{2003Ta13}}}}\\
\vspace{4pt}
\vspace{8pt}
\parbox[b][0.3cm]{17.7cm}{\addtolength{\parindent}{-0.2in}\href{https://www.nndc.bnl.gov/nsr/nsrlink.jsp?1968Gi09,B}{1968Gi09}: \ensuremath{^{\textnormal{16}}}O(\ensuremath{^{\textnormal{3}}}He,n\ensuremath{\gamma}) E=9.2, 12 and 13.2 MeV; measured n\ensuremath{\gamma} coincidences using a Ge(Li) at \ensuremath{\theta}\ensuremath{_{\textnormal{lab}}}=90\ensuremath{^\circ} and an NE-213 liquid}\\
\parbox[b][0.3cm]{17.7cm}{scintillator at \ensuremath{\theta}\ensuremath{_{\textnormal{lab}}}=0\ensuremath{^\circ}; measured \ensuremath{\gamma}-rays and their angular correlations from the de-exciting \ensuremath{^{\textnormal{18}}}Ne* levels at 1887-, 3376-, 3576-,}\\
\parbox[b][0.3cm]{17.7cm}{and 3616 keV; deduced lifetimes of these states using the Doppler shift attenuation method; deduced J\ensuremath{^{\ensuremath{\pi}}} assignments using the}\\
\parbox[b][0.3cm]{17.7cm}{mirror level analysis.}\\
\parbox[b][0.3cm]{17.7cm}{\addtolength{\parindent}{-0.2in}\href{https://www.nndc.bnl.gov/nsr/nsrlink.jsp?1969Ro08,B}{1969Ro08}: \ensuremath{^{\textnormal{16}}}O(\ensuremath{^{\textnormal{3}}}He,n\ensuremath{\gamma}) E=8.5-13.15 MeV; part of the results of this experiment was published in (\href{https://www.nndc.bnl.gov/nsr/nsrlink.jsp?1968Gi09,B}{1968Gi09}). Measured n\ensuremath{\gamma} and}\\
\parbox[b][0.3cm]{17.7cm}{\ensuremath{\gamma}\ensuremath{\gamma} coincidences using rotatable Ge(Li) detectors (resolution=8 keV at 1.84 MeV (FWHM)), a NaI detector, and a NE-213 liquid}\\
\parbox[b][0.3cm]{17.7cm}{scintillator at \ensuremath{\theta}\ensuremath{_{\textnormal{lab}}}=0\ensuremath{^\circ}; measured \ensuremath{\sigma}(E\ensuremath{_{\ensuremath{\gamma}}},\ensuremath{\theta}(n\ensuremath{\gamma})). Measured E\ensuremath{_{\ensuremath{\gamma}}}=1887.3 keV \textit{2}, 1488.9 keV \textit{3}, 1689 keV \textit{2}, 1729.2 keV \textit{5}. Measured}\\
\parbox[b][0.3cm]{17.7cm}{lifetimes of the 1887-, 3376-, 3576-, and 3616-keV states using the Doppler shift attenuation method. Measured \ensuremath{\gamma}-branching ratios}\\
\parbox[b][0.3cm]{17.7cm}{using n\ensuremath{\gamma} coincidences with the NaI or a Ge(Li) detector at 55\ensuremath{^\circ} and at E\ensuremath{_{\textnormal{lab}}}=12 MeV, \ensuremath{\gamma}-ray angular distributions at 8.5 MeV, and}\\
\parbox[b][0.3cm]{17.7cm}{\ensuremath{\gamma}-ray angular correlations at 8.5 and 12 MeV; deduced \ensuremath{^{\textnormal{18}}}Ne level energies, J, and \ensuremath{\delta} (mixing ratios); calculated transition rates in}\\
\parbox[b][0.3cm]{17.7cm}{\ensuremath{^{\textnormal{18}}}Ne based on shell model.}\\
\parbox[b][0.3cm]{17.7cm}{\addtolength{\parindent}{-0.2in}\href{https://www.nndc.bnl.gov/nsr/nsrlink.jsp?1969Ro22,B}{1969Ro22}: \ensuremath{^{\textnormal{16}}}O(\ensuremath{^{\textnormal{3}}}He,n\ensuremath{\gamma}) E=9.5-11.5 MeV; measured n\ensuremath{\gamma} coincidences using 3 Ge(Li) detectors and a NE-213 scintillator at}\\
\parbox[b][0.3cm]{17.7cm}{\ensuremath{\theta}\ensuremath{_{\textnormal{lab}}}=0\ensuremath{^\circ}; measured the \ensuremath{\gamma}-ray from the decay of the 3616-keV state directly to the ground state; measured n\ensuremath{\gamma}\ensuremath{\gamma} coincidences using}\\
\parbox[b][0.3cm]{17.7cm}{a Ge(Li) and 4 NaI detectors; measured \ensuremath{\sigma}(E;E\ensuremath{_{\textnormal{n}}},E\ensuremath{_{\ensuremath{\gamma}_{\textnormal{1}}}},E\ensuremath{_{\ensuremath{\gamma}_{\textnormal{2}}}},\ensuremath{\theta}(n\ensuremath{\gamma})); deduced \ensuremath{^{\textnormal{18}}}Ne levels at 1887-, 3376-, 3576-, and 3616-keV;}\\
\parbox[b][0.3cm]{17.7cm}{deduced \ensuremath{\gamma}-branching, J, \ensuremath{\pi}, and \ensuremath{\delta} (mixing ratios); confirmed the results of (\href{https://www.nndc.bnl.gov/nsr/nsrlink.jsp?1968Gi09,B}{1968Gi09}, \href{https://www.nndc.bnl.gov/nsr/nsrlink.jsp?1969Ro08,B}{1969Ro08}) regarding to the existence of the}\\
\parbox[b][0.3cm]{17.7cm}{level at 3576 keV and its 1689-keV \ensuremath{\gamma}-ray from the decay to the 1887-keV state.}\\
\parbox[b][0.3cm]{17.7cm}{\addtolength{\parindent}{-0.2in}\href{https://www.nndc.bnl.gov/nsr/nsrlink.jsp?1969Be31,B}{1969Be31}: \ensuremath{^{\textnormal{24}}}Mg(\ensuremath{^{\textnormal{3}}}He,n\ensuremath{\gamma}) E=5.5, 7.8, and 10 MeV. The first excited state of \ensuremath{^{\textnormal{18}}}Ne was also populated due to oxygen}\\
\parbox[b][0.3cm]{17.7cm}{contamination of the enriched \ensuremath{^{\textnormal{24}}}Mg self-supporting target. The \ensuremath{\gamma}-ray from the de-excitation of the 1887-keV state in \ensuremath{^{\textnormal{18}}}Ne to the}\\
\parbox[b][0.3cm]{17.7cm}{ground state was observed.}\\
\parbox[b][0.3cm]{17.7cm}{\addtolength{\parindent}{-0.2in}\href{https://www.nndc.bnl.gov/nsr/nsrlink.jsp?1970Sh04,B}{1970Sh04}: \ensuremath{^{\textnormal{16}}}O(\ensuremath{^{\textnormal{3}}}He,n\ensuremath{\gamma}) E=9 and 10 MeV; measured \ensuremath{\sigma}(E;E\ensuremath{_{\textnormal{n}}}), n\ensuremath{\gamma} coincidences, n\ensuremath{\gamma}(\ensuremath{\theta}) using a \ensuremath{\gamma}-ray time-of-flight technique}\\
\parbox[b][0.3cm]{17.7cm}{employing a Ge(Li) (and a NaI(Tl)) counter at \ensuremath{\theta}\ensuremath{_{\textnormal{lab}}}=90\ensuremath{^\circ} and a NE-102 plastic scintillator at \ensuremath{\theta}\ensuremath{_{\textnormal{lab}}}=0\ensuremath{^\circ} and 40\ensuremath{^\circ}, measured n\ensuremath{\gamma}}\\
\parbox[b][0.3cm]{17.7cm}{correlations from the \ensuremath{^{\textnormal{18}}}Ne*(3626 keV, 2\ensuremath{^{\textnormal{+}}_{\textnormal{2}}}) and \ensuremath{^{\textnormal{18}}}Ne*(3383 keV, 4\ensuremath{^{\textnormal{+}}_{\textnormal{1}}}) states using a WO\ensuremath{_{\textnormal{3}}} target and the Ge(Li) at \ensuremath{\theta}\ensuremath{_{\textnormal{lab}}}=55\ensuremath{^\circ},}\\
\parbox[b][0.3cm]{17.7cm}{90\ensuremath{^\circ}, 105\ensuremath{^\circ}, 125\ensuremath{^\circ}, 135\ensuremath{^\circ} and 145\ensuremath{^\circ}; combined this information with lifetime measurements of (\href{https://www.nndc.bnl.gov/nsr/nsrlink.jsp?1968Gi09,B}{1968Gi09}, \href{https://www.nndc.bnl.gov/nsr/nsrlink.jsp?1969Ro08,B}{1969Ro08}) to deduce the}\\
\parbox[b][0.3cm]{17.7cm}{isoscalar and isovector matrix elements involved in the decay of 2\ensuremath{^{\textnormal{+}}_{\textnormal{2}}} levels of the A=18 triad; measured \ensuremath{^{\textnormal{18}}}Ne \ensuremath{\gamma}-rays at 1890 keV}\\
\parbox[b][0.3cm]{17.7cm}{\textit{2}, 1466 keV \textit{2}, 1493 keV \textit{2}, 1689 keV \textit{2}, 1736 keV \textit{2}, and 3630 keV \textit{2}; deduced \ensuremath{^{\textnormal{18}}}Ne levels at 1890 keV \textit{2}, 3383 keV \textit{4}, 3576}\\
\parbox[b][0.3cm]{17.7cm}{keV \textit{4}, and 3623 keV \textit{4}; deduced J, \ensuremath{\pi}, \ensuremath{\gamma}-branching ratios, and \ensuremath{\delta} (mixing ratios) for the observed transitions.}\\
\parbox[b][0.3cm]{17.7cm}{\addtolength{\parindent}{-0.2in}\href{https://www.nndc.bnl.gov/nsr/nsrlink.jsp?1971Ro18,B}{1971Ro18}: \ensuremath{^{\textnormal{16}}}O(\ensuremath{^{\textnormal{3}}}He,n\ensuremath{\gamma}) E=10.3 MeV; measured n\ensuremath{\gamma} coincidences for the triplet states at E\ensuremath{_{\textnormal{x}}}\ensuremath{\sim}3.5 MeV using the Oxford \ensuremath{\gamma}-neutron}\\
\parbox[b][0.3cm]{17.7cm}{time-of-flight spectrometer consisting of a large Ge(Li) and a large scintillator; measured \ensuremath{\sigma}(E\ensuremath{_{\textnormal{n}}},E\ensuremath{_{\ensuremath{\gamma}}}); found no evidence for the 1466}\\
\parbox[b][0.3cm]{17.7cm}{keV \ensuremath{\gamma}-ray observed in (\href{https://www.nndc.bnl.gov/nsr/nsrlink.jsp?1970Sh04,B}{1970Sh04}) and attributed the origin of this \ensuremath{\gamma}-ray to the Compton scattering from the 1736-keV \ensuremath{\gamma}-ray;}\\
\parbox[b][0.3cm]{17.7cm}{deduced the \ensuremath{\gamma}-decay of the triplet excited states of \ensuremath{^{\textnormal{18}}}Ne at E\ensuremath{_{\textnormal{x}}}\ensuremath{\sim}3.5 MeV; confirmed that the 3576-keV state is the 0\ensuremath{^{\textnormal{+}}_{\textnormal{2}}} state;}\\
\parbox[b][0.3cm]{17.7cm}{deduced the branching ratio and mixing ratio of the 3616\ensuremath{\rightarrow}1887 keV transition but these are not published here.}\\
\parbox[b][0.3cm]{17.7cm}{\addtolength{\parindent}{-0.2in}\href{https://www.nndc.bnl.gov/nsr/nsrlink.jsp?1972Gi01,B}{1972Gi01}: \ensuremath{^{\textnormal{16}}}O(\ensuremath{^{\textnormal{3}}}He,n\ensuremath{\gamma}) E=10.3-10.8 MeV; measured n\ensuremath{\gamma} coincidences using a Ge(Li) detector with 2.3 keV resolution at 1.33}\\
\parbox[b][0.3cm]{17.7cm}{MeV and by neutron time-of-flight method using a NE-213 liquid scintillator at \ensuremath{\theta}\ensuremath{_{\textnormal{lab}}}=0\ensuremath{^\circ}; The Ge(Li) detector was placed at}\\
\parbox[b][0.3cm]{17.7cm}{\ensuremath{\theta}\ensuremath{_{\textnormal{lab}}}=90\ensuremath{^\circ}, 115\ensuremath{^\circ}, 125\ensuremath{^\circ}, 135\ensuremath{^\circ}, 140\ensuremath{^\circ}, and 145\ensuremath{^\circ} to measure the n\ensuremath{\gamma} angular correlations and the \ensuremath{\gamma}-ray decay scheme; measured}\\
\parbox[b][0.3cm]{17.7cm}{lifetimes of the 3376- and 3576-keV states via the Doppler-shift recoil distance method with the Ge(Li) detector at \ensuremath{\theta}\ensuremath{_{\textnormal{lab}}}=0\ensuremath{^\circ} and the}\\
\parbox[b][0.3cm]{17.7cm}{scintillator at \ensuremath{\theta}\ensuremath{_{\textnormal{lab}}}=30\ensuremath{^\circ}; measured \ensuremath{\sigma}(\ensuremath{\theta}) for the 1887-, 1729-, and 3616-keV \ensuremath{\gamma}-rays; deduced \ensuremath{^{\textnormal{18}}}Ne levels, branching ratios, and}\\
\parbox[b][0.3cm]{17.7cm}{\ensuremath{\gamma}-mixing ratios for the \ensuremath{^{\textnormal{18}}}Ne*(1887, 3376, 3576, 3616 keV) states.}\\
\parbox[b][0.3cm]{17.7cm}{\addtolength{\parindent}{-0.2in}\href{https://www.nndc.bnl.gov/nsr/nsrlink.jsp?1974Mc17,B}{1974Mc17}: \ensuremath{^{\textnormal{16}}}O(\ensuremath{^{\textnormal{3}}}He,n\ensuremath{\gamma}) E=7.5 and 9 MeV; measured E\ensuremath{_{\ensuremath{\gamma}}}=1490-, 1690-, 1730-, 1890-, and 3620-keV and I\ensuremath{_{\ensuremath{\gamma}}} a Ge(Li) detector}\\
\parbox[b][0.3cm]{17.7cm}{placed at \ensuremath{\theta}\ensuremath{_{\textnormal{lab}}}=0\ensuremath{^\circ}, 90\ensuremath{^\circ}, and 135\ensuremath{^\circ}; measured n\ensuremath{\gamma} and \ensuremath{\gamma}\ensuremath{\gamma} coincidences using a NE-213 liquid scintillator at \ensuremath{\theta}\ensuremath{_{\textnormal{lab}}}=0\ensuremath{^\circ} and 2 Ge(Li)}\\
\parbox[b][0.3cm]{17.7cm}{detectors at \ensuremath{\theta}\ensuremath{_{\textnormal{lab}}}=45\ensuremath{^\circ} and 135\ensuremath{^\circ}; Ge(Li) detectors with resolution of 3.5 keV at 1.88 MeV; measured the lifetime of the 1.89-MeV}\\
\parbox[b][0.3cm]{17.7cm}{state of \ensuremath{^{\textnormal{18}}}Ne using the Doppler shift attenuation method using the stopping powers of (\href{https://www.nndc.bnl.gov/nsr/nsrlink.jsp?1963Li17,B}{1963Li17}); transition rates in the \ensuremath{^{\textnormal{18}}}Ne and}\\
\parbox[b][0.3cm]{17.7cm}{\ensuremath{^{\textnormal{18}}}O mirror nuclei are compared with theoretical predictions.}\\
\parbox[b][0.3cm]{17.7cm}{\addtolength{\parindent}{-0.2in}\href{https://www.nndc.bnl.gov/nsr/nsrlink.jsp?1976Mc02,B}{1976Mc02}: \ensuremath{^{\textnormal{3}}}He(\ensuremath{^{\textnormal{16}}}O,n\ensuremath{\gamma}) E=38 MeV; measured n\ensuremath{\gamma} coincidences using a Ge(Li) and a NE-213 liquid scintillator both at \ensuremath{\theta}\ensuremath{_{\textnormal{lab}}}=0\ensuremath{^\circ};}\\
\parbox[b][0.3cm]{17.7cm}{remeasured the lifetime of the \ensuremath{^{\textnormal{18}}}Ne*(1890 keV) state using Doppler broadened line shape analysis; used inverse kinematics to}\\
\parbox[b][0.3cm]{17.7cm}{produce high velocity (v/c=6\%) recoils, for which accurate stopping powers had been measured by (\href{https://www.nndc.bnl.gov/nsr/nsrlink.jsp?1976Fo20,B}{1976Fo20}). Deduced level}\\
\parbox[b][0.3cm]{17.7cm}{energy, lifetime and B(E2) for the 1890 keV state. The lifetime was determined using a method which is insensitive to the}\\
\parbox[b][0.3cm]{17.7cm}{systematic uncertainties arising from stopping powers.}\\
\parbox[b][0.3cm]{17.7cm}{\addtolength{\parindent}{-0.2in}\href{https://www.nndc.bnl.gov/nsr/nsrlink.jsp?2003Ri08,B}{2003Ri08}: \ensuremath{^{\textnormal{3}}}He(\ensuremath{^{\textnormal{16}}}O,n\ensuremath{\gamma}) E=38 MeV. The goal was to resolve the discrepancy between B(E2:0\ensuremath{_{\textnormal{g.s.}}^{\textnormal{+}}}\ensuremath{\rightarrow}2\ensuremath{^{\textnormal{+}}_{\textnormal{1}}}) deduced from}\\
\parbox[b][0.3cm]{17.7cm}{(\href{https://www.nndc.bnl.gov/nsr/nsrlink.jsp?2000Ri15,B}{2000Ri15}: see section \ensuremath{^{\textnormal{197}}}Au(\ensuremath{^{\textnormal{18}}}Ne,\ensuremath{^{\textnormal{18}}}Ne\ensuremath{'}):COULEX) and the B(E2:2\ensuremath{^{\textnormal{+}}_{\textnormal{1}}}\ensuremath{\rightarrow}0\ensuremath{^{\textnormal{+}}_{\textnormal{g.s.}}}) deduced by (\href{https://www.nndc.bnl.gov/nsr/nsrlink.jsp?1976Mc02,B}{1976Mc02}). Measured E\ensuremath{_{\ensuremath{\gamma}}}, I\ensuremath{_{\ensuremath{\gamma}}}, n\ensuremath{\gamma}}\\
\parbox[b][0.3cm]{17.7cm}{coincidences for the \ensuremath{^{\textnormal{18}}}Ne*(2\ensuremath{^{\textnormal{+}}_{\textnormal{1}}}) state using 2 large clover Ge detectors with Compton suppression both at \ensuremath{\theta}\ensuremath{_{\textnormal{lab}}}=135\ensuremath{^\circ} and a}\\
\parbox[b][0.3cm]{17.7cm}{NE-213 liquid scintillator at \ensuremath{\theta}\ensuremath{_{\textnormal{lab}}}=0\ensuremath{^\circ}; measured the lifetime of the \ensuremath{^{\textnormal{18}}}Ne(2\ensuremath{^{\textnormal{+}}_{\textnormal{1}}}) state using Doppler shift attenuation method;}\\
\parbox[b][0.3cm]{17.7cm}{deduced B(E2: 0\ensuremath{^{\textnormal{+}}_{\textnormal{g.s.}}}\ensuremath{\rightarrow}2\ensuremath{^{\textnormal{+}}_{\textnormal{1}}}); performed a semi-microscopic reanalysis of the intermediate energy \ensuremath{^{\textnormal{18}}}Ne+\ensuremath{^{\textnormal{197}}}Au scattering data of}\\
\parbox[b][0.3cm]{17.7cm}{(\href{https://www.nndc.bnl.gov/nsr/nsrlink.jsp?2000Ri15,B}{2000Ri15}) to resolve the aforementioned discrepancy; discussed comparison with previous results.}\\
\clearpage
\vspace{0.3cm}
{\bf \small \underline{\ensuremath{^{\textnormal{16}}}O(\ensuremath{^{\textnormal{3}}}He,n\ensuremath{\gamma})\hspace{0.2in}\href{https://www.nndc.bnl.gov/nsr/nsrlink.jsp?1968Gi09,B}{1968Gi09},\href{https://www.nndc.bnl.gov/nsr/nsrlink.jsp?2003Ta13,B}{2003Ta13} (continued)}}\\
\vspace{0.3cm}
\parbox[b][0.3cm]{17.7cm}{\addtolength{\parindent}{-0.2in}\href{https://www.nndc.bnl.gov/nsr/nsrlink.jsp?2003Ta13,B}{2003Ta13}: \ensuremath{^{\textnormal{16}}}O(\ensuremath{^{\textnormal{3}}}He,n\ensuremath{\gamma}) E=3.7-36 MeV; measured E\ensuremath{_{\ensuremath{\gamma}}}=1489 keV and 1887 keV; measured I\ensuremath{_{\ensuremath{\gamma}}}, \ensuremath{\sigma}(\ensuremath{\theta}) using 4 large Compton}\\
\parbox[b][0.3cm]{17.7cm}{suppressed HPGe detectors at \ensuremath{\theta}\ensuremath{_{\textnormal{lab}}}=90\ensuremath{^\circ}, 112.5\ensuremath{^\circ}, 135\ensuremath{^\circ}, and 157.5\ensuremath{^\circ}; deduced \ensuremath{\gamma}-ray production \ensuremath{\sigma}(E) of the 1887-keV \ensuremath{\gamma}-ray from the}\\
\parbox[b][0.3cm]{17.7cm}{de-excitation of \ensuremath{^{\textnormal{18}}}Ne*(1.89 MeV) state. These cross sections are given in numerical format as a function of laboratory energy. The}\\
\parbox[b][0.3cm]{17.7cm}{astrophysical implications for solar flares and \ensuremath{\gamma}-ray astronomy are discussed.}\\
\vspace{12pt}
\underline{$^{18}$Ne Levels}\\
\vspace{0.34cm}
\parbox[b][0.3cm]{17.7cm}{\addtolength{\parindent}{-0.254cm}Cross sections measured by (\href{https://www.nndc.bnl.gov/nsr/nsrlink.jsp?2003Ta13,B}{2003Ta13}) are uncertain by 13-17\%.}\\
\parbox[b][0.3cm]{17.7cm}{\addtolength{\parindent}{-0.254cm}The \ensuremath{\gamma}-ray energies (and thus the deduced excitation energies) from (\href{https://www.nndc.bnl.gov/nsr/nsrlink.jsp?1970Sh04,B}{1970Sh04}) are consistently higher than all other \ensuremath{\gamma}-ray}\\
\parbox[b][0.3cm]{17.7cm}{measurements except for the E\ensuremath{_{\textnormal{x}}}=3576 keV. Thus, these results are not used due to potentially unknown systematic uncertainties.}\\
\vspace{0.34cm}
\begin{longtable}{cccccc@{\extracolsep{\fill}}c}
\multicolumn{2}{c}{E(level)$^{}$}&J$^{\pi}$$^{}$&\multicolumn{2}{c}{T$_{1/2}$$^{}$}&Comments&\\[-.2cm]
\multicolumn{2}{c}{\hrulefill}&\hrulefill&\multicolumn{2}{c}{\hrulefill}&\hrulefill&
\endfirsthead
\multicolumn{1}{r@{}}{0}&\multicolumn{1}{@{}l}{}&\multicolumn{1}{l}{0\ensuremath{^{+}}}&&&\parbox[t][0.3cm]{12.962581cm}{\raggedright E(level): The ground state was populated indirectly via the \ensuremath{\gamma}-decay of the first excited state in\vspace{0.1cm}}&\\
&&&&&\parbox[t][0.3cm]{12.962581cm}{\raggedright {\ }{\ }{\ }(\href{https://www.nndc.bnl.gov/nsr/nsrlink.jsp?1968Gi09,B}{1968Gi09}, \href{https://www.nndc.bnl.gov/nsr/nsrlink.jsp?1969Be31,B}{1969Be31}, \href{https://www.nndc.bnl.gov/nsr/nsrlink.jsp?1969Ro08,B}{1969Ro08}, \href{https://www.nndc.bnl.gov/nsr/nsrlink.jsp?1969Ro22,B}{1969Ro22}, \href{https://www.nndc.bnl.gov/nsr/nsrlink.jsp?1970Sh04,B}{1970Sh04}, \href{https://www.nndc.bnl.gov/nsr/nsrlink.jsp?1971Ro18,B}{1971Ro18}, \href{https://www.nndc.bnl.gov/nsr/nsrlink.jsp?1971Gi02,B}{1971Gi02}, \href{https://www.nndc.bnl.gov/nsr/nsrlink.jsp?1974Mc17,B}{1974Mc17},\vspace{0.1cm}}&\\
&&&&&\parbox[t][0.3cm]{12.962581cm}{\raggedright {\ }{\ }{\ }\href{https://www.nndc.bnl.gov/nsr/nsrlink.jsp?1976Mc02,B}{1976Mc02}, \href{https://www.nndc.bnl.gov/nsr/nsrlink.jsp?2003Ri08,B}{2003Ri08}, \href{https://www.nndc.bnl.gov/nsr/nsrlink.jsp?2003Ta13,B}{2003Ta13}).\vspace{0.1cm}}&\\
&&&&&\parbox[t][0.3cm]{12.962581cm}{\raggedright J\ensuremath{^{\pi}}: From the \ensuremath{^{\textnormal{18}}}Ne Adopted Levels.\vspace{0.1cm}}&\\
&&&&&\parbox[t][0.3cm]{12.962581cm}{\raggedright T\ensuremath{_{\textnormal{z}}}={\textminus}1 (\href{https://www.nndc.bnl.gov/nsr/nsrlink.jsp?1968Gi09,B}{1968Gi09}).\vspace{0.1cm}}&\\
\multicolumn{1}{r@{}}{1887}&\multicolumn{1}{@{.}l}{4\ensuremath{^{{\hyperlink{NE16LEVEL0}{a}}}} {\it 2}}&\multicolumn{1}{l}{2\ensuremath{^{+}}\ensuremath{^{{\hyperlink{NE16LEVEL2}{c}}}}}&\multicolumn{1}{r@{}}{0}&\multicolumn{1}{@{.}l}{47 ps {\it 4}}&\parbox[t][0.3cm]{12.962581cm}{\raggedright T=1 (\href{https://www.nndc.bnl.gov/nsr/nsrlink.jsp?1976Mc02,B}{1976Mc02})\vspace{0.1cm}}&\\
&&&&&\parbox[t][0.3cm]{12.962581cm}{\raggedright E(level): From recoil correction of E\ensuremath{_{\ensuremath{\gamma}}}=1887.3 keV \textit{2} (\href{https://www.nndc.bnl.gov/nsr/nsrlink.jsp?1968Gi09,B}{1968Gi09}, \href{https://www.nndc.bnl.gov/nsr/nsrlink.jsp?1969Ro08,B}{1969Ro08}). See also\vspace{0.1cm}}&\\
&&&&&\parbox[t][0.3cm]{12.962581cm}{\raggedright {\ }{\ }{\ }excitation energies deduced that appear to neglect the \ensuremath{^{\textnormal{18}}}Ne recoil corrections: E\ensuremath{_{\textnormal{x}}}=1887.3\vspace{0.1cm}}&\\
&&&&&\parbox[t][0.3cm]{12.962581cm}{\raggedright {\ }{\ }{\ }keV \textit{2} (\href{https://www.nndc.bnl.gov/nsr/nsrlink.jsp?1968Gi09,B}{1968Gi09}, \href{https://www.nndc.bnl.gov/nsr/nsrlink.jsp?1969Ro08,B}{1969Ro08}); E\ensuremath{_{\textnormal{x}}}=1887 keV (\href{https://www.nndc.bnl.gov/nsr/nsrlink.jsp?1969Ro22,B}{1969Ro22}, \href{https://www.nndc.bnl.gov/nsr/nsrlink.jsp?1971Ro18,B}{1971Ro18}, \href{https://www.nndc.bnl.gov/nsr/nsrlink.jsp?1972Gi01,B}{1972Gi01}, \href{https://www.nndc.bnl.gov/nsr/nsrlink.jsp?2003Ri08,B}{2003Ri08},\vspace{0.1cm}}&\\
&&&&&\parbox[t][0.3cm]{12.962581cm}{\raggedright {\ }{\ }{\ }\href{https://www.nndc.bnl.gov/nsr/nsrlink.jsp?2003Ta13,B}{2003Ta13}); E\ensuremath{_{\textnormal{x}}}=1890 keV \textit{2} (\href{https://www.nndc.bnl.gov/nsr/nsrlink.jsp?1970Sh04,B}{1970Sh04}); and E\ensuremath{_{\textnormal{x}}}=1890 keV (\href{https://www.nndc.bnl.gov/nsr/nsrlink.jsp?1974Mc17,B}{1974Mc17}, \href{https://www.nndc.bnl.gov/nsr/nsrlink.jsp?1976Mc02,B}{1976Mc02}).\vspace{0.1cm}}&\\
&&&&&\parbox[t][0.3cm]{12.962581cm}{\raggedright T\ensuremath{_{1/2}}: From \ensuremath{\tau}=0.68 ps \textit{6}, which is the weighted average (with external errors) of (1) \ensuremath{\tau}=0.49 ps\vspace{0.1cm}}&\\
&&&&&\parbox[t][0.3cm]{12.962581cm}{\raggedright {\ }{\ }{\ }\textit{+17{\textminus}9} (\href{https://www.nndc.bnl.gov/nsr/nsrlink.jsp?1968Gi09,B}{1968Gi09}): deduced \ensuremath{\tau}=0.45 ps \textit{+22{\textminus}15} and \ensuremath{\tau}=0.37 ps {14} using thick SrO and CdO\vspace{0.1cm}}&\\
&&&&&\parbox[t][0.3cm]{12.962581cm}{\raggedright {\ }{\ }{\ }targets, respectively, and recommended \ensuremath{\tau}=0.49 ps \textit{+17{\textminus}9} (see the footnote of Table 1 in\vspace{0.1cm}}&\\
&&&&&\parbox[t][0.3cm]{12.962581cm}{\raggedright {\ }{\ }{\ }(\href{https://www.nndc.bnl.gov/nsr/nsrlink.jsp?1968Gi09,B}{1968Gi09})). A year later, (\href{https://www.nndc.bnl.gov/nsr/nsrlink.jsp?1969Ro08,B}{1969Ro08}) obtained \ensuremath{\tau}=0.44 ps \textit{+24{\textminus}16} and \ensuremath{\tau}=0.54 ps \textit{+24{\textminus}10}\vspace{0.1cm}}&\\
&&&&&\parbox[t][0.3cm]{12.962581cm}{\raggedright {\ }{\ }{\ }using a thin WO\ensuremath{_{\textnormal{3}}} target on gold and lead backings, respectively, and again recommended\vspace{0.1cm}}&\\
&&&&&\parbox[t][0.3cm]{12.962581cm}{\raggedright {\ }{\ }{\ }\ensuremath{\tau}=0.49 ps \textit{+17{\textminus}9}; (2) \ensuremath{\tau}=0.67 ps \textit{6} (\href{https://www.nndc.bnl.gov/nsr/nsrlink.jsp?1976Mc02,B}{1976Mc02}); and (3) \ensuremath{\tau}=0.77 ps \textit{+9{\textminus}7} (\href{https://www.nndc.bnl.gov/nsr/nsrlink.jsp?2003Ri08,B}{2003Ri08}).\vspace{0.1cm}}&\\
&&&&&\parbox[t][0.3cm]{12.962581cm}{\raggedright T\ensuremath{_{1/2}}: See also \ensuremath{\tau}=0.63 ps \textit{13} (sys.) (\href{https://www.nndc.bnl.gov/nsr/nsrlink.jsp?1974Mc17,B}{1974Mc17}): they measured the lifetime using Doppler shift\vspace{0.1cm}}&\\
&&&&&\parbox[t][0.3cm]{12.962581cm}{\raggedright {\ }{\ }{\ }attenuation method. They acknowledged that for low recoil velocities (v/c\ensuremath{\leq}1\%, which was\vspace{0.1cm}}&\\
&&&&&\parbox[t][0.3cm]{12.962581cm}{\raggedright {\ }{\ }{\ }the case), the lifetime was uncertain to the extent of about \ensuremath{\pm}20\%, and this uncertainty was\vspace{0.1cm}}&\\
&&&&&\parbox[t][0.3cm]{12.962581cm}{\raggedright {\ }{\ }{\ }systematic arising from lack of experimental stopping powers and the inadequacies of the\vspace{0.1cm}}&\\
&&&&&\parbox[t][0.3cm]{12.962581cm}{\raggedright {\ }{\ }{\ }theoretical stopping powers deduced by (\href{https://www.nndc.bnl.gov/nsr/nsrlink.jsp?1963Li17,B}{1963Li17}). (\href{https://www.nndc.bnl.gov/nsr/nsrlink.jsp?1976Mc02,B}{1976Mc02}) re-measured this lifetime\vspace{0.1cm}}&\\
&&&&&\parbox[t][0.3cm]{12.962581cm}{\raggedright {\ }{\ }{\ }using inverse kinematics to increase the velocity of the recoils, for which more accurate\vspace{0.1cm}}&\\
&&&&&\parbox[t][0.3cm]{12.962581cm}{\raggedright {\ }{\ }{\ }stopping powers were measured. This measurement used Doppler broadened line shape\vspace{0.1cm}}&\\
&&&&&\parbox[t][0.3cm]{12.962581cm}{\raggedright {\ }{\ }{\ }analysis, which is insensitive to the systematic uncertainties arising from stopping powers.\vspace{0.1cm}}&\\
&&&&&\parbox[t][0.3cm]{12.962581cm}{\raggedright {\ }{\ }{\ }Consequently, (\href{https://www.nndc.bnl.gov/nsr/nsrlink.jsp?1976Mc02,B}{1976Mc02}) were able to remeasure the lifetime of this state more accurately.\vspace{0.1cm}}&\\
&&&&&\parbox[t][0.3cm]{12.962581cm}{\raggedright {\ }{\ }{\ }We therefore excluded \ensuremath{\tau}=0.63 ps \textit{13} (sys.) (\href{https://www.nndc.bnl.gov/nsr/nsrlink.jsp?1974Mc17,B}{1974Mc17}).\vspace{0.1cm}}&\\
&&&&&\parbox[t][0.3cm]{12.962581cm}{\raggedright J\ensuremath{^{\pi}}: From (\href{https://www.nndc.bnl.gov/nsr/nsrlink.jsp?1969Ro22,B}{1969Ro22}: J=2 uniquely determined from the coefficients of the Legendre\vspace{0.1cm}}&\\
&&&&&\parbox[t][0.3cm]{12.962581cm}{\raggedright {\ }{\ }{\ }polynomial fits to the \ensuremath{\gamma}-ray angular correlations data). The positive parity is preferred by\vspace{0.1cm}}&\\
&&&&&\parbox[t][0.3cm]{12.962581cm}{\raggedright {\ }{\ }{\ }(\href{https://www.nndc.bnl.gov/nsr/nsrlink.jsp?1969Ro22,B}{1969Ro22}) from the lifetime measurement of (\href{https://www.nndc.bnl.gov/nsr/nsrlink.jsp?1968Gi09,B}{1968Gi09}). This assignment is also supported\vspace{0.1cm}}&\\
&&&&&\parbox[t][0.3cm]{12.962581cm}{\raggedright {\ }{\ }{\ }by the mirror levels analysis of (\href{https://www.nndc.bnl.gov/nsr/nsrlink.jsp?1970Sh04,B}{1970Sh04}). (\href{https://www.nndc.bnl.gov/nsr/nsrlink.jsp?1971Ro18,B}{1971Ro18}) also recommended the J\ensuremath{^{\ensuremath{\pi}}}=2\ensuremath{^{\textnormal{+}}}\vspace{0.1cm}}&\\
&&&&&\parbox[t][0.3cm]{12.962581cm}{\raggedright {\ }{\ }{\ }assignment for this state.\vspace{0.1cm}}&\\
\multicolumn{1}{r@{}}{3376}&\multicolumn{1}{@{.}l}{4\ensuremath{^{{\hyperlink{NE16LEVEL0}{a}}}} {\it 4}}&\multicolumn{1}{l}{4\ensuremath{^{+}}\ensuremath{^{{\hyperlink{NE16LEVEL2}{c}}}}}&\multicolumn{1}{r@{}}{3}&\multicolumn{1}{@{.}l}{05 ps {\it 42}}&\parbox[t][0.3cm]{12.962581cm}{\raggedright E(level): From recoil correction of E\ensuremath{_{\ensuremath{\gamma}}}=1488.9 keV \textit{3} (\href{https://www.nndc.bnl.gov/nsr/nsrlink.jsp?1968Gi09,B}{1968Gi09}, \href{https://www.nndc.bnl.gov/nsr/nsrlink.jsp?1969Ro08,B}{1969Ro08}). See also\vspace{0.1cm}}&\\
&&&&&\parbox[t][0.3cm]{12.962581cm}{\raggedright {\ }{\ }{\ }excitation energies deduced that appear to neglect the \ensuremath{^{\textnormal{18}}}Ne recoil corrections: 3376.2 keV \textit{4}\vspace{0.1cm}}&\\
&&&&&\parbox[t][0.3cm]{12.962581cm}{\raggedright {\ }{\ }{\ }(\href{https://www.nndc.bnl.gov/nsr/nsrlink.jsp?1968Gi09,B}{1968Gi09}, \href{https://www.nndc.bnl.gov/nsr/nsrlink.jsp?1969Ro08,B}{1969Ro08}); 3376 keV (\href{https://www.nndc.bnl.gov/nsr/nsrlink.jsp?1969Ro22,B}{1969Ro22}, \href{https://www.nndc.bnl.gov/nsr/nsrlink.jsp?1971Ro18,B}{1971Ro18}, \href{https://www.nndc.bnl.gov/nsr/nsrlink.jsp?1972Gi01,B}{1972Gi01}, \href{https://www.nndc.bnl.gov/nsr/nsrlink.jsp?2003Ta13,B}{2003Ta13}); and 3383\vspace{0.1cm}}&\\
&&&&&\parbox[t][0.3cm]{12.962581cm}{\raggedright {\ }{\ }{\ }keV \textit{4} (\href{https://www.nndc.bnl.gov/nsr/nsrlink.jsp?1970Sh04,B}{1970Sh04}).\vspace{0.1cm}}&\\
&&&&&\parbox[t][0.3cm]{12.962581cm}{\raggedright E(level): (\href{https://www.nndc.bnl.gov/nsr/nsrlink.jsp?1968Gi09,B}{1968Gi09}) did not find any evidence to support the suggestion of (Shapiro et al.,\vspace{0.1cm}}&\\
&&&&&\parbox[t][0.3cm]{12.962581cm}{\raggedright {\ }{\ }{\ }Bull. Amer. Phys. Soc. 13 (1968) 698) concerning the doublet nature of the 3376 keV state.\vspace{0.1cm}}&\\
&&&&&\parbox[t][0.3cm]{12.962581cm}{\raggedright T\ensuremath{_{1/2}}: From \ensuremath{\tau}=4.4 ps \textit{6} (\href{https://www.nndc.bnl.gov/nsr/nsrlink.jsp?1972Gi01,B}{1972Gi01}). Note that this lifetime is consistent with \ensuremath{\tau}=4.3 ps\vspace{0.1cm}}&\\
&&&&&\parbox[t][0.3cm]{12.962581cm}{\raggedright {\ }{\ }{\ }(\href{https://www.nndc.bnl.gov/nsr/nsrlink.jsp?1966Be29,B}{1966Be29}: calculated); 4.1 ps; and 4.7 ps (private communication of T. Engeland and P. J.\vspace{0.1cm}}&\\
&&&&&\parbox[t][0.3cm]{12.962581cm}{\raggedright {\ }{\ }{\ }Ellis with the authors of (\href{https://www.nndc.bnl.gov/nsr/nsrlink.jsp?1972Gi01,B}{1972Gi01}): lifetimes were calculated assuming the two-particle\vspace{0.1cm}}&\\
&&&&&\parbox[t][0.3cm]{12.962581cm}{\raggedright {\ }{\ }{\ }configuration model).\vspace{0.1cm}}&\\
&&&&&\parbox[t][0.3cm]{12.962581cm}{\raggedright T\ensuremath{_{1/2}}: See also \ensuremath{\tau}=1.9 ps \textit{+7{\textminus}4} (\href{https://www.nndc.bnl.gov/nsr/nsrlink.jsp?1969Ro08,B}{1969Ro08}) and \ensuremath{\tau}=1.9 ps \textit{+10{\textminus}4} (\href{https://www.nndc.bnl.gov/nsr/nsrlink.jsp?1968Gi09,B}{1968Gi09}: the preliminary\vspace{0.1cm}}&\\
&&&&&\parbox[t][0.3cm]{12.962581cm}{\raggedright {\ }{\ }{\ }result published by the same authors before (\href{https://www.nndc.bnl.gov/nsr/nsrlink.jsp?1969Ro08,B}{1969Ro08})). Both of these values are measured\vspace{0.1cm}}&\\
\end{longtable}
\begin{textblock}{29}(0,27.3)
Continued on next page (footnotes at end of table)
\end{textblock}
\clearpage
\begin{longtable}{cccccc@{\extracolsep{\fill}}c}
\\[-.4cm]
\multicolumn{7}{c}{{\bf \small \underline{\ensuremath{^{\textnormal{16}}}O(\ensuremath{^{\textnormal{3}}}He,n\ensuremath{\gamma})\hspace{0.2in}\href{https://www.nndc.bnl.gov/nsr/nsrlink.jsp?1968Gi09,B}{1968Gi09},\href{https://www.nndc.bnl.gov/nsr/nsrlink.jsp?2003Ta13,B}{2003Ta13} (continued)}}}\\
\multicolumn{7}{c}{~}\\
\multicolumn{7}{c}{\underline{\ensuremath{^{18}}Ne Levels (continued)}}\\
\multicolumn{7}{c}{~}\\
\multicolumn{2}{c}{E(level)$^{}$}&J$^{\pi}$$^{}$&\multicolumn{2}{c}{T$_{1/2}$$^{}$}&Comments&\\[-.2cm]
\multicolumn{2}{c}{\hrulefill}&\hrulefill&\multicolumn{2}{c}{\hrulefill}&\hrulefill&
\endhead
&&&&&\parbox[t][0.3cm]{12.196461cm}{\raggedright {\ }{\ }{\ }via Doppler shift attenuation method. (\href{https://www.nndc.bnl.gov/nsr/nsrlink.jsp?1972Gi01,B}{1972Gi01}) raised concern about the validity of\vspace{0.1cm}}&\\
&&&&&\parbox[t][0.3cm]{12.196461cm}{\raggedright {\ }{\ }{\ }the (\href{https://www.nndc.bnl.gov/nsr/nsrlink.jsp?1963Li17,B}{1963Li17}) stopping theory for light nuclei recoiling into high-Z materials with low\vspace{0.1cm}}&\\
&&&&&\parbox[t][0.3cm]{12.196461cm}{\raggedright {\ }{\ }{\ }velocities (used in deducing the aforementioned lifetimes), which cast doubt on the\vspace{0.1cm}}&\\
&&&&&\parbox[t][0.3cm]{12.196461cm}{\raggedright {\ }{\ }{\ }results of (\href{https://www.nndc.bnl.gov/nsr/nsrlink.jsp?1968Gi09,B}{1968Gi09}, \href{https://www.nndc.bnl.gov/nsr/nsrlink.jsp?1969Ro08,B}{1969Ro08}) and may explain the inconsistencies between those\vspace{0.1cm}}&\\
&&&&&\parbox[t][0.3cm]{12.196461cm}{\raggedright {\ }{\ }{\ }values and the lifetime deduced by (\href{https://www.nndc.bnl.gov/nsr/nsrlink.jsp?1972Gi01,B}{1972Gi01}).\vspace{0.1cm}}&\\
&&&&&\parbox[t][0.3cm]{12.196461cm}{\raggedright \ensuremath{\Gamma}=0.15 meV: from the lifetime deduced by (\href{https://www.nndc.bnl.gov/nsr/nsrlink.jsp?1972Gi01,B}{1972Gi01}: see Table 1). See also similar\vspace{0.1cm}}&\\
&&&&&\parbox[t][0.3cm]{12.196461cm}{\raggedright {\ }{\ }{\ }calculated values of: \ensuremath{\Gamma}=0.15 meV (\href{https://www.nndc.bnl.gov/nsr/nsrlink.jsp?1966Be29,B}{1966Be29}) as cited in (\href{https://www.nndc.bnl.gov/nsr/nsrlink.jsp?1969Ro08,B}{1969Ro08}); \ensuremath{\Gamma}=0.14 meV\vspace{0.1cm}}&\\
&&&&&\parbox[t][0.3cm]{12.196461cm}{\raggedright {\ }{\ }{\ }(\href{https://www.nndc.bnl.gov/nsr/nsrlink.jsp?1968Ar02,B}{1968Ar02}) as cited in (\href{https://www.nndc.bnl.gov/nsr/nsrlink.jsp?1969Ro08,B}{1969Ro08}); and \ensuremath{\Gamma}=0.16 meV (private communication with T.\vspace{0.1cm}}&\\
&&&&&\parbox[t][0.3cm]{12.196461cm}{\raggedright {\ }{\ }{\ }Engeland and P. J. Ellis and the authors of (\href{https://www.nndc.bnl.gov/nsr/nsrlink.jsp?1972Gi01,B}{1972Gi01})).\vspace{0.1cm}}&\\
&&&&&\parbox[t][0.3cm]{12.196461cm}{\raggedright J\ensuremath{^{\pi}}: J=2,4 from the measured angular correlations in (\href{https://www.nndc.bnl.gov/nsr/nsrlink.jsp?1969Ro08,B}{1969Ro08}, \href{https://www.nndc.bnl.gov/nsr/nsrlink.jsp?1969Ro22,B}{1969Ro22}). Both of\vspace{0.1cm}}&\\
&&&&&\parbox[t][0.3cm]{12.196461cm}{\raggedright {\ }{\ }{\ }these assignments are compatible with the lifetime measurement of (\href{https://www.nndc.bnl.gov/nsr/nsrlink.jsp?1968Gi09,B}{1968Gi09},\vspace{0.1cm}}&\\
&&&&&\parbox[t][0.3cm]{12.196461cm}{\raggedright {\ }{\ }{\ }\href{https://www.nndc.bnl.gov/nsr/nsrlink.jsp?1969Ro08,B}{1969Ro08}). However, (\href{https://www.nndc.bnl.gov/nsr/nsrlink.jsp?1969Ro08,B}{1969Ro08}) selected J\ensuremath{^{\ensuremath{\pi}}}=4\ensuremath{^{\textnormal{+}}} based on their mirror levels analysis.\vspace{0.1cm}}&\\
&&&&&\parbox[t][0.3cm]{12.196461cm}{\raggedright {\ }{\ }{\ }(\href{https://www.nndc.bnl.gov/nsr/nsrlink.jsp?1969Ro22,B}{1969Ro22}) ruled out J=2 and made the J\ensuremath{^{\ensuremath{\pi}}}=4\ensuremath{^{\textnormal{+}}} assignment tentative. Furthermore, the\vspace{0.1cm}}&\\
&&&&&\parbox[t][0.3cm]{12.196461cm}{\raggedright {\ }{\ }{\ }n\ensuremath{\gamma} angular correlation measurements of (\href{https://www.nndc.bnl.gov/nsr/nsrlink.jsp?1970Sh04,B}{1970Sh04}) yielded a unique spin-parity\vspace{0.1cm}}&\\
&&&&&\parbox[t][0.3cm]{12.196461cm}{\raggedright {\ }{\ }{\ }assignment of J\ensuremath{^{\ensuremath{\pi}}}=4\ensuremath{^{\textnormal{+}}} for this level. The angular correlation data were fitted\vspace{0.1cm}}&\\
&&&&&\parbox[t][0.3cm]{12.196461cm}{\raggedright {\ }{\ }{\ }simultaneously for the 1493-keV and 1890-keV \ensuremath{\gamma}-rays observed by (\href{https://www.nndc.bnl.gov/nsr/nsrlink.jsp?1970Sh04,B}{1970Sh04}) using\vspace{0.1cm}}&\\
&&&&&\parbox[t][0.3cm]{12.196461cm}{\raggedright {\ }{\ }{\ }the code of Warburton (\href{https://www.nndc.bnl.gov/nsr/nsrlink.jsp?1965Po01,B}{1965Po01}). (\href{https://www.nndc.bnl.gov/nsr/nsrlink.jsp?1971Ro18,B}{1971Ro18}) also recommended the J\ensuremath{^{\ensuremath{\pi}}}=4\ensuremath{^{\textnormal{+}}}\vspace{0.1cm}}&\\
&&&&&\parbox[t][0.3cm]{12.196461cm}{\raggedright {\ }{\ }{\ }assignment for this state.\vspace{0.1cm}}&\\
\multicolumn{1}{r@{}}{3576}&\multicolumn{1}{@{.}l}{5\ensuremath{^{{\hyperlink{NE16LEVEL0}{a}}{\hyperlink{NE16LEVEL1}{b}}}} {\it 20}}&\multicolumn{1}{l}{0\ensuremath{^{+}}\ensuremath{^{{\hyperlink{NE16LEVEL2}{c}}}}}&\multicolumn{1}{r@{}}{2}&\multicolumn{1}{@{.}l}{8 ps {\it 14}}&\parbox[t][0.3cm]{12.196461cm}{\raggedright E(level): From recoil correction of E\ensuremath{_{\ensuremath{\gamma}}}=1689 keV \textit{2} (\href{https://www.nndc.bnl.gov/nsr/nsrlink.jsp?1968Gi09,B}{1968Gi09}, \href{https://www.nndc.bnl.gov/nsr/nsrlink.jsp?1969Ro08,B}{1969Ro08}). See also\vspace{0.1cm}}&\\
&&&&&\parbox[t][0.3cm]{12.196461cm}{\raggedright {\ }{\ }{\ }excitation energies deduced that appear to neglect the \ensuremath{^{\textnormal{18}}}Ne recoil corrections: 3576.3\vspace{0.1cm}}&\\
&&&&&\parbox[t][0.3cm]{12.196461cm}{\raggedright {\ }{\ }{\ }keV \textit{20} (\href{https://www.nndc.bnl.gov/nsr/nsrlink.jsp?1968Gi09,B}{1968Gi09}, \href{https://www.nndc.bnl.gov/nsr/nsrlink.jsp?1969Ro08,B}{1969Ro08}); 3576 keV (\href{https://www.nndc.bnl.gov/nsr/nsrlink.jsp?1969Ro22,B}{1969Ro22}, \href{https://www.nndc.bnl.gov/nsr/nsrlink.jsp?1971Ro18,B}{1971Ro18}, \href{https://www.nndc.bnl.gov/nsr/nsrlink.jsp?1972Gi01,B}{1972Gi01}); and 3576\vspace{0.1cm}}&\\
&&&&&\parbox[t][0.3cm]{12.196461cm}{\raggedright {\ }{\ }{\ }keV \textit{4} (\href{https://www.nndc.bnl.gov/nsr/nsrlink.jsp?1970Sh04,B}{1970Sh04}).\vspace{0.1cm}}&\\
&&&&&\parbox[t][0.3cm]{12.196461cm}{\raggedright T\ensuremath{_{1/2}}: From \ensuremath{\tau}=4 ps \textit{2} (\href{https://www.nndc.bnl.gov/nsr/nsrlink.jsp?1972Gi01,B}{1972Gi01}). (\href{https://www.nndc.bnl.gov/nsr/nsrlink.jsp?1968Gi09,B}{1968Gi09}, \href{https://www.nndc.bnl.gov/nsr/nsrlink.jsp?1969Ro08,B}{1969Ro08}) deduced \ensuremath{\tau}\ensuremath{>}2 ps. This lower\vspace{0.1cm}}&\\
&&&&&\parbox[t][0.3cm]{12.196461cm}{\raggedright {\ }{\ }{\ }limit comes from the lack of observation of Doppler shift for the 1689 keV \ensuremath{\gamma} ray from\vspace{0.1cm}}&\\
&&&&&\parbox[t][0.3cm]{12.196461cm}{\raggedright {\ }{\ }{\ }this state when either a CdO or SrO thick target was used to populate this state.\vspace{0.1cm}}&\\
&&&&&\parbox[t][0.3cm]{12.196461cm}{\raggedright {\ }{\ }{\ }(\href{https://www.nndc.bnl.gov/nsr/nsrlink.jsp?1972Gi01,B}{1972Gi01}) obtained \ensuremath{\tau}\ensuremath{<}6 ps, combined their result with the \ensuremath{\tau}\ensuremath{>}2 ps from (\href{https://www.nndc.bnl.gov/nsr/nsrlink.jsp?1968Gi09,B}{1968Gi09},\vspace{0.1cm}}&\\
&&&&&\parbox[t][0.3cm]{12.196461cm}{\raggedright {\ }{\ }{\ }\href{https://www.nndc.bnl.gov/nsr/nsrlink.jsp?1969Ro08,B}{1969Ro08}), and deduced \ensuremath{\tau}=4 ps \textit{2} for this state. Note that this value is inconsistent\vspace{0.1cm}}&\\
&&&&&\parbox[t][0.3cm]{12.196461cm}{\raggedright {\ }{\ }{\ }with the calculated values of \ensuremath{\tau}=13 ps (\href{https://www.nndc.bnl.gov/nsr/nsrlink.jsp?1966Be29,B}{1966Be29}) or 16 ps (private communication of\vspace{0.1cm}}&\\
&&&&&\parbox[t][0.3cm]{12.196461cm}{\raggedright {\ }{\ }{\ }T. Engeland and P. J. Ellis with the authors of (\href{https://www.nndc.bnl.gov/nsr/nsrlink.jsp?1972Gi01,B}{1972Gi01})). \ensuremath{\tau}=4 ps \textit{2} (\href{https://www.nndc.bnl.gov/nsr/nsrlink.jsp?1972Gi01,B}{1972Gi01}) was\vspace{0.1cm}}&\\
&&&&&\parbox[t][0.3cm]{12.196461cm}{\raggedright {\ }{\ }{\ }historically accepted by the previous ENSDF evaluators, and thus it is also adopted\vspace{0.1cm}}&\\
&&&&&\parbox[t][0.3cm]{12.196461cm}{\raggedright {\ }{\ }{\ }here.\vspace{0.1cm}}&\\
&&&&&\parbox[t][0.3cm]{12.196461cm}{\raggedright \ensuremath{\Gamma}=0.16 meV: from (\href{https://www.nndc.bnl.gov/nsr/nsrlink.jsp?1972Gi01,B}{1972Gi01}: see Table 1). This value should be compared to \ensuremath{\Gamma}=0.05\vspace{0.1cm}}&\\
&&&&&\parbox[t][0.3cm]{12.196461cm}{\raggedright {\ }{\ }{\ }meV (\href{https://www.nndc.bnl.gov/nsr/nsrlink.jsp?1966Be29,B}{1966Be29}) as cited in (\href{https://www.nndc.bnl.gov/nsr/nsrlink.jsp?1969Ro08,B}{1969Ro08}); \ensuremath{\Gamma}=0.03 meV (\href{https://www.nndc.bnl.gov/nsr/nsrlink.jsp?1968Ar02,B}{1968Ar02}) as cited in\vspace{0.1cm}}&\\
&&&&&\parbox[t][0.3cm]{12.196461cm}{\raggedright {\ }{\ }{\ }(\href{https://www.nndc.bnl.gov/nsr/nsrlink.jsp?1969Ro08,B}{1969Ro08}); and \ensuremath{\Gamma}=0.04 meV (private communication of T. Engeland and P. J. Ellis\vspace{0.1cm}}&\\
&&&&&\parbox[t][0.3cm]{12.196461cm}{\raggedright {\ }{\ }{\ }with the authors of (\href{https://www.nndc.bnl.gov/nsr/nsrlink.jsp?1972Gi01,B}{1972Gi01})).\vspace{0.1cm}}&\\
&&&&&\parbox[t][0.3cm]{12.196461cm}{\raggedright J\ensuremath{^{\pi}}: Out of J\ensuremath{\leq}4 deduced from the measured angular correlations in (\href{https://www.nndc.bnl.gov/nsr/nsrlink.jsp?1969Ro08,B}{1969Ro08}), J\ensuremath{^{\ensuremath{\pi}}}=0\ensuremath{^{\textnormal{+}}} is\vspace{0.1cm}}&\\
&&&&&\parbox[t][0.3cm]{12.196461cm}{\raggedright {\ }{\ }{\ }selected by (\href{https://www.nndc.bnl.gov/nsr/nsrlink.jsp?1969Ro08,B}{1969Ro08}) based on the mirror levels analysis by (\href{https://www.nndc.bnl.gov/nsr/nsrlink.jsp?1969Ro08,B}{1969Ro08}). The n\ensuremath{\gamma}\vspace{0.1cm}}&\\
&&&&&\parbox[t][0.3cm]{12.196461cm}{\raggedright {\ }{\ }{\ }angular correlation measured by (\href{https://www.nndc.bnl.gov/nsr/nsrlink.jsp?1969Ro22,B}{1969Ro22}) for the 3576 keV\ensuremath{\rightarrow}1887 keV \ensuremath{\gamma}-ray\vspace{0.1cm}}&\\
&&&&&\parbox[t][0.3cm]{12.196461cm}{\raggedright {\ }{\ }{\ }transition was isotropic, and thus not in disagreement with the suggested J\ensuremath{^{\ensuremath{\pi}}}=0\ensuremath{^{\textnormal{+}}}\vspace{0.1cm}}&\\
&&&&&\parbox[t][0.3cm]{12.196461cm}{\raggedright {\ }{\ }{\ }assignment for the 3576 keV state (\href{https://www.nndc.bnl.gov/nsr/nsrlink.jsp?1968Gi09,B}{1968Gi09}, \href{https://www.nndc.bnl.gov/nsr/nsrlink.jsp?1969Ro08,B}{1969Ro08}). (\href{https://www.nndc.bnl.gov/nsr/nsrlink.jsp?1971Ro18,B}{1971Ro18}) also\vspace{0.1cm}}&\\
&&&&&\parbox[t][0.3cm]{12.196461cm}{\raggedright {\ }{\ }{\ }recommended the J\ensuremath{^{\ensuremath{\pi}}}=0\ensuremath{^{\textnormal{+}}} assignment for this state.\vspace{0.1cm}}&\\
\multicolumn{1}{r@{}}{3616}&\multicolumn{1}{@{.}l}{6\ensuremath{^{{\hyperlink{NE16LEVEL0}{a}}}} {\it 6}}&\multicolumn{1}{l}{2\ensuremath{^{+}}\ensuremath{^{{\hyperlink{NE16LEVEL2}{c}}}}}&\multicolumn{1}{r@{}}{44}&\multicolumn{1}{@{ }l}{fs {\it +21\textminus14}}&\parbox[t][0.3cm]{12.196461cm}{\raggedright T=1 (\href{https://www.nndc.bnl.gov/nsr/nsrlink.jsp?1970Sh04,B}{1970Sh04})\vspace{0.1cm}}&\\
&&&&&\parbox[t][0.3cm]{12.196461cm}{\raggedright E(level): From the least-squares fit to E\ensuremath{_{\ensuremath{\gamma}}}, which includes nuclear recoil corrections. See\vspace{0.1cm}}&\\
&&&&&\parbox[t][0.3cm]{12.196461cm}{\raggedright {\ }{\ }{\ }also excitation energies that appear to not be corrected for the recoil energy: 3616.4\vspace{0.1cm}}&\\
&&&&&\parbox[t][0.3cm]{12.196461cm}{\raggedright {\ }{\ }{\ }keV \textit{6} (\href{https://www.nndc.bnl.gov/nsr/nsrlink.jsp?1968Gi09,B}{1968Gi09}, \href{https://www.nndc.bnl.gov/nsr/nsrlink.jsp?1969Ro08,B}{1969Ro08}: simply adding E\ensuremath{_{\ensuremath{\gamma}}}=1887.3 keV \textit{2} and E\ensuremath{_{\ensuremath{\gamma}}}=1729.2 keV \textit{5}\vspace{0.1cm}}&\\
&&&&&\parbox[t][0.3cm]{12.196461cm}{\raggedright {\ }{\ }{\ }to follow what was done by these studies to deduce E\ensuremath{_{\textnormal{x}}}, one would obtain E\ensuremath{_{\textnormal{x}}}=3616.5\vspace{0.1cm}}&\\
&&&&&\parbox[t][0.3cm]{12.196461cm}{\raggedright {\ }{\ }{\ }keV \textit{5}. It is not clear why the energy was reported in (\href{https://www.nndc.bnl.gov/nsr/nsrlink.jsp?1968Gi09,B}{1968Gi09}, \href{https://www.nndc.bnl.gov/nsr/nsrlink.jsp?1969Ro08,B}{1969Ro08}) as\vspace{0.1cm}}&\\
&&&&&\parbox[t][0.3cm]{12.196461cm}{\raggedright {\ }{\ }{\ }E\ensuremath{_{\textnormal{x}}}=3616.4 keV {6}); 3616 keV (\href{https://www.nndc.bnl.gov/nsr/nsrlink.jsp?1969Ro22,B}{1969Ro22}); 3623 keV \textit{4} (\href{https://www.nndc.bnl.gov/nsr/nsrlink.jsp?1970Sh04,B}{1970Sh04}); and 3616 keV\vspace{0.1cm}}&\\
&&&&&\parbox[t][0.3cm]{12.196461cm}{\raggedright {\ }{\ }{\ }(\href{https://www.nndc.bnl.gov/nsr/nsrlink.jsp?1971Ro18,B}{1971Ro18}).\vspace{0.1cm}}&\\
&&&&&\parbox[t][0.3cm]{12.196461cm}{\raggedright T\ensuremath{_{1/2}}: From \ensuremath{\tau}=0.063 ps \textit{+30{\textminus}20} (\href{https://www.nndc.bnl.gov/nsr/nsrlink.jsp?1969Ro08,B}{1969Ro08}: the analysis of the 1.73 MeV \ensuremath{\gamma}-ray observed\vspace{0.1cm}}&\\
&&&&&\parbox[t][0.3cm]{12.196461cm}{\raggedright {\ }{\ }{\ }using the thick SrO and CdO targets yielded lifetimes of \ensuremath{\tau}=0.074 ps \textit{+60{\textminus}40} and\vspace{0.1cm}}&\\
&&&&&\parbox[t][0.3cm]{12.196461cm}{\raggedright {\ }{\ }{\ }\ensuremath{\tau}=0.059 ps \textit{+40{\textminus}30}, respectively. (\href{https://www.nndc.bnl.gov/nsr/nsrlink.jsp?1969Ro08,B}{1969Ro08}) then recommended the lifetime of \ensuremath{\tau}=0.063\vspace{0.1cm}}&\\
&&&&&\parbox[t][0.3cm]{12.196461cm}{\raggedright {\ }{\ }{\ }ps \textit{+30{\textminus}20}). See also a similar value of \ensuremath{\tau}=0.06 ps \textit{+3{\textminus}2} from (\href{https://www.nndc.bnl.gov/nsr/nsrlink.jsp?1968Gi09,B}{1968Gi09}).\vspace{0.1cm}}&\\
&&&&&\parbox[t][0.3cm]{12.196461cm}{\raggedright \ensuremath{\Gamma}=1.05\ensuremath{\times}10\ensuremath{^{\textnormal{$-$2}}} from the recommended lifetime deduced by (\href{https://www.nndc.bnl.gov/nsr/nsrlink.jsp?1969Ro08,B}{1969Ro08}).\vspace{0.1cm}}&\\
&&&&&\parbox[t][0.3cm]{12.196461cm}{\raggedright J\ensuremath{^{\pi}}: From the n\ensuremath{\gamma} angular correlation measurements of (\href{https://www.nndc.bnl.gov/nsr/nsrlink.jsp?1970Sh04,B}{1970Sh04}), which yielded a\vspace{0.1cm}}&\\
&&&&&\parbox[t][0.3cm]{12.196461cm}{\raggedright {\ }{\ }{\ }unique spin-parity assignment of J\ensuremath{^{\ensuremath{\pi}}}=2\ensuremath{^{\textnormal{+}}} for this level. The angular correlation data were\vspace{0.1cm}}&\\
&&&&&\parbox[t][0.3cm]{12.196461cm}{\raggedright {\ }{\ }{\ }fitted simultaneously for the 1736-keV and 1890-keV \ensuremath{\gamma} rays observed by (\href{https://www.nndc.bnl.gov/nsr/nsrlink.jsp?1970Sh04,B}{1970Sh04})\vspace{0.1cm}}&\\
\end{longtable}
\begin{textblock}{29}(0,27.3)
Continued on next page (footnotes at end of table)
\end{textblock}
\clearpage
\begin{longtable}{cccccc@{\extracolsep{\fill}}c}
\\[-.4cm]
\multicolumn{7}{c}{{\bf \small \underline{\ensuremath{^{\textnormal{16}}}O(\ensuremath{^{\textnormal{3}}}He,n\ensuremath{\gamma})\hspace{0.2in}\href{https://www.nndc.bnl.gov/nsr/nsrlink.jsp?1968Gi09,B}{1968Gi09},\href{https://www.nndc.bnl.gov/nsr/nsrlink.jsp?2003Ta13,B}{2003Ta13} (continued)}}}\\
\multicolumn{7}{c}{~}\\
\multicolumn{7}{c}{\underline{\ensuremath{^{18}}Ne Levels (continued)}}\\
\multicolumn{7}{c}{~}\\
\multicolumn{2}{c}{E(level)$^{}$}&J$^{\pi}$$^{}$&\multicolumn{2}{c}{T$_{1/2}$$^{}$}&Comments&\\[-.2cm]
\multicolumn{2}{c}{\hrulefill}&\hrulefill&\multicolumn{2}{c}{\hrulefill}&\hrulefill&
\endhead
&&&&&\parbox[t][0.3cm]{14.559021cm}{\raggedright {\ }{\ }{\ }using the code of Warburton (\href{https://www.nndc.bnl.gov/nsr/nsrlink.jsp?1965Po01,B}{1965Po01}).\vspace{0.1cm}}&\\
&&&&&\parbox[t][0.3cm]{14.559021cm}{\raggedright J\ensuremath{^{\pi}}: See similar assignments from (\href{https://www.nndc.bnl.gov/nsr/nsrlink.jsp?1969Ro08,B}{1969Ro08}: J\ensuremath{^{\ensuremath{\pi}}}=2\ensuremath{^{\textnormal{+}}} based on the E2 sum rules of (\href{https://www.nndc.bnl.gov/nsr/nsrlink.jsp?1968Ar02,B}{1968Ar02}). The\vspace{0.1cm}}&\\
&&&&&\parbox[t][0.3cm]{14.559021cm}{\raggedright {\ }{\ }{\ }lifetime measurement together with the ground state decay from this state exclude J=4); (\href{https://www.nndc.bnl.gov/nsr/nsrlink.jsp?1969Ro22,B}{1969Ro22}: J=2,4\vspace{0.1cm}}&\\
&&&&&\parbox[t][0.3cm]{14.559021cm}{\raggedright {\ }{\ }{\ }but J=4 is excluded as mentioned previously. A positive parity is selected based on lifetime measurement\vspace{0.1cm}}&\\
&&&&&\parbox[t][0.3cm]{14.559021cm}{\raggedright {\ }{\ }{\ }by (\href{https://www.nndc.bnl.gov/nsr/nsrlink.jsp?1969Ro08,B}{1969Ro08}) and mixing ratio from (\href{https://www.nndc.bnl.gov/nsr/nsrlink.jsp?1969Ro22,B}{1969Ro22})); and (\href{https://www.nndc.bnl.gov/nsr/nsrlink.jsp?1972Gi01,B}{1972Gi01}: J\ensuremath{^{\ensuremath{\pi}}}=2\ensuremath{^{\textnormal{+}}}).\vspace{0.1cm}}&\\
&&&&&\parbox[t][0.3cm]{14.559021cm}{\raggedright (\href{https://www.nndc.bnl.gov/nsr/nsrlink.jsp?1972Gi01,B}{1972Gi01}) reports a population parameter p\ensuremath{>}0.75 at 10.3 MeV bombarding energy. It is not clear what\vspace{0.1cm}}&\\
&&&&&\parbox[t][0.3cm]{14.559021cm}{\raggedright {\ }{\ }{\ }this parameter refers to.\vspace{0.1cm}}&\\
\end{longtable}
\parbox[b][0.3cm]{17.7cm}{\makebox[1ex]{\ensuremath{^{\hypertarget{NE16LEVEL0}{a}}}} It appears that none of the above mentioned experiments considered the \ensuremath{^{\textnormal{18}}}Ne recoil energies when calculating the excitation}\\
\parbox[b][0.3cm]{17.7cm}{{\ }{\ }energies. Instead, it appears that the excitation energies were calculated simply by adding the Doppler shift corrected \ensuremath{\gamma}-ray}\\
\parbox[b][0.3cm]{17.7cm}{{\ }{\ }energies from the decay cascades to deduce the excitation energies. This is true for all the previous evaluations of \ensuremath{^{\textnormal{18}}}Ne.}\\
\parbox[b][0.3cm]{17.7cm}{{\ }{\ }Therefore, the evaluator corrected the excitation energies of all the bound states of \ensuremath{^{\textnormal{18}}}Ne by performing a least-squares fit to E\ensuremath{_{\ensuremath{\gamma}}},}\\
\parbox[b][0.3cm]{17.7cm}{{\ }{\ }which also takes into account the nuclear recoil energies.}\\
\parbox[b][0.3cm]{17.7cm}{\makebox[1ex]{\ensuremath{^{\hypertarget{NE16LEVEL1}{b}}}} This state was observed in (\href{https://www.nndc.bnl.gov/nsr/nsrlink.jsp?1968Gi09,B}{1968Gi09}) for the first time via measuring the \ensuremath{\gamma}-ray decay to the first excited state confirmed by the}\\
\parbox[b][0.3cm]{17.7cm}{{\ }{\ }n\ensuremath{\gamma} coincidences measurement in (\href{https://www.nndc.bnl.gov/nsr/nsrlink.jsp?1968Gi09,B}{1968Gi09}).}\\
\parbox[b][0.3cm]{17.7cm}{\makebox[1ex]{\ensuremath{^{\hypertarget{NE16LEVEL2}{c}}}} Supported by the mirror analysis of (\href{https://www.nndc.bnl.gov/nsr/nsrlink.jsp?1968Gi09,B}{1968Gi09}); and the measured \ensuremath{\gamma}-ray angular correlations by (\href{https://www.nndc.bnl.gov/nsr/nsrlink.jsp?1969Ro08,B}{1969Ro08}). The latter were}\\
\parbox[b][0.3cm]{17.7cm}{{\ }{\ }fitted with the theoretical distributions taken from (O. H\"{a}usser, J. S. Lopes, H. J. Rose and R. D. Gill, University of Oxford}\\
\parbox[b][0.3cm]{17.7cm}{{\ }{\ }Laboratory Report (1966)) for a number of assumed spins involved in the \ensuremath{\gamma}-ray transitions. (\href{https://www.nndc.bnl.gov/nsr/nsrlink.jsp?1969Ro08,B}{1969Ro08}) deduced a\ensuremath{_{\textnormal{2}}} and a\ensuremath{_{\textnormal{4}}}}\\
\parbox[b][0.3cm]{17.7cm}{{\ }{\ }coefficients of Legendre polynomials and used them together with the sum rules of (\href{https://www.nndc.bnl.gov/nsr/nsrlink.jsp?1968Ar02,B}{1968Ar02}) for the E2 transitions and the}\\
\parbox[b][0.3cm]{17.7cm}{{\ }{\ }mirror levels in \ensuremath{^{\textnormal{18}}}O to deduced one final spin for the excited states, which are reported here.}\\
\vspace{0.5cm}
\underline{$\gamma$($^{18}$Ne)}\\
\vspace{0.34cm}
\parbox[b][0.3cm]{17.7cm}{\addtolength{\parindent}{-0.254cm}In (\href{https://www.nndc.bnl.gov/nsr/nsrlink.jsp?1970Sh04,B}{1970Sh04}), \ensuremath{\gamma}-ray de-excitation data were quoted in terms of the relative intensities for \ensuremath{\theta}\ensuremath{_{\ensuremath{\gamma}}}=90\ensuremath{^\circ} and \ensuremath{\theta}\ensuremath{_{\textnormal{n}}}=0\ensuremath{^\circ} whenever a}\\
\parbox[b][0.3cm]{17.7cm}{transition was observed definitely.}\\
\parbox[b][0.3cm]{17.7cm}{\addtolength{\parindent}{-0.254cm}The results from (\href{https://www.nndc.bnl.gov/nsr/nsrlink.jsp?1976Mc02,B}{1976Mc02}) for the B(E2: 2\ensuremath{^{\textnormal{+}}_{\textnormal{1}}}\ensuremath{\rightarrow}0\ensuremath{^{\textnormal{+}}_{\textnormal{g.s.}}}) transitions confirm the presence of two-body contributions to the effective}\\
\parbox[b][0.3cm]{17.7cm}{E2 transition operator. (\href{https://www.nndc.bnl.gov/nsr/nsrlink.jsp?1970Ha75,B}{1970Ha75}, \href{https://www.nndc.bnl.gov/nsr/nsrlink.jsp?1970Ha49,B}{1970Ha49}) predicted two-body contributions of the order of 25\% to the E2 transition strengths}\\
\parbox[b][0.3cm]{17.7cm}{in \ensuremath{^{\textnormal{18}}}Ne.}\\
\vspace{0.34cm}
\begin{longtable}{ccccccccc@{}ccc@{\extracolsep{\fill}}c}
\multicolumn{2}{c}{E\ensuremath{_{i}}(level)}&J\ensuremath{^{\pi}_{i}}&\multicolumn{2}{c}{E\ensuremath{_{\gamma}}\ensuremath{^{\hyperlink{NE16GAMMA0}{a}}}}&\multicolumn{2}{c}{I\ensuremath{_{\gamma}}}&\multicolumn{2}{c}{E\ensuremath{_{f}}}&J\ensuremath{^{\pi}_{f}}&Mult.&Comments&\\[-.2cm]
\multicolumn{2}{c}{\hrulefill}&\hrulefill&\multicolumn{2}{c}{\hrulefill}&\multicolumn{2}{c}{\hrulefill}&\multicolumn{2}{c}{\hrulefill}&\hrulefill&\hrulefill&\hrulefill&
\endfirsthead
\multicolumn{1}{r@{}}{1887}&\multicolumn{1}{@{.}l}{4}&\multicolumn{1}{l}{2\ensuremath{^{+}}}&\multicolumn{1}{r@{}}{1887}&\multicolumn{1}{@{.}l}{3 {\it 2}}&\multicolumn{1}{r@{}}{100}&\multicolumn{1}{@{}l}{}&\multicolumn{1}{r@{}}{0}&\multicolumn{1}{@{}l}{}&\multicolumn{1}{@{}l}{0\ensuremath{^{+}}}&\multicolumn{1}{l}{E2}&\parbox[t][0.3cm]{10.441801cm}{\raggedright B(E2)(W.u.)=18.0 \textit{+17{\textminus}14}\vspace{0.1cm}}&\\
&&&&&&&&&&&\parbox[t][0.3cm]{10.441801cm}{\raggedright E\ensuremath{_{\gamma}}: From (\href{https://www.nndc.bnl.gov/nsr/nsrlink.jsp?1968Gi09,B}{1968Gi09}, \href{https://www.nndc.bnl.gov/nsr/nsrlink.jsp?1969Ro08,B}{1969Ro08}). See also 1890 keV (\href{https://www.nndc.bnl.gov/nsr/nsrlink.jsp?1969Be31,B}{1969Be31},\vspace{0.1cm}}&\\
&&&&&&&&&&&\parbox[t][0.3cm]{10.441801cm}{\raggedright {\ }{\ }{\ }\href{https://www.nndc.bnl.gov/nsr/nsrlink.jsp?1974Mc17,B}{1974Mc17}, \href{https://www.nndc.bnl.gov/nsr/nsrlink.jsp?1976Mc02,B}{1976Mc02}); 1887 keV (\href{https://www.nndc.bnl.gov/nsr/nsrlink.jsp?1969Ro22,B}{1969Ro22}, \href{https://www.nndc.bnl.gov/nsr/nsrlink.jsp?1971Ro18,B}{1971Ro18}, \href{https://www.nndc.bnl.gov/nsr/nsrlink.jsp?1972Gi01,B}{1972Gi01},\vspace{0.1cm}}&\\
&&&&&&&&&&&\parbox[t][0.3cm]{10.441801cm}{\raggedright {\ }{\ }{\ }\href{https://www.nndc.bnl.gov/nsr/nsrlink.jsp?2003Ta13,B}{2003Ta13}); and 1890 keV \textit{2} (\href{https://www.nndc.bnl.gov/nsr/nsrlink.jsp?1970Sh04,B}{1970Sh04}).\vspace{0.1cm}}&\\
&&&&&&&&&&&\parbox[t][0.3cm]{10.441801cm}{\raggedright I\ensuremath{_{\gamma}}: From (\href{https://www.nndc.bnl.gov/nsr/nsrlink.jsp?1968Gi09,B}{1968Gi09}, \href{https://www.nndc.bnl.gov/nsr/nsrlink.jsp?1969Ro08,B}{1969Ro08}, \href{https://www.nndc.bnl.gov/nsr/nsrlink.jsp?1969Ro22,B}{1969Ro22}).\vspace{0.1cm}}&\\
&&&&&&&&&&&\parbox[t][0.3cm]{10.441801cm}{\raggedright Mult.: From (\href{https://www.nndc.bnl.gov/nsr/nsrlink.jsp?1974Mc17,B}{1974Mc17}).\vspace{0.1cm}}&\\
&&&&&&&&&&&\parbox[t][0.3cm]{10.441801cm}{\raggedright \ensuremath{\delta}=0 (mixing ratio) from (\href{https://www.nndc.bnl.gov/nsr/nsrlink.jsp?1969Ro08,B}{1969Ro08}).\vspace{0.1cm}}&\\
&&&&&&&&&&&\parbox[t][0.3cm]{10.441801cm}{\raggedright (\href{https://www.nndc.bnl.gov/nsr/nsrlink.jsp?1969Ro08,B}{1969Ro08}): deduced coefficients of Legendre polynomials from the \ensuremath{\gamma}-ray\vspace{0.1cm}}&\\
&&&&&&&&&&&\parbox[t][0.3cm]{10.441801cm}{\raggedright {\ }{\ }{\ }angular distributions. These are: a\ensuremath{_{\textnormal{2}}}=0.44 \textit{3} and a\ensuremath{_{\textnormal{4}}}={\textminus}0.33 \textit{5} (E\ensuremath{_{\textnormal{lab}}} not\vspace{0.1cm}}&\\
&&&&&&&&&&&\parbox[t][0.3cm]{10.441801cm}{\raggedright {\ }{\ }{\ }given).\vspace{0.1cm}}&\\
&&&&&&&&&&&\parbox[t][0.3cm]{10.441801cm}{\raggedright (\href{https://www.nndc.bnl.gov/nsr/nsrlink.jsp?1969Ro22,B}{1969Ro22}): deduced a\ensuremath{_{\textnormal{2}}}=0.31 \textit{9} and a\ensuremath{_{\textnormal{4}}}={\textminus}0.92 \textit{7} from the measured \ensuremath{\gamma}-ray\vspace{0.1cm}}&\\
&&&&&&&&&&&\parbox[t][0.3cm]{10.441801cm}{\raggedright {\ }{\ }{\ }angular correlations at E\ensuremath{_{\textnormal{lab}}}=9.5 MeV. These coefficients uniquely\vspace{0.1cm}}&\\
&&&&&&&&&&&\parbox[t][0.3cm]{10.441801cm}{\raggedright {\ }{\ }{\ }determine J=2 for the \ensuremath{^{\textnormal{18}}}Ne*(1887 keV) level.\vspace{0.1cm}}&\\
&&&&&&&&&&&\parbox[t][0.3cm]{10.441801cm}{\raggedright (\href{https://www.nndc.bnl.gov/nsr/nsrlink.jsp?2003Ta13,B}{2003Ta13}): deduced values of a\ensuremath{_{\textnormal{2}}}/a\ensuremath{_{\textnormal{0}}}=0.40 and a\ensuremath{_{\textnormal{4}}}/a\ensuremath{_{\textnormal{0}}}={\textminus}0.15 \textit{23} from the\vspace{0.1cm}}&\\
&&&&&&&&&&&\parbox[t][0.3cm]{10.441801cm}{\raggedright {\ }{\ }{\ }measured angular distribution, for the 1887 keV\ensuremath{\rightarrow}g.s. transition, averaged\vspace{0.1cm}}&\\
&&&&&&&&&&&\parbox[t][0.3cm]{10.441801cm}{\raggedright {\ }{\ }{\ }over E\ensuremath{_{\textnormal{lab}}}=6-30 MeV.\vspace{0.1cm}}&\\
&&&&&&&&&&&\parbox[t][0.3cm]{10.441801cm}{\raggedright (\href{https://www.nndc.bnl.gov/nsr/nsrlink.jsp?1976Mc02,B}{1976Mc02}) calculated an experimental value of B(E2:2\ensuremath{^{\textnormal{+}}_{\textnormal{1}}}\ensuremath{\rightarrow}0\ensuremath{^{\textnormal{+}}_{\textnormal{1}}})=52 e\ensuremath{^{\textnormal{2}}}fm\ensuremath{^{\textnormal{4}}}\vspace{0.1cm}}&\\
&&&&&&&&&&&\parbox[t][0.3cm]{10.441801cm}{\raggedright {\ }{\ }{\ }\textit{5} for the 1887 keV\ensuremath{\rightarrow}g.s. transition using \ensuremath{\tau}=0.66 ps \textit{6}. [This value was\vspace{0.1cm}}&\\
&&&&&&&&&&&\parbox[t][0.3cm]{10.441801cm}{\raggedright {\ }{\ }{\ }deduced by (\href{https://www.nndc.bnl.gov/nsr/nsrlink.jsp?1976Mc02,B}{1976Mc02}) as the weighted average of \ensuremath{\tau}=0.63 ps \textit{13}\vspace{0.1cm}}&\\
\end{longtable}
\begin{textblock}{29}(0,27.3)
Continued on next page (footnotes at end of table)
\end{textblock}
\clearpage
\begin{longtable}{ccccccccc@{}ccc@{\extracolsep{\fill}}c}
\\[-.4cm]
\multicolumn{13}{c}{{\bf \small \underline{\ensuremath{^{\textnormal{16}}}O(\ensuremath{^{\textnormal{3}}}He,n\ensuremath{\gamma})\hspace{0.2in}\href{https://www.nndc.bnl.gov/nsr/nsrlink.jsp?1968Gi09,B}{1968Gi09},\href{https://www.nndc.bnl.gov/nsr/nsrlink.jsp?2003Ta13,B}{2003Ta13} (continued)}}}\\
\multicolumn{13}{c}{~}\\
\multicolumn{13}{c}{\underline{$\gamma$($^{18}$Ne) (continued)}}\\
\multicolumn{13}{c}{~~~}\\
\multicolumn{2}{c}{E\ensuremath{_{i}}(level)}&J\ensuremath{^{\pi}_{i}}&\multicolumn{2}{c}{E\ensuremath{_{\gamma}}\ensuremath{^{\hyperlink{NE16GAMMA0}{a}}}}&\multicolumn{2}{c}{I\ensuremath{_{\gamma}}}&\multicolumn{2}{c}{E\ensuremath{_{f}}}&J\ensuremath{^{\pi}_{f}}&Mult.&Comments&\\[-.2cm]
\multicolumn{2}{c}{\hrulefill}&\hrulefill&\multicolumn{2}{c}{\hrulefill}&\multicolumn{2}{c}{\hrulefill}&\multicolumn{2}{c}{\hrulefill}&\hrulefill&\hrulefill&\hrulefill&
\endhead
&&&&&&&&&&&\parbox[t][0.3cm]{9.864cm}{\raggedright {\ }{\ }{\ }(\href{https://www.nndc.bnl.gov/nsr/nsrlink.jsp?1974Mc17,B}{1974Mc17}); \ensuremath{\tau}=0.67 ps \textit{6} (\href{https://www.nndc.bnl.gov/nsr/nsrlink.jsp?1976Mc02,B}{1976Mc02}); and \ensuremath{\tau}=0.49 ps \textit{+17{\textminus}9}\vspace{0.1cm}}&\\
&&&&&&&&&&&\parbox[t][0.3cm]{9.864cm}{\raggedright {\ }{\ }{\ }(\href{https://www.nndc.bnl.gov/nsr/nsrlink.jsp?1969Ro08,B}{1969Ro08}). Evaluator notes that even though the uncertainty in\vspace{0.1cm}}&\\
&&&&&&&&&&&\parbox[t][0.3cm]{9.864cm}{\raggedright {\ }{\ }{\ }\ensuremath{\tau}=0.63 ps \textit{13} (\href{https://www.nndc.bnl.gov/nsr/nsrlink.jsp?1974Mc17,B}{1974Mc17}) was systematic, (\href{https://www.nndc.bnl.gov/nsr/nsrlink.jsp?1976Mc02,B}{1976Mc02}) appears to\vspace{0.1cm}}&\\
&&&&&&&&&&&\parbox[t][0.3cm]{9.864cm}{\raggedright {\ }{\ }{\ }have ignored that, and we cannot reproduce the weighted average\vspace{0.1cm}}&\\
&&&&&&&&&&&\parbox[t][0.3cm]{9.864cm}{\raggedright {\ }{\ }{\ }obtained by (\href{https://www.nndc.bnl.gov/nsr/nsrlink.jsp?1976Mc02,B}{1976Mc02}).] The deduced B(E2) value should be\vspace{0.1cm}}&\\
&&&&&&&&&&&\parbox[t][0.3cm]{9.864cm}{\raggedright {\ }{\ }{\ }compared with calculated values of B(E2)=32 e\ensuremath{^{\textnormal{2}}}fm\ensuremath{^{\textnormal{4}}} (T. Engeland and\vspace{0.1cm}}&\\
&&&&&&&&&&&\parbox[t][0.3cm]{9.864cm}{\raggedright {\ }{\ }{\ }P. J. Ellis, Nucl. Phys. A 181 (1972) 368: from shell model);\vspace{0.1cm}}&\\
&&&&&&&&&&&\parbox[t][0.3cm]{9.864cm}{\raggedright {\ }{\ }{\ }B(E2)=40 e\ensuremath{^{\textnormal{2}}}fm\ensuremath{^{\textnormal{4}}} (\href{https://www.nndc.bnl.gov/nsr/nsrlink.jsp?1976Mc02,B}{1976Mc02}: calculated assuming that the wave\vspace{0.1cm}}&\\
&&&&&&&&&&&\parbox[t][0.3cm]{9.864cm}{\raggedright {\ }{\ }{\ }functions for the states in \ensuremath{^{\textnormal{18}}}Ne have the same structure as the\vspace{0.1cm}}&\\
&&&&&&&&&&&\parbox[t][0.3cm]{9.864cm}{\raggedright {\ }{\ }{\ }corresponding T=1 states in \ensuremath{^{\textnormal{18}}}O); B(E2)=52 e\ensuremath{^{\textnormal{2}}}fm\ensuremath{^{\textnormal{4}}} (\href{https://www.nndc.bnl.gov/nsr/nsrlink.jsp?1976Mc02,B}{1976Mc02}: from\vspace{0.1cm}}&\\
&&&&&&&&&&&\parbox[t][0.3cm]{9.864cm}{\raggedright {\ }{\ }{\ }the phenomenological estimates of (\href{https://www.nndc.bnl.gov/nsr/nsrlink.jsp?1970Ha75,B}{1970Ha75}, \href{https://www.nndc.bnl.gov/nsr/nsrlink.jsp?1970Ha49,B}{1970Ha49}) for the\vspace{0.1cm}}&\\
&&&&&&&&&&&\parbox[t][0.3cm]{9.864cm}{\raggedright {\ }{\ }{\ }two-body part of the effective transition operator); and B(E2)=48\vspace{0.1cm}}&\\
&&&&&&&&&&&\parbox[t][0.3cm]{9.864cm}{\raggedright {\ }{\ }{\ }e\ensuremath{^{\textnormal{2}}}fm\ensuremath{^{\textnormal{4}}} (\href{https://www.nndc.bnl.gov/nsr/nsrlink.jsp?1976Mc02,B}{1976Mc02}: from the realistic reaction matrix elements of (F.\vspace{0.1cm}}&\\
&&&&&&&&&&&\parbox[t][0.3cm]{9.864cm}{\raggedright {\ }{\ }{\ }C. Khanna, M. Harvey, D. W. L. Sprung and A. Jopko, The two body\vspace{0.1cm}}&\\
&&&&&&&&&&&\parbox[t][0.3cm]{9.864cm}{\raggedright {\ }{\ }{\ }force in nuclei, ed. S. M. Austin and G. M. Crawley (Plenum Press,\vspace{0.1cm}}&\\
&&&&&&&&&&&\parbox[t][0.3cm]{9.864cm}{\raggedright {\ }{\ }{\ }New York, 1972) p. 229)).\vspace{0.1cm}}&\\
&&&&&&&&&&&\parbox[t][0.3cm]{9.864cm}{\raggedright (\href{https://www.nndc.bnl.gov/nsr/nsrlink.jsp?2003Ri08,B}{2003Ri08}): deduced, using the measured lifetime of 0.77 ps \textit{+9{\textminus}7},\vspace{0.1cm}}&\\
&&&&&&&&&&&\parbox[t][0.3cm]{9.864cm}{\raggedright {\ }{\ }{\ }B(E2: 0\ensuremath{^{\textnormal{+}}_{\textnormal{g.s.}}}\ensuremath{\rightarrow}2\ensuremath{^{\textnormal{+}}_{\textnormal{1}}})=222 e\ensuremath{^{\textnormal{2}}}fm\ensuremath{^{\textnormal{4}}} \textit{20}, which is consistent with B(E2:\vspace{0.1cm}}&\\
&&&&&&&&&&&\parbox[t][0.3cm]{9.864cm}{\raggedright {\ }{\ }{\ }0\ensuremath{^{\textnormal{+}}_{\textnormal{g.s.}}}\ensuremath{\rightarrow}2\ensuremath{^{\textnormal{+}}_{\textnormal{1}}})=260 e\ensuremath{^{\textnormal{2}}}fm\ensuremath{^{\textnormal{4}}} \textit{25} (\href{https://www.nndc.bnl.gov/nsr/nsrlink.jsp?1976Mc02,B}{1976Mc02}). Both of these values are\vspace{0.1cm}}&\\
&&&&&&&&&&&\parbox[t][0.3cm]{9.864cm}{\raggedright {\ }{\ }{\ }inconsistent with B(E2: 0\ensuremath{^{\textnormal{+}}_{\textnormal{g.s.}}}\ensuremath{\rightarrow}2\ensuremath{^{\textnormal{+}}_{\textnormal{1}}})=113 e\ensuremath{^{\textnormal{2}}}fm\ensuremath{^{\textnormal{4}}} \textit{18} (\href{https://www.nndc.bnl.gov/nsr/nsrlink.jsp?2000Ri15,B}{2000Ri15}) and\vspace{0.1cm}}&\\
&&&&&&&&&&&\parbox[t][0.3cm]{9.864cm}{\raggedright {\ }{\ }{\ }B(E2: 0\ensuremath{^{\textnormal{+}}_{\textnormal{g.s.}}}\ensuremath{\rightarrow}2\ensuremath{^{\textnormal{+}}_{\textnormal{1}}})=137 e\ensuremath{^{\textnormal{2}}}fm\ensuremath{^{\textnormal{4}}} \textit{22} (\href{https://www.nndc.bnl.gov/nsr/nsrlink.jsp?2000Ri15,B}{2000Ri15}).\vspace{0.1cm}}&\\
&&&&&&&&&&&\parbox[t][0.3cm]{9.864cm}{\raggedright Note that the B(E2: 0\ensuremath{^{\textnormal{+}}_{\textnormal{g.s.}}}\ensuremath{\rightarrow}2\ensuremath{^{\textnormal{+}}_{\textnormal{1}}})=222 e\ensuremath{^{\textnormal{2}}}fm\ensuremath{^{\textnormal{4}}} \textit{20} from (\href{https://www.nndc.bnl.gov/nsr/nsrlink.jsp?2003Ri08,B}{2003Ri08}), in\vspace{0.1cm}}&\\
&&&&&&&&&&&\parbox[t][0.3cm]{9.864cm}{\raggedright {\ }{\ }{\ }turn, yields a proton matrix element M\ensuremath{_{\textnormal{p}}}(\ensuremath{^{\textnormal{18}}}Ne)=0.149 \textit{7} (\href{https://www.nndc.bnl.gov/nsr/nsrlink.jsp?2006ChZY,B}{2006ChZY}).\vspace{0.1cm}}&\\
&&&&&&&&&&&\parbox[t][0.3cm]{9.864cm}{\raggedright By adopting the mirror hypothesis, (\href{https://www.nndc.bnl.gov/nsr/nsrlink.jsp?2003Ri08,B}{2003Ri08}) deduced B(E2:\vspace{0.1cm}}&\\
&&&&&&&&&&&\parbox[t][0.3cm]{9.864cm}{\raggedright {\ }{\ }{\ }0\ensuremath{^{\textnormal{+}}_{\textnormal{g.s.}}}\ensuremath{\rightarrow}2\ensuremath{^{\textnormal{+}}_{\textnormal{1}}})=190 e\ensuremath{^{\textnormal{2}}}fm\ensuremath{^{\textnormal{4}}} \textit{14} from the \ensuremath{^{\textnormal{18}}}O empirical neutron transition\vspace{0.1cm}}&\\
&&&&&&&&&&&\parbox[t][0.3cm]{9.864cm}{\raggedright {\ }{\ }{\ }density. This calculation yields an integrated cross section of 52.7 mb\vspace{0.1cm}}&\\
&&&&&&&&&&&\parbox[t][0.3cm]{9.864cm}{\raggedright {\ }{\ }{\ }for the excitation of the first 2\ensuremath{^{\textnormal{+}}} state of \ensuremath{^{\textnormal{18}}}Ne. This cross section is\vspace{0.1cm}}&\\
&&&&&&&&&&&\parbox[t][0.3cm]{9.864cm}{\raggedright {\ }{\ }{\ }very close to the measured value of 40 mb \textit{11} from (\href{https://www.nndc.bnl.gov/nsr/nsrlink.jsp?2000Ri15,B}{2000Ri15}).\vspace{0.1cm}}&\\
&&&&&&&&&&&\parbox[t][0.3cm]{9.864cm}{\raggedright (\href{https://www.nndc.bnl.gov/nsr/nsrlink.jsp?2003Ta13,B}{2003Ta13}): a broad resonance like structure dominates the\vspace{0.1cm}}&\\
&&&&&&&&&&&\parbox[t][0.3cm]{9.864cm}{\raggedright {\ }{\ }{\ }\ensuremath{^{\textnormal{16}}}O(\ensuremath{^{\textnormal{3}}}He,n\ensuremath{\gamma}\ensuremath{_{\textnormal{1887}}})\ensuremath{^{\textnormal{18}}}Ne* excitation function at E\ensuremath{_{\textnormal{lab}}}\ensuremath{\sim}9.5 MeV, which\vspace{0.1cm}}&\\
&&&&&&&&&&&\parbox[t][0.3cm]{9.864cm}{\raggedright {\ }{\ }{\ }corresponds to an excitation energy of \ensuremath{\sim}16.4 MeV in the compound\vspace{0.1cm}}&\\
&&&&&&&&&&&\parbox[t][0.3cm]{9.864cm}{\raggedright {\ }{\ }{\ }nucleus \ensuremath{^{\textnormal{19}}}Ne. A narrower resonance was also observed at E\ensuremath{_{\textnormal{lab}}}\ensuremath{\sim}6.5\vspace{0.1cm}}&\\
&&&&&&&&&&&\parbox[t][0.3cm]{9.864cm}{\raggedright {\ }{\ }{\ }MeV, which could correspond to the 13.8-MeV level of \ensuremath{^{\textnormal{19}}}Ne\vspace{0.1cm}}&\\
&&&&&&&&&&&\parbox[t][0.3cm]{9.864cm}{\raggedright {\ }{\ }{\ }(\href{https://www.nndc.bnl.gov/nsr/nsrlink.jsp?1995Ti07,B}{1995Ti07}).\vspace{0.1cm}}&\\
&&&&&&&&&&&\parbox[t][0.3cm]{9.864cm}{\raggedright Maximum production cross section for this transition was reported as\vspace{0.1cm}}&\\
&&&&&&&&&&&\parbox[t][0.3cm]{9.864cm}{\raggedright {\ }{\ }{\ }\ensuremath{\sim}65 mb at E\ensuremath{_{\textnormal{lab}}}=10 MeV (\href{https://www.nndc.bnl.gov/nsr/nsrlink.jsp?2003Ta13,B}{2003Ta13}).\vspace{0.1cm}}&\\
\multicolumn{1}{r@{}}{3376}&\multicolumn{1}{@{.}l}{4}&\multicolumn{1}{l}{4\ensuremath{^{+}}}&\multicolumn{1}{r@{}}{1488}&\multicolumn{1}{@{.}l}{9 {\it 3}}&\multicolumn{1}{r@{}}{100}&\multicolumn{1}{@{}l}{}&\multicolumn{1}{r@{}}{1887}&\multicolumn{1}{@{.}l}{4 }&\multicolumn{1}{@{}l}{2\ensuremath{^{+}}}&\multicolumn{1}{l}{E2}&\parbox[t][0.3cm]{9.864cm}{\raggedright B(E2)(W.u.)=9.1 \textit{+15{\textminus}11}\vspace{0.1cm}}&\\
&&&&&&&&&&&\parbox[t][0.3cm]{9.864cm}{\raggedright E\ensuremath{_{\gamma}}: From (\href{https://www.nndc.bnl.gov/nsr/nsrlink.jsp?1968Gi09,B}{1968Gi09}, \href{https://www.nndc.bnl.gov/nsr/nsrlink.jsp?1969Ro08,B}{1969Ro08}). See also 1489 keV (\href{https://www.nndc.bnl.gov/nsr/nsrlink.jsp?1969Ro22,B}{1969Ro22},\vspace{0.1cm}}&\\
&&&&&&&&&&&\parbox[t][0.3cm]{9.864cm}{\raggedright {\ }{\ }{\ }\href{https://www.nndc.bnl.gov/nsr/nsrlink.jsp?1971Ro18,B}{1971Ro18}, \href{https://www.nndc.bnl.gov/nsr/nsrlink.jsp?1972Gi01,B}{1972Gi01}, \href{https://www.nndc.bnl.gov/nsr/nsrlink.jsp?2003Ta13,B}{2003Ta13}); and 1493 keV \textit{2} (\href{https://www.nndc.bnl.gov/nsr/nsrlink.jsp?1970Sh04,B}{1970Sh04}).\vspace{0.1cm}}&\\
&&&&&&&&&&&\parbox[t][0.3cm]{9.864cm}{\raggedright Mult.: From (\href{https://www.nndc.bnl.gov/nsr/nsrlink.jsp?1972Gi01,B}{1972Gi01}). See also E2/M3 (\href{https://www.nndc.bnl.gov/nsr/nsrlink.jsp?1969Ro08,B}{1969Ro08}, \href{https://www.nndc.bnl.gov/nsr/nsrlink.jsp?1969Ro22,B}{1969Ro22},\vspace{0.1cm}}&\\
&&&&&&&&&&&\parbox[t][0.3cm]{9.864cm}{\raggedright {\ }{\ }{\ }\href{https://www.nndc.bnl.gov/nsr/nsrlink.jsp?1970Sh04,B}{1970Sh04}) with a mixing ratio of \ensuremath{\delta}=0.04 \textit{3} determined from the\vspace{0.1cm}}&\\
&&&&&&&&&&&\parbox[t][0.3cm]{9.864cm}{\raggedright {\ }{\ }{\ }weighted average of +0.06 \textit{7} (\href{https://www.nndc.bnl.gov/nsr/nsrlink.jsp?1969Ro08,B}{1969Ro08}: see Table 2 assuming J=4);\vspace{0.1cm}}&\\
&&&&&&&&&&&\parbox[t][0.3cm]{9.864cm}{\raggedright {\ }{\ }{\ }+0.00 \textit{4} (\href{https://www.nndc.bnl.gov/nsr/nsrlink.jsp?1969Ro22,B}{1969Ro22}: see Fig. 3 assuming J=4); and +0.12 \textit{7}\vspace{0.1cm}}&\\
&&&&&&&&&&&\parbox[t][0.3cm]{9.864cm}{\raggedright {\ }{\ }{\ }(\href{https://www.nndc.bnl.gov/nsr/nsrlink.jsp?1970Sh04,B}{1970Sh04}: see Fig. 1). See also \ensuremath{\delta}=+8 \textit{3} (\href{https://www.nndc.bnl.gov/nsr/nsrlink.jsp?1969Ro08,B}{1969Ro08}: see Table 2\vspace{0.1cm}}&\\
&&&&&&&&&&&\parbox[t][0.3cm]{9.864cm}{\raggedright {\ }{\ }{\ }assuming J=4). (\href{https://www.nndc.bnl.gov/nsr/nsrlink.jsp?1969Ro08,B}{1969Ro08}) mentions that the latter value may be\vspace{0.1cm}}&\\
&&&&&&&&&&&\parbox[t][0.3cm]{9.864cm}{\raggedright {\ }{\ }{\ }excluded because this value together with the measured lifetime by\vspace{0.1cm}}&\\
&&&&&&&&&&&\parbox[t][0.3cm]{9.864cm}{\raggedright {\ }{\ }{\ }(\href{https://www.nndc.bnl.gov/nsr/nsrlink.jsp?1969Ro08,B}{1969Ro08}) would imply an enormously enhanced M3 matrix element\vspace{0.1cm}}&\\
&&&&&&&&&&&\parbox[t][0.3cm]{9.864cm}{\raggedright {\ }{\ }{\ }(\ensuremath{\approx}10\ensuremath{^{\textnormal{4}}}). Note that (\href{https://www.nndc.bnl.gov/nsr/nsrlink.jsp?1970Sh04,B}{1970Sh04}) and (\href{https://www.nndc.bnl.gov/nsr/nsrlink.jsp?1972Gi01,B}{1972Gi01}) both used the phase\vspace{0.1cm}}&\\
&&&&&&&&&&&\parbox[t][0.3cm]{9.864cm}{\raggedright {\ }{\ }{\ }convention of (\href{https://www.nndc.bnl.gov/nsr/nsrlink.jsp?1967Ro21,B}{1967Ro21}) to deduce their mixing ratios. The\vspace{0.1cm}}&\\
&&&&&&&&&&&\parbox[t][0.3cm]{9.864cm}{\raggedright {\ }{\ }{\ }uncertainty in the mixing ratio of (\href{https://www.nndc.bnl.gov/nsr/nsrlink.jsp?1970Sh04,B}{1970Sh04}) is statistical.\vspace{0.1cm}}&\\
&&&&&&&&&&&\parbox[t][0.3cm]{9.864cm}{\raggedright I\ensuremath{_{\gamma}}: From (\href{https://www.nndc.bnl.gov/nsr/nsrlink.jsp?1968Gi09,B}{1968Gi09}, \href{https://www.nndc.bnl.gov/nsr/nsrlink.jsp?1969Ro08,B}{1969Ro08}, \href{https://www.nndc.bnl.gov/nsr/nsrlink.jsp?1969Ro22,B}{1969Ro22}: see Fig. 3), and (\href{https://www.nndc.bnl.gov/nsr/nsrlink.jsp?1970Sh04,B}{1970Sh04}:\vspace{0.1cm}}&\\
&&&&&&&&&&&\parbox[t][0.3cm]{9.864cm}{\raggedright {\ }{\ }{\ }see Fig. 1).\vspace{0.1cm}}&\\
&&&&&&&&&&&\parbox[t][0.3cm]{9.864cm}{\raggedright (\href{https://www.nndc.bnl.gov/nsr/nsrlink.jsp?1969Ro08,B}{1969Ro08}) deduced coefficients of Legendre polynomials from the\vspace{0.1cm}}&\\
&&&&&&&&&&&\parbox[t][0.3cm]{9.864cm}{\raggedright {\ }{\ }{\ }\ensuremath{\gamma}-ray angular distributions: a\ensuremath{_{\textnormal{2}}}=0.37 \textit{6} and a\ensuremath{_{\textnormal{4}}}={\textminus}0.28 \textit{9}.\vspace{0.1cm}}&\\
&&&&&&&&&&&\parbox[t][0.3cm]{9.864cm}{\raggedright (\href{https://www.nndc.bnl.gov/nsr/nsrlink.jsp?1969Ro22,B}{1969Ro22}) deduced a\ensuremath{_{\textnormal{2}}}=0.22 \textit{9} and a\ensuremath{_{\textnormal{4}}}={\textminus}0.20 \textit{10} at E\ensuremath{_{\textnormal{lab}}}=10 MeV;\vspace{0.1cm}}&\\
&&&&&&&&&&&\parbox[t][0.3cm]{9.864cm}{\raggedright {\ }{\ }{\ }a\ensuremath{_{\textnormal{2}}}=0.61 \textit{10} and a\ensuremath{_{\textnormal{4}}}={\textminus}0.11 \textit{11} at E\ensuremath{_{\textnormal{lab}}}=10.2 MeV; and a\ensuremath{_{\textnormal{2}}}=0.38 \textit{10} and\vspace{0.1cm}}&\\
\end{longtable}
\begin{textblock}{29}(0,27.3)
Continued on next page (footnotes at end of table)
\end{textblock}
\clearpage
\begin{longtable}{ccccccccc@{}ccc@{\extracolsep{\fill}}c}
\\[-.4cm]
\multicolumn{13}{c}{{\bf \small \underline{\ensuremath{^{\textnormal{16}}}O(\ensuremath{^{\textnormal{3}}}He,n\ensuremath{\gamma})\hspace{0.2in}\href{https://www.nndc.bnl.gov/nsr/nsrlink.jsp?1968Gi09,B}{1968Gi09},\href{https://www.nndc.bnl.gov/nsr/nsrlink.jsp?2003Ta13,B}{2003Ta13} (continued)}}}\\
\multicolumn{13}{c}{~}\\
\multicolumn{13}{c}{\underline{$\gamma$($^{18}$Ne) (continued)}}\\
\multicolumn{13}{c}{~~~}\\
\multicolumn{2}{c}{E\ensuremath{_{i}}(level)}&J\ensuremath{^{\pi}_{i}}&\multicolumn{2}{c}{E\ensuremath{_{\gamma}}\ensuremath{^{\hyperlink{NE16GAMMA0}{a}}}}&\multicolumn{2}{c}{I\ensuremath{_{\gamma}}}&\multicolumn{2}{c}{E\ensuremath{_{f}}}&J\ensuremath{^{\pi}_{f}}&Mult.&Comments&\\[-.2cm]
\multicolumn{2}{c}{\hrulefill}&\hrulefill&\multicolumn{2}{c}{\hrulefill}&\multicolumn{2}{c}{\hrulefill}&\multicolumn{2}{c}{\hrulefill}&\hrulefill&\hrulefill&\hrulefill&
\endhead
&&&&&&&&&&&\parbox[t][0.3cm]{10.103681cm}{\raggedright {\ }{\ }{\ }a\ensuremath{_{\textnormal{4}}}={\textminus}0.40 \textit{12} at E\ensuremath{_{\textnormal{lab}}}=11.5 MeV from the measured \ensuremath{\gamma}-ray angular\vspace{0.1cm}}&\\
&&&&&&&&&&&\parbox[t][0.3cm]{10.103681cm}{\raggedright {\ }{\ }{\ }correlations for the 1488.9-keV transition.\vspace{0.1cm}}&\\
&&&&&&&&&&&\parbox[t][0.3cm]{10.103681cm}{\raggedright The fit to the n\ensuremath{\gamma} angular correlation data of (\href{https://www.nndc.bnl.gov/nsr/nsrlink.jsp?1970Sh04,B}{1970Sh04}) with J=4 is the\vspace{0.1cm}}&\\
&&&&&&&&&&&\parbox[t][0.3cm]{10.103681cm}{\raggedright {\ }{\ }{\ }best fit with 0.1\% confidence level (\href{https://www.nndc.bnl.gov/nsr/nsrlink.jsp?1970Sh04,B}{1970Sh04}: see Fig. 9).\vspace{0.1cm}}&\\
&&&&&&&&&&&\parbox[t][0.3cm]{10.103681cm}{\raggedright Using \ensuremath{\tau}=4.4 ps \textit{6} (see \href{https://www.nndc.bnl.gov/nsr/nsrlink.jsp?1974Mc17,B}{1974Mc17}) for the \ensuremath{^{\textnormal{18}}}Ne*(3380 keV) state,\vspace{0.1cm}}&\\
&&&&&&&&&&&\parbox[t][0.3cm]{10.103681cm}{\raggedright {\ }{\ }{\ }(\href{https://www.nndc.bnl.gov/nsr/nsrlink.jsp?1976Mc02,B}{1976Mc02}) calculated an experimental value of B(E2: 4\ensuremath{^{\textnormal{+}}_{\textnormal{1}}}\ensuremath{\rightarrow}2\ensuremath{^{\textnormal{+}}_{\textnormal{1}}})=25\vspace{0.1cm}}&\\
&&&&&&&&&&&\parbox[t][0.3cm]{10.103681cm}{\raggedright {\ }{\ }{\ }e\ensuremath{^{\textnormal{2}}}fm\ensuremath{^{\textnormal{4}}} \textit{3} for this transition. This value should be compared with\vspace{0.1cm}}&\\
&&&&&&&&&&&\parbox[t][0.3cm]{10.103681cm}{\raggedright {\ }{\ }{\ }calculated values of B(E2)=26 e\ensuremath{^{\textnormal{2}}}fm\ensuremath{^{\textnormal{4}}} (T. Engeland and P. J. Ellis, Nucl.\vspace{0.1cm}}&\\
&&&&&&&&&&&\parbox[t][0.3cm]{10.103681cm}{\raggedright {\ }{\ }{\ }Phys. A 181 (1972) 368: from shell model); B(E2)=26 e\ensuremath{^{\textnormal{2}}}fm\ensuremath{^{\textnormal{4}}}\vspace{0.1cm}}&\\
&&&&&&&&&&&\parbox[t][0.3cm]{10.103681cm}{\raggedright {\ }{\ }{\ }(\href{https://www.nndc.bnl.gov/nsr/nsrlink.jsp?1976Mc02,B}{1976Mc02}: calculated assuming that the wave functions for the states\vspace{0.1cm}}&\\
&&&&&&&&&&&\parbox[t][0.3cm]{10.103681cm}{\raggedright {\ }{\ }{\ }in \ensuremath{^{\textnormal{18}}}Ne have the same structure as the corresponding T=1 states in\vspace{0.1cm}}&\\
&&&&&&&&&&&\parbox[t][0.3cm]{10.103681cm}{\raggedright {\ }{\ }{\ }\ensuremath{^{\textnormal{18}}}O); B(E2)=33 e\ensuremath{^{\textnormal{2}}}fm\ensuremath{^{\textnormal{4}}} (\href{https://www.nndc.bnl.gov/nsr/nsrlink.jsp?1976Mc02,B}{1976Mc02}: from the phenomenological\vspace{0.1cm}}&\\
&&&&&&&&&&&\parbox[t][0.3cm]{10.103681cm}{\raggedright {\ }{\ }{\ }estimates of (\href{https://www.nndc.bnl.gov/nsr/nsrlink.jsp?1970Ha75,B}{1970Ha75}, \href{https://www.nndc.bnl.gov/nsr/nsrlink.jsp?1970Ha49,B}{1970Ha49}) for the two-body part of the\vspace{0.1cm}}&\\
&&&&&&&&&&&\parbox[t][0.3cm]{10.103681cm}{\raggedright {\ }{\ }{\ }effective transition operator); and B(E2)=26 e\ensuremath{^{\textnormal{2}}}fm\ensuremath{^{\textnormal{4}}} (\href{https://www.nndc.bnl.gov/nsr/nsrlink.jsp?1976Mc02,B}{1976Mc02}: from the\vspace{0.1cm}}&\\
&&&&&&&&&&&\parbox[t][0.3cm]{10.103681cm}{\raggedright {\ }{\ }{\ }realistic reaction matrix elements of (F. C. Khanna, M. Harvey, D. W.\vspace{0.1cm}}&\\
&&&&&&&&&&&\parbox[t][0.3cm]{10.103681cm}{\raggedright {\ }{\ }{\ }L. Sprung and A. Jopko, The two body force in nuclei, ed. S. M.\vspace{0.1cm}}&\\
&&&&&&&&&&&\parbox[t][0.3cm]{10.103681cm}{\raggedright {\ }{\ }{\ }Austin and G. M. Crawley (Plenum Press, New York, 1972) p. 229)).\vspace{0.1cm}}&\\
&&&&&&&&&&&\parbox[t][0.3cm]{10.103681cm}{\raggedright (\href{https://www.nndc.bnl.gov/nsr/nsrlink.jsp?1970Sh04,B}{1970Sh04}) found that at \ensuremath{\theta}\ensuremath{_{\ensuremath{\gamma}\textnormal{,lab}}}=90\ensuremath{^\circ} the ratio of the yield of the\vspace{0.1cm}}&\\
&&&&&&&&&&&\parbox[t][0.3cm]{10.103681cm}{\raggedright {\ }{\ }{\ }1.87-MeV to the 1.49-MeV \ensuremath{\gamma} rays is 1.01 \textit{4}. This is consistent with\vspace{0.1cm}}&\\
&&&&&&&&&&&\parbox[t][0.3cm]{10.103681cm}{\raggedright {\ }{\ }{\ }J\ensuremath{^{\ensuremath{\pi}}}=4\ensuremath{^{\textnormal{+}}} assignment for the 3376.4-keV level because both of the \ensuremath{\gamma}-ray\vspace{0.1cm}}&\\
&&&&&&&&&&&\parbox[t][0.3cm]{10.103681cm}{\raggedright {\ }{\ }{\ }transitions involved in the 4\ensuremath{^{\textnormal{+}}_{\textnormal{1}}}\ensuremath{\rightarrow}2\ensuremath{^{\textnormal{+}}_{\textnormal{1}}}\ensuremath{\rightarrow}0\ensuremath{^{\textnormal{+}}_{\textnormal{1}}} cascade have the same\vspace{0.1cm}}&\\
&&&&&&&&&&&\parbox[t][0.3cm]{10.103681cm}{\raggedright {\ }{\ }{\ }angular correlation with respect to the corresponding neutron groups\vspace{0.1cm}}&\\
&&&&&&&&&&&\parbox[t][0.3cm]{10.103681cm}{\raggedright {\ }{\ }{\ }independent of the reaction mechanism.\vspace{0.1cm}}&\\
&&&&&&&&&&&\parbox[t][0.3cm]{10.103681cm}{\raggedright The ground state decay branch (3376 keV \ensuremath{\rightarrow} g.s.) was not observed. The\vspace{0.1cm}}&\\
&&&&&&&&&&&\parbox[t][0.3cm]{10.103681cm}{\raggedright {\ }{\ }{\ }branching ratio for this unobserved branch was estimated to be \ensuremath{<}1\%\vspace{0.1cm}}&\\
&&&&&&&&&&&\parbox[t][0.3cm]{10.103681cm}{\raggedright {\ }{\ }{\ }(\href{https://www.nndc.bnl.gov/nsr/nsrlink.jsp?1969Ro22,B}{1969Ro22}: see Fig. 3); \ensuremath{<}1\% (\href{https://www.nndc.bnl.gov/nsr/nsrlink.jsp?1970Sh04,B}{1970Sh04}: see Fig. 1); and \ensuremath{<}4\%\vspace{0.1cm}}&\\
&&&&&&&&&&&\parbox[t][0.3cm]{10.103681cm}{\raggedright {\ }{\ }{\ }(\href{https://www.nndc.bnl.gov/nsr/nsrlink.jsp?1968Gi09,B}{1968Gi09}, \href{https://www.nndc.bnl.gov/nsr/nsrlink.jsp?1969Ro08,B}{1969Ro08}).\vspace{0.1cm}}&\\
\multicolumn{1}{r@{}}{3576}&\multicolumn{1}{@{.}l}{5}&\multicolumn{1}{l}{0\ensuremath{^{+}}}&\multicolumn{1}{r@{}}{1689}&\multicolumn{1}{@{ }l}{{\it 2}}&\multicolumn{1}{r@{}}{100}&\multicolumn{1}{@{}l}{}&\multicolumn{1}{r@{}}{1887}&\multicolumn{1}{@{.}l}{4 }&\multicolumn{1}{@{}l}{2\ensuremath{^{+}}}&\multicolumn{1}{l}{E2}&\parbox[t][0.3cm]{10.103681cm}{\raggedright B(E2)(W.u.)=5.3 \textit{+46{\textminus}18}\vspace{0.1cm}}&\\
&&&&&&&&&&&\parbox[t][0.3cm]{10.103681cm}{\raggedright E\ensuremath{_{\gamma}}: From (\href{https://www.nndc.bnl.gov/nsr/nsrlink.jsp?1968Gi09,B}{1968Gi09}, \href{https://www.nndc.bnl.gov/nsr/nsrlink.jsp?1969Ro08,B}{1969Ro08}). See also 1689 keV (\href{https://www.nndc.bnl.gov/nsr/nsrlink.jsp?1969Ro22,B}{1969Ro22},\vspace{0.1cm}}&\\
&&&&&&&&&&&\parbox[t][0.3cm]{10.103681cm}{\raggedright {\ }{\ }{\ }\href{https://www.nndc.bnl.gov/nsr/nsrlink.jsp?1971Ro18,B}{1971Ro18}, \href{https://www.nndc.bnl.gov/nsr/nsrlink.jsp?1972Gi01,B}{1972Gi01}); and 1689 keV \textit{2} (\href{https://www.nndc.bnl.gov/nsr/nsrlink.jsp?1970Sh04,B}{1970Sh04}: which is most\vspace{0.1cm}}&\\
&&&&&&&&&&&\parbox[t][0.3cm]{10.103681cm}{\raggedright {\ }{\ }{\ }likely what was reported by (\href{https://www.nndc.bnl.gov/nsr/nsrlink.jsp?1969Ro08,B}{1969Ro08})).\vspace{0.1cm}}&\\
&&&&&&&&&&&\parbox[t][0.3cm]{10.103681cm}{\raggedright E\ensuremath{_{\gamma}}: This \ensuremath{\gamma} ray was first observed by (\href{https://www.nndc.bnl.gov/nsr/nsrlink.jsp?1968Gi09,B}{1968Gi09}, \href{https://www.nndc.bnl.gov/nsr/nsrlink.jsp?1969Ro08,B}{1969Ro08}). (\href{https://www.nndc.bnl.gov/nsr/nsrlink.jsp?1969Ro22,B}{1969Ro22})\vspace{0.1cm}}&\\
&&&&&&&&&&&\parbox[t][0.3cm]{10.103681cm}{\raggedright {\ }{\ }{\ }confirmed that this \ensuremath{\gamma} ray belongs to \ensuremath{^{\textnormal{18}}}Ne by using n\ensuremath{\gamma}\ensuremath{\gamma} coincidences\vspace{0.1cm}}&\\
&&&&&&&&&&&\parbox[t][0.3cm]{10.103681cm}{\raggedright {\ }{\ }{\ }gating on the 1887-keV \ensuremath{\gamma} ray.\vspace{0.1cm}}&\\
&&&&&&&&&&&\parbox[t][0.3cm]{10.103681cm}{\raggedright I\ensuremath{_{\gamma}}: From (\href{https://www.nndc.bnl.gov/nsr/nsrlink.jsp?1968Gi09,B}{1968Gi09}, \href{https://www.nndc.bnl.gov/nsr/nsrlink.jsp?1969Ro08,B}{1969Ro08}, \href{https://www.nndc.bnl.gov/nsr/nsrlink.jsp?1969Ro22,B}{1969Ro22}: see Fig. 3).\vspace{0.1cm}}&\\
&&&&&&&&&&&\parbox[t][0.3cm]{10.103681cm}{\raggedright Mult.: From (\href{https://www.nndc.bnl.gov/nsr/nsrlink.jsp?1972Gi01,B}{1972Gi01}).\vspace{0.1cm}}&\\
&&&&&&&&&&&\parbox[t][0.3cm]{10.103681cm}{\raggedright (\href{https://www.nndc.bnl.gov/nsr/nsrlink.jsp?1969Ro08,B}{1969Ro08}) deduced coefficients of Legendre polynomials from the \ensuremath{\gamma}-ray\vspace{0.1cm}}&\\
&&&&&&&&&&&\parbox[t][0.3cm]{10.103681cm}{\raggedright {\ }{\ }{\ }angular distributions: a\ensuremath{_{\textnormal{2}}}=0.0 \textit{3}, a\ensuremath{_{\textnormal{4}}}=0.0 \textit{5}.\vspace{0.1cm}}&\\
&&&&&&&&&&&\parbox[t][0.3cm]{10.103681cm}{\raggedright (\href{https://www.nndc.bnl.gov/nsr/nsrlink.jsp?1969Ro22,B}{1969Ro22}) deduced a\ensuremath{_{\textnormal{2}}}=0.0 \textit{3} and a\ensuremath{_{\textnormal{4}}}={\textminus}0.1 \textit{6} at E\ensuremath{_{\textnormal{lab}}}=11.5 MeV from the\vspace{0.1cm}}&\\
&&&&&&&&&&&\parbox[t][0.3cm]{10.103681cm}{\raggedright {\ }{\ }{\ }measured \ensuremath{\gamma}-ray angular correlations.\vspace{0.1cm}}&\\
&&&&&&&&&&&\parbox[t][0.3cm]{10.103681cm}{\raggedright Assuming \ensuremath{\tau}=4 ps \textit{2} (\href{https://www.nndc.bnl.gov/nsr/nsrlink.jsp?1972Gi01,B}{1972Gi01}), (\href{https://www.nndc.bnl.gov/nsr/nsrlink.jsp?1974Mc17,B}{1974Mc17}) calculated B(E2:\vspace{0.1cm}}&\\
&&&&&&&&&&&\parbox[t][0.3cm]{10.103681cm}{\raggedright {\ }{\ }{\ }0\ensuremath{^{\textnormal{+}}_{\textnormal{2}}}\ensuremath{\rightarrow}2\ensuremath{^{\textnormal{+}}_{\textnormal{1}}})=14.5 e\ensuremath{^{\textnormal{2}}}fm\ensuremath{^{\textnormal{4}}} \textit{74} for this transition. This value should be\vspace{0.1cm}}&\\
&&&&&&&&&&&\parbox[t][0.3cm]{10.103681cm}{\raggedright {\ }{\ }{\ }compared with B(E2)=3.36 e\ensuremath{^{\textnormal{2}}}fm\ensuremath{^{\textnormal{4}}} calculated by (T. Engeland and P. J.\vspace{0.1cm}}&\\
&&&&&&&&&&&\parbox[t][0.3cm]{10.103681cm}{\raggedright {\ }{\ }{\ }Ellis, Nucl. Phys. A 181 (1972) 368).\vspace{0.1cm}}&\\
&&&&&&&&&&&\parbox[t][0.3cm]{10.103681cm}{\raggedright The decay branch from the 3576-keV state to the ground state was not\vspace{0.1cm}}&\\
&&&&&&&&&&&\parbox[t][0.3cm]{10.103681cm}{\raggedright {\ }{\ }{\ }observed. The branching ratio for this unobserved branch was estimated\vspace{0.1cm}}&\\
&&&&&&&&&&&\parbox[t][0.3cm]{10.103681cm}{\raggedright {\ }{\ }{\ }to be \ensuremath{<}5\% (\href{https://www.nndc.bnl.gov/nsr/nsrlink.jsp?1969Ro22,B}{1969Ro22}: see Fig. 3); and \ensuremath{<}17\% (\href{https://www.nndc.bnl.gov/nsr/nsrlink.jsp?1968Gi09,B}{1968Gi09}, \href{https://www.nndc.bnl.gov/nsr/nsrlink.jsp?1969Ro08,B}{1969Ro08}).\vspace{0.1cm}}&\\
\end{longtable}
\begin{textblock}{29}(0,27.3)
Continued on next page (footnotes at end of table)
\end{textblock}
\clearpage
\begin{longtable}{ccccccccc@{}ccccc@{\extracolsep{\fill}}c}
\\[-.4cm]
\multicolumn{15}{c}{{\bf \small \underline{\ensuremath{^{\textnormal{16}}}O(\ensuremath{^{\textnormal{3}}}He,n\ensuremath{\gamma})\hspace{0.2in}\href{https://www.nndc.bnl.gov/nsr/nsrlink.jsp?1968Gi09,B}{1968Gi09},\href{https://www.nndc.bnl.gov/nsr/nsrlink.jsp?2003Ta13,B}{2003Ta13} (continued)}}}\\
\multicolumn{15}{c}{~}\\
\multicolumn{15}{c}{\underline{$\gamma$($^{18}$Ne) (continued)}}\\
\multicolumn{15}{c}{~~~}\\
\multicolumn{2}{c}{E\ensuremath{_{i}}(level)}&J\ensuremath{^{\pi}_{i}}&\multicolumn{2}{c}{E\ensuremath{_{\gamma}}\ensuremath{^{\hyperlink{NE16GAMMA0}{a}}}}&\multicolumn{2}{c}{I\ensuremath{_{\gamma}}}&\multicolumn{2}{c}{E\ensuremath{_{f}}}&J\ensuremath{^{\pi}_{f}}&Mult.&\multicolumn{2}{c}{\ensuremath{\delta}}&Comments&\\[-.2cm]
\multicolumn{2}{c}{\hrulefill}&\hrulefill&\multicolumn{2}{c}{\hrulefill}&\multicolumn{2}{c}{\hrulefill}&\multicolumn{2}{c}{\hrulefill}&\hrulefill&\hrulefill&\multicolumn{2}{c}{\hrulefill}&\hrulefill&
\endhead
\multicolumn{1}{r@{}}{3616}&\multicolumn{1}{@{.}l}{6}&\multicolumn{1}{l}{2\ensuremath{^{+}}}&\multicolumn{1}{r@{}}{1729}&\multicolumn{1}{@{.}l}{2 {\it 5}}&\multicolumn{1}{r@{}}{91}&\multicolumn{1}{@{ }l}{{\it 3}}&\multicolumn{1}{r@{}}{1887}&\multicolumn{1}{@{.}l}{4 }&\multicolumn{1}{@{}l}{2\ensuremath{^{+}}}&\multicolumn{1}{l}{M1+E2}&\multicolumn{1}{r@{}}{0}&\multicolumn{1}{@{.}l}{05 {\it 7}}&\parbox[t][0.3cm]{8.156461cm}{\raggedright B(M1)(W.u.)=0.088 \textit{+45{\textminus}31}; B(E2)(W.u.)\ensuremath{<}5.8\vspace{0.1cm}}&\\
&&&&&&&&&&&&&\parbox[t][0.3cm]{8.156461cm}{\raggedright E\ensuremath{_{\gamma}}: From (\href{https://www.nndc.bnl.gov/nsr/nsrlink.jsp?1968Gi09,B}{1968Gi09}, \href{https://www.nndc.bnl.gov/nsr/nsrlink.jsp?1969Ro08,B}{1969Ro08}). See also 1729 keV\vspace{0.1cm}}&\\
&&&&&&&&&&&&&\parbox[t][0.3cm]{8.156461cm}{\raggedright {\ }{\ }{\ }(\href{https://www.nndc.bnl.gov/nsr/nsrlink.jsp?1969Ro22,B}{1969Ro22}, \href{https://www.nndc.bnl.gov/nsr/nsrlink.jsp?1971Ro18,B}{1971Ro18}, \href{https://www.nndc.bnl.gov/nsr/nsrlink.jsp?1972Gi01,B}{1972Gi01}); and 1736 keV \textit{2}\vspace{0.1cm}}&\\
&&&&&&&&&&&&&\parbox[t][0.3cm]{8.156461cm}{\raggedright {\ }{\ }{\ }(\href{https://www.nndc.bnl.gov/nsr/nsrlink.jsp?1970Sh04,B}{1970Sh04}).\vspace{0.1cm}}&\\
&&&&&&&&&&&&&\parbox[t][0.3cm]{8.156461cm}{\raggedright I\ensuremath{_{\gamma}}: Weighted average (rounded to the nearest integer) of\vspace{0.1cm}}&\\
&&&&&&&&&&&&&\parbox[t][0.3cm]{8.156461cm}{\raggedright {\ }{\ }{\ }93\% \textit{2} (\href{https://www.nndc.bnl.gov/nsr/nsrlink.jsp?1969Ro22,B}{1969Ro22}: see Fig. 3) and 87.5\% \textit{25} (\href{https://www.nndc.bnl.gov/nsr/nsrlink.jsp?1972Gi01,B}{1972Gi01}).\vspace{0.1cm}}&\\
&&&&&&&&&&&&&\parbox[t][0.3cm]{8.156461cm}{\raggedright {\ }{\ }{\ }See also 100\% (\href{https://www.nndc.bnl.gov/nsr/nsrlink.jsp?1968Gi09,B}{1968Gi09}, \href{https://www.nndc.bnl.gov/nsr/nsrlink.jsp?1969Ro08,B}{1969Ro08}, \href{https://www.nndc.bnl.gov/nsr/nsrlink.jsp?1970Sh04,B}{1970Sh04}: see Fig.\vspace{0.1cm}}&\\
&&&&&&&&&&&&&\parbox[t][0.3cm]{8.156461cm}{\raggedright {\ }{\ }{\ }1).\vspace{0.1cm}}&\\
&&&&&&&&&&&&&\parbox[t][0.3cm]{8.156461cm}{\raggedright \ensuremath{\delta}: Weighted average (with external errors) of {\textminus}1.1 \textit{+10{\textminus}3}\vspace{0.1cm}}&\\
&&&&&&&&&&&&&\parbox[t][0.3cm]{8.156461cm}{\raggedright {\ }{\ }{\ }(\href{https://www.nndc.bnl.gov/nsr/nsrlink.jsp?1969Ro08,B}{1969Ro08}: see Table 2 assuming J=2); {\textminus}0.9 \textit{7}\vspace{0.1cm}}&\\
&&&&&&&&&&&&&\parbox[t][0.3cm]{8.156461cm}{\raggedright {\ }{\ }{\ }(\href{https://www.nndc.bnl.gov/nsr/nsrlink.jsp?1969Ro22,B}{1969Ro22}: see Fig. 3); +0.09 \textit{7} (\href{https://www.nndc.bnl.gov/nsr/nsrlink.jsp?1970Sh04,B}{1970Sh04}: see Fig. 1;\vspace{0.1cm}}&\\
&&&&&&&&&&&&&\parbox[t][0.3cm]{8.156461cm}{\raggedright {\ }{\ }{\ }used the phase convention of (\href{https://www.nndc.bnl.gov/nsr/nsrlink.jsp?1967Ro21,B}{1967Ro21}) to deduce the\vspace{0.1cm}}&\\
&&&&&&&&&&&&&\parbox[t][0.3cm]{8.156461cm}{\raggedright {\ }{\ }{\ }mixing ratio. The uncertainty in this mixing ratio is\vspace{0.1cm}}&\\
&&&&&&&&&&&&&\parbox[t][0.3cm]{8.156461cm}{\raggedright {\ }{\ }{\ }statistical); and +0.03 \textit{9} (\href{https://www.nndc.bnl.gov/nsr/nsrlink.jsp?1972Gi01,B}{1972Gi01}: the mixing ratio was\vspace{0.1cm}}&\\
&&&&&&&&&&&&&\parbox[t][0.3cm]{8.156461cm}{\raggedright {\ }{\ }{\ }determined following the convention of (\href{https://www.nndc.bnl.gov/nsr/nsrlink.jsp?1967Ro21,B}{1967Ro21})). See\vspace{0.1cm}}&\\
&&&&&&&&&&&&&\parbox[t][0.3cm]{8.156461cm}{\raggedright {\ }{\ }{\ }also \ensuremath{\delta}=0.06 \textit{6} deduced by (\href{https://www.nndc.bnl.gov/nsr/nsrlink.jsp?1972Gi01,B}{1972Gi01}) as the weighted\vspace{0.1cm}}&\\
&&&&&&&&&&&&&\parbox[t][0.3cm]{8.156461cm}{\raggedright {\ }{\ }{\ }average of the same values mentioned above.\vspace{0.1cm}}&\\
&&&&&&&&&&&&&\parbox[t][0.3cm]{8.156461cm}{\raggedright Mult.: From (\href{https://www.nndc.bnl.gov/nsr/nsrlink.jsp?1969Ro08,B}{1969Ro08}, \href{https://www.nndc.bnl.gov/nsr/nsrlink.jsp?1970Sh04,B}{1970Sh04}, \href{https://www.nndc.bnl.gov/nsr/nsrlink.jsp?1972Gi01,B}{1972Gi01}).\vspace{0.1cm}}&\\
&&&&&&&&&&&&&\parbox[t][0.3cm]{8.156461cm}{\raggedright {\ }{\ }{\ }(\href{https://www.nndc.bnl.gov/nsr/nsrlink.jsp?1969Ro08,B}{1969Ro08}) calculated the matrix elements for E2 and\vspace{0.1cm}}&\\
&&&&&&&&&&&&&\parbox[t][0.3cm]{8.156461cm}{\raggedright {\ }{\ }{\ }M1 transitions as \ensuremath{\vert}M\ensuremath{\vert}\ensuremath{^{\textnormal{2}}}\ensuremath{<}190 W.u. and 0.03\ensuremath{<}\ensuremath{\vert}M\ensuremath{\vert}\ensuremath{^{\textnormal{2}}}\ensuremath{<}0.1,\vspace{0.1cm}}&\\
&&&&&&&&&&&&&\parbox[t][0.3cm]{8.156461cm}{\raggedright {\ }{\ }{\ }respectively.\vspace{0.1cm}}&\\
&&&&&&&&&&&&&\parbox[t][0.3cm]{8.156461cm}{\raggedright (\href{https://www.nndc.bnl.gov/nsr/nsrlink.jsp?1969Ro08,B}{1969Ro08}) deduced coefficients of Legendre polynomials\vspace{0.1cm}}&\\
&&&&&&&&&&&&&\parbox[t][0.3cm]{8.156461cm}{\raggedright {\ }{\ }{\ }from the \ensuremath{\gamma}-ray angular distributions: a\ensuremath{_{\textnormal{2}}}=0.74 \textit{11},\vspace{0.1cm}}&\\
&&&&&&&&&&&&&\parbox[t][0.3cm]{8.156461cm}{\raggedright {\ }{\ }{\ }a\ensuremath{_{\textnormal{4}}}={\textminus}0.13 \textit{17}.\vspace{0.1cm}}&\\
&&&&&&&&&&&&&\parbox[t][0.3cm]{8.156461cm}{\raggedright (\href{https://www.nndc.bnl.gov/nsr/nsrlink.jsp?1969Ro22,B}{1969Ro22}) deduced a\ensuremath{_{\textnormal{2}}}=0.48 \textit{22} and a\ensuremath{_{\textnormal{4}}}=0.02 \textit{26} at\vspace{0.1cm}}&\\
&&&&&&&&&&&&&\parbox[t][0.3cm]{8.156461cm}{\raggedright {\ }{\ }{\ }E\ensuremath{_{\textnormal{lab}}}=10 MeV; a\ensuremath{_{\textnormal{2}}}=0.35 \textit{18} and a\ensuremath{_{\textnormal{4}}}={\textminus}0.69 \textit{19} at E\ensuremath{_{\textnormal{lab}}}=10.2\vspace{0.1cm}}&\\
&&&&&&&&&&&&&\parbox[t][0.3cm]{8.156461cm}{\raggedright {\ }{\ }{\ }MeV; and a\ensuremath{_{\textnormal{2}}}=0.62 \textit{17} and a\ensuremath{_{\textnormal{4}}}={\textminus}0.34 \textit{25} at E\ensuremath{_{\textnormal{lab}}}=11.5\vspace{0.1cm}}&\\
&&&&&&&&&&&&&\parbox[t][0.3cm]{8.156461cm}{\raggedright {\ }{\ }{\ }MeV from the measured \ensuremath{\gamma}-ray angular correlations.\vspace{0.1cm}}&\\
&&&&&&&&&&&&&\parbox[t][0.3cm]{8.156461cm}{\raggedright \ensuremath{\Gamma}=9.1 meV was calculated by (\href{https://www.nndc.bnl.gov/nsr/nsrlink.jsp?1972Gi01,B}{1972Gi01}) for this\vspace{0.1cm}}&\\
&&&&&&&&&&&&&\parbox[t][0.3cm]{8.156461cm}{\raggedright {\ }{\ }{\ }transition using the mixing ratio of +0.06 \textit{6} (\href{https://www.nndc.bnl.gov/nsr/nsrlink.jsp?1972Gi01,B}{1972Gi01}:\vspace{0.1cm}}&\\
&&&&&&&&&&&&&\parbox[t][0.3cm]{8.156461cm}{\raggedright {\ }{\ }{\ }see the comment on \ensuremath{\delta}) and the lifetime of the 3616-keV\vspace{0.1cm}}&\\
&&&&&&&&&&&&&\parbox[t][0.3cm]{8.156461cm}{\raggedright {\ }{\ }{\ }state from (\href{https://www.nndc.bnl.gov/nsr/nsrlink.jsp?1972Gi01,B}{1972Gi01}). This width should be compared\vspace{0.1cm}}&\\
&&&&&&&&&&&&&\parbox[t][0.3cm]{8.156461cm}{\raggedright {\ }{\ }{\ }with \ensuremath{\Gamma}=9.4 meV (\href{https://www.nndc.bnl.gov/nsr/nsrlink.jsp?1966Be29,B}{1966Be29}) as cited in (\href{https://www.nndc.bnl.gov/nsr/nsrlink.jsp?1969Ro08,B}{1969Ro08}) and\vspace{0.1cm}}&\\
&&&&&&&&&&&&&\parbox[t][0.3cm]{8.156461cm}{\raggedright {\ }{\ }{\ }\ensuremath{\Gamma}=6.7 meV (private communication of T. Engeland and\vspace{0.1cm}}&\\
&&&&&&&&&&&&&\parbox[t][0.3cm]{8.156461cm}{\raggedright {\ }{\ }{\ }P. J. Ellis with the authors of (\href{https://www.nndc.bnl.gov/nsr/nsrlink.jsp?1972Gi01,B}{1972Gi01})).\vspace{0.1cm}}&\\
&&&&&&&&&&&&&\parbox[t][0.3cm]{8.156461cm}{\raggedright Assuming \ensuremath{\delta}=+0.06 \textit{6} and BR=87.5\% \textit{25} deduced from the\vspace{0.1cm}}&\\
&&&&&&&&&&&&&\parbox[t][0.3cm]{8.156461cm}{\raggedright {\ }{\ }{\ }results of (\href{https://www.nndc.bnl.gov/nsr/nsrlink.jsp?1972Gi01,B}{1972Gi01}), (\href{https://www.nndc.bnl.gov/nsr/nsrlink.jsp?1974Mc17,B}{1974Mc17}) calculated B(E2:\vspace{0.1cm}}&\\
&&&&&&&&&&&&&\parbox[t][0.3cm]{8.156461cm}{\raggedright {\ }{\ }{\ }2\ensuremath{^{\textnormal{+}}_{\textnormal{2}}}\ensuremath{\rightarrow}2\ensuremath{^{\textnormal{+}}_{\textnormal{1}}})=2.6 e\ensuremath{^{\textnormal{2}}}fm\ensuremath{^{\textnormal{4}}} \textit{+77{\textminus}27} for this transition. This\vspace{0.1cm}}&\\
&&&&&&&&&&&&&\parbox[t][0.3cm]{8.156461cm}{\raggedright {\ }{\ }{\ }value should be compared with the calculated values of\vspace{0.1cm}}&\\
&&&&&&&&&&&&&\parbox[t][0.3cm]{8.156461cm}{\raggedright {\ }{\ }{\ }B(E2)=30.5 e\ensuremath{^{\textnormal{2}}}fm\ensuremath{^{\textnormal{4}}} (\href{https://www.nndc.bnl.gov/nsr/nsrlink.jsp?1970Ha49,B}{1970Ha49}: based on wave functions\vspace{0.1cm}}&\\
&&&&&&&&&&&&&\parbox[t][0.3cm]{8.156461cm}{\raggedright {\ }{\ }{\ }of (\href{https://www.nndc.bnl.gov/nsr/nsrlink.jsp?1969Be94,B}{1969Be94}) without two-body contributions),\vspace{0.1cm}}&\\
&&&&&&&&&&&&&\parbox[t][0.3cm]{8.156461cm}{\raggedright {\ }{\ }{\ }B(E2)=22.14 e\ensuremath{^{\textnormal{2}}}fm\ensuremath{^{\textnormal{4}}} (T. Engeland and P. J. Ellis, Nucl.\vspace{0.1cm}}&\\
&&&&&&&&&&&&&\parbox[t][0.3cm]{8.156461cm}{\raggedright {\ }{\ }{\ }Phys. A 181 (1972) 368), and B(E2)=36 e\ensuremath{^{\textnormal{2}}}fm\ensuremath{^{\textnormal{4}}}\vspace{0.1cm}}&\\
&&&&&&&&&&&&&\parbox[t][0.3cm]{8.156461cm}{\raggedright {\ }{\ }{\ }(\href{https://www.nndc.bnl.gov/nsr/nsrlink.jsp?1970Ha49,B}{1970Ha49}).\vspace{0.1cm}}&\\
&&&&&&&&&&&&&\parbox[t][0.3cm]{8.156461cm}{\raggedright Using \ensuremath{\tau}=0.063 ps \textit{+30{\textminus}20} (\href{https://www.nndc.bnl.gov/nsr/nsrlink.jsp?1968Gi09,B}{1968Gi09}), (\href{https://www.nndc.bnl.gov/nsr/nsrlink.jsp?1974Mc17,B}{1974Mc17})\vspace{0.1cm}}&\\
&&&&&&&&&&&&&\parbox[t][0.3cm]{8.156461cm}{\raggedright {\ }{\ }{\ }calculated B(M1)=0.00170 e\ensuremath{^{\textnormal{2}}}fm\ensuremath{^{\textnormal{2}}} \textit{+77{\textminus}53}, which should\vspace{0.1cm}}&\\
&&&&&&&&&&&&&\parbox[t][0.3cm]{8.156461cm}{\raggedright {\ }{\ }{\ }be compared with B(M1)=0.0012 e\ensuremath{^{\textnormal{2}}}fm\ensuremath{^{\textnormal{2}}} calculated by (T.\vspace{0.1cm}}&\\
&&&&&&&&&&&&&\parbox[t][0.3cm]{8.156461cm}{\raggedright {\ }{\ }{\ }Engeland and P. J. Ellis, Nucl. Phys. A 181 (1972) 368).\vspace{0.1cm}}&\\
&&&\multicolumn{1}{r@{}}{3614}&\multicolumn{1}{@{ }l}{{\it 3}}&\multicolumn{1}{r@{}}{9}&\multicolumn{1}{@{ }l}{{\it 3}}&\multicolumn{1}{r@{}}{0}&\multicolumn{1}{@{}l}{}&\multicolumn{1}{@{}l}{0\ensuremath{^{+}}}&\multicolumn{1}{l}{E2}&&&\parbox[t][0.3cm]{8.156461cm}{\raggedright B(E2)(W.u.)=0.68 \textit{+37{\textminus}29}\vspace{0.1cm}}&\\
&&&&&&&&&&&&&\parbox[t][0.3cm]{8.156461cm}{\raggedright E\ensuremath{_{\gamma}}: From (\href{https://www.nndc.bnl.gov/nsr/nsrlink.jsp?1969Ro22,B}{1969Ro22}). See also 3630 keV \textit{2} (\href{https://www.nndc.bnl.gov/nsr/nsrlink.jsp?1970Sh04,B}{1970Sh04});\vspace{0.1cm}}&\\
&&&&&&&&&&&&&\parbox[t][0.3cm]{8.156461cm}{\raggedright {\ }{\ }{\ }and 3616 keV (\href{https://www.nndc.bnl.gov/nsr/nsrlink.jsp?1972Gi01,B}{1972Gi01}, \href{https://www.nndc.bnl.gov/nsr/nsrlink.jsp?1971Ro18,B}{1971Ro18}: this \ensuremath{\gamma}-ray was\vspace{0.1cm}}&\\
&&&&&&&&&&&&&\parbox[t][0.3cm]{8.156461cm}{\raggedright {\ }{\ }{\ }observed in this experiment. This is confirmed in the text\vspace{0.1cm}}&\\
&&&&&&&&&&&&&\parbox[t][0.3cm]{8.156461cm}{\raggedright {\ }{\ }{\ }but not shown on the Fig. 2).\vspace{0.1cm}}&\\
&&&&&&&&&&&&&\parbox[t][0.3cm]{8.156461cm}{\raggedright E\ensuremath{_{\gamma}}: (\href{https://www.nndc.bnl.gov/nsr/nsrlink.jsp?1969Ro22,B}{1969Ro22}): confirmed that this \ensuremath{\gamma}-ray belongs to the\vspace{0.1cm}}&\\
&&&&&&&&&&&&&\parbox[t][0.3cm]{8.156461cm}{\raggedright {\ }{\ }{\ }3617-keV \ensuremath{^{\textnormal{18}}}Ne state because the energy of the\vspace{0.1cm}}&\\
\end{longtable}
\begin{textblock}{29}(0,27.3)
Continued on next page (footnotes at end of table)
\end{textblock}
\clearpage
\begin{longtable}{ccccc@{\extracolsep{\fill}}c}
\\[-.4cm]
\multicolumn{6}{c}{{\bf \small \underline{\ensuremath{^{\textnormal{16}}}O(\ensuremath{^{\textnormal{3}}}He,n\ensuremath{\gamma})\hspace{0.2in}\href{https://www.nndc.bnl.gov/nsr/nsrlink.jsp?1968Gi09,B}{1968Gi09},\href{https://www.nndc.bnl.gov/nsr/nsrlink.jsp?2003Ta13,B}{2003Ta13} (continued)}}}\\
\multicolumn{6}{c}{~}\\
\multicolumn{6}{c}{\underline{$\gamma$($^{18}$Ne) (continued)}}\\
\multicolumn{6}{c}{~~~}\\
\multicolumn{2}{c}{E\ensuremath{_{i}}(level)}&\multicolumn{2}{c}{E\ensuremath{_{\gamma}}\ensuremath{^{\hyperlink{NE16GAMMA0}{a}}}}&Comments&\\[-.2cm]
\multicolumn{2}{c}{\hrulefill}&\multicolumn{2}{c}{\hrulefill}&\hrulefill&
\endhead
&&&&\parbox[t][0.3cm]{15.170881cm}{\raggedright {\ }{\ }{\ }photopeak and double-escape peak from the decay of this state fit the excitation energy of this level within the\vspace{0.1cm}}&\\
&&&&\parbox[t][0.3cm]{15.170881cm}{\raggedright {\ }{\ }{\ }errors.\vspace{0.1cm}}&\\
&&&&\parbox[t][0.3cm]{15.170881cm}{\raggedright I\ensuremath{_{\gamma}}: Weighted average (rounded to the nearest integer) of 7\% \textit{2} (\href{https://www.nndc.bnl.gov/nsr/nsrlink.jsp?1969Ro22,B}{1969Ro22}: see Fig. 3, deduced from\vspace{0.1cm}}&\\
&&&&\parbox[t][0.3cm]{15.170881cm}{\raggedright {\ }{\ }{\ }measurement of angular correlations using a NaI counter at 10.2 MeV); and 12.5\% \textit{25} (\href{https://www.nndc.bnl.gov/nsr/nsrlink.jsp?1972Gi01,B}{1972Gi01}). See also\vspace{0.1cm}}&\\
&&&&\parbox[t][0.3cm]{15.170881cm}{\raggedright {\ }{\ }{\ }BR\ensuremath{<}9\% (\href{https://www.nndc.bnl.gov/nsr/nsrlink.jsp?1968Gi09,B}{1968Gi09}, \href{https://www.nndc.bnl.gov/nsr/nsrlink.jsp?1969Ro22,B}{1969Ro22}); BR\ensuremath{<}3\% (\href{https://www.nndc.bnl.gov/nsr/nsrlink.jsp?1970Sh04,B}{1970Sh04}: see Fig. 1). Note that (\href{https://www.nndc.bnl.gov/nsr/nsrlink.jsp?1969Ro22,B}{1969Ro22}) deduced branching\vspace{0.1cm}}&\\
&&&&\parbox[t][0.3cm]{15.170881cm}{\raggedright {\ }{\ }{\ }ratio of 8\% \textit{2} from n\ensuremath{\gamma}-coincidence measurement using Ge(Li) detector. They recommended the value of 7\% \textit{2}\vspace{0.1cm}}&\\
&&&&\parbox[t][0.3cm]{15.170881cm}{\raggedright {\ }{\ }{\ }deduced from angular correlations. However, later on, (\href{https://www.nndc.bnl.gov/nsr/nsrlink.jsp?1972Gi01,B}{1972Gi01}) argued that the 7\% \textit{2} branching ratio\vspace{0.1cm}}&\\
&&&&\parbox[t][0.3cm]{15.170881cm}{\raggedright {\ }{\ }{\ }reported by (\href{https://www.nndc.bnl.gov/nsr/nsrlink.jsp?1969Ro22,B}{1969Ro22}) is a calculated value and not from measurement. It is not clear to the evaluator if this\vspace{0.1cm}}&\\
&&&&\parbox[t][0.3cm]{15.170881cm}{\raggedright {\ }{\ }{\ }opinion is correct.\vspace{0.1cm}}&\\
&&&&\parbox[t][0.3cm]{15.170881cm}{\raggedright Mult.: From (\href{https://www.nndc.bnl.gov/nsr/nsrlink.jsp?1972Gi01,B}{1972Gi01}).\vspace{0.1cm}}&\\
&&&&\parbox[t][0.3cm]{15.170881cm}{\raggedright Using the branching ratio for this transition and the lifetime of the 3616-keV state from (\href{https://www.nndc.bnl.gov/nsr/nsrlink.jsp?1972Gi01,B}{1972Gi01}), they\vspace{0.1cm}}&\\
&&&&\parbox[t][0.3cm]{15.170881cm}{\raggedright {\ }{\ }{\ }calculated \ensuremath{\Gamma}=1.4 meV for this transition. This value should be compared with the theoretical values of \ensuremath{\Gamma}=0.4\vspace{0.1cm}}&\\
&&&&\parbox[t][0.3cm]{15.170881cm}{\raggedright {\ }{\ }{\ }meV from (\href{https://www.nndc.bnl.gov/nsr/nsrlink.jsp?1966Be29,B}{1966Be29}) as cited in (\href{https://www.nndc.bnl.gov/nsr/nsrlink.jsp?1969Ro08,B}{1969Ro08}); \ensuremath{\Gamma}=0.6 meV from (\href{https://www.nndc.bnl.gov/nsr/nsrlink.jsp?1968Ar02,B}{1968Ar02}) as cited in (\href{https://www.nndc.bnl.gov/nsr/nsrlink.jsp?1969Ro08,B}{1969Ro08}); and\vspace{0.1cm}}&\\
&&&&\parbox[t][0.3cm]{15.170881cm}{\raggedright {\ }{\ }{\ }\ensuremath{\Gamma}=0.64 meV (private communication of T. Engeland and P. J. Ellis with the authors of (\href{https://www.nndc.bnl.gov/nsr/nsrlink.jsp?1972Gi01,B}{1972Gi01})).\vspace{0.1cm}}&\\
&&&&\parbox[t][0.3cm]{15.170881cm}{\raggedright Assuming BR=12.5 \% \textit{25} (\href{https://www.nndc.bnl.gov/nsr/nsrlink.jsp?1972Gi01,B}{1972Gi01}), (\href{https://www.nndc.bnl.gov/nsr/nsrlink.jsp?1974Mc17,B}{1974Mc17}) calculated B(E2: 2\ensuremath{^{\textnormal{+}}_{\textnormal{2}}}\ensuremath{\rightarrow}0\ensuremath{^{\textnormal{+}}_{\textnormal{1}}})=2.5 e\ensuremath{^{\textnormal{2}}}fm\ensuremath{^{\textnormal{4}}} \textit{+13{\textminus}10} for this\vspace{0.1cm}}&\\
&&&&\parbox[t][0.3cm]{15.170881cm}{\raggedright {\ }{\ }{\ }transition. This value should be compared with the calculated values of B(E2)=1.68 e\ensuremath{^{\textnormal{2}}}fm\ensuremath{^{\textnormal{4}}} (\href{https://www.nndc.bnl.gov/nsr/nsrlink.jsp?1970Ha49,B}{1970Ha49}: based\vspace{0.1cm}}&\\
&&&&\parbox[t][0.3cm]{15.170881cm}{\raggedright {\ }{\ }{\ }on wave functions of (\href{https://www.nndc.bnl.gov/nsr/nsrlink.jsp?1969Be94,B}{1969Be94}) without two-body contributions), B(E2)=1.26 e\ensuremath{^{\textnormal{2}}}fm\ensuremath{^{\textnormal{4}}} (T. Engeland and P. J.\vspace{0.1cm}}&\\
&&&&\parbox[t][0.3cm]{15.170881cm}{\raggedright {\ }{\ }{\ }Ellis, Nucl. Phys. A 181 (1972) 368), and B(E2)=2.77 e\ensuremath{^{\textnormal{2}}}fm\ensuremath{^{\textnormal{4}}} (\href{https://www.nndc.bnl.gov/nsr/nsrlink.jsp?1970Ha49,B}{1970Ha49}).\vspace{0.1cm}}&\\
\end{longtable}
\parbox[b][0.3cm]{17.7cm}{\makebox[1ex]{\ensuremath{^{\hypertarget{NE16GAMMA0}{a}}}} The \ensuremath{\gamma}-ray energies (and thus the deduced excitation energies) from (\href{https://www.nndc.bnl.gov/nsr/nsrlink.jsp?1970Sh04,B}{1970Sh04}) are consistently higher than all other \ensuremath{\gamma}-ray}\\
\parbox[b][0.3cm]{17.7cm}{{\ }{\ }measurements reported here. Due to potentially unknown systematic uncertainties, the evaluator did not consider the \ensuremath{\gamma}-ray}\\
\parbox[b][0.3cm]{17.7cm}{{\ }{\ }energies reported by (\href{https://www.nndc.bnl.gov/nsr/nsrlink.jsp?1970Sh04,B}{1970Sh04}) in finding the adopted E\ensuremath{_{\ensuremath{\gamma}}} and E\ensuremath{_{\textnormal{x}}} energies.}\\
\vspace{0.5cm}
\clearpage
\begin{figure}[h]
\begin{center}
\includegraphics{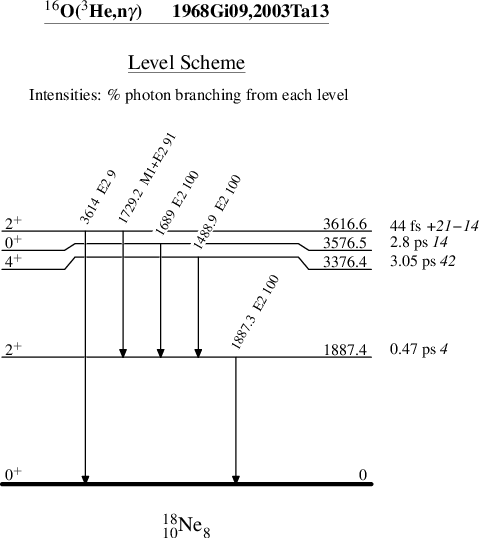}\\
\end{center}
\end{figure}
\clearpage
\subsection[\hspace{-0.2cm}\ensuremath{^{\textnormal{16}}}O(\ensuremath{^{\textnormal{10}}}B,\ensuremath{^{\textnormal{8}}}Li)]{ }
\vspace{-27pt}
\vspace{0.3cm}
\hypertarget{NE17}{{\bf \small \underline{\ensuremath{^{\textnormal{16}}}O(\ensuremath{^{\textnormal{10}}}B,\ensuremath{^{\textnormal{8}}}Li)\hspace{0.2in}\href{https://www.nndc.bnl.gov/nsr/nsrlink.jsp?1977HaYB,B}{1977HaYB},\href{https://www.nndc.bnl.gov/nsr/nsrlink.jsp?1983Os07,B}{1983Os07}}}}\\
\vspace{4pt}
\vspace{8pt}
\parbox[b][0.3cm]{17.7cm}{\addtolength{\parindent}{-0.2in}\href{https://www.nndc.bnl.gov/nsr/nsrlink.jsp?1977HaYB,B}{1977HaYB}: \ensuremath{^{\textnormal{16}}}O(\ensuremath{^{\textnormal{10}}}B,\ensuremath{^{\textnormal{8}}}Li) E=100 MeV; measured yields.}\\
\parbox[b][0.3cm]{17.7cm}{\addtolength{\parindent}{-0.2in}\href{https://www.nndc.bnl.gov/nsr/nsrlink.jsp?1978Ha10,B}{1978Ha10}: \ensuremath{^{\textnormal{16}}}O(\ensuremath{^{\textnormal{10}}}B,\ensuremath{^{\textnormal{8}}}Li) E=100 MeV; measured the reaction products using a \ensuremath{\Delta}E-E Si telescope with an energy resolution of}\\
\parbox[b][0.3cm]{17.7cm}{\ensuremath{\Delta}E(FWHM)=300-400 keV. Measured \ensuremath{\sigma}(\ensuremath{\theta}) and a few excited levels of \ensuremath{^{\textnormal{18}}}Ne at 1890 keV \textit{19}, 3380 keV \textit{34}, 7200 keV \textit{72}, and}\\
\parbox[b][0.3cm]{17.7cm}{8200 keV \textit{82}. Measured the angular distribution of the \ensuremath{^{\textnormal{18}}}Ne*(3380 keV) state at \ensuremath{\theta}\ensuremath{_{\textnormal{c.m.}}}\ensuremath{\sim}17\ensuremath{^\circ}{\textminus}35\ensuremath{^\circ}, which was analyzed using finite}\\
\parbox[b][0.3cm]{17.7cm}{range DWBA analysis via the SATURN and MARS codes. Deduced the spectroscopic factor and compared the result with that}\\
\parbox[b][0.3cm]{17.7cm}{obtained using shell model calculations. The authors concluded that the (\ensuremath{^{\textnormal{10}}}B,\ensuremath{^{\textnormal{8}}}Li) reaction selectively populates the high spin states.}\\
\vspace{0.385cm}
\parbox[b][0.3cm]{17.7cm}{\addtolength{\parindent}{-0.2in}\textit{Theory}:}\\
\parbox[b][0.3cm]{17.7cm}{\addtolength{\parindent}{-0.2in}\href{https://www.nndc.bnl.gov/nsr/nsrlink.jsp?1976ToZV,B}{1976ToZV}: \ensuremath{^{\textnormal{16}}}O(\ensuremath{^{\textnormal{10}}}B,\ensuremath{^{\textnormal{8}}}Li) E=100 MeV.}\\
\parbox[b][0.3cm]{17.7cm}{\addtolength{\parindent}{-0.2in}\href{https://www.nndc.bnl.gov/nsr/nsrlink.jsp?1983Os07,B}{1983Os07}: \ensuremath{^{\textnormal{16}}}O(\ensuremath{^{\textnormal{10}}}B,\ensuremath{^{\textnormal{8}}}Li) E=100 MeV; developed an expression for the DWBA differential cross sections of heavy ion reactions}\\
\parbox[b][0.3cm]{17.7cm}{with two nucleon transfer taking into account the finite range effects. Analyzed the experimental \ensuremath{\sigma}(\ensuremath{\theta}) (\href{https://www.nndc.bnl.gov/nsr/nsrlink.jsp?1978Ha10,B}{1978Ha10}) for}\\
\parbox[b][0.3cm]{17.7cm}{\ensuremath{^{\textnormal{16}}}O(\ensuremath{^{\textnormal{10}}}B,\ensuremath{^{\textnormal{8}}}Li) at 100 MeV. Deduced optical model parameters for the finite range DWBA calculations performed using the}\\
\parbox[b][0.3cm]{17.7cm}{LAJOLLA code. Deduced the spectroscopic factors. The theoretical calculations reproduced the experimental measurements of the}\\
\parbox[b][0.3cm]{17.7cm}{differential cross sections both in shape and in magnitude.}\\
\vspace{12pt}
\underline{$^{18}$Ne Levels}\\
\begin{longtable}{cccccc@{\extracolsep{\fill}}c}
\multicolumn{2}{c}{E(level)$^{{\hyperlink{NE17LEVEL0}{a}}}$}&J$^{\pi}$$^{}$&\multicolumn{2}{c}{C\ensuremath{^{\textnormal{2}}}S$^{{\hyperlink{NE17LEVEL2}{c}}}$}&Comments&\\[-.2cm]
\multicolumn{2}{c}{\hrulefill}&\hrulefill&\multicolumn{2}{c}{\hrulefill}&\hrulefill&
\endfirsthead
\multicolumn{1}{r@{}}{1890}&\multicolumn{1}{@{}l}{\ensuremath{^{{\hyperlink{NE17LEVEL1}{b}}}} {\it 19}}&\multicolumn{1}{l}{2\ensuremath{^{+}}}&&&\parbox[t][0.3cm]{13.78862cm}{\raggedright J\ensuremath{^{\pi}}: From Fig. 2 of (\href{https://www.nndc.bnl.gov/nsr/nsrlink.jsp?1978Ha10,B}{1978Ha10}), measured at \ensuremath{\theta}\ensuremath{_{\textnormal{lab}}}=10.8\ensuremath{^\circ}. J\ensuremath{^{\ensuremath{\pi}}} is determined from the exact finite\vspace{0.1cm}}&\\
&&&&&\parbox[t][0.3cm]{13.78862cm}{\raggedright {\ }{\ }{\ }range DWBA calculation performed in (\href{https://www.nndc.bnl.gov/nsr/nsrlink.jsp?1978Ha10,B}{1978Ha10}) but L is not given.\vspace{0.1cm}}&\\
\multicolumn{1}{r@{}}{3380}&\multicolumn{1}{@{ }l}{{\it 34}}&\multicolumn{1}{l}{4\ensuremath{^{+}}}&\multicolumn{1}{r@{}}{0}&\multicolumn{1}{@{.}l}{50 {\it 13}}&\parbox[t][0.3cm]{13.78862cm}{\raggedright J\ensuremath{^{\pi}},C\ensuremath{^{\textnormal{2}}}S: Determined from the exact finite range DWBA calculation performed in (\href{https://www.nndc.bnl.gov/nsr/nsrlink.jsp?1978Ha10,B}{1978Ha10}). The\vspace{0.1cm}}&\\
&&&&&\parbox[t][0.3cm]{13.78862cm}{\raggedright {\ }{\ }{\ }agreement with the data is very good except at \ensuremath{\theta}\ensuremath{_{\textnormal{c.m.}}}=35\ensuremath{^\circ} (see Fig. 5). The orbital angular\vspace{0.1cm}}&\\
&&&&&\parbox[t][0.3cm]{13.78862cm}{\raggedright {\ }{\ }{\ }momentum for the transfer is not provided. This state is strongly populated (see Fig. 2), and it is\vspace{0.1cm}}&\\
&&&&&\parbox[t][0.3cm]{13.78862cm}{\raggedright {\ }{\ }{\ }mentioned in (\href{https://www.nndc.bnl.gov/nsr/nsrlink.jsp?1978Ha10,B}{1978Ha10}) that the results of the exact finite-range DWBA showed that only those\vspace{0.1cm}}&\\
&&&&&\parbox[t][0.3cm]{13.78862cm}{\raggedright {\ }{\ }{\ }states which have shell model configuration of the target nucleus coupled to an \textit{s}=0, T=1 proton\vspace{0.1cm}}&\\
&&&&&\parbox[t][0.3cm]{13.78862cm}{\raggedright {\ }{\ }{\ }pair are strongly populated for A=18 nuclei. (\href{https://www.nndc.bnl.gov/nsr/nsrlink.jsp?1978Ha10,B}{1978Ha10}) deduced the ratio between experimental to\vspace{0.1cm}}&\\
&&&&&\parbox[t][0.3cm]{13.78862cm}{\raggedright {\ }{\ }{\ }theoretical cluster spectroscopic factor for this state and found that S\ensuremath{_{\textnormal{exp}}}/S\ensuremath{_{\textnormal{theo}}}=2.50 \textit{63}, where the\vspace{0.1cm}}&\\
&&&&&\parbox[t][0.3cm]{13.78862cm}{\raggedright {\ }{\ }{\ }theoretical spectroscopic factor is calculated to be S\ensuremath{_{\textnormal{theo}}}=0.199 using shell model wave functions\vspace{0.1cm}}&\\
&&&&&\parbox[t][0.3cm]{13.78862cm}{\raggedright {\ }{\ }{\ }from (\href{https://www.nndc.bnl.gov/nsr/nsrlink.jsp?1976La13,B}{1976La13}). (\href{https://www.nndc.bnl.gov/nsr/nsrlink.jsp?1978Ha10,B}{1978Ha10}) reported that this rather large ratio may be caused by an interference\vspace{0.1cm}}&\\
&&&&&\parbox[t][0.3cm]{13.78862cm}{\raggedright {\ }{\ }{\ }between \textit{s}=0 and \textit{s}=1 transfer [the proton pair may be transferred in a relative \textit{s}=1 state], which\vspace{0.1cm}}&\\
&&&&&\parbox[t][0.3cm]{13.78862cm}{\raggedright {\ }{\ }{\ }could enhance the population strength of this state (\href{https://www.nndc.bnl.gov/nsr/nsrlink.jsp?1978Ha10,B}{1978Ha10}). A similar theoretical finite range\vspace{0.1cm}}&\\
&&&&&\parbox[t][0.3cm]{13.78862cm}{\raggedright {\ }{\ }{\ }DWBA calculation performed by (\href{https://www.nndc.bnl.gov/nsr/nsrlink.jsp?1983Os07,B}{1983Os07}) reports \ensuremath{\Delta}L=3,4. Evaluator notes that \ensuremath{\Delta}L=3 most\vspace{0.1cm}}&\\
&&&&&\parbox[t][0.3cm]{13.78862cm}{\raggedright {\ }{\ }{\ }likely indicates that the two protons are transferred from two different shells with a relative L=1,\vspace{0.1cm}}&\\
&&&&&\parbox[t][0.3cm]{13.78862cm}{\raggedright {\ }{\ }{\ }whereas \ensuremath{\Delta}L=4 refers to a diproton transfer in a relative \textit{s}=0 state.\vspace{0.1cm}}&\\
&&&&&\parbox[t][0.3cm]{13.78862cm}{\raggedright C\ensuremath{^{\textnormal{2}}}S: (\href{https://www.nndc.bnl.gov/nsr/nsrlink.jsp?1978Ha10,B}{1978Ha10}): \ensuremath{\sigma}\ensuremath{_{\textnormal{exp}}}(\ensuremath{\theta})=C\ensuremath{_{\textnormal{1}}^{\textnormal{2}}}S\ensuremath{_{\textnormal{1}}}C\ensuremath{_{\textnormal{2}}^{\textnormal{2}}}S\ensuremath{_{\textnormal{2}}}\ensuremath{\sigma}\ensuremath{_{\textnormal{DWBA}}} for 2p transfer. Therefore, the normalization factor\vspace{0.1cm}}&\\
&&&&&\parbox[t][0.3cm]{13.78862cm}{\raggedright {\ }{\ }{\ }is N=C\ensuremath{_{\textnormal{1}}^{\textnormal{2}}}S\ensuremath{_{\textnormal{1}}}C\ensuremath{_{\textnormal{2}}^{\textnormal{2}}}S\ensuremath{_{\textnormal{2}}}. N=0.095 \textit{24} (\href{https://www.nndc.bnl.gov/nsr/nsrlink.jsp?1978Ha10,B}{1978Ha10}). Assuming C\ensuremath{_{\textnormal{1}}^{\textnormal{2}}}S\ensuremath{_{\textnormal{1}}}=0.191 (\href{https://www.nndc.bnl.gov/nsr/nsrlink.jsp?1978Ha10,B}{1978Ha10}), C\ensuremath{_{\textnormal{2}}^{\textnormal{2}}}S\ensuremath{^{\textnormal{2}}}=0.50 \textit{13}\vspace{0.1cm}}&\\
&&&&&\parbox[t][0.3cm]{13.78862cm}{\raggedright {\ }{\ }{\ }(\href{https://www.nndc.bnl.gov/nsr/nsrlink.jsp?1978Ha10,B}{1978Ha10}).\vspace{0.1cm}}&\\
&&&&&\parbox[t][0.3cm]{13.78862cm}{\raggedright C\ensuremath{^{\textnormal{2}}}S: (\href{https://www.nndc.bnl.gov/nsr/nsrlink.jsp?1978Ha10,B}{1978Ha10}): shell mode1 wave functions from (\href{https://www.nndc.bnl.gov/nsr/nsrlink.jsp?1976La13,B}{1976La13}) (constrained-II set) assumed for this\vspace{0.1cm}}&\\
&&&&&\parbox[t][0.3cm]{13.78862cm}{\raggedright {\ }{\ }{\ }state in \ensuremath{^{\textnormal{18}}}Ne is: 0.986\ensuremath{\vert}\textit{d}\ensuremath{_{\textnormal{5/2}}^{\textnormal{2}}}\ensuremath{\rangle}+0.151\ensuremath{\vert}\textit{d}\ensuremath{_{\textnormal{5/2}}}\textit{d}\ensuremath{~_{\textnormal{3/2}}}\ensuremath{\rangle}. The corresponding amplitude of the leading SU\ensuremath{_{\textnormal{3}}}\vspace{0.1cm}}&\\
&&&&&\parbox[t][0.3cm]{13.78862cm}{\raggedright {\ }{\ }{\ }term in the transformed wave function is 0.576 (\href{https://www.nndc.bnl.gov/nsr/nsrlink.jsp?1978Ha10,B}{1978Ha10}).\vspace{0.1cm}}&\\
&&&&&\parbox[t][0.3cm]{13.78862cm}{\raggedright C\ensuremath{^{\textnormal{2}}}S: See also S\ensuremath{_{\textnormal{theoretical}}}=0.6531 (\href{https://www.nndc.bnl.gov/nsr/nsrlink.jsp?1983Os07,B}{1983Os07}).\vspace{0.1cm}}&\\
\multicolumn{1}{r@{}}{7200}&\multicolumn{1}{@{}l}{\ensuremath{^{{\hyperlink{NE17LEVEL1}{b}}}} {\it 72}}&&&&\parbox[t][0.3cm]{13.78862cm}{\raggedright E(level): Due to insufficient information, this state was not considered for the \ensuremath{^{\textnormal{18}}}Ne Adopted Levels.\vspace{0.1cm}}&\\
&&&&&\parbox[t][0.3cm]{13.78862cm}{\raggedright {\ }{\ }{\ }However, based on the excitation energy, we paired this level with the 7120-keV state in the\vspace{0.1cm}}&\\
&&&&&\parbox[t][0.3cm]{13.78862cm}{\raggedright {\ }{\ }{\ }Adopted Levels.\vspace{0.1cm}}&\\
\multicolumn{1}{r@{}}{8200}&\multicolumn{1}{@{}l}{\ensuremath{^{{\hyperlink{NE17LEVEL1}{b}}}} {\it 82}}&&&&&\\
\end{longtable}
\parbox[b][0.3cm]{17.7cm}{\makebox[1ex]{\ensuremath{^{\hypertarget{NE17LEVEL0}{a}}}} From (\href{https://www.nndc.bnl.gov/nsr/nsrlink.jsp?1978Ha10,B}{1978Ha10}). An uncertainty of 1\% due to calibration is added to the excitation energies based on what is reported in the}\\
\parbox[b][0.3cm]{17.7cm}{{\ }{\ }text of (\href{https://www.nndc.bnl.gov/nsr/nsrlink.jsp?1978Ha10,B}{1978Ha10}).}\\
\parbox[b][0.3cm]{17.7cm}{\makebox[1ex]{\ensuremath{^{\hypertarget{NE17LEVEL1}{b}}}} From Fig. 2 of (\href{https://www.nndc.bnl.gov/nsr/nsrlink.jsp?1978Ha10,B}{1978Ha10}), measured at \ensuremath{\theta}\ensuremath{_{\textnormal{lab}}}=10.8\ensuremath{^\circ}.}\\
\parbox[b][0.3cm]{17.7cm}{\makebox[1ex]{\ensuremath{^{\hypertarget{NE17LEVEL2}{c}}}} Cluster spectroscopic factor from (\href{https://www.nndc.bnl.gov/nsr/nsrlink.jsp?1978Ha10,B}{1978Ha10}).}\\
\vspace{0.5cm}
\clearpage
\subsection[\hspace{-0.2cm}\ensuremath{^{\textnormal{16}}}O(\ensuremath{^{\textnormal{11}}}B,\ensuremath{^{\textnormal{9}}}Li)]{ }
\vspace{-27pt}
\vspace{0.3cm}
\hypertarget{NE18}{{\bf \small \underline{\ensuremath{^{\textnormal{16}}}O(\ensuremath{^{\textnormal{11}}}B,\ensuremath{^{\textnormal{9}}}Li)\hspace{0.2in}\href{https://www.nndc.bnl.gov/nsr/nsrlink.jsp?1979Ra10,B}{1979Ra10}}}}\\
\vspace{4pt}
\vspace{8pt}
\parbox[b][0.3cm]{17.7cm}{\addtolength{\parindent}{-0.2in}\href{https://www.nndc.bnl.gov/nsr/nsrlink.jsp?1979Ra10,B}{1979Ra10}: \ensuremath{^{\textnormal{16}}}O(\ensuremath{^{\textnormal{11}}}B,\ensuremath{^{\textnormal{9}}}Li) E=115 MeV; measured the \ensuremath{^{\textnormal{9}}}Li ejectiles using a \ensuremath{\Delta}E-\ensuremath{\Delta}E-E telescope with an overall energy resolution of}\\
\parbox[b][0.3cm]{17.7cm}{\ensuremath{\sim}250 keV. This two-proton transfer reaction populated only one state of \ensuremath{^{\textnormal{18}}}Ne at 3.38 MeV.}\\
\vspace{12pt}
\underline{$^{18}$Ne Levels}\\
\begin{longtable}{ccc@{\extracolsep{\fill}}c}
\multicolumn{2}{c}{E(level)$^{{\hyperlink{NE18LEVEL0}{a}}}$}&J$^{\pi}$$^{{\hyperlink{NE18LEVEL0}{a}}}$&\\[-.2cm]
\multicolumn{2}{c}{\hrulefill}&\hrulefill&
\endfirsthead
\multicolumn{1}{r@{}}{3376}&\multicolumn{1}{@{.}l}{4}&\multicolumn{1}{l}{4\ensuremath{^{+}}}&\\
\end{longtable}
\parbox[b][0.3cm]{17.7cm}{\makebox[1ex]{\ensuremath{^{\hypertarget{NE18LEVEL0}{a}}}} From the \ensuremath{^{\textnormal{18}}}Ne Adopted Levels.}\\
\vspace{0.5cm}
\clearpage
\subsection[\hspace{-0.2cm}\ensuremath{^{\textnormal{16}}}O(\ensuremath{^{\textnormal{12}}}C,\ensuremath{^{\textnormal{10}}}Be)]{ }
\vspace{-27pt}
\vspace{0.3cm}
\hypertarget{NE19}{{\bf \small \underline{\ensuremath{^{\textnormal{16}}}O(\ensuremath{^{\textnormal{12}}}C,\ensuremath{^{\textnormal{10}}}Be)\hspace{0.2in}\href{https://www.nndc.bnl.gov/nsr/nsrlink.jsp?1972Sc21,B}{1972Sc21},\href{https://www.nndc.bnl.gov/nsr/nsrlink.jsp?1988Me10,B}{1988Me10}}}}\\
\vspace{4pt}
\vspace{8pt}
\parbox[b][0.3cm]{17.7cm}{\addtolength{\parindent}{-0.2in}\href{https://www.nndc.bnl.gov/nsr/nsrlink.jsp?1972Sc21,B}{1972Sc21}: \ensuremath{^{\textnormal{16}}}O(\ensuremath{^{\textnormal{12}}}C,\ensuremath{^{\textnormal{10}}}B) E=114 MeV; measured the reaction products, from \ensuremath{^{\textnormal{6}}}Li to \ensuremath{^{\textnormal{13}}}C, using a solid state \ensuremath{\Delta}E-E telescope}\\
\parbox[b][0.3cm]{17.7cm}{covering \ensuremath{\theta}\ensuremath{_{\textnormal{lab}}}=7\ensuremath{^\circ}{\textminus}35\ensuremath{^\circ}. Energy resolution was \ensuremath{\Delta}E(FWHM)=300 keV. Measured charged particle spectra. The mechanism of the}\\
\parbox[b][0.3cm]{17.7cm}{reactions appears to be dominated by a surface interaction. The experimenters concluded that the reactions caused by heavy ions}\\
\parbox[b][0.3cm]{17.7cm}{with incident energies of \ensuremath{\sim}10 MeV/nucleon tend to populate high-spin excited states with stretched particle configurations. In the}\\
\parbox[b][0.3cm]{17.7cm}{case of two-proton transfer, the exclusion principle allows only states of even spin with T=1.}\\
\parbox[b][0.3cm]{17.7cm}{\addtolength{\parindent}{-0.2in}\href{https://www.nndc.bnl.gov/nsr/nsrlink.jsp?1974An36,B}{1974An36}: \ensuremath{^{\textnormal{16}}}O(\ensuremath{^{\textnormal{12}}}C,\ensuremath{^{\textnormal{10}}}B) E=114 MeV; measured the reaction products using a telescope that consisted of \ensuremath{\Delta}E-\ensuremath{\Delta}E-E fully depleted}\\
\parbox[b][0.3cm]{17.7cm}{Si surface barrier detectors in anti-coincidence with a final Si veto detector. The experimental energy resolution was 400 keV}\\
\parbox[b][0.3cm]{17.7cm}{(FWHM). Angular distributions were measured at \ensuremath{\theta}\ensuremath{_{\textnormal{lab}}}=6\ensuremath{^\circ}{\textminus}30\ensuremath{^\circ}. The observed cross section for a transfer of two protons was on the}\\
\parbox[b][0.3cm]{17.7cm}{order of 0.02 mb/sr. Reaction mechanism, and spectroscopic amplitudes are discussed. The experimenters report that the excitation}\\
\parbox[b][0.3cm]{17.7cm}{of \ensuremath{^{\textnormal{10}}}Be is mostly obscured and conclude that this heavy-ion transfer reaction selectively populates high spin states by transferring}\\
\parbox[b][0.3cm]{17.7cm}{particles into the \textit{sd} shell coupling to the maximum possible spin.}\\
\parbox[b][0.3cm]{17.7cm}{\addtolength{\parindent}{-0.2in}\href{https://www.nndc.bnl.gov/nsr/nsrlink.jsp?1988Kr11,B}{1988Kr11}, \href{https://www.nndc.bnl.gov/nsr/nsrlink.jsp?1988Me10,B}{1988Me10}: \ensuremath{^{\textnormal{16}}}O(\ensuremath{^{\textnormal{12}}}C,\ensuremath{^{\textnormal{10}}}B) E=480 MeV; measured the reaction products using the SPEG spectrometer and its associated}\\
\parbox[b][0.3cm]{17.7cm}{focal plane equipment resulting in an energy resolution of 200 keV. The angular distributions of the most strongly populated excited}\\
\parbox[b][0.3cm]{17.7cm}{states in various residual nuclei, including \ensuremath{^{\textnormal{18}}}Ne, were measured (for the case of \ensuremath{^{\textnormal{18}}}Ne states, at \ensuremath{\theta}\ensuremath{_{\textnormal{lab}}}=2\ensuremath{^\circ}{\textminus}6\ensuremath{^\circ}). For the residual}\\
\parbox[b][0.3cm]{17.7cm}{nuclei other than \ensuremath{^{\textnormal{18}}}Ne, these distributions were analyzed using the exact finite range DWBA with the PTOLEMY code. The \ensuremath{^{\textnormal{18}}}Ne}\\
\parbox[b][0.3cm]{17.7cm}{states were unresolved from the \ensuremath{^{\textnormal{30}}}S excited states populated by the (\ensuremath{^{\textnormal{12}}}C,\ensuremath{^{\textnormal{10}}}Be) reaction on the Si content of the target. Shell}\\
\parbox[b][0.3cm]{17.7cm}{model calculations were performed to explain the observed levels.}\\
\vspace{12pt}
\underline{$^{18}$Ne Levels}\\
\begin{longtable}{cccc@{\extracolsep{\fill}}c}
\multicolumn{2}{c}{E(level)$^{}$}&J$^{\pi}$$^{}$&Comments&\\[-.2cm]
\multicolumn{2}{c}{\hrulefill}&\hrulefill&\hrulefill&
\endfirsthead
\multicolumn{1}{r@{}}{0}&\multicolumn{1}{@{}l}{\ensuremath{^{{\hyperlink{NE19LEVEL0}{a}}}}}&\multicolumn{1}{l}{0\ensuremath{^{+}}\ensuremath{^{{\hyperlink{NE19LEVEL1}{b}}}}}&\parbox[t][0.3cm]{15.023221cm}{\raggedright E(level),J\ensuremath{^{\pi}}: From Fig. 3(c) of (\href{https://www.nndc.bnl.gov/nsr/nsrlink.jsp?1972Sc21,B}{1972Sc21}), Fig. 5(b) of (\href{https://www.nndc.bnl.gov/nsr/nsrlink.jsp?1974An36,B}{1974An36}), and Fig. 3 of (\href{https://www.nndc.bnl.gov/nsr/nsrlink.jsp?1988Kr11,B}{1988Kr11}).\vspace{0.1cm}}&\\
\multicolumn{1}{r@{}}{1.8\ensuremath{\times10^{3}}}&\multicolumn{1}{@{}l}{\ensuremath{^{{\hyperlink{NE19LEVEL0}{a}}}}}&\multicolumn{1}{l}{2\ensuremath{^{+}}\ensuremath{^{{\hyperlink{NE19LEVEL1}{b}}}}}&\parbox[t][0.3cm]{15.023221cm}{\raggedright E(level),J\ensuremath{^{\pi}}: From Fig. 3 of (\href{https://www.nndc.bnl.gov/nsr/nsrlink.jsp?1988Kr11,B}{1988Kr11}). It is not clear if the J\ensuremath{^{\ensuremath{\pi}}} assignment was independently determined.\vspace{0.1cm}}&\\
\multicolumn{1}{r@{}}{3380}&\multicolumn{1}{@{}l}{\ensuremath{^{{\hyperlink{NE19LEVEL0}{a}}}}}&\multicolumn{1}{l}{4\ensuremath{^{+}}}&\parbox[t][0.3cm]{15.023221cm}{\raggedright T=1 (\href{https://www.nndc.bnl.gov/nsr/nsrlink.jsp?1974An36,B}{1974An36})\vspace{0.1cm}}&\\
&&&\parbox[t][0.3cm]{15.023221cm}{\raggedright E(level),J\ensuremath{^{\pi}}: From Fig. 3(c) of (\href{https://www.nndc.bnl.gov/nsr/nsrlink.jsp?1972Sc21,B}{1972Sc21}), Fig. 5(b) of (\href{https://www.nndc.bnl.gov/nsr/nsrlink.jsp?1974An36,B}{1974An36}), and Fig. 3 of (\href{https://www.nndc.bnl.gov/nsr/nsrlink.jsp?1988Kr11,B}{1988Kr11}).\vspace{0.1cm}}&\\
&&&\parbox[t][0.3cm]{15.023221cm}{\raggedright E(level): See also 3.4 MeV (\href{https://www.nndc.bnl.gov/nsr/nsrlink.jsp?1988Kr11,B}{1988Kr11}, \href{https://www.nndc.bnl.gov/nsr/nsrlink.jsp?1988Me10,B}{1988Me10}).\vspace{0.1cm}}&\\
&&&\parbox[t][0.3cm]{15.023221cm}{\raggedright E(level),J\ensuremath{^{\pi}}: The shell model calculation of (\href{https://www.nndc.bnl.gov/nsr/nsrlink.jsp?1988Kr11,B}{1988Kr11}) predicted the 4\ensuremath{^{\textnormal{+}}_{\textnormal{1}}} state in \ensuremath{^{\textnormal{18}}}Ne to be at 3.32 MeV\vspace{0.1cm}}&\\
&&&\parbox[t][0.3cm]{15.023221cm}{\raggedright {\ }{\ }{\ }with a configuration of 1\textit{d}\ensuremath{^{\textnormal{2}}_{\textnormal{5/2}}}. The evaluator assumed the J\ensuremath{^{\ensuremath{\pi}}} assignment is from the shell model calculation\vspace{0.1cm}}&\\
&&&\parbox[t][0.3cm]{15.023221cm}{\raggedright {\ }{\ }{\ }in (\href{https://www.nndc.bnl.gov/nsr/nsrlink.jsp?1988Kr11,B}{1988Kr11}).\vspace{0.1cm}}&\\
&&&\parbox[t][0.3cm]{15.023221cm}{\raggedright This state, formed by 2p transfer on \ensuremath{^{\textnormal{16}}}O, has a configuration of (\textit{d}\ensuremath{_{\textnormal{5/2}}})\ensuremath{^{\textnormal{2}}} (\href{https://www.nndc.bnl.gov/nsr/nsrlink.jsp?1972Sc21,B}{1972Sc21}, \href{https://www.nndc.bnl.gov/nsr/nsrlink.jsp?1970Ad02,B}{1970Ad02}).\vspace{0.1cm}}&\\
&&&\parbox[t][0.3cm]{15.023221cm}{\raggedright (\href{https://www.nndc.bnl.gov/nsr/nsrlink.jsp?1988Kr11,B}{1988Kr11}): the \ensuremath{^{\textnormal{18}}}Ne state at 3.4 MeV appears to be wide (see Fig. 3). This is because in the data analysis,\vspace{0.1cm}}&\\
&&&\parbox[t][0.3cm]{15.023221cm}{\raggedright {\ }{\ }{\ }the position of the focal plane is reconstructed for the Si target and not for the O target, which causes the\vspace{0.1cm}}&\\
&&&\parbox[t][0.3cm]{15.023221cm}{\raggedright {\ }{\ }{\ }\ensuremath{^{\textnormal{18}}}Ne states to suffer from kinematics broadening (out of focus).\vspace{0.1cm}}&\\
\multicolumn{1}{r@{}}{7.9\ensuremath{\times10^{3}}}&\multicolumn{1}{@{}l}{}&\multicolumn{1}{l}{(4\ensuremath{^{+}})}&\parbox[t][0.3cm]{15.023221cm}{\raggedright E(level),J\ensuremath{^{\pi}}: From Fig. 3 of (\href{https://www.nndc.bnl.gov/nsr/nsrlink.jsp?1988Kr11,B}{1988Kr11}).\vspace{0.1cm}}&\\
&&&\parbox[t][0.3cm]{15.023221cm}{\raggedright E(level),J\ensuremath{^{\pi}}: The shell model calculation of (\href{https://www.nndc.bnl.gov/nsr/nsrlink.jsp?1988Kr11,B}{1988Kr11}) predicted the 4\ensuremath{^{\textnormal{+}}_{\textnormal{2}}} state in \ensuremath{^{\textnormal{18}}}Ne to be at 8.32 MeV\vspace{0.1cm}}&\\
&&&\parbox[t][0.3cm]{15.023221cm}{\raggedright {\ }{\ }{\ }with a configuration of 1\textit{d}\ensuremath{^{\textnormal{1}}_{\textnormal{5/2}}}1\textit{d}\ensuremath{^{\textnormal{1}}_{\textnormal{3/2}}}. So, the evaluator assumed the J\ensuremath{^{\ensuremath{\pi}}} assignment is from the shell model\vspace{0.1cm}}&\\
&&&\parbox[t][0.3cm]{15.023221cm}{\raggedright {\ }{\ }{\ }calculation in (\href{https://www.nndc.bnl.gov/nsr/nsrlink.jsp?1988Kr11,B}{1988Kr11}).\vspace{0.1cm}}&\\
\end{longtable}
\parbox[b][0.3cm]{17.7cm}{\makebox[1ex]{\ensuremath{^{\hypertarget{NE19LEVEL0}{a}}}} From (\href{https://www.nndc.bnl.gov/nsr/nsrlink.jsp?1988Kr11,B}{1988Kr11}): these states were unresolved and mixed with a \ensuremath{^{\textnormal{30}}}S* excited state with E\ensuremath{_{\textnormal{x}}}\ensuremath{\sim}5.1-8.3 MeV. The (\ensuremath{^{\textnormal{12}}}C,\ensuremath{^{\textnormal{10}}}Be)}\\
\parbox[b][0.3cm]{17.7cm}{{\ }{\ }reactions on the Si and O contents of the SiO\ensuremath{_{\textnormal{2}}} target, as well as poor energy resolution made the spectra of \ensuremath{^{\textnormal{18}}}Ne and \ensuremath{^{\textnormal{30}}}S in}\\
\parbox[b][0.3cm]{17.7cm}{{\ }{\ }(\href{https://www.nndc.bnl.gov/nsr/nsrlink.jsp?1988Kr11,B}{1988Kr11}) complicated. However, different angular kinematic shifts of the \ensuremath{^{\textnormal{10}}}Be reaction products, arising either from a reaction}\\
\parbox[b][0.3cm]{17.7cm}{{\ }{\ }on the O or on the Si, allowed partial disentanglement of the two reactions.}\\
\parbox[b][0.3cm]{17.7cm}{\makebox[1ex]{\ensuremath{^{\hypertarget{NE19LEVEL1}{b}}}} It is not clear from (\href{https://www.nndc.bnl.gov/nsr/nsrlink.jsp?1972Sc21,B}{1972Sc21}) and (\href{https://www.nndc.bnl.gov/nsr/nsrlink.jsp?1974An36,B}{1974An36}) if the reported J\ensuremath{^{\ensuremath{\pi}}} assignment is based on the shell model calculations, or from}\\
\parbox[b][0.3cm]{17.7cm}{{\ }{\ }the literature. In (\href{https://www.nndc.bnl.gov/nsr/nsrlink.jsp?1988Kr11,B}{1988Kr11}), finite range DWBA was used but no angular distribution data are presented or discussed for the}\\
\parbox[b][0.3cm]{17.7cm}{{\ }{\ }populated \ensuremath{^{\textnormal{18}}}Ne states. So, it is not clear if the reported J\ensuremath{^{\ensuremath{\pi}}} assignments in (\href{https://www.nndc.bnl.gov/nsr/nsrlink.jsp?1988Kr11,B}{1988Kr11}) are from the DWBA calculations or from}\\
\parbox[b][0.3cm]{17.7cm}{{\ }{\ }shell model calculations.}\\
\vspace{0.5cm}
\clearpage
\subsection[\hspace{-0.2cm}\ensuremath{^{\textnormal{18}}}O(\ensuremath{\pi}\ensuremath{^{\textnormal{+}}},\ensuremath{\pi}\ensuremath{^{-}})]{ }
\vspace{-27pt}
\vspace{0.3cm}
\hypertarget{NE20}{{\bf \small \underline{\ensuremath{^{\textnormal{18}}}O(\ensuremath{\pi}\ensuremath{^{\textnormal{+}}},\ensuremath{\pi}\ensuremath{^{-}})\hspace{0.2in}\href{https://www.nndc.bnl.gov/nsr/nsrlink.jsp?1974LiZR,B}{1974LiZR},\href{https://www.nndc.bnl.gov/nsr/nsrlink.jsp?2007Ke07,B}{2007Ke07}}}}\\
\vspace{4pt}
\vspace{8pt}
\parbox[b][0.3cm]{17.7cm}{\addtolength{\parindent}{-0.2in}\href{https://www.nndc.bnl.gov/nsr/nsrlink.jsp?1968Ch46,B}{1968Ch46}: \ensuremath{^{\textnormal{18}}}O(\ensuremath{\pi}\ensuremath{^{\ensuremath{\pm}}},\ensuremath{\pi}\ensuremath{^{\ensuremath{\mp}}}) E=80-280 MeV; measured the activation cross sections for reactions of pions with light nuclie, including}\\
\parbox[b][0.3cm]{17.7cm}{\ensuremath{^{\textnormal{18}}}O over the energy range of 80 to 280 MeV. No \ensuremath{^{\textnormal{18}}}Ne state was observed. An upper limit on the \ensuremath{^{\textnormal{18}}}O(\ensuremath{\pi}\ensuremath{^{\textnormal{+}}},\ensuremath{\pi}\ensuremath{^{-}})\ensuremath{^{\textnormal{18}}}Ne\ensuremath{_{\textnormal{g.s.}}} cross}\\
\parbox[b][0.3cm]{17.7cm}{section was set to \ensuremath{\sigma}\ensuremath{<}0.1 mb.}\\
\parbox[b][0.3cm]{17.7cm}{\addtolength{\parindent}{-0.2in}\href{https://www.nndc.bnl.gov/nsr/nsrlink.jsp?1974LiZR,B}{1974LiZR}: \ensuremath{^{\textnormal{18}}}O(\ensuremath{\pi}\ensuremath{^{\textnormal{+}}},\ensuremath{\pi}\ensuremath{^{-}}); measured \ensuremath{\sigma}.}\\
\parbox[b][0.3cm]{17.7cm}{\addtolength{\parindent}{-0.2in}\href{https://www.nndc.bnl.gov/nsr/nsrlink.jsp?1977Ma02,B}{1977Ma02}, \href{https://www.nndc.bnl.gov/nsr/nsrlink.jsp?1977MaYB,B}{1977MaYB}: \ensuremath{^{\textnormal{18}}}O(\ensuremath{\pi}\ensuremath{^{\textnormal{+}}},\ensuremath{\pi}\ensuremath{^{-}}) E=95-139 MeV (\href{https://www.nndc.bnl.gov/nsr/nsrlink.jsp?1977MaYB,B}{1977MaYB}), E=139 MeV (\href{https://www.nndc.bnl.gov/nsr/nsrlink.jsp?1977Ma02,B}{1977Ma02}); measured the \ensuremath{\pi}\ensuremath{^{-}} particles using the}\\
\parbox[b][0.3cm]{17.7cm}{focal plane detection system of the low-energy pion beamline of the Los Alamos Meson Physics Facility, which was used as a}\\
\parbox[b][0.3cm]{17.7cm}{spectrometer. The detection system consisted of 3 helical wire proportional chambers, 3 plastic Cherenkov counters, and 5}\\
\parbox[b][0.3cm]{17.7cm}{scintillator counters to measure the \ensuremath{\pi}\ensuremath{^{-}} trajectories, energies and time-of-flight. The intrinsic experimental resolution was 4 MeV.}\\
\parbox[b][0.3cm]{17.7cm}{The \ensuremath{^{\textnormal{18}}}Ne\ensuremath{_{\textnormal{g.s.}}} was observed at E\ensuremath{_{\ensuremath{\pi}^{-}}}=130 MeV. The differential cross section at \ensuremath{\theta}\ensuremath{_{\textnormal{c.m.}}}=0\ensuremath{^\circ} was deduced to be d\ensuremath{\sigma}/d\ensuremath{\Omega}\ensuremath{_{\textnormal{lab}}}(0\ensuremath{^\circ})=1.78}\\
\parbox[b][0.3cm]{17.7cm}{\ensuremath{\mu}b/sr \textit{30}. Comparison between this result and theoretical calculations are presented.}\\
\parbox[b][0.3cm]{17.7cm}{\addtolength{\parindent}{-0.2in}\href{https://www.nndc.bnl.gov/nsr/nsrlink.jsp?1977Pe12,B}{1977Pe12}: \ensuremath{^{\textnormal{18}}}O(\ensuremath{\pi}\ensuremath{^{\textnormal{+}}},\ensuremath{\pi}\ensuremath{^{-}}) E=148 and 187 MeV; used the SIN \ensuremath{\pi}M1 beamline for tuning \ensuremath{\pi}\ensuremath{^{\textnormal{+}}} beam and pion spectrometer for tuning}\\
\parbox[b][0.3cm]{17.7cm}{the \ensuremath{\pi}\ensuremath{^{-}} ejectiles; resolutions were 0.7 MeV at 148 MeV, and 1.1 MeV at 187 MeV; the experiments were performed at \ensuremath{\theta}\ensuremath{_{\textnormal{lab}}}=18\ensuremath{^\circ}; a}\\
\parbox[b][0.3cm]{17.7cm}{series of multi-wire proportional counters and scintillators were used to measure the momenta and scattering angles; measured}\\
\parbox[b][0.3cm]{17.7cm}{differential cross sections normalized to the measured \ensuremath{^{\textnormal{12}}}C elastic scattering data; deduced d\ensuremath{\sigma}/d\ensuremath{\Omega}\ensuremath{_{\textnormal{lab}}}(\ensuremath{\theta}\ensuremath{_{\textnormal{lab}}}=18\ensuremath{^\circ}, E=148 MeV)=0.30}\\
\parbox[b][0.3cm]{17.7cm}{\ensuremath{\mu}b/sr \textit{10} and d\ensuremath{\sigma}/d\ensuremath{\Omega}\ensuremath{_{\textnormal{lab}}}(\ensuremath{\theta}\ensuremath{_{\textnormal{lab}}}=18\ensuremath{^\circ}, E=187 MeV)=0.21 \ensuremath{\mu}b/sr \textit{8} for the \ensuremath{^{\textnormal{18}}}O(\ensuremath{\pi}\ensuremath{^{\textnormal{+}}},\ensuremath{\pi}\ensuremath{^{-}})\ensuremath{^{\textnormal{18}}}Ne\ensuremath{_{\textnormal{g.s.}}}; upper limits were set for the cross}\\
\parbox[b][0.3cm]{17.7cm}{section of the first excited state of \ensuremath{^{\textnormal{18}}}Ne (unresolved from \ensuremath{^{\textnormal{18}}}Ne\ensuremath{_{\textnormal{g.s.}}}): d\ensuremath{\sigma}/d\ensuremath{\Omega}\ensuremath{_{\textnormal{lab}}}(\ensuremath{\theta}\ensuremath{_{\textnormal{lab}}}=18\ensuremath{^\circ}, E=148 MeV)\ensuremath{<}70 nb/sr and}\\
\parbox[b][0.3cm]{17.7cm}{d\ensuremath{\sigma}/d\ensuremath{\Omega}\ensuremath{_{\textnormal{lab}}}(\ensuremath{\theta}\ensuremath{_{\textnormal{lab}}}=18\ensuremath{^\circ}, E=187 MeV)\ensuremath{\sim}120 nb/sr for \ensuremath{^{\textnormal{18}}}Ne*(1.89 MeV) level. The authors deduced the laboratory differential cross}\\
\parbox[b][0.3cm]{17.7cm}{section by integrating over the double charge exchange continuum and the excited states of \ensuremath{^{\textnormal{18}}}Ne up to 20 MeV excitation energy}\\
\parbox[b][0.3cm]{17.7cm}{and obtained d\ensuremath{\sigma}/d\ensuremath{\Omega}\ensuremath{_{\textnormal{lab}}}=3.8 \ensuremath{\mu}b/sr \textit{70} and 3.0 \ensuremath{\mu}b/sr \textit{5} for \ensuremath{\theta}\ensuremath{_{\textnormal{lab}}}=18\ensuremath{^\circ} and E=148 MeV and 187 MeV, respectively.}\\
\parbox[b][0.3cm]{17.7cm}{\addtolength{\parindent}{-0.2in}\href{https://www.nndc.bnl.gov/nsr/nsrlink.jsp?1978Bu09,B}{1978Bu09}: \ensuremath{^{\textnormal{18}}}O(\ensuremath{\pi}\ensuremath{^{\textnormal{+}}},\ensuremath{\pi}\ensuremath{^{-}}) E=95, 126, and 139 MeV; used the low energy pion channel at the Los Alamos Meson Physics Facility.}\\
\parbox[b][0.3cm]{17.7cm}{The experimental setup and detection system was identical to that of (\href{https://www.nndc.bnl.gov/nsr/nsrlink.jsp?1977Ma02,B}{1977Ma02}). The authors state that the (\href{https://www.nndc.bnl.gov/nsr/nsrlink.jsp?1977Ma02,B}{1977Ma02})}\\
\parbox[b][0.3cm]{17.7cm}{publication is the initial results of this experiment. Energy resolution was 4 MeV (FWHM). Laboratory differential cross sections}\\
\parbox[b][0.3cm]{17.7cm}{for the \ensuremath{^{\textnormal{18}}}O(\ensuremath{\pi}\ensuremath{^{\textnormal{+}}},\ensuremath{\pi}\ensuremath{^{-}})\ensuremath{^{\textnormal{18}}}Ne\ensuremath{_{\textnormal{g.s.}}} reaction at \ensuremath{\theta}\ensuremath{_{\textnormal{lab}}}=0\ensuremath{^\circ} were deduced to be d\ensuremath{\sigma}/d\ensuremath{\Omega}\ensuremath{_{\textnormal{lab}}}(\ensuremath{\theta}\ensuremath{_{\textnormal{lab}}}=0\ensuremath{^\circ})=2.00 \ensuremath{\mu}b/sr \textit{34} at E=139 MeV,}\\
\parbox[b][0.3cm]{17.7cm}{d\ensuremath{\sigma}/d\ensuremath{\Omega}\ensuremath{_{\textnormal{lab}}}(\ensuremath{\theta}\ensuremath{_{\textnormal{lab}}}=0\ensuremath{^\circ})=2.19 \ensuremath{\mu}b/sr \textit{44} at E=126 MeV, and d\ensuremath{\sigma}/d\ensuremath{\Omega}\ensuremath{_{\textnormal{lab}}}(\ensuremath{\theta}\ensuremath{_{\textnormal{lab}}}=0\ensuremath{^\circ})=1.67 \ensuremath{\mu}b/sr \textit{38} at E=95 MeV. The result at 139 MeV}\\
\parbox[b][0.3cm]{17.7cm}{is consistent with that of (\href{https://www.nndc.bnl.gov/nsr/nsrlink.jsp?1977Ma02,B}{1977Ma02}). Comparison between the deduced ground state cross section and various theoretical models}\\
\parbox[b][0.3cm]{17.7cm}{are presented and discussed. A conclusion is made that the calculations based on first-order optical equations are an incomplete}\\
\parbox[b][0.3cm]{17.7cm}{representation for \ensuremath{^{\textnormal{18}}}Ne\ensuremath{_{\textnormal{g.s.}}}, and that higher-order processes contribute significantly to the \ensuremath{^{\textnormal{18}}}O(\ensuremath{\pi}\ensuremath{^{\textnormal{+}}},\ensuremath{\pi}\ensuremath{^{-}})\ensuremath{^{\textnormal{18}}}Ne\ensuremath{_{\textnormal{g.s.}}} reaction. The}\\
\parbox[b][0.3cm]{17.7cm}{(\href{https://www.nndc.bnl.gov/nsr/nsrlink.jsp?1977Ma02,B}{1977Ma02}) presents the initial results of (\href{https://www.nndc.bnl.gov/nsr/nsrlink.jsp?1978Bu09,B}{1978Bu09}).}\\
\parbox[b][0.3cm]{17.7cm}{\addtolength{\parindent}{-0.2in}\href{https://www.nndc.bnl.gov/nsr/nsrlink.jsp?1978SeZU,B}{1978SeZU}: \ensuremath{^{\textnormal{18}}}O(\ensuremath{\pi}\ensuremath{^{\textnormal{+}}},\ensuremath{\pi}\ensuremath{^{-}}) E not given; measured absolute \ensuremath{\sigma}.}\\
\parbox[b][0.3cm]{17.7cm}{\addtolength{\parindent}{-0.2in}\href{https://www.nndc.bnl.gov/nsr/nsrlink.jsp?1979Gr18,B}{1979Gr18}, \href{https://www.nndc.bnl.gov/nsr/nsrlink.jsp?1979GrZG,B}{1979GrZG}: \ensuremath{^{\textnormal{18}}}O(\ensuremath{\pi}\ensuremath{^{\textnormal{+}}},\ensuremath{\pi}\ensuremath{^{-}}) E=164 and 292 MeV; measured the ground state angular distribution at 164 MeV and}\\
\parbox[b][0.3cm]{17.7cm}{\ensuremath{\theta}\ensuremath{_{\textnormal{lab}}}=5\ensuremath{^\circ}{\textminus}33\ensuremath{^\circ}; measured the excitation function of the \ensuremath{^{\textnormal{18}}}O(\ensuremath{\pi}\ensuremath{^{\textnormal{+}}},\ensuremath{\pi}\ensuremath{^{-}})\ensuremath{^{\textnormal{18}}}Ne reaction at \ensuremath{\theta}\ensuremath{_{\textnormal{lab}}}=5\ensuremath{^\circ} using the EPICS pion spectrometer}\\
\parbox[b][0.3cm]{17.7cm}{facility at LAMPF and a time-of-flight and a freon gas Cherenkov detector to reject electrons in the spectrometer$'$s focal plane; The}\\
\parbox[b][0.3cm]{17.7cm}{excitation function shows the ground and first excited states of \ensuremath{^{\textnormal{18}}}Ne. The first excited state is populated with relatively high}\\
\parbox[b][0.3cm]{17.7cm}{statistics. This excitation function covers, for the first time, the full region of the (3,3) \ensuremath{\pi}-nucleon resonance (aka \ensuremath{\Delta}\ensuremath{_{\textnormal{33}}} resonance).}\\
\parbox[b][0.3cm]{17.7cm}{In addition, d\ensuremath{\sigma}/d\ensuremath{\Omega}\ensuremath{_{\textnormal{lab}}}(\ensuremath{\theta}\ensuremath{_{\textnormal{lab}}}=5\ensuremath{^\circ}) was measured as a function of incident energy from 80 to 292 MeV. Cross section varied between}\\
\parbox[b][0.3cm]{17.7cm}{0.8 \ensuremath{\mu}b/sr at 80 MeV to 2.5 \ensuremath{\mu}b/sr at 292 MeV. These data were compared with calculations using a local Laplacian potential}\\
\parbox[b][0.3cm]{17.7cm}{(\href{https://www.nndc.bnl.gov/nsr/nsrlink.jsp?1974Mi22,B}{1974Mi22}), and a fixed scatterer model (W. R. Gibbs, B. F. Gibson, A. T. Hess and G. J. Stephenson, private communication).}\\
\parbox[b][0.3cm]{17.7cm}{The agreement between these calculations and the data is poor but the former model predicts the overall shape better.}\\
\parbox[b][0.3cm]{17.7cm}{\addtolength{\parindent}{-0.2in}\href{https://www.nndc.bnl.gov/nsr/nsrlink.jsp?1979Se08,B}{1979Se08}, Los Alamos Scientific Laboratory Report No. LA-7892-C, 1979, p. 201 (unpublished): \ensuremath{^{\textnormal{18}}}O(\ensuremath{\pi}\ensuremath{^{\textnormal{+}}},\ensuremath{\pi}\ensuremath{^{-}}) E=164 MeV; used}\\
\parbox[b][0.3cm]{17.7cm}{the EPICS pion spectrometer facility at LAMPF; energy resolution: 600 {\textminus} 800 keV (FWHM); (\href{https://www.nndc.bnl.gov/nsr/nsrlink.jsp?1979Se08,B}{1979Se08}) is the first measurement}\\
\parbox[b][0.3cm]{17.7cm}{ever of angular distributions for double charge exchange transitions to a discrete excited nuclear state; measured angular}\\
\parbox[b][0.3cm]{17.7cm}{distributions at \ensuremath{\theta}\ensuremath{_{\textnormal{lab}}}=13\ensuremath{^\circ}, 18\ensuremath{^\circ}, 23\ensuremath{^\circ}, 30\ensuremath{^\circ}, and 45\ensuremath{^\circ}. The angular distribution for the transition to the \ensuremath{^{\textnormal{18}}}Ne\ensuremath{_{\textnormal{g.s.}}} monotonically decrease}\\
\parbox[b][0.3cm]{17.7cm}{from 5.8 \ensuremath{\mu}b/sr around \ensuremath{\theta}\ensuremath{_{\textnormal{c.m.}}}=13\ensuremath{^\circ} (see the errata) to 2.5 \ensuremath{\mu}b/sr at \ensuremath{\theta}\ensuremath{_{\textnormal{c.m.}}}=45\ensuremath{^\circ}. Measured \ensuremath{\pi}\ensuremath{^{\textnormal{+}}} elastic scattering yields between 13\ensuremath{^\circ}}\\
\parbox[b][0.3cm]{17.7cm}{and 45\ensuremath{^\circ}; used these data and those of (\href{https://www.nndc.bnl.gov/nsr/nsrlink.jsp?1978Iv01,B}{1978Iv01}, \href{https://www.nndc.bnl.gov/nsr/nsrlink.jsp?1979Iv03,B}{1979Iv03}) to deduce an average normalization for the absolute cross section;}\\
\parbox[b][0.3cm]{17.7cm}{measured \ensuremath{\sigma}(\ensuremath{\theta}) for the ground and the 2\ensuremath{^{\textnormal{+}}_{\textnormal{1}}} states in \ensuremath{^{\textnormal{18}}}Ne; placed an upper limit of \ensuremath{\sim}200 nb/sr for populating the \ensuremath{^{\textnormal{18}}}Ne*(2\ensuremath{^{\textnormal{+}}_{\textnormal{1}}}) state}\\
\parbox[b][0.3cm]{17.7cm}{at \ensuremath{\theta}\ensuremath{_{\textnormal{c.m.}}}=0\ensuremath{^\circ}; discussed strengths and shortcomings of theoretical models used (\href{https://www.nndc.bnl.gov/nsr/nsrlink.jsp?1976Mi23,B}{1976Mi23}, \href{https://www.nndc.bnl.gov/nsr/nsrlink.jsp?1978Sp07,B}{1978Sp07}, \href{https://www.nndc.bnl.gov/nsr/nsrlink.jsp?1978Os02,B}{1978Os02}) to describe}\\
\parbox[b][0.3cm]{17.7cm}{measured angular distributions; the agreement between the data and these models were poor; argued that a large radius (4.8 fm) is}\\
\parbox[b][0.3cm]{17.7cm}{required for \ensuremath{^{\textnormal{18}}}O to theoretically produce the first minima of the measured angular distributions at \ensuremath{\theta}\ensuremath{_{\textnormal{c.m.}}}\ensuremath{\sim}21\ensuremath{^\circ}; populated analog and}\\
\parbox[b][0.3cm]{17.7cm}{non-analog \ensuremath{^{\textnormal{18}}}Ne states.}\\
\parbox[b][0.3cm]{17.7cm}{\addtolength{\parindent}{-0.2in}\href{https://www.nndc.bnl.gov/nsr/nsrlink.jsp?1980GrZZ,B}{1980GrZZ}: \ensuremath{^{\textnormal{18}}}O(\ensuremath{\pi}\ensuremath{^{\textnormal{+}}},\ensuremath{\pi}\ensuremath{^{-}}) E=164 and 292 MeV; measured \ensuremath{\sigma}(\ensuremath{\theta}). Deduced double-isobaric analog transitions relative to non-analog}\\
\parbox[b][0.3cm]{17.7cm}{transitions in \ensuremath{^{\textnormal{18}}}Ne and investigated their energy dependence.}\\
\parbox[b][0.3cm]{17.7cm}{\addtolength{\parindent}{-0.2in}\href{https://www.nndc.bnl.gov/nsr/nsrlink.jsp?1981GrZS,B}{1981GrZS}: \ensuremath{^{\textnormal{18}}}O(\ensuremath{\pi}\ensuremath{^{\textnormal{+}}},\ensuremath{\pi}\ensuremath{^{-}}) E=100-290 MeV; measured d\ensuremath{\sigma}/d\ensuremath{\Omega}(\ensuremath{^{\textnormal{18}}}Ne\ensuremath{_{\textnormal{g.s.}}}) and d\ensuremath{\sigma}/d\ensuremath{\Omega}(\ensuremath{^{\textnormal{18}}}Ne(2\ensuremath{^{\textnormal{+}}_{\textnormal{1}}})) at \ensuremath{\theta}\ensuremath{_{\textnormal{lab}}}=5\ensuremath{^\circ} as a function of incident}\\
\parbox[b][0.3cm]{17.7cm}{energy for E=80-310 MeV using the EPICS pion spectrometer facility. Deduced reaction mechanism and structure effects.}\\
\clearpage
\vspace{0.3cm}
{\bf \small \underline{\ensuremath{^{\textnormal{18}}}O(\ensuremath{\pi}\ensuremath{^{\textnormal{+}}},\ensuremath{\pi}\ensuremath{^{-}})\hspace{0.2in}\href{https://www.nndc.bnl.gov/nsr/nsrlink.jsp?1974LiZR,B}{1974LiZR},\href{https://www.nndc.bnl.gov/nsr/nsrlink.jsp?2007Ke07,B}{2007Ke07} (continued)}}\\
\vspace{0.3cm}
\parbox[b][0.3cm]{17.7cm}{\addtolength{\parindent}{-0.2in}\href{https://www.nndc.bnl.gov/nsr/nsrlink.jsp?1982Gr02,B}{1982Gr02},\href{https://www.nndc.bnl.gov/nsr/nsrlink.jsp?1982GrZV,B}{1982GrZV}: \ensuremath{^{\textnormal{18}}}O(\ensuremath{\pi}\ensuremath{^{\textnormal{+}}},\ensuremath{\pi}\ensuremath{^{-}}) E=80-292 MeV; used the EPICS spectrometer facility; measured angular distributions for}\\
\parbox[b][0.3cm]{17.7cm}{\ensuremath{^{\textnormal{18}}}Ne\ensuremath{_{\textnormal{g.s.}}} at \ensuremath{\theta}\ensuremath{_{\textnormal{lab}}}=5\ensuremath{^\circ}{\textminus}35\ensuremath{^\circ}; measured \ensuremath{\sigma}(\ensuremath{\theta}) vs. E for \ensuremath{^{\textnormal{18}}}Ne\ensuremath{_{\textnormal{g.s.}}} and observed that the cross section peaks near 130 MeV, dips near 170}\\
\parbox[b][0.3cm]{17.7cm}{MeV, and shows a smooth rise above that energy, which is in contrast with the theoretical predictions (\href{https://www.nndc.bnl.gov/nsr/nsrlink.jsp?1975Li04,B}{1975Li04}, \href{https://www.nndc.bnl.gov/nsr/nsrlink.jsp?1980Ge09,B}{1980Ge09}, G. A.}\\
\parbox[b][0.3cm]{17.7cm}{Miller, Bull. Am. Phys. Soc. 25 (1981) 731). (\href{https://www.nndc.bnl.gov/nsr/nsrlink.jsp?1982Gr02,B}{1982Gr02}) showed that the \ensuremath{\sigma}(\ensuremath{\theta}) vs. E for the ground state of \ensuremath{^{\textnormal{18}}}Ne can be}\\
\parbox[b][0.3cm]{17.7cm}{explained by adding the direct double-isobaric analog transition (DIAT) amplitude (quasi-elastic amplitude) and non-DIAT}\\
\parbox[b][0.3cm]{17.7cm}{amplitude, which is the sum of all processes that can change any two neutrons into any two protons without violating the Pauli}\\
\parbox[b][0.3cm]{17.7cm}{principle. The interference of these two amplitudes gives a good account of the \ensuremath{\sigma}(\ensuremath{\theta}) vs. E data throughout the energy range. This}\\
\parbox[b][0.3cm]{17.7cm}{model also explains the \ensuremath{^{\textnormal{18}}}Ne\ensuremath{_{\textnormal{g.s.}}} angular distributions at 164 and 292 MeV: strong interference with non-DIAT amplitude changes}\\
\parbox[b][0.3cm]{17.7cm}{the location of the minimum to lower angles at 164 MeV incident energy.}\\
\parbox[b][0.3cm]{17.7cm}{\addtolength{\parindent}{-0.2in}\href{https://www.nndc.bnl.gov/nsr/nsrlink.jsp?1982Gr28,B}{1982Gr28}: \ensuremath{^{\textnormal{18}}}O(\ensuremath{\pi}\ensuremath{^{\textnormal{+}}},\ensuremath{\pi}\ensuremath{^{-}}) E=100-300 MeV; a systematic investigation via (\ensuremath{\pi}\ensuremath{^{\textnormal{+}}},\ensuremath{\pi}\ensuremath{^{-}}) on \ensuremath{^{\textnormal{9}}}Be, \ensuremath{^{\textnormal{12,13}}}C, \ensuremath{^{\textnormal{16,18}}}O, \ensuremath{^{\textnormal{24,26}}}Mg, \ensuremath{^{\textnormal{32}}}S, and}\\
\parbox[b][0.3cm]{17.7cm}{\ensuremath{^{\textnormal{209}}}Bi. The EPICS pion facility and its associated detection system at LAMPF were used. Energy resolution was \ensuremath{\sim}150 keV.}\\
\parbox[b][0.3cm]{17.7cm}{Angular distributions were measured at \ensuremath{\theta}\ensuremath{_{\textnormal{lab}}}=5\ensuremath{^\circ}{\textminus}33\ensuremath{^\circ} for \ensuremath{^{\textnormal{18}}}Ne\ensuremath{_{\textnormal{g.s.}}} (at E=164 and 292 MeV) and the \ensuremath{^{\textnormal{18}}}Ne(2\ensuremath{^{\textnormal{+}}_{\textnormal{1}}}) state (at 164 MeV).}\\
\parbox[b][0.3cm]{17.7cm}{At the latter incident energy, the angular distribution of \ensuremath{^{\textnormal{18}}}Ne\ensuremath{_{\textnormal{g.s.}}} shows a variation in the cross section from \ensuremath{\sim}1 \ensuremath{\mu}b/sr at 5\ensuremath{^\circ} falling}\\
\parbox[b][0.3cm]{17.7cm}{into a minimum (40 nb/sr) at 20\ensuremath{^\circ} followed by a rise to \ensuremath{\sim}250 nb/sr at 33\ensuremath{^\circ}. At 292 MeV, the position of the minimum is shifted}\\
\parbox[b][0.3cm]{17.7cm}{outwards to near 25\ensuremath{^\circ}. The angular distribution of the 2\ensuremath{^{\textnormal{+}}_{\textnormal{1}}} state has a shape consistent with a \ensuremath{\Delta}L=2 transition, in agreement with}\\
\parbox[b][0.3cm]{17.7cm}{the result of (\href{https://www.nndc.bnl.gov/nsr/nsrlink.jsp?1979Se08,B}{1979Se08}). The missing mass spectrum of the \ensuremath{^{\textnormal{18}}}Ne was deduced. The excitation functions (d\ensuremath{\sigma}/d\ensuremath{\Omega}\ensuremath{_{\textnormal{lab}}}(E)) were}\\
\parbox[b][0.3cm]{17.7cm}{measured for \ensuremath{^{\textnormal{18}}}Ne(g.s., and 2\ensuremath{^{\textnormal{+}}_{\textnormal{1}}}) levels at \ensuremath{\theta}\ensuremath{_{\textnormal{lab}}}=5\ensuremath{^\circ} and E=80-320 MeV. The authors discuss the result of a second-order optical}\\
\parbox[b][0.3cm]{17.7cm}{model with isovector and isotensor terms, developed by (M. B. Johnson and E. R. Siciliano, Bull. Am. Phys. Soc. 25 (1980) 741),}\\
\parbox[b][0.3cm]{17.7cm}{which fits very well the angular distribution of \ensuremath{^{\textnormal{18}}}Ne\ensuremath{_{\textnormal{g.s.}}} at 164 MeV.}\\
\parbox[b][0.3cm]{17.7cm}{\addtolength{\parindent}{-0.2in}\href{https://www.nndc.bnl.gov/nsr/nsrlink.jsp?1983AnZT,B}{1983AnZT}: \ensuremath{^{\textnormal{18}}}O(\ensuremath{\pi}\ensuremath{^{\textnormal{+}}},\ensuremath{\pi}\ensuremath{^{-}}) E=65 MeV; measured \ensuremath{\sigma}(E\ensuremath{_{\ensuremath{\pi}}}); deduced reaction \ensuremath{\sigma}(\ensuremath{\theta}).}\\
\parbox[b][0.3cm]{17.7cm}{\addtolength{\parindent}{-0.2in}\href{https://www.nndc.bnl.gov/nsr/nsrlink.jsp?1984MoZU,B}{1984MoZU}: \ensuremath{^{\textnormal{18}}}O(\ensuremath{\pi}\ensuremath{^{\textnormal{+}}},\ensuremath{\pi}\ensuremath{^{-}}) E=140, 200 MeV; measured \ensuremath{\sigma}(\ensuremath{\theta}). Deduced \ensuremath{^{\textnormal{18}}}Ne analog double charge exchange excitation energy}\\
\parbox[b][0.3cm]{17.7cm}{dependence.}\\
\parbox[b][0.3cm]{17.7cm}{\addtolength{\parindent}{-0.2in}\href{https://www.nndc.bnl.gov/nsr/nsrlink.jsp?1985Al15,B}{1985Al15},\href{https://www.nndc.bnl.gov/nsr/nsrlink.jsp?1985AlZV,B}{1985AlZV},\href{https://www.nndc.bnl.gov/nsr/nsrlink.jsp?1985GiZX,B}{1985GiZX}: \ensuremath{^{\textnormal{18}}}O(\ensuremath{\pi}\ensuremath{^{\textnormal{+}}},\ensuremath{\pi}\ensuremath{^{-}}) E=50 MeV; measured d\ensuremath{\sigma}/d\ensuremath{\Omega}(\ensuremath{\theta}) for the double isobaric analog state (DIAS) using}\\
\parbox[b][0.3cm]{17.7cm}{the M13 pion channel and the QQD spectrometer at TRIUMF. The differential cross sections were measured at \ensuremath{\theta}\ensuremath{_{\textnormal{lab}}}=22\ensuremath{^\circ}, 32\ensuremath{^\circ}, 42\ensuremath{^\circ},}\\
\parbox[b][0.3cm]{17.7cm}{52\ensuremath{^\circ}, 92\ensuremath{^\circ}, and 122\ensuremath{^\circ}. The detection system consisted of 4 two-dimensional multi-wire proportional chambers with which the momenta}\\
\parbox[b][0.3cm]{17.7cm}{of the \ensuremath{\pi}\ensuremath{^{-}} ejectiles were measured. The resolution was 1 MeV. The cross section for populating \ensuremath{^{\textnormal{18}}}Ne\ensuremath{_{\textnormal{g.s.}}} at \ensuremath{\theta}\ensuremath{_{\textnormal{c.m.}}}=0\ensuremath{^\circ} at 50 MeV}\\
\parbox[b][0.3cm]{17.7cm}{was deduced to be 5.3 \ensuremath{\mu}b/sr \textit{5}, and a total angle integrated cross section was deduced as 16.7 \ensuremath{\mu}b \textit{12}. Comparison with}\\
\parbox[b][0.3cm]{17.7cm}{\ensuremath{^{\textnormal{14}}}C(\ensuremath{\pi}\ensuremath{^{\textnormal{+}}},\ensuremath{\pi}\ensuremath{^{-}})\ensuremath{^{\textnormal{14}}}O\ensuremath{_{\textnormal{g.s.}}} at 50 MeV, and the A-dependence of the cross section at 50, 164 and 292 MeV are discussed.}\\
\parbox[b][0.3cm]{17.7cm}{\addtolength{\parindent}{-0.2in}\href{https://www.nndc.bnl.gov/nsr/nsrlink.jsp?1985Se08,B}{1985Se08}, \href{https://www.nndc.bnl.gov/nsr/nsrlink.jsp?1985SeZZ,B}{1985SeZZ}: \ensuremath{^{\textnormal{18}}}O(\ensuremath{\pi}\ensuremath{^{\textnormal{+}}},\ensuremath{\pi}\ensuremath{^{-}}) E=100-292 MeV; used the EPICS facility at LAMPF and its associated detection system.}\\
\parbox[b][0.3cm]{17.7cm}{Measured the differential cross sections of the \ensuremath{^{\textnormal{18}}}O(\ensuremath{\pi}\ensuremath{^{\textnormal{+}}},\ensuremath{\pi}\ensuremath{^{-}}) reaction populating the \ensuremath{^{\textnormal{18}}}Ne\ensuremath{_{\textnormal{g.s.}}} and \ensuremath{^{\textnormal{18}}}Ne(2\ensuremath{^{\textnormal{+}}_{\textnormal{1}}}) states as a function of}\\
\parbox[b][0.3cm]{17.7cm}{pion incident energy and momentum transfer. Performed calculations representing the lowest-order sequential charge exchange}\\
\parbox[b][0.3cm]{17.7cm}{through the intermediate analog state using the PIESDEX code for E\ensuremath{_{\ensuremath{\pi}}}=180-292 MeV. This model describes the measured angular}\\
\parbox[b][0.3cm]{17.7cm}{distribution of the differential cross sections of \ensuremath{^{\textnormal{18}}}Ne\ensuremath{_{\textnormal{g.s.}}} very well at E\ensuremath{_{\ensuremath{\pi}}}=230 and 292 MeV. But it fails to explain the data at}\\
\parbox[b][0.3cm]{17.7cm}{E\ensuremath{_{\ensuremath{\pi}}}=180 and 200 MeV. For these lower energies, the angular distributions exhibit a minima at forward angles, which is the}\\
\parbox[b][0.3cm]{17.7cm}{evidence for a different reaction mechanism involving higher-order dynamic effects.}\\
\parbox[b][0.3cm]{17.7cm}{\addtolength{\parindent}{-0.2in}\href{https://www.nndc.bnl.gov/nsr/nsrlink.jsp?1986AnZX,B}{1986AnZX}: \ensuremath{^{\textnormal{18}}}O(\ensuremath{\pi}\ensuremath{^{\textnormal{+}}},\ensuremath{\pi}\ensuremath{^{-}}) E=24-80 MeV; measured the differential cross sections of the DIAT from the \ensuremath{^{\textnormal{18}}}O(\ensuremath{\pi}\ensuremath{^{\textnormal{+}}},\ensuremath{\pi}\ensuremath{^{-}}) reaction at}\\
\parbox[b][0.3cm]{17.7cm}{forward angles for different energies using the QQD-spectrometer at the pion channels M13 and M11 of TRIUMF. The results}\\
\parbox[b][0.3cm]{17.7cm}{together with those of (\href{https://www.nndc.bnl.gov/nsr/nsrlink.jsp?1982Gr28,B}{1982Gr28}, \href{https://www.nndc.bnl.gov/nsr/nsrlink.jsp?1985Se08,B}{1985Se08}) indicate that this cross section peaks at forward angles at low energies and that the}\\
\parbox[b][0.3cm]{17.7cm}{shape of its angular distribution differs for E\ensuremath{_{\ensuremath{\pi}}}\ensuremath{<}65 MeV and E\ensuremath{_{\ensuremath{\pi}}}\ensuremath{>}65 MeV.}\\
\parbox[b][0.3cm]{17.7cm}{\addtolength{\parindent}{-0.2in}\href{https://www.nndc.bnl.gov/nsr/nsrlink.jsp?1986AnZY,B}{1986AnZY}: \ensuremath{^{\textnormal{18}}}O(\ensuremath{\pi}\ensuremath{^{\textnormal{+}}},\ensuremath{\pi}\ensuremath{^{-}}) E=22, 33, 64 MeV; measured \ensuremath{\sigma}(\ensuremath{\theta}); deduced reaction mechanism and energy dependence of the}\\
\parbox[b][0.3cm]{17.7cm}{\ensuremath{^{\textnormal{18}}}O(\ensuremath{\pi}\ensuremath{^{\textnormal{+}}},\ensuremath{\pi}\ensuremath{^{-}}) reaction populating the DIAS (\ensuremath{^{\textnormal{18}}}Ne\ensuremath{_{\textnormal{g.s.}}}).}\\
\parbox[b][0.3cm]{17.7cm}{\addtolength{\parindent}{-0.2in}\href{https://www.nndc.bnl.gov/nsr/nsrlink.jsp?1989Wi02,B}{1989Wi02}: \ensuremath{^{\textnormal{18}}}O(\ensuremath{\pi}\ensuremath{^{\textnormal{+}}},\ensuremath{\pi}\ensuremath{^{-}}) E=300, 400, 500 MeV; used the Large Acceptance Spectrometer at the P\ensuremath{^{\textnormal{3}}} high energy channel of LAMPF.}\\
\parbox[b][0.3cm]{17.7cm}{The energy resolution was 2.8 MeV (FWHM) at 500 MeV. Measured the differential cross sections of the \ensuremath{^{\textnormal{18}}}O(\ensuremath{\pi}\ensuremath{^{\textnormal{+}}},\ensuremath{\pi}\ensuremath{^{-}}) reaction as}\\
\parbox[b][0.3cm]{17.7cm}{a function of energy at \ensuremath{\theta}\ensuremath{_{\textnormal{lab}}}=5\ensuremath{^\circ} and as a function of center-of-mass angle. Due to poor energy resolution, \ensuremath{^{\textnormal{18}}}Ne\ensuremath{_{\textnormal{g.s.}}} could not be}\\
\parbox[b][0.3cm]{17.7cm}{resolved from the contributions from the 2\ensuremath{^{\textnormal{+}}_{\textnormal{1}}} state at 1.89 MeV and the next triple excited states at \ensuremath{\sim}3.5 MeV. The angular}\\
\parbox[b][0.3cm]{17.7cm}{distribution of the cross section was measured, which was affected by the contributions of the \ensuremath{^{\textnormal{18}}}Ne excited states. Comparison with}\\
\parbox[b][0.3cm]{17.7cm}{literature is presented, and the results are discussed.}\\
\parbox[b][0.3cm]{17.7cm}{\addtolength{\parindent}{-0.2in}\href{https://www.nndc.bnl.gov/nsr/nsrlink.jsp?1992JoZZ,B}{1992JoZZ}, \href{https://www.nndc.bnl.gov/nsr/nsrlink.jsp?1993Jo03,B}{1993Jo03}: \ensuremath{^{\textnormal{18}}}O(\ensuremath{\pi}\ensuremath{^{\textnormal{+}}},\ensuremath{\pi}\ensuremath{^{-}}) E=350-440 MeV; measured the excitation function of the \ensuremath{^{\textnormal{18}}}O(\ensuremath{\pi}\ensuremath{^{\textnormal{+}}},\ensuremath{\pi}\ensuremath{^{-}}) reaction, populating the}\\
\parbox[b][0.3cm]{17.7cm}{DIAS, for momentum transfers of q=0, 105 and 210 MeV/c. The experiment was performed using the Large Acceptance}\\
\parbox[b][0.3cm]{17.7cm}{Spectrometer at the P\ensuremath{^{\textnormal{3}}} high energy channel of LAMPF with configuration, target and detection system identical to the ones used in}\\
\parbox[b][0.3cm]{17.7cm}{(\href{https://www.nndc.bnl.gov/nsr/nsrlink.jsp?1989Wi02,B}{1989Wi02}). The energy resolution was 2.3 MeV (FWHM). \ensuremath{^{\textnormal{18}}}O(\ensuremath{\pi}\ensuremath{^{\textnormal{+}}},\ensuremath{\pi}\ensuremath{^{-}})\ensuremath{^{\textnormal{18}}}Ne(DIAS) was measured at \ensuremath{\theta}\ensuremath{_{\textnormal{lab}}}=5\ensuremath{^\circ} and for fixed}\\
\parbox[b][0.3cm]{17.7cm}{momentum transfers of q=105 and 210 MeV/c. The DIAS state could not be resolved from the next 4 excited states up to 3.6 MeV}\\
\parbox[b][0.3cm]{17.7cm}{in excitation energy. The missing mass spectrum was deduced. Presented evidence for the observation of a structure near the \ensuremath{\eta}}\\
\clearpage
\vspace{0.3cm}
{\bf \small \underline{\ensuremath{^{\textnormal{18}}}O(\ensuremath{\pi}\ensuremath{^{\textnormal{+}}},\ensuremath{\pi}\ensuremath{^{-}})\hspace{0.2in}\href{https://www.nndc.bnl.gov/nsr/nsrlink.jsp?1974LiZR,B}{1974LiZR},\href{https://www.nndc.bnl.gov/nsr/nsrlink.jsp?2007Ke07,B}{2007Ke07} (continued)}}\\
\vspace{0.3cm}
\parbox[b][0.3cm]{17.7cm}{production threshold. However, poor statistics prevented the authors to make a conclusive argument.}\\
\vspace{0.385cm}
\parbox[b][0.3cm]{17.7cm}{\addtolength{\parindent}{-0.2in}\textit{Theory}:}\\
\parbox[b][0.3cm]{17.7cm}{\addtolength{\parindent}{-0.2in}\href{https://www.nndc.bnl.gov/nsr/nsrlink.jsp?1965Pa21,B}{1965Pa21}: \ensuremath{^{\textnormal{18}}}O(\ensuremath{\pi}\ensuremath{^{\textnormal{+}}},\ensuremath{\pi}\ensuremath{^{-}}); calculated d\ensuremath{\sigma}/d\ensuremath{\Omega} at 0\ensuremath{^\circ}, for \ensuremath{^{\textnormal{18}}}Ne\ensuremath{_{\textnormal{g.s.}}}, as a function of incident pion momentum (in units of m\ensuremath{_{\ensuremath{\pi}}}) using}\\
\parbox[b][0.3cm]{17.7cm}{shell model without the spin-orbit coupling effects. The authors used plane wave Born approximation and Chew-Low model.}\\
\parbox[b][0.3cm]{17.7cm}{\addtolength{\parindent}{-0.2in}\href{https://www.nndc.bnl.gov/nsr/nsrlink.jsp?1965Ko25,B}{1965Ko25}: \ensuremath{^{\textnormal{18}}}O(\ensuremath{\pi}\ensuremath{^{\textnormal{+}}},\ensuremath{\pi}\ensuremath{^{-}}) E=20-40 MeV; calculated the double charge exchange cross sections for pions with low incident energies.}\\
\parbox[b][0.3cm]{17.7cm}{The authors considered the pion scattering on nucleon pairs from light targets using the second Born approximation via a}\\
\parbox[b][0.3cm]{17.7cm}{semi-phenomenological \textit{s}-wave pion-nucleon interaction. The resultant cross sections were deduced to be \ensuremath{\sim}7 \ensuremath{\mu}b/sr.}\\
\parbox[b][0.3cm]{17.7cm}{\addtolength{\parindent}{-0.2in}\href{https://www.nndc.bnl.gov/nsr/nsrlink.jsp?1968Da32,B}{1968Da32}: \ensuremath{^{\textnormal{18}}}O(\ensuremath{\pi}\ensuremath{^{\textnormal{+}}},\ensuremath{\pi}\ensuremath{^{-}}); calculated the differential cross section of the double charge exchange process as a function of the}\\
\parbox[b][0.3cm]{17.7cm}{momentum transfer and of the excitation energy of the residual nucleus.}\\
\parbox[b][0.3cm]{17.7cm}{\addtolength{\parindent}{-0.2in}\href{https://www.nndc.bnl.gov/nsr/nsrlink.jsp?1974Mi22,B}{1974Mi22}, \href{https://www.nndc.bnl.gov/nsr/nsrlink.jsp?1976Mi23,B}{1976Mi23}: \ensuremath{^{\textnormal{18}}}O(\ensuremath{\pi}\ensuremath{^{\textnormal{+}}},\ensuremath{\pi}\ensuremath{^{-}}); calculated two nucleon short range correlations and angle transformation related to the}\\
\parbox[b][0.3cm]{17.7cm}{Kisslinger and local Laplacian potentials (\href{https://www.nndc.bnl.gov/nsr/nsrlink.jsp?1974Mi22,B}{1974Mi22}: off-shell), which are used for the framework of a coupled-channel optical}\\
\parbox[b][0.3cm]{17.7cm}{model. This model only considered analog transitions between pure \textit{d}\ensuremath{_{\textnormal{5/2}}^{\textnormal{2}}} states.}\\
\parbox[b][0.3cm]{17.7cm}{\addtolength{\parindent}{-0.2in}\href{https://www.nndc.bnl.gov/nsr/nsrlink.jsp?1974Ka07,B}{1974Ka07}: \ensuremath{^{\textnormal{18}}}O(\ensuremath{\pi}\ensuremath{^{\textnormal{+}}},\ensuremath{\pi}\ensuremath{^{-}}) E=180 MeV; calculated \ensuremath{\sigma}(E) using a fixed scatterer model, which included elastic multiple scattering}\\
\parbox[b][0.3cm]{17.7cm}{corrections to infinite order.}\\
\parbox[b][0.3cm]{17.7cm}{\addtolength{\parindent}{-0.2in}\href{https://www.nndc.bnl.gov/nsr/nsrlink.jsp?1975Li04,B}{1975Li04}: \ensuremath{^{\textnormal{18}}}O(\ensuremath{\pi}\ensuremath{^{\textnormal{+}}},\ensuremath{\pi}\ensuremath{^{-}}) E\ensuremath{<}550 MeV; calculated \ensuremath{\sigma}(E), \ensuremath{\sigma}(E,\ensuremath{\theta}=0\ensuremath{^\circ}) and \ensuremath{\sigma}(\ensuremath{\theta}) using an Eikonalized distorted wave formulation}\\
\parbox[b][0.3cm]{17.7cm}{obtained from the Glauber calculations. The calculated \ensuremath{\sigma}(E) shows a minimum near 130 MeV and a maximum near the (3,3)}\\
\parbox[b][0.3cm]{17.7cm}{resonance energy. The \ensuremath{\sigma}(\ensuremath{\theta}) distributions are forward peaked.}\\
\parbox[b][0.3cm]{17.7cm}{\addtolength{\parindent}{-0.2in}\href{https://www.nndc.bnl.gov/nsr/nsrlink.jsp?1976HeZU,B}{1976HeZU}: \ensuremath{^{\textnormal{18}}}O(\ensuremath{\pi}\ensuremath{^{\textnormal{+}}},\ensuremath{\pi}\ensuremath{^{-}}); calculated \ensuremath{\sigma}.}\\
\parbox[b][0.3cm]{17.7cm}{\addtolength{\parindent}{-0.2in}\href{https://www.nndc.bnl.gov/nsr/nsrlink.jsp?1977Le16,B}{1977Le16}: \ensuremath{^{\textnormal{18}}}O(\ensuremath{\pi}\ensuremath{^{\textnormal{+}}},\ensuremath{\pi}\ensuremath{^{-}}); investigated the effects of nuclear structure, such as pairing correlations in the nuclear wave functions, on}\\
\parbox[b][0.3cm]{17.7cm}{the total double charge exchange cross section using Glauber calculations. This was an attempt to explain the experimental ratio}\\
\parbox[b][0.3cm]{17.7cm}{deduced for the d\ensuremath{\sigma}/d\ensuremath{\Omega}(\ensuremath{^{\textnormal{18}}}Ne\ensuremath{_{\textnormal{g.s.}}}) to d\ensuremath{\sigma}/d\ensuremath{\Omega}(\ensuremath{^{\textnormal{16}}}Ne\ensuremath{_{\textnormal{g.s.}}}), both of which were populated by the (\ensuremath{\pi}\ensuremath{^{\textnormal{+}}},\ensuremath{\pi}\ensuremath{^{-}}) reactions.}\\
\parbox[b][0.3cm]{17.7cm}{\addtolength{\parindent}{-0.2in}\href{https://www.nndc.bnl.gov/nsr/nsrlink.jsp?1978Os02,B}{1978Os02}: \ensuremath{^{\textnormal{18}}}O(\ensuremath{\pi}\ensuremath{^{\textnormal{+}}},\ensuremath{\pi}\ensuremath{^{-}}) E=187-190 MeV; calculated \ensuremath{\sigma}(\ensuremath{\theta}) using the Glauber model formalism with a \textit{d}\ensuremath{_{\textnormal{5/2}}}, \textit{s}\ensuremath{_{\textnormal{1/2}}} basis for A=18}\\
\parbox[b][0.3cm]{17.7cm}{wave functions. The results were sensitive to the anti-symmetry of the nuclear wave function and details of the nuclear surface.}\\
\parbox[b][0.3cm]{17.7cm}{\addtolength{\parindent}{-0.2in}\href{https://www.nndc.bnl.gov/nsr/nsrlink.jsp?1978Sp07,B}{1978Sp07}: \ensuremath{^{\textnormal{18}}}O(\ensuremath{\pi}\ensuremath{^{\textnormal{+}}},\ensuremath{\pi}\ensuremath{^{-}}) E=100, 139, 187 MeV; calculated \ensuremath{\sigma}(\ensuremath{\theta}) for the ground state of \ensuremath{^{\textnormal{18}}}Ne using optical model assuming that a}\\
\parbox[b][0.3cm]{17.7cm}{single intermediate state in \ensuremath{^{\textnormal{18}}}F contributes. These singles particle states were considered to be the 0\ensuremath{^{\textnormal{+}}_{\textnormal{1}}}, 2\ensuremath{^{\textnormal{+}}_{\textnormal{1}}} and 4\ensuremath{^{\textnormal{+}}_{\textnormal{1}}} states of \ensuremath{^{\textnormal{18}}}F,}\\
\parbox[b][0.3cm]{17.7cm}{whose contributions seem to be important at E=100-200 MeV. The authors emphasized on the importance of non-analog}\\
\parbox[b][0.3cm]{17.7cm}{contributions to the double charge exchange scattering amplitude for populating a double isobaric analog state. However, their}\\
\parbox[b][0.3cm]{17.7cm}{model did not explain the slope of the measured cross sections from 0\ensuremath{^\circ}{\textminus}18\ensuremath{^\circ} (\href{https://www.nndc.bnl.gov/nsr/nsrlink.jsp?1977Ma02,B}{1977Ma02}, \href{https://www.nndc.bnl.gov/nsr/nsrlink.jsp?1977Pe12,B}{1977Pe12}, \href{https://www.nndc.bnl.gov/nsr/nsrlink.jsp?1978Bu09,B}{1978Bu09}).}\\
\parbox[b][0.3cm]{17.7cm}{\addtolength{\parindent}{-0.2in}\href{https://www.nndc.bnl.gov/nsr/nsrlink.jsp?1979RoZN,B}{1979RoZN}: \ensuremath{^{\textnormal{18}}}O(\ensuremath{\pi}\ensuremath{^{\textnormal{+}}},\ensuremath{\pi}\ensuremath{^{-}}) E\ensuremath{\approx}100 MeV; calculated \ensuremath{\sigma}(\ensuremath{\theta}) using pion-nucleus optical model and first-order multiple scattering theory.}\\
\parbox[b][0.3cm]{17.7cm}{\addtolength{\parindent}{-0.2in}Los Alamos Scientific Laboratory Report No. LA-7892-C, 1979, p. 343 (unpublished), \href{https://www.nndc.bnl.gov/nsr/nsrlink.jsp?1980Jo06,B}{1980Jo06}: \ensuremath{^{\textnormal{18}}}O(\ensuremath{\pi}\ensuremath{^{\textnormal{+}}},\ensuremath{\pi}\ensuremath{^{-}}) E=180, 164 MeV;}\\
\parbox[b][0.3cm]{17.7cm}{calculated \ensuremath{\sigma}(\ensuremath{\theta}) of pion single and double charge exchange to analog states. Calculations were performed using the geometrical}\\
\parbox[b][0.3cm]{17.7cm}{Eikonal approximation. The theoretical angular distributions were deduced using the Bessel functions with a cross section scaling}\\
\parbox[b][0.3cm]{17.7cm}{factor of A\ensuremath{^{\textnormal{$-$10/3}}}. This model was designed to consider T\ensuremath{\geq}1 nuclei and was limited to the region of \textit{p}-wave dominance. This model}\\
\parbox[b][0.3cm]{17.7cm}{failed to reproduce the observed angular distribution of the \ensuremath{^{\textnormal{18}}}O(\ensuremath{\pi}\ensuremath{^{\textnormal{+}}},\ensuremath{\pi}\ensuremath{^{-}})\ensuremath{^{\textnormal{18}}}Ne\ensuremath{_{\textnormal{g.s.}}} reaction.}\\
\parbox[b][0.3cm]{17.7cm}{\addtolength{\parindent}{-0.2in}\href{https://www.nndc.bnl.gov/nsr/nsrlink.jsp?1981Li04,B}{1981Li04}: \ensuremath{^{\textnormal{18}}}O(\ensuremath{\pi}\ensuremath{^{\textnormal{+}}},\ensuremath{\pi}\ensuremath{^{-}}) E=162 MeV; calculated \ensuremath{\sigma}(\ensuremath{\theta}) using the closure approximation, following the procedure of (\href{https://www.nndc.bnl.gov/nsr/nsrlink.jsp?1977Le16,B}{1977Le16}), to}\\
\parbox[b][0.3cm]{17.7cm}{sum the two-step contributions from all the intermediate states of \ensuremath{^{\textnormal{18}}}F. This study was performed to try to describe the measured}\\
\parbox[b][0.3cm]{17.7cm}{results of (\href{https://www.nndc.bnl.gov/nsr/nsrlink.jsp?1979Se08,B}{1979Se08}: a minimum was observed around \ensuremath{\theta}\ensuremath{_{\textnormal{c.m.}}}=13\ensuremath{^\circ} in the angular distribution of \ensuremath{^{\textnormal{18}}}Ne\ensuremath{_{\textnormal{g.s.}}}). However, the calculated}\\
\parbox[b][0.3cm]{17.7cm}{first minimum of the differential cross section of the \ensuremath{^{\textnormal{18}}}O(\ensuremath{\pi}\ensuremath{^{\textnormal{+}}},\ensuremath{\pi}\ensuremath{^{-}})\ensuremath{^{\textnormal{18}}}Ne\ensuremath{_{\textnormal{g.s.}}} is at 30\ensuremath{^\circ}, which is much smaller than that produced by}\\
\parbox[b][0.3cm]{17.7cm}{the optical-model calculation of (\href{https://www.nndc.bnl.gov/nsr/nsrlink.jsp?1974Mi22,B}{1974Mi22}).}\\
\parbox[b][0.3cm]{17.7cm}{\addtolength{\parindent}{-0.2in}X. Liu, Z. Wu, Z. Huang, and Y. Li, Sci. Sin. 24 (1981) 789: \ensuremath{^{\textnormal{18}}}O(\ensuremath{\pi}\ensuremath{^{\textnormal{+}}},\ensuremath{\pi}\ensuremath{^{-}}); another study using Glauber calculation.}\\
\parbox[b][0.3cm]{17.7cm}{\addtolength{\parindent}{-0.2in}M. B. Johnson and E. R. Siciliano, Bull. Am. Phys. Soc. 25 (1980) 741: \ensuremath{^{\textnormal{18}}}O(\ensuremath{\pi}\ensuremath{^{\textnormal{+}}},\ensuremath{\pi}\ensuremath{^{-}}); demonstrated that the pion-nucleus elastic,}\\
\parbox[b][0.3cm]{17.7cm}{single, and double charge exchange scattering to isobaric analog states can be systematically described by including isoscalar,}\\
\parbox[b][0.3cm]{17.7cm}{isovector, and isotensor correlation terms in the optical potential. Calculations were performed using the Kisslinger optical potential,}\\
\parbox[b][0.3cm]{17.7cm}{which assumes a zero range for the pion-nucleon interaction. The inclusion of an isotensor component in the optical potential}\\
\parbox[b][0.3cm]{17.7cm}{significantly altered the shape of the resultant angular distribution for \ensuremath{^{\textnormal{18}}}O(\ensuremath{\pi}\ensuremath{^{\textnormal{+}}},\ensuremath{\pi}\ensuremath{^{-}})\ensuremath{^{\textnormal{18}}}Ne\ensuremath{_{\textnormal{g.s.}}}. The result agrees very well with the}\\
\parbox[b][0.3cm]{17.7cm}{shape of the measured angular distribution of (\href{https://www.nndc.bnl.gov/nsr/nsrlink.jsp?1979Se08,B}{1979Se08}); however, the minimum deduced by this calculation is \ensuremath{\sim}20\ensuremath{^\circ} higher. The}\\
\parbox[b][0.3cm]{17.7cm}{authors concluded that this measurement suggests the presence of an isotensor interaction, and the effects of including finite-range}\\
\parbox[b][0.3cm]{17.7cm}{pion-nucleon form factors must also be understood.}\\
\parbox[b][0.3cm]{17.7cm}{\addtolength{\parindent}{-0.2in}\href{https://www.nndc.bnl.gov/nsr/nsrlink.jsp?1981McZU,B}{1981McZU}: \ensuremath{^{\textnormal{18}}}O(\ensuremath{\pi}\ensuremath{^{\textnormal{+}}},\ensuremath{\pi}\ensuremath{^{-}}) E=164 MeV; calculated \ensuremath{\sigma}(\ensuremath{\theta}); investigated the finite-range effects suggested by (M. B. Johnson and E. R.}\\
\parbox[b][0.3cm]{17.7cm}{Siciliano, Bull. Am. Phys. Soc. 25 (1980) 741) by performing the lowest-order optical potential calculations in momentum space for}\\
\parbox[b][0.3cm]{17.7cm}{a number of different off-shell pion-nucleon models; deduced nuclear density pion-nucleon interaction dependence. The theoretical}\\
\parbox[b][0.3cm]{17.7cm}{angular distribution deduced from this study for \ensuremath{^{\textnormal{18}}}O(\ensuremath{\pi}\ensuremath{^{\textnormal{+}}},\ensuremath{\pi}\ensuremath{^{-}})\ensuremath{^{\textnormal{18}}}Ne\ensuremath{_{\textnormal{g.s.}}} at 164 MeV shows no qualitative difference between the}\\
\clearpage
\vspace{0.3cm}
{\bf \small \underline{\ensuremath{^{\textnormal{18}}}O(\ensuremath{\pi}\ensuremath{^{\textnormal{+}}},\ensuremath{\pi}\ensuremath{^{-}})\hspace{0.2in}\href{https://www.nndc.bnl.gov/nsr/nsrlink.jsp?1974LiZR,B}{1974LiZR},\href{https://www.nndc.bnl.gov/nsr/nsrlink.jsp?2007Ke07,B}{2007Ke07} (continued)}}\\
\vspace{0.3cm}
\parbox[b][0.3cm]{17.7cm}{zero-range calculation and the lowest-order finite-range calculation performed in this study. They both miss the observed minimum}\\
\parbox[b][0.3cm]{17.7cm}{by \ensuremath{\sim}20\ensuremath{^\circ}. The authors concluded that some important piece of physics, such as higher-order terms in the optical potential, must have}\\
\parbox[b][0.3cm]{17.7cm}{been ignored in all the theoretical work up to then.}\\
\parbox[b][0.3cm]{17.7cm}{\addtolength{\parindent}{-0.2in}\href{https://www.nndc.bnl.gov/nsr/nsrlink.jsp?1981Mi09,B}{1981Mi09}: \ensuremath{^{\textnormal{18}}}O(\ensuremath{\pi}\ensuremath{^{\textnormal{+}}},\ensuremath{\pi}\ensuremath{^{-}}) E=80-300 MeV; analyzed \ensuremath{\sigma}(\ensuremath{\theta},E) using a simple sequential single charge exchange model. They concluded}\\
\parbox[b][0.3cm]{17.7cm}{that the measured cross sections of double charge exchange reactions at relatively high energies (E\ensuremath{>}250 MeV) can be confidently}\\
\parbox[b][0.3cm]{17.7cm}{described by two subsequent single charge exchange. However, at energies in the vicinity of the pion-nucleon \ensuremath{\Delta}\ensuremath{_{\textnormal{33}}} resonance, this}\\
\parbox[b][0.3cm]{17.7cm}{simple model fails, which is a strong evidence that the double charge exchange reaction at these energies proceed via two nucleon}\\
\parbox[b][0.3cm]{17.7cm}{processes (such as true pion absorption).}\\
\parbox[b][0.3cm]{17.7cm}{\addtolength{\parindent}{-0.2in}\href{https://www.nndc.bnl.gov/nsr/nsrlink.jsp?1982LiZP,B}{1982LiZP}: \ensuremath{^{\textnormal{18}}}O(\ensuremath{\pi}\ensuremath{^{\textnormal{+}}},\ensuremath{\pi}\ensuremath{^{-}}) E=164, 292 MeV; analyzed \ensuremath{\sigma}(\ensuremath{\theta}) using sequential single charge exchange, which did not reproduce the data}\\
\parbox[b][0.3cm]{17.7cm}{observed for \ensuremath{^{\textnormal{18}}}Ne\ensuremath{_{\textnormal{g.s.}}} (\href{https://www.nndc.bnl.gov/nsr/nsrlink.jsp?1979Se08,B}{1979Se08}, \href{https://www.nndc.bnl.gov/nsr/nsrlink.jsp?1979Gr18,B}{1979Gr18}). These data showed a minimum at around \ensuremath{\theta}\ensuremath{_{\textnormal{c.m.}}}=13\ensuremath{^\circ}. But up to this point, none of}\\
\parbox[b][0.3cm]{17.7cm}{the above mentioned theories could produce a minimum at an angle lower than \ensuremath{\theta}\ensuremath{_{\textnormal{c.m.}}}=30\ensuremath{^\circ}. Using the same model that was used in}\\
\parbox[b][0.3cm]{17.7cm}{(\href{https://www.nndc.bnl.gov/nsr/nsrlink.jsp?1981Li02,B}{1981Li02}) involving coupled-channel formalism, the authors included two-nucleon processes such as true pion absorption. As a}\\
\parbox[b][0.3cm]{17.7cm}{result, the theoretical angular distributions of the \ensuremath{^{\textnormal{18}}}O(\ensuremath{\pi}\ensuremath{^{\textnormal{+}}},\ensuremath{\pi}\ensuremath{^{-}})\ensuremath{^{\textnormal{18}}}Ne\ensuremath{_{\textnormal{g.s.}}} reaction displayed minima at 23\ensuremath{^\circ} at 164 MeV and at 22\ensuremath{^\circ} at}\\
\parbox[b][0.3cm]{17.7cm}{292 MeV. The results were in much better agreement with the data of (\href{https://www.nndc.bnl.gov/nsr/nsrlink.jsp?1979Se08,B}{1979Se08}, \href{https://www.nndc.bnl.gov/nsr/nsrlink.jsp?1979Gr18,B}{1979Gr18}). Using an inert \ensuremath{^{\textnormal{16}}}O core could not}\\
\parbox[b][0.3cm]{17.7cm}{move the minimum at 164 MeV closer to the measured value of \ensuremath{\theta}\ensuremath{_{\textnormal{c.m.}}}=13\ensuremath{^\circ} (\href{https://www.nndc.bnl.gov/nsr/nsrlink.jsp?1979Se08,B}{1979Se08}). The authors then assumed an \ensuremath{^{\textnormal{18}}}O core}\\
\parbox[b][0.3cm]{17.7cm}{polarization and introduced into the valence neutron wave function of \ensuremath{^{\textnormal{18}}}O a collective state component arising from 4p-2h}\\
\parbox[b][0.3cm]{17.7cm}{excitation. Consequently, the minimum in the resultant angular distributions moved closer to the measured data, and the theoretical}\\
\parbox[b][0.3cm]{17.7cm}{\ensuremath{\sigma}(\ensuremath{\theta}) described the data well.}\\
\parbox[b][0.3cm]{17.7cm}{\addtolength{\parindent}{-0.2in}\href{https://www.nndc.bnl.gov/nsr/nsrlink.jsp?1983FoZX,B}{1983FoZX}: \ensuremath{^{\textnormal{18}}}O(\ensuremath{\pi}\ensuremath{^{\textnormal{+}}},\ensuremath{\pi}\ensuremath{^{-}}) E=164 MeV; analyzed \ensuremath{\sigma}(\ensuremath{\theta}) using the two-amplitude model.}\\
\parbox[b][0.3cm]{17.7cm}{\addtolength{\parindent}{-0.2in}\href{https://www.nndc.bnl.gov/nsr/nsrlink.jsp?1983Ho02,B}{1983Ho02}: \ensuremath{^{\textnormal{18}}}O(\ensuremath{\pi}\ensuremath{^{\textnormal{+}}},\ensuremath{\pi}\ensuremath{^{-}}) E=164 MeV; calculated \ensuremath{\sigma}(\ensuremath{\theta}) using the optical model and isobar dynamics. This model included a Lane}\\
\parbox[b][0.3cm]{17.7cm}{potential term, which incorporates the effects of \ensuremath{\Delta} charge exchange to account for processes ignored by the sequential charge}\\
\parbox[b][0.3cm]{17.7cm}{exchange models. The resultant theoretical cross sections did not satisfactorily reproduce the data of (\href{https://www.nndc.bnl.gov/nsr/nsrlink.jsp?1979Se08,B}{1979Se08}, \href{https://www.nndc.bnl.gov/nsr/nsrlink.jsp?1979Gr18,B}{1979Gr18}). The}\\
\parbox[b][0.3cm]{17.7cm}{authors concluded that the double charge exchange reaction mechanism must include effects such as short range correlations,}\\
\parbox[b][0.3cm]{17.7cm}{spin-flip terms, core polarization, inclusion of recoils and delta-particles in the nuclear wave function.}\\
\parbox[b][0.3cm]{17.7cm}{\addtolength{\parindent}{-0.2in}\href{https://www.nndc.bnl.gov/nsr/nsrlink.jsp?1983Jo06,B}{1983Jo06}: \ensuremath{^{\textnormal{18}}}O(\ensuremath{\pi}\ensuremath{^{\textnormal{+}}},\ensuremath{\pi}\ensuremath{^{-}}) E=164, 180 MeV; calculated \ensuremath{\sigma}(\ensuremath{\theta}) using second-order optical potential, Klein-Gordon equation, and Eikonal}\\
\parbox[b][0.3cm]{17.7cm}{theory in an isospin invariant framework. The authors emphasized the necessity of the distinction between nuclear structure and}\\
\parbox[b][0.3cm]{17.7cm}{reaction dynamics in the optical potential so that these effects manifest themselves differently in the deduced cross section.}\\
\parbox[b][0.3cm]{17.7cm}{Discussed the effects of the inclusion of isotensor term to properly account for the reaction dynamics. This model agrees with the}\\
\parbox[b][0.3cm]{17.7cm}{experimental data on the relative variation of the zero-degree cross section for the single and double charge exchange reactions}\\
\parbox[b][0.3cm]{17.7cm}{throughout the periodic table.}\\
\parbox[b][0.3cm]{17.7cm}{\addtolength{\parindent}{-0.2in}\href{https://www.nndc.bnl.gov/nsr/nsrlink.jsp?1983Li08,B}{1983Li08}: \ensuremath{^{\textnormal{18}}}O(\ensuremath{\pi}\ensuremath{^{\textnormal{+}}},\ensuremath{\pi}\ensuremath{^{-}}) E=50-300 MeV; calculated \ensuremath{\sigma}(\ensuremath{\theta}) vs. E using coupled-channel diffractive scattering theory with the inclusion}\\
\parbox[b][0.3cm]{17.7cm}{of the higher-order processes. The authors deduced reaction mechanism, and discussed the roles that nuclear structure and}\\
\parbox[b][0.3cm]{17.7cm}{two-nucleon processes play in the double charge exchange reaction mechanism. This model considers core-excitation and reflects}\\
\parbox[b][0.3cm]{17.7cm}{the effects of true pion absorption. It therefore provides a good prediction of the measured angular distributions at 164 MeV, and}\\
\parbox[b][0.3cm]{17.7cm}{the measured excitation function of \ensuremath{^{\textnormal{18}}}Ne\ensuremath{_{\textnormal{g.s.}}}.}\\
\parbox[b][0.3cm]{17.7cm}{\addtolength{\parindent}{-0.2in}\href{https://www.nndc.bnl.gov/nsr/nsrlink.jsp?1983Os09,B}{1983Os09}: \ensuremath{^{\textnormal{18}}}O(\ensuremath{\pi}\ensuremath{^{\textnormal{+}}},\ensuremath{\pi}\ensuremath{^{-}}) E=130-250 MeV; described reaction mechanism by taking into account the virtual \ensuremath{\Delta} mesons in the nuclear}\\
\parbox[b][0.3cm]{17.7cm}{medium and investigating the effects of meson exchange currents; deduced \ensuremath{\sigma}(E,\ensuremath{\theta}) for the DIAS at 130-250 MeV. It appears that at}\\
\parbox[b][0.3cm]{17.7cm}{E\ensuremath{<}130 MeV and E\ensuremath{>}250 MeV, the corrections to the cross section from meson exchange current can be larger than 50\%, and thus}\\
\parbox[b][0.3cm]{17.7cm}{these effects have to be included to enable accurate calculations of the double charge exchange amplitudes.}\\
\parbox[b][0.3cm]{17.7cm}{\addtolength{\parindent}{-0.2in}\href{https://www.nndc.bnl.gov/nsr/nsrlink.jsp?1984Gr27,B}{1984Gr27}: \ensuremath{^{\textnormal{18}}}O(\ensuremath{\pi}\ensuremath{^{\textnormal{+}}},\ensuremath{\pi}\ensuremath{^{-}}) E=165 MeV; calculated \ensuremath{\sigma}(\ensuremath{\theta},A) using phenomenological analysis and second-order optical model}\\
\parbox[b][0.3cm]{17.7cm}{calculations. According to this study, the anomalous position of the minima in the angular distribution of the \ensuremath{^{\textnormal{18}}}Ne\ensuremath{_{\textnormal{g.s.}}} is attributed}\\
\parbox[b][0.3cm]{17.7cm}{to the interference with higher-order terms (due to two-nucleon processes) in the pion-nucleus optical potential. In this approach,}\\
\parbox[b][0.3cm]{17.7cm}{the magnitudes of the isovector and isotensor terms were adjusted to fit the available data on the single charge exchange reactions}\\
\parbox[b][0.3cm]{17.7cm}{at \ensuremath{\theta}\ensuremath{_{\textnormal{c.m.}}}=0\ensuremath{^\circ} and on the double charge exchange reaction populating \ensuremath{^{\textnormal{18}}}Ne\ensuremath{_{\textnormal{g.s.}}} at \ensuremath{\theta}\ensuremath{_{\textnormal{c.m.}}}=5\ensuremath{^\circ}. The resulting optical model parameters}\\
\parbox[b][0.3cm]{17.7cm}{provide good predictions of the measured angular-distribution shapes.}\\
\parbox[b][0.3cm]{17.7cm}{\addtolength{\parindent}{-0.2in}\href{https://www.nndc.bnl.gov/nsr/nsrlink.jsp?1984Jo01,B}{1984Jo01}: \ensuremath{^{\textnormal{18}}}O(\ensuremath{\pi}\ensuremath{^{\textnormal{+}}},\ensuremath{\pi}\ensuremath{^{-}}) E=100-300 MeV; calculated \ensuremath{\sigma}(\ensuremath{\theta}=0\ensuremath{^\circ}) vs. E; constructed double charge exchange cross sections from the}\\
\parbox[b][0.3cm]{17.7cm}{solution of the Klein-Gordon equation for the pion; investigated the effects of direct meson-isobar interactions and found out}\\
\parbox[b][0.3cm]{17.7cm}{conclusively that the inclusion of the \ensuremath{\Delta}-particles in the wave functions has a negligible effect on the energy region near the \ensuremath{\Delta}\ensuremath{_{\textnormal{33}}}}\\
\parbox[b][0.3cm]{17.7cm}{resonance in the double charge exchange reactions populating the DIAS. Inclusion of these terms improves agreement with the}\\
\parbox[b][0.3cm]{17.7cm}{experimental data below the \ensuremath{\Delta}\ensuremath{_{\textnormal{33}}} resonance. Data at the higher energies cannot be described by a simple combination of sequential}\\
\parbox[b][0.3cm]{17.7cm}{pion-nucleon scattering plus a direct meson-nucleus interaction.}\\
\parbox[b][0.3cm]{17.7cm}{\addtolength{\parindent}{-0.2in}\href{https://www.nndc.bnl.gov/nsr/nsrlink.jsp?1984Ka26,B}{1984Ka26}: \ensuremath{^{\textnormal{18}}}O(\ensuremath{\pi}\ensuremath{^{\textnormal{+}}},\ensuremath{\pi}\ensuremath{^{-}}) E=164 MeV; calculated \ensuremath{\sigma}(\ensuremath{\theta}) using \ensuremath{\Delta}-hole and non-local \ensuremath{\pi}-nucleon T-matrix formalisms and closure}\\
\parbox[b][0.3cm]{17.7cm}{approximation; discussed \ensuremath{\Delta}-nucleon interaction and the role of \ensuremath{\Delta} recoil and the non-analog intermediate states in \ensuremath{^{\textnormal{18}}}F. The}\\
\clearpage
\vspace{0.3cm}
{\bf \small \underline{\ensuremath{^{\textnormal{18}}}O(\ensuremath{\pi}\ensuremath{^{\textnormal{+}}},\ensuremath{\pi}\ensuremath{^{-}})\hspace{0.2in}\href{https://www.nndc.bnl.gov/nsr/nsrlink.jsp?1974LiZR,B}{1974LiZR},\href{https://www.nndc.bnl.gov/nsr/nsrlink.jsp?2007Ke07,B}{2007Ke07} (continued)}}\\
\vspace{0.3cm}
\parbox[b][0.3cm]{17.7cm}{deduced angular distribution has more or less a similar shape to the measured data of (\href{https://www.nndc.bnl.gov/nsr/nsrlink.jsp?1979Gr18,B}{1979Gr18}, \href{https://www.nndc.bnl.gov/nsr/nsrlink.jsp?1979Se08,B}{1979Se08}) but the location of the}\\
\parbox[b][0.3cm]{17.7cm}{minimum is not exactly reproduced.}\\
\parbox[b][0.3cm]{17.7cm}{\addtolength{\parindent}{-0.2in}\href{https://www.nndc.bnl.gov/nsr/nsrlink.jsp?1985Gi01,B}{1985Gi01}: \ensuremath{^{\textnormal{18}}}O(\ensuremath{\pi}\ensuremath{^{\textnormal{+}}},\ensuremath{\pi}\ensuremath{^{-}}) E=164 MeV; calculated \ensuremath{\sigma}(\ensuremath{\theta}) using the two-amplitude model to describe the anomalous excitation function}\\
\parbox[b][0.3cm]{17.7cm}{measured by (\href{https://www.nndc.bnl.gov/nsr/nsrlink.jsp?1982Gr28,B}{1982Gr28}) near 160 MeV and the angular distribution for \ensuremath{^{\textnormal{18}}}Ne\ensuremath{_{\textnormal{g.s.}}}. The authors suggested that the unusual behavior}\\
\parbox[b][0.3cm]{17.7cm}{of these measured distributions arises from non-double-analog processes. To describe the measured angular distribution of \ensuremath{^{\textnormal{18}}}Ne\ensuremath{_{\textnormal{g.s.}}}:}\\
\parbox[b][0.3cm]{17.7cm}{the authors parameterized the non-analog amplitude using Bessel functions, and Legendre polynomial expansion. They then added}\\
\parbox[b][0.3cm]{17.7cm}{the analog amplitude produced by Eikonal approximation of (\href{https://www.nndc.bnl.gov/nsr/nsrlink.jsp?1980Jo06,B}{1980Jo06}) to each of the parameterized non-analog amplitudes. These}\\
\parbox[b][0.3cm]{17.7cm}{were used to fit the measured (by \href{https://www.nndc.bnl.gov/nsr/nsrlink.jsp?1982Gr28,B}{1982Gr28}) angular distribution. The fit consisting of the non-analog parameterization using}\\
\parbox[b][0.3cm]{17.7cm}{Legendre polynomials describes the data very well (see Fig. 5 of (\href{https://www.nndc.bnl.gov/nsr/nsrlink.jsp?1985Gi01,B}{1985Gi01})). To describe the measured excitation function at}\\
\parbox[b][0.3cm]{17.7cm}{\ensuremath{\theta}\ensuremath{_{\textnormal{lab}}}=5\ensuremath{^\circ} for \ensuremath{^{\textnormal{18}}}Ne\ensuremath{_{\textnormal{g.s.}}}, the authors achieved an excellent fit of the data using the sum of double charge exchange analog amplitude}\\
\parbox[b][0.3cm]{17.7cm}{(produced by the simple sequential single charge exchange (\href{https://www.nndc.bnl.gov/nsr/nsrlink.jsp?1981Mi09,B}{1981Mi09})) and the analog amplitude produced by the Eikonal}\\
\parbox[b][0.3cm]{17.7cm}{approximation of (\href{https://www.nndc.bnl.gov/nsr/nsrlink.jsp?1980Jo06,B}{1980Jo06}) together with either of the non-analog parameterized amplitudes. The fits with both parameterizations}\\
\parbox[b][0.3cm]{17.7cm}{describe the data very well.}\\
\parbox[b][0.3cm]{17.7cm}{\addtolength{\parindent}{-0.2in}\href{https://www.nndc.bnl.gov/nsr/nsrlink.jsp?1985Gm01,B}{1985Gm01}: \ensuremath{^{\textnormal{18}}}O(\ensuremath{\pi}\ensuremath{^{\textnormal{+}}},\ensuremath{\pi}\ensuremath{^{-}}) E\ensuremath{\approx}100-340 MeV; calculated \ensuremath{\sigma}(\ensuremath{\theta}) vs. E using coupled-channel and distorted-wave impulse approximation}\\
\parbox[b][0.3cm]{17.7cm}{(DWIA). Both these calculations are performed under the same kinematical and dynamical assumptions in constructing the}\\
\parbox[b][0.3cm]{17.7cm}{corresponding transition matrix elements. This model is used to explain the d\ensuremath{\sigma}/d\ensuremath{\Omega}(E) for \ensuremath{^{\textnormal{18}}}O(\ensuremath{\pi}\ensuremath{^{\textnormal{+}}},\ensuremath{\pi}\ensuremath{^{-}})\ensuremath{^{\textnormal{18}}}Ne\ensuremath{_{\textnormal{g.s.}}}. However, the}\\
\parbox[b][0.3cm]{17.7cm}{model cannot describe the experimentally observed (at \ensuremath{\theta}\ensuremath{_{\textnormal{lab}}}=5\ensuremath{^\circ}) minimum in the cross section. The minimum that this model}\\
\parbox[b][0.3cm]{17.7cm}{produces occurs at much larger angles. The authors concluded that core excitation is important to be able to understand the}\\
\parbox[b][0.3cm]{17.7cm}{double-charge exchange reaction mechanism populating the double analog states.}\\
\parbox[b][0.3cm]{17.7cm}{\addtolength{\parindent}{-0.2in}\href{https://www.nndc.bnl.gov/nsr/nsrlink.jsp?1986Fo06,B}{1986Fo06}: \ensuremath{^{\textnormal{18}}}O(\ensuremath{\pi}\ensuremath{^{\textnormal{+}}},\ensuremath{\pi}\ensuremath{^{-}}) E=292 MeV, \ensuremath{\theta}\ensuremath{_{\textnormal{lab}}}=5\ensuremath{^\circ}; analyzed \ensuremath{\sigma}(E) for \ensuremath{^{\textnormal{18}}}Ne\ensuremath{_{\textnormal{g.s.}}}; deduced reaction mechanism using two-amplitude}\\
\parbox[b][0.3cm]{17.7cm}{model. The authors assume that the double charge exchange amplitude for a T=1 nucleus is the sum of a double isobaric analog}\\
\parbox[b][0.3cm]{17.7cm}{transition amplitude (DIAT) and a non-DIAT amplitude. The calculations are insensitive to the relative phase. They conclude that}\\
\parbox[b][0.3cm]{17.7cm}{at 292 MeV, the cross section of the \ensuremath{^{\textnormal{18}}}O(\ensuremath{\pi}\ensuremath{^{\textnormal{+}}},\ensuremath{\pi}\ensuremath{^{-}})\ensuremath{^{\textnormal{18}}}Ne\ensuremath{_{\textnormal{g.s.}}} reaction contains contributions from two distinct amplitudes, one varying}\\
\parbox[b][0.3cm]{17.7cm}{as A\ensuremath{^{\textnormal{$-$5/3}}}, the other as A\ensuremath{^{\textnormal{$-$2/3}}}. The relative phase between the two amplitude is \ensuremath{\sim}95\ensuremath{^\circ} at 292 MeV and seems to be independent of A.}\\
\parbox[b][0.3cm]{17.7cm}{For the double charge exchange on \ensuremath{^{\textnormal{18}}}O, the ratio of analog to nonanalog amplitudes is 1.087 at 292 MeV. This calculation predicts}\\
\parbox[b][0.3cm]{17.7cm}{the cross section for the \ensuremath{^{\textnormal{18}}}O(\ensuremath{\pi}\ensuremath{^{\textnormal{+}}},\ensuremath{\pi}\ensuremath{^{-}})\ensuremath{^{\textnormal{18}}}Ne\ensuremath{_{\textnormal{g.s.}}} reaction at 292 MeV and \ensuremath{\theta}\ensuremath{_{\textnormal{lab}}}=5\ensuremath{^\circ} to be 2.47 \ensuremath{\mu}b/sr, which is consistent with the}\\
\parbox[b][0.3cm]{17.7cm}{experimentally deduced value of 2.40 \ensuremath{\mu}b/sr \textit{19} (\href{https://www.nndc.bnl.gov/nsr/nsrlink.jsp?1982Gr28,B}{1982Gr28}).}\\
\parbox[b][0.3cm]{17.7cm}{\addtolength{\parindent}{-0.2in}\href{https://www.nndc.bnl.gov/nsr/nsrlink.jsp?1986Ge06,B}{1986Ge06}: \ensuremath{^{\textnormal{18}}}O(\ensuremath{\pi}\ensuremath{^{\textnormal{+}}},\ensuremath{\pi}\ensuremath{^{-}}) E=50 MeV; calculated \ensuremath{\sigma}(\ensuremath{\theta}); deduced \ensuremath{\sigma}(E) for \ensuremath{^{\textnormal{18}}}Ne\ensuremath{_{\textnormal{g.s.}}} using coupled-channels approach with the}\\
\parbox[b][0.3cm]{17.7cm}{inclusion of the long range shape correlations and quadrupole coupling to the first 2\ensuremath{^{\textnormal{+}}} state in \ensuremath{^{\textnormal{18}}}F. The authors used a Kisslinger}\\
\parbox[b][0.3cm]{17.7cm}{optical potential model with further corrections for \textit{s}-wave pion absorption and for Fermi averaging of the \ensuremath{\pi}-nucleus amplitudes.}\\
\parbox[b][0.3cm]{17.7cm}{The integrated cross sections for mass-18 were deduced. The authors conclude that by adjusting the pion-nucleus exchange currents}\\
\parbox[b][0.3cm]{17.7cm}{when the quadrupole couplings are included, the agreement between the calculated and observed double charge exchange dataset}\\
\parbox[b][0.3cm]{17.7cm}{improves.}\\
\parbox[b][0.3cm]{17.7cm}{\addtolength{\parindent}{-0.2in}\href{https://www.nndc.bnl.gov/nsr/nsrlink.jsp?1986Jo04,B}{1986Jo04}: \ensuremath{^{\textnormal{18}}}O(\ensuremath{\pi}\ensuremath{^{\textnormal{+}}},\ensuremath{\pi}\ensuremath{^{-}}) E\ensuremath{\approx}20-320 MeV; calculated double isobaric analog state excitation$'$s \ensuremath{\sigma}(\ensuremath{\theta}) vs. E using the six-quark cluster}\\
\parbox[b][0.3cm]{17.7cm}{model to calculate their contributions on pion double charge exchange. It is assumed that the two excess neutrons in \ensuremath{^{\textnormal{18}}}Ne\ensuremath{_{\textnormal{g.s.}}} on}\\
\parbox[b][0.3cm]{17.7cm}{which the double charge exchange takes place reside in a (1\textit{d}\ensuremath{_{\textnormal{5/2}}})\ensuremath{^{\textnormal{2}}} configuration. The resulting cross sections are roughly a factor}\\
\parbox[b][0.3cm]{17.7cm}{of 2 smaller than the experimentally deduced ones. The authors conclude that the direct \ensuremath{\pi} coupling to the interior region quarks}\\
\parbox[b][0.3cm]{17.7cm}{and long term mesonic corrections should be included in the quark cluster models in order to improve the agreement between the}\\
\parbox[b][0.3cm]{17.7cm}{data and theoretical calculations.}\\
\parbox[b][0.3cm]{17.7cm}{\addtolength{\parindent}{-0.2in}\href{https://www.nndc.bnl.gov/nsr/nsrlink.jsp?1986Os06,B}{1986Os06}: \ensuremath{^{\textnormal{18}}}O(\ensuremath{\pi}\ensuremath{^{\textnormal{+}}},\ensuremath{\pi}\ensuremath{^{-}}); discussed pion pole and pion contact terms for the double charge-exchange reactions in the context of}\\
\parbox[b][0.3cm]{17.7cm}{distorted wave born approximation.}\\
\parbox[b][0.3cm]{17.7cm}{\addtolength{\parindent}{-0.2in}\href{https://www.nndc.bnl.gov/nsr/nsrlink.jsp?1987Ka39,B}{1987Ka39}: \ensuremath{^{\textnormal{18}}}O(\ensuremath{\pi}\ensuremath{^{\textnormal{+}}},\ensuremath{\pi}\ensuremath{^{-}}) E=50, 164 MeV; calculated \ensuremath{\sigma}(\ensuremath{\theta}) using a semi-microscopic model for the double charge exchange}\\
\parbox[b][0.3cm]{17.7cm}{reactions. The double charge-exchange amplitude is separated into sequential and non-sequential amplitudes that interfere with one}\\
\parbox[b][0.3cm]{17.7cm}{another. The pion distortion is based on delta-hole optical potential. This model is able to describe the experimentally deduced}\\
\parbox[b][0.3cm]{17.7cm}{(\href{https://www.nndc.bnl.gov/nsr/nsrlink.jsp?1985Al15,B}{1985Al15}) \ensuremath{^{\textnormal{18}}}O(\ensuremath{\pi}\ensuremath{^{\textnormal{+}}},\ensuremath{\pi}\ensuremath{^{-}})\ensuremath{^{\textnormal{18}}}Ne\ensuremath{_{\textnormal{g.s.}}} cross section at 50 MeV but fails to reproduce the energy and angular dependence of this}\\
\parbox[b][0.3cm]{17.7cm}{cross-section in the medium energy region.}\\
\parbox[b][0.3cm]{17.7cm}{\addtolength{\parindent}{-0.2in}\href{https://www.nndc.bnl.gov/nsr/nsrlink.jsp?1987Mi02,B}{1987Mi02}: \ensuremath{^{\textnormal{18}}}O(\ensuremath{\pi}\ensuremath{^{\textnormal{+}}},\ensuremath{\pi}\ensuremath{^{-}}); computed the effects of a mechanism in which the double charge exchange reaction proceeds via pion}\\
\parbox[b][0.3cm]{17.7cm}{absorption and emission on a six-quark bag.}\\
\parbox[b][0.3cm]{17.7cm}{\addtolength{\parindent}{-0.2in}\href{https://www.nndc.bnl.gov/nsr/nsrlink.jsp?1987Ha29,B}{1987Ha29}: \ensuremath{^{\textnormal{18}}}O(\ensuremath{\pi}\ensuremath{^{\textnormal{+}}},\ensuremath{\pi}\ensuremath{^{-}}); predicted a narrow (\ensuremath{\Gamma}\ensuremath{\sim}10 MeV) resonance structure, related to the strongly bound \ensuremath{\eta}-nucleus system, in}\\
\parbox[b][0.3cm]{17.7cm}{the excitation function of the (\ensuremath{\pi}\ensuremath{^{\textnormal{+}}},\ensuremath{\pi}\ensuremath{^{-}}) double charge exchange reactions at a pion kinetic energy of \ensuremath{\sim}419 MeV.}\\
\parbox[b][0.3cm]{17.7cm}{\addtolength{\parindent}{-0.2in}\href{https://www.nndc.bnl.gov/nsr/nsrlink.jsp?1988Os02,B}{1988Os02}: \ensuremath{^{\textnormal{18}}}O(\ensuremath{\pi}\ensuremath{^{\textnormal{+}}},\ensuremath{\pi}\ensuremath{^{-}}) E=130-230 MeV; calculated \ensuremath{\sigma}(\ensuremath{\theta}). The authors investigated the contribution of the \ensuremath{\Delta} interaction to the}\\
\parbox[b][0.3cm]{17.7cm}{pion-induced double-charge exchange reaction to isobaric analogue states at energies around the \ensuremath{\Delta}\ensuremath{_{\textnormal{33}}} resonance region and}\\
\parbox[b][0.3cm]{17.7cm}{compared it to the conventional mechanisms involving two sequential single charge exchange steps.}\\
\clearpage
\vspace{0.3cm}
{\bf \small \underline{\ensuremath{^{\textnormal{18}}}O(\ensuremath{\pi}\ensuremath{^{\textnormal{+}}},\ensuremath{\pi}\ensuremath{^{-}})\hspace{0.2in}\href{https://www.nndc.bnl.gov/nsr/nsrlink.jsp?1974LiZR,B}{1974LiZR},\href{https://www.nndc.bnl.gov/nsr/nsrlink.jsp?2007Ke07,B}{2007Ke07} (continued)}}\\
\vspace{0.3cm}
\parbox[b][0.3cm]{17.7cm}{\addtolength{\parindent}{-0.2in}\href{https://www.nndc.bnl.gov/nsr/nsrlink.jsp?1988Yu04,B}{1988Yu04}, Yu, Cai and Ma, Sao Paulo (1989) 9: \ensuremath{^{\textnormal{18}}}O(\ensuremath{\pi}\ensuremath{^{\textnormal{+}}},\ensuremath{\pi}\ensuremath{^{-}}) E=164, 292 MeV; calculated \ensuremath{\sigma}(\ensuremath{\theta}) by including the effect of the}\\
\parbox[b][0.3cm]{17.7cm}{second kind of meson exchange currents. The authors found out that meson exchange currents decrease the \ensuremath{^{\textnormal{18}}}O(\ensuremath{\pi}\ensuremath{^{\textnormal{+}}},\ensuremath{\pi}\ensuremath{^{-}})\ensuremath{^{\textnormal{18}}}Ne\ensuremath{_{\textnormal{g.s.}}}}\\
\parbox[b][0.3cm]{17.7cm}{cross sections at low energies and increase it at the energies higher than 180 MeV. The theoretical results has a minimum at 30\ensuremath{^\circ},}\\
\parbox[b][0.3cm]{17.7cm}{which is far away from the experimentally observed minimum at 13\ensuremath{^\circ} (\href{https://www.nndc.bnl.gov/nsr/nsrlink.jsp?1979Se08,B}{1979Se08}). The authors concluded that even though these}\\
\parbox[b][0.3cm]{17.7cm}{effects are important, the second kind of meson exchange currents do not provide a better understanding of the location of the}\\
\parbox[b][0.3cm]{17.7cm}{minimum in the experimental differential cross section at 164 MeV.}\\
\parbox[b][0.3cm]{17.7cm}{\addtolength{\parindent}{-0.2in}Ching \textit{et al}., Commun. Theor. Phys. 11 (1989) 171: \ensuremath{^{\textnormal{18}}}O(\ensuremath{\pi}\ensuremath{^{\textnormal{+}}},\ensuremath{\pi}\ensuremath{^{-}}) E=50 MeV; calculated \ensuremath{\sigma}(\ensuremath{\theta}). The authors used two-nucleon pion}\\
\parbox[b][0.3cm]{17.7cm}{absorption-emission mechanism to describe the pion double charge exchange reaction at low energies.}\\
\parbox[b][0.3cm]{17.7cm}{\addtolength{\parindent}{-0.2in}\href{https://www.nndc.bnl.gov/nsr/nsrlink.jsp?1989Ch21,B}{1989Ch21}: \ensuremath{^{\textnormal{18}}}O(\ensuremath{\pi}\ensuremath{^{\textnormal{+}}},\ensuremath{\pi}\ensuremath{^{-}}) E=24-79 MeV; calculated \ensuremath{\sigma}(\ensuremath{\theta}) using quark-antiquark annihilation mechanism to calculate the short-range}\\
\parbox[b][0.3cm]{17.7cm}{contribution to the pion double charge exchange reaction. The authors found that this process contributes significantly to the double}\\
\parbox[b][0.3cm]{17.7cm}{charge exchange reaction and can account for most of the cross section of the \ensuremath{^{\textnormal{18}}}Ne\ensuremath{_{\textnormal{g.s.}}} double isobaric analog state in the forward}\\
\parbox[b][0.3cm]{17.7cm}{angles at low pion incident energies. However, their model did not well describe the experimental data (provided by R. R. Johnson}\\
\parbox[b][0.3cm]{17.7cm}{via private communication).}\\
\parbox[b][0.3cm]{17.7cm}{\addtolength{\parindent}{-0.2in}\href{https://www.nndc.bnl.gov/nsr/nsrlink.jsp?1989Fo02,B}{1989Fo02}: \ensuremath{^{\textnormal{18}}}O(\ensuremath{\pi}\ensuremath{^{\textnormal{+}}},\ensuremath{\pi}\ensuremath{^{-}}) E\ensuremath{\approx}100-300 MeV; calculated \ensuremath{\sigma}(\ensuremath{\theta}) using a two-amplitude model. The author discussed the relative}\\
\parbox[b][0.3cm]{17.7cm}{magnitudes, the relative phase between them, and their interferences at different pion energies and scattering angles. The model}\\
\parbox[b][0.3cm]{17.7cm}{used in this study reproduces the experimental data of (\href{https://www.nndc.bnl.gov/nsr/nsrlink.jsp?1982Gr28,B}{1982Gr28}, \href{https://www.nndc.bnl.gov/nsr/nsrlink.jsp?1985Se08,B}{1985Se08}) very well.}\\
\parbox[b][0.3cm]{17.7cm}{\addtolength{\parindent}{-0.2in}\href{https://www.nndc.bnl.gov/nsr/nsrlink.jsp?1989Wi20,B}{1989Wi20}: \ensuremath{^{\textnormal{18}}}O(\ensuremath{\pi}\ensuremath{^{\textnormal{+}}},\ensuremath{\pi}\ensuremath{^{-}}) E=164 MeV; calculated \ensuremath{\sigma}(\ensuremath{\theta}). They investigated in detail the \ensuremath{\Delta}-nucleus interactions contributing to the}\\
\parbox[b][0.3cm]{17.7cm}{double charge exchange reaction populating the double isobaric analog state. They studied the model dependence and nuclear}\\
\parbox[b][0.3cm]{17.7cm}{structure sensitivity of various processes which are responsible for the \ensuremath{\Delta}-nucleus interaction.}\\
\parbox[b][0.3cm]{17.7cm}{\addtolength{\parindent}{-0.2in}\href{https://www.nndc.bnl.gov/nsr/nsrlink.jsp?1990Ch14,B}{1990Ch14}, Ching, Ho and Zou, Panic XII (1990) Paper III-77: \ensuremath{^{\textnormal{18}}}O(\ensuremath{\pi}\ensuremath{^{\textnormal{+}}},\ensuremath{\pi}\ensuremath{^{-}}) E\ensuremath{\leq}300 MeV; calculated d\ensuremath{\sigma}/d\ensuremath{\Omega}(E) at \ensuremath{\theta}\ensuremath{_{\textnormal{lab}}}=0\ensuremath{^\circ} by}\\
\parbox[b][0.3cm]{17.7cm}{combining the contribution of two-nucleon pion absorption-emission mechanism (in the framework of distorted wave impulse}\\
\parbox[b][0.3cm]{17.7cm}{approximation) with the conventional two sequential single charge exchange mechanism to describe the pion induced double charge}\\
\parbox[b][0.3cm]{17.7cm}{exchange reaction. As a result, the agreement between the data and theoretical results is improved at low energy. Deduced new pion}\\
\parbox[b][0.3cm]{17.7cm}{absorption-emission mechanism role.}\\
\parbox[b][0.3cm]{17.7cm}{\addtolength{\parindent}{-0.2in}\href{https://www.nndc.bnl.gov/nsr/nsrlink.jsp?1990Ch27,B}{1990Ch27}: \ensuremath{^{\textnormal{18}}}O(\ensuremath{\pi}\ensuremath{^{\textnormal{+}}},\ensuremath{\pi}\ensuremath{^{-}}) E\ensuremath{\approx}20-300 MeV; calculated d\ensuremath{\sigma}/d\ensuremath{\Omega}(E) at \ensuremath{\theta}\ensuremath{_{\textnormal{lab}}}=0\ensuremath{^\circ} by using operator expansion approach for the pion}\\
\parbox[b][0.3cm]{17.7cm}{absorption-emission mechanism. This method is used to avoid truncating higher order terms. Therefore, this work was the}\\
\parbox[b][0.3cm]{17.7cm}{continuation of that of (\href{https://www.nndc.bnl.gov/nsr/nsrlink.jsp?1990Ch14,B}{1990Ch14}) to investigate the influence of all highly excited states in the intermediate \ensuremath{^{\textnormal{18}}}F nucleus. The}\\
\parbox[b][0.3cm]{17.7cm}{authors concluded that these excited states do not play an important role in the double charge exchange reaction. They stated that}\\
\parbox[b][0.3cm]{17.7cm}{the two-nucleon pion absorption-emission mechanism together with the two sequential single charge exchange can account for the}\\
\parbox[b][0.3cm]{17.7cm}{anomalous, large, forward cross section of \ensuremath{^{\textnormal{18}}}O(\ensuremath{\pi}\ensuremath{^{\textnormal{+}}},\ensuremath{\pi}\ensuremath{^{-}})\ensuremath{^{\textnormal{18}}}Ne\ensuremath{_{\textnormal{g.s.}}} at the low energies. Their theoretical cross section distortion}\\
\parbox[b][0.3cm]{17.7cm}{reproduces the shape of the observed data at low energy but the magnitude of their cross section does not describe the}\\
\parbox[b][0.3cm]{17.7cm}{measurements.}\\
\parbox[b][0.3cm]{17.7cm}{\addtolength{\parindent}{-0.2in}\href{https://www.nndc.bnl.gov/nsr/nsrlink.jsp?1990MaZW,B}{1990MaZW}: \ensuremath{^{\textnormal{18}}}O(\ensuremath{\pi}\ensuremath{^{\textnormal{+}}},\ensuremath{\pi}\ensuremath{^{-}}) E in the region of \ensuremath{\Delta} resonance; deduced the influence of quark effects on double charge exchange}\\
\parbox[b][0.3cm]{17.7cm}{mechanism. Analyzed the measurements using a hybrid quark hadron model.}\\
\parbox[b][0.3cm]{17.7cm}{\addtolength{\parindent}{-0.2in}\href{https://www.nndc.bnl.gov/nsr/nsrlink.jsp?1992Ma46,B}{1992Ma46}: \ensuremath{^{\textnormal{18}}}O(\ensuremath{\pi}\ensuremath{^{\textnormal{+}}},\ensuremath{\pi}\ensuremath{^{-}}); E=164 MeV; calculated \ensuremath{\sigma}(\ensuremath{\theta}); deduced dibaryon effects and its role in double charge exchange mechanism.}\\
\parbox[b][0.3cm]{17.7cm}{\addtolength{\parindent}{-0.2in}\href{https://www.nndc.bnl.gov/nsr/nsrlink.jsp?1992Os05,B}{1992Os05}: \ensuremath{^{\textnormal{18}}}O(\ensuremath{\pi}\ensuremath{^{\textnormal{+}}},\ensuremath{\pi}\ensuremath{^{-}}) E=164 MeV; calculated \ensuremath{\sigma}(\ensuremath{\theta}) by evaluating the pion absorption contribution near E\ensuremath{_{\ensuremath{\pi}}}=50 MeV. Near the}\\
\parbox[b][0.3cm]{17.7cm}{\ensuremath{\Delta}\ensuremath{_{\textnormal{33}}} resonance region, the corrections due to pion absorption are small. The most noticeable effect from this contribution is a small}\\
\parbox[b][0.3cm]{17.7cm}{shift of the minimum in the cross section at smaller angles. This shift is however not big enough to reproduce the experimental data}\\
\parbox[b][0.3cm]{17.7cm}{(\href{https://www.nndc.bnl.gov/nsr/nsrlink.jsp?1979Gr18,B}{1979Gr18}, \href{https://www.nndc.bnl.gov/nsr/nsrlink.jsp?1979Se08,B}{1979Se08}).}\\
\parbox[b][0.3cm]{17.7cm}{\addtolength{\parindent}{-0.2in}\href{https://www.nndc.bnl.gov/nsr/nsrlink.jsp?1993Bi10,B}{1993Bi10}: \ensuremath{^{\textnormal{18}}}O(\ensuremath{\pi}\ensuremath{^{\textnormal{+}}},\ensuremath{\pi}\ensuremath{^{-}}) E\ensuremath{\approx}20-300 MeV; analyzed \ensuremath{\sigma}(\ensuremath{\theta}) vs. E. The authors discussed that the anomalous energy dependence of the}\\
\parbox[b][0.3cm]{17.7cm}{cross section of \ensuremath{^{\textnormal{18}}}O(\ensuremath{\pi}\ensuremath{^{\textnormal{+}}},\ensuremath{\pi}\ensuremath{^{-}})\ensuremath{^{\textnormal{18}}}Ne\ensuremath{_{\textnormal{g.s.}}} at E\ensuremath{_{\ensuremath{\pi}}}=50 MeV may be explained by a narrow resonance in the \ensuremath{\pi}NN subsystem with J\ensuremath{^{\ensuremath{\pi}}}=0\ensuremath{^{-}}}\\
\parbox[b][0.3cm]{17.7cm}{and a mass of 2.065 GeV.}\\
\parbox[b][0.3cm]{17.7cm}{\addtolength{\parindent}{-0.2in}\href{https://www.nndc.bnl.gov/nsr/nsrlink.jsp?1993Gi03,B}{1993Gi03}: \ensuremath{^{\textnormal{18}}}O(\ensuremath{\pi}\ensuremath{^{\textnormal{+}}},\ensuremath{\pi}\ensuremath{^{-}}); E=100-300 MeV; calculated \ensuremath{\sigma}(\ensuremath{\theta}); the contribution of sequential charge exchange and delta-nucleon charge}\\
\parbox[b][0.3cm]{17.7cm}{exchange is examined. It is found that the nonanalog double charge exchange cannot be quantitatively understood in terms of these}\\
\parbox[b][0.3cm]{17.7cm}{two reaction mechanisms. The authors recommended that the contributions of the double spin-flip to the sequential single charge}\\
\parbox[b][0.3cm]{17.7cm}{exchange be considered.}\\
\parbox[b][0.3cm]{17.7cm}{\addtolength{\parindent}{-0.2in}\href{https://www.nndc.bnl.gov/nsr/nsrlink.jsp?1993Os01,B}{1993Os01}: \ensuremath{^{\textnormal{18}}}O(\ensuremath{\pi}\ensuremath{^{\textnormal{+}}},\ensuremath{\pi}\ensuremath{^{-}}) E=200-1400 MeV; calculated differential cross sections at \ensuremath{\theta}\ensuremath{_{\textnormal{lab}}}=0\ensuremath{^\circ} and excitation function at \ensuremath{\theta}\ensuremath{_{\textnormal{lab}}}=5\ensuremath{^\circ} for}\\
\parbox[b][0.3cm]{17.7cm}{double charge exchange on \ensuremath{^{\textnormal{14}}}C and \ensuremath{^{\textnormal{18}}}O using a zero-parameter Glauber theory that includes spin-flip and pion absorption. The}\\
\parbox[b][0.3cm]{17.7cm}{wave functions are calculated using the Glasgow shell model code. The authors compared their theoretical results with the}\\
\parbox[b][0.3cm]{17.7cm}{experimental data of (\href{https://www.nndc.bnl.gov/nsr/nsrlink.jsp?1989Wi02,B}{1989Wi02}) at E\ensuremath{_{\ensuremath{\pi}}}=300-525 MeV. The authors found that theory reproduces the shape of the excitation}\\
\parbox[b][0.3cm]{17.7cm}{function but is about a factor of 3 too large compared with experiment. The effects of medium polarization on the isovector pion}\\
\parbox[b][0.3cm]{17.7cm}{operator were calculated. As a result of isovector normalization, the theoretical center-of-mass differential cross sections at \ensuremath{\theta}\ensuremath{_{\textnormal{lab}}}=5\ensuremath{^\circ}}\\
\parbox[b][0.3cm]{17.7cm}{reproduces the experimental results of (\href{https://www.nndc.bnl.gov/nsr/nsrlink.jsp?1989Wi02,B}{1989Wi02}) very well.}\\
\clearpage
\vspace{0.3cm}
{\bf \small \underline{\ensuremath{^{\textnormal{18}}}O(\ensuremath{\pi}\ensuremath{^{\textnormal{+}}},\ensuremath{\pi}\ensuremath{^{-}})\hspace{0.2in}\href{https://www.nndc.bnl.gov/nsr/nsrlink.jsp?1974LiZR,B}{1974LiZR},\href{https://www.nndc.bnl.gov/nsr/nsrlink.jsp?2007Ke07,B}{2007Ke07} (continued)}}\\
\vspace{0.3cm}
\parbox[b][0.3cm]{17.7cm}{\addtolength{\parindent}{-0.2in}\href{https://www.nndc.bnl.gov/nsr/nsrlink.jsp?1993Os07,B}{1993Os07}: \ensuremath{^{\textnormal{18}}}O(\ensuremath{\pi}\ensuremath{^{\textnormal{+}}},\ensuremath{\pi}\ensuremath{^{-}}) E=400 MeV; calculated \ensuremath{\sigma}(\ensuremath{\theta}) using a microscopic, parameter free Glauber approach. The authors}\\
\parbox[b][0.3cm]{17.7cm}{considered corrections in the single and double charge exchange amplitudes due to the medium polarization from an isospin-flip}\\
\parbox[b][0.3cm]{17.7cm}{spin-nonflip source. This kind of mechanism dominates these reactions at energies near \ensuremath{\Delta}(3/2,3/2) resonance. The authors discuss}\\
\parbox[b][0.3cm]{17.7cm}{their theoretical calculations for the \ensuremath{^{\textnormal{18}}}O(\ensuremath{\pi}\ensuremath{^{\textnormal{+}}},\ensuremath{\pi}\ensuremath{^{-}}) reaction populating the non-analog excited states of \ensuremath{^{\textnormal{18}}}Ne at 1.89 MeV and 3.56}\\
\parbox[b][0.3cm]{17.7cm}{MeV.}\\
\parbox[b][0.3cm]{17.7cm}{\addtolength{\parindent}{-0.2in}\href{https://www.nndc.bnl.gov/nsr/nsrlink.jsp?1993Os09,B}{1993Os09}: \ensuremath{^{\textnormal{18}}}O(\ensuremath{\pi}\ensuremath{^{\textnormal{+}}},\ensuremath{\pi}\ensuremath{^{-}}) E=300-525 MeV; compiled and reviewed theoretical \ensuremath{\sigma}(\ensuremath{\theta}).}\\
\parbox[b][0.3cm]{17.7cm}{\addtolength{\parindent}{-0.2in}\href{https://www.nndc.bnl.gov/nsr/nsrlink.jsp?1993Wa30,B}{1993Wa30}: \ensuremath{^{\textnormal{18}}}O(\ensuremath{\pi}\ensuremath{^{\textnormal{+}}},\ensuremath{\pi}\ensuremath{^{-}}) E\ensuremath{\leq}300 MeV; analyzed \ensuremath{\sigma}(\ensuremath{\theta}) available data and deduced \ensuremath{\pi}NN-system resonance parameters.}\\
\parbox[b][0.3cm]{17.7cm}{\addtolength{\parindent}{-0.2in}\href{https://www.nndc.bnl.gov/nsr/nsrlink.jsp?1995Ma58,B}{1995Ma58}: \ensuremath{^{\textnormal{18}}}O(\ensuremath{\pi}\ensuremath{^{\textnormal{+}}},\ensuremath{\pi}\ensuremath{^{-}}) E=164 MeV; calculated \ensuremath{\sigma}(\ensuremath{\theta}) using a hybrid quark hadron model. In this model, the dominant mechanisms}\\
\parbox[b][0.3cm]{17.7cm}{contributing to the double charge exchange reactions are the six-quark cluster mechanisms (short range) in the interior quark region,}\\
\parbox[b][0.3cm]{17.7cm}{and the conventional two-nucleon mechanisms (long range) in the exterior hadronic region. This model seems to be able to}\\
\parbox[b][0.3cm]{17.7cm}{reproduce the experimental data of (\href{https://www.nndc.bnl.gov/nsr/nsrlink.jsp?1979Se08,B}{1979Se08}, \href{https://www.nndc.bnl.gov/nsr/nsrlink.jsp?1979Gr18,B}{1979Gr18}).}\\
\parbox[b][0.3cm]{17.7cm}{\addtolength{\parindent}{-0.2in}\href{https://www.nndc.bnl.gov/nsr/nsrlink.jsp?1996Al15,B}{1996Al15}: \ensuremath{^{\textnormal{18}}}O(\ensuremath{\pi}\ensuremath{^{\textnormal{+}}},\ensuremath{\pi}\ensuremath{^{-}}) E=0.4-1.4 GeV; calculated \ensuremath{\sigma}(\ensuremath{\theta}) vs. E by evaluating the sequential single charge-exchange and meson}\\
\parbox[b][0.3cm]{17.7cm}{exchange currents. The authors state that the contribution of the meson-exchange currents becomes relevant at E\ensuremath{_{\ensuremath{\pi}}}\ensuremath{>}600 MeV.}\\
\parbox[b][0.3cm]{17.7cm}{\addtolength{\parindent}{-0.2in}\href{https://www.nndc.bnl.gov/nsr/nsrlink.jsp?2003Wu09,B}{2003Wu09}: \ensuremath{^{\textnormal{18}}}O(\ensuremath{\pi}\ensuremath{^{\textnormal{+}}},\ensuremath{\pi}\ensuremath{^{-}}) E=20-220 MeV; calculated \ensuremath{\sigma}(\ensuremath{\theta}) for double isobaric analog state transition in \ensuremath{^{\textnormal{18}}}Ne and the ground states}\\
\parbox[b][0.3cm]{17.7cm}{transitions in \ensuremath{^{\textnormal{16}}}O(\ensuremath{\pi}\ensuremath{^{\textnormal{+}}},\ensuremath{\pi}\ensuremath{^{-}})\ensuremath{^{\textnormal{16}}}Ne and \ensuremath{^{\textnormal{40}}}Ca(\ensuremath{\pi}\ensuremath{^{\textnormal{+}}},\ensuremath{\pi}\ensuremath{^{-}})\ensuremath{^{\textnormal{40}}}Ti ; deduced configuration mixing effects. The authors emphasized on the}\\
\parbox[b][0.3cm]{17.7cm}{importance of nuclear structure effects on the double charge exchange reactions.}\\
\parbox[b][0.3cm]{17.7cm}{\addtolength{\parindent}{-0.2in}\href{https://www.nndc.bnl.gov/nsr/nsrlink.jsp?2007Ke07,B}{2007Ke07}: \ensuremath{^{\textnormal{18}}}O(\ensuremath{\pi}\ensuremath{^{\textnormal{+}}},\ensuremath{\pi}\ensuremath{^{-}}) E=600-1400 MeV; calculated \ensuremath{\sigma}(E,\ensuremath{\theta}) using a composite-meson model. The authors discussed the}\\
\parbox[b][0.3cm]{17.7cm}{contribution of meson exchange current in double charge exchange reaction mechanism.}\\
\vspace{0.385cm}
\parbox[b][0.3cm]{17.7cm}{\addtolength{\parindent}{-0.2in}\textit{Informative Reviews of Experiments and Theory}:}\\
\parbox[b][0.3cm]{17.7cm}{\addtolength{\parindent}{-0.2in}\href{https://www.nndc.bnl.gov/nsr/nsrlink.jsp?1979Al35,B}{1979Al35}, S. J. Greene, The Positive-Pion Double Charge Exchange Reaction, Los Alamos National Laboratory Report No.}\\
\parbox[b][0.3cm]{17.7cm}{LA8891-T (1981), R. A. Gilman, Systematics of Pion Double Charge Exchange, Thesis, Los Alamos National Laboratory Report}\\
\parbox[b][0.3cm]{17.7cm}{No. LA-10524-T (1985), \href{https://www.nndc.bnl.gov/nsr/nsrlink.jsp?1983Os09,B}{1983Os09}, \href{https://www.nndc.bnl.gov/nsr/nsrlink.jsp?1988Se13,B}{1988Se13}, \href{https://www.nndc.bnl.gov/nsr/nsrlink.jsp?1993Jo16,B}{1993Jo16}.}\\
\vspace{0.385cm}
\parbox[b][0.3cm]{17.7cm}{\addtolength{\parindent}{-0.2in}\textit{See also}:}\\
\parbox[b][0.3cm]{17.7cm}{\addtolength{\parindent}{-0.2in}E. Aslanides, T. Bressani, M. Caria, \textit{et al}., Proc. Int. Conf. on Nucleus-Nucleus Collisions, 26 September-1 October (1982), East}\\
\parbox[b][0.3cm]{17.7cm}{Lansing, MI, USA (1982) 2.}\\
\parbox[b][0.3cm]{17.7cm}{\addtolength{\parindent}{-0.2in}X.-H. Liu and Y.-G. Li, Phys. Energ. Fortis Phys. Nucl. 7 (1983) 197.}\\
\parbox[b][0.3cm]{17.7cm}{\addtolength{\parindent}{-0.2in}Bauer \textit{et al}., 10\ensuremath{^{\textnormal{th}}} Int. Conf. on Paricles and Nuclei, 30 July-3 August (1984), Heidelberg (1984) F21.}\\
\parbox[b][0.3cm]{17.7cm}{\addtolength{\parindent}{-0.2in}H. W. Baer and G. A. Miller, Comments Nucl. Part. Phys. 15 (1986) 269.}\\
\parbox[b][0.3cm]{17.7cm}{\addtolength{\parindent}{-0.2in}B. Parker, K. Seth and R. Soundranayagam, Panic (1987) 356.}\\
\parbox[b][0.3cm]{17.7cm}{\addtolength{\parindent}{-0.2in}Baer, Bull. Amer. Phys. Soc. 34 (1989) 1210.}\\
\parbox[b][0.3cm]{17.7cm}{\addtolength{\parindent}{-0.2in}Strottman, Fund. Symm. and Nucl. Struct., Eds. Ginocchio and Rosen, in Santa Fe, NM 1988 (World Scientific: 1989) 247.}\\
\vspace{12pt}
\underline{$^{18}$Ne Levels}\\
\vspace{0.34cm}
\parbox[b][0.3cm]{17.7cm}{\addtolength{\parindent}{-0.254cm}(\href{https://www.nndc.bnl.gov/nsr/nsrlink.jsp?1977Pe12,B}{1977Pe12}): the laboratory differential cross section was deduced by integrating over the double charge exchange continuum and the}\\
\parbox[b][0.3cm]{17.7cm}{excited states of \ensuremath{^{\textnormal{18}}}Ne up to 20 MeV excitation energy. The results are: d\ensuremath{\sigma}/d\ensuremath{\Omega}\ensuremath{_{\textnormal{lab}}}=3.8 \ensuremath{\mu}b/sr \textit{70} and 3.0 \ensuremath{\mu}b/sr \textit{5} for E-148 MeV}\\
\parbox[b][0.3cm]{17.7cm}{and 187 MeV, respectively, and at \ensuremath{\theta}\ensuremath{_{\textnormal{lab}}}=18\ensuremath{^\circ}.}\\
\vspace{0.34cm}
\begin{longtable}{cccc@{\extracolsep{\fill}}c}
\multicolumn{2}{c}{E(level)$^{{\hyperlink{NE20LEVEL0}{a}}}$}&J$^{\pi}$$^{{\hyperlink{NE20LEVEL2}{c}}}$&Comments&\\[-.2cm]
\multicolumn{2}{c}{\hrulefill}&\hrulefill&\hrulefill&
\endfirsthead
\multicolumn{1}{r@{}}{0}&\multicolumn{1}{@{}l}{}&\multicolumn{1}{l}{0\ensuremath{^{+}}}&\parbox[t][0.3cm]{15.102401cm}{\raggedright T=1 (\href{https://www.nndc.bnl.gov/nsr/nsrlink.jsp?1979Se08,B}{1979Se08})\vspace{0.1cm}}&\\
&&&\parbox[t][0.3cm]{15.102401cm}{\raggedright E(level): In the \ensuremath{^{\textnormal{18}}}O(\ensuremath{\pi}\ensuremath{^{\textnormal{+}}},\ensuremath{\pi}\ensuremath{^{-}})\ensuremath{^{\textnormal{18}}}Ne\ensuremath{_{\textnormal{g.s.}}} reaction, the final state is the \ensuremath{\Delta}T\ensuremath{_{\textnormal{z}}}=2, isobaric analog state of \ensuremath{^{\textnormal{18}}}O\ensuremath{_{\textnormal{g.s.}}}.\vspace{0.1cm}}&\\
&&&\parbox[t][0.3cm]{15.102401cm}{\raggedright E(level): Populated in (\href{https://www.nndc.bnl.gov/nsr/nsrlink.jsp?1977Ma02,B}{1977Ma02}, \href{https://www.nndc.bnl.gov/nsr/nsrlink.jsp?1977Pe12,B}{1977Pe12}, \href{https://www.nndc.bnl.gov/nsr/nsrlink.jsp?1978Bu09,B}{1978Bu09}, \href{https://www.nndc.bnl.gov/nsr/nsrlink.jsp?1979Gr18,B}{1979Gr18}, \href{https://www.nndc.bnl.gov/nsr/nsrlink.jsp?1979Se08,B}{1979Se08}, \href{https://www.nndc.bnl.gov/nsr/nsrlink.jsp?1982Gr02,B}{1982Gr02}, \href{https://www.nndc.bnl.gov/nsr/nsrlink.jsp?1982Gr28,B}{1982Gr28},\vspace{0.1cm}}&\\
&&&\parbox[t][0.3cm]{15.102401cm}{\raggedright {\ }{\ }{\ }\href{https://www.nndc.bnl.gov/nsr/nsrlink.jsp?1985Al15,B}{1985Al15}, \href{https://www.nndc.bnl.gov/nsr/nsrlink.jsp?1985Se08,B}{1985Se08}, \href{https://www.nndc.bnl.gov/nsr/nsrlink.jsp?1989Wi02,B}{1989Wi02}, \href{https://www.nndc.bnl.gov/nsr/nsrlink.jsp?1992JoZZ,B}{1992JoZZ}, and \href{https://www.nndc.bnl.gov/nsr/nsrlink.jsp?1993Jo03,B}{1993Jo03}).\vspace{0.1cm}}&\\
&&&\parbox[t][0.3cm]{15.102401cm}{\raggedright J\ensuremath{^{\pi}}: This state, when populated via the \ensuremath{^{\textnormal{18}}}O(\ensuremath{\pi}\ensuremath{^{\textnormal{+}}},\ensuremath{\pi}\ensuremath{^{-}}) reaction, is the double isobaric analog state, which makes\vspace{0.1cm}}&\\
&&&\parbox[t][0.3cm]{15.102401cm}{\raggedright {\ }{\ }{\ }it a 0\ensuremath{^{\textnormal{+}}} state. Moreover, (\href{https://www.nndc.bnl.gov/nsr/nsrlink.jsp?1979Se08,B}{1979Se08}) stated that the measured angular distribution of this state populated via\vspace{0.1cm}}&\\
&&&\parbox[t][0.3cm]{15.102401cm}{\raggedright {\ }{\ }{\ }the \ensuremath{^{\textnormal{18}}}O(\ensuremath{\pi}\ensuremath{^{\textnormal{+}}},\ensuremath{\pi}\ensuremath{^{-}}) reaction shows an apparent diffractive shape characteristic of L=0 transfer, which is evidence\vspace{0.1cm}}&\\
&&&\parbox[t][0.3cm]{15.102401cm}{\raggedright {\ }{\ }{\ }for the J\ensuremath{^{\ensuremath{\pi}}}=0\ensuremath{^{\textnormal{+}}} assignment.\vspace{0.1cm}}&\\
\multicolumn{1}{r@{}}{1886}&\multicolumn{1}{@{}l}{}&\multicolumn{1}{l}{2\ensuremath{^{+}}}&\parbox[t][0.3cm]{15.102401cm}{\raggedright E(level): From an unweighted average of the energies reported by (\href{https://www.nndc.bnl.gov/nsr/nsrlink.jsp?1977Pe12,B}{1977Pe12}: 1890 keV); (\href{https://www.nndc.bnl.gov/nsr/nsrlink.jsp?1979Gr18,B}{1979Gr18}: 1880\vspace{0.1cm}}&\\
&&&\parbox[t][0.3cm]{15.102401cm}{\raggedright {\ }{\ }{\ }keV); (\href{https://www.nndc.bnl.gov/nsr/nsrlink.jsp?1979Se08,B}{1979Se08}: 1890 keV); (\href{https://www.nndc.bnl.gov/nsr/nsrlink.jsp?1981GrZS,B}{1981GrZS}: 1880 keV); (\href{https://www.nndc.bnl.gov/nsr/nsrlink.jsp?1982Gr28,B}{1982Gr28}: 1880 keV); (\href{https://www.nndc.bnl.gov/nsr/nsrlink.jsp?1985Al15,B}{1985Al15}: 1890 keV); and\vspace{0.1cm}}&\\
&&&\parbox[t][0.3cm]{15.102401cm}{\raggedright {\ }{\ }{\ }(\href{https://www.nndc.bnl.gov/nsr/nsrlink.jsp?1985Se08,B}{1985Se08}: 1890 keV) and rounded to the nearest integer.\vspace{0.1cm}}&\\
\end{longtable}
\begin{textblock}{29}(0,27.3)
Continued on next page (footnotes at end of table)
\end{textblock}
\clearpage
\begin{longtable}{cccc@{\extracolsep{\fill}}c}
\\[-.4cm]
\multicolumn{5}{c}{{\bf \small \underline{\ensuremath{^{\textnormal{18}}}O(\ensuremath{\pi}\ensuremath{^{\textnormal{+}}},\ensuremath{\pi}\ensuremath{^{-}})\hspace{0.2in}\href{https://www.nndc.bnl.gov/nsr/nsrlink.jsp?1974LiZR,B}{1974LiZR},\href{https://www.nndc.bnl.gov/nsr/nsrlink.jsp?2007Ke07,B}{2007Ke07} (continued)}}}\\
\multicolumn{5}{c}{~}\\
\multicolumn{5}{c}{\underline{\ensuremath{^{18}}Ne Levels (continued)}}\\
\multicolumn{5}{c}{~}\\
\multicolumn{2}{c}{E(level)$^{{\hyperlink{NE20LEVEL0}{a}}}$}&J$^{\pi}$$^{{\hyperlink{NE20LEVEL2}{c}}}$&Comments&\\[-.2cm]
\multicolumn{2}{c}{\hrulefill}&\hrulefill&\hrulefill&
\endhead
&&&\parbox[t][0.3cm]{14.869141cm}{\raggedright J\ensuremath{^{\pi}}: From (\href{https://www.nndc.bnl.gov/nsr/nsrlink.jsp?1979Gr18,B}{1979Gr18}, \href{https://www.nndc.bnl.gov/nsr/nsrlink.jsp?1979Se08,B}{1979Se08}, \href{https://www.nndc.bnl.gov/nsr/nsrlink.jsp?1982Gr28,B}{1982Gr28}, and \href{https://www.nndc.bnl.gov/nsr/nsrlink.jsp?1985Se08,B}{1985Se08}). Note that none of these studies actually\vspace{0.1cm}}&\\
&&&\parbox[t][0.3cm]{14.869141cm}{\raggedright {\ }{\ }{\ }measured the J\ensuremath{^{\ensuremath{\pi}}} value for this state. The evaluator speculates that the reported J\ensuremath{^{\ensuremath{\pi}}}=2\ensuremath{^{\textnormal{+}}} for this state in these\vspace{0.1cm}}&\\
&&&\parbox[t][0.3cm]{14.869141cm}{\raggedright {\ }{\ }{\ }studies come from the established \ensuremath{^{\textnormal{18}}}Ne Adopted Levels in ENSDF (\href{https://www.nndc.bnl.gov/nsr/nsrlink.jsp?1978Aj03,B}{1978Aj03} and \href{https://www.nndc.bnl.gov/nsr/nsrlink.jsp?1983Aj01,B}{1983Aj01}). (\href{https://www.nndc.bnl.gov/nsr/nsrlink.jsp?1979Se08,B}{1979Se08})\vspace{0.1cm}}&\\
&&&\parbox[t][0.3cm]{14.869141cm}{\raggedright {\ }{\ }{\ }stated that the measured angular distribution of this state populated via the \ensuremath{^{\textnormal{18}}}O(\ensuremath{\pi}\ensuremath{^{\textnormal{+}}},\ensuremath{\pi}\ensuremath{^{-}}) reaction shows an\vspace{0.1cm}}&\\
&&&\parbox[t][0.3cm]{14.869141cm}{\raggedright {\ }{\ }{\ }apparent diffractive shape characteristic of L=2 transfer, which is evidence for the J\ensuremath{^{\ensuremath{\pi}}}=2\ensuremath{^{\textnormal{+}}} assignment. Also,\vspace{0.1cm}}&\\
&&&\parbox[t][0.3cm]{14.869141cm}{\raggedright {\ }{\ }{\ }the measured angular distribution (by \href{https://www.nndc.bnl.gov/nsr/nsrlink.jsp?1982Gr28,B}{1982Gr28}) of this state has a shape consistent with a \ensuremath{\Delta}L=2\vspace{0.1cm}}&\\
&&&\parbox[t][0.3cm]{14.869141cm}{\raggedright {\ }{\ }{\ }transition, in agreement with the result of (\href{https://www.nndc.bnl.gov/nsr/nsrlink.jsp?1979Se08,B}{1979Se08}).\vspace{0.1cm}}&\\
\multicolumn{1}{r@{}}{3.62\ensuremath{\times10^{3}}}&\multicolumn{1}{@{}l}{}&\multicolumn{1}{l}{2\ensuremath{^{+}}\ensuremath{^{{\hyperlink{NE20LEVEL3}{d}}}}}&\parbox[t][0.3cm]{14.869141cm}{\raggedright E(level): From (\href{https://www.nndc.bnl.gov/nsr/nsrlink.jsp?1979Se08,B}{1979Se08}).\vspace{0.1cm}}&\\
\multicolumn{1}{r@{}}{5.09\ensuremath{\times10^{3}}}&\multicolumn{1}{@{}l}{\ensuremath{^{{\hyperlink{NE20LEVEL1}{b}}}}}&\multicolumn{1}{l}{3\ensuremath{^{-}}\ensuremath{^{{\hyperlink{NE20LEVEL3}{d}}}}}&\parbox[t][0.3cm]{14.869141cm}{\raggedright E(level): From (\href{https://www.nndc.bnl.gov/nsr/nsrlink.jsp?1979Se08,B}{1979Se08}).\vspace{0.1cm}}&\\
\multicolumn{1}{r@{}}{5.14\ensuremath{\times10^{3}}}&\multicolumn{1}{@{}l}{\ensuremath{^{{\hyperlink{NE20LEVEL1}{b}}}}}&\multicolumn{1}{l}{3\ensuremath{^{-}}\ensuremath{^{{\hyperlink{NE20LEVEL3}{d}}}}}&\parbox[t][0.3cm]{14.869141cm}{\raggedright E(level): From (\href{https://www.nndc.bnl.gov/nsr/nsrlink.jsp?1979Se08,B}{1979Se08}).\vspace{0.1cm}}&\\
\end{longtable}
\parbox[b][0.3cm]{17.7cm}{\makebox[1ex]{\ensuremath{^{\hypertarget{NE20LEVEL0}{a}}}} (\href{https://www.nndc.bnl.gov/nsr/nsrlink.jsp?1979Se08,B}{1979Se08}): the uncertainty in the excitation energies reported in this work may be 30\%.}\\
\parbox[b][0.3cm]{17.7cm}{\makebox[1ex]{\ensuremath{^{\hypertarget{NE20LEVEL1}{b}}}} (\href{https://www.nndc.bnl.gov/nsr/nsrlink.jsp?1979Se08,B}{1979Se08}): these states are not resolved.}\\
\parbox[b][0.3cm]{17.7cm}{\makebox[1ex]{\ensuremath{^{\hypertarget{NE20LEVEL2}{c}}}} (\href{https://www.nndc.bnl.gov/nsr/nsrlink.jsp?1979Se08,B}{1979Se08}): the measured angular distributions (at 164 MeV) for the double isobaric analog state (\ensuremath{^{\textnormal{18}}}Ne\ensuremath{_{\textnormal{g.s.}}}) and for the}\\
\parbox[b][0.3cm]{17.7cm}{{\ }{\ }\ensuremath{^{\textnormal{18}}}Ne(2\ensuremath{^{\textnormal{+}}_{\textnormal{1}}}) state showed apparent diffractive shapes, characteristic of L=0 and L=2 transfers, respectively, observed in surface}\\
\parbox[b][0.3cm]{17.7cm}{{\ }{\ }dominated direct reactions. These are in contrast to the flat, featureless distributions expected from two consecutive uncorrelated}\\
\parbox[b][0.3cm]{17.7cm}{{\ }{\ }steps of single charge exchange.}\\
\parbox[b][0.3cm]{17.7cm}{\makebox[1ex]{\ensuremath{^{\hypertarget{NE20LEVEL3}{d}}}} (\href{https://www.nndc.bnl.gov/nsr/nsrlink.jsp?1979Se08,B}{1979Se08}) did not measure the J\ensuremath{^{\ensuremath{\pi}}} assignment of this state, so the evaluator speculates that the reported J\ensuremath{^{\ensuremath{\pi}}}=2\ensuremath{^{\textnormal{+}}} and 3\ensuremath{^{-}}}\\
\parbox[b][0.3cm]{17.7cm}{{\ }{\ }assignments most likely come from the \ensuremath{^{\textnormal{18}}}Ne Adopted Levels established in (\href{https://www.nndc.bnl.gov/nsr/nsrlink.jsp?1978Aj03,B}{1978Aj03}).}\\
\vspace{0.5cm}
\clearpage
\subsection[\hspace{-0.2cm}\ensuremath{^{\textnormal{19}}}F(p,2n)]{ }
\vspace{-27pt}
\vspace{0.3cm}
\hypertarget{NE21}{{\bf \small \underline{\ensuremath{^{\textnormal{19}}}F(p,2n)\hspace{0.2in}\href{https://www.nndc.bnl.gov/nsr/nsrlink.jsp?1954Go17,B}{1954Go17}}}}\\
\vspace{4pt}
\vspace{8pt}
\parbox[b][0.3cm]{17.7cm}{\addtolength{\parindent}{-0.2in}\href{https://www.nndc.bnl.gov/nsr/nsrlink.jsp?1954Go17,B}{1954Go17}: \ensuremath{^{\textnormal{19}}}F(p,2n) E not given. This study reports the first observation of \ensuremath{^{\textnormal{18}}}Ne (see also \href{https://www.nndc.bnl.gov/nsr/nsrlink.jsp?2012Th01,B}{2012Th01}). A proton beam (E not}\\
\parbox[b][0.3cm]{17.7cm}{given) impinged on a teflon and a LiF crystal target. Beam was on for a ``\textit{short time}'', and positrons were counted during beam off}\\
\parbox[b][0.3cm]{17.7cm}{time using a 180\ensuremath{^\circ} magnetic spectrograph to deflect the positrons 180\ensuremath{^\circ} away from the target. Measured positrons coincidences using}\\
\parbox[b][0.3cm]{17.7cm}{two proportional counters facing each other. Measured a single \ensuremath{\beta}\ensuremath{^{\textnormal{+}}} branch of the maximum energy of 3.2 MeV \textit{2}, which was}\\
\parbox[b][0.3cm]{17.7cm}{attributed to the positron decay of \ensuremath{^{\textnormal{18}}}Ne to the ground state of \ensuremath{^{\textnormal{18}}}F. Measured \ensuremath{^{\textnormal{18}}}Ne\ensuremath{_{\textnormal{g.s.}}} decay curve and half-life (T\ensuremath{_{\textnormal{1/2}}}=1.6 s \textit{2}).}\\
\parbox[b][0.3cm]{17.7cm}{Deduced a log\textit{ft} value of 2.9 \textit{2}.}\\
\parbox[b][0.3cm]{17.7cm}{\addtolength{\parindent}{-0.2in}\href{https://www.nndc.bnl.gov/nsr/nsrlink.jsp?2003An02,B}{2003An02}, \href{https://www.nndc.bnl.gov/nsr/nsrlink.jsp?2003An28,B}{2003An28}, \href{https://www.nndc.bnl.gov/nsr/nsrlink.jsp?2004An28,B}{2004An28}, \href{https://www.nndc.bnl.gov/nsr/nsrlink.jsp?2006AcZY,B}{2006AcZY}, \href{https://www.nndc.bnl.gov/nsr/nsrlink.jsp?2008Pe02,B}{2008Pe02}: \ensuremath{^{\textnormal{19}}}F(p,2n) E=21, 23.5, 25, 28 MeV; \ensuremath{^{\textnormal{1}}}H, \ensuremath{^{\textnormal{nat}}}C(\ensuremath{^{\textnormal{18}}}Ne,p) E=66 MeV; and}\\
\parbox[b][0.3cm]{17.7cm}{\ensuremath{^{\textnormal{1}}}H(\ensuremath{^{\textnormal{18}}}Ne,\ensuremath{^{\textnormal{18}}}Ne),\ensuremath{^{\textnormal{1}}}H(\ensuremath{^{\textnormal{18}}}Ne,\ensuremath{^{\textnormal{18}}}Ne\ensuremath{'}) E=66 MeV. These studies focused on \ensuremath{^{\textnormal{19}}}Na states and their proton decay to \ensuremath{^{\textnormal{18}}}Ne. In order to}\\
\parbox[b][0.3cm]{17.7cm}{carry out these experiments, a beam of \ensuremath{^{\textnormal{18}}}Ne\ensuremath{_{\textnormal{g.s.}}} was produced using the \ensuremath{^{\textnormal{19}}}F(p,2n) reaction.}\\
\vspace{12pt}
\underline{$^{18}$Ne Levels}\\
\begin{longtable}{cccc@{\extracolsep{\fill}}c}
\multicolumn{2}{c}{E(level)$^{{\hyperlink{NE21LEVEL0}{a}}}$}&\multicolumn{2}{c}{T$_{1/2}$$^{{\hyperlink{NE21LEVEL0}{a}}}$}&\\[-.2cm]
\multicolumn{2}{c}{\hrulefill}&\multicolumn{2}{c}{\hrulefill}&
\endfirsthead
\multicolumn{1}{r@{}}{0}&\multicolumn{1}{@{}l}{}&\multicolumn{1}{r@{}}{1}&\multicolumn{1}{@{.}l}{6 s {\it 2}}&\\
\end{longtable}
\parbox[b][0.3cm]{17.7cm}{\makebox[1ex]{\ensuremath{^{\hypertarget{NE21LEVEL0}{a}}}} From (\href{https://www.nndc.bnl.gov/nsr/nsrlink.jsp?1954Go17,B}{1954Go17}).}\\
\vspace{0.5cm}
\clearpage
\subsection[\hspace{-0.2cm}\ensuremath{^{\textnormal{20}}}Ne(p,t)]{ }
\vspace{-27pt}
\vspace{0.3cm}
\hypertarget{NE22}{{\bf \small \underline{\ensuremath{^{\textnormal{20}}}Ne(p,t)\hspace{0.2in}\href{https://www.nndc.bnl.gov/nsr/nsrlink.jsp?1969Ha38,B}{1969Ha38},\href{https://www.nndc.bnl.gov/nsr/nsrlink.jsp?2017Ch32,B}{2017Ch32}}}}\\
\vspace{4pt}
\vspace{8pt}
\parbox[b][0.3cm]{17.7cm}{\addtolength{\parindent}{-0.2in}D. K. Olsen and R. E. Brown, John H. Williams Laboratory of Nuclear Physics, University of Minnesota Report No.}\\
\parbox[b][0.3cm]{17.7cm}{COO-1265-67, 1968, p. 86 (unpublished): \ensuremath{^{\textnormal{20}}}Ne(p,t) E=40 MeV; measured tritons using 32 surface barrier detectors placed on the}\\
\parbox[b][0.3cm]{17.7cm}{focal plane of a spectrometer (no information is given on its type). The energy resolution was 150-200 keV (FWHM). The ground}\\
\parbox[b][0.3cm]{17.7cm}{state of \ensuremath{^{\textnormal{18}}}Ne was observed. The triton angular distribution corresponding to \ensuremath{^{\textnormal{18}}}Ne\ensuremath{_{\textnormal{g.s.}}} was measured at \ensuremath{\theta}\ensuremath{_{\textnormal{lab}}}=8\ensuremath{^\circ}{\textminus}75\ensuremath{^\circ}.}\\
\parbox[b][0.3cm]{17.7cm}{\addtolength{\parindent}{-0.2in}\href{https://www.nndc.bnl.gov/nsr/nsrlink.jsp?1969Ha38,B}{1969Ha38}: \ensuremath{^{\textnormal{20}}}Ne(p,t) E=45 MeV; measured the reaction products using two telescopes consisting of a phosphorus diffused silicon}\\
\parbox[b][0.3cm]{17.7cm}{\ensuremath{\Delta}E counter and a Si(Li) E counter mounted on opposite sides of the scattering chamber operating in coincidence mode. The energy}\\
\parbox[b][0.3cm]{17.7cm}{resolution was 100-130 keV (FWHM). Tritons were measured at \ensuremath{\theta}\ensuremath{_{\textnormal{lab}}}=22.3\ensuremath{^\circ}, 26.8\ensuremath{^\circ}, and 41\ensuremath{^\circ}. \ensuremath{^{\textnormal{18}}}Ne levels were deduced at 1890}\\
\parbox[b][0.3cm]{17.7cm}{keV \textit{20}, 3375 keV \textit{30}, 3588 keV \textit{25}, 4580 keV \textit{30}, and 5115 keV \textit{25}. Comparisons with the reported energies from previous}\\
\parbox[b][0.3cm]{17.7cm}{experiments are presented. The angular distributions of tritons corresponding to \ensuremath{^{\textnormal{18}}}Ne\ensuremath{_{\textnormal{g.s.}}} were measured at \ensuremath{\theta}\ensuremath{_{\textnormal{c.m.}}}=10\ensuremath{^\circ}{\textminus}40\ensuremath{^\circ}. Angular}\\
\parbox[b][0.3cm]{17.7cm}{momentum transfers were deduced by DWBA analysis. Coulomb displacement energies for the A=18 multiplet were calculated}\\
\parbox[b][0.3cm]{17.7cm}{using two sets of equations: one derived in low-seniority j-j coupling limit, and the other in the Wigner supermultiplet scheme. The}\\
\parbox[b][0.3cm]{17.7cm}{results are discussed and compared with the previous theoretical calculations in (\href{https://www.nndc.bnl.gov/nsr/nsrlink.jsp?1968Be82,B}{1968Be82}).}\\
\parbox[b][0.3cm]{17.7cm}{\addtolength{\parindent}{-0.2in}\href{https://www.nndc.bnl.gov/nsr/nsrlink.jsp?1970Fa17,B}{1970Fa17}: \ensuremath{^{\textnormal{20}}}Ne(p,t) E=42.6 MeV. The reaction products$'$ detection system was comprised of a \ensuremath{\Delta}E-E telescope consisting of a}\\
\parbox[b][0.3cm]{17.7cm}{silicon surface barrier \ensuremath{\Delta}E and a Si(Li) E detector. The energy resolution was 140 keV. The excitation energy spectrum of \ensuremath{^{\textnormal{18}}}Ne}\\
\parbox[b][0.3cm]{17.7cm}{(up to E\ensuremath{_{\textnormal{x}}}=9170 keV \textit{30}) and the tritons angular distributions at \ensuremath{\theta}\ensuremath{_{\textnormal{c.m.}}}=10\ensuremath{^\circ}{\textminus}80\ensuremath{^\circ} were measured. A finite range DWBA analysis was}\\
\parbox[b][0.3cm]{17.7cm}{performed to deduce L, J\ensuremath{^{\ensuremath{\pi}}}, and the two-nucleon spectroscopic amplitudes. Comparison with \ensuremath{^{\textnormal{18}}}O mirror levels are presented. The}\\
\parbox[b][0.3cm]{17.7cm}{spectroscopic amplitudes deduced by (\href{https://www.nndc.bnl.gov/nsr/nsrlink.jsp?1967Ku09,B}{1967Ku09}: includes 1\textit{d}\ensuremath{_{\textnormal{5/2}}} and 2\textit{s}\ensuremath{_{\textnormal{1/2}}} configurations) used in the DWBA calculations of}\\
\parbox[b][0.3cm]{17.7cm}{(\href{https://www.nndc.bnl.gov/nsr/nsrlink.jsp?1970Fa17,B}{1970Fa17}) resulted in good agreement with all the five lowest \ensuremath{^{\textnormal{18}}}Ne states except for the 4\ensuremath{^{\textnormal{+}}_{\textnormal{1}}} level.}\\
\parbox[b][0.3cm]{17.7cm}{\addtolength{\parindent}{-0.2in}\href{https://www.nndc.bnl.gov/nsr/nsrlink.jsp?1970Le08,B}{1970Le08}: \ensuremath{^{\textnormal{20}}}Ne(p,t) E=50 MeV; momentum analyzed the reaction products using a \textit{n}=1/2 magnetic spectrometer (J. Bonn \textit{et al.},}\\
\parbox[b][0.3cm]{17.7cm}{Rutherford Laboratory Report No. RHEL/R136, 1966 (unpublished), p. 141). The focal plane detector consisted of a sonic spark}\\
\parbox[b][0.3cm]{17.7cm}{chamber backed by a \ensuremath{\Delta}E-E scintillator telescope. Energy resolution was 120 keV (FWHM). The \ensuremath{^{\textnormal{18}}}Ne excitation energy spectrum}\\
\parbox[b][0.3cm]{17.7cm}{up to E\ensuremath{_{\textnormal{x}}}=6.34 MeV was measured at \ensuremath{\theta}\ensuremath{_{\textnormal{lab}}}=15\ensuremath{^\circ}{\textminus}50\ensuremath{^\circ}. The triton angular distributions were measured at \ensuremath{\theta}\ensuremath{_{\textnormal{lab}}}=15\ensuremath{^\circ}{\textminus}50\ensuremath{^\circ}. Finite-range}\\
\parbox[b][0.3cm]{17.7cm}{DWBA calculations were performed to deduce the transferred angular momenta and the J\ensuremath{^{\ensuremath{\pi}}} values. The spectroscopic amplitudes}\\
\parbox[b][0.3cm]{17.7cm}{were deduced for the \ensuremath{^{\textnormal{20}}}Ne(p,t) reaction at 50 MeV.}\\
\parbox[b][0.3cm]{17.7cm}{\addtolength{\parindent}{-0.2in}\href{https://www.nndc.bnl.gov/nsr/nsrlink.jsp?1971PaZX,B}{1971PaZX}, \href{https://www.nndc.bnl.gov/nsr/nsrlink.jsp?1972Pa02,B}{1972Pa02}: \ensuremath{^{\textnormal{20}}}Ne(p,t) E=45 MeV; identified the reaction products using a \ensuremath{\Delta}E-E telescope consisting of silicon surface}\\
\parbox[b][0.3cm]{17.7cm}{barrier detectors. The energy resolution was 90 keV (FWHM). Measured the excitation energy spectrum of \ensuremath{^{\textnormal{18}}}Ne (up to E\ensuremath{_{\textnormal{x}}}=9215}\\
\parbox[b][0.3cm]{17.7cm}{MeV) and triton angular distributions of the strongly populated levels at \ensuremath{\theta}\ensuremath{_{\textnormal{c.m.}}}\ensuremath{\sim}10\ensuremath{^\circ}{\textminus}120\ensuremath{^\circ}. The transferred angular momenta, and J\ensuremath{^{\ensuremath{\pi}}}}\\
\parbox[b][0.3cm]{17.7cm}{values were deduced by finite-range DWBA calculations using the JULIE code. Comparison with the previously assigned J\ensuremath{^{\ensuremath{\pi}}} values}\\
\parbox[b][0.3cm]{17.7cm}{and known level energies are presented.}\\
\parbox[b][0.3cm]{17.7cm}{\addtolength{\parindent}{-0.2in}\href{https://www.nndc.bnl.gov/nsr/nsrlink.jsp?1974Ne04,B}{1974Ne04}: \ensuremath{^{\textnormal{20}}}Ne(p,t) E=41.8 MeV; detected tritons at \ensuremath{\theta}\ensuremath{_{\textnormal{lab}}}=10\ensuremath{^\circ}{\textminus}45\ensuremath{^\circ} using a solid state telescope. The experimental resolution was}\\
\parbox[b][0.3cm]{17.7cm}{50 keV. The authors were interested to study a doublet in \ensuremath{^{\textnormal{18}}}Ne at E\ensuremath{_{\textnormal{x}}}\ensuremath{\sim}4.5 MeV that consisted of a 0\ensuremath{^{\textnormal{+}}} state and a 1\ensuremath{^{-}} state based on}\\
\parbox[b][0.3cm]{17.7cm}{mirror analysis (\href{https://www.nndc.bnl.gov/nsr/nsrlink.jsp?1970Ad02,B}{1970Ad02}). The \ensuremath{^{\textnormal{18}}}Ne excitation energy spectrum was measured up to E\ensuremath{_{\textnormal{x}}}=9198 MeV, and two states at 4522 keV}\\
\parbox[b][0.3cm]{17.7cm}{\textit{10} and 4592 keV \textit{10} were observed among others. The triton angular distributions corresponding to these two states were measured.}\\
\parbox[b][0.3cm]{17.7cm}{Level energies, J, \ensuremath{\pi}, and L were deduced based on comparisons of the shapes of triton angular distributions and the previously}\\
\parbox[b][0.3cm]{17.7cm}{published results.}\\
\parbox[b][0.3cm]{17.7cm}{\addtolength{\parindent}{-0.2in}\href{https://www.nndc.bnl.gov/nsr/nsrlink.jsp?1974OlZQ,B}{1974OlZQ}, \href{https://www.nndc.bnl.gov/nsr/nsrlink.jsp?1975Ol03,B}{1975Ol03}: \ensuremath{^{\textnormal{20}}}Ne(p,t) E=39.8 MeV; momentum analyzed the reaction products using a magnetic spectrometer with an}\\
\parbox[b][0.3cm]{17.7cm}{array of 32 Si surface barrier detectors on the focal plane. Deduced \ensuremath{^{\textnormal{18}}}Ne excitation energies for states up to 5140 keV. Measured}\\
\parbox[b][0.3cm]{17.7cm}{triton angular distributions at \ensuremath{\theta}\ensuremath{_{\textnormal{c.m.}}}\ensuremath{\sim}10\ensuremath{^\circ}{\textminus}85\ensuremath{^\circ}. The energy resolution was 100 keV. Deduced the two-neutron spectroscopic}\\
\parbox[b][0.3cm]{17.7cm}{amplitudes for the 2\textit{s}\ensuremath{_{\textnormal{1/2}}}, 1\textit{d}\ensuremath{_{\textnormal{3/2}}} and 1\textit{d}\ensuremath{_{\textnormal{5/2}}} sub shells. Performed zero-range coupled-channel Born approximation calculations using}\\
\parbox[b][0.3cm]{17.7cm}{the JUPITOR-1 and MARS codes; and DWBA calculations (using the MARS code). Deduced the rotational and vibrational}\\
\parbox[b][0.3cm]{17.7cm}{deformation parameters for the T=1 excited states in mass-18.}\\
\parbox[b][0.3cm]{17.7cm}{\addtolength{\parindent}{-0.2in}\href{https://www.nndc.bnl.gov/nsr/nsrlink.jsp?1981Ne09,B}{1981Ne09}: \ensuremath{^{\textnormal{20}}}Ne(p,t) E=41.8 MeV; measured the reaction products using a \ensuremath{\Delta}E-E telescope; measured the angular distributions of}\\
\parbox[b][0.3cm]{17.7cm}{tritons at \ensuremath{\theta}\ensuremath{_{\textnormal{lab}}}=10\ensuremath{^\circ}{\textminus}40\ensuremath{^\circ}; deduced \ensuremath{^{\textnormal{18}}}Ne excitation energies up to E\ensuremath{_{\textnormal{x}}}=9198 keV. Comparison with previous \ensuremath{^{\textnormal{20}}}Ne(p,t) and}\\
\parbox[b][0.3cm]{17.7cm}{\ensuremath{^{\textnormal{16}}}O(\ensuremath{^{\textnormal{3}}}He,n) experiments, and the mirror levels in \ensuremath{^{\textnormal{18}}}O and \ensuremath{^{\textnormal{18}}}Ne are discussed. The authors calculated the Coulomb shifts for the}\\
\parbox[b][0.3cm]{17.7cm}{\ensuremath{^{\textnormal{18}}}Ne states from the ground state to the 0\ensuremath{^{\textnormal{+}}_{\textnormal{3}}} state. The authors recommended wave functions that place most of the \textit{s}\ensuremath{^{\textnormal{2}}_{\textnormal{1/2}}} strength}\\
\parbox[b][0.3cm]{17.7cm}{in the 0\ensuremath{^{\textnormal{+}}_{\textnormal{3}}} state observed at 4.5 MeV in \ensuremath{^{\textnormal{18}}}Ne, such as the wave functions of (\href{https://www.nndc.bnl.gov/nsr/nsrlink.jsp?1969Be94,B}{1969Be94}, \href{https://www.nndc.bnl.gov/nsr/nsrlink.jsp?1972En03,B}{1972En03}, and \href{https://www.nndc.bnl.gov/nsr/nsrlink.jsp?1970El23,B}{1970El23}). The authors}\\
\parbox[b][0.3cm]{17.7cm}{concluded that the existence of three 0\ensuremath{^{\textnormal{+}}} and three 2\ensuremath{^{\textnormal{+}}} states at low excitation energy and the observation of enhanced E2 transition}\\
\parbox[b][0.3cm]{17.7cm}{rates in \ensuremath{^{\textnormal{18}}}Ne suggest the presence of deformed configurations in this region. They assigned the deformed strength to the 0\ensuremath{^{\textnormal{+}}_{\textnormal{2}}} state.}\\
\parbox[b][0.3cm]{17.7cm}{As a by-product of the remeasurement of the \ensuremath{^{\textnormal{18}}}Ne mass (F. P. Calaprice, S. J. Freedman, and A. V. Nero, private communication,}\\
\parbox[b][0.3cm]{17.7cm}{unpublished), the doublets at E\ensuremath{_{\textnormal{x}}}(\ensuremath{^{\textnormal{18}}}Ne)=4.5 MeV and 5.1 MeV, as well as the triplet at E\ensuremath{_{\textnormal{x}}}\ensuremath{\sim}3.5 MeV were remeasured using the}\\
\parbox[b][0.3cm]{17.7cm}{\ensuremath{^{\textnormal{20}}}Ne(p,t) reaction at 41.8 MeV beam energy. The tritons were measured using the high resolution Princeton Q3D spectrograph.}\\
\parbox[b][0.3cm]{17.7cm}{With this latter measurement, the members of the 3.5-MeV triplet and those of the 4.5-MeV and 5.1-MeV doublets were resolved}\\
\parbox[b][0.3cm]{17.7cm}{(see Fig. 13 of \href{https://www.nndc.bnl.gov/nsr/nsrlink.jsp?1981Ne09,B}{1981Ne09}). The authors deduced E\ensuremath{_{\textnormal{x}}}=4522 keV \textit{10} and E\ensuremath{_{\textnormal{x}}}=4592 keV \textit{10} states and \ensuremath{\Gamma}=40 keV \textit{20} and \ensuremath{\Gamma}=25 keV}\\
\clearpage
\vspace{0.3cm}
{\bf \small \underline{\ensuremath{^{\textnormal{20}}}Ne(p,t)\hspace{0.2in}\href{https://www.nndc.bnl.gov/nsr/nsrlink.jsp?1969Ha38,B}{1969Ha38},\href{https://www.nndc.bnl.gov/nsr/nsrlink.jsp?2017Ch32,B}{2017Ch32} (continued)}}\\
\vspace{0.3cm}
\parbox[b][0.3cm]{17.7cm}{\textit{15} for the 5099 keV \textit{10} and 5151 keV \textit{10} states, respectively.}\\
\parbox[b][0.3cm]{17.7cm}{\addtolength{\parindent}{-0.2in}\href{https://www.nndc.bnl.gov/nsr/nsrlink.jsp?1995La27,B}{1995La27}: \ensuremath{^{\textnormal{nat}}}Ne(p,X) E=25-67.5 MeV. This study focused on the commercial production of \ensuremath{^{\textnormal{18}}}F as a source for positron emission}\\
\parbox[b][0.3cm]{17.7cm}{tomography purposes. Several reaction channels contributed to the total production cross section of \ensuremath{^{\textnormal{18}}}F\ensuremath{_{\textnormal{g.s.}}}. As a by-product,}\\
\parbox[b][0.3cm]{17.7cm}{\ensuremath{^{\textnormal{18}}}Ne\ensuremath{_{\textnormal{g.s.}}} was produced via the (p,t), (p,dn), and (p,p2n) reactions on \ensuremath{^{\textnormal{20}}}Ne; the (p,4n), (p,p3n), (p,d2n), and (p,tn) reactions on}\\
\parbox[b][0.3cm]{17.7cm}{\ensuremath{^{\textnormal{21}}}Ne; and the (p,d3n), (p,p4n), (p,t2n), and (p,\ensuremath{\alpha}n) reactions on \ensuremath{^{\textnormal{22}}}Ne. The results and recommendations for production of \ensuremath{^{\textnormal{18}}}F are}\\
\parbox[b][0.3cm]{17.7cm}{discussed. Single, cumulative and saturation yields of \ensuremath{^{\textnormal{18}}}F were discussed.}\\
\parbox[b][0.3cm]{17.7cm}{\addtolength{\parindent}{-0.2in}\href{https://www.nndc.bnl.gov/nsr/nsrlink.jsp?1996Ha26,B}{1996Ha26}: \ensuremath{^{\textnormal{20}}}Ne(p,t) E=88.4 and 40 MeV. Performed two separate experiments: in the first experiment, a 88.4-MeV proton beam}\\
\parbox[b][0.3cm]{17.7cm}{from the Indiana University Cyclotron Facility bombarded a \ensuremath{^{\textnormal{20}}}Ne implanted target. The tritons were measured at \ensuremath{\theta}\ensuremath{_{\textnormal{lab}}}=6\ensuremath{^\circ}{\textminus}11\ensuremath{^\circ}}\\
\parbox[b][0.3cm]{17.7cm}{using the high resolution K600 spectrometer, and its associated focal plane detectors, in dispersion matching mode. The energy}\\
\parbox[b][0.3cm]{17.7cm}{resolution was 20-25 keV. The \ensuremath{^{\textnormal{18}}}Ne states with E\ensuremath{_{\textnormal{x}}}=0\ensuremath{\sim}8 MeV were observed in this experiment. In the second experiment, a}\\
\parbox[b][0.3cm]{17.7cm}{40-MeV proton beam from the Princeton AVF cyclotron bombarded the \ensuremath{^{\textnormal{20}}}Ne implanted target. Tritons were measured at \ensuremath{\theta}\ensuremath{_{\textnormal{lab}}}=10\ensuremath{^\circ}}\\
\parbox[b][0.3cm]{17.7cm}{and 20\ensuremath{^\circ} using the Princeton Q3D spectrograph, with an energy resolution of 15 keV. The \ensuremath{^{\textnormal{18}}}Ne excited states with E\ensuremath{_{\textnormal{x}}}\ensuremath{<}6 MeV were}\\
\parbox[b][0.3cm]{17.7cm}{observed. With the Q3D spectrograph, the authors easily resolved the doublets at 4.5 MeV and 5.1 MeV (see Fig. 11). There was}\\
\parbox[b][0.3cm]{17.7cm}{no indication in any of these experiments of a missing state with J\ensuremath{^{\ensuremath{\pi}}}=3\ensuremath{^{\textnormal{+}}} proposed by (\href{https://www.nndc.bnl.gov/nsr/nsrlink.jsp?1988Wi08,B}{1988Wi08}). With these two experiments, the}\\
\parbox[b][0.3cm]{17.7cm}{\ensuremath{^{\textnormal{18}}}Ne level energies and widths were deduced.}\\
\parbox[b][0.3cm]{17.7cm}{\addtolength{\parindent}{-0.2in}\href{https://www.nndc.bnl.gov/nsr/nsrlink.jsp?1996Ro11,B}{1996Ro11}: \ensuremath{^{\textnormal{nat}}}Ne(p,X) E=19-38 and 41 MeV. This study also focused on the production cross section of \ensuremath{^{\textnormal{18}}}F for medical purposes.}\\
\parbox[b][0.3cm]{17.7cm}{The authors measured the residual \ensuremath{^{\textnormal{18}}}F yields and calculated thick target saturation yields. Comparison with the results of}\\
\parbox[b][0.3cm]{17.7cm}{(\href{https://www.nndc.bnl.gov/nsr/nsrlink.jsp?1995La27,B}{1995La27}) are discussed. The authors concluded that \ensuremath{^{\textnormal{18}}}F can be produced with high yield using proton energies above 40 MeV.}\\
\parbox[b][0.3cm]{17.7cm}{\addtolength{\parindent}{-0.2in}\href{https://www.nndc.bnl.gov/nsr/nsrlink.jsp?1998PaZZ,B}{1998PaZZ}, \href{https://www.nndc.bnl.gov/nsr/nsrlink.jsp?1998PaZR,B}{1998PaZR}, \href{https://www.nndc.bnl.gov/nsr/nsrlink.jsp?1998KuZX,B}{1998KuZX}, \href{https://www.nndc.bnl.gov/nsr/nsrlink.jsp?1999Pa07,B}{1999Pa07}: \ensuremath{^{\textnormal{20}}}Ne(p,t) E=35 MeV; measured the tritons using a high resolution Q2D magnetic}\\
\parbox[b][0.3cm]{17.7cm}{spectrograph. The energy resolution was 12 keV. Measured the excitation energies of \ensuremath{^{\textnormal{18}}}Ne from E\ensuremath{_{\textnormal{x}}}=4520 keV to E\ensuremath{_{\textnormal{x}}}=6358 keV.}\\
\parbox[b][0.3cm]{17.7cm}{Owing to the high resolution achieved in this study, the member states of the 3 doublets near 4.5 MeV, 5.1 MeV and 6.3 MeV}\\
\parbox[b][0.3cm]{17.7cm}{were all fully resolved. Measured the triton angular distributions for the members of the 5.1 MeV doublet at \ensuremath{\theta}\ensuremath{_{\textnormal{c.m.}}}=15\ensuremath{^\circ}{\textminus}90\ensuremath{^\circ}.}\\
\parbox[b][0.3cm]{17.7cm}{DWBA calculations were performed using DWUCK4; and J, \ensuremath{\pi}, and L were deduced for these two states. The widths of the}\\
\parbox[b][0.3cm]{17.7cm}{unbound members of these 3 closely spaced doublets were deduced. Comparison with the previous results are discussed. The}\\
\parbox[b][0.3cm]{17.7cm}{authors searched for the missing 3\ensuremath{^{\textnormal{+}}} state proposed by (\href{https://www.nndc.bnl.gov/nsr/nsrlink.jsp?1988Wi08,B}{1988Wi08}) at 4.33 MeV and found no conclusive evidence for such a state.}\\
\parbox[b][0.3cm]{17.7cm}{\addtolength{\parindent}{-0.2in}\href{https://www.nndc.bnl.gov/nsr/nsrlink.jsp?2017Ch32,B}{2017Ch32}: \ensuremath{^{\textnormal{20}}}Ne(p,t) E not given. A proton beam impinged upon a neon jet gas target using the JENSA instrument (\href{https://www.nndc.bnl.gov/nsr/nsrlink.jsp?2014Ch56,B}{2014Ch56}).}\\
\parbox[b][0.3cm]{17.7cm}{The preliminary experimental program is outlined.}\\
\vspace{0.385cm}
\parbox[b][0.3cm]{17.7cm}{\addtolength{\parindent}{-0.2in}\textit{Theory}:}\\
\parbox[b][0.3cm]{17.7cm}{\addtolength{\parindent}{-0.2in}\href{https://www.nndc.bnl.gov/nsr/nsrlink.jsp?1969So08,B}{1969So08}: \ensuremath{^{\textnormal{20}}}Ne(p,t); investigated pairing type correlations in the structure of even-N nuclei; calculated neutron pairing strength,}\\
\parbox[b][0.3cm]{17.7cm}{collective boson Hamiltonian, and neutron pairing type states (zero seniority) for \ensuremath{^{\textnormal{18}}}Ne and suggested that \ensuremath{^{\textnormal{18}}}Ne may be slightly}\\
\parbox[b][0.3cm]{17.7cm}{deformed due to the observation of a weak 0\ensuremath{^{\textnormal{+}}} state populated (via the \ensuremath{^{\textnormal{20}}}Ne(p,t) reaction: \href{https://www.nndc.bnl.gov/nsr/nsrlink.jsp?1970Le08,B}{1970Le08}) at approximately twice the}\\
\parbox[b][0.3cm]{17.7cm}{2\ensuremath{^{\textnormal{+}}_{\textnormal{1}}} energy.}\\
\parbox[b][0.3cm]{17.7cm}{\addtolength{\parindent}{-0.2in}\href{https://www.nndc.bnl.gov/nsr/nsrlink.jsp?1973OlZU,B}{1973OlZU}: \ensuremath{^{\textnormal{20}}}Ne(p,t); calculated \ensuremath{\sigma}(\ensuremath{\theta}).}\\
\vspace{12pt}
\underline{$^{18}$Ne Levels}\\
\vspace{0.34cm}
\parbox[b][0.3cm]{17.7cm}{\addtolength{\parindent}{-0.254cm}T: From (\href{https://www.nndc.bnl.gov/nsr/nsrlink.jsp?1974Ne04,B}{1974Ne04}, \href{https://www.nndc.bnl.gov/nsr/nsrlink.jsp?1981Ne09,B}{1981Ne09}) unless otherwise noted.}\\
\parbox[b][0.3cm]{17.7cm}{\addtolength{\parindent}{-0.254cm}(\href{https://www.nndc.bnl.gov/nsr/nsrlink.jsp?1972Pa02,B}{1972Pa02}) deduced their excitation energies using the \ensuremath{^{\textnormal{18}}}Ne mass excess of 5319.3 keV \textit{47} from (\href{https://www.nndc.bnl.gov/nsr/nsrlink.jsp?1965Ma54,B}{1965Ma54}). This could explain}\\
\parbox[b][0.3cm]{17.7cm}{why the excitation energies reported by (\href{https://www.nndc.bnl.gov/nsr/nsrlink.jsp?1972Pa02,B}{1972Pa02}) are on average \ensuremath{\sim}10 keV higher than those deduced by (\href{https://www.nndc.bnl.gov/nsr/nsrlink.jsp?1974Ne04,B}{1974Ne04}).}\\
\parbox[b][0.3cm]{17.7cm}{\addtolength{\parindent}{-0.254cm}The overall normalization uncertainty for the measured cross sections in (\href{https://www.nndc.bnl.gov/nsr/nsrlink.jsp?1972Pa02,B}{1972Pa02}) is about \ensuremath{\pm}5\%.}\\
\parbox[b][0.3cm]{17.7cm}{\addtolength{\parindent}{-0.254cm}The overall normalization uncertainty of about \ensuremath{\pm}10\% should be added in quadrature to the error bars shown on the experimental}\\
\parbox[b][0.3cm]{17.7cm}{data points in Fig. 4 of (\href{https://www.nndc.bnl.gov/nsr/nsrlink.jsp?1970Fa17,B}{1970Fa17}).}\\
\parbox[b][0.3cm]{17.7cm}{\addtolength{\parindent}{-0.254cm}The differential cross sections measured in (\href{https://www.nndc.bnl.gov/nsr/nsrlink.jsp?1975Ol03,B}{1975Ol03}) have a relative uncertainty of \ensuremath{\pm}4\%. The absolute cross sections have a 5\%}\\
\parbox[b][0.3cm]{17.7cm}{uncertainty (standard deviation).}\\
\parbox[b][0.3cm]{17.7cm}{\addtolength{\parindent}{-0.254cm}(\href{https://www.nndc.bnl.gov/nsr/nsrlink.jsp?1981Ne09,B}{1981Ne09}): measured cross sections have 15\% uncertainty.}\\
\parbox[b][0.3cm]{17.7cm}{\addtolength{\parindent}{-0.254cm}(\href{https://www.nndc.bnl.gov/nsr/nsrlink.jsp?1981Ne09,B}{1981Ne09}): from the measured triton angular distributions, the relative intensities deduced by (\href{https://www.nndc.bnl.gov/nsr/nsrlink.jsp?1981Ne09,B}{1981Ne09}) are: \ensuremath{\sigma}(0\ensuremath{^{\textnormal{+}}_{\textnormal{2}}})/\ensuremath{\sigma}(0\ensuremath{^{\textnormal{+}}_{\textnormal{1}}})\ensuremath{\sim}12}\\
\parbox[b][0.3cm]{17.7cm}{and \ensuremath{\sigma}(2\ensuremath{^{\textnormal{+}}_{\textnormal{2}}})/\ensuremath{\sigma}(2\ensuremath{^{\textnormal{+}}_{\textnormal{1}}})\ensuremath{\sim}4.}\\
\vspace{0.34cm}
\begin{textblock}{29}(0,27.3)
Continued on next page (footnotes at end of table)
\end{textblock}
\clearpage
\vspace{0.3cm}
{\bf \small \underline{\ensuremath{^{\textnormal{20}}}Ne(p,t)\hspace{0.2in}\href{https://www.nndc.bnl.gov/nsr/nsrlink.jsp?1969Ha38,B}{1969Ha38},\href{https://www.nndc.bnl.gov/nsr/nsrlink.jsp?2017Ch32,B}{2017Ch32} (continued)}}\\
\vspace{0.3cm}
\underline{$^{18}$Ne Levels (continued)}\\
\begin{longtable}{ccccc@{\extracolsep{\fill}}c}
\multicolumn{2}{c}{E(level)$^{}$}&J$^{\pi}$$^{{\hyperlink{NE22LEVEL5}{f}}}$&L$^{}$&Comments&\\[-.2cm]
\multicolumn{2}{c}{\hrulefill}&\hrulefill&\hrulefill&\hrulefill&
\endfirsthead
\multicolumn{1}{r@{}}{0}&\multicolumn{1}{@{}l}{\ensuremath{^{{\hyperlink{NE22LEVEL0}{a}}}}}&\multicolumn{1}{l}{0\ensuremath{^{+}}\ensuremath{^{{\hyperlink{NE22LEVEL3}{d}}{\hyperlink{NE22LEVEL4}{e}}}}}&\multicolumn{1}{l}{0$^{{\hyperlink{NE22LEVEL6}{g}}}$}&\parbox[t][0.3cm]{13.732981cm}{\raggedright T=1\vspace{0.1cm}}&\\
&&&&\parbox[t][0.3cm]{13.732981cm}{\raggedright E(level): Populated in (\href{https://www.nndc.bnl.gov/nsr/nsrlink.jsp?1969Ha38,B}{1969Ha38}, \href{https://www.nndc.bnl.gov/nsr/nsrlink.jsp?1970Fa17,B}{1970Fa17}, \href{https://www.nndc.bnl.gov/nsr/nsrlink.jsp?1970Le08,B}{1970Le08}, \href{https://www.nndc.bnl.gov/nsr/nsrlink.jsp?1972Pa02,B}{1972Pa02}, \href{https://www.nndc.bnl.gov/nsr/nsrlink.jsp?1974Ne04,B}{1974Ne04}, \href{https://www.nndc.bnl.gov/nsr/nsrlink.jsp?1975Ol03,B}{1975Ol03}, \href{https://www.nndc.bnl.gov/nsr/nsrlink.jsp?1981Ne09,B}{1981Ne09},\vspace{0.1cm}}&\\
&&&&\parbox[t][0.3cm]{13.732981cm}{\raggedright {\ }{\ }{\ }\href{https://www.nndc.bnl.gov/nsr/nsrlink.jsp?1995La27,B}{1995La27}, and \href{https://www.nndc.bnl.gov/nsr/nsrlink.jsp?1996Ro11,B}{1996Ro11}).\vspace{0.1cm}}&\\
&&&&\parbox[t][0.3cm]{13.732981cm}{\raggedright J\ensuremath{^{\pi}}: From the DWBA analyses of (\href{https://www.nndc.bnl.gov/nsr/nsrlink.jsp?1969Ha38,B}{1969Ha38}, \href{https://www.nndc.bnl.gov/nsr/nsrlink.jsp?1970Fa17,B}{1970Fa17}, \href{https://www.nndc.bnl.gov/nsr/nsrlink.jsp?1970Le08,B}{1970Le08}, \href{https://www.nndc.bnl.gov/nsr/nsrlink.jsp?1972Pa02,B}{1972Pa02}: finite-range DWBA),\vspace{0.1cm}}&\\
&&&&\parbox[t][0.3cm]{13.732981cm}{\raggedright {\ }{\ }{\ }and the zero-range coupled-channel Born approximation calculations of (\href{https://www.nndc.bnl.gov/nsr/nsrlink.jsp?1975Ol03,B}{1975Ol03}).\vspace{0.1cm}}&\\
&&&&\parbox[t][0.3cm]{13.732981cm}{\raggedright L: From the DWBA analyses of (\href{https://www.nndc.bnl.gov/nsr/nsrlink.jsp?1969Ha38,B}{1969Ha38}, \href{https://www.nndc.bnl.gov/nsr/nsrlink.jsp?1970Fa17,B}{1970Fa17}, \href{https://www.nndc.bnl.gov/nsr/nsrlink.jsp?1970Le08,B}{1970Le08}: finite-range DWBA).\vspace{0.1cm}}&\\
&&&&\parbox[t][0.3cm]{13.732981cm}{\raggedright S\ensuremath{_{\textnormal{2n}}}=28.50 MeV (\href{https://www.nndc.bnl.gov/nsr/nsrlink.jsp?1970Fa17,B}{1970Fa17}).\vspace{0.1cm}}&\\
&&&&\parbox[t][0.3cm]{13.732981cm}{\raggedright (\href{https://www.nndc.bnl.gov/nsr/nsrlink.jsp?1975Ol03,B}{1975Ol03}) measured d\ensuremath{\sigma}/d\ensuremath{\Omega}\ensuremath{_{\textnormal{c.m.}}}=747 \ensuremath{\mu}b/sr for \ensuremath{^{\textnormal{18}}}Ne\ensuremath{_{\textnormal{g.s.}}} at the first maximum near \ensuremath{\theta}\ensuremath{_{\textnormal{c.m.}}}=30\ensuremath{^\circ} and\vspace{0.1cm}}&\\
&&&&\parbox[t][0.3cm]{13.732981cm}{\raggedright {\ }{\ }{\ }at E\ensuremath{_{\textnormal{p}}}=39.8 MeV. This result is in agreement with the 650-750 \ensuremath{\mu}b/sr cross sections measured at\vspace{0.1cm}}&\\
&&&&\parbox[t][0.3cm]{13.732981cm}{\raggedright {\ }{\ }{\ }E\ensuremath{_{\textnormal{p}}}=42.6 MeV (\href{https://www.nndc.bnl.gov/nsr/nsrlink.jsp?1970Fa17,B}{1970Fa17}), 45 MeV (\href{https://www.nndc.bnl.gov/nsr/nsrlink.jsp?1969Ha38,B}{1969Ha38}, \href{https://www.nndc.bnl.gov/nsr/nsrlink.jsp?1972Pa02,B}{1972Pa02}), and 50 MeV (\href{https://www.nndc.bnl.gov/nsr/nsrlink.jsp?1970Le08,B}{1970Le08}).\vspace{0.1cm}}&\\
\multicolumn{1}{r@{}}{1889}&\multicolumn{1}{@{}l}{\ensuremath{^{{\hyperlink{NE22LEVEL0}{a}}}} {\it 7}}&\multicolumn{1}{l}{2\ensuremath{^{+}}\ensuremath{^{{\hyperlink{NE22LEVEL4}{e}}}}}&\multicolumn{1}{l}{2$^{{\hyperlink{NE22LEVEL6}{g}}}$}&\parbox[t][0.3cm]{13.732981cm}{\raggedright T=1 (\href{https://www.nndc.bnl.gov/nsr/nsrlink.jsp?1981Ne09,B}{1981Ne09})\vspace{0.1cm}}&\\
&&&&\parbox[t][0.3cm]{13.732981cm}{\raggedright E(level): Weighted average of 1890 keV 20 (\href{https://www.nndc.bnl.gov/nsr/nsrlink.jsp?1969Ha38,B}{1969Ha38}); 1830 keV \textit{50} (\href{https://www.nndc.bnl.gov/nsr/nsrlink.jsp?1970Le08,B}{1970Le08}); 1894 keV \textit{10}\vspace{0.1cm}}&\\
&&&&\parbox[t][0.3cm]{13.732981cm}{\raggedright {\ }{\ }{\ }(\href{https://www.nndc.bnl.gov/nsr/nsrlink.jsp?1972Pa02,B}{1972Pa02}); and 1886 keV \textit{10} (\href{https://www.nndc.bnl.gov/nsr/nsrlink.jsp?1974Ne04,B}{1974Ne04}, \href{https://www.nndc.bnl.gov/nsr/nsrlink.jsp?1981Ne09,B}{1981Ne09}). See also E\ensuremath{_{\textnormal{x}}}=1887 keV (\href{https://www.nndc.bnl.gov/nsr/nsrlink.jsp?1970Fa17,B}{1970Fa17}) and\vspace{0.1cm}}&\\
&&&&\parbox[t][0.3cm]{13.732981cm}{\raggedright {\ }{\ }{\ }1890 keV (\href{https://www.nndc.bnl.gov/nsr/nsrlink.jsp?1975Ol03,B}{1975Ol03}).\vspace{0.1cm}}&\\
&&&&\parbox[t][0.3cm]{13.732981cm}{\raggedright J\ensuremath{^{\pi}}: From the DWBA analyses of (\href{https://www.nndc.bnl.gov/nsr/nsrlink.jsp?1970Fa17,B}{1970Fa17}, \href{https://www.nndc.bnl.gov/nsr/nsrlink.jsp?1970Le08,B}{1970Le08}, \href{https://www.nndc.bnl.gov/nsr/nsrlink.jsp?1972Pa02,B}{1972Pa02}: finite-range DWBA) and the\vspace{0.1cm}}&\\
&&&&\parbox[t][0.3cm]{13.732981cm}{\raggedright {\ }{\ }{\ }zero-range coupled-channel Born approximation calculations of (\href{https://www.nndc.bnl.gov/nsr/nsrlink.jsp?1975Ol03,B}{1975Ol03}).\vspace{0.1cm}}&\\
&&&&\parbox[t][0.3cm]{13.732981cm}{\raggedright L: From the finite-range DWBA analyses of (\href{https://www.nndc.bnl.gov/nsr/nsrlink.jsp?1970Fa17,B}{1970Fa17}, \href{https://www.nndc.bnl.gov/nsr/nsrlink.jsp?1970Le08,B}{1970Le08}).\vspace{0.1cm}}&\\
&&&&\parbox[t][0.3cm]{13.732981cm}{\raggedright \ensuremath{\sigma}/\ensuremath{\sigma}\ensuremath{_{\textnormal{g.s.}}}(\ensuremath{\theta}\ensuremath{_{\textnormal{lab}}}=30\ensuremath{^\circ})=0.42 (\href{https://www.nndc.bnl.gov/nsr/nsrlink.jsp?1970Le08,B}{1970Le08}) at E\ensuremath{_{\textnormal{p}}}=50 MeV.\vspace{0.1cm}}&\\
\multicolumn{1}{r@{}}{3379}&\multicolumn{1}{@{}l}{\ensuremath{^{{\hyperlink{NE22LEVEL0}{a}}{\hyperlink{NE22LEVEL1}{b}}}} {\it 8}}&\multicolumn{1}{l}{4\ensuremath{^{+}}\ensuremath{^{{\hyperlink{NE22LEVEL4}{e}}}}}&\multicolumn{1}{l}{4$^{{\hyperlink{NE22LEVEL6}{g}}}$}&\parbox[t][0.3cm]{13.732981cm}{\raggedright T=1\vspace{0.1cm}}&\\
&&&&\parbox[t][0.3cm]{13.732981cm}{\raggedright E(level): Weighted average of 3375 keV \textit{30} (\href{https://www.nndc.bnl.gov/nsr/nsrlink.jsp?1969Ha38,B}{1969Ha38}); 3360 keV \textit{50} (\href{https://www.nndc.bnl.gov/nsr/nsrlink.jsp?1970Le08,B}{1970Le08}); 3390 keV \textit{14}\vspace{0.1cm}}&\\
&&&&\parbox[t][0.3cm]{13.732981cm}{\raggedright {\ }{\ }{\ }(\href{https://www.nndc.bnl.gov/nsr/nsrlink.jsp?1972Pa02,B}{1972Pa02}); and 3375 keV \textit{10} (\href{https://www.nndc.bnl.gov/nsr/nsrlink.jsp?1974Ne04,B}{1974Ne04}, \href{https://www.nndc.bnl.gov/nsr/nsrlink.jsp?1981Ne09,B}{1981Ne09}). See also E\ensuremath{_{\textnormal{x}}}=3376 keV (\href{https://www.nndc.bnl.gov/nsr/nsrlink.jsp?1970Fa17,B}{1970Fa17}); 3380\vspace{0.1cm}}&\\
&&&&\parbox[t][0.3cm]{13.732981cm}{\raggedright {\ }{\ }{\ }keV (\href{https://www.nndc.bnl.gov/nsr/nsrlink.jsp?1975Ol03,B}{1975Ol03}); and 3376 keV (see Fig. 13 of (\href{https://www.nndc.bnl.gov/nsr/nsrlink.jsp?1981Ne09,B}{1981Ne09}): from the remeasurement of this state\vspace{0.1cm}}&\\
&&&&\parbox[t][0.3cm]{13.732981cm}{\raggedright {\ }{\ }{\ }using the Princeton Q3D spectrograph).\vspace{0.1cm}}&\\
&&&&\parbox[t][0.3cm]{13.732981cm}{\raggedright E(level): (\href{https://www.nndc.bnl.gov/nsr/nsrlink.jsp?1972Pa02,B}{1972Pa02}) paired their level observed at 3390 keV \textit{14} to the level in \ensuremath{^{\textnormal{18}}}Ne with\vspace{0.1cm}}&\\
&&&&\parbox[t][0.3cm]{13.732981cm}{\raggedright {\ }{\ }{\ }E\ensuremath{_{\textnormal{x}}}=3376.2 keV \textit{4}, which is the weighted average of the energies reported for this state in\vspace{0.1cm}}&\\
&&&&\parbox[t][0.3cm]{13.732981cm}{\raggedright {\ }{\ }{\ }(\href{https://www.nndc.bnl.gov/nsr/nsrlink.jsp?1970Fa17,B}{1970Fa17}, \href{https://www.nndc.bnl.gov/nsr/nsrlink.jsp?1968To09,B}{1968To09}, \href{https://www.nndc.bnl.gov/nsr/nsrlink.jsp?1970Le08,B}{1970Le08}, and \href{https://www.nndc.bnl.gov/nsr/nsrlink.jsp?1969Ro08,B}{1969Ro08}).\vspace{0.1cm}}&\\
&&&&\parbox[t][0.3cm]{13.732981cm}{\raggedright J\ensuremath{^{\pi}}: From the DWBA analyses of (\href{https://www.nndc.bnl.gov/nsr/nsrlink.jsp?1970Fa17,B}{1970Fa17}, \href{https://www.nndc.bnl.gov/nsr/nsrlink.jsp?1970Le08,B}{1970Le08}, \href{https://www.nndc.bnl.gov/nsr/nsrlink.jsp?1972Pa02,B}{1972Pa02}: J\ensuremath{^{\ensuremath{\pi}}}=(4\ensuremath{^{\textnormal{+}}}) from finite-range\vspace{0.1cm}}&\\
&&&&\parbox[t][0.3cm]{13.732981cm}{\raggedright {\ }{\ }{\ }DWBA) and the zero-range coupled-channel Born approximation calculations of (\href{https://www.nndc.bnl.gov/nsr/nsrlink.jsp?1975Ol03,B}{1975Ol03}). Note\vspace{0.1cm}}&\\
&&&&\parbox[t][0.3cm]{13.732981cm}{\raggedright {\ }{\ }{\ }that the DWBA fits performed by (\href{https://www.nndc.bnl.gov/nsr/nsrlink.jsp?1970Le08,B}{1970Le08}) and (\href{https://www.nndc.bnl.gov/nsr/nsrlink.jsp?1972Pa02,B}{1972Pa02}) with L=4 for J\ensuremath{^{\ensuremath{\pi}}}=4\ensuremath{^{\textnormal{+}}} do not well\vspace{0.1cm}}&\\
&&&&\parbox[t][0.3cm]{13.732981cm}{\raggedright {\ }{\ }{\ }describe the measured triton angular distribution data corresponding to this state.\vspace{0.1cm}}&\\
&&&&\parbox[t][0.3cm]{13.732981cm}{\raggedright J\ensuremath{^{\pi}}: (\href{https://www.nndc.bnl.gov/nsr/nsrlink.jsp?1972Pa02,B}{1972Pa02}) considered the 1\textit{d}\ensuremath{_{\textnormal{5/2}}} sub-shell for the two transferred neutrons.\vspace{0.1cm}}&\\
&&&&\parbox[t][0.3cm]{13.732981cm}{\raggedright L: From the finite-range DWBA analyses of (\href{https://www.nndc.bnl.gov/nsr/nsrlink.jsp?1970Fa17,B}{1970Fa17}, \href{https://www.nndc.bnl.gov/nsr/nsrlink.jsp?1970Le08,B}{1970Le08}).\vspace{0.1cm}}&\\
&&&&\parbox[t][0.3cm]{13.732981cm}{\raggedright \ensuremath{\sigma}/\ensuremath{\sigma}\ensuremath{_{\textnormal{g.s.}}}(\ensuremath{\theta}\ensuremath{_{\textnormal{lab}}}=30\ensuremath{^\circ})=0.06 (\href{https://www.nndc.bnl.gov/nsr/nsrlink.jsp?1970Le08,B}{1970Le08}) at E\ensuremath{_{\textnormal{p}}}=50 MeV.\vspace{0.1cm}}&\\
\multicolumn{1}{r@{}}{3580}&\multicolumn{1}{@{}l}{\ensuremath{^{{\hyperlink{NE22LEVEL0}{a}}{\hyperlink{NE22LEVEL1}{b}}}} {\it 10}}&\multicolumn{1}{l}{(0\ensuremath{^{+}})\ensuremath{^{{\hyperlink{NE22LEVEL4}{e}}}}}&\multicolumn{1}{l}{0$^{{\hyperlink{NE22LEVEL6}{g}}}$}&\parbox[t][0.3cm]{13.732981cm}{\raggedright T=1\vspace{0.1cm}}&\\
&&&&\parbox[t][0.3cm]{13.732981cm}{\raggedright E(level): From (\href{https://www.nndc.bnl.gov/nsr/nsrlink.jsp?1974Ne04,B}{1974Ne04}, \href{https://www.nndc.bnl.gov/nsr/nsrlink.jsp?1981Ne09,B}{1981Ne09}). See also E\ensuremath{_{\textnormal{x}}}=3588 keV \textit{25} (\href{https://www.nndc.bnl.gov/nsr/nsrlink.jsp?1969Ha38,B}{1969Ha38}: unresolved doublet\vspace{0.1cm}}&\\
&&&&\parbox[t][0.3cm]{13.732981cm}{\raggedright {\ }{\ }{\ }consisting of a (0\ensuremath{^{\textnormal{+}}}) state at 3576.3 keV and the 2\ensuremath{^{\textnormal{(+)}}} state at 3616.4 keV with the energies from\vspace{0.1cm}}&\\
&&&&\parbox[t][0.3cm]{13.732981cm}{\raggedright {\ }{\ }{\ }(\href{https://www.nndc.bnl.gov/nsr/nsrlink.jsp?1968Gi09,B}{1968Gi09})); 3576 keV (\href{https://www.nndc.bnl.gov/nsr/nsrlink.jsp?1970Fa17,B}{1970Fa17}: an unresolved doublet); 3580 keV \textit{50} (\href{https://www.nndc.bnl.gov/nsr/nsrlink.jsp?1970Le08,B}{1970Le08}: an\vspace{0.1cm}}&\\
&&&&\parbox[t][0.3cm]{13.732981cm}{\raggedright {\ }{\ }{\ }unresolved doublet consisting of a 0\ensuremath{^{\textnormal{+}}} and a 2\ensuremath{^{\textnormal{+}}} state at 3.59 MeV and 3.63 MeV, respectively);\vspace{0.1cm}}&\\
&&&&\parbox[t][0.3cm]{13.732981cm}{\raggedright {\ }{\ }{\ }3614 keV \textit{13} (\href{https://www.nndc.bnl.gov/nsr/nsrlink.jsp?1972Pa02,B}{1972Pa02}: see below); 3580 keV (\href{https://www.nndc.bnl.gov/nsr/nsrlink.jsp?1975Ol03,B}{1975Ol03}); and 3576 keV (see Fig. 13 of\vspace{0.1cm}}&\\
&&&&\parbox[t][0.3cm]{13.732981cm}{\raggedright {\ }{\ }{\ }(\href{https://www.nndc.bnl.gov/nsr/nsrlink.jsp?1981Ne09,B}{1981Ne09}): from the remeasurement of this state using the Princeton Q3D spectrograph).\vspace{0.1cm}}&\\
&&&&\parbox[t][0.3cm]{13.732981cm}{\raggedright E(level): There were known, 0\ensuremath{^{\textnormal{+}}} and 2\ensuremath{^{\textnormal{+}}} states at E\ensuremath{_{\textnormal{x}}}=3576.3 keV \textit{20} and 3616.4 keV \textit{6} (see\vspace{0.1cm}}&\\
&&&&\parbox[t][0.3cm]{13.732981cm}{\raggedright {\ }{\ }{\ }\href{https://www.nndc.bnl.gov/nsr/nsrlink.jsp?1968Gi09,B}{1968Gi09}). (\href{https://www.nndc.bnl.gov/nsr/nsrlink.jsp?1972Pa02,B}{1972Pa02}) observed a level at 3614 keV \textit{13}, which was thought to be an unresolved\vspace{0.1cm}}&\\
&&&&\parbox[t][0.3cm]{13.732981cm}{\raggedright {\ }{\ }{\ }doublet consisting of the 0\ensuremath{^{\textnormal{+}}} and 2\ensuremath{^{\textnormal{+}}} levels mentioned above. The evaluator notes that the level\vspace{0.1cm}}&\\
&&&&\parbox[t][0.3cm]{13.732981cm}{\raggedright {\ }{\ }{\ }energies measured by (\href{https://www.nndc.bnl.gov/nsr/nsrlink.jsp?1972Pa02,B}{1972Pa02}) seem to be on average \ensuremath{\sim}10 keV higher than those of\vspace{0.1cm}}&\\
&&&&\parbox[t][0.3cm]{13.732981cm}{\raggedright {\ }{\ }{\ }(\href{https://www.nndc.bnl.gov/nsr/nsrlink.jsp?1974Ne04,B}{1974Ne04}), so there may be \ensuremath{\sim}10 keV systematic uncertainty that was not considered by\vspace{0.1cm}}&\\
&&&&\parbox[t][0.3cm]{13.732981cm}{\raggedright {\ }{\ }{\ }(\href{https://www.nndc.bnl.gov/nsr/nsrlink.jsp?1972Pa02,B}{1972Pa02}). They paired the E\ensuremath{_{\textnormal{x}}}=3614 keV \textit{13} level to the known 0\ensuremath{^{\textnormal{+}}} state at E\ensuremath{_{\textnormal{x}}}=3576 keV and\vspace{0.1cm}}&\\
&&&&\parbox[t][0.3cm]{13.732981cm}{\raggedright {\ }{\ }{\ }stated that the level observed at 3614 keV may have been the 0\ensuremath{^{\textnormal{+}}} state populated more strongly\vspace{0.1cm}}&\\
&&&&\parbox[t][0.3cm]{13.732981cm}{\raggedright {\ }{\ }{\ }because the DWBA fit (by \href{https://www.nndc.bnl.gov/nsr/nsrlink.jsp?1972Pa02,B}{1972Pa02}) with L=0 and \textit{s}\ensuremath{_{\textnormal{1/2}}^{\textnormal{2}}} shell configuration appeared to describe\vspace{0.1cm}}&\\
&&&&\parbox[t][0.3cm]{13.732981cm}{\raggedright {\ }{\ }{\ }the triton angular distribution of the 3614-keV level better that that with L=2 and \textit{d}\ensuremath{_{\textnormal{5/2}}^{\textnormal{2}}} shell\vspace{0.1cm}}&\\
&&&&\parbox[t][0.3cm]{13.732981cm}{\raggedright {\ }{\ }{\ }configuration (see Fig. 8).\vspace{0.1cm}}&\\
&&&&\parbox[t][0.3cm]{13.732981cm}{\raggedright J\ensuremath{^{\pi}},L: From the finite-range DWBA analyses of (\href{https://www.nndc.bnl.gov/nsr/nsrlink.jsp?1970Fa17,B}{1970Fa17}: J\ensuremath{^{\ensuremath{\pi}}}=0\ensuremath{^{\textnormal{+}}}+2\ensuremath{^{\textnormal{+}}} (an unresolved doublet) with\vspace{0.1cm}}&\\
&&&&\parbox[t][0.3cm]{13.732981cm}{\raggedright {\ }{\ }{\ }L=0+2); (\href{https://www.nndc.bnl.gov/nsr/nsrlink.jsp?1970Le08,B}{1970Le08}: J\ensuremath{^{\ensuremath{\pi}}}=0\ensuremath{^{\textnormal{+}}}+2\ensuremath{^{\textnormal{+}}} (an unresolved doublet) but L=0 was preferred due to a better\vspace{0.1cm}}&\\
&&&&\parbox[t][0.3cm]{13.732981cm}{\raggedright {\ }{\ }{\ }DWBA fit); (\href{https://www.nndc.bnl.gov/nsr/nsrlink.jsp?1972Pa02,B}{1972Pa02}: J\ensuremath{^{\ensuremath{\pi}}}=(0\ensuremath{^{\textnormal{+}}}) with L=0 preferred due to a better DWBA fit); and the\vspace{0.1cm}}&\\
&&&&\parbox[t][0.3cm]{13.732981cm}{\raggedright {\ }{\ }{\ }zero-range coupled-channel Born approximation calculations of (\href{https://www.nndc.bnl.gov/nsr/nsrlink.jsp?1975Ol03,B}{1975Ol03}: J\ensuremath{^{\ensuremath{\pi}}}=0\ensuremath{^{\textnormal{+}}} with L=0).\vspace{0.1cm}}&\\
&&&&\parbox[t][0.3cm]{13.732981cm}{\raggedright (\href{https://www.nndc.bnl.gov/nsr/nsrlink.jsp?1970Le08,B}{1970Le08}): \ensuremath{\sigma}/\ensuremath{\sigma}\ensuremath{_{\textnormal{g.s.}}}(\ensuremath{\theta}\ensuremath{_{\textnormal{lab}}}=30\ensuremath{^\circ})=0.11 for L=0 and 0.05 for L=2 at E\ensuremath{_{\textnormal{p}}}=50 MeV.\vspace{0.1cm}}&\\
\end{longtable}
\begin{textblock}{29}(0,27.3)
Continued on next page (footnotes at end of table)
\end{textblock}
\clearpage
\begin{longtable}{ccccccc@{\extracolsep{\fill}}c}
\\[-.4cm]
\multicolumn{8}{c}{{\bf \small \underline{\ensuremath{^{\textnormal{20}}}Ne(p,t)\hspace{0.2in}\href{https://www.nndc.bnl.gov/nsr/nsrlink.jsp?1969Ha38,B}{1969Ha38},\href{https://www.nndc.bnl.gov/nsr/nsrlink.jsp?2017Ch32,B}{2017Ch32} (continued)}}}\\
\multicolumn{8}{c}{~}\\
\multicolumn{8}{c}{\underline{\ensuremath{^{18}}Ne Levels (continued)}}\\
\multicolumn{8}{c}{~}\\
\multicolumn{2}{c}{E(level)$^{}$}&J$^{\pi}$$^{{\hyperlink{NE22LEVEL5}{f}}}$&\multicolumn{2}{c}{\ensuremath{\Gamma} (keV)$^{{\hyperlink{NE22LEVEL2}{c}}}$}&L$^{}$&Comments&\\[-.2cm]
\multicolumn{2}{c}{\hrulefill}&\hrulefill&\multicolumn{2}{c}{\hrulefill}&\hrulefill&\hrulefill&
\endhead
\multicolumn{1}{r@{}}{3612}&\multicolumn{1}{@{}l}{\ensuremath{^{{\hyperlink{NE22LEVEL1}{b}}}} {\it 10}}&\multicolumn{1}{l}{(2\ensuremath{^{+}})\ensuremath{^{{\hyperlink{NE22LEVEL4}{e}}}}}&&&\multicolumn{1}{l}{$^{{\hyperlink{NE22LEVEL6}{g}}}$}&\parbox[t][0.3cm]{11.338501cm}{\raggedright T=1\vspace{0.1cm}}&\\
&&&&&&\parbox[t][0.3cm]{11.338501cm}{\raggedright E(level): From (\href{https://www.nndc.bnl.gov/nsr/nsrlink.jsp?1974Ne04,B}{1974Ne04}, \href{https://www.nndc.bnl.gov/nsr/nsrlink.jsp?1981Ne09,B}{1981Ne09}: state resolved using the high resolution\vspace{0.1cm}}&\\
&&&&&&\parbox[t][0.3cm]{11.338501cm}{\raggedright {\ }{\ }{\ }Princeton Q3D spectrograph). See also E\ensuremath{_{\textnormal{x}}}=3616 keV (\href{https://www.nndc.bnl.gov/nsr/nsrlink.jsp?1970Fa17,B}{1970Fa17}: unresolved\vspace{0.1cm}}&\\
&&&&&&\parbox[t][0.3cm]{11.338501cm}{\raggedright {\ }{\ }{\ }doublet); 3620 keV (\href{https://www.nndc.bnl.gov/nsr/nsrlink.jsp?1975Ol03,B}{1975Ol03}); and 3616 keV (\href{https://www.nndc.bnl.gov/nsr/nsrlink.jsp?1981Ne09,B}{1981Ne09}: Fig. 13 from a\vspace{0.1cm}}&\\
&&&&&&\parbox[t][0.3cm]{11.338501cm}{\raggedright {\ }{\ }{\ }remeasurement of this state using the Princeton spectrograph).\vspace{0.1cm}}&\\
&&&&&&\parbox[t][0.3cm]{11.338501cm}{\raggedright J\ensuremath{^{\pi}}: From (\href{https://www.nndc.bnl.gov/nsr/nsrlink.jsp?1975Ol03,B}{1975Ol03}), who assigned a J\ensuremath{^{\ensuremath{\pi}}}=(2\ensuremath{^{\textnormal{+}}}) to this state based on comparison\vspace{0.1cm}}&\\
&&&&&&\parbox[t][0.3cm]{11.338501cm}{\raggedright {\ }{\ }{\ }of the cross section with a zero-range coupled-channel Born approximation\vspace{0.1cm}}&\\
&&&&&&\parbox[t][0.3cm]{11.338501cm}{\raggedright {\ }{\ }{\ }calculation.\vspace{0.1cm}}&\\
&&&&&&\parbox[t][0.3cm]{11.338501cm}{\raggedright S\ensuremath{_{\textnormal{2n}}}=32.12 MeV (\href{https://www.nndc.bnl.gov/nsr/nsrlink.jsp?1970Fa17,B}{1970Fa17}).\vspace{0.1cm}}&\\
\multicolumn{1}{r@{}}{4522}&\multicolumn{1}{@{}l}{\ensuremath{^{{\hyperlink{NE22LEVEL1}{b}}}} {\it 9}}&\multicolumn{1}{l}{1\ensuremath{^{-}}\ensuremath{^{{\hyperlink{NE22LEVEL3}{d}}{\hyperlink{NE22LEVEL4}{e}}}}}&\multicolumn{1}{r@{}}{9}&\multicolumn{1}{@{ }l}{keV {\it 6}}&\multicolumn{1}{l}{1$^{{\hyperlink{NE22LEVEL6}{g}}}$}&\parbox[t][0.3cm]{11.338501cm}{\raggedright E(level): Weighted average of 4530 keV \textit{20} (\href{https://www.nndc.bnl.gov/nsr/nsrlink.jsp?1970Fa17,B}{1970Fa17}); 4460 keV \textit{50} (\href{https://www.nndc.bnl.gov/nsr/nsrlink.jsp?1970Le08,B}{1970Le08});\vspace{0.1cm}}&\\
&&&&&&\parbox[t][0.3cm]{11.338501cm}{\raggedright {\ }{\ }{\ }and 4522 keV \textit{10} (\href{https://www.nndc.bnl.gov/nsr/nsrlink.jsp?1974Ne04,B}{1974Ne04}, \href{https://www.nndc.bnl.gov/nsr/nsrlink.jsp?1981Ne09,B}{1981Ne09}). See also E\ensuremath{_{\textnormal{x}}}=4576 keV (\href{https://www.nndc.bnl.gov/nsr/nsrlink.jsp?1972Pa02,B}{1972Pa02}: a\vspace{0.1cm}}&\\
&&&&&&\parbox[t][0.3cm]{11.338501cm}{\raggedright {\ }{\ }{\ }(0\ensuremath{^{\textnormal{+}}},1\ensuremath{^{-}}) unresolved doublet with L=0,1); 4560 (\href{https://www.nndc.bnl.gov/nsr/nsrlink.jsp?1975Ol03,B}{1975Ol03}); 4519 keV (\href{https://www.nndc.bnl.gov/nsr/nsrlink.jsp?1981Ne09,B}{1981Ne09}:\vspace{0.1cm}}&\\
&&&&&&\parbox[t][0.3cm]{11.338501cm}{\raggedright {\ }{\ }{\ }remeasured by the Princeton Q3D spectrograph, see Fig. 13); and 4520\vspace{0.1cm}}&\\
&&&&&&\parbox[t][0.3cm]{11.338501cm}{\raggedright {\ }{\ }{\ }(\href{https://www.nndc.bnl.gov/nsr/nsrlink.jsp?1999Pa07,B}{1999Pa07}).\vspace{0.1cm}}&\\
&&&&&&\parbox[t][0.3cm]{11.338501cm}{\raggedright \ensuremath{\Gamma}: From (\href{https://www.nndc.bnl.gov/nsr/nsrlink.jsp?1999Pa07,B}{1999Pa07}). See also \ensuremath{\Gamma}\ensuremath{\leq}20 keV (\href{https://www.nndc.bnl.gov/nsr/nsrlink.jsp?1981Ne09,B}{1981Ne09}), which supersedes \ensuremath{\Gamma}\ensuremath{\leq}40\vspace{0.1cm}}&\\
&&&&&&\parbox[t][0.3cm]{11.338501cm}{\raggedright {\ }{\ }{\ }keV (\href{https://www.nndc.bnl.gov/nsr/nsrlink.jsp?1974Ne04,B}{1974Ne04}).\vspace{0.1cm}}&\\
&&&&&&\parbox[t][0.3cm]{11.338501cm}{\raggedright J\ensuremath{^{\pi}}: From the DWBA analyses of (\href{https://www.nndc.bnl.gov/nsr/nsrlink.jsp?1970Fa17,B}{1970Fa17}) and (\href{https://www.nndc.bnl.gov/nsr/nsrlink.jsp?1972Pa02,B}{1972Pa02}: finite-range DWBA).\vspace{0.1cm}}&\\
&&&&&&\parbox[t][0.3cm]{11.338501cm}{\raggedright (\href{https://www.nndc.bnl.gov/nsr/nsrlink.jsp?1972Pa02,B}{1972Pa02}): the DWBA fit was obtained considering the two transferred neutrons\vspace{0.1cm}}&\\
&&&&&&\parbox[t][0.3cm]{11.338501cm}{\raggedright {\ }{\ }{\ }from the 1\textit{p}\ensuremath{_{\textnormal{1/2}}} and 2\textit{s}\ensuremath{_{\textnormal{1/2}}} sub shells (see Fig. 8).\vspace{0.1cm}}&\\
&&&&&&\parbox[t][0.3cm]{11.338501cm}{\raggedright L: From (\href{https://www.nndc.bnl.gov/nsr/nsrlink.jsp?1970Fa17,B}{1970Fa17}) and (\href{https://www.nndc.bnl.gov/nsr/nsrlink.jsp?1970Le08,B}{1970Le08}), where L=1 was deduced. (\href{https://www.nndc.bnl.gov/nsr/nsrlink.jsp?1970Fa17,B}{1970Fa17}) ruled\vspace{0.1cm}}&\\
&&&&&&\parbox[t][0.3cm]{11.338501cm}{\raggedright {\ }{\ }{\ }out L=0 transfer for population of this state due to an observed minimum at\vspace{0.1cm}}&\\
&&&&&&\parbox[t][0.3cm]{11.338501cm}{\raggedright {\ }{\ }{\ }\ensuremath{\theta}\ensuremath{_{\textnormal{c.m.}}}=25\ensuremath{^\circ} in the triton angular distribution corresponding to this state, which\vspace{0.1cm}}&\\
&&&&&&\parbox[t][0.3cm]{11.338501cm}{\raggedright {\ }{\ }{\ }would have not been seen if this state had J=0. Furthermore, note that the angular\vspace{0.1cm}}&\\
&&&&&&\parbox[t][0.3cm]{11.338501cm}{\raggedright {\ }{\ }{\ }distribution for the tritons corresponding to the population of this state agrees\vspace{0.1cm}}&\\
&&&&&&\parbox[t][0.3cm]{11.338501cm}{\raggedright {\ }{\ }{\ }well with an L=1 transfer (\href{https://www.nndc.bnl.gov/nsr/nsrlink.jsp?1974Ne04,B}{1974Ne04}, see Fig. 2). A minimum is observed at\vspace{0.1cm}}&\\
&&&&&&\parbox[t][0.3cm]{11.338501cm}{\raggedright {\ }{\ }{\ }\ensuremath{\theta}\ensuremath{_{\textnormal{c.m.}}}=25\ensuremath{^\circ}, in agreement with what was observed in (\href{https://www.nndc.bnl.gov/nsr/nsrlink.jsp?1970Fa17,B}{1970Fa17}).\vspace{0.1cm}}&\\
&&&&&&\parbox[t][0.3cm]{11.338501cm}{\raggedright S\ensuremath{_{\textnormal{2n}}}=33.03 MeV (\href{https://www.nndc.bnl.gov/nsr/nsrlink.jsp?1970Fa17,B}{1970Fa17}).\vspace{0.1cm}}&\\
&&&&&&\parbox[t][0.3cm]{11.338501cm}{\raggedright \ensuremath{\sigma}/\ensuremath{\sigma}\ensuremath{_{\textnormal{g.s.}}}(\ensuremath{\theta}\ensuremath{_{\textnormal{lab}}}=30\ensuremath{^\circ})=0.16 (\href{https://www.nndc.bnl.gov/nsr/nsrlink.jsp?1970Le08,B}{1970Le08}) at E\ensuremath{_{\textnormal{p}}}=50 MeV.\vspace{0.1cm}}&\\
\multicolumn{1}{r@{}}{4592}&\multicolumn{1}{@{ }l}{{\it 10}}&\multicolumn{1}{l}{0\ensuremath{^{+}}\ensuremath{^{{\hyperlink{NE22LEVEL3}{d}}{\hyperlink{NE22LEVEL4}{e}}}}}&\multicolumn{1}{r@{}}{2}&\multicolumn{1}{@{ }l}{keV {\it +6\textminus2}}&\multicolumn{1}{l}{0$^{{\hyperlink{NE22LEVEL6}{g}}}$}&\parbox[t][0.3cm]{11.338501cm}{\raggedright T=1\vspace{0.1cm}}&\\
&&&&&&\parbox[t][0.3cm]{11.338501cm}{\raggedright E(level): From (\href{https://www.nndc.bnl.gov/nsr/nsrlink.jsp?1974Ne04,B}{1974Ne04}, \href{https://www.nndc.bnl.gov/nsr/nsrlink.jsp?1981Ne09,B}{1981Ne09}). See also E\ensuremath{_{\textnormal{x}}}=4580 keV \textit{30} (\href{https://www.nndc.bnl.gov/nsr/nsrlink.jsp?1969Ha38,B}{1969Ha38}: an\vspace{0.1cm}}&\\
&&&&&&\parbox[t][0.3cm]{11.338501cm}{\raggedright {\ }{\ }{\ }unresolved state); 4576 keV (\href{https://www.nndc.bnl.gov/nsr/nsrlink.jsp?1972Pa02,B}{1972Pa02}: (0\ensuremath{^{\textnormal{+}}},1\ensuremath{^{-}}) unresolved doublet with L=0,1);\vspace{0.1cm}}&\\
&&&&&&\parbox[t][0.3cm]{11.338501cm}{\raggedright {\ }{\ }{\ }5120 keV \textit{50} (\href{https://www.nndc.bnl.gov/nsr/nsrlink.jsp?1970Le08,B}{1970Le08}: an unresolved doublet with J\ensuremath{^{\ensuremath{\pi}}}=0\ensuremath{^{\textnormal{+}}},2\ensuremath{^{\textnormal{+}}}); 4590\vspace{0.1cm}}&\\
&&&&&&\parbox[t][0.3cm]{11.338501cm}{\raggedright {\ }{\ }{\ }(\href{https://www.nndc.bnl.gov/nsr/nsrlink.jsp?1981Ne09,B}{1981Ne09}: see Fig. 13, from the remeasurement of this state using the Princeton\vspace{0.1cm}}&\\
&&&&&&\parbox[t][0.3cm]{11.338501cm}{\raggedright {\ }{\ }{\ }Q3D spectrograph); and 4589 keV (\href{https://www.nndc.bnl.gov/nsr/nsrlink.jsp?1999Pa07,B}{1999Pa07}).\vspace{0.1cm}}&\\
&&&&&&\parbox[t][0.3cm]{11.338501cm}{\raggedright \ensuremath{\Gamma}: From (\href{https://www.nndc.bnl.gov/nsr/nsrlink.jsp?1999Pa07,B}{1999Pa07}), which reported \ensuremath{\Gamma}=2 keV \textit{6}. The uncertainty was changed to\vspace{0.1cm}}&\\
&&&&&&\parbox[t][0.3cm]{11.338501cm}{\raggedright {\ }{\ }{\ }\textit{+6{\textminus}2} keV by the evaluator to avoid having a negative width. See also \ensuremath{\Gamma}\ensuremath{\leq}20 keV\vspace{0.1cm}}&\\
&&&&&&\parbox[t][0.3cm]{11.338501cm}{\raggedright {\ }{\ }{\ }(\href{https://www.nndc.bnl.gov/nsr/nsrlink.jsp?1981Ne09,B}{1981Ne09}), which supersedes \ensuremath{\Gamma}\ensuremath{\leq}40 keV (\href{https://www.nndc.bnl.gov/nsr/nsrlink.jsp?1974Ne04,B}{1974Ne04}).\vspace{0.1cm}}&\\
&&&&&&\parbox[t][0.3cm]{11.338501cm}{\raggedright J\ensuremath{^{\pi}},L: From (\href{https://www.nndc.bnl.gov/nsr/nsrlink.jsp?1974Ne04,B}{1974Ne04}): triton angular distribution corresponding to the 4.59-MeV\vspace{0.1cm}}&\\
&&&&&&\parbox[t][0.3cm]{11.338501cm}{\raggedright {\ }{\ }{\ }state has the characteristics of L=0 in that a mimimum occurs at \ensuremath{\theta}\ensuremath{_{\textnormal{c.m.}}}\ensuremath{<}25\ensuremath{^\circ}.\vspace{0.1cm}}&\\
&&&&&&\parbox[t][0.3cm]{11.338501cm}{\raggedright A predominantly \textit{s}\ensuremath{^{\textnormal{2}}_{\textnormal{1/2}}} configuration was assigned to this state based on its large\vspace{0.1cm}}&\\
&&&&&&\parbox[t][0.3cm]{11.338501cm}{\raggedright {\ }{\ }{\ }downward shift with respect to the analog state in \ensuremath{^{\textnormal{18}}}O (\href{https://www.nndc.bnl.gov/nsr/nsrlink.jsp?1974Ne04,B}{1974Ne04}).\vspace{0.1cm}}&\\
\multicolumn{1}{r@{}}{5101}&\multicolumn{1}{@{ }l}{{\it 9}}&\multicolumn{1}{l}{2\ensuremath{^{+}}}&\multicolumn{1}{r@{}}{46}&\multicolumn{1}{@{ }l}{keV {\it 3}}&\multicolumn{1}{l}{2}&\parbox[t][0.3cm]{11.338501cm}{\raggedright T=1\vspace{0.1cm}}&\\
&&&&&&\parbox[t][0.3cm]{11.338501cm}{\raggedright E(level): Weighted average of 5115 keV \textit{25} (\href{https://www.nndc.bnl.gov/nsr/nsrlink.jsp?1969Ha38,B}{1969Ha38}); and 5099 keV \textit{10}\vspace{0.1cm}}&\\
&&&&&&\parbox[t][0.3cm]{11.338501cm}{\raggedright {\ }{\ }{\ }(\href{https://www.nndc.bnl.gov/nsr/nsrlink.jsp?1981Ne09,B}{1981Ne09}), which supersedes the 5095 keV \textit{15} result by (\href{https://www.nndc.bnl.gov/nsr/nsrlink.jsp?1974Ne04,B}{1974Ne04}). See also\vspace{0.1cm}}&\\
&&&&&&\parbox[t][0.3cm]{11.338501cm}{\raggedright {\ }{\ }{\ }E\ensuremath{_{\textnormal{x}}}=5100 keV \textit{20} (\href{https://www.nndc.bnl.gov/nsr/nsrlink.jsp?1970Fa17,B}{1970Fa17}: an unresolved doublet with J\ensuremath{^{\ensuremath{\pi}}}=2\ensuremath{^{\textnormal{+}}}, 3\ensuremath{^{-}} members);\vspace{0.1cm}}&\\
&&&&&&\parbox[t][0.3cm]{11.338501cm}{\raggedright {\ }{\ }{\ }5085 keV (\href{https://www.nndc.bnl.gov/nsr/nsrlink.jsp?1981Ne09,B}{1981Ne09}: see Fig. 13, remeasurement by the Princeton Q3D\vspace{0.1cm}}&\\
&&&&&&\parbox[t][0.3cm]{11.338501cm}{\raggedright {\ }{\ }{\ }spectrograph); 5095 keV (\href{https://www.nndc.bnl.gov/nsr/nsrlink.jsp?1996Ha26,B}{1996Ha26}: used as energy calibration); and 5106 keV\vspace{0.1cm}}&\\
&&&&&&\parbox[t][0.3cm]{11.338501cm}{\raggedright {\ }{\ }{\ }(\href{https://www.nndc.bnl.gov/nsr/nsrlink.jsp?1999Pa07,B}{1999Pa07}).\vspace{0.1cm}}&\\
&&&&&&\parbox[t][0.3cm]{11.338501cm}{\raggedright \ensuremath{\Gamma}: Weighted average of 49 keV \textit{6} (\href{https://www.nndc.bnl.gov/nsr/nsrlink.jsp?1996Ha26,B}{1996Ha26}: the (p,t) reaction using the K600\vspace{0.1cm}}&\\
&&&&&&\parbox[t][0.3cm]{11.338501cm}{\raggedright {\ }{\ }{\ }spectrograph at the Indiana University); 45 keV \textit{5} (\href{https://www.nndc.bnl.gov/nsr/nsrlink.jsp?1996Ha26,B}{1996Ha26}: the (p,t) reaction\vspace{0.1cm}}&\\
&&&&&&\parbox[t][0.3cm]{11.338501cm}{\raggedright {\ }{\ }{\ }study using the Q3D spectrograph at Princeton University); 40 keV \textit{20}\vspace{0.1cm}}&\\
&&&&&&\parbox[t][0.3cm]{11.338501cm}{\raggedright {\ }{\ }{\ }(\href{https://www.nndc.bnl.gov/nsr/nsrlink.jsp?1981Ne09,B}{1981Ne09}: from the remeasurement of this state using the Princeton Q3D\vspace{0.1cm}}&\\
&&&&&&\parbox[t][0.3cm]{11.338501cm}{\raggedright {\ }{\ }{\ }spectrograph); and 45 keV \textit{7} (\href{https://www.nndc.bnl.gov/nsr/nsrlink.jsp?1999Pa07,B}{1999Pa07}). See also \ensuremath{\Gamma}\ensuremath{\leq}80 keV (\href{https://www.nndc.bnl.gov/nsr/nsrlink.jsp?1974Ne04,B}{1974Ne04}).\vspace{0.1cm}}&\\
&&&&&&\parbox[t][0.3cm]{11.338501cm}{\raggedright J\ensuremath{^{\pi}},L: From the DWBA analysis of (\href{https://www.nndc.bnl.gov/nsr/nsrlink.jsp?1999Pa07,B}{1999Pa07}). See also (\href{https://www.nndc.bnl.gov/nsr/nsrlink.jsp?1970Le08,B}{1970Le08}), where the\vspace{0.1cm}}&\\
\end{longtable}
\begin{textblock}{29}(0,27.3)
Continued on next page (footnotes at end of table)
\end{textblock}
\clearpage
\begin{longtable}{ccccccc@{\extracolsep{\fill}}c}
\\[-.4cm]
\multicolumn{8}{c}{{\bf \small \underline{\ensuremath{^{\textnormal{20}}}Ne(p,t)\hspace{0.2in}\href{https://www.nndc.bnl.gov/nsr/nsrlink.jsp?1969Ha38,B}{1969Ha38},\href{https://www.nndc.bnl.gov/nsr/nsrlink.jsp?2017Ch32,B}{2017Ch32} (continued)}}}\\
\multicolumn{8}{c}{~}\\
\multicolumn{8}{c}{\underline{\ensuremath{^{18}}Ne Levels (continued)}}\\
\multicolumn{8}{c}{~}\\
\multicolumn{2}{c}{E(level)$^{}$}&J$^{\pi}$$^{{\hyperlink{NE22LEVEL5}{f}}}$&\multicolumn{2}{c}{\ensuremath{\Gamma} (keV)$^{{\hyperlink{NE22LEVEL2}{c}}}$}&L$^{}$&Comments&\\[-.2cm]
\multicolumn{2}{c}{\hrulefill}&\hrulefill&\multicolumn{2}{c}{\hrulefill}&\hrulefill&\hrulefill&
\endhead
&&&&&&\parbox[t][0.3cm]{11.84691cm}{\raggedright {\ }{\ }{\ }triton angular distribution corresponding to the population of this state could be fitted\vspace{0.1cm}}&\\
&&&&&&\parbox[t][0.3cm]{11.84691cm}{\raggedright {\ }{\ }{\ }equally well with L=0, L=2, and L=3 momentum transfers; and (\href{https://www.nndc.bnl.gov/nsr/nsrlink.jsp?1970Fa17,B}{1970Fa17}), where\vspace{0.1cm}}&\\
&&&&&&\parbox[t][0.3cm]{11.84691cm}{\raggedright {\ }{\ }{\ }the triton angular distribution was best fitted with a sum of L=2+3.\vspace{0.1cm}}&\\
\multicolumn{1}{r@{}}{5151}&\multicolumn{1}{@{}l}{\ensuremath{^{{\hyperlink{NE22LEVEL0}{a}}}} {\it 8}}&\multicolumn{1}{l}{3\ensuremath{^{-}}}&\multicolumn{1}{r@{}}{10}&\multicolumn{1}{@{ }l}{keV {\it 5}}&\multicolumn{1}{l}{3}&\parbox[t][0.3cm]{11.84691cm}{\raggedright T=1\vspace{0.1cm}}&\\
&&&&&&\parbox[t][0.3cm]{11.84691cm}{\raggedright E(level): Weighted average of 5150 keV \textit{14} (\href{https://www.nndc.bnl.gov/nsr/nsrlink.jsp?1972Pa02,B}{1972Pa02}); and 5151 keV {10} (\href{https://www.nndc.bnl.gov/nsr/nsrlink.jsp?1981Ne09,B}{1981Ne09}),\vspace{0.1cm}}&\\
&&&&&&\parbox[t][0.3cm]{11.84691cm}{\raggedright {\ }{\ }{\ }which supersedes the 5149 keV \textit{5} (\href{https://www.nndc.bnl.gov/nsr/nsrlink.jsp?1974Ne04,B}{1974Ne04}). See also E\ensuremath{_{\textnormal{x}}}=5120 keV \textit{50} (\href{https://www.nndc.bnl.gov/nsr/nsrlink.jsp?1970Le08,B}{1970Le08}:\vspace{0.1cm}}&\\
&&&&&&\parbox[t][0.3cm]{11.84691cm}{\raggedright {\ }{\ }{\ }this level was believed to be composed of more than one states); 5140 (\href{https://www.nndc.bnl.gov/nsr/nsrlink.jsp?1975Ol03,B}{1975Ol03}:\vspace{0.1cm}}&\\
&&&&&&\parbox[t][0.3cm]{11.84691cm}{\raggedright {\ }{\ }{\ }see Fig. 1); 5140 keV (\href{https://www.nndc.bnl.gov/nsr/nsrlink.jsp?1981Ne09,B}{1981Ne09}: see Fig. 13 from remeasurement using the\vspace{0.1cm}}&\\
&&&&&&\parbox[t][0.3cm]{11.84691cm}{\raggedright {\ }{\ }{\ }Princeton Q3D spectrograph); 5150 keV (\href{https://www.nndc.bnl.gov/nsr/nsrlink.jsp?1996Ha26,B}{1996Ha26}: used as energy calibration); and\vspace{0.1cm}}&\\
&&&&&&\parbox[t][0.3cm]{11.84691cm}{\raggedright {\ }{\ }{\ }5153 keV (\href{https://www.nndc.bnl.gov/nsr/nsrlink.jsp?1999Pa07,B}{1999Pa07}).\vspace{0.1cm}}&\\
&&&&&&\parbox[t][0.3cm]{11.84691cm}{\raggedright (\href{https://www.nndc.bnl.gov/nsr/nsrlink.jsp?1972Pa02,B}{1972Pa02}) paired the level observed at 5150 keV \textit{14} to the level in \ensuremath{^{\textnormal{18}}}Ne with\vspace{0.1cm}}&\\
&&&&&&\parbox[t][0.3cm]{11.84691cm}{\raggedright {\ }{\ }{\ }E\ensuremath{_{\textnormal{x}}}=5120 keV reported in (\href{https://www.nndc.bnl.gov/nsr/nsrlink.jsp?1970Le08,B}{1970Le08}, \href{https://www.nndc.bnl.gov/nsr/nsrlink.jsp?1970Ad02,B}{1970Ad02}).\vspace{0.1cm}}&\\
&&&&&&\parbox[t][0.3cm]{11.84691cm}{\raggedright \ensuremath{\Gamma}: Weighted average of 25 keV \textit{15} (\href{https://www.nndc.bnl.gov/nsr/nsrlink.jsp?1981Ne09,B}{1981Ne09}); and 8 keV \textit{5} (\href{https://www.nndc.bnl.gov/nsr/nsrlink.jsp?1999Pa07,B}{1999Pa07}). See also\vspace{0.1cm}}&\\
&&&&&&\parbox[t][0.3cm]{11.84691cm}{\raggedright {\ }{\ }{\ }\ensuremath{\Gamma}\ensuremath{\leq}50 keV (\href{https://www.nndc.bnl.gov/nsr/nsrlink.jsp?1974Ne04,B}{1974Ne04}); \ensuremath{\Gamma}\ensuremath{\leq}20 keV (\href{https://www.nndc.bnl.gov/nsr/nsrlink.jsp?1996Ha26,B}{1996Ha26}: from the (p,t) study using the K600\vspace{0.1cm}}&\\
&&&&&&\parbox[t][0.3cm]{11.84691cm}{\raggedright {\ }{\ }{\ }spectrograph in Indiana University, see Table V); and \ensuremath{\Gamma}\ensuremath{\leq}15 (\href{https://www.nndc.bnl.gov/nsr/nsrlink.jsp?1996Ha26,B}{1996Ha26}: from the\vspace{0.1cm}}&\\
&&&&&&\parbox[t][0.3cm]{11.84691cm}{\raggedright {\ }{\ }{\ }(p,t) study using the Princeton Q3D spectrograph, see Table V).\vspace{0.1cm}}&\\
&&&&&&\parbox[t][0.3cm]{11.84691cm}{\raggedright J\ensuremath{^{\pi}},L: From the DWBA analysis by (\href{https://www.nndc.bnl.gov/nsr/nsrlink.jsp?1999Pa07,B}{1999Pa07}). See also (\href{https://www.nndc.bnl.gov/nsr/nsrlink.jsp?1970Le08,B}{1970Le08}), where \ensuremath{\sigma}(E\ensuremath{_{\textnormal{t}}},\ensuremath{\theta})\vspace{0.1cm}}&\\
&&&&&&\parbox[t][0.3cm]{11.84691cm}{\raggedright {\ }{\ }{\ }could be fitted, using finite-range DWBA calculations, equally well with L=0, L=2,\vspace{0.1cm}}&\\
&&&&&&\parbox[t][0.3cm]{11.84691cm}{\raggedright {\ }{\ }{\ }and L=3 momentum transfers.\vspace{0.1cm}}&\\
&&&&&&\parbox[t][0.3cm]{11.84691cm}{\raggedright \ensuremath{\sigma}/\ensuremath{\sigma}\ensuremath{_{\textnormal{g.s.}}}(\ensuremath{\theta}\ensuremath{_{\textnormal{lab}}}=30\ensuremath{^\circ})=0.24 (\href{https://www.nndc.bnl.gov/nsr/nsrlink.jsp?1970Le08,B}{1970Le08}) at E\ensuremath{_{\textnormal{p}}}=50 MeV.\vspace{0.1cm}}&\\
\multicolumn{1}{r@{}}{5464}&\multicolumn{1}{@{ }l}{{\it 6}}&&\multicolumn{1}{r@{}}{6}&\multicolumn{1}{@{ }l}{keV {\it 6}}&&\parbox[t][0.3cm]{11.84691cm}{\raggedright E(level): Weighted average (with external errors) of 5453 keV \textit{10} (\href{https://www.nndc.bnl.gov/nsr/nsrlink.jsp?1974Ne04,B}{1974Ne04},\vspace{0.1cm}}&\\
&&&&&&\parbox[t][0.3cm]{11.84691cm}{\raggedright {\ }{\ }{\ }\href{https://www.nndc.bnl.gov/nsr/nsrlink.jsp?1981Ne09,B}{1981Ne09}) and 5467 keV \textit{5} (\href{https://www.nndc.bnl.gov/nsr/nsrlink.jsp?1999Pa07,B}{1999Pa07}). See also E\ensuremath{_{\textnormal{x}}}\ensuremath{\approx}5360 keV (\href{https://www.nndc.bnl.gov/nsr/nsrlink.jsp?1970Fa17,B}{1970Fa17}: very\vspace{0.1cm}}&\\
&&&&&&\parbox[t][0.3cm]{11.84691cm}{\raggedright {\ }{\ }{\ }weakly populated state).\vspace{0.1cm}}&\\
&&&&&&\parbox[t][0.3cm]{11.84691cm}{\raggedright \ensuremath{\Gamma}: From (\href{https://www.nndc.bnl.gov/nsr/nsrlink.jsp?1999Pa07,B}{1999Pa07}). See also \ensuremath{\Gamma}\ensuremath{\leq}50 keV (\href{https://www.nndc.bnl.gov/nsr/nsrlink.jsp?1974Ne04,B}{1974Ne04}, \href{https://www.nndc.bnl.gov/nsr/nsrlink.jsp?1981Ne09,B}{1981Ne09}).\vspace{0.1cm}}&\\
\multicolumn{1}{r@{}}{6302}&\multicolumn{1}{@{}l}{\ensuremath{^{{\hyperlink{NE22LEVEL0}{a}}}} {\it 4}}&\multicolumn{1}{l}{(4\ensuremath{^{+}})}&\multicolumn{1}{r@{}}{8}&\multicolumn{1}{@{ }l}{keV {\it 7}}&\multicolumn{1}{l}{(4)}&\parbox[t][0.3cm]{11.84691cm}{\raggedright E(level): Weighted average (with external errors) of 6280 keV \textit{20} (\href{https://www.nndc.bnl.gov/nsr/nsrlink.jsp?1970Fa17,B}{1970Fa17}); 6326\vspace{0.1cm}}&\\
&&&&&&\parbox[t][0.3cm]{11.84691cm}{\raggedright {\ }{\ }{\ }keV \textit{18} (\href{https://www.nndc.bnl.gov/nsr/nsrlink.jsp?1972Pa02,B}{1972Pa02}); 6297 keV \textit{10} (\href{https://www.nndc.bnl.gov/nsr/nsrlink.jsp?1974Ne04,B}{1974Ne04}, \href{https://www.nndc.bnl.gov/nsr/nsrlink.jsp?1981Ne09,B}{1981Ne09}); 6286 keV \textit{10} (\href{https://www.nndc.bnl.gov/nsr/nsrlink.jsp?1996Ha26,B}{1996Ha26}:\vspace{0.1cm}}&\\
&&&&&&\parbox[t][0.3cm]{11.84691cm}{\raggedright {\ }{\ }{\ }from the (p,t) study using the K600 spectrograph at Indiana University); and 6305\vspace{0.1cm}}&\\
&&&&&&\parbox[t][0.3cm]{11.84691cm}{\raggedright {\ }{\ }{\ }keV \textit{4} (\href{https://www.nndc.bnl.gov/nsr/nsrlink.jsp?1999Pa07,B}{1999Pa07}). See also 6340 keV \textit{50} (\href{https://www.nndc.bnl.gov/nsr/nsrlink.jsp?1970Le08,B}{1970Le08}: thought to be composed of\vspace{0.1cm}}&\\
&&&&&&\parbox[t][0.3cm]{11.84691cm}{\raggedright {\ }{\ }{\ }more than one states).\vspace{0.1cm}}&\\
&&&&&&\parbox[t][0.3cm]{11.84691cm}{\raggedright \ensuremath{\Gamma}: From (\href{https://www.nndc.bnl.gov/nsr/nsrlink.jsp?1999Pa07,B}{1999Pa07}). See also \ensuremath{\Gamma}\ensuremath{\leq}60 keV (\href{https://www.nndc.bnl.gov/nsr/nsrlink.jsp?1974Ne04,B}{1974Ne04}, \href{https://www.nndc.bnl.gov/nsr/nsrlink.jsp?1981Ne09,B}{1981Ne09}); and \ensuremath{\Gamma}\ensuremath{\leq}20 keV\vspace{0.1cm}}&\\
&&&&&&\parbox[t][0.3cm]{11.84691cm}{\raggedright {\ }{\ }{\ }(\href{https://www.nndc.bnl.gov/nsr/nsrlink.jsp?1996Ha26,B}{1996Ha26}: from the (p,t) reaction study using the K600 spectrograph at Indiana\vspace{0.1cm}}&\\
&&&&&&\parbox[t][0.3cm]{11.84691cm}{\raggedright {\ }{\ }{\ }University).\vspace{0.1cm}}&\\
&&&&&&\parbox[t][0.3cm]{11.84691cm}{\raggedright J\ensuremath{^{\pi}},L: From (\href{https://www.nndc.bnl.gov/nsr/nsrlink.jsp?1970Fa17,B}{1970Fa17}: deduced from a finite-range DWBA analysis; however, the\vspace{0.1cm}}&\\
&&&&&&\parbox[t][0.3cm]{11.84691cm}{\raggedright {\ }{\ }{\ }DWBA curve does not really describe the data well. Therefore, the evaluator made\vspace{0.1cm}}&\\
&&&&&&\parbox[t][0.3cm]{11.84691cm}{\raggedright {\ }{\ }{\ }the L value tentative).\vspace{0.1cm}}&\\
\multicolumn{1}{r@{}}{6355}&\multicolumn{1}{@{ }l}{{\it 4}}&&\multicolumn{1}{r@{}}{30}&\multicolumn{1}{@{ }l}{keV {\it 13}}&&\parbox[t][0.3cm]{11.84691cm}{\raggedright E(level): Weighted average of 6353 keV \textit{10} (\href{https://www.nndc.bnl.gov/nsr/nsrlink.jsp?1974Ne04,B}{1974Ne04}, \href{https://www.nndc.bnl.gov/nsr/nsrlink.jsp?1981Ne09,B}{1981Ne09}); 6343 keV \textit{20}\vspace{0.1cm}}&\\
&&&&&&\parbox[t][0.3cm]{11.84691cm}{\raggedright {\ }{\ }{\ }(\href{https://www.nndc.bnl.gov/nsr/nsrlink.jsp?1996Ha26,B}{1996Ha26}: measured at \ensuremath{\theta}\ensuremath{_{\textnormal{lab}}}=6\ensuremath{^\circ} using the K600 spectrograph at Indiana\vspace{0.1cm}}&\\
&&&&&&\parbox[t][0.3cm]{11.84691cm}{\raggedright {\ }{\ }{\ }University); 6346 keV {10} (\href{https://www.nndc.bnl.gov/nsr/nsrlink.jsp?1996Ha26,B}{1996Ha26}: measured at \ensuremath{\theta}\ensuremath{_{\textnormal{lab}}}=11\ensuremath{^\circ} using the K600\vspace{0.1cm}}&\\
&&&&&&\parbox[t][0.3cm]{11.84691cm}{\raggedright {\ }{\ }{\ }spectrograph at Indiana University); and 6358 keV \textit{5} (\href{https://www.nndc.bnl.gov/nsr/nsrlink.jsp?1999Pa07,B}{1999Pa07}). See also 6340 keV\vspace{0.1cm}}&\\
&&&&&&\parbox[t][0.3cm]{11.84691cm}{\raggedright {\ }{\ }{\ }\textit{50} (\href{https://www.nndc.bnl.gov/nsr/nsrlink.jsp?1970Le08,B}{1970Le08}: this level was believed to be composed of more than one states).\vspace{0.1cm}}&\\
&&&&&&\parbox[t][0.3cm]{11.84691cm}{\raggedright \ensuremath{\Gamma}: Weighted average of 45 keV \textit{10} (\href{https://www.nndc.bnl.gov/nsr/nsrlink.jsp?1996Ha26,B}{1996Ha26}: from the (p,t) reaction study using the\vspace{0.1cm}}&\\
&&&&&&\parbox[t][0.3cm]{11.84691cm}{\raggedright {\ }{\ }{\ }K600 spectrograph at Indiana University); and 18 keV \textit{9} (\href{https://www.nndc.bnl.gov/nsr/nsrlink.jsp?1999Pa07,B}{1999Pa07}). See also \ensuremath{\Gamma}\ensuremath{\leq}60\vspace{0.1cm}}&\\
&&&&&&\parbox[t][0.3cm]{11.84691cm}{\raggedright {\ }{\ }{\ }keV (\href{https://www.nndc.bnl.gov/nsr/nsrlink.jsp?1974Ne04,B}{1974Ne04}, \href{https://www.nndc.bnl.gov/nsr/nsrlink.jsp?1981Ne09,B}{1981Ne09}).\vspace{0.1cm}}&\\
&&&&&&\parbox[t][0.3cm]{11.84691cm}{\raggedright \ensuremath{\sigma}/\ensuremath{\sigma}\ensuremath{_{\textnormal{g.s.}}}(\ensuremath{\theta}\ensuremath{_{\textnormal{lab}}}=30\ensuremath{^\circ})=0.06 (\href{https://www.nndc.bnl.gov/nsr/nsrlink.jsp?1970Le08,B}{1970Le08}) at E\ensuremath{_{\textnormal{p}}}=50 MeV.\vspace{0.1cm}}&\\
\multicolumn{1}{r@{}}{7713}&\multicolumn{1}{@{ }l}{{\it 10}}&&\multicolumn{1}{r@{}}{$\leq$60}&\multicolumn{1}{@{ }l}{keV}&&\parbox[t][0.3cm]{11.84691cm}{\raggedright E(level),\ensuremath{\Gamma}: From (\href{https://www.nndc.bnl.gov/nsr/nsrlink.jsp?1974Ne04,B}{1974Ne04}, \href{https://www.nndc.bnl.gov/nsr/nsrlink.jsp?1981Ne09,B}{1981Ne09}).\vspace{0.1cm}}&\\
\multicolumn{1}{r@{}}{7942}&\multicolumn{1}{@{ }l}{{\it 8}}&&\multicolumn{1}{r@{}}{70}&\multicolumn{1}{@{ }l}{keV {\it 20}}&&\parbox[t][0.3cm]{11.84691cm}{\raggedright E(level): Weighted average of 7957 keV \textit{25} (\href{https://www.nndc.bnl.gov/nsr/nsrlink.jsp?1972Pa02,B}{1972Pa02}); 7949 keV \textit{10} (\href{https://www.nndc.bnl.gov/nsr/nsrlink.jsp?1974Ne04,B}{1974Ne04},\vspace{0.1cm}}&\\
&&&&&&\parbox[t][0.3cm]{11.84691cm}{\raggedright {\ }{\ }{\ }\href{https://www.nndc.bnl.gov/nsr/nsrlink.jsp?1981Ne09,B}{1981Ne09}); 7924 keV \textit{20} (\href{https://www.nndc.bnl.gov/nsr/nsrlink.jsp?1996Ha26,B}{1996Ha26}: measured at \ensuremath{\theta}\ensuremath{_{\textnormal{lab}}}=6\ensuremath{^\circ} using the K600\vspace{0.1cm}}&\\
&&&&&&\parbox[t][0.3cm]{11.84691cm}{\raggedright {\ }{\ }{\ }spectrograph at Indiana University); and 7920 keV \textit{20} (\href{https://www.nndc.bnl.gov/nsr/nsrlink.jsp?1996Ha26,B}{1996Ha26}: measured at\vspace{0.1cm}}&\\
&&&&&&\parbox[t][0.3cm]{11.84691cm}{\raggedright {\ }{\ }{\ }\ensuremath{\theta}\ensuremath{_{\textnormal{lab}}}=11\ensuremath{^\circ} using the K600 spectrograph in Indiana University). See also, 7920 keV \textit{20}\vspace{0.1cm}}&\\
&&&&&&\parbox[t][0.3cm]{11.84691cm}{\raggedright {\ }{\ }{\ }(\href{https://www.nndc.bnl.gov/nsr/nsrlink.jsp?1996Ha26,B}{1996Ha26}: Table V).\vspace{0.1cm}}&\\
&&&&&&\parbox[t][0.3cm]{11.84691cm}{\raggedright \ensuremath{\Gamma}: From (\href{https://www.nndc.bnl.gov/nsr/nsrlink.jsp?1996Ha26,B}{1996Ha26}: from the (p,t) reaction study using the K600 spectrograph at\vspace{0.1cm}}&\\
&&&&&&\parbox[t][0.3cm]{11.84691cm}{\raggedright {\ }{\ }{\ }Indiana University). See also \ensuremath{\Gamma}\ensuremath{\leq}60 keV (\href{https://www.nndc.bnl.gov/nsr/nsrlink.jsp?1974Ne04,B}{1974Ne04}, \href{https://www.nndc.bnl.gov/nsr/nsrlink.jsp?1981Ne09,B}{1981Ne09}).\vspace{0.1cm}}&\\
\multicolumn{1}{r@{}}{9199}&\multicolumn{1}{@{ }l}{{\it 9}}&&\multicolumn{1}{r@{}}{$\leq$60}&\multicolumn{1}{@{ }l}{keV}&&\parbox[t][0.3cm]{11.84691cm}{\raggedright E(level): Weighted average of 9170 keV \textit{30} (\href{https://www.nndc.bnl.gov/nsr/nsrlink.jsp?1970Fa17,B}{1970Fa17}); 9215 keV \textit{20} (\href{https://www.nndc.bnl.gov/nsr/nsrlink.jsp?1972Pa02,B}{1972Pa02});\vspace{0.1cm}}&\\
&&&&&&\parbox[t][0.3cm]{11.84691cm}{\raggedright {\ }{\ }{\ }and 9198 keV \textit{10} (\href{https://www.nndc.bnl.gov/nsr/nsrlink.jsp?1974Ne04,B}{1974Ne04}, \href{https://www.nndc.bnl.gov/nsr/nsrlink.jsp?1981Ne09,B}{1981Ne09}).\vspace{0.1cm}}&\\
&&&&&&\parbox[t][0.3cm]{11.84691cm}{\raggedright \ensuremath{\Gamma}: From (\href{https://www.nndc.bnl.gov/nsr/nsrlink.jsp?1981Ne09,B}{1981Ne09}), which supersedes \ensuremath{\Gamma}\ensuremath{\leq}50 keV (\href{https://www.nndc.bnl.gov/nsr/nsrlink.jsp?1974Ne04,B}{1974Ne04}).\vspace{0.1cm}}&\\
\end{longtable}
\begin{textblock}{29}(0,27.3)
Continued on next page (footnotes at end of table)
\end{textblock}
\clearpage
\vspace*{-0.5cm}
{\bf \small \underline{\ensuremath{^{\textnormal{20}}}Ne(p,t)\hspace{0.2in}\href{https://www.nndc.bnl.gov/nsr/nsrlink.jsp?1969Ha38,B}{1969Ha38},\href{https://www.nndc.bnl.gov/nsr/nsrlink.jsp?2017Ch32,B}{2017Ch32} (continued)}}\\
\vspace{0.3cm}
\underline{$^{18}$Ne Levels (continued)}\\
\vspace{0.3cm}
\parbox[b][0.3cm]{17.7cm}{\makebox[1ex]{\ensuremath{^{\hypertarget{NE22LEVEL0}{a}}}} This level was strongly populated in (\href{https://www.nndc.bnl.gov/nsr/nsrlink.jsp?1972Pa02,B}{1972Pa02}), and \ensuremath{\sigma}(E\ensuremath{_{\textnormal{t}}},\ensuremath{\theta}), where t=triton, was measured at \ensuremath{\theta}\ensuremath{_{\textnormal{c.m.}}}\ensuremath{\sim}10\ensuremath{^\circ}{\textminus}120\ensuremath{^\circ}.}\\
\parbox[b][0.3cm]{17.7cm}{\makebox[1ex]{\ensuremath{^{\hypertarget{NE22LEVEL1}{b}}}} (\href{https://www.nndc.bnl.gov/nsr/nsrlink.jsp?1981Ne09,B}{1981Ne09}): this state was remeasured by (F. P. Calaprice, S. J. Freedman, and A. V. Nero, private communication) using the}\\
\parbox[b][0.3cm]{17.7cm}{{\ }{\ }\ensuremath{^{\textnormal{20}}}Ne(p,t) reaction with a 41.8 MeV proton beam. In this additional experiment, tritons were measured using the high resolution}\\
\parbox[b][0.3cm]{17.7cm}{{\ }{\ }Princeton Q3D spectrograph. As a result, the constituent states in the 4.5-MeV and 5.1-MeV doublets, as well as those of the}\\
\parbox[b][0.3cm]{17.7cm}{{\ }{\ }triplet around 3.5 MeV were resolved (\href{https://www.nndc.bnl.gov/nsr/nsrlink.jsp?1981Ne09,B}{1981Ne09}: Fig. 13). The widths of the members of the doublet at 5.1 MeV reported by}\\
\parbox[b][0.3cm]{17.7cm}{{\ }{\ }(\href{https://www.nndc.bnl.gov/nsr/nsrlink.jsp?1981Ne09,B}{1981Ne09}) are from this additional measurement, which is published together with (\href{https://www.nndc.bnl.gov/nsr/nsrlink.jsp?1981Ne09,B}{1981Ne09}).}\\
\parbox[b][0.3cm]{17.7cm}{\makebox[1ex]{\ensuremath{^{\hypertarget{NE22LEVEL2}{c}}}} (\href{https://www.nndc.bnl.gov/nsr/nsrlink.jsp?1999Pa07,B}{1999Pa07}): the instrumental width was extracted from a particle bound state in \ensuremath{^{\textnormal{18}}}Ne at E\ensuremath{_{\textnormal{x}}}=3616 keV. It was assumed that the}\\
\parbox[b][0.3cm]{17.7cm}{{\ }{\ }instrumental spreads are Gaussian distributions, but the proton resonance structures are Lorentzian distributions. Hence, these two}\\
\parbox[b][0.3cm]{17.7cm}{{\ }{\ }different functions were convoluted for each state, in order to deduce their intrinsic widths.}\\
\parbox[b][0.3cm]{17.7cm}{\makebox[1ex]{\ensuremath{^{\hypertarget{NE22LEVEL3}{d}}}} See similar results in (\href{https://www.nndc.bnl.gov/nsr/nsrlink.jsp?1974Ne04,B}{1974Ne04}), where the triton angular distributions were analyzed without a DWBA analysis. These authors}\\
\parbox[b][0.3cm]{17.7cm}{{\ }{\ }assigned J\ensuremath{^{\ensuremath{\pi}}} values and inferred the transferred angular momentum, L, based on arguments on whether or not the shape of}\\
\parbox[b][0.3cm]{17.7cm}{{\ }{\ }distributions have the characteristics of a certain angular momentum transfer. These educated guesses were guided by the results}\\
\parbox[b][0.3cm]{17.7cm}{{\ }{\ }of the previous studies of these states.}\\
\parbox[b][0.3cm]{17.7cm}{\makebox[1ex]{\ensuremath{^{\hypertarget{NE22LEVEL4}{e}}}} See also (\href{https://www.nndc.bnl.gov/nsr/nsrlink.jsp?1981Ne09,B}{1981Ne09}), where similar J\ensuremath{^{\ensuremath{\pi}}} was deduced through comparison with the known states, mirror level and two-nucleon}\\
\parbox[b][0.3cm]{17.7cm}{{\ }{\ }Coulomb shift analysis.}\\
\parbox[b][0.3cm]{17.7cm}{\makebox[1ex]{\ensuremath{^{\hypertarget{NE22LEVEL5}{f}}}} (\href{https://www.nndc.bnl.gov/nsr/nsrlink.jsp?1970Le08,B}{1970Le08}) specified that to deduce J from L, they assumed J=L+S (see footnote (a) under Table IV).}\\
\parbox[b][0.3cm]{17.7cm}{\makebox[1ex]{\ensuremath{^{\hypertarget{NE22LEVEL6}{g}}}} See also (\href{https://www.nndc.bnl.gov/nsr/nsrlink.jsp?1981Ne09,B}{1981Ne09}), which deduced similar L-values from comparison with the data of (\href{https://www.nndc.bnl.gov/nsr/nsrlink.jsp?1970Le08,B}{1970Le08}, \href{https://www.nndc.bnl.gov/nsr/nsrlink.jsp?1970Fa17,B}{1970Fa17}). The authors of}\\
\parbox[b][0.3cm]{17.7cm}{{\ }{\ }(\href{https://www.nndc.bnl.gov/nsr/nsrlink.jsp?1981Ne09,B}{1981Ne09}) appear to have been guided by the same DWBA calculations performed in (\href{https://www.nndc.bnl.gov/nsr/nsrlink.jsp?1970Fa17,B}{1970Fa17}) due to the proximity of the}\\
\parbox[b][0.3cm]{17.7cm}{{\ }{\ }beam energies in these two experiments.}\\
\vspace{0.5cm}
\clearpage
\subsection[\hspace{-0.2cm}\ensuremath{^{\textnormal{27}}}Al(\ensuremath{\pi}\ensuremath{^{-}},\ensuremath{^{\textnormal{18}}}Ne\ensuremath{\gamma})]{ }
\vspace{-27pt}
\vspace{0.3cm}
\hypertarget{NE23}{{\bf \small \underline{\ensuremath{^{\textnormal{27}}}Al(\ensuremath{\pi}\ensuremath{^{-}},\ensuremath{^{\textnormal{18}}}Ne\ensuremath{\gamma})\hspace{0.2in}\href{https://www.nndc.bnl.gov/nsr/nsrlink.jsp?1976Li18,B}{1976Li18},\href{https://www.nndc.bnl.gov/nsr/nsrlink.jsp?1978Li18,B}{1978Li18}}}}\\
\vspace{4pt}
\vspace{8pt}
\parbox[b][0.3cm]{17.7cm}{\addtolength{\parindent}{-0.2in}\href{https://www.nndc.bnl.gov/nsr/nsrlink.jsp?1976Li18,B}{1976Li18}, \href{https://www.nndc.bnl.gov/nsr/nsrlink.jsp?1978Li18,B}{1978Li18}: \ensuremath{^{\textnormal{27}}}Al(\ensuremath{\pi}\ensuremath{^{-}},X\ensuremath{\gamma}) E=230-235 MeV; measured prompt \ensuremath{\gamma} rays from the de-exciting reaction products using a}\\
\parbox[b][0.3cm]{17.7cm}{Ge(Li) detector at \ensuremath{\theta}\ensuremath{_{\textnormal{lab}}}=90\ensuremath{^\circ} surrounded by a cup-shaped Compton supressed scintillator. Measured beam-\ensuremath{\gamma} coincidences; resolution}\\
\parbox[b][0.3cm]{17.7cm}{was 5 keV at 1 MeV in (\href{https://www.nndc.bnl.gov/nsr/nsrlink.jsp?1976Li18,B}{1976Li18}) and 4 keV for E\ensuremath{_{\ensuremath{\gamma}}}=0.3-6.4 MeV in (\href{https://www.nndc.bnl.gov/nsr/nsrlink.jsp?1978Li18,B}{1978Li18}). Measured \ensuremath{\gamma} yields. Deduced production \ensuremath{\sigma} for}\\
\parbox[b][0.3cm]{17.7cm}{multi-nucleon removal, \ensuremath{\sigma}(E\ensuremath{_{\ensuremath{\gamma}}}) and nuclear recoil momenta.}\\
\vspace{12pt}
\underline{$^{18}$Ne Levels}\\
\begin{longtable}{cccccc@{\extracolsep{\fill}}c}
\multicolumn{2}{c}{E(level)$^{{\hyperlink{NE23LEVEL0}{a}}}$}&J$^{\pi}$$^{{\hyperlink{NE23LEVEL0}{a}}}$&\multicolumn{2}{c}{\ensuremath{\sigma} (mb)$^{{\hyperlink{NE23LEVEL1}{b}}}$}&Comments&\\[-.2cm]
\multicolumn{2}{c}{\hrulefill}&\hrulefill&\multicolumn{2}{c}{\hrulefill}&\hrulefill&
\endfirsthead
\multicolumn{1}{r@{}}{0}&\multicolumn{1}{@{}l}{}&&&&&\\
\multicolumn{1}{r@{}}{1887}&\multicolumn{1}{@{.}l}{4}&\multicolumn{1}{l}{2\ensuremath{^{+}}}&\multicolumn{1}{r@{}}{$<$3}&\multicolumn{1}{@{.}l}{3}&\parbox[t][0.3cm]{13.5254cm}{\raggedright The \ensuremath{\gamma}-ray transition from 1887 keV\ensuremath{\rightarrow}g.s. was observed in (\href{https://www.nndc.bnl.gov/nsr/nsrlink.jsp?1976Li18,B}{1976Li18}) (see Table II) but the \ensuremath{\gamma}-ray\vspace{0.1cm}}&\\
&&&&&\parbox[t][0.3cm]{13.5254cm}{\raggedright {\ }{\ }{\ }energy is not reported.\vspace{0.1cm}}&\\
\multicolumn{1}{r@{}}{3376}&\multicolumn{1}{@{.}l}{4}&\multicolumn{1}{l}{4\ensuremath{^{+}}}&\multicolumn{1}{r@{}}{2}&\multicolumn{1}{@{.}l}{1 {\it 5}}&\parbox[t][0.3cm]{13.5254cm}{\raggedright The \ensuremath{\gamma}-ray transition from 3376 keV\ensuremath{\rightarrow}1887 keV was observed in (\href{https://www.nndc.bnl.gov/nsr/nsrlink.jsp?1976Li18,B}{1976Li18}) (see Table II) but the\vspace{0.1cm}}&\\
&&&&&\parbox[t][0.3cm]{13.5254cm}{\raggedright {\ }{\ }{\ }\ensuremath{\gamma}-ray energy is not reported.\vspace{0.1cm}}&\\
&&&&&\parbox[t][0.3cm]{13.5254cm}{\raggedright \ensuremath{\sigma} (mb): The production cross section was corrected for \ensuremath{\gamma} feeding from higher states known to be\vspace{0.1cm}}&\\
&&&&&\parbox[t][0.3cm]{13.5254cm}{\raggedright {\ }{\ }{\ }excited in (\href{https://www.nndc.bnl.gov/nsr/nsrlink.jsp?1976Li18,B}{1976Li18}). These higher energy states are not reported in (\href{https://www.nndc.bnl.gov/nsr/nsrlink.jsp?1976Li18,B}{1976Li18}).\vspace{0.1cm}}&\\
&&&&&\parbox[t][0.3cm]{13.5254cm}{\raggedright \ensuremath{\sigma} (mb): See also 2.1 mb \textit{5} (\href{https://www.nndc.bnl.gov/nsr/nsrlink.jsp?1978Li18,B}{1978Li18}).\vspace{0.1cm}}&\\
\end{longtable}
\parbox[b][0.3cm]{17.7cm}{\makebox[1ex]{\ensuremath{^{\hypertarget{NE23LEVEL0}{a}}}} From the \ensuremath{^{\textnormal{18}}}Ne Adopted Levels.}\\
\parbox[b][0.3cm]{17.7cm}{\makebox[1ex]{\ensuremath{^{\hypertarget{NE23LEVEL1}{b}}}} This is the cross section for production of a particular state by direct excitation and/or by feeding from higher states (\href{https://www.nndc.bnl.gov/nsr/nsrlink.jsp?1976Li18,B}{1976Li18}).}\\
\vspace{0.5cm}
\underline{$\gamma$($^{18}$Ne)}\\
\begin{longtable}{ccccccc@{}c@{\extracolsep{\fill}}c}
\multicolumn{2}{c}{E\ensuremath{_{\gamma}}\ensuremath{^{\hyperlink{NE23GAMMA0}{a}}}}&\multicolumn{2}{c}{E\ensuremath{_{i}}(level)}&J\ensuremath{^{\pi}_{i}}&\multicolumn{2}{c}{E\ensuremath{_{f}}}&J\ensuremath{^{\pi}_{f}}&\\[-.2cm]
\multicolumn{2}{c}{\hrulefill}&\multicolumn{2}{c}{\hrulefill}&\hrulefill&\multicolumn{2}{c}{\hrulefill}&\hrulefill&
\endfirsthead
\multicolumn{1}{r@{}}{1488}&\multicolumn{1}{@{.}l}{9}&\multicolumn{1}{r@{}}{3376}&\multicolumn{1}{@{.}l}{4}&\multicolumn{1}{l}{4\ensuremath{^{+}}}&\multicolumn{1}{r@{}}{1887}&\multicolumn{1}{@{.}l}{4}&\multicolumn{1}{@{}l}{2\ensuremath{^{+}}}&\\
\multicolumn{1}{r@{}}{1887}&\multicolumn{1}{@{.}l}{3}&\multicolumn{1}{r@{}}{1887}&\multicolumn{1}{@{.}l}{4}&\multicolumn{1}{l}{2\ensuremath{^{+}}}&\multicolumn{1}{r@{}}{0}&\multicolumn{1}{@{}l}{}&&\\
\end{longtable}
\parbox[b][0.3cm]{17.7cm}{\makebox[1ex]{\ensuremath{^{\hypertarget{NE23GAMMA0}{a}}}} From the \ensuremath{^{\textnormal{18}}}Ne Adopted Gammas.}\\
\vspace{0.5cm}
\begin{figure}[h]
\begin{center}
\includegraphics{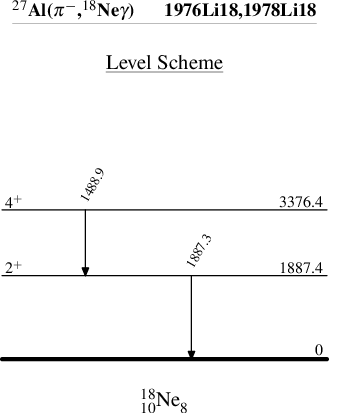}\\
\end{center}
\end{figure}
\clearpage
\subsection[\hspace{-0.2cm}\ensuremath{^{\textnormal{27}}}Al(p,X\ensuremath{\gamma})]{ }
\vspace{-27pt}
\vspace{0.3cm}
\hypertarget{NE24}{{\bf \small \underline{\ensuremath{^{\textnormal{27}}}Al(p,X\ensuremath{\gamma})\hspace{0.2in}\href{https://www.nndc.bnl.gov/nsr/nsrlink.jsp?1997Vo03,B}{1997Vo03}}}}\\
\vspace{4pt}
\vspace{8pt}
\parbox[b][0.3cm]{17.7cm}{\addtolength{\parindent}{-0.2in}\href{https://www.nndc.bnl.gov/nsr/nsrlink.jsp?1997Vo03,B}{1997Vo03}: \ensuremath{^{\textnormal{27}}}Al(p,X\ensuremath{\gamma}) E=800 MeV; measured the production cross sections for \ensuremath{^{\textnormal{18}}}Ne and 47 other nuclei; measured the prompt}\\
\parbox[b][0.3cm]{17.7cm}{and delayed \ensuremath{\gamma} rays from the decay of these nuclei using a HPGe detector (energy resolution=3 keV at FWHM) placed 30 m away}\\
\parbox[b][0.3cm]{17.7cm}{from the target at \ensuremath{\theta}\ensuremath{_{\textnormal{lab}}}=150\ensuremath{^\circ}. Compared the data with literature and semi-empirical systematics and quantum molecular dynamics}\\
\parbox[b][0.3cm]{17.7cm}{calculations. Deduced upper limits of \ensuremath{\sigma}\ensuremath{<}1.7 mb and \ensuremath{\sigma}\ensuremath{<}0.05 mb for the \ensuremath{^{\textnormal{18}}}Ne\ensuremath{_{\textnormal{g.s.}}} production from \ensuremath{^{\textnormal{27}}}Al+p at E\ensuremath{_{\textnormal{p}}}=800 MeV, and for}\\
\parbox[b][0.3cm]{17.7cm}{the \ensuremath{^{\textnormal{18}}}Ne*(1887 keV) state by considering the prompt \ensuremath{^{\textnormal{18}}}Ne(2\ensuremath{^{\textnormal{+}}_{\textnormal{1}}}\ensuremath{\rightarrow}g.s.) transition, respectively. This study assumes that at E\ensuremath{_{\textnormal{p}}}=800}\\
\parbox[b][0.3cm]{17.7cm}{MeV, well over 90\% of all \ensuremath{\gamma}-ray cascades in even-even nuclei proceed through the 2\ensuremath{^{\textnormal{+}}_{\textnormal{1}}}\ensuremath{\rightarrow}g.s. transition because of the large}\\
\parbox[b][0.3cm]{17.7cm}{average spin of the residual nuclei cascading downwards. It is also assumed that the 2\ensuremath{^{\textnormal{+}}_{\textnormal{1}}}\ensuremath{\rightarrow}g.s. transition is isotropic at E\ensuremath{_{\textnormal{p}}}=800}\\
\parbox[b][0.3cm]{17.7cm}{MeV.}\\
\vspace{12pt}
\underline{$^{18}$Ne Levels}\\
\begin{longtable}{cccccc@{\extracolsep{\fill}}c}
\multicolumn{2}{c}{E(level)$^{{\hyperlink{NE24LEVEL0}{a}}}$}&J$^{\pi}$$^{{\hyperlink{NE24LEVEL0}{a}}}$&\multicolumn{2}{c}{\ensuremath{\sigma} (mb)$^{{\hyperlink{NE24LEVEL1}{b}}}$}&Comments&\\[-.2cm]
\multicolumn{2}{c}{\hrulefill}&\hrulefill&\multicolumn{2}{c}{\hrulefill}&\hrulefill&
\endfirsthead
\multicolumn{1}{r@{}}{0?}&\multicolumn{1}{@{}l}{}&\multicolumn{1}{l}{0\ensuremath{^{+}}}&\multicolumn{1}{r@{}}{$<$1}&\multicolumn{1}{@{.}l}{7}&\parbox[t][0.3cm]{13.5254cm}{\raggedright \ensuremath{\sigma} (mb): Deduced from \ensuremath{^{\textnormal{18}}}Ne(\ensuremath{\beta}\ensuremath{^{\textnormal{+}}})\ensuremath{^{\textnormal{18}}}F*(1042 keV)\ensuremath{\rightarrow}\ensuremath{^{\textnormal{18}}}F\ensuremath{_{\textnormal{g.s.}}}: E\ensuremath{_{\ensuremath{\gamma}}}=1041 keV and I\ensuremath{_{\ensuremath{\gamma}}}=7.9\%\vspace{0.1cm}}&\\
&&&&&\parbox[t][0.3cm]{13.5254cm}{\raggedright {\ }{\ }{\ }(\href{https://www.nndc.bnl.gov/nsr/nsrlink.jsp?1997Vo03,B}{1997Vo03}).\vspace{0.1cm}}&\\
\multicolumn{1}{r@{}}{1887}&\multicolumn{1}{@{.}l}{4?}&\multicolumn{1}{l}{2\ensuremath{^{+}}}&\multicolumn{1}{r@{}}{$<$0}&\multicolumn{1}{@{.}l}{05}&\parbox[t][0.3cm]{13.5254cm}{\raggedright E(level): Since the cross sections given here are upper limits, the evaluator assumed the level\vspace{0.1cm}}&\\
&&&&&\parbox[t][0.3cm]{13.5254cm}{\raggedright {\ }{\ }{\ }energy and its \ensuremath{\gamma}-ray transition to the \ensuremath{^{\textnormal{18}}}Ne\ensuremath{_{\textnormal{g.s.}}} tentative.\vspace{0.1cm}}&\\
&&&&&\parbox[t][0.3cm]{13.5254cm}{\raggedright \ensuremath{\sigma} (mb): Deduced from intensity of the prompt 2\ensuremath{^{\textnormal{+}}_{\textnormal{1}}}\ensuremath{\rightarrow}g.s. transition in \ensuremath{^{\textnormal{18}}}Ne (\href{https://www.nndc.bnl.gov/nsr/nsrlink.jsp?1997Vo03,B}{1997Vo03}).\vspace{0.1cm}}&\\
\end{longtable}
\parbox[b][0.3cm]{17.7cm}{\makebox[1ex]{\ensuremath{^{\hypertarget{NE24LEVEL0}{a}}}} From the \ensuremath{^{\textnormal{18}}}Ne Adopted Levels.}\\
\parbox[b][0.3cm]{17.7cm}{\makebox[1ex]{\ensuremath{^{\hypertarget{NE24LEVEL1}{b}}}} \ensuremath{^{\textnormal{18}}}Ne production cross section from (\href{https://www.nndc.bnl.gov/nsr/nsrlink.jsp?1997Vo03,B}{1997Vo03}).}\\
\vspace{0.5cm}
\underline{$\gamma$($^{18}$Ne)}\\
\begin{longtable}{ccccccc@{}c@{\extracolsep{\fill}}c}
\multicolumn{2}{c}{E\ensuremath{_{\gamma}}\ensuremath{^{\hyperlink{NE24GAMMA0}{a}}}}&\multicolumn{2}{c}{E\ensuremath{_{i}}(level)}&J\ensuremath{^{\pi}_{i}}&\multicolumn{2}{c}{E\ensuremath{_{f}}}&J\ensuremath{^{\pi}_{f}}&\\[-.2cm]
\multicolumn{2}{c}{\hrulefill}&\multicolumn{2}{c}{\hrulefill}&\hrulefill&\multicolumn{2}{c}{\hrulefill}&\hrulefill&
\endfirsthead
\multicolumn{1}{r@{}}{1887}&\multicolumn{1}{@{.}l}{3\ensuremath{^{\hyperlink{NE24GAMMA1}{b}}}}&\multicolumn{1}{r@{}}{1887}&\multicolumn{1}{@{.}l}{4?}&\multicolumn{1}{l}{2\ensuremath{^{+}}}&\multicolumn{1}{r@{}}{0?}&\multicolumn{1}{@{}l}{}&\multicolumn{1}{@{}l}{0\ensuremath{^{+}}}&\\
\end{longtable}
\parbox[b][0.3cm]{17.7cm}{\makebox[1ex]{\ensuremath{^{\hypertarget{NE24GAMMA0}{a}}}} From the \ensuremath{^{\textnormal{18}}}Ne Adopted Gammas.}\\
\parbox[b][0.3cm]{17.7cm}{\makebox[1ex]{\ensuremath{^{\hypertarget{NE24GAMMA1}{b}}}} Placement of transition in the level scheme is uncertain.}\\
\vspace{0.5cm}
\clearpage
\begin{figure}[h]
\begin{center}
\includegraphics{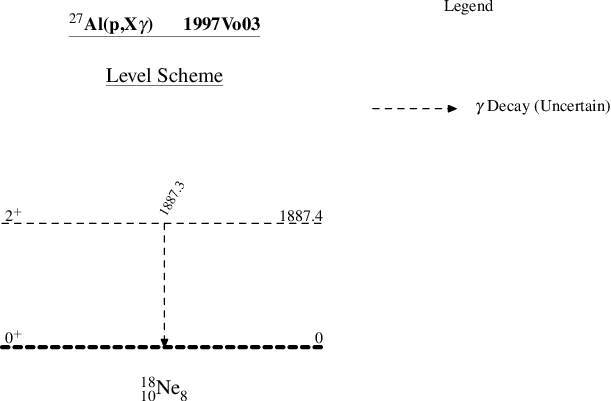}\\
\end{center}
\end{figure}
\clearpage
\subsection[\hspace{-0.2cm}\ensuremath{^{\textnormal{27}}}Al(\ensuremath{\alpha},\ensuremath{^{\textnormal{18}}}Ne\ensuremath{\gamma}),\ensuremath{^{\textnormal{28}}}Si(\ensuremath{\alpha},\ensuremath{^{\textnormal{18}}}Ne\ensuremath{\gamma})]{ }
\vspace{-27pt}
\vspace{0.3cm}
\hypertarget{NE25}{{\bf \small \underline{\ensuremath{^{\textnormal{27}}}Al(\ensuremath{\alpha},\ensuremath{^{\textnormal{18}}}Ne\ensuremath{\gamma}),\ensuremath{^{\textnormal{28}}}Si(\ensuremath{\alpha},\ensuremath{^{\textnormal{18}}}Ne\ensuremath{\gamma})\hspace{0.2in}\href{https://www.nndc.bnl.gov/nsr/nsrlink.jsp?1979Gl01,B}{1979Gl01},\href{https://www.nndc.bnl.gov/nsr/nsrlink.jsp?2001Na02,B}{2001Na02}}}}\\
\vspace{4pt}
\vspace{8pt}
\parbox[b][0.3cm]{17.7cm}{\addtolength{\parindent}{-0.2in}\href{https://www.nndc.bnl.gov/nsr/nsrlink.jsp?1979Gl01,B}{1979Gl01}: Inclusive \ensuremath{^{\textnormal{27}}}Al(\ensuremath{\alpha},X\ensuremath{\gamma}) E=140 MeV; target mounted at 45\ensuremath{^\circ}; performed in-beam \ensuremath{\gamma}-ray spectroscopy to study prompt and}\\
\parbox[b][0.3cm]{17.7cm}{\ensuremath{\beta}-delayed \ensuremath{\gamma} rays using a Ge(Li) detector (energy resolution of 2 keV at 1 MeV) located at \ensuremath{\theta}\ensuremath{_{\textnormal{lab}}}=90\ensuremath{^\circ}. Measured lowest energy \ensuremath{\gamma}}\\
\parbox[b][0.3cm]{17.7cm}{rays using a planar intrinsic Ge detector (energy resolution of \ensuremath{<}1 keV at \ensuremath{<}0.5 MeV). Identified the nuclei produced through}\\
\parbox[b][0.3cm]{17.7cm}{measuring E\ensuremath{_{\ensuremath{\gamma}}}, and determined production \ensuremath{\sigma} for specific levels in the residual nuclei. Measured Doppler broadening \ensuremath{\sigma}(\ensuremath{\theta},E) for}\\
\parbox[b][0.3cm]{17.7cm}{discrete energy groups corresponding to p, d, t, \ensuremath{^{\textnormal{3}}}He, and \ensuremath{\alpha} ejectiles assuming isotropic \ensuremath{\gamma}-ray emission. Also measured light}\\
\parbox[b][0.3cm]{17.7cm}{reaction products using rotatable Si \ensuremath{\Delta}E-\ensuremath{\Delta}E and NaI-E, and Si \ensuremath{\Delta}E-E telescopes, respectively.}\\
\parbox[b][0.3cm]{17.7cm}{\addtolength{\parindent}{-0.2in}\href{https://www.nndc.bnl.gov/nsr/nsrlink.jsp?1980Li14,B}{1980Li14}: \ensuremath{^{\textnormal{27}}}Al(\ensuremath{\alpha},X\ensuremath{\gamma}),\ensuremath{^{\textnormal{28}}}Si(\ensuremath{\alpha},X\ensuremath{\gamma}) E=180 MeV/nucleon; targets mounted at \ensuremath{\theta}\ensuremath{_{\textnormal{lab}}}=45\ensuremath{^\circ}; measured prompt \ensuremath{\gamma} rays from the decay of}\\
\parbox[b][0.3cm]{17.7cm}{the populated residual nuclei, E\ensuremath{_{\ensuremath{\gamma}}} and I\ensuremath{_{\ensuremath{\gamma}}} from the decaying nuclei, and \ensuremath{\gamma}-recoil coincidences using a Compton suppressed Ge(Li)}\\
\parbox[b][0.3cm]{17.7cm}{detector at \ensuremath{\theta}\ensuremath{_{\textnormal{lab}}}=90\ensuremath{^\circ} and a scintillator \ensuremath{\Delta}E-E telescope. Deduced the momenta of the residual nuclei from Doppler broadening.}\\
\parbox[b][0.3cm]{17.7cm}{Deduced \ensuremath{^{\textnormal{18}}}Ne levels and the production \ensuremath{\sigma} assuming that the \ensuremath{\gamma} rays were emitted isotropically. Comparison between the}\\
\parbox[b][0.3cm]{17.7cm}{interactions of \ensuremath{\alpha} and pion beams with \ensuremath{^{\textnormal{27}}}Al and \ensuremath{^{\textnormal{28}}}Si targets, as well as the calculation of the spallation yield are discussed.}\\
\vspace{0.385cm}
\parbox[b][0.3cm]{17.7cm}{\addtolength{\parindent}{-0.2in}\textit{Theory}:}\\
\parbox[b][0.3cm]{17.7cm}{\addtolength{\parindent}{-0.2in}\href{https://www.nndc.bnl.gov/nsr/nsrlink.jsp?2001Na02,B}{2001Na02}: \ensuremath{^{\textnormal{nat}}}Si(p,X)\ensuremath{^{\textnormal{18}}}Ne E=20{\textminus}110 MeV; calculated excitation functions of several nuclei from p+\ensuremath{^{\textnormal{nat}}}Si interactions using the}\\
\parbox[b][0.3cm]{17.7cm}{ALICE code. The aim was to investigate radiation damage to Si (including quartz beam viewers) from structural changes inflicted}\\
\parbox[b][0.3cm]{17.7cm}{by the residual nuclei produced from the activation of Si by energetic protons. Comparisons with experimental data are discussed}\\
\parbox[b][0.3cm]{17.7cm}{for a few of these nuclei, for which data are available. The formation cross section of \ensuremath{^{\textnormal{18}}}Ne from the \ensuremath{^{\textnormal{nat}}}Si(p,X) reaction was}\\
\parbox[b][0.3cm]{17.7cm}{estimated to be less than 0.1 mb for proton incident energies of 70-110 MeV (see Fig. 8 of (\href{https://www.nndc.bnl.gov/nsr/nsrlink.jsp?2001Na02,B}{2001Na02})).}\\
\vspace{12pt}
\underline{$^{18}$Ne Levels}\\
\begin{longtable}{cccccc@{\extracolsep{\fill}}c}
\multicolumn{2}{c}{E(level)$^{{\hyperlink{NE25LEVEL1}{b}}}$}&J$^{\pi}$$^{{\hyperlink{NE25LEVEL1}{b}}}$&\multicolumn{2}{c}{Total Production Cross Section \ensuremath{\sigma}\ensuremath{_{\textnormal{total}}} (mb)$^{{\hyperlink{NE25LEVEL0}{a}}}$}&Comments&\\[-.2cm]
\multicolumn{2}{c}{\hrulefill}&\hrulefill&\multicolumn{2}{c}{\hrulefill}&\hrulefill&
\endfirsthead
\multicolumn{1}{r@{}}{0}&\multicolumn{1}{@{}l}{}&&&&&\\
\multicolumn{1}{r@{}}{1887?}&\multicolumn{1}{@{ }l}{{\it 1}}&\multicolumn{1}{l}{2\ensuremath{^{+}}}&\multicolumn{1}{r@{}}{$<$0}&\multicolumn{1}{@{.}l}{8}&\parbox[t][0.3cm]{8.727519cm}{\raggedright E(level): Since the cross sections given here are upper limits,\vspace{0.1cm}}&\\
&&&&&\parbox[t][0.3cm]{8.727519cm}{\raggedright {\ }{\ }{\ }the evaluator assumed the level energy and its \ensuremath{\gamma}-ray\vspace{0.1cm}}&\\
&&&&&\parbox[t][0.3cm]{8.727519cm}{\raggedright {\ }{\ }{\ }transition to the \ensuremath{^{\textnormal{18}}}Ne\ensuremath{_{\textnormal{g.s.}}} tentative.\vspace{0.1cm}}&\\
&&&&&\parbox[t][0.3cm]{8.727519cm}{\raggedright Total Production Cross Section \ensuremath{\sigma}\ensuremath{_{\textnormal{total}}} (mb): From (\href{https://www.nndc.bnl.gov/nsr/nsrlink.jsp?1980Li14,B}{1980Li14}):\vspace{0.1cm}}&\\
&&&&&\parbox[t][0.3cm]{8.727519cm}{\raggedright {\ }{\ }{\ }\ensuremath{\sigma}\ensuremath{_{\textnormal{ex}}}\ensuremath{<}0.8 mb for \ensuremath{\alpha}+\ensuremath{^{\textnormal{27}}}Al. For \ensuremath{\alpha}+\ensuremath{^{\textnormal{28}}}Si: \ensuremath{\sigma}\ensuremath{_{\textnormal{tot}}}\ensuremath{<}0.6 mb and\vspace{0.1cm}}&\\
&&&&&\parbox[t][0.3cm]{8.727519cm}{\raggedright {\ }{\ }{\ }\ensuremath{\sigma}\ensuremath{_{\textnormal{ex}}}\ensuremath{<}0.6 mb.\vspace{0.1cm}}&\\
&&&&&\parbox[t][0.3cm]{8.727519cm}{\raggedright Total Production Cross Section \ensuremath{\sigma}\ensuremath{_{\textnormal{total}}} (mb): See also \ensuremath{\sigma}\ensuremath{_{\textnormal{x}}}\ensuremath{\leq}0.4\vspace{0.1cm}}&\\
&&&&&\parbox[t][0.3cm]{8.727519cm}{\raggedright {\ }{\ }{\ }mb (\href{https://www.nndc.bnl.gov/nsr/nsrlink.jsp?1979Gl01,B}{1979Gl01}), where \ensuremath{\sigma}\ensuremath{_{\textnormal{x}}} is the cross section for production\vspace{0.1cm}}&\\
&&&&&\parbox[t][0.3cm]{8.727519cm}{\raggedright {\ }{\ }{\ }of a particular state by direct excitation and/or by \ensuremath{\gamma} feeding\vspace{0.1cm}}&\\
&&&&&\parbox[t][0.3cm]{8.727519cm}{\raggedright {\ }{\ }{\ }from unidentified higher lying states. \ensuremath{\gamma} feeding by transitions\vspace{0.1cm}}&\\
&&&&&\parbox[t][0.3cm]{8.727519cm}{\raggedright {\ }{\ }{\ }from identified higher lying states was subtracted. In\vspace{0.1cm}}&\\
&&&&&\parbox[t][0.3cm]{8.727519cm}{\raggedright {\ }{\ }{\ }(\href{https://www.nndc.bnl.gov/nsr/nsrlink.jsp?1979Gl01,B}{1979Gl01}), no \ensuremath{\gamma} rays from the decay of \ensuremath{^{\textnormal{18}}}Ne, produced by\vspace{0.1cm}}&\\
&&&&&\parbox[t][0.3cm]{8.727519cm}{\raggedright {\ }{\ }{\ }the \ensuremath{^{\textnormal{27}}}Al(\ensuremath{\alpha},X\ensuremath{\gamma}) reaction, were observed. Therefore, only an\vspace{0.1cm}}&\\
&&&&&\parbox[t][0.3cm]{8.727519cm}{\raggedright {\ }{\ }{\ }upper limit for the production of the reported excited state\vspace{0.1cm}}&\\
&&&&&\parbox[t][0.3cm]{8.727519cm}{\raggedright {\ }{\ }{\ }was determined.\vspace{0.1cm}}&\\
&&&&&\parbox[t][0.3cm]{8.727519cm}{\raggedright (\href{https://www.nndc.bnl.gov/nsr/nsrlink.jsp?2001Na02,B}{2001Na02}) calculated the formation cross section of \ensuremath{^{\textnormal{18}}}Ne\vspace{0.1cm}}&\\
&&&&&\parbox[t][0.3cm]{8.727519cm}{\raggedright {\ }{\ }{\ }from the \ensuremath{^{\textnormal{nat}}}Si(p,X) reaction. The result was less than 0.1 mb\vspace{0.1cm}}&\\
&&&&&\parbox[t][0.3cm]{8.727519cm}{\raggedright {\ }{\ }{\ }for proton incident energies of 70-110 MeV (see Fig. 8 of\vspace{0.1cm}}&\\
&&&&&\parbox[t][0.3cm]{8.727519cm}{\raggedright {\ }{\ }{\ }(\href{https://www.nndc.bnl.gov/nsr/nsrlink.jsp?2001Na02,B}{2001Na02})).\vspace{0.1cm}}&\\
\end{longtable}
\parbox[b][0.3cm]{17.7cm}{\makebox[1ex]{\ensuremath{^{\hypertarget{NE25LEVEL0}{a}}}} From (\href{https://www.nndc.bnl.gov/nsr/nsrlink.jsp?1980Li14,B}{1980Li14}): \ensuremath{\sigma}\ensuremath{_{\textnormal{tot}}} is the cross section for production of the excited state, including the \ensuremath{\gamma} ray feeding from higher excited}\\
\parbox[b][0.3cm]{17.7cm}{{\ }{\ }states. \ensuremath{\sigma}\ensuremath{_{\textnormal{ex}}} is \ensuremath{\sigma}\ensuremath{_{\textnormal{total}}} corrected for the \ensuremath{\gamma} ray feeding from the populated, known higher excited states.}\\
\parbox[b][0.3cm]{17.7cm}{\makebox[1ex]{\ensuremath{^{\hypertarget{NE25LEVEL1}{b}}}} Taken by (\href{https://www.nndc.bnl.gov/nsr/nsrlink.jsp?1979Gl01,B}{1979Gl01}, \href{https://www.nndc.bnl.gov/nsr/nsrlink.jsp?1980Li14,B}{1980Li14}) from the Adopted Levels of \ensuremath{^{\textnormal{18}}}Ne in (\href{https://www.nndc.bnl.gov/nsr/nsrlink.jsp?1978Aj03,B}{1978Aj03}).}\\
\vspace{0.5cm}
\clearpage
\vspace{0.3cm}
\vspace*{-0.5cm}
{\bf \small \underline{\ensuremath{^{\textnormal{27}}}Al(\ensuremath{\alpha},\ensuremath{^{\textnormal{18}}}Ne\ensuremath{\gamma}),\ensuremath{^{\textnormal{28}}}Si(\ensuremath{\alpha},\ensuremath{^{\textnormal{18}}}Ne\ensuremath{\gamma})\hspace{0.2in}\href{https://www.nndc.bnl.gov/nsr/nsrlink.jsp?1979Gl01,B}{1979Gl01},\href{https://www.nndc.bnl.gov/nsr/nsrlink.jsp?2001Na02,B}{2001Na02} (continued)}}\\
\vspace{0.3cm}
\underline{$\gamma$($^{18}$Ne)}\\
\begin{longtable}{ccccccc@{\extracolsep{\fill}}c}
\multicolumn{2}{c}{E\ensuremath{_{\gamma}}\ensuremath{^{\hyperlink{NE25GAMMA0}{a}}}}&\multicolumn{2}{c}{E\ensuremath{_{i}}(level)}&J\ensuremath{^{\pi}_{i}}&\multicolumn{2}{c}{E\ensuremath{_{f}}}&\\[-.2cm]
\multicolumn{2}{c}{\hrulefill}&\multicolumn{2}{c}{\hrulefill}&\hrulefill&\multicolumn{2}{c}{\hrulefill}&
\endfirsthead
\multicolumn{1}{r@{}}{1887}&\multicolumn{1}{@{.}l}{3\ensuremath{^{\hyperlink{NE25GAMMA1}{b}}}}&\multicolumn{1}{r@{}}{1887?}&\multicolumn{1}{@{}l}{}&\multicolumn{1}{l}{2\ensuremath{^{+}}}&\multicolumn{1}{r@{}}{0}&\multicolumn{1}{@{}l}{}&\\
\end{longtable}
\parbox[b][0.3cm]{17.7cm}{\makebox[1ex]{\ensuremath{^{\hypertarget{NE25GAMMA0}{a}}}} From the \ensuremath{^{\textnormal{18}}}Ne Adopted Gammas.}\\
\parbox[b][0.3cm]{17.7cm}{\makebox[1ex]{\ensuremath{^{\hypertarget{NE25GAMMA1}{b}}}} Placement of transition in the level scheme is uncertain.}\\
\vspace{0.5cm}
\begin{figure}[h]
\begin{center}
\includegraphics{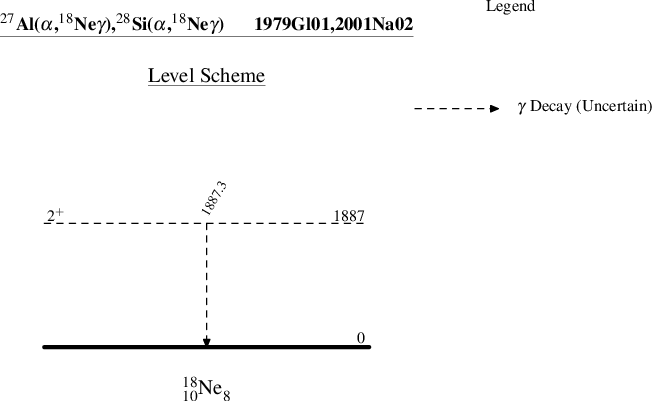}\\
\end{center}
\end{figure}
\clearpage
\subsection[\hspace{-0.2cm}\ensuremath{^{\textnormal{197}}}Au(\ensuremath{^{\textnormal{18}}}Ne,\ensuremath{^{\textnormal{18}}}Ne\ensuremath{'}):coulex]{ }
\vspace{-27pt}
\vspace{0.3cm}
\hypertarget{NE26}{{\bf \small \underline{\ensuremath{^{\textnormal{197}}}Au(\ensuremath{^{\textnormal{18}}}Ne,\ensuremath{^{\textnormal{18}}}Ne\ensuremath{'}):coulex\hspace{0.2in}\href{https://www.nndc.bnl.gov/nsr/nsrlink.jsp?2000Ri15,B}{2000Ri15},\href{https://www.nndc.bnl.gov/nsr/nsrlink.jsp?2016Li45,B}{2016Li45}}}}\\
\vspace{4pt}
\vspace{8pt}
\parbox[b][0.3cm]{17.7cm}{\addtolength{\parindent}{-0.2in}\href{https://www.nndc.bnl.gov/nsr/nsrlink.jsp?2000Ri15,B}{2000Ri15}: \ensuremath{^{\textnormal{197}}}Au(\ensuremath{^{\textnormal{18}}}Ne,\ensuremath{^{\textnormal{18}}}Ne\ensuremath{'}), \ensuremath{^{\textnormal{197}}}Au(\ensuremath{^{\textnormal{18}}}O,\ensuremath{^{\textnormal{18}}}O\ensuremath{'}) E=60 and 46 MeV/nucleon, respectively. The \ensuremath{^{\textnormal{18}}}Ne and \ensuremath{^{\textnormal{18}}}O beams were slowed}\\
\parbox[b][0.3cm]{17.7cm}{down in the gold target and stopped in a cylindrical fast/slow plastic phoswich detector located at \ensuremath{\theta}\ensuremath{_{\textnormal{lab}}}=0\ensuremath{^\circ} subtending \ensuremath{\theta}\ensuremath{_{\textnormal{lab}}}=0\ensuremath{^\circ}{\textminus}4\ensuremath{^\circ}.}\\
\parbox[b][0.3cm]{17.7cm}{Measured \ensuremath{\gamma} rays from decays of the Coulomb excited states in coincidence with beam particles using an array of 10 position}\\
\parbox[b][0.3cm]{17.7cm}{sensitive NaI(Tl) detectors subtending \ensuremath{\theta}\ensuremath{_{\textnormal{lab}}}=56.5\ensuremath{^\circ}{\textminus}123.5\ensuremath{^\circ}. Measured E\ensuremath{_{\ensuremath{\gamma}}} and I\ensuremath{_{\ensuremath{\gamma}}}. Deduced the integrated \ensuremath{\sigma} for the excitation of the}\\
\parbox[b][0.3cm]{17.7cm}{2\ensuremath{^{\textnormal{+}}_{\textnormal{1}}} state in \ensuremath{^{\textnormal{18}}}Ne in the forward angle. Analyzed data using a macroscopic model that included both Coulomb and nuclear}\\
\parbox[b][0.3cm]{17.7cm}{excitation mechanisms. Deduced B(E2:0\ensuremath{^{\textnormal{+}}_{\textnormal{1}}}\ensuremath{\rightarrow}2\ensuremath{^{\textnormal{+}}_{\textnormal{1}}}), as well as the Coulomb and nuclear deformation parameters of \ensuremath{\beta}\ensuremath{_{\textnormal{C}}}=0.450 \textit{36} and}\\
\parbox[b][0.3cm]{17.7cm}{\ensuremath{\beta}\ensuremath{_{\textnormal{N}}}=0.481 \textit{39}, respectively.}\\
\parbox[b][0.3cm]{17.7cm}{\addtolength{\parindent}{-0.2in}\href{https://www.nndc.bnl.gov/nsr/nsrlink.jsp?2009Ji02,B}{2009Ji02}, \href{https://www.nndc.bnl.gov/nsr/nsrlink.jsp?2010Li33,B}{2010Li33}, \href{https://www.nndc.bnl.gov/nsr/nsrlink.jsp?2011LiZV,B}{2011LiZV}, \href{https://www.nndc.bnl.gov/nsr/nsrlink.jsp?2016Li45,B}{2016Li45}: \ensuremath{^{\textnormal{197}}}Au(\ensuremath{^{\textnormal{18}}}Ne,\ensuremath{^{\textnormal{18}}}Ne$'$\ensuremath{\rightarrow}\ensuremath{^{\textnormal{16}}}O+2p) E\ensuremath{\approx}65 MeV/A (deduced from LISE++); studied the}\\
\parbox[b][0.3cm]{17.7cm}{in-flight decay of \ensuremath{^{\textnormal{18}}}Ne excited states populated via Coulomb excitation and detected in coincidence with the heavy fragments and}\\
\parbox[b][0.3cm]{17.7cm}{light decay products. These were measured using a position sensitive telescope with 6 Si \ensuremath{\Delta}E detectors followed by an array of}\\
\parbox[b][0.3cm]{17.7cm}{stopping CsI crystals that covered a maximum opening angle of \ensuremath{\pm}13.2\ensuremath{^\circ}. Measured the relative momenta, angular and energy}\\
\parbox[b][0.3cm]{17.7cm}{correlations of the proton pairs (from \ensuremath{^{\textnormal{18}}}Ne*\ensuremath{\rightarrow}\ensuremath{^{\textnormal{16}}}O+2p decay) in the center{\textminus}of{\textminus}mass system. The experimental energy resolution was}\\
\parbox[b][0.3cm]{17.7cm}{200-400 keV. Deduced the invariant mass of the \ensuremath{^{\textnormal{16}}}O\ensuremath{_{\textnormal{g.s.}}}+2p from the complete kinematics reconstruction of these decay events.}\\
\parbox[b][0.3cm]{17.7cm}{Simulations and data support diproton decay of the \ensuremath{^{\textnormal{18}}}Ne*(6.15 MeV) state. No obvious diproton emission was found for}\\
\parbox[b][0.3cm]{17.7cm}{higher-lying \ensuremath{^{\textnormal{18}}}Ne states.}\\
\parbox[b][0.3cm]{17.7cm}{\addtolength{\parindent}{-0.2in}The \ensuremath{^{\textnormal{18}}}Ne beam energy of 65 MeV/nucleon is estimated by the evaluator from the incident \ensuremath{^{\textnormal{20}}}Ne primary beam energy of 78.24}\\
\parbox[b][0.3cm]{17.7cm}{MeV/nucleon (\href{https://www.nndc.bnl.gov/nsr/nsrlink.jsp?2009Ji02,B}{2009Ji02}) on the \ensuremath{^{\textnormal{9}}}Be target using LISE++ computer code.}\\
\parbox[b][0.3cm]{17.7cm}{\addtolength{\parindent}{-0.2in}\href{https://www.nndc.bnl.gov/nsr/nsrlink.jsp?2010Xu11,B}{2010Xu11}: \ensuremath{^{\textnormal{197}}}Au(\ensuremath{^{\textnormal{18}}}Ne,\ensuremath{^{\textnormal{18}}}Ne\ensuremath{'}) E=51.8 MeV/nucleon; measured momenta of \ensuremath{^{\textnormal{18}}}Ne* decay products using a position sensitive}\\
\parbox[b][0.3cm]{17.7cm}{telescope along \ensuremath{\theta}\ensuremath{_{\textnormal{lab}}}=0\ensuremath{^\circ} that covered \ensuremath{\theta}\ensuremath{_{\textnormal{lab}}}=\ensuremath{\pm}11\ensuremath{^\circ}. The array comprised several position sensitive \ensuremath{\Delta}E detectors and a stopping E}\\
\parbox[b][0.3cm]{17.7cm}{layer of CsI. Deduced the excitation function of \ensuremath{^{\textnormal{18}}}Ne. The resolutions of the experimental setup were \ensuremath{\sim}500 keV at FWHM for E\ensuremath{_{\textnormal{x}}},}\\
\parbox[b][0.3cm]{17.7cm}{5\ensuremath{^\circ} for \ensuremath{\theta}\ensuremath{_{\textnormal{c.m.}}^{\ensuremath{\alpha}\ensuremath{\alpha}}}, and 5 MeV/c for the c.m. relative momentum of the two \ensuremath{\alpha}-particles. Deduced the c.m. relative angular correlations}\\
\parbox[b][0.3cm]{17.7cm}{and momenta between the \ensuremath{\alpha} pairs from the decay of the \ensuremath{^{\textnormal{18}}}Ne* states. The decay mechanism that best describes the data is a}\\
\parbox[b][0.3cm]{17.7cm}{sequential decay via the \ensuremath{^{\textnormal{14}}}O* intermediate states.}\\
\vspace{12pt}
\underline{$^{18}$Ne Levels}\\
\begin{longtable}{cccc@{\extracolsep{\fill}}c}
\multicolumn{2}{c}{E(level)$^{}$}&J$^{\pi}$$^{}$&Comments&\\[-.2cm]
\multicolumn{2}{c}{\hrulefill}&\hrulefill&\hrulefill&
\endfirsthead
\multicolumn{1}{r@{}}{0}&\multicolumn{1}{@{}l}{\ensuremath{^{{\hyperlink{NE26LEVEL0}{a}}}}}&\multicolumn{1}{l}{0\ensuremath{^{+}}\ensuremath{^{{\hyperlink{NE26LEVEL0}{a}}}}}&&\\
\multicolumn{1}{r@{}}{1887}&\multicolumn{1}{@{.}l}{4\ensuremath{^{{\hyperlink{NE26LEVEL0}{a}}}}}&\multicolumn{1}{l}{2\ensuremath{^{+}}\ensuremath{^{{\hyperlink{NE26LEVEL0}{a}}}}}&\parbox[t][0.3cm]{15.2243805cm}{\raggedright (\href{https://www.nndc.bnl.gov/nsr/nsrlink.jsp?2000Ri15,B}{2000Ri15}): deduced \ensuremath{\sigma}=45 mb \textit{6} for the production of the 1887 keV \ensuremath{\gamma} ray in \ensuremath{^{\textnormal{18}}}Ne assuming a \ensuremath{\gamma}-ray angular\vspace{0.1cm}}&\\
&&&\parbox[t][0.3cm]{15.2243805cm}{\raggedright {\ }{\ }{\ }distribution corresponding to a pure E2 transition and by integrating over \ensuremath{\theta}\ensuremath{_{\textnormal{lab}}}=0\ensuremath{^\circ}{\textminus}4\ensuremath{^\circ}. The authors noted that\vspace{0.1cm}}&\\
&&&\parbox[t][0.3cm]{15.2243805cm}{\raggedright {\ }{\ }{\ }this cross section may not be identical to that for directly exciting the \ensuremath{^{\textnormal{18}}}Ne*(2\ensuremath{^{\textnormal{+}}_{\textnormal{1}}}) state via scattering. This is\vspace{0.1cm}}&\\
&&&\parbox[t][0.3cm]{15.2243805cm}{\raggedright {\ }{\ }{\ }because the latter state can be fed by \ensuremath{\gamma} decays from higher-lying states. A total cross section of 40 mb \textit{11}\vspace{0.1cm}}&\\
&&&\parbox[t][0.3cm]{15.2243805cm}{\raggedright {\ }{\ }{\ }was deduced, using coupled channels calculations via the ECIS88 code, for directly populating the 2\ensuremath{^{\textnormal{+}}_{\textnormal{1}}} state in\vspace{0.1cm}}&\\
&&&\parbox[t][0.3cm]{15.2243805cm}{\raggedright {\ }{\ }{\ }\ensuremath{^{\textnormal{18}}}Ne.\vspace{0.1cm}}&\\
&&&\parbox[t][0.3cm]{15.2243805cm}{\raggedright (\href{https://www.nndc.bnl.gov/nsr/nsrlink.jsp?2000Ri15,B}{2000Ri15}): using the optical model parameters of (\href{https://www.nndc.bnl.gov/nsr/nsrlink.jsp?1987Me05,B}{1987Me05}), the Coulomb and nuclear deformation\vspace{0.1cm}}&\\
&&&\parbox[t][0.3cm]{15.2243805cm}{\raggedright {\ }{\ }{\ }parameters were deduced as \ensuremath{\beta}\ensuremath{_{\textnormal{C}}}=0.450 \textit{36} and \ensuremath{\beta}\ensuremath{_{\textnormal{N}}}=0.481 \textit{39}, respectively. These resulted in\vspace{0.1cm}}&\\
&&&\parbox[t][0.3cm]{15.2243805cm}{\raggedright {\ }{\ }{\ }B(E2;0\ensuremath{^{\textnormal{+}}_{\textnormal{1}}}\ensuremath{\rightarrow}2\ensuremath{^{\textnormal{+}}_{\textnormal{1}}})=113 e\ensuremath{^{\textnormal{2}}}fm\ensuremath{^{\textnormal{4}}} \textit{18} and M\ensuremath{_{\textnormal{p}}}=10.6 fm\ensuremath{^{\textnormal{2}}} \textit{9}.\vspace{0.1cm}}&\\
&&&\parbox[t][0.3cm]{15.2243805cm}{\raggedright (\href{https://www.nndc.bnl.gov/nsr/nsrlink.jsp?2000Ri15,B}{2000Ri15}): using the optical model parameters of (\href{https://www.nndc.bnl.gov/nsr/nsrlink.jsp?1988Ba39,B}{1988Ba39}), the above-mentioned values were deduced to be:\vspace{0.1cm}}&\\
&&&\parbox[t][0.3cm]{15.2243805cm}{\raggedright {\ }{\ }{\ }\ensuremath{\beta}\ensuremath{_{\textnormal{C}}}=0.496 \textit{40}, \ensuremath{\beta}\ensuremath{_{\textnormal{N}}}=0.503 \textit{40}, B(E2;0\ensuremath{^{\textnormal{+}}_{\textnormal{1}}}\ensuremath{\rightarrow}2\ensuremath{^{\textnormal{+}}_{\textnormal{1}}})=137 e\ensuremath{^{\textnormal{2}}}fm\ensuremath{^{\textnormal{4}}} \textit{22}, and M\ensuremath{_{\textnormal{p}}}=11.7 fm\ensuremath{^{\textnormal{2}}} \textit{9}.\vspace{0.1cm}}&\\
&&&\parbox[t][0.3cm]{15.2243805cm}{\raggedright Both the above-mentioned groups of results are under the assumption that b\ensuremath{_{\textnormal{n}}^{\textnormal{F}}}/b\ensuremath{_{\textnormal{p}}^{\textnormal{F}}}=0.820 for \ensuremath{^{\textnormal{197}}}Au, where\vspace{0.1cm}}&\\
&&&\parbox[t][0.3cm]{15.2243805cm}{\raggedright {\ }{\ }{\ }b\ensuremath{_{\textnormal{n/p}}^{\textnormal{F}}} is the external field interaction strength of the probe F (which is \ensuremath{^{\textnormal{197}}}Au) with neutrons or protons in\vspace{0.1cm}}&\\
&&&\parbox[t][0.3cm]{15.2243805cm}{\raggedright {\ }{\ }{\ }\ensuremath{^{\textnormal{18}}}Ne. If \ensuremath{^{\textnormal{197}}}Au is assumed to be an isoscalar probe (b\ensuremath{_{\textnormal{n}}^{\textnormal{F}}}/b\ensuremath{_{\textnormal{p}}^{\textnormal{F}}}=1), then B(E2;0\ensuremath{^{\textnormal{+}}_{\textnormal{g.s.}}}\ensuremath{\rightarrow}2\ensuremath{^{\textnormal{+}}_{\textnormal{1}}}) increases by 0.4\%\vspace{0.1cm}}&\\
&&&\parbox[t][0.3cm]{15.2243805cm}{\raggedright {\ }{\ }{\ }(\href{https://www.nndc.bnl.gov/nsr/nsrlink.jsp?2000Ri15,B}{2000Ri15}).\vspace{0.1cm}}&\\
\multicolumn{1}{r@{}}{3616}&\multicolumn{1}{@{.}l}{5?\ensuremath{^{{\hyperlink{NE26LEVEL0}{a}}}}}&\multicolumn{1}{l}{2\ensuremath{^{+}}\ensuremath{^{{\hyperlink{NE26LEVEL0}{a}}}}}&\parbox[t][0.3cm]{15.2243805cm}{\raggedright (\href{https://www.nndc.bnl.gov/nsr/nsrlink.jsp?2000Ri15,B}{2000Ri15}) speculated that even though the 1729-keV \ensuremath{\gamma} ray was not observed in the projectile frame spectrum,\vspace{0.1cm}}&\\
&&&\parbox[t][0.3cm]{15.2243805cm}{\raggedright {\ }{\ }{\ }it would have been possible that the \ensuremath{^{\textnormal{18}}}Ne*(3616.5 keV, 2\ensuremath{^{\textnormal{+}}_{\textnormal{2}}}) state would be significantly populated in their\vspace{0.1cm}}&\\
&&&\parbox[t][0.3cm]{15.2243805cm}{\raggedright {\ }{\ }{\ }measurement specially when the \ensuremath{^{\textnormal{18}}}O*(2\ensuremath{^{\textnormal{+}}_{\textnormal{2}}}) state was strongly populated. Using GEANT simulations, an upper\vspace{0.1cm}}&\\
&&&\parbox[t][0.3cm]{15.2243805cm}{\raggedright {\ }{\ }{\ }limit of 10 mb was deduced for production \ensuremath{\sigma} of the 1729-keV \ensuremath{\gamma} ray from the decay of the \ensuremath{^{\textnormal{18}}}Ne*(2\ensuremath{^{\textnormal{+}}_{\textnormal{2}}}) state\vspace{0.1cm}}&\\
&&&\parbox[t][0.3cm]{15.2243805cm}{\raggedright {\ }{\ }{\ }to the \ensuremath{^{\textnormal{18}}}Ne*(2\ensuremath{^{\textnormal{+}}_{\textnormal{1}}}) state.\vspace{0.1cm}}&\\
\multicolumn{1}{r@{}}{5150}&\multicolumn{1}{@{}l}{\ensuremath{^{{\hyperlink{NE26LEVEL1}{b}}}}}&&&\\
\multicolumn{1}{r@{}}{6150}&\multicolumn{1}{@{}l}{\ensuremath{^{{\hyperlink{NE26LEVEL1}{b}}}}}&&\parbox[t][0.3cm]{15.2243805cm}{\raggedright Mode of decay: \ensuremath{^{\textnormal{2}}}He (diproton) decay to \ensuremath{^{\textnormal{16}}}O\ensuremath{_{\textnormal{g.s.}}} (\href{https://www.nndc.bnl.gov/nsr/nsrlink.jsp?2009Ji02,B}{2009Ji02}, \href{https://www.nndc.bnl.gov/nsr/nsrlink.jsp?2011LiZV,B}{2011LiZV}, \href{https://www.nndc.bnl.gov/nsr/nsrlink.jsp?2016Li45,B}{2016Li45}).\vspace{0.1cm}}&\\
&&&\parbox[t][0.3cm]{15.2243805cm}{\raggedright (\href{https://www.nndc.bnl.gov/nsr/nsrlink.jsp?2016Li45,B}{2016Li45}) performed a Hanbury-Brown Twiss interferometry analysis to probe the space-time character of\vspace{0.1cm}}&\\
&&&\parbox[t][0.3cm]{15.2243805cm}{\raggedright {\ }{\ }{\ }\ensuremath{^{\textnormal{18}}}Ne assuming a Gaussian shape. As a result, the p-p distances inside \ensuremath{^{\textnormal{18}}}Ne nucleus (for the valence proton\vspace{0.1cm}}&\\
\end{longtable}
\begin{textblock}{29}(0,27.3)
Continued on next page (footnotes at end of table)
\end{textblock}
\clearpage
\begin{longtable}{ccc@{\extracolsep{\fill}}c}
\\[-.4cm]
\multicolumn{4}{c}{{\bf \small \underline{\ensuremath{^{\textnormal{197}}}Au(\ensuremath{^{\textnormal{18}}}Ne,\ensuremath{^{\textnormal{18}}}Ne\ensuremath{'}):coulex\hspace{0.2in}\href{https://www.nndc.bnl.gov/nsr/nsrlink.jsp?2000Ri15,B}{2000Ri15},\href{https://www.nndc.bnl.gov/nsr/nsrlink.jsp?2016Li45,B}{2016Li45} (continued)}}}\\
\multicolumn{4}{c}{~}\\
\multicolumn{4}{c}{\underline{\ensuremath{^{18}}Ne Levels (continued)}}\\
\multicolumn{4}{c}{~}\\
\multicolumn{2}{c}{E(level)$^{}$}&Comments&\\[-.2cm]
\multicolumn{2}{c}{\hrulefill}&\hrulefill&
\endhead
&&\parbox[t][0.3cm]{16.24712cm}{\raggedright {\ }{\ }{\ }pair that are emitted in the 2p-decay) was determined to be 5.44 fm \textit{+19{\textminus}17}. This result may indicate a small opening\vspace{0.1cm}}&\\
&&\parbox[t][0.3cm]{16.24712cm}{\raggedright {\ }{\ }{\ }angle between the proton pair, a signature of the crossover of Bardeen-Cooper-Schrieffer (BCS) pairing correlation to\vspace{0.1cm}}&\\
&&\parbox[t][0.3cm]{16.24712cm}{\raggedright {\ }{\ }{\ }the Bose-Einstein Condensation (BEC) pairing correlation in the dilute nuclear matter, i.e., a 2p halo nucleus.\vspace{0.1cm}}&\\
\multicolumn{1}{r@{}}{7060}&\multicolumn{1}{@{}l}{\ensuremath{^{{\hyperlink{NE26LEVEL1}{b}}}}}&&\\
\multicolumn{1}{r@{}}{7910}&\multicolumn{1}{@{}l}{\ensuremath{^{{\hyperlink{NE26LEVEL1}{b}}}}}&&\\
\multicolumn{1}{r@{}}{8500}&\multicolumn{1}{@{}l}{\ensuremath{^{{\hyperlink{NE26LEVEL1}{b}}}}}&&\\
\multicolumn{1}{r@{}}{9600}&\multicolumn{1}{@{}l}{\ensuremath{^{{\hyperlink{NE26LEVEL1}{b}}}}}&&\\
\multicolumn{1}{r@{}}{10900}&\multicolumn{1}{@{}l}{\ensuremath{^{{\hyperlink{NE26LEVEL1}{b}}}}}&&\\
\multicolumn{1}{r@{}}{11500}&\multicolumn{1}{@{}l}{\ensuremath{^{{\hyperlink{NE26LEVEL1}{b}}}}}&&\\
\multicolumn{1}{r@{}}{12100}&\multicolumn{1}{@{}l}{\ensuremath{^{{\hyperlink{NE26LEVEL1}{b}}}}}&&\\
\multicolumn{1}{r@{}}{12700}&\multicolumn{1}{@{}l}{\ensuremath{^{{\hyperlink{NE26LEVEL1}{b}}}}}&&\\
\multicolumn{1}{r@{}}{20700}&\multicolumn{1}{@{}l}{\ensuremath{^{{\hyperlink{NE26LEVEL2}{c}}}}}&\parbox[t][0.3cm]{16.24712cm}{\raggedright (\href{https://www.nndc.bnl.gov/nsr/nsrlink.jsp?2010Xu11,B}{2010Xu11}): mode of decay is most likely via \ensuremath{^{\textnormal{14}}}O* + 2\ensuremath{\alpha} (\href{https://www.nndc.bnl.gov/nsr/nsrlink.jsp?2010Xu11,B}{2010Xu11}).\vspace{0.1cm}}&\\
\multicolumn{1}{r@{}}{23300}&\multicolumn{1}{@{}l}{\ensuremath{^{{\hyperlink{NE26LEVEL2}{c}}}}}&\parbox[t][0.3cm]{16.24712cm}{\raggedright (\href{https://www.nndc.bnl.gov/nsr/nsrlink.jsp?2010Xu11,B}{2010Xu11}): mode of decay is most likely via \ensuremath{^{\textnormal{14}}}O* + 2\ensuremath{\alpha} (\href{https://www.nndc.bnl.gov/nsr/nsrlink.jsp?2010Xu11,B}{2010Xu11}).\vspace{0.1cm}}&\\
\multicolumn{1}{r@{}}{26200}&\multicolumn{1}{@{}l}{\ensuremath{^{{\hyperlink{NE26LEVEL2}{c}}}}}&\parbox[t][0.3cm]{16.24712cm}{\raggedright (\href{https://www.nndc.bnl.gov/nsr/nsrlink.jsp?2010Xu11,B}{2010Xu11}): mode of decay is most likely via \ensuremath{^{\textnormal{14}}}O* + 2\ensuremath{\alpha} (\href{https://www.nndc.bnl.gov/nsr/nsrlink.jsp?2010Xu11,B}{2010Xu11}).\vspace{0.1cm}}&\\
\end{longtable}
\parbox[b][0.3cm]{17.7cm}{\makebox[1ex]{\ensuremath{^{\hypertarget{NE26LEVEL0}{a}}}} From the \ensuremath{^{\textnormal{18}}}Ne Adopted Levels.}\\
\parbox[b][0.3cm]{17.7cm}{\makebox[1ex]{\ensuremath{^{\hypertarget{NE26LEVEL1}{b}}}} From the complete reconstruction of the invariant mass of the \ensuremath{^{\textnormal{16}}}O+2p events observed in coincidence in (\href{https://www.nndc.bnl.gov/nsr/nsrlink.jsp?2009Ji02,B}{2009Ji02}).}\\
\parbox[b][0.3cm]{17.7cm}{\makebox[1ex]{\ensuremath{^{\hypertarget{NE26LEVEL2}{c}}}} From (\href{https://www.nndc.bnl.gov/nsr/nsrlink.jsp?2010Xu11,B}{2010Xu11}).}\\
\vspace{0.5cm}
\underline{$\gamma$($^{18}$Ne)}\\
\begin{longtable}{ccccccc@{}c@{\extracolsep{\fill}}c}
\multicolumn{2}{c}{E\ensuremath{_{\gamma}}}&\multicolumn{2}{c}{E\ensuremath{_{i}}(level)}&J\ensuremath{^{\pi}_{i}}&\multicolumn{2}{c}{E\ensuremath{_{f}}}&J\ensuremath{^{\pi}_{f}}&\\[-.2cm]
\multicolumn{2}{c}{\hrulefill}&\multicolumn{2}{c}{\hrulefill}&\hrulefill&\multicolumn{2}{c}{\hrulefill}&\hrulefill&
\endfirsthead
\multicolumn{1}{r@{}}{1729}&\multicolumn{1}{@{.}l}{2\ensuremath{^{\hyperlink{NE26GAMMA0}{a}\hyperlink{NE26GAMMA1}{b}}}}&\multicolumn{1}{r@{}}{3616}&\multicolumn{1}{@{.}l}{5?}&\multicolumn{1}{l}{2\ensuremath{^{+}}}&\multicolumn{1}{r@{}}{1887}&\multicolumn{1}{@{.}l}{4}&\multicolumn{1}{@{}l}{2\ensuremath{^{+}}}&\\
\multicolumn{1}{r@{}}{1887}&\multicolumn{1}{@{.}l}{3\ensuremath{^{\hyperlink{NE26GAMMA0}{a}}}}&\multicolumn{1}{r@{}}{1887}&\multicolumn{1}{@{.}l}{4}&\multicolumn{1}{l}{2\ensuremath{^{+}}}&\multicolumn{1}{r@{}}{0}&\multicolumn{1}{@{}l}{}&\multicolumn{1}{@{}l}{0\ensuremath{^{+}}}&\\
\end{longtable}
\parbox[b][0.3cm]{17.7cm}{\makebox[1ex]{\ensuremath{^{\hypertarget{NE26GAMMA0}{a}}}} From the \ensuremath{^{\textnormal{18}}}Ne Adopted Gammas.}\\
\parbox[b][0.3cm]{17.7cm}{\makebox[1ex]{\ensuremath{^{\hypertarget{NE26GAMMA1}{b}}}} Placement of transition in the level scheme is uncertain.}\\
\vspace{0.5cm}
\clearpage
\begin{figure}[h]
\begin{center}
\includegraphics{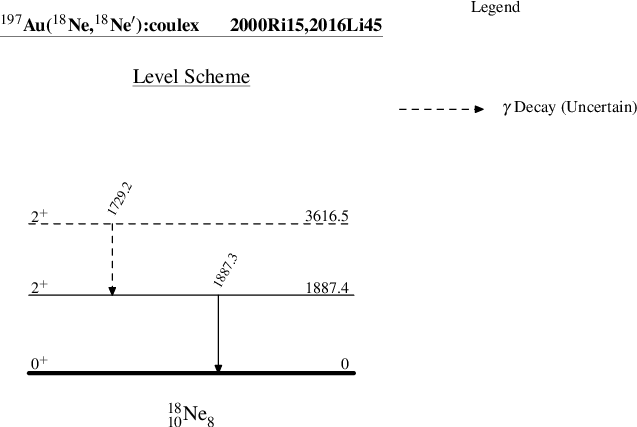}\\
\end{center}
\end{figure}
\clearpage
\subsection[\hspace{-0.2cm}Si(p,\ensuremath{^{\textnormal{18}}}Ne)]{ }
\vspace{-27pt}
\vspace{0.3cm}
\hypertarget{NE27}{{\bf \small \underline{Si(p,\ensuremath{^{\textnormal{18}}}Ne)\hspace{0.2in}\href{https://www.nndc.bnl.gov/nsr/nsrlink.jsp?2005Ba87,B}{2005Ba87},\href{https://www.nndc.bnl.gov/nsr/nsrlink.jsp?2017DuZU,B}{2017DuZU}}}}\\
\vspace{4pt}
\vspace{8pt}
\parbox[b][0.3cm]{17.7cm}{\addtolength{\parindent}{-0.2in}\href{https://www.nndc.bnl.gov/nsr/nsrlink.jsp?2005Ba87,B}{2005Ba87}, \href{https://www.nndc.bnl.gov/nsr/nsrlink.jsp?2007Gr18,B}{2007Gr18}, \href{https://www.nndc.bnl.gov/nsr/nsrlink.jsp?2013Gr03,B}{2013Gr03}: SiC(p,\ensuremath{^{\textnormal{18}}}Ne) E=30 keV (\href{https://www.nndc.bnl.gov/nsr/nsrlink.jsp?2005Ba87,B}{2005Ba87}, \href{https://www.nndc.bnl.gov/nsr/nsrlink.jsp?2007Gr18,B}{2007Gr18}) and E=60 keV (\href{https://www.nndc.bnl.gov/nsr/nsrlink.jsp?2013Gr03,B}{2013Gr03}); implanted the \ensuremath{^{\textnormal{18}}}Ne}\\
\parbox[b][0.3cm]{17.7cm}{beam into a movable mylar backed aluminum tape placed in the mutual center of 8\ensuremath{\pi} spectrometer and Scintillating}\\
\parbox[b][0.3cm]{17.7cm}{Electron-Positron Tagging Array (SCEPTAR). Measured \ensuremath{\beta}-\ensuremath{\gamma} coincidences. Measured the level of \ensuremath{^{\textnormal{18}}}F and H\ensuremath{^{\textnormal{17}}}F contaminants,}\\
\parbox[b][0.3cm]{17.7cm}{whose decays are not followed by a prompt \ensuremath{\gamma} ray, using SCEPTAR. Measured the 1042-keV \ensuremath{\gamma} rays following the superallowed}\\
\parbox[b][0.3cm]{17.7cm}{\ensuremath{\beta}-decay of \ensuremath{^{\textnormal{18}}}Ne*. In (\href{https://www.nndc.bnl.gov/nsr/nsrlink.jsp?2005Ba87,B}{2005Ba87}, \href{https://www.nndc.bnl.gov/nsr/nsrlink.jsp?2007Gr18,B}{2007Gr18}), the implantation took 7 seconds followed by 40 s \ensuremath{\gamma}-ray counting before the tape was}\\
\parbox[b][0.3cm]{17.7cm}{moved 1.5 m away to stop counting, and the cycle repeated. In (\href{https://www.nndc.bnl.gov/nsr/nsrlink.jsp?2013Gr03,B}{2013Gr03}), such cycles consisted of 2.5, 5, or 7 s of background}\\
\parbox[b][0.3cm]{17.7cm}{counting, 120 s of implantation, 40 s counting followed by 1 s for moving the tape outside the detection array. Deduced the relative}\\
\parbox[b][0.3cm]{17.7cm}{\ensuremath{\gamma}-ray intensities, \ensuremath{\beta}-decay branching ratios, and the half-life of \ensuremath{^{\textnormal{18}}}Ne. The result of (\href{https://www.nndc.bnl.gov/nsr/nsrlink.jsp?2013Gr03,B}{2013Gr03}) supersedes that of (\href{https://www.nndc.bnl.gov/nsr/nsrlink.jsp?2007Gr18,B}{2007Gr18}). The}\\
\parbox[b][0.3cm]{17.7cm}{results of (\href{https://www.nndc.bnl.gov/nsr/nsrlink.jsp?2005Ba87,B}{2005Ba87}) were preliminary and should not be used.}\\
\parbox[b][0.3cm]{17.7cm}{\addtolength{\parindent}{-0.2in}\href{https://www.nndc.bnl.gov/nsr/nsrlink.jsp?2014LaZV,B}{2014LaZV}, \href{https://www.nndc.bnl.gov/nsr/nsrlink.jsp?2015La19,B}{2015La19}, \href{https://www.nndc.bnl.gov/nsr/nsrlink.jsp?2017DuZU,B}{2017DuZU}: SiC(p,\ensuremath{^{\textnormal{18}}}Ne) E=30 keV; measured the half-life of \ensuremath{^{\textnormal{18}}}Ne in two independent \ensuremath{\beta}\ensuremath{^{\textnormal{+}}}-counting}\\
\parbox[b][0.3cm]{17.7cm}{measurements:}\\
\parbox[b][0.3cm]{17.7cm}{\addtolength{\parindent}{-0.2in}A 30-keV \ensuremath{^{\textnormal{18}}}Ne beam was implanted in the thick aluminum layer of a single-sided aluminized Mylar tape (\href{https://www.nndc.bnl.gov/nsr/nsrlink.jsp?2014LaZV,B}{2014LaZV}). Once a}\\
\parbox[b][0.3cm]{17.7cm}{sufficient amount of \ensuremath{^{\textnormal{18}}}Ne activity was implanted, the beam was deflected, and the tape was transferred to the center of a 4\ensuremath{\pi}}\\
\parbox[b][0.3cm]{17.7cm}{continuous-flow gas proportional counter for counting. Ions were implanted in 440 beam-on, beam-off cycles, each consisting of a}\\
\parbox[b][0.3cm]{17.7cm}{0.5 s of implantation followed by 1.5-4.5 s for cooling, 2 s for transfer to the counting station and 40 s (\ensuremath{\approx}24 half-lives) of}\\
\parbox[b][0.3cm]{17.7cm}{\ensuremath{\beta}-counting. In this experiment, a cold-transfer line was used between the spallation target and the ion source to reduce the \ensuremath{^{\textnormal{18}}}F}\\
\parbox[b][0.3cm]{17.7cm}{beam contaminant to undetectable levels.}\\
\parbox[b][0.3cm]{17.7cm}{\addtolength{\parindent}{-0.2in}In the second experiment, the cold-transfer line was not utilized. This resulted in a higher intensity \ensuremath{\approx}4.8\ensuremath{\times}10\ensuremath{^{\textnormal{5}}} \ensuremath{^{\textnormal{18}}}Ne ions/s beam}\\
\parbox[b][0.3cm]{17.7cm}{with a 1\ensuremath{\times}10\ensuremath{^{\textnormal{6}}} \ensuremath{^{\textnormal{18}}}F ions/s contaminant. In this case, the beam was implanted in the same kind of tape but at 20 keV during a total}\\
\parbox[b][0.3cm]{17.7cm}{of 813 beam-on beam-off cycles. The \ensuremath{\beta}-counting was carried out with a different 4\ensuremath{\pi} gas proportional counter at lower operating}\\
\parbox[b][0.3cm]{17.7cm}{voltage. The \ensuremath{^{\textnormal{18}}}Ne half life was measured independently in both experiments.}\\
\vspace{12pt}
\underline{$^{18}$Ne Levels}\\
\begin{longtable}{cccccc@{\extracolsep{\fill}}c}
\multicolumn{2}{c}{E(level)$^{{\hyperlink{NE27LEVEL0}{a}}}$}&J$^{\pi}$$^{{\hyperlink{NE27LEVEL0}{a}}}$&\multicolumn{2}{c}{T$_{1/2}$$^{{\hyperlink{NE27LEVEL1}{b}}}$}&Comments&\\[-.2cm]
\multicolumn{2}{c}{\hrulefill}&\hrulefill&\multicolumn{2}{c}{\hrulefill}&\hrulefill&
\endfirsthead
\multicolumn{1}{r@{}}{0}&\multicolumn{1}{@{}l}{}&\multicolumn{1}{l}{0\ensuremath{^{+}}}&\multicolumn{1}{r@{}}{1}&\multicolumn{1}{@{.}l}{66415 s {\it +57\textminus48}}&\parbox[t][0.3cm]{12.24354cm}{\raggedright \%\ensuremath{\varepsilon}+\%\ensuremath{\beta}\ensuremath{^{\textnormal{+}}}=100 (\href{https://www.nndc.bnl.gov/nsr/nsrlink.jsp?1981Ad01,B}{1981Ad01})\vspace{0.1cm}}&\\
&&&&&\parbox[t][0.3cm]{12.24354cm}{\raggedright Decays to the \ensuremath{^{\textnormal{18}}}F(g.s., 1042-, 1081-, and 1701-keV) states (\href{https://www.nndc.bnl.gov/nsr/nsrlink.jsp?2013Gr03,B}{2013Gr03}).\vspace{0.1cm}}&\\
\end{longtable}
\parbox[b][0.3cm]{17.7cm}{\makebox[1ex]{\ensuremath{^{\hypertarget{NE27LEVEL0}{a}}}} From the \ensuremath{^{\textnormal{18}}}Ne Adopted Levels.}\\
\parbox[b][0.3cm]{17.7cm}{\makebox[1ex]{\ensuremath{^{\hypertarget{NE27LEVEL1}{b}}}} From the weighted average of T\ensuremath{_{\textnormal{1/2}}}=1.6648 s \textit{11} (\href{https://www.nndc.bnl.gov/nsr/nsrlink.jsp?2013Gr03,B}{2013Gr03}) and T\ensuremath{_{\textnormal{1/2}}}=1.66400 s \textit{+57{\textminus}48} (\href{https://www.nndc.bnl.gov/nsr/nsrlink.jsp?2015La19,B}{2015La19}). The evaluator assumed that}\\
\parbox[b][0.3cm]{17.7cm}{{\ }{\ }the uncertainties are systematic, and therefore cannot decrease.}\\
\vspace{0.5cm}
\clearpage
\subsection[\hspace{-0.2cm}Pb(\ensuremath{^{\textnormal{18}}}Ne,\ensuremath{^{\textnormal{18}}}Ne\ensuremath{'}):coulex]{ }
\vspace{-27pt}
\vspace{0.3cm}
\hypertarget{NE28}{{\bf \small \underline{Pb(\ensuremath{^{\textnormal{18}}}Ne,\ensuremath{^{\textnormal{18}}}Ne\ensuremath{'}):coulex\hspace{0.2in}\href{https://www.nndc.bnl.gov/nsr/nsrlink.jsp?2007Ra36,B}{2007Ra36},\href{https://www.nndc.bnl.gov/nsr/nsrlink.jsp?2010Gi05,B}{2010Gi05}}}}\\
\vspace{4pt}
\vspace{8pt}
\parbox[b][0.3cm]{17.7cm}{\addtolength{\parindent}{-0.2in}\href{https://www.nndc.bnl.gov/nsr/nsrlink.jsp?2007Ra36,B}{2007Ra36}, \href{https://www.nndc.bnl.gov/nsr/nsrlink.jsp?2007CaZT,B}{2007CaZT}, \href{https://www.nndc.bnl.gov/nsr/nsrlink.jsp?2008SfZZ,B}{2008SfZZ}, \href{https://www.nndc.bnl.gov/nsr/nsrlink.jsp?2008Ra12,B}{2008Ra12}, \href{https://www.nndc.bnl.gov/nsr/nsrlink.jsp?2010Ra14,B}{2010Ra14}, \href{https://www.nndc.bnl.gov/nsr/nsrlink.jsp?2010RaZZ,B}{2010RaZZ}, \href{https://www.nndc.bnl.gov/nsr/nsrlink.jsp?2010Gi05,B}{2010Gi05}: Pb(\ensuremath{^{\textnormal{18}}}Ne,\ensuremath{^{\textnormal{17}}}F+p) and Pb(\ensuremath{^{\textnormal{18}}}Ne,\ensuremath{^{\textnormal{16}}}O+2p) E=32{\textminus}35}\\
\parbox[b][0.3cm]{17.7cm}{MeV/nucleon; targets: \ensuremath{^{\textnormal{207}}}Pb (\href{https://www.nndc.bnl.gov/nsr/nsrlink.jsp?2007Ra36,B}{2007Ra36}) and \ensuremath{^{\textnormal{nat}}}Pb (other studies); measured the angles, energies and velocities of heavy reaction}\\
\parbox[b][0.3cm]{17.7cm}{and light decay products in coincidence using two Si-CsI hodoscopes covering an angular range of \ensuremath{\theta}\ensuremath{_{\textnormal{lab}}}=\ensuremath{\pm}4.5\ensuremath{^\circ} to \ensuremath{\pm}16.5\ensuremath{^\circ}}\\
\parbox[b][0.3cm]{17.7cm}{(\href{https://www.nndc.bnl.gov/nsr/nsrlink.jsp?2007Ra36,B}{2007Ra36}), \ensuremath{\theta}\ensuremath{_{\textnormal{lab}}}=0\ensuremath{^\circ}{\textminus}20\ensuremath{^\circ} in (\href{https://www.nndc.bnl.gov/nsr/nsrlink.jsp?2007CaZT,B}{2007CaZT}), and \ensuremath{\theta}\ensuremath{_{\textnormal{lab}}}=\ensuremath{\pm}5\ensuremath{^\circ} to \ensuremath{\pm}21.5\ensuremath{^\circ} in (\href{https://www.nndc.bnl.gov/nsr/nsrlink.jsp?2008SfZZ,B}{2008SfZZ}, \href{https://www.nndc.bnl.gov/nsr/nsrlink.jsp?2008Ra12,B}{2008Ra12}, \href{https://www.nndc.bnl.gov/nsr/nsrlink.jsp?2010Gi05,B}{2010Gi05}, and \href{https://www.nndc.bnl.gov/nsr/nsrlink.jsp?2010Ra14,B}{2010Ra14},}\\
\parbox[b][0.3cm]{17.7cm}{\href{https://www.nndc.bnl.gov/nsr/nsrlink.jsp?2010RaZZ,B}{2010RaZZ}). Deduced the invariant mass from a reconstruction of the \ensuremath{^{\textnormal{17}}}F+p and \ensuremath{^{\textnormal{16}}}O+2p complete decay kinematics in the}\\
\parbox[b][0.3cm]{17.7cm}{center-of-mass frame. The experimental energy resolution varied between 250 keV (\href{https://www.nndc.bnl.gov/nsr/nsrlink.jsp?2007Ra36,B}{2007Ra36}) to 500 keV (other studies). The}\\
\parbox[b][0.3cm]{17.7cm}{particle decay of the observed states are discussed.}\\
\vspace{0.385cm}
\parbox[b][0.3cm]{17.7cm}{\addtolength{\parindent}{-0.2in}\textit{Theory}:}\\
\parbox[b][0.3cm]{17.7cm}{\addtolength{\parindent}{-0.2in}\href{https://www.nndc.bnl.gov/nsr/nsrlink.jsp?2007Be54,B}{2007Be54}: \ensuremath{^{\textnormal{208}}}Pb(\ensuremath{^{\textnormal{18}}}Ne,\ensuremath{^{\textnormal{18}}}Ne\ensuremath{'}); Coulomb excitation of low-lying states of unstable, light and medium heavy nuclei in intermediate}\\
\parbox[b][0.3cm]{17.7cm}{collision energies of interest to radioactive beam facilities is investigated. Coulomb excitation cross sections of numerous projectiles}\\
\parbox[b][0.3cm]{17.7cm}{incident on Pb and Au targets at laboratory bombarding energies of 10, 20, 30, 50, 100, 200, 500 MeV/nucleon and retardation}\\
\parbox[b][0.3cm]{17.7cm}{effects are calculated. B(E1:J\ensuremath{_{\textnormal{g.s.}}}\ensuremath{\rightarrow}J\ensuremath{_{\textnormal{f}}}), B(E2:J\ensuremath{_{\textnormal{g.s.}}}\ensuremath{\rightarrow}J\ensuremath{_{\textnormal{f}}}), or B(M1:J\ensuremath{_{\textnormal{g.s.}}}\ensuremath{\rightarrow}J\ensuremath{_{\textnormal{f}}}) are obtained for the lowest-lying transitions of various}\\
\parbox[b][0.3cm]{17.7cm}{reaction products, including \ensuremath{^{\textnormal{18}}}Ne. Comparison with data are discussed. The calculated Coulomb excitation cross section for}\\
\parbox[b][0.3cm]{17.7cm}{\ensuremath{^{\textnormal{18}}}Ne*(1.89 MeV) is very large (615 mb at 10 MeV/nucleon, reducing to 22.1 mb at 500 MeV/nucleon). B(E2:0\ensuremath{^{\textnormal{+}}_{\textnormal{1}}}\ensuremath{\rightarrow}2\ensuremath{^{\textnormal{+}}_{\textnormal{1}}})=248}\\
\parbox[b][0.3cm]{17.7cm}{e\ensuremath{^{\textnormal{2}}}fm\ensuremath{^{\textnormal{4}}} is calculated by (\href{https://www.nndc.bnl.gov/nsr/nsrlink.jsp?2007Be54,B}{2007Be54}).}\\
\parbox[b][0.3cm]{17.7cm}{\addtolength{\parindent}{-0.2in}N. Yu, E. Maglione and L. S. Ferreira, Nucl. Sci. Tech., 24 (2013) 050517 and \href{https://www.nndc.bnl.gov/nsr/nsrlink.jsp?2024Fe02,B}{2024Fe02}: Used the shell model interaction of}\\
\parbox[b][0.3cm]{17.7cm}{(\href{https://www.nndc.bnl.gov/nsr/nsrlink.jsp?2011Bo20,B}{2011Bo20}) to fit the experimentally obtained excitation energies of (\href{https://www.nndc.bnl.gov/nsr/nsrlink.jsp?2008Ra12,B}{2008Ra12}). Assuming \ensuremath{^{\textnormal{18}}}Ne as a spherical nucleus, in 2013,}\\
\parbox[b][0.3cm]{17.7cm}{the authors deduced excited states with negative parity, some of which are narrow resonances with excitation energies above 10}\\
\parbox[b][0.3cm]{17.7cm}{MeV. These theoretically calculated states decay preferentially by one proton emission to \ensuremath{^{\textnormal{17}}}F*, and are therefore possible}\\
\parbox[b][0.3cm]{17.7cm}{candidates for a sequential two-proton decay.}\\
\vspace{12pt}
\underline{$^{18}$Ne Levels}\\
\vspace{0.34cm}
\parbox[b][0.3cm]{17.7cm}{\addtolength{\parindent}{-0.254cm}(\href{https://www.nndc.bnl.gov/nsr/nsrlink.jsp?2008Ra12,B}{2008Ra12}) determined branching ratios for the decay of the \ensuremath{^{\textnormal{18}}}Ne*(\ensuremath{>}6.5 MeV) states: 64\% \textit{7} for 3-body direct break up involving}\\
\parbox[b][0.3cm]{17.7cm}{an uncorrelated emission of two protons (usually referred to as democratic emission), 30\% \textit{4} for true sequential decay via the}\\
\parbox[b][0.3cm]{17.7cm}{\ensuremath{^{\textnormal{17}}}F(3.1 MeV, 1/2\ensuremath{^{-}}) state, and 6\% \textit{2} for \ensuremath{^{\textnormal{2}}}He decay.}\\
\vspace{0.34cm}
\begin{longtable}{cccc@{\extracolsep{\fill}}c}
\multicolumn{2}{c}{E(level)$^{{\hyperlink{NE28LEVEL0}{a}}}$}&J$^{\pi}$$^{{\hyperlink{NE28LEVEL1}{b}}}$&Comments&\\[-.2cm]
\multicolumn{2}{c}{\hrulefill}&\hrulefill&\hrulefill&
\endfirsthead
\multicolumn{1}{r@{}}{5090}&\multicolumn{1}{@{}l}{\ensuremath{^{{\hyperlink{NE28LEVEL2}{c}}}}}&\multicolumn{1}{l}{2\ensuremath{^{+}}}&\parbox[t][0.3cm]{14.640161cm}{\raggedright E(level): From (\href{https://www.nndc.bnl.gov/nsr/nsrlink.jsp?2008SfZZ,B}{2008SfZZ}, \href{https://www.nndc.bnl.gov/nsr/nsrlink.jsp?2008Ra12,B}{2008Ra12}, \href{https://www.nndc.bnl.gov/nsr/nsrlink.jsp?2010Gi05,B}{2010Gi05}, and \href{https://www.nndc.bnl.gov/nsr/nsrlink.jsp?2010Ra14,B}{2010Ra14}). See also 5110 keV (\href{https://www.nndc.bnl.gov/nsr/nsrlink.jsp?2007Ra36,B}{2007Ra36}).\vspace{0.1cm}}&\\
\multicolumn{1}{r@{}}{5150}&\multicolumn{1}{@{}l}{\ensuremath{^{{\hyperlink{NE28LEVEL2}{c}}}}}&\multicolumn{1}{l}{2\ensuremath{^{+}}}&\parbox[t][0.3cm]{14.640161cm}{\raggedright E(level): From (\href{https://www.nndc.bnl.gov/nsr/nsrlink.jsp?2007Ra36,B}{2007Ra36}, \href{https://www.nndc.bnl.gov/nsr/nsrlink.jsp?2008SfZZ,B}{2008SfZZ}, \href{https://www.nndc.bnl.gov/nsr/nsrlink.jsp?2008Ra12,B}{2008Ra12}, \href{https://www.nndc.bnl.gov/nsr/nsrlink.jsp?2010Gi05,B}{2010Gi05}, and \href{https://www.nndc.bnl.gov/nsr/nsrlink.jsp?2010Ra14,B}{2010Ra14}).\vspace{0.1cm}}&\\
\multicolumn{1}{r@{}}{6150}&\multicolumn{1}{@{}l}{\ensuremath{^{{\hyperlink{NE28LEVEL2}{c}}{\hyperlink{NE28LEVEL3}{d}}}}}&\multicolumn{1}{l}{1\ensuremath{^{-}}}&\parbox[t][0.3cm]{14.640161cm}{\raggedright E(level): Observed in both \ensuremath{^{\textnormal{17}}}F+p and \ensuremath{^{\textnormal{16}}}O+2p decay channels (\href{https://www.nndc.bnl.gov/nsr/nsrlink.jsp?2007Ra36,B}{2007Ra36}, \href{https://www.nndc.bnl.gov/nsr/nsrlink.jsp?2008SfZZ,B}{2008SfZZ}, \href{https://www.nndc.bnl.gov/nsr/nsrlink.jsp?2008Ra12,B}{2008Ra12}, \href{https://www.nndc.bnl.gov/nsr/nsrlink.jsp?2010Gi05,B}{2010Gi05},\vspace{0.1cm}}&\\
&&&\parbox[t][0.3cm]{14.640161cm}{\raggedright {\ }{\ }{\ }and \href{https://www.nndc.bnl.gov/nsr/nsrlink.jsp?2010Ra14,B}{2010Ra14}) but not resolved from the neighboring states.\vspace{0.1cm}}&\\
&&&\parbox[t][0.3cm]{14.640161cm}{\raggedright The cross section of the \ensuremath{^{\textnormal{17}}}F+p decay channel in the excitation energy region of 6 MeV was extracted in\vspace{0.1cm}}&\\
&&&\parbox[t][0.3cm]{14.640161cm}{\raggedright {\ }{\ }{\ }(\href{https://www.nndc.bnl.gov/nsr/nsrlink.jsp?2007CaZT,B}{2007CaZT}) to be of the order of 20 mb \textit{10}.\vspace{0.1cm}}&\\
&&&\parbox[t][0.3cm]{14.640161cm}{\raggedright This state has various decay modes (see the Adopted Levels), including 2p decay mode. The 2-proton decay\vspace{0.1cm}}&\\
&&&\parbox[t][0.3cm]{14.640161cm}{\raggedright {\ }{\ }{\ }mode proceeds through (1) a \ensuremath{^{\textnormal{2}}}He (diproton) resonance leading to the \ensuremath{^{\textnormal{16}}}O\ensuremath{_{\textnormal{g.s.}}} 31\% \textit{7} of the time\vspace{0.1cm}}&\\
&&&\parbox[t][0.3cm]{14.640161cm}{\raggedright {\ }{\ }{\ }(\href{https://www.nndc.bnl.gov/nsr/nsrlink.jsp?2008Ra12,B}{2008Ra12}, \href{https://www.nndc.bnl.gov/nsr/nsrlink.jsp?2010Gi05,B}{2010Gi05}, \href{https://www.nndc.bnl.gov/nsr/nsrlink.jsp?2010Ra14,B}{2010Ra14}); (2) 2p democratic three-body decay 66\% \textit{9} of the time (\href{https://www.nndc.bnl.gov/nsr/nsrlink.jsp?2008Ra12,B}{2008Ra12},\vspace{0.1cm}}&\\
&&&\parbox[t][0.3cm]{14.640161cm}{\raggedright {\ }{\ }{\ }\href{https://www.nndc.bnl.gov/nsr/nsrlink.jsp?2010Gi05,B}{2010Gi05}, \href{https://www.nndc.bnl.gov/nsr/nsrlink.jsp?2010Ra14,B}{2010Ra14}); and (3) 2p virtual sequential decay via the \ensuremath{^{\textnormal{17}}}F*(3.1 MeV, 1/2\ensuremath{^{-}}) state 3\% \textit{2} of the\vspace{0.1cm}}&\\
&&&\parbox[t][0.3cm]{14.640161cm}{\raggedright {\ }{\ }{\ }time (\href{https://www.nndc.bnl.gov/nsr/nsrlink.jsp?2008Ra12,B}{2008Ra12}, \href{https://www.nndc.bnl.gov/nsr/nsrlink.jsp?2010Gi05,B}{2010Gi05}, \href{https://www.nndc.bnl.gov/nsr/nsrlink.jsp?2010Ra14,B}{2010Ra14}).\vspace{0.1cm}}&\\
&&&\parbox[t][0.3cm]{14.640161cm}{\raggedright The first experimental proof for diproton emission from this state comes from a combination of (\href{https://www.nndc.bnl.gov/nsr/nsrlink.jsp?2008SfZZ,B}{2008SfZZ},\vspace{0.1cm}}&\\
&&&\parbox[t][0.3cm]{14.640161cm}{\raggedright {\ }{\ }{\ }\href{https://www.nndc.bnl.gov/nsr/nsrlink.jsp?2008Ra12,B}{2008Ra12}) analyses.\vspace{0.1cm}}&\\
\multicolumn{1}{r@{}}{7060}&\multicolumn{1}{@{}l}{\ensuremath{^{{\hyperlink{NE28LEVEL3}{d}}{\hyperlink{NE28LEVEL4}{e}}{\hyperlink{NE28LEVEL6}{g}}}}}&\multicolumn{1}{l}{(1\ensuremath{^{-}},2\ensuremath{^{+}})}&\parbox[t][0.3cm]{14.640161cm}{\raggedright E(level): From (\href{https://www.nndc.bnl.gov/nsr/nsrlink.jsp?2007Ra36,B}{2007Ra36}, \href{https://www.nndc.bnl.gov/nsr/nsrlink.jsp?2008SfZZ,B}{2008SfZZ}, \href{https://www.nndc.bnl.gov/nsr/nsrlink.jsp?2008Ra12,B}{2008Ra12}, \href{https://www.nndc.bnl.gov/nsr/nsrlink.jsp?2010Gi05,B}{2010Gi05}, \href{https://www.nndc.bnl.gov/nsr/nsrlink.jsp?2010Ra14,B}{2010Ra14}). Note that this state is mistakenly\vspace{0.1cm}}&\\
&&&\parbox[t][0.3cm]{14.640161cm}{\raggedright {\ }{\ }{\ }labeled as 7.59 MeV (instead of 7.059 MeV) in (\href{https://www.nndc.bnl.gov/nsr/nsrlink.jsp?2007Ra36,B}{2007Ra36}). In this study, it is shown on Fig. 7 that this\vspace{0.1cm}}&\\
&&&\parbox[t][0.3cm]{14.640161cm}{\raggedright {\ }{\ }{\ }state is populated via the \ensuremath{^{\textnormal{17}}}F+p decay channel. The higher statistics of the studies later carried out by\vspace{0.1cm}}&\\
&&&\parbox[t][0.3cm]{14.640161cm}{\raggedright {\ }{\ }{\ }(\href{https://www.nndc.bnl.gov/nsr/nsrlink.jsp?2008Ra12,B}{2008Ra12}, \href{https://www.nndc.bnl.gov/nsr/nsrlink.jsp?2010Gi05,B}{2010Gi05}) indicate that this state is also populated in the \ensuremath{^{\textnormal{16}}}O+2p channel.\vspace{0.1cm}}&\\
\multicolumn{1}{r@{}}{7910}&\multicolumn{1}{@{}l}{\ensuremath{^{{\hyperlink{NE28LEVEL3}{d}}{\hyperlink{NE28LEVEL4}{e}}{\hyperlink{NE28LEVEL6}{g}}}}}&\multicolumn{1}{l}{(1\ensuremath{^{-}},2\ensuremath{^{+}})}&\parbox[t][0.3cm]{14.640161cm}{\raggedright E(level): From (\href{https://www.nndc.bnl.gov/nsr/nsrlink.jsp?2008SfZZ,B}{2008SfZZ}, \href{https://www.nndc.bnl.gov/nsr/nsrlink.jsp?2008Ra12,B}{2008Ra12}, \href{https://www.nndc.bnl.gov/nsr/nsrlink.jsp?2010Gi05,B}{2010Gi05}, \href{https://www.nndc.bnl.gov/nsr/nsrlink.jsp?2010Ra14,B}{2010Ra14}).\vspace{0.1cm}}&\\
\multicolumn{1}{r@{}}{8500}&\multicolumn{1}{@{}l}{\ensuremath{^{{\hyperlink{NE28LEVEL3}{d}}{\hyperlink{NE28LEVEL4}{e}}{\hyperlink{NE28LEVEL5}{f}}}}}&\multicolumn{1}{l}{(1\ensuremath{^{-}},2\ensuremath{^{+}})}&\parbox[t][0.3cm]{14.640161cm}{\raggedright E(level): From (\href{https://www.nndc.bnl.gov/nsr/nsrlink.jsp?2007Ra36,B}{2007Ra36}, \href{https://www.nndc.bnl.gov/nsr/nsrlink.jsp?2008SfZZ,B}{2008SfZZ}, \href{https://www.nndc.bnl.gov/nsr/nsrlink.jsp?2008Ra12,B}{2008Ra12}, \href{https://www.nndc.bnl.gov/nsr/nsrlink.jsp?2010Gi05,B}{2010Gi05}, \href{https://www.nndc.bnl.gov/nsr/nsrlink.jsp?2010Ra14,B}{2010Ra14}). In (\href{https://www.nndc.bnl.gov/nsr/nsrlink.jsp?2007Ra36,B}{2007Ra36}), it is shown on Fig.\vspace{0.1cm}}&\\
&&&\parbox[t][0.3cm]{14.640161cm}{\raggedright {\ }{\ }{\ }7 that this state is populated via the \ensuremath{^{\textnormal{17}}}F+p channel. The higher statistics of the studies by (\href{https://www.nndc.bnl.gov/nsr/nsrlink.jsp?2008Ra12,B}{2008Ra12},\vspace{0.1cm}}&\\
&&&\parbox[t][0.3cm]{14.640161cm}{\raggedright {\ }{\ }{\ }\href{https://www.nndc.bnl.gov/nsr/nsrlink.jsp?2010Gi05,B}{2010Gi05}) carried out later indicate that this state was populated in the \ensuremath{^{\textnormal{16}}}O+2p channel. (\href{https://www.nndc.bnl.gov/nsr/nsrlink.jsp?2008Ra12,B}{2008Ra12})\vspace{0.1cm}}&\\
&&&\parbox[t][0.3cm]{14.640161cm}{\raggedright {\ }{\ }{\ }specifically mentions that \textit{this} \textit{state is not observed in the \ensuremath{^{17}}F+p decay channel}.\vspace{0.1cm}}&\\
&&&\parbox[t][0.3cm]{14.640161cm}{\raggedright J\ensuremath{^{\pi}}: From (\href{https://www.nndc.bnl.gov/nsr/nsrlink.jsp?2007Ra36,B}{2007Ra36}, \href{https://www.nndc.bnl.gov/nsr/nsrlink.jsp?2008Ra12,B}{2008Ra12}).\vspace{0.1cm}}&\\
\end{longtable}
\begin{textblock}{29}(0,27.3)
Continued on next page (footnotes at end of table)
\end{textblock}
\clearpage
\begin{longtable}{cccc@{\extracolsep{\fill}}c}
\\[-.4cm]
\multicolumn{5}{c}{{\bf \small \underline{Pb(\ensuremath{^{\textnormal{18}}}Ne,\ensuremath{^{\textnormal{18}}}Ne\ensuremath{'}):coulex\hspace{0.2in}\href{https://www.nndc.bnl.gov/nsr/nsrlink.jsp?2007Ra36,B}{2007Ra36},\href{https://www.nndc.bnl.gov/nsr/nsrlink.jsp?2010Gi05,B}{2010Gi05} (continued)}}}\\
\multicolumn{5}{c}{~}\\
\multicolumn{5}{c}{\underline{\ensuremath{^{18}}Ne Levels (continued)}}\\
\multicolumn{5}{c}{~}\\
\multicolumn{2}{c}{E(level)$^{{\hyperlink{NE28LEVEL0}{a}}}$}&J$^{\pi}$$^{{\hyperlink{NE28LEVEL1}{b}}}$&Comments&\\[-.2cm]
\multicolumn{2}{c}{\hrulefill}&\hrulefill&\hrulefill&
\endhead
&&&\parbox[t][0.3cm]{14.250681cm}{\raggedright This state is a candidate for diproton decay (\href{https://www.nndc.bnl.gov/nsr/nsrlink.jsp?2008Ra12,B}{2008Ra12}).\vspace{0.1cm}}&\\
\multicolumn{1}{r@{}}{10700}&\multicolumn{1}{@{}l}{\ensuremath{^{{\hyperlink{NE28LEVEL3}{d}}{\hyperlink{NE28LEVEL4}{e}}{\hyperlink{NE28LEVEL5}{f}}{\hyperlink{NE28LEVEL6}{g}}}}}&\multicolumn{1}{l}{(1\ensuremath{^{-}},2\ensuremath{^{+}})}&\parbox[t][0.3cm]{14.250681cm}{\raggedright E(level): From (\href{https://www.nndc.bnl.gov/nsr/nsrlink.jsp?2008SfZZ,B}{2008SfZZ}, \href{https://www.nndc.bnl.gov/nsr/nsrlink.jsp?2008Ra12,B}{2008Ra12}, \href{https://www.nndc.bnl.gov/nsr/nsrlink.jsp?2010Gi05,B}{2010Gi05}, \href{https://www.nndc.bnl.gov/nsr/nsrlink.jsp?2010Ra14,B}{2010Ra14}).\vspace{0.1cm}}&\\
&&&\parbox[t][0.3cm]{14.250681cm}{\raggedright J\ensuremath{^{\pi}}: From (\href{https://www.nndc.bnl.gov/nsr/nsrlink.jsp?2008Ra12,B}{2008Ra12}).\vspace{0.1cm}}&\\
\multicolumn{1}{r@{}}{12500}&\multicolumn{1}{@{}l}{\ensuremath{^{{\hyperlink{NE28LEVEL3}{d}}{\hyperlink{NE28LEVEL4}{e}}{\hyperlink{NE28LEVEL5}{f}}}}}&\multicolumn{1}{l}{(1\ensuremath{^{-}},2\ensuremath{^{+}})}&\parbox[t][0.3cm]{14.250681cm}{\raggedright E(level): From (\href{https://www.nndc.bnl.gov/nsr/nsrlink.jsp?2008SfZZ,B}{2008SfZZ}, \href{https://www.nndc.bnl.gov/nsr/nsrlink.jsp?2008Ra12,B}{2008Ra12}, \href{https://www.nndc.bnl.gov/nsr/nsrlink.jsp?2010Gi05,B}{2010Gi05}, \href{https://www.nndc.bnl.gov/nsr/nsrlink.jsp?2010Ra14,B}{2010Ra14}).\vspace{0.1cm}}&\\
&&&\parbox[t][0.3cm]{14.250681cm}{\raggedright J\ensuremath{^{\pi}}: From (\href{https://www.nndc.bnl.gov/nsr/nsrlink.jsp?2008Ra12,B}{2008Ra12}).\vspace{0.1cm}}&\\
\multicolumn{1}{r@{}}{13700}&\multicolumn{1}{@{}l}{\ensuremath{^{{\hyperlink{NE28LEVEL3}{d}}{\hyperlink{NE28LEVEL4}{e}}{\hyperlink{NE28LEVEL5}{f}}{\hyperlink{NE28LEVEL6}{g}}}}}&\multicolumn{1}{l}{(1\ensuremath{^{-}},2\ensuremath{^{+}})}&\parbox[t][0.3cm]{14.250681cm}{\raggedright E(level): From (\href{https://www.nndc.bnl.gov/nsr/nsrlink.jsp?2008SfZZ,B}{2008SfZZ}, \href{https://www.nndc.bnl.gov/nsr/nsrlink.jsp?2008Ra12,B}{2008Ra12}, \href{https://www.nndc.bnl.gov/nsr/nsrlink.jsp?2010Gi05,B}{2010Gi05}, \href{https://www.nndc.bnl.gov/nsr/nsrlink.jsp?2010Ra14,B}{2010Ra14}).\vspace{0.1cm}}&\\
&&&\parbox[t][0.3cm]{14.250681cm}{\raggedright J\ensuremath{^{\pi}}: From (\href{https://www.nndc.bnl.gov/nsr/nsrlink.jsp?2008Ra12,B}{2008Ra12}).\vspace{0.1cm}}&\\
\end{longtable}
\parbox[b][0.3cm]{17.7cm}{\makebox[1ex]{\ensuremath{^{\hypertarget{NE28LEVEL0}{a}}}} From invariant mass spectra of \ensuremath{^{\textnormal{18}}}Ne created from the \ensuremath{^{\textnormal{16}}}O+2p and \ensuremath{^{\textnormal{17}}}F+p events (\href{https://www.nndc.bnl.gov/nsr/nsrlink.jsp?2007Ra36,B}{2007Ra36}, \href{https://www.nndc.bnl.gov/nsr/nsrlink.jsp?2007CaZT,B}{2007CaZT}, \href{https://www.nndc.bnl.gov/nsr/nsrlink.jsp?2008SfZZ,B}{2008SfZZ}, \href{https://www.nndc.bnl.gov/nsr/nsrlink.jsp?2008Ra12,B}{2008Ra12},}\\
\parbox[b][0.3cm]{17.7cm}{{\ }{\ }\href{https://www.nndc.bnl.gov/nsr/nsrlink.jsp?2010Gi05,B}{2010Gi05}, and \href{https://www.nndc.bnl.gov/nsr/nsrlink.jsp?2010Ra14,B}{2010Ra14}).}\\
\parbox[b][0.3cm]{17.7cm}{\makebox[1ex]{\ensuremath{^{\hypertarget{NE28LEVEL1}{b}}}} It is assumed by the evaluator that the reported J\ensuremath{^{\ensuremath{\pi}}} in (\href{https://www.nndc.bnl.gov/nsr/nsrlink.jsp?2007Ra36,B}{2007Ra36}, \href{https://www.nndc.bnl.gov/nsr/nsrlink.jsp?2007CaZT,B}{2007CaZT}, \href{https://www.nndc.bnl.gov/nsr/nsrlink.jsp?2008SfZZ,B}{2008SfZZ}, \href{https://www.nndc.bnl.gov/nsr/nsrlink.jsp?2008Ra12,B}{2008Ra12}, \href{https://www.nndc.bnl.gov/nsr/nsrlink.jsp?2010Gi05,B}{2010Gi05}, and \href{https://www.nndc.bnl.gov/nsr/nsrlink.jsp?2010Ra14,B}{2010Ra14})}\\
\parbox[b][0.3cm]{17.7cm}{{\ }{\ }are based on the Coulomb excitation selection rules for spin zero targets (J\ensuremath{^{\ensuremath{\pi}}}=1\ensuremath{^{-}} and 2\ensuremath{^{\textnormal{+}}} for E1 and E2 transitions, respectively).}\\
\parbox[b][0.3cm]{17.7cm}{\makebox[1ex]{\ensuremath{^{\hypertarget{NE28LEVEL2}{c}}}} This state is populated in the \ensuremath{^{\textnormal{17}}}F+p decay channel (\href{https://www.nndc.bnl.gov/nsr/nsrlink.jsp?2007Ra36,B}{2007Ra36}, \href{https://www.nndc.bnl.gov/nsr/nsrlink.jsp?2008SfZZ,B}{2008SfZZ}, \href{https://www.nndc.bnl.gov/nsr/nsrlink.jsp?2008Ra12,B}{2008Ra12}, \href{https://www.nndc.bnl.gov/nsr/nsrlink.jsp?2010Gi05,B}{2010Gi05}, and \href{https://www.nndc.bnl.gov/nsr/nsrlink.jsp?2010Ra14,B}{2010Ra14}).}\\
\parbox[b][0.3cm]{17.7cm}{\makebox[1ex]{\ensuremath{^{\hypertarget{NE28LEVEL3}{d}}}} This state is populated in the \ensuremath{^{\textnormal{16}}}O+2p decay channel (\href{https://www.nndc.bnl.gov/nsr/nsrlink.jsp?2007Ra36,B}{2007Ra36}, \href{https://www.nndc.bnl.gov/nsr/nsrlink.jsp?2008SfZZ,B}{2008SfZZ}, \href{https://www.nndc.bnl.gov/nsr/nsrlink.jsp?2008Ra12,B}{2008Ra12}, \href{https://www.nndc.bnl.gov/nsr/nsrlink.jsp?2010Gi05,B}{2010Gi05}, and \href{https://www.nndc.bnl.gov/nsr/nsrlink.jsp?2010Ra14,B}{2010Ra14}).}\\
\parbox[b][0.3cm]{17.7cm}{\makebox[1ex]{\ensuremath{^{\hypertarget{NE28LEVEL4}{e}}}} The sequential 2p-decay via an intermediate \ensuremath{^{\textnormal{17}}}F state is energetically possible for this state (\href{https://www.nndc.bnl.gov/nsr/nsrlink.jsp?2008Ra12,B}{2008Ra12}, \href{https://www.nndc.bnl.gov/nsr/nsrlink.jsp?2010Ra14,B}{2010Ra14}).}\\
\parbox[b][0.3cm]{17.7cm}{\makebox[1ex]{\ensuremath{^{\hypertarget{NE28LEVEL5}{f}}}} The 1p decay branching ratio should be negligible for this state as it is not observed in the \ensuremath{^{\textnormal{17}}}F+p channel (\href{https://www.nndc.bnl.gov/nsr/nsrlink.jsp?2008Ra12,B}{2008Ra12}).}\\
\parbox[b][0.3cm]{17.7cm}{\makebox[1ex]{\ensuremath{^{\hypertarget{NE28LEVEL6}{g}}}} Mode of 2p-decay from (\href{https://www.nndc.bnl.gov/nsr/nsrlink.jsp?2008Ra12,B}{2008Ra12}): either sequential decay or 3-body democratic decay.}\\
\vspace{0.5cm}
\end{center}
\clearpage
\newpage
\pagestyle{plain}
\section[References]{ }
\vspace{-30pt}
\begin{longtable}{l@{\hskip 0.9cm}l}
\multicolumn{2}{c}{REFERENCES FOR A=18}\\
&\endfirsthead
\multicolumn{2}{c}{REFERENCES FOR A=18(CONTINUED)}\\
&\endhead
\href{https://www.nndc.bnl.gov/nsr/nsrlink.jsp?1953Ku08,B}{1953Ku08}&\parbox[t]{6in}{\addtolength{\parindent}{-0.25cm}D.N.Kundu, T.W.Donaven, M.L.Pool, J.K.Long - Phys.Rev. 89, 1200 (1953).}\\
&\parbox[t]{6in}{\addtolength{\parindent}{-0.25cm} \textit{Nuclear Reactions with 21-MeV He\ensuremath{^{\textnormal{3}}} in a Cyclotron.}}\\
\href{https://www.nndc.bnl.gov/nsr/nsrlink.jsp?1954Go17,B}{1954Go17}&\parbox[t]{6in}{\addtolength{\parindent}{-0.25cm}J.D.Gow, L.W.Alvarez - Phys.Rev. 94, 365 (1954).}\\
&\parbox[t]{6in}{\addtolength{\parindent}{-0.25cm} \textit{Neon-18.}}\\
\href{https://www.nndc.bnl.gov/nsr/nsrlink.jsp?1959Aj76,B}{1959Aj76}&\parbox[t]{6in}{\addtolength{\parindent}{-0.25cm}F.Ajzenberg-Selove, T.Lauritsen - Nuclear Phys. 11, 1 (1959).}\\
&\parbox[t]{6in}{\addtolength{\parindent}{-0.25cm} \textit{Energy Levels of Light Nuclei VI.}}\\
\href{https://www.nndc.bnl.gov/nsr/nsrlink.jsp?1959Du81,B}{1959Du81}&\parbox[t]{6in}{\addtolength{\parindent}{-0.25cm}K.L.Dunning, J.W.Butler - Bull.Am.Phys.Soc. 4, No.8, 444, C9 (1959).}\\
&\parbox[t]{6in}{\addtolength{\parindent}{-0.25cm} \textit{O\ensuremath{^{\textnormal{16}}}(He\ensuremath{^{\textnormal{3}}},n)Ne\ensuremath{^{\textnormal{18}}} Reaction.}}\\
\end{longtable}
\end{document}